\pdfoutput=1
\documentclass[runningheads,openany]{svmult}

\usepackage{palatino}

\usepackage{euler}

\usepackage{helvet}

\usepackage{graphicx}  

\usepackage{physprbb}  

\usepackage{mathrsfs}

\usepackage{proc}  

\usepackage[figuresright]{rotating}

\makeindex             

\usepackage{microtype}

\usepackage{amsmath}   
\usepackage{amssymb}
\usepackage{multicol}
\usepackage{bm}

\usepackage[dvips,cam]{crop}

\let\tocauthor\authorrunning

\begin{document}

\frontmatter


\thispagestyle{empty}
\parindent=0pt

{\Large\sc Blejske delavnice iz fizike \hfill Letnik~11, \v{s}t. 2}

\smallskip

{\large\sc Bled Workshops in Physics \hfill Vol.~11, No.~2}

\smallskip

\hrule

\hrule

\hrule

\vspace{0.5mm}

\hrule

\medskip
{\sc ISSN 1580-4992}

\vfill

\bigskip\bigskip
\begin{center}

{\bfseries 
{\Large  Proceedings to the $13^\textrm{th}$ Workshop}\\
{\Huge What Comes Beyond the Standard Models\\}
\bigskip
{\Large Bled, July 12--22, 2010}\\
\bigskip
}

\vspace{5mm}

\vfill

{\bfseries\large
Edited by

\vspace{5mm}
Norma Susana Manko\v c Bor\v stnik

\smallskip

Holger Bech Nielsen

\smallskip

Dragan Lukman

\bigskip


\vspace{12pt}

\vspace{3mm}

\vrule height 1pt depth 0pt width 54 mm}

\vspace*{3cm}

{\large {\sc  DMFA -- zalo\v{z}ni\v{s}tvo} \\[6pt]
{\sc Ljubljana, december 2010}}
\end{center}
\newpage

\thispagestyle{empty}
\parindent=0pt
\begin{flushright}
{\parskip 6pt
{\bfseries\large
                  The 13th Workshop \textit{What Comes Beyond  
                  the Standard Models}, 12.-- 22. July 2010, Bled}

\bigskip\bigskip

{\bfseries\large was organized by}

{\parindent8pt
\textit{Department of Physics, Faculty of Mathematics and Physics,
University of Ljubljana}

}

\bigskip

{\bfseries\large and sponsored by}

{\parindent8pt
\textit{Slovenian Research Agency}

\textit{Department of Physics, Faculty of Mathematics and Physics,
University of Ljubljana}


\textit{Society of Mathematicians, Physicists and Astronomers
of Slovenia}}}
\bigskip
\medskip

{\bfseries\large Organizing Committee}

\medskip

{\parindent9pt
\textit{Norma Susana Manko\v c Bor\v stnik}

\textit{Holger Bech Nielsen}

\textit{Maxim Yu. Khlopov}}

\bigskip

\medskip

\medskip

The Members of the Organizing Committee of the International Workshop 
``What Comes Beyond the Standard Models'', Bled, Slovenia, state that the articles 
published in the Proceedings to the $13^{\textrm{th}}$ Workshop 
``What Comes Beyond the Standard Models'', Bled, Slovenia are refereed at the Workshop 
in intense in-depth discussions. 
\end{flushright}

\setcounter{tocdepth}{0}

\tableofcontents

\cleardoublepage

\chapter*{Preface}
\addcontentsline{toc}{chapter}{Preface in English and SlovenianLanguage}

The series of workshops on "What Comes Beyond the Standard Models?" started
in 1998 with the idea of Norma and Holger for organizing a real workshop, in which participants
would spend most of the time in discussions, confronting different
approaches and ideas. The picturesque town of Bled by the lake of the
same name, surrounded by beautiful mountains and offering pleasant walks and 
mountaineering, was chosen to stimulate the discussions. 
The workshops take place in the house gifted to  the Society of
Mathematicians, Physicists and Astronomers of Slovenia by the Slovenian mathematician 
Josip Plemelj, well known to the 
participants by his work in complex algebra.

The idea was successful and has developed into an annual workshop, which is
taking place every year since 1998. This year the thirteenth workshop took place.
Very open-minded and fruitful discussions
have become the trade-mark of our workshop, producing several new ideas and 
clarifying the proposed ones. The first versions of published works appeared in the
proceedings to the workshop.

In this thirteenth  workshop, which took place from $12^{th}$ to $22^{nd}$ of July
2010, we were discussing several topics, most of them presented in
this Proceedings  and in the discussion section. 

One of the main topics was this time the "approach unifying spin and charges 
and predicting families" (the {\it spin-charge-family-theory} shortly), 
as the new way beyond the {\it standard model of the 
electroweak and colour interactions}, accompanied by the critical 
discussions about all the traps, which the theory does and might in future 
confront, before being accepted as a theory which answers the open questions which the 
{\it standard model} leaves unanswered. 
Proposing the mechanism for generating families, this theory is 
predicting the fourth family which is waiting to be observed  
and the stable fifth family which have a chance to form the dark matter. 
There were discussions of the questions like: To which extent can the {\it spin-charge-family-theory}
answer the open questions of both standard models --  the elementary  particle one and the 
cosmological one?   Are the clusters of the fifth family members 
alone what constitute the dark matter? Can the fifth family baryons and neutrinos 
explain the observed properties of the dark matter with the direct measurements included? 
How do fermions and gauge bosons of this theory behave in phase transitions, through which 
the primordial plasma went?  
How do very heavy  fifth family neutrinos and colourless fifth family baryons behave in the 
electroweak phase transitions? 
How  do (very heavy) fifth family quarks behave in the colour phase transition? 
Why does the colour phase transition occur at $\approx 1$ GeV? 
What does trigger the fermion-antifermion asymmetry in this theory and how does the existence of 
two stable families influences the matter-antimatter asymmetry?
Although the theory predicts the mass matrices, it also connects strongly 
the mass matrix properties of the members of families on the tree level. Does the 
coherent contribution of fields beyond the tree level explains the great 
difference in masses and mixing matrices among the (so far measured) families?  
Can a complex action  function as a cutoff in loop diagrams?

We  discuss the model  with the complex action and its 
application to the presently observed properties of fermions and bosons as well as 
about the possibility that this model would lead to improvement in the sense of interpretation 
of quantum mechanic, since it includes  the DeBroglie-Bohm-particle approach to quantum mechanics.

We discuss also the dark matter direct measurements and possible explanations of the experimental data, 
if a very heavy stable family, as the fifth family is, constitutes the dark matter as neutral baryons 
and neutrinos or as  neutral nuclei (both with respect to the colour and electromagnetic charge). 
There were also the talk and discussions afterwards about what signals 
from the dark matter are expected to be seen at the LHC .

Talks and discussions in our workshop are not at
all talks in the usual way. Each talk or discussions lasted several
hours, divided in two hours blocks, with a lot of questions,
explanations, trials to agree or disagree from the audience or a
speaker side. 
Most of talks are "unusual" in the sense that they are trying to find out new 
ways of understanding and describing the observed  
phenomena. Although we always hope that the progress made in discussions 
will reflect in the same year proceedings, it 
happened  many a time that the topics appear in the next or after the next year proceedings. 
This is happening also in this year. There fore neither the discussion section nor the 
talks published in this proceedings, manifest all the discussions and the work 
done in this workshop.  

Several teleconferences were taking place during the Workshop on various
topics. It was organized by the Virtual Institute for Astrophysics \\
(wwww.cosmovia.org) of Maxim  with able support by Didier Rouable.
We managed to have ample discussions and we thank all the participants, those 
presenting a talk and those contributing in discussions.
The reader can find the talks  delivered by John Ellis,
N.S. Manko\v c Bor\v stnik and H.B. Nielsen on www.cosmovia.org,\\[2mm]
http://viavca.in2p3.fr/what\_comes\_beyond\_the\_standard\_models\_xiii.html\\[2mm]
Let us thanks cordially  all the participants, those present really
and those present virtually,  for their presentations and in particular
for really fruitful discussions
and the good working atmosphere. \\[5mm]

\parbox[b]{\textwidth}{%
  \textit{Norma Manko\v c Bor\v stnik, Holger Bech Nielsen, Maxim Y. Khlopov,}\\
   \textit{(the Organizing comittee)}\\[2mm]
   \textit{Norma Manko\v c Bor\v stnik, Holger Bech Nielsen, Dragan Lukman,} \\
   \textit{(the Editors)}

\hfill\textit{Ljubljana, December 2010}}

\newpage

\chapter*{Predgovor (Preface in Slovenian Language)}

Serija delavnic "Kako prese\v ci oba standardna modela, kozmolo\v skega in 
elektro\v sibkega" ("What Comes Beyond the Standard Models?") se je za\v cela
leta 1998 z idejo, da bi organizirali delavnice, v katerih bi udele\v zenci 
posvetili veliko \v casa diskusijam, ki bi kriti\v cno soo\v cile razli\v cne 
ideje in teorije. Mesto Bled ob slikovitem jezeru je za take delavnice zelo  primerno,  
ker prijetni sprehodi in pohodi na \v cudovite gore, ki kipijo nad mestom, ponujajo 
prilo\v znosti in vzpodbudo za diskusije. Delavnica poteka v hi\v si, ki jo je  
Dru\v stvu matematikov, fizikov in astronomov Slovenije zapustil v last slovenski matematik 
Josip Plemelj, udele\v zencem delavnic, ki prihajajo iz razli\v cnih koncev sveta, dobro poznan
po svojem delu v kompleksni algebri.

Ideja je za\v zivela, rodila se je serija letnih delavnic, ki potekajo vsako   
leto od 1998 naprej. To leto je potekala trinajsti\v c. Zelo odprte in
plodne diskusije so postale zna\v cilnost na\v sih delavnic, porodile so marsikatero 
novo idejo in pomagale razjasniti in narediti naslednji korak predlaganim idejam in teorijam. 
Povzetki prvih novih korakov in dognanj so 
iz\v sle v zbornikih delavnic. 

Na leto\v snji, trinajsti, delavnici, ki je potekala od 12. do 22. malega srpana (julija) 2010, smo 
razpravljali o ve\v c temah, ve\v cina je predstavljena v tem zborniku. 

Ena od osnovnih tem je bila tokrat "teorija enotnih spinov in nabojev, ki napoveduje
dru\v zine" (na kratko: {\it teorija spina-nabojev-dru\v zin}) kot novo pot za raz\v siritev 
{\it standardnega modela elektro\v sibke in barvne interakcije}. Spremljale so jo
kriti\v cne razprave o pasteh, s katerimi se ta teorija soo\v ca in se bo soo\v cala 
v prihodnje, preden bo lahko sprejeta kot odgovor na odprta vpra\v sanja {\it standardnega modela}.
Teorija predlaga mehanizem za nastanek dru\v zin, napoveduje \v cetrto 
dru\v zino, ki jo utegnejo opaziti, in peto dru\v zino, iz katere bi utegnila biti temna snov.
Razpravljali smo o vpra\v sanjih kot so: V kolik\v sni meri lahko 
{\it teorija spina-nabojev-dru\v zin} odgovori na odprta vpra\v sanja obeh 
standardnih modelov -- standardnega modela za osnovne  delce  in polja in standardnega 
kozmolo\v skega modela? Ali so gru\v ce iz \v clanov pete dru\v zine edina sestavina temne snovi?
Ali lahko barioni in nevtrini pete dru\v zine pojasnijo opa\v zene lastnosti 
temne snovi, vklju\v cno z direktnimi meritvami? Kako se fermioni in umeritvena polja  
te teorije obna\v sajo pri faznih prehodih, skozi katere je \v sla prvotna plazma?
Kako se zelo te\v zki nevtrini in brezbarvni barioni pete dru\v zine obna\v sajo pri 
elektro\v sibkih faznih prehodih? Kako se obna\v sajo (zelo te\v zki) kvarki pete dru\v zine
pri barvnem faznem prehodu? Zakaj se barvni fazni prehod zgodi pri $\approx 1$ GeV? 
Kaj v tej teoriji spro\v zi asimetrijo fermionov in antifermionov in kako obstoj dveh stabilnih
dru\v zin vpliva na asimetrijo snov-antisnov? \v Ceprav teorija napove masne matrike, 
obenem krepko pove\v ze lastnosti masnih matrik \v clanov dru\v zin na drevesnem nivoju.
Ali koherentni prispevek polj pod drevesnim nivojem pojasni velike razlike v masah in
me\v salnih matrikah med (\v ze izmerjenimi) dru\v zinami? Ali lahko kompleksna akcija 
u\v cinkuje kot zgornja meja  v diagramih zank?

Razpravljali  smo o modelu s kompleksno akcijo in o mo\v znosti, da kompleksna akcija 
pojasni nekatere  lastnosti  fermionov in bozonov, denimo  skalo elektro\v sibkega prehoda pri 
200 GeV, pa tudi o  izbol\v sani interpretacijo kvantne mehanike, ker vklju\v cuje 
 DeBroglie-Bohmov pristop k kvantni mehaniki. 

Veliko \v casa smo posvetili  direktnim meritvam temne snovi in mo\v znim razlagam doslej zbranih 
podatkov, \v ce gradi temno snov te\v zka stabilna  dru\v zina kvarkov in leptonov, denimo peta dru\v zina.   
Je temna snov iz nevtralnih barionov (nevtralnih glede na barvni in 
elektromagnetni naboj) in nevtrinov ali iz elektromagnetno nevtralnih jeder, ki jih sestavljajo 
barvno nevtralni te\v zki barioni in brezbarvni barioni  prve dru\v zine. 

Imeli smo predavanje in nato \v zivahno diskusijo o tem, kak\v sne signale temne snovi lahko pri\v cakujemo
na pospe\v sevalniku LHC.

Predavanja in razprave na na\v si delavnici niso predavanja v obi\v cajnem smislu. Vsako
predavanje ali razprava je trajala ve\v c ur, razdelejnih na bloke po dve uri, 
z veliko vpra\v sanji, pojasnili, poskusi, da bi  predavatelj  in ob\v cinstvo razumeli trditve, 
kritike in se na koncu strinjali ali pa tudi ne. Ve\v cina predavanj je 'neobi\v cajnih' v tem smislu, da
posku\v sajo najti nove matemati\v cne na\v cine opisa, pa tudi razumevanja doslej opa\v zenih pojavov.
\v Ceprav vedno upamo, da bomo vsako leto uspeli zapisati vsa nova dognanja, nastala v ali ob diskusijah, 
se vseeno mnogokrat zgodi, da se prvi zapisi o napredku pojavijo \v sele v kasnej\v sih zbornikh.
Tako tudi leto\v snji zbornik ne vsebuje povzetkov vseh uspe\v snih 
razprav ter napredka pri temah, predstavljenih v predavanjih.

Med delavnico smo imeli ve\v c spletnih konferenc na razli\v cne teme. Organiziral jih
je Virtualni institut za astrofiziko iz Pariza (www.cosmovia.org, vodi ga Maxim) ob
spretni podpori Didierja Rouablea. Uspelo nam je odprto diskutirati kar z nekaj laboratoriji 
po svetu.  
Toplo se zahvaljujemo  vsem udele\v zencem, tako tistim, ki so imeli predavanje, 
kot tistim, ki so sodelovali v razpravi, na Bledu ali preko spleta. 
Bralec lahko najde posnetke predavanj, ki
so jih imeli John Ellis, N.S. Manko\v c Bor\v stnik in H.B. Nielsen na spletni povezavi\\[2mm]
http://viavca.in2p3.fr/what\_comes\_beyond\_the\_standard\_models\_xiii.html\\[2mm] 
Prisr\v cno se zahvaljujemo vsem udele\v zencem, ki so bili prisotni, tako fizi\v cno kot 
virtualno, za njihova predavanja, za zelo plodne razprave in za delovno vzdu\v sje.\\[5mm]

\parbox[b]{\textwidth}{%
  \textit{Norma Manko\v c Bor\v stnik, Holger Bech Nielsen, Maxim Y. Khlopov,}\\
   \textit{(Organizacijski odbor)}\\[2mm]
   \textit{Norma Manko\v c Bor\v stnik, Holger Bech Nielsen, Dragan Lukman,} \\
   \textit{(uredniki)}

\hfill\textit{Ljubljana, grudna (decembra) 2010}}

\newpage

\cleardoublepage


\mainmatter

\cleardoublepage
\thispagestyle{empty}
\vspace*{5cm}
{\bfseries \Large Talk Section}\\[1cm]
\addcontentsline{toc}{chapter}{Talk Section}
All talk contributions are arranged alphabetically with respect to the first author's name.

\newpage

\cleardoublepage
\parindent=20pt

\setcounter{page}{1}

\newcommand {\hmtr}{{\rm tr\,}}
\title{Noncommutativity and Topology within Lattice Field Theories}
\author{A. Ali Khan${}^1$ and H. Markum${}^2$\thanks{Thanks to the organizers of the Workshop 2010
	'What Comes Beyond the Standard Models' in Bled}}
\institute{%
${}^1$ Faculty of Applied Sciences, University of Taiz, Yemen\\
        arifa.ali-khan@physik.uni-regensburg.de\\
${}^2$ Atominstitut, Vienna University of Technology, Austria\\
        markum@tuwien.ac.at}

\authorrunning{A. Ali Khan and H. Markum}
\titlerunning{Noncommutativity and Topology within Lattice Field Theories}
\maketitle

\begin{abstract}
Theories with noncommutative space-time coordinates represent alternative candidates of grand unified theories. We discuss U(1) gauge theory in 2 and 4 dimensions on a lattice with N sites. The mapping to a U(N) plaquette model in the sense of Eguchi and Kawai can be used for computer simulations. In 2D it turns out that the formulation of the topological charge leads to the imaginary part of the plaquette. Concerning 4D, the definition of instantons seems straightforward. One can transcribe the plaquette and hypercube formulation to the matrix theory. The transcription of a monopole observable seems to be difficult. The analogy to commutative U(1) theory of summing up the phases over an elementary cube does not obviously transfer to the U(N) theory in the matrix model. It would be interesting to measure the topological charge on a noncommutative hypercube. It would even be more interesting to find arguments and evidence for realization of noncommutative space-time in nature.
\end{abstract}

\section{Motivation}
\label{contribution:hmak}
In noncommutative geometry, where the coordinate operators $\hat{x}_\mu$
satisfy the commutation relation 
$[\hat{x}_\mu , \hat{x}_\nu] = i \theta_{\mu\nu}$, 
a mixing between ultraviolet and infrared degrees
of freedom takes place \cite{hmakszabo}. 
So lattice simulations are a promising tool to get
deeper insight into noncommutative quantum field theories.
For noncommutative U($1$) gauge there exists an equivalent matrix model
which makes numerical calculations feasible \cite{hmakhofheinz}. 

The main topic of the underlying contribution is to discuss the 
topological charge of noncommutative U($1$) gauge theory in two and 
in higher dimensions. In two dimensions the instanton configurations 
carry a topological charge $q$ which was shown being non-integer 
\cite{hmaknishi,hmakDAfrisch}. We work out the definition of 
instantons in four and higher dimensions. 

\section{Topology and Instantons in QCD}
The Lagragian of pure gluodynamics (the Yang-Mills theory with no matter fields) in Euclidean spacetime can be written as
\begin{equation}
\mathcal{L}=\frac{1}{4g^2}G_{\mu \nu}^a G_{\mu \nu}^a
\end{equation}
where $G_{\mu \nu}^a$ is the gluon field strength tensor
\begin{equation}
G_{\mu \nu}^a=\partial_{\mu}A_{\nu}^{a}-\partial_{\nu}A_{\mu}^{a}+f^{abs}A_{\mu}^{b}A_{\nu}^{c}
\end{equation}
and $f^{abc}$ are structure constants of the gauge group considered. The classical action of the Yang-Mills fields can be identically rewritten as 
\begin{equation}
S=\frac{1}{8g^2}\int dx^4 (G_{\mu \nu}^a \pm \tilde{G}_{\mu \nu}^a)^2 \mp \frac{8 \pi^2}{g^2}Q
\end{equation}
where Q denotes the topological charge
\begin{equation}
Q=\frac{1}{32 \pi^2} \int dx^4 G_{\mu \nu}^a \tilde{G}_{\mu \nu}^a
\end{equation}
with 
\begin{equation}
\tilde{G}_{\mu \nu}^a=\frac{1}{2} \epsilon_{\mu \nu \alpha \beta} G_{\alpha \beta}^a
\end{equation}

\section{Definition of the Topological Charge in Two Dimensions}

\subsection{Lattice Regularization of Noncommutative  Two-Dimensional U(1) Gauge Theory}
The lattice regularized version of the theory can be defined by an
analog of Wilson's plaquette action 
\begin{equation}
S= - \beta \sum_{x} \sum_{\mu < \nu}
U_\mu (x) \star U_\nu (x + a \hat{\mu}) \star U_\mu (x + a \hat{\nu})^{\dagger} \star U_\nu (x)^{\dagger}
 + \mbox{c.c.} 
\label{hmaklat-action}
\end{equation}
where the symbol $\hat{\mu}$ represents 
a unit vector in the $\mu$-direction
and we have introduced the lattice spacing $a$.
The link variables $U_\mu(x)$ 
are complex fields on the lattice 
satisfying the star-unitarity condition.
The star-product \cite{hmakszabo} on the lattice can be obtained by
rewriting its definition within noncommutatiuve derivatives 
in terms of Fourier modes
and restricting the momenta to the Brillouin zone.

Let us define the topological charge for a gauge field configuration on the discretized two-dimensional torus.
In the language of fields, we define the topological charge as
\begin{equation}
q = \frac{1}{4 \pi i}
\sum_{x} \sum_{\mu \nu} \epsilon_{\mu\nu}
U_\mu (x) \star U_\nu (x + a \hat{\mu}) \star
U_\mu (x + a \hat{\nu})^{\dagger} \star U_\nu (x)^{\dagger}
\label{hmakdef-q}
\end{equation}
which reduces to the usual definition of the topological charge
in 2d gauge theory
\begin{equation}
q = \frac{1}{4 \pi} \int d^2 x \,
\epsilon_{\mu\nu} G_{\mu\nu}
\end{equation}
in the continuum limit.

\subsection{Matrix-Model Formulation}
It is much more convenient for computer simulations  to use
an equivalent formulation, in which one maps functions on a noncommutative space to operators
so that the star-product becomes nothing but the usual operator product, which is noncommutative.
The action (\ref{hmaklat-action}) can then be written as 
\begin{eqnarray}
S &=& - N \beta \, \sum_{\mu \ne \nu} 
\hmtr ~\Bigl\{\hat{U}_\mu\, (\Gamma_\mu 
\hat{U}_\nu \Gamma_\mu^\dag)\,
(\Gamma_\nu \hat{U}_\mu^\dag \Gamma_\nu^\dag)
\, \hat{U}_\nu^\dag \Bigr\}  + 2 \beta N^2 \nonumber\\
&=& - N \beta \, \sum_{\mu \ne \nu} 
{\cal  Z}_{\nu\mu}
\hmtr ~\Bigl(V_\mu\,V_\nu\,V_\mu^\dag\,V_\nu^\dag\Bigr) 
 + 2 \beta N^2
\label{hmakTEK-action}
\end{eqnarray}
where $V_\mu \equiv \hat{U}_\mu \Gamma_\mu$ is a U($N$) matrix 
and $N$ is the linear extent of the original lattice.
An explicit representation of $\Gamma_\mu$
in the $d=2$ case shall be given
in Sec. \ref{hmaksection:class_sol}. 
This is the twisted Eguchi-Kawai (TEK) model \cite{hmakTEK},
which appeared in history as a matrix model equivalent to the 
large $N$ gauge theory \cite{hmakEK}.
We have added the constant term $2\beta N^2$ to
what we would obtain from (\ref{hmaklat-action})
in order to make the absolute minimum of the action zero.
%

By using the map between fields and matrices,
the topological charge (\ref{hmakdef-q}) can be represented
in terms of matrices as
\begin{eqnarray}
q&=& \frac{1}{4 \pi i} N \, \sum_{\mu \nu} \epsilon_{\mu\nu}
\hmtr ~\Bigl\{ \hat{U}_\mu\, (\Gamma_\mu 
\hat{U}_\nu \Gamma_\mu^\dag ) \,
(\Gamma_\nu \hat{U}_\mu^\dag \Gamma_\nu^\dag)
\, \hat{U}_\nu^\dag \Bigr\} \nonumber\\
&=& \frac{1}{4 \pi i} N \, \sum_{\mu \nu} \epsilon_{\mu\nu}
{\cal  Z}_{\nu\mu}
\hmtr ~\Bigl(V_\mu\,V_\nu\,V_\mu^\dag\,V_\nu^\dag\Bigr)
\label{hmakdef-q-mat}
\end{eqnarray}

\subsection{Classical Solutions}
\label{hmaksection:class_sol}
The classical equation of motion was worked out in the literature 
\cite{hmakGriguolo:2003kq,hmaknishi} for the action (\ref{hmakTEK-action}) 
\begin{equation}
V_\mu ^\dag (W - W^\dag ) V_\mu = W - W^\dag  
\end{equation}
with the unitary matrix $W$ 
\begin{equation}
W = {\cal  Z}_{\nu\mu} V_\mu\,V_\nu\,V_\mu^\dag\,V_\nu^\dag 
\end{equation}
General solutions
to this equation can be written 
in a block-diagonal form \cite{hmakGriguolo:2003kq}
\begin{eqnarray}
V_\mu = 
\begin{pmatrix}
\Gamma_{\mu} ^{(1)} & & & \cr 
& \Gamma_{\mu} ^{(2)} & & \cr
& & \ddots & \cr 
& & & \Gamma_{\mu} ^{(k)} \cr  
\end{pmatrix}  
\label{hmakgeneral_sol}
\end{eqnarray}
by an appropriate SU($N$) transformation, where $\Gamma_{\mu} ^{(j)}$
are $n_j \times n_j$ unitary matrices, $j = 1,\ldots, k$,
satisfying the 't~Hooft-Weyl algebra
\begin{eqnarray}
\Gamma_\mu ^{(j)} \Gamma_\nu ^{(j)} &=& Z_{\mu\nu}^{(j)}
\Gamma_\nu ^{(j)} \Gamma_\mu ^{(j)}        \\
Z_{12}^{(j)} &=& Z_{21}^{(j)*} = \exp 
\left( 2\pi i \frac{ m_j }{n_j}  \right)\\
      m_j&=&\frac{n_j+1}{2}
\end{eqnarray}
An explicit representation is given by the clock and shift operators, 
$Q$ and $P$
\begin{equation}
\Gamma_1 ^{(j)}  = P_{n_j} \ , 
\quad \Gamma_2  ^{(j)} = (Q_{n_j})^{m_j}
\label{hmakexplicit-Gamma}
\end{equation}
An example is shown in the appendix. Refs.~\cite{hmaknishi,hmakGriguolo:2003kq} 
quote expressions for the action and the topological charge
\begin{eqnarray}
\label{hmakaction-classical}
S &=& 4 N \beta
\sum_j n_j \sin ^2 \left\{ \pi \left( 
\frac{m_j}{n_j} - \frac{M}{N}
\right) \right\}     \\
q &=& \frac{N}{2\pi} \sum_j n_j \sin \left\{ 2 \pi  \left( 
\frac{m_j}{n_j} - \frac{M}{N} \right) \right\}
\label{hmakq-classical}
\end{eqnarray}
In general, the topological charge $q$ is not an integer.
If we require the action to be less than of order $N$ 
the argument of the sine has to vanish for all $j$.
In that case the topological charge approaches an integer
\begin{equation}
q \simeq N \left( \sum_j m_j  - M \right) 
\label{hmakq-integer}
\end{equation}
being a multiple of $N$.
\section{Definition of the Topological Charge in Four and Higher Dimensions}
The lattice action (\ref{hmaklat-action}) taking into account the star-product 
can be used in any dimension. The field-theoretic definition of the 
topological charge (\ref{hmakdef-q}) can be extended in two ways. 

One can rely on the so-called plaquette definition which then yields a 
product of two plaquettes
\begin{eqnarray}
  q^{(P)} = \frac{-1}{32 \pi^2}
  \sum_{x} \sum_{\mu \nu\rho\sigma} \epsilon_{\mu\nu\rho\sigma}
  U_\mu (x) \star U_\nu (x + a \hat{\mu}) \star
  U_\mu (x + a \hat{\nu})^{\dagger} \star U_\nu (x)^{\dagger} \nonumber\\
  \star U_\rho (x) \star U_\sigma (x + a \hat{\rho}) \star
  U_\rho (x + a \hat{\sigma})^{\dagger} \star U_\sigma (x)^{\dagger}
  \label{hmakdef-q_p-4d}
\end{eqnarray}
which reduces to the definition of the topological charge
in 4d gauge theory
\begin{equation}
  q^{(P)} = \frac{1}{32 \pi^2} \int d^4 x \,
  \epsilon_{\mu\nu\rho\sigma} G_{\mu\nu} \star G_{\rho\sigma}
\end{equation}
in the continuum limit.

By using the map between fields and matrices,
the topological charge (\ref{hmakdef-q_p-4d}) can be represented
in terms of matrices as
\begin{eqnarray}
  q^{(P)}= \frac{-1}{32 \pi^2} N \, \sum_{\mu \nu \rho \sigma} \epsilon_{\mu\nu\rho\sigma}
  \hmtr ~\Bigl\{ \hat{U}_\mu\, (\Gamma_\mu
  \hat{U}_\nu \Gamma_\mu^\dag ) \,
  (\Gamma_\nu \hat{U}_\mu^\dag \Gamma_\nu^\dag)
  \, \hat{U}_\nu^\dag \Bigr\} \nonumber\\ 
  \Bigl\{ \hat{U}_\rho\, (\Gamma_\rho
  \hat{U}_\sigma \Gamma_\rho^\dag ) \,
  (\Gamma_\sigma \hat{U}_\rho^\dag \Gamma_\sigma^\dag)
  \, \hat{U}_\sigma^\dag \Bigr\} \nonumber\\
  = \frac{-1}{32 \pi^2} N \, \sum_{\mu \nu \rho \sigma} \epsilon_{\mu\nu\rho\sigma}
  {\cal  Z}_{\nu\mu} {\cal  Z}_{\rho\sigma}
  \hmtr ~\Bigl(V_\mu\,V_\nu\,V_\mu^\dag\,V_\nu^\dag
  V_\rho\,V_\sigma\,V_\rho^\dag\,V_\sigma^\dag\Bigr)
  \label{hmakdef-q_p-mat-4d}
\end{eqnarray}

Alternatively, one can rely on the so-called hybercube definition 
which leads to a star-product of matrices winding along the edges 
of the hybercube
\begin{eqnarray}
  q^{(H)} = \frac{-1}{32 \pi^2}
  \sum_{x} \sum_{\mu \nu\rho\sigma} \epsilon_{\mu\nu\rho\sigma}
  U_\mu (x) \star U_\nu (x + a \hat{\mu}) \star
  U_\rho (x + a \hat{\mu} + a \hat{\nu}) \nonumber\\
  \star U_\sigma (x + a \hat{\mu} + a \hat{\nu} + a \hat{\rho})
  \star U_\mu (x + a \hat{\nu} + a \hat{\rho} + a \hat{\sigma})^{\dagger}) 
  \nonumber\\
  \star U_\nu (x + a \hat{\rho} + a \hat{\sigma})^{\dagger}) \star
  U_\rho (x + a \hat{\sigma})^{\dagger} \star U_\sigma (x)^{\dagger})
  \label{hmakdef-q_h-4d}
\end{eqnarray}
which reduces to the definition of the topological charge
in 4d gauge theory
\begin{equation}
  q^{(H)} = \frac{1}{32 \pi^2} \int d^4 x \,
  \epsilon_{\mu\nu\rho\sigma} G_{\mu\nu} \star G_{\rho\sigma}
\end{equation}
in the continuum limit.

By using the map between fields and matrices,
the topological charge (\ref{hmakdef-q_p-4d}) can be represented
in terms of matrices as
\begin{eqnarray}
  q^{(H)}= \frac{-1}{32 \pi^2} N \, \sum_{\mu \nu \rho \sigma} 
  \epsilon_{\mu\nu\rho\sigma}
  \hmtr ~\Bigl\{ \hat{U}_\mu \, (\Gamma_\mu \hat{U}_\nu \Gamma_\mu^\dag ) \,
  (\Gamma_\nu \Gamma_\mu \ \hat{U}_\rho \Gamma_\mu^\dag \Gamma_\nu^\dag) \,
  \nonumber\\
  (\Gamma_\rho \Gamma_\nu \Gamma_\mu \ \hat{U}_\sigma 
  \Gamma_\mu^\dag \Gamma_\nu^\dag \Gamma_\rho^\dag) \,
  (\Gamma_\sigma \Gamma_\rho \Gamma_\nu \ \hat{U}_\mu^\dag 
  \Gamma_\nu^\dag \Gamma_\rho^\dag \Gamma_\sigma^\dag) \,
  (\Gamma_\sigma \Gamma_\rho \hat{U}_\nu^\dag 
  \Gamma_\rho^\dag \Gamma_\sigma^\dag) \, 
  (\Gamma_\sigma \hat{U}_\rho^\dag \Gamma_\sigma^\dag) \, 
  \hat{U}_\sigma^\dag \Bigr\} \nonumber\\
  = \frac{-1}{32 \pi^2} N \, \sum_{\mu \nu \rho \sigma} 
  \epsilon_{\mu\nu\rho\sigma}
  {\cal  Z}_{\nu\mu} {\cal  Z}_{\rho\nu} 
  {\cal  Z}_{\sigma\rho} {\cal  Z}_{\rho\mu}
  {\cal  Z}_{\sigma\nu} {\cal  Z}_{\sigma\mu}
  \hmtr ~\Bigl(V_\mu\,V_\nu\,V_\rho\,V_\sigma
  V_\mu^\dag\,V_\nu^\dag\,V_\rho^\dag\,V_\sigma^\dag\Bigr) \nonumber\\
  \label{hmakdef-q_h-mat-4d}
\end{eqnarray}
The extension to higher dimensions is straight-forward.
In practical studies, one can choose one or more planes noncommutative 
while leaving the others commutative \cite{hmakbietenholz}.

\section{Conclusion and Outlook}
Today there exist several investigations of the topological sector of 
the two-dimens\-ional noncommutative U($1$) theory \cite{hmaknishi,hmakDAfrisch}. 
Also classical solutions are available. 
The situation with the field-theoretic definition of instantons is 
reminiscent of lattice QCD where quantum gauge field configurations 
are topologically trivial and one needs to apply some smoothing procedure 
onto the gauge fields to unhide instantons. 

In this contribution we worked out the field-theoretic definition to four 
and higher dimensions. We demonstrated that both the plaquette and hybercube 
definition can be taken over from the commutative gauge theory by 
respecting the star-multiplication and applying the map to the matrix model. 

It would be interesting to adapt cooling techniques from QCD to
the four-dimensional noncommutative U($1$) theory \cite{hmakbietenholz}. 
At present we are working on this. It would be disirable to send 
the noncommutativity parameter $\theta$ of the four-dimens\-ional 
noncommutative gauge theory to zero in order to obtain
a realistic comparison of its topological content with the well-studied
topological objects like instantons and monopoles in QCD.        

Unfortunately, the transcription of a monopole observable seems to be 
difficult. The analogy to commutative U(1) theory of summing up the 
phases of the field over an elementary cube does not obviously transfer 
to the U(N) theory in the matrix model. Finding a reasonable definition 
one could be able to measure the monopole number on a noncommutative 
hypercube.

\section*{Appendix: Example for Calculation of Topological Charge}
\begin{eqnarray}
q &=& \frac{N}{4\pi i}\sum_{\mu\nu}\epsilon_{\mu\nu} Z_{\nu\mu}\mathrm{tr}
\left(V_\mu V_\nu V_\mu^\dagger V_\nu^\dagger\right) \nonumber\\
& = & \frac{N}{4\pi i}\left[
\epsilon_{12} Z_{21}\mathrm{tr}\left(V_1 V_2 V_1^\dagger V_2^\dagger\right)
+\epsilon_{21} Z_{21}^*\mathrm{tr}\left(V_1 V_2 V_1^\dagger V_2^\dagger\right)^\dagger
\right] \nonumber\\
& = & \frac{N}{4\pi i}\left\{e^{-\pi i\frac{N+1}{N}}
\mathrm{tr}\left(V_1 V_2 V_1^\dagger V_2^\dagger\right) -
e^{\pi i\frac{N+1}{N}} \mathrm{tr}\left(V_1 V_2 V_1^\dagger V_2^\dagger
\right)^\dagger \right\}
\end{eqnarray}
For demonstration we choose $N =5$ with a decomposition $n_1 = 2$ and 
$n_2 = 3$. This leads to $5\times 5$ matrices of the form
\begin{equation}
V_1  = \left(\begin{array}{ccccc}
0 & 1 & 0 & 0 & 0 \\
1 & 0 & 0 & 0 & 0 \\
0 & 0 & 0 & 1 & 0 \\
0 & 0 & 0 & 0 & 1 \\
0 & 0 & 1 & 0 & 0 \\
\end{array} \right)
\end{equation}
\begin{equation}
V_2  = \left(\begin{array}{ccccc}
1 & 0 & 0 & 0 & 0 \\
0 & e^{\frac{3i\pi}{2}} & 0 & 0 & 0 \\
0 & 0 & 1 & 0 & 0 \\
0 & 0 & 0 & e^{\frac{4i\pi}{3}} & 0 \\
0 & 0 & 0 & 0 &  e^{\frac{8i\pi}{3}}\\
\end{array} \right)
\end{equation}

\begin{equation}
V_1 V_2  = \left(\begin{array}{ccccc}
0 & e^{\frac{3i\pi}{2}} & 0 & 0 & 0 \\
1 & 0 & 0 & 0 & 0 \\
0 & 0 & 0 & e^{\frac{4i\pi}{3}} & 0 \\
0 & 0 & 0 & 0 & e^{\frac{8i\pi}{3}} \\
0 & 0 & 1 & 0 &  0\\
\end{array} \right)
\end{equation}

\begin{equation}
V_2 V_1  = \left(\begin{array}{ccccc}
0 & 1 & 0 & 0 & 0 \\
e^{\frac{3i\pi}{2}} & 0 & 0 & 0 & 0 \\
0 & 0 & 0 & 1 & 0 \\
0 & 0 & 0 & 0 & e^{\frac{4i\pi}{3}} \\
0 & 0 & e^{\frac{8i\pi}{3}} & 0 &  0\\
\end{array} \right)
\end{equation}

\begin{equation}
V_1^\dagger V_2^\dagger = (V_2 V_1)^\dagger  = \left(\begin{array}{ccccc}
0 & e^{-\frac{3i\pi}{2}} & 0 & 0 & 0 \\
1 & 0 & 0 & 0 & 0 \\
0 & 0 & 0 & 0 &  e^{-\frac{8i\pi}{3}}\\
0 & 0 & 1 & 0 & 0 \\
0 & 0 & 0 & e^{-\frac{4i\pi}{3}} &  0\\
\end{array} \right)
\end{equation}
\begin{equation}
\mathrm{tr}(V_1V_2V_1^\dagger V_2^\dagger) =
\mathrm{tr}  \left(\begin{array}{ccccc}
e^{\frac{3i\pi}{2}} & 0 & 0 & 0 & 0 \\
0 & e^{-\frac{3i\pi}{2}} & 0 & 0 & 0 \\
0 & 0 & e^{\frac{4i\pi}{3}} & 0 & 0 \\
0 & 0 & 0 & e^{\frac{4i\pi}{3}} & 0 \\
0 & 0 & 0 & 0 &  e^{-\frac{8i\pi}{3}}\\
\end{array} \right)
\end{equation}
So we obtain for the classical topological charge
\begin{eqnarray}
q &=& \frac{5}{4\pi i} 2i\mathrm{Im}\left[e^{-\frac{6i\pi}{5}}\left(
e^{\frac{3i\pi}{2}} + e^{\frac{-3i\pi}{2}} + 2e^{\frac{4i\pi}{3}}
+ e^{\frac{-8i\pi}{3}}  \right)\right] \\
&=& \frac{5}{2\pi } \mathrm{Im}\left[
e^{i\pi(\frac{3}{2}-\frac 6 5) } +  e^{i\pi(-\frac{3}{2}-\frac 6 5) }  +2
e^{i\pi(\frac{4}{3}-\frac 6 5) } +e^{i\pi(-\frac{8}{3}-\frac 6 5) } 
\right] \nonumber\\
&=& \frac{5}{2\pi }\left[\sin[\pi(\frac 3 2 -\frac 6 5 )]+
\sin[\pi(-\frac 3 2 -\frac 6 5 )]+2\sin[\pi(\frac 4 3 -\frac 6 5 )] +
\sin[\pi(-\frac 8 3  -\frac 6 5 )]\right] \nonumber
\end{eqnarray}
This result is in agreement with the relation
\begin{equation}
  q =  \frac{N}{2\pi}\sum_j \left\{(n_j -
  1)\sin\left[\pi(\frac{m_j}{n_j}-\frac{M}{N})\right] +
  \sin\left[\pi(-\frac{(n_j-1)m_j}{n_j}-\frac{M}{N})\right]\right\}
\end{equation}


\title{{The Construction of Quantum Field Operators: Something of Interest}\thanks{The invited talks at the VIII International Workshop "Applied Category Theory. Graph-Operad-Logic", San Blas, Nayarit, M\'exico, January 9-16, 2010, and at the 6th International Conference on the Dark Side of the Universe (DSU2010), Leon, Gto, M\'exico, June 1-6, 2010.}}
\author{V.V. Dvoeglazov}
\institute{%
Universidad de Zacatecas\\Ap. Postal 636, Suc. 3 Cruces, C. P. 98062\\Zacatecas, Zac., M\'exico}

\titlerunning{The Construction of Quantum Field Operators: Something of Interest}
\authorrunning{V.V. Dvoeglazov}
\maketitle

\begin{abstract}
We  draw attention to some tune problems in constructions of the quantum-field operators for spins 1/2 and 1. They are related to the existence of negative-energy and acausal solutions of relativistic wave equations. Particular attention is paid to the chiral theories, and to the method of the Lorentz boosts.
\end{abstract}

\section{The Dirac Equation}

First of all, I would like to remind you some basic things in the quantum field theory.
The Dirac equation has been considered in detail in a pedagogical way~\cite{vd1Sakurai,vd1Ryder}:
\begin{equation}
[i\gamma^\mu \partial_\mu -m]\Psi (x) =0\,.\label{vd1Dirac}
\end{equation}
At least, 3 methods of its derivation exist:
\begin{itemize}
  \item the Dirac one (the Hamiltonian should be linear in $\partial/\partial x^\mu$, and be compatible with $E^2 -{\bf p}^2 c^2 =m^2 c^4$);
  \item the Sakurai one (based on the equation $(E- {\bf \sigma} \cdot {\bf p}) (E+ {\bf \sigma} \cdot {\bf p}) \phi =m^2 \phi$);
  \item the Ryder one (the relation between  2-spinors at rest is $\phi_R ({\bf 0}) = \pm \phi_L ({\bf 0})$).
\end{itemize}
The $\gamma^\mu$ are the Clifford algebra matrices: 
\begin{equation}
\gamma^\mu \gamma^\nu +\gamma^\nu \gamma^\mu = 2g^{\mu\nu}\,.
\end{equation}
Usually, everybody uses the following definition of the field operator~\cite{vd1Itzykson}:
\begin{equation}
\Psi (x) = \frac{1}{(2\pi)^3}\sum_\sigma \int \frac{d^3 {\bf p}}{2E_p} [ u_\sigma ({\bf p}) a_\sigma ({\bf p}) e^{-ip\cdot x}
+ v_\sigma ({\bf p}) b_\sigma^\dagger ({\bf p})] e^{+ip\cdot x}]\,,
\end{equation}
as given {\it ab initio}.

I studied in the previous 
works~\cite{vd1Dvoeglazov1,vd1Dvoeglazov2,vd1Dvoeglazov3}:
\begin{itemize}
  \item $\sigma \rightarrow h$  (the helicity basis);
  \item  the modified Sakurai derivation (the additional $m_2 \gamma^5$ term in the Dirac equation);
  \item  the derivation of the Barut equation~\cite{vd1Barut} from the first principles, namely 
  based on the generalized Ryder relation, ($\phi_L^h ({\bf 0}) = \hat A \phi_L^{-h\,\ast} ({\bf 0}) + \hat B \phi_L^{h\,\ast} ({\bf 0})$). In fact, we have the second mass state ($\mu$-meson)  from that equation:
  \begin{equation}[i\gamma^\mu \partial_\mu - \alpha \partial_\mu \partial^\mu  /m -\beta] \psi =0\,;\end{equation}  
\item the self/anti-self charge-conjugate Majorana 4-spinors
\cite{vd1Majorana,vd1Bilenky} in the momentum representation.
  \end{itemize}
The Wigner rules~\cite{vd1Wigner} of the Lorentz transformations
for the $(0,S)$ left- $\phi_L ({\bf p})$ and  the $(S,0)$ right-
$\phi_R ({\bf p})$ spinors are:
\begin{eqnarray}
(S,0):&&\phi_R ({\bf p})= \Lambda_R ({\bf p} \leftarrow
{\bf 0})\,\phi_R ({\bf 0})  =  \exp (+\,{\bf S} \cdot
{\bf \varphi}) \,\phi_R ({\bf 0}),\label{vd1boost0a}\\
(0,S):&&\phi_L ({\bf p}) = \Lambda_L ({\bf p} \leftarrow
{\bf 0})\,\phi_L
({\bf 0})  =  \exp (-\,{\bf S} \cdot {\bf \varphi})\,\phi_L
({\bf 0}),\label{vd1boost0}
\end{eqnarray}
with ${\bf \varphi} = {\bf n} \varphi$ being the boost parameters:
\begin{eqnarray}
&&cosh (\varphi ) =\gamma = \frac{1}{\sqrt{1-v^2/c^2}}, 
sinh (\varphi ) =\beta \gamma =\frac{v/c}{\sqrt{1-v^2/c^2}}\\
&&tanh (\varphi ) =v/c\,.
\end{eqnarray} 
They are well known and given, {\it e.g.}, in~\cite{vd1Wigner,vd1Faustov,vd1Ryder}. 

On using the Wigner rules and the Ryder relations we can recover the Dirac equation in the matrix form:
\begin{eqnarray}
\begin{pmatrix}\mp m \, 1 & p_0 + {\bf \sigma}\cdot {\bf p}\\
p_0 - {\bf \sigma}\cdot {\bf p} & \mp m \, 1\end{pmatrix} \psi (p^\mu)\,
=\, 0\,,
\end{eqnarray}
or
$(\gamma\cdot p - m) u ({\bf p})=0$ and $(\gamma\cdot p + m) v ({\bf p})=0$.
We have used the property $\left [\Lambda_{L,R} ({\bf p}
\leftarrow {\bf 0})\right ]^{-1} =
\left [\Lambda_{R,L} ({\bf p} \leftarrow {\bf 0})\right ]^\dagger$ above,
and that both ${\bf S}$ and $\Lambda_{R,L}$ are Hermitian
for the finite $(S=1/2,0)\oplus (0,S=1/2)$ representation
of the Lorentz group.
Introducing $\psi (x) \equiv \psi (p)  \exp (\mp ip\cdot x)$
and letting $p_\mu \rightarrow i\partial_\mu$, the above equation
becomes the Dirac equation (\ref{vd1Dirac}).

The solutions of the Dirac equation are denoted by 
$$u ({\bf p}) = \begin{pmatrix} \phi_R ({\bf p})\\ \phi_L ({\bf p})\end{pmatrix}$$ 
and $v ({\bf p}) =\gamma^5 u ({\bf p})$.  Let me remind
that the boosted 4-spinors in the common-used basis  (the standard representation of $\gamma$ matrices) are
\begin{eqnarray}
u_{{1\over 2},{1\over 2}} &=& \sqrt{\frac{(E+m)}{2m}}
\begin{pmatrix}1\\ 0\\ p_z/(E+m)\\ p_r/(E+m)\end{pmatrix}\,,\nonumber\\
u_{{1\over 2},-{1\over 2}} &=&\sqrt{\frac{(E+m)}{2m}}
\begin{pmatrix}0\\ 1\\ p_l/(E+m)\\ -p_z/(E+m)\end{pmatrix}\,,\\
v_{{1\over 2},{1\over 2}} &=& \sqrt{\frac{(E+m)}{2m}}\begin{pmatrix}p_z/(E+m)\\ p_r/(E+m)\\
1\\ 0\end{pmatrix}\,,\nonumber\\
v_{{1\over 2},-{1\over 2}} &=&\sqrt{\frac{(E+m)}{2m}} \begin{pmatrix}p_l/(E+m)\\ -p_z/(E+m)\\ 0\\ 
1\end{pmatrix}\,.
\end{eqnarray}
$E=\sqrt{{\bf p}^2 +m^2}>0$, $p_0=\pm E$, $p^\pm = E\pm p_z$, $p_{r,l}= p_x\pm ip_y$.
They  are the parity eigenstates with the eigenvalues of $\pm 1$. In
the parity operator the matrix $\gamma_0=\begin{pmatrix}1&0\\ 0 &
-1\end{pmatrix}$ was used as usual. They also describe
eigenstates of the charge operator, $Q$, if at rest
\begin{equation}\label{vd1rb}
\phi_R ({\bf 0})
=\pm \phi_L ({\bf 0})
\end{equation}
(otherwise the corresponding physical states are no longer
the charge eigenstates). 
Their normalizations are:
\begin{eqnarray}
&&\bar u_\sigma ({\bf p})u_{\sigma^\prime} ({\bf p}) =+\delta_{\sigma\sigma^\prime}\,,\\
&&\bar v_\sigma ({\bf p})v_{\sigma^\prime} ({\bf p}) =-\delta_{\sigma\sigma^\prime}\,,\\
&&\bar u_\sigma ({\bf p})v_{\sigma^\prime} ({\bf p}) = 0\,.
\end{eqnarray}
The bar over the 4-spinors signifies the Dirac conjugation.

Thus in this Section we have used the basis for charged particles in the $(S,0)\oplus (0,S)$ representation (in general) 
\begin{eqnarray}
u_{+\sigma} ({\bf 0}) &=& N(\sigma)\begin{pmatrix}1\\ 0\\ . \\ .
\\ . \\ 0\end{pmatrix},\,
u_{\sigma-1} ({\bf 0})=N(\sigma) \begin{pmatrix}0\\
1\\ . \\ .  \\ . \\ 0\end{pmatrix},\ldots
v_{-\sigma} ({\bf 0})=N(\sigma) \begin{pmatrix}0\\ 0\\ . \\ . \\ . \\ 1\end{pmatrix}\nonumber\\
&&
\end{eqnarray}
Sometimes, the normalization factor is convenient to choose $N(\sigma)=m^\sigma$ in order
the rest spinors to vanish in the massless limit. 

However, other constructs are possible in the $(1/2,0)\oplus (0,1/2)$ representation.

\section{Majorana Spinors in the Momentum Representation}

During the 20th century various authors introduced {\it self/anti-self} charge-conjugate 4-spinors
(including in the momentum representation), see~\cite{vd1Majorana,vd1Bilenky,vd1Ziino,vd1Ahluwalia}. \\
Later~\cite{vd1Lounesto,vd1Dvoeglazov1,vd1Dvoeglazov2,vd1Kirchbach} {\it etc} studied these spinors, they found corresponding dynamical equations, gauge transformations 
and other specific features of them.
The definitions are:
\begin{equation}
C= e^{i\theta} \begin{pmatrix}0&0&0&-i\\
0&0&i&0\\
0&i&0&0\\
-i&0&0&0\end{pmatrix} {\cal K} = -e^{i\theta} \gamma^2 {\cal K}
\end{equation}
is the anti-linear operator of charge conjugation. ${\cal K}$ is the complex conjugation operator. We  define the {\it self/anti-self} charge-conjugate 4-spinors 
in the momentum space
\begin{eqnarray}
C\lambda^{S,A} ({\bf p}) &=& \pm \lambda^{S,A} ({\bf p})\,,\\
C\rho^{S,A} ({\bf p}) &=& \pm \rho^{S,A} ({\bf p})\,.
\end{eqnarray}
Thus,
\begin{equation}
\lambda^{S,A} (p^\mu)=\begin{pmatrix}\pm i\Theta \phi^\ast_L ({\bf p})\\
\phi_L ({\bf p})\end{pmatrix}\,,
\end{equation}
and
\begin{equation}
\rho^{S,A} ({\bf p})=\begin{pmatrix}\phi_R ({\bf p})\\ \mp i\Theta \phi^\ast_R ({\bf p})\end{pmatrix}\,.
\end{equation}
The Wigner matrix is
\begin{equation}
\Theta_{[1/2]}=-i\sigma_2=\begin{pmatrix}0&-1\\
1&0\end{pmatrix}\,,
\end{equation}
and $\phi_L$, $\phi_R$ can be boosted with $\Lambda_{L,R}$ 
matrices.\footnote{Such definitions of 4-spinors differ, of course, from the original Majorana definition in x-representation:
\begin{equation}
\nu (x) = \frac{1}{\sqrt{2}} (\Psi_D (x) + \Psi_D^c (x))\,,
\end{equation}
$C \nu (x) = \nu (x)$ that represents the positive real $C-$ parity field operator. However, the momentum-space Majorana-like spinors 
open various possibilities for description of neutral  particles 
(with experimental consequences, see~\cite{vd1Kirchbach}). For instance, "for imaginary $C$ parities, the neutrino mass 
can drop out from the single $\beta $ decay trace and 
reappear in $0\nu \beta\beta $, a curious and in principle  
experimentally testable signature for a  non-trivial impact of 
Majorana framework in experiments with polarized sources."}

The rest $\lambda$ and $\rho$ spinors are:
\begin{eqnarray}
\lambda^S_\uparrow ({\bf 0}) &=& \sqrt{\frac{m}{2}}
\begin{pmatrix}0\\ i \\ 1\\ 0\end{pmatrix}\,,\,
\lambda^S_\downarrow ({\bf 0})= \sqrt{\frac{m}{2}}
\begin{pmatrix}-i \\ 0\\ 0\\ 1\end{pmatrix}\,,\,\\
\lambda^A_\uparrow ({\bf 0}) &=& \sqrt{\frac{m}{2}}
\begin{pmatrix}0\\ -i\\ 1\\ 0\end{pmatrix}\,,\,
\lambda^A_\downarrow ({\bf 0}) = \sqrt{\frac{m}{2}}
\begin{pmatrix}i\\ 0\\ 0\\ 1\end{pmatrix}\,,\,\\
\rho^S_\uparrow ({\bf 0}) &=& \sqrt{\frac{m}{2}}
\begin{pmatrix}1\\ 0\\ 0\\ -i\end{pmatrix}\,,\,
\rho^S_\downarrow ({\bf 0}) = \sqrt{\frac{m}{2}}
\begin{pmatrix}0\\ 1\\ i\\ 0\end{pmatrix}\,,\,\\
\rho^A_\uparrow ({\bf 0}) &=& \sqrt{\frac{m}{2}}
\begin{pmatrix}1\\ 0\\ 0\\ i\end{pmatrix}\,,\,
\rho^A_\downarrow ({\bf 0}) = \sqrt{\frac{m}{2}}
\begin{pmatrix}0\\ 1\\ -i\\ 0\end{pmatrix}\,.
\end{eqnarray}
Thus, in this basis the explicite forms of the 4-spinors of the second kind  $\lambda^{S,A}_{\uparrow\downarrow}
({\bf p})$ and $\rho^{S,A}_{\uparrow\downarrow} ({\bf p})$
are
\begin{eqnarray}
\lambda^S_\uparrow ({\bf p}) &=& \frac{1}{2\sqrt{E+m}}
\begin{pmatrix}ip_l\\ i (p^- +m)\\ p^- +m\\ -p_r\end{pmatrix},
\lambda^S_\downarrow ({\bf p})= \frac{1}{2\sqrt{E+m}}
\begin{pmatrix}-i (p^+ +m)\\ -ip_r\\ -p_l\\ (p^+ +m)\end{pmatrix}\nonumber\\
\\
\lambda^A_\uparrow ({\bf p}) &=& \frac{1}{2\sqrt{E+m}}
\begin{pmatrix}-ip_l\\ -i(p^- +m)\\ (p^- +m)\\ -p_r\end{pmatrix},
\lambda^A_\downarrow ({\bf p}) = \frac{1}{2\sqrt{E+m}}
\begin{pmatrix}i(p^+ +m)\\ ip_r\\ -p_l\\ (p^+ +m)\end{pmatrix}\nonumber\\
\\
\rho^S_\uparrow ({\bf p}) &=& \frac{1}{2\sqrt{E+m}}
\begin{pmatrix}p^+ +m\\ p_r\\ ip_l\\ -i(p^+ +m)\end{pmatrix},
\rho^S_\downarrow ({\bf p}) = \frac{1}{2\sqrt{E+m}}
\begin{pmatrix}p_l\\ (p^- +m)\\ i(p^- +m)\\ -ip_r\end{pmatrix}\nonumber\\
\\
\rho^A_\uparrow ({\bf p}) &=& \frac{1}{2\sqrt{E+m}}
\begin{pmatrix}p^+ +m\\ p_r\\ -ip_l\\ i (p^+ +m)\end{pmatrix},
\rho^A_\downarrow ({\bf p}) = \frac{1}{2\sqrt{E+m}}
\begin{pmatrix}p_l\\ (p^- +m)\\ -i(p^- +m)\\ ip_r\end{pmatrix}.\nonumber
\\
\end{eqnarray}
As we showed $\lambda$ and $\rho$ 4-spinors are NOT the eigenspinors of the helicity. Moreover, 
$\lambda$ and $\rho$ are NOT the eigenspinors of the parity (in this
representation 
$P=\begin{pmatrix}0&1\\ 1&0\end{pmatrix}R$), as opposed to the Dirac case.
The indices $\uparrow\downarrow$ should be referred to the chiral helicity 
quantum number introduced 
in the 60s, $\eta=-\gamma^5 h$.
While 
\begin{equation}
Pu_\sigma ({\bf p}) = + u_\sigma ({\bf p})\,,
Pv_\sigma ({\bf p}) = - v_\sigma ({\bf p})\,,
\end{equation}
we have
\begin{equation}
P\lambda^{S,A} ({\bf p}) = \rho^{A,S} ({\bf p})\,,
P \rho^{S,A} ({\bf p}) = \lambda^{A,S} ({\bf p})\,,
\end{equation}
for the Majorana-like momentum-space 4-spinors on the first quantization level.
In this basis one has
\begin{eqnarray}
\rho^S_\uparrow ({\bf p}) \,&=&\, - i \lambda^A_\downarrow ({\bf p})\,,\,
\rho^S_\downarrow ({\bf p}) \,=\, + i \lambda^A_\uparrow ({\bf p})\,,\,\\
\rho^A_\uparrow ({\bf p}) \,&=&\, + i \lambda^S_\downarrow ({\bf p})\,,\,
\rho^A_\downarrow ({\bf p}) \,=\, - i \lambda^S_\uparrow ({\bf p})\,.
\end{eqnarray}

The normalization of the spinors $\lambda^{S,A}_{\uparrow\downarrow}
({\bf p})$ and $\rho^{S,A}_{\uparrow\downarrow} ({\bf p})$ are the following ones:
\begin{eqnarray}
\overline \lambda^S_\uparrow ({\bf p}) \lambda^S_\downarrow ({\bf p}) \,&=&\,
- i m \quad,\quad
\overline \lambda^S_\downarrow ({\bf p}) \lambda^S_\uparrow ({\bf p}) \,= \,
+ i m \quad,\quad\\
\overline \lambda^A_\uparrow ({\bf p}) \lambda^A_\downarrow ({\bf p}) \,&=&\,
+ i m \quad,\quad
\overline \lambda^A_\downarrow ({\bf p}) \lambda^A_\uparrow ({\bf p}) \,=\,
- i m \quad,\quad\\
\overline \rho^S_\uparrow ({\bf p}) \rho^S_\downarrow ({\bf p}) \, &=&  \,
+ i m\quad,\quad
\overline \rho^S_\downarrow ({\bf p}) \rho^S_\uparrow ({\bf p})  \, =  \,
- i m\quad,\quad\\
\overline \rho^A_\uparrow ({\bf p}) \rho^A_\downarrow ({\bf p})  \,&=&\,
- i m\quad,\quad
\overline \rho^A_\downarrow ({\bf p}) \rho^A_\uparrow ({\bf p}) \,=\,
+ i m\quad.
\end{eqnarray}
All other conditions are equal to zero.

The dynamical coordinate-space equations are:
\begin{eqnarray}
i \gamma^\mu \partial_\mu \lambda^S (x) - m \rho^A (x) &=& 0 \,,
\label{vd111}\\
i \gamma^\mu \partial_\mu \rho^A (x) - m \lambda^S (x) &=& 0 \,,
\label{vd112}\\
i \gamma^\mu \partial_\mu \lambda^A (x) + m \rho^S (x) &=& 0\,,
\label{vd113}\\
i \gamma^\mu \partial_\mu \rho^S (x) + m \lambda^A (x) &=& 0\,.
\label{vd114}
\end{eqnarray}
These are NOT the Dirac equation.
However, they can be written in the 8-component form as follows:
\begin{eqnarray}
\left [i \Gamma^\mu \partial_\mu - m\right ] \Psi_{_{(+)}} (x) &=& 0\,,
\label{vd1psi1}\\
\left [i \Gamma^\mu \partial_\mu + m\right ] \Psi_{_{(-)}} (x) &=& 0\,,
\label{vd1psi2}
\end{eqnarray}
with
\begin{eqnarray}
\Psi_{(+)} (x) = \begin{pmatrix}\rho^A (x)\\
\lambda^S (x)\end{pmatrix},
\Psi_{(-)} (x) = \begin{pmatrix}\rho^S (x)\\
\lambda^A (x)\end{pmatrix}, \mbox{and}\,\Gamma^\mu =\begin{pmatrix}0 & \gamma^\mu\\
\gamma^\mu & 0\end{pmatrix}
\end{eqnarray}
One can also re-write the equations into the two-component form.
Similar formulations have been presented by M. Markov~\cite{vd1Markov}, and
A. Barut and G. Ziino~\cite{vd1Ziino}. The group-theoretical basis for such doubling has been given
in the papers by Gelfand, Tsetlin and Sokolik~\cite{vd1Gelfand}.

The Lagrangian is
\begin{eqnarray}
&&{\cal L}= \frac{i}{2} \left[\bar \lambda^S \gamma^\mu \partial_\mu \lambda^S - (\partial_\mu \bar \lambda^S ) \gamma^\mu \lambda^S +
\bar \rho^A \gamma^\mu \partial_\mu \rho^A - (\partial_\mu \bar \rho^A ) \gamma^\mu \rho^A +\right.\nonumber\\
&&\left.\bar \lambda^A \gamma^\mu \partial_\mu \lambda^A - (\partial_\mu \bar \lambda^A ) \gamma^\mu \lambda^A +
\bar \rho^S
\gamma^\mu \partial_\mu \rho^S - (\partial_\mu \bar \rho^S ) \gamma^\mu \rho^S -\right.\nonumber\\
&&\left. - m (\bar\lambda^S \rho^A +\bar \lambda^S \rho^A -\bar\lambda^S \rho^A -\bar\lambda^S \rho^A )
\right ]
\end{eqnarray}

The connection with the Dirac spinors has been found. 
For instance,
\begin{eqnarray}
\begin{pmatrix}\lambda^S_\uparrow ({\bf p}) \\ \lambda^S_\downarrow ({\bf p}) \\
\lambda^A_\uparrow ({\bf p}) \\ \lambda^A_\downarrow ({\bf p})\end{pmatrix} = {1\over
2} \begin{pmatrix}1 & i & -1 & i\\ -i & 1 & -i & -1\\ 1 & -i & -1 & -i\\ i&
1& i& -1\end{pmatrix} \begin{pmatrix}u_{+1/2} ({\bf p}) \\ u_{-1/2} ({\bf p}) \\
v_{+1/2} ({\bf p}) \\ v_{-1/2} ({\bf p})\end{pmatrix}.\label{vd1connect}
\end{eqnarray}
See also ref.~\cite{vd1Gelfand,vd1Ziino}.

The sets of $\lambda$ spinors and of $\rho$ spinors are claimed to be
{\it bi-orthonormal} sets each in the mathematical sense~\cite{vd1Ahluwalia},  provided
that overall phase factors of 2-spinors $\theta_1 +\theta_2 = 0$ or $\pi$.
For instance, on the classical level $\bar \lambda^S_\uparrow
\lambda^S_\downarrow = 2iN^2 \cos ( \theta_1 + \theta_2 )$.\footnote{We used above 
$\theta_1=\theta_2 =0$.} 

Few remarks have been given in the previous works:
\begin{itemize}
\item
While in the massive case there are four $\lambda$-type spinors, two
$\lambda^S$ and two $\lambda^A$ (the $\rho$ spinors are connected by
certain relations with the $\lambda$ spinors for any spin case),  in a
massless case $\lambda^S_\uparrow$ and $\lambda^A_\uparrow$ identically
vanish, provided that one takes into account that $\phi_L^{\pm 1/2}$ are
 eigenspinors of ${\bf \sigma}\cdot \hat {\bf n}$, the
$2\times 2$ helicity operator.

\item
It was noted the possibility of the generalization of the concept of the
Fock space, which leads to the ``doubling" Fock space~\cite{vd1Gelfand,vd1Ziino}.

\end{itemize}

It was shown~\cite{vd1Dvoeglazov1} that the covariant derivative (and, hence, the
 interaction) can be introduced in this construct in the following way:
\begin{equation}
\partial_\mu \rightarrow \nabla_\mu = \partial_\mu - ig \mbox{\L}^5 A_\mu\quad,
\end{equation}
where $\mbox{\L}^5 = \mbox{diag} (\gamma^5 \quad -\gamma^5)$, the $8\times 8$
matrix. With respect to the transformations
\begin{eqnarray}
\lambda^\prime (x)
\rightarrow (\cos \alpha -i\gamma^5 \sin\alpha) \lambda
(x)\quad,\label{vd1g10}\\
\overline \lambda^{\,\prime} (x) \rightarrow
\overline \lambda (x) (\cos \alpha - i\gamma^5
\sin\alpha)\quad,\label{vd1g20}\\
\rho^\prime (x) \rightarrow  (\cos \alpha +
i\gamma^5 \sin\alpha) \rho (x) \quad,\label{vd1g30}\\
\overline \rho^{\,\prime} (x) \rightarrow  \overline \rho (x)
(\cos \alpha + i\gamma^5 \sin\alpha)\quad\label{vd1g40}
\end{eqnarray}
the spinors retain their properties to be self/anti-self charge conjugate
spinors and the proposed Lagrangian~\cite[p.1472]{vd1Dvoeglazov1} remains to be invariant.
This tells us that while self/anti-self charge conjugate states have
zero eigenvalues of the ordinary (scalar) charge operator but they can
possess the axial charge (cf.  with the discussion of~\cite{vd1Ziino} and
the old idea of R. E. Marshak).

In fact, from this consideration one can recover the Feynman-Gell-Mann
equation (and its charge-conjugate equation). It is re-written in the
two-component form
\begin{eqnarray} 
\begin{cases}\left [\pi_\mu^- \pi^{\mu\,-}
-m^2 -{g\over 2} \sigma^{\mu\nu} F_{\mu\nu} \right ] \chi (x)=0\,, &\\
\left [\pi_\mu^+ \pi^{\mu\,+} -m^2
+{g\over 2} \widetilde\sigma^{\mu\nu} F_{\mu\nu} \right ] \phi (x)
=0\,, &\end{cases}\label{vd1iii}
\end{eqnarray}
where already one has $\pi_\mu^\pm =
i\partial_\mu \pm gA_\mu$, \, $\sigma^{0i} = -\widetilde\sigma^{0i} =
i\sigma^i$, $\sigma^{ij} = \widetilde\sigma^{ij} = \epsilon_{ijk}
\sigma^k$ and $\nu^{^{DL}} (x) =\mbox{column} (\chi \quad \phi )$.

Next, because the transformations
\begin{eqnarray}
\lambda_S^\prime ({\bf p}) &=& \begin{pmatrix}\Xi &0\\ 0&\Xi\end{pmatrix} \lambda_S ({\bf p})
\equiv \lambda_A^\ast ({\bf p}),\\
\lambda_S^{\prime\prime} ({\bf p}) &=& \begin{pmatrix}i\Xi &0\\ 0&-i\Xi\end{pmatrix} \lambda_S
({\bf p}) \equiv -i\lambda_S^\ast ({\bf p}),\\
\lambda_S^{\prime\prime\prime} ({\bf p}) &=& \begin{pmatrix}0& i\Xi\\
i\Xi &0\end{pmatrix} \lambda_S ({\bf p}) \equiv i\gamma^0 \lambda_A^\ast
({\bf p}),\\
\lambda_S^{IV} ({\bf p}) &=& \begin{pmatrix}0& \Xi\\
-\Xi&0\end{pmatrix} \lambda_S ({\bf p}) \equiv \gamma^0\lambda_S^\ast
({\bf p})
\end{eqnarray}
with the $2\times 2$ matrix $\Xi$ defined as ($\phi$ is the azimuthal
angle  related with ${\bf p} \rightarrow {\bf 0}$)
\begin{equation}
\Xi = \begin{pmatrix}e^{i\phi} & 0\\ 0 &
e^{-i\phi}\end{pmatrix}\quad,\quad \Xi \Lambda_{R,L} ({\bf p} \leftarrow
{\bf 0}) \Xi^{-1} = \Lambda_{R,L}^\ast ({\bf p} \leftarrow
 {\bf 0})\,\,\, ,
\end{equation}
and corresponding transformations for
$\lambda^A$ do {\it not} change the properties of bispi\-nors to be in the
self/anti-self charge conjugate spaces, the Majorana-like field operator
($b^\dagger \equiv a^\dagger$) admits additional phase (and, in general,
normalization) transformations:
\begin{equation} \nu^{ML\,\,\prime}
(x^\mu) = \left [ c_0 + i({\bf \tau}\cdot  {\bf c}) \right
]\nu^{ML\,\,\dagger} (x^\mu) \,, 
\end{equation} 
where $c_\alpha$ are
arbitrary parameters. The ${\bf \tau}$ matrices are defined over the
field of $2\times 2$ matrices and the Hermitian
conjugation operation is assumed to act on the $c$- numbers as the complex
conjugation. One can parametrize $c_0 = \cos\phi$ and ${\bf c} = {\bf n}
\sin\phi$ and, thus, define the $SU(2)$ group of phase transformations.
One can select the Lagrangian which is composed from the both field
operators (with $\lambda$ spinors and $\rho$ spinors)
and which remains to be
invariant with respect to this kind of transformations.  The conclusion
is: it is permitted a non-Abelian construct which is based on
the spinors of the Lorentz group only (cf. with the old ideas of T. W.
Kibble and R. Utiyama) .  This is not surprising because both the $SU(2)$
group and $U(1)$ group are  the sub-groups of the extended Poincar\'e group
(cf.~\cite{vd1Ryder}).

The Dirac-like and Majorana-like field operators can
be built from both $\lambda^{S,A} ({\bf p})$ and $\rho^{S,A} ({\bf p})$,
or their combinations. For 
instance,
\begin{eqnarray}
&&\Psi (x^\mu) \equiv \int {d^3 {\bf p}\over (2\pi)^3} {1\over 2E_p}
\sum_\eta \left [ \lambda^S_\eta ({\bf p}) \, a_\eta ({\bf p}) \,\exp
(-ip\cdot x) +\right.\nonumber\\
&+&\left.\lambda^A_\eta ({\bf p})\, b^\dagger_\eta ({\bf p}) \,\exp
(+ip\cdot x)\right ].\label{vd1oper}
\end{eqnarray}

The anticommutation relations are the following ones (due to the {\it bi-orthonor\-mality}):
\begin{eqnarray}
[a_{\eta{\prime}} ({\bf p}^{\prime}), a_\eta^\dagger ({\bf p}) ]_\pm = (2\pi)^3 2E_p \delta ({\bf p} -{\bf p}^\prime) \delta_{\eta,-\eta^\prime}
\end{eqnarray}
and 
\begin{eqnarray}
[b_{\eta{\prime}} ({\bf p}^{\prime}), b_\eta^\dagger ({\bf p}) ]_\pm = (2\pi)^3 2E_p \delta ({\bf p} -{\bf p}^\prime) \delta_{\eta,-\eta^\prime}
\end{eqnarray}
Other (anti)commutators are equal to zero: ($[ a_{\eta^\prime} ({\bf p}^{\prime}), 
b_\eta^\dagger ({\bf p}) ]=0$).

In the Fock space operations of the charge conjugation and space
inversions can be defined through unitary operators such that:
\begin{eqnarray}
U^c_{[1/2]} \Psi (x^\mu) (U^c_{[1/2]})^{-1} &=& {\cal C}_{[1/2]}
\Psi^\dagger_{[1/2]} (x^\mu),\\
U^s_{[1/2]} \Psi (x^\mu) (U^s_{[1/2]})^{-1} &=& \gamma^0
\Psi (x^{\prime^{\,\mu}}),
\end{eqnarray}
the time reversal operation, through {\it an antiunitary}
operator\footnote{Let us remind that the operator of hermitian conjugation does not act on $c$-numbers on the left side of the equation (\ref{vd1tr}).
This fact is conected with the properties of an antiunitary operator:
$\left [ V^{^T} \lambda A (V^{^T})^{-1}\right ]^\dagger =
\left [\lambda^\ast V^{^T} A (V^{^T})^{-1}\right ]^\dagger =
\lambda \left [ V^{^T} A^\dagger (V^{^T})^{-1} \right ]$.}
\begin{equation}
\left [V^{^T}_{[1/2]}  \Psi (x^\mu)
(V^{^T}_{[1/2]})^{-1} \right ]^\dagger = S(T) \Psi^\dagger
(x^{{\prime\prime}^\mu}) \quad,\label{vd1tr}
\end{equation}
with
$x^{\prime^{\,\mu}} \equiv (x^0, -{\bf x})$ and $x^{{\prime\prime}^{\,\mu}}
=(-x^0,{\bf x})$.  We  further assume the vacuum state to be assigned an
even $P$- and $C$-eigenvalue and, then, proceed as in ref.~\cite{vd1Itzykson}.

As a result we have the following properties of creation (annihilation)
operators in the Fock space:
\begin{eqnarray}
U^s_{[1/2]} a_\uparrow ({\bf p}) (U^s_{[1/2]})^{-1} &=& - ia_\downarrow
(-  {\bf p}),\nonumber\\
U^s_{[1/2]} a_\downarrow ({\bf p}) (U^s_{[1/2]})^{-1} &=& + ia_\uparrow
(- {\bf p})\,.\\
U^s_{[1/2]} b_\uparrow^\dagger ({\bf p}) (U^s_{[1/2]})^{-1} &=&
+ i b_\downarrow^\dagger (- {\bf p}),\nonumber\\
U^s_{[1/2]} b_\downarrow^\dagger ({\bf p}) (U^s_{[1/2]})^{-1} &=&
- i b_\uparrow (- {\bf p}),
\end{eqnarray} 
what signifies that the states created by the operators $a^\dagger
({\bf p})$ and $b^\dagger ({\bf p})$ have very different properties
with respect to the space inversion operation, comparing with
Dirac states (the case also regarded in~\cite{vd1Ziino}):
\begin{eqnarray}
U^s_{[1/2]} \vert {\bf p},\,\uparrow >^+ &=& + i \vert -{\bf p},\,
\downarrow >^+ ,
U^s_{[1/2]} \vert {\bf p},\,\uparrow >^- = + i
\vert -{\bf p},\, \downarrow >^-\nonumber\\
\\
U^s_{[1/2]} \vert {\bf p},\,\downarrow >^+ &=& - i \vert -{\bf p},\,
\uparrow >^+,
U^s_{[1/2]} \vert {\bf p},\,\downarrow >^- =  - i
\vert -{\bf p},\, \uparrow >^- .\nonumber\\
\end{eqnarray}

For the charge conjugation operation in the Fock space we have
two physically different possibilities. The first one, {\it e.g.},
\begin{eqnarray}
U^c_{[1/2]} a_\uparrow ({\bf p}) (U^c_{[1/2]})^{-1} &=& + b_\uparrow
({\bf p})\,,\,
U^c_{[1/2]} a_\downarrow ({\bf p}) (U^c_{[1/2]})^{-1} = + b_\downarrow
({\bf p}),\nonumber\\
\\
U^c_{[1/2]} b_\uparrow^\dagger ({\bf p}) (U^c_{[1/2]})^{-1} &=&
-a_\uparrow^\dagger ({\bf p})\,,\,
U^c_{[1/2]} b_\downarrow^\dagger ({\bf p})
(U^c_{[1/2]})^{-1} = -a_\downarrow^\dagger ({\bf p})\,,\nonumber\\
\end{eqnarray}
in fact, has some similarities with the Dirac construct.
However, the action of this operator on the physical states are
\begin{eqnarray}
U^c_{[1/2]} \vert {\bf p}, \uparrow >^+ &=& + \vert {\bf p},
\uparrow >^- ,\,
U^c_{[1/2]} \vert {\bf p}, \downarrow >^+ = + \vert {\bf p},
\downarrow >^- ,\\
U^c_{[1/2]} \vert {\bf p}, \uparrow >^-
&=&  - \, \vert {\bf p}, \uparrow >^+ ,\,
U^c_{[1/2]} \vert
{\bf p}, \, \downarrow >^- = - \vert {\bf p}, \downarrow >^+ .
\end{eqnarray}
But, one can also construct the charge conjugation operator in the
Fock space which acts, {\it e.g.}, in the following manner:
\begin{eqnarray}
\widetilde U^c_{[1/2]} a_\uparrow ({\bf p}) (\widetilde U^c_{[1/2]})^{-1}
&=& - b_\downarrow ({\bf p})\,,\, \widetilde U^c_{[1/2]}
a_\downarrow ({\bf p}) (\widetilde U^c_{[1/2]})^{-1} = - b_\uparrow
({\bf p}),\nonumber\\
\\
\widetilde U^c_{[1/2]} b_\uparrow^\dagger ({\bf p})
(\widetilde U^c_{[1/2]})^{-1} &=& + a_\downarrow^\dagger ({\bf
p})\,,\,
\widetilde U^c_{[1/2]} b_\downarrow^\dagger ({\bf p})
(\widetilde U^c_{[1/2]})^{-1} = + a_\uparrow^\dagger ({\bf p}),\nonumber
\\
\end{eqnarray}
and, therefore,
\begin{eqnarray}
\widetilde U^c_{[1/2]} \vert {\bf p}, \, \uparrow >^+ &=& - \,\vert {\bf
p},\, \downarrow >^- \,,\,
\widetilde U^c_{[1/2]} \vert {\bf p}, \, \downarrow
>^+ = - \, \vert {\bf p},\, \uparrow >^- \,,\\
\widetilde U^c_{[1/2]} \vert
{\bf p}, \, \uparrow >^- &=& + \, \vert {\bf p},\, \downarrow >^+
\,,\,
\widetilde U^c_{[1/2]} \vert {\bf p}, \, \downarrow >^- = + \, \vert {\bf
p},\, \uparrow >^+ \,.
\end{eqnarray}

Investigations of several important cases, which are different from the
 above ones, are required a separate paper to. Next, it is
 possible a situation when the operators of the space inversion and 
charge conjugation commute each other in the Fock space~\cite{vd1Foldy}. For instance,
\begin{eqnarray}
U^c_{[1/2]} U^s_{[1/2]} \vert {\bf
p}, \uparrow >^+ &=& + i U^c_{[1/2]}\vert -{\bf p},\, \downarrow >^+ =
+ i \vert -{\bf p}, \downarrow >^- \\
U^s_{[1/2]} U^c_{[1/2]} \vert {\bf
p}, \uparrow >^+ &=& U^s_{[1/2]}\vert {\bf p}, \uparrow >^- = + i
\vert -{\bf p}, \downarrow >^- .
\end{eqnarray}
The second choice of the charge conjugation operator answers for the case
when the $\widetilde U^c_{[1/2]}$ and $U^s_{[1/2]}$ operations
anticommute:
\begin{eqnarray}
\widetilde U^c_{[1/2]} U^s_{[1/2]} \vert {\bf p},\, \uparrow >^+ &=&
+ i \widetilde U^c_{[1/2]}\vert -{\bf
p},\, \downarrow >^+ = -i \, \vert -{\bf p},\, \uparrow >^- \\
U^s_{[1/2]} \widetilde U^c_{[1/2]} \vert {\bf p},\, \uparrow >^+ &=& -
U^s_{[1/2]}\vert {\bf p},\, \downarrow >^- = + i \, \vert -{\bf p},\,
\uparrow >^- \,.
\end{eqnarray}

Next, one can compose states which would have somewhat similar
properties to those which we have become accustomed.
The states $\vert {\bf p}, \,\uparrow >^+ \pm
i\vert {\bf p},\, \downarrow >^+$ answer for positive (negative) parity,
respectively.  But, what is important, {\it the antiparticle states}
(moving backward in time) have the same properties with respect to the
operation of space inversion as the corresponding {\it particle states}
(as opposed to $j=1/2$ Dirac particles).  
The states which are 
eigenstates of the charge conjugation operator in the Fock space are
\begin{equation}
U^c_{[1/2]} \left ( \vert {\bf p},\, \uparrow >^+ \pm i\,
\vert {\bf p},\, \uparrow >^- \right ) = \mp i\,  \left ( \vert {\bf p},\,
\uparrow >^+ \pm i\, \vert {\bf p},\, \uparrow >^- \right ) \,.
\end{equation}
There is no any simultaneous set of states which would be eigenstates of the 
operator of the space inversion and of the charge conjugation 
$U^c_{[1/2]}$.

Finally, the time reversal {\it anti-unitary} operator in 
the Fock space should be defined in such a way that the formalism to be
 compatible with the $CPT$ theorem. If we wish the Dirac states to transform 
as 
$V(T) \vert {\bf p}, \pm 1/2 > = \pm \,\vert -{\bf p}, \mp 1/2 >$ we
 have to choose (within a phase factor), ref.~\cite{vd1Itzykson}:
\begin{equation}
S(T) = \begin{pmatrix}\Theta_{[1/2]} &0\\ 0 &
\Theta_{[1/2]}\end{pmatrix}\,.
\end{equation}
Thus, in the first relevant case we obtain for the $\Psi
(x^\mu)$ field, Eq.  (\ref{vd1oper}):
\begin{eqnarray}
V^{^T} a^\dagger_\uparrow ({\bf p}) (V^{^T})^{-1} &=& a^\dagger_\downarrow
(-{\bf p}),\,
V^{^T} a^\dagger_\downarrow ({\bf p}) (V^{^T})^{-1} = -
a^\dagger_\uparrow (-{\bf p}) \\
V^{^T} b_\uparrow ({\bf p}) (V^{^T})^{-1} &=& b_\downarrow
(-{\bf p}),\,
V^{^T} b_\downarrow ({\bf p}) (V^{^T})^{-1} = -
b_\uparrow (-{\bf p}).
\end{eqnarray}
Thus, this construct has very different properties with respect to $C,P$ and $T$ comparing 
with the Dirac construct. 

{\bf But, at least for mathematicians, the dependence of the physical results on the choice 
of the basis is a bit strange thing. Somewhat similar things have been presented in~\cite{vd1Dvoeglazov3} when compared the Dirac-like constructs in the parity and helicity bases. It was shown that the helicity eigenstates
$({\bf \sigma} \cdot {\bf n})\otimes I$) are NOT the parity eigenstates (and the 
${\bf S_3}$ eigenstates), and vice versa, in the helicity basis (cf. with [Berestetskii,Lifshitz, Pitaevskii]), while they obey the same Dirac equation.  The bases are connected by the unitary transformation. And, the both sets of 4-spinors form the complete system in a mathematical sense.}

\section{The Spin 1}

\subsection{Maxwell Equations as Quantum Equations}

In refs.~\cite{vd1Gersten,vd1Dvoeglazov4}
the Maxwell-like equations have been derived\footnote{I call them "Maxwell-like"  because an additional gradient of a scalar field $\chi$ can be introduced therein.} from the Klein-Gordon equation. Here they are:
\begin{eqnarray}
&&{\bf \nabla}\times {\bf
E}=-\frac{1}{c}\frac{\partial {\bf B}}{\partial t} + {\bf
\nabla} {\Im} \chi \,, \label{vd11A}\\
&&{\bf \nabla }\times {\bf B}=\frac{1}{c}\frac{\partial {\bf
E}}{\partial t}  +{\bf \nabla} {\Re} \chi\,,\label{vd12A}\\
&&{\bf \nabla}\cdot {\bf E}=-{1\over c} {\partial \over \partial
t} {\Re}\chi \,,\label{vd13}\\
&&{\bf \nabla }\cdot {\bf B}= {1\over
c} {\partial \over \partial t} {\Im} \chi \,.  \label{vd14}
\end{eqnarray}
Of course, similar equations can be obtained 
in the massive case $m\neq 0$, i.e., within the Proca-like theory.
We should then consider
\begin{equation}
(E^2 -c^2 {\bf p}^2 - m^2 c^4 ) \Psi^{(3)} =0\, .\label{vd15}
\end{equation}
In the spin-1/2 case the equation (\ref{vd15}) can be written 
for the two-component spinor ($c=\hbar =1$)
\begin{equation}
(E I^{(2)} - {\bf\sigma}\cdot {\bf p})
(E I^{(2)} + {\bf\sigma}\cdot {\bf p})\Psi^{(2)} = m^2 \Psi^{(2)}\,,
\end{equation}
or, in the 4-component form
\begin{equation}
[i\gamma_\mu \partial_\mu +m_1 +m_2 \gamma^5 ] \Psi^{(4)} = 0\,.
\end{equation}
In the spin-1 case  we have
\begin{equation}
(E I^{(3)} - {\bf S}\cdot {\bf p})
(E I^{(3)} + {\bf S}\cdot {\bf p}){\bf \Psi}^{(3)} 
- {\bf p} ({\bf p}\cdot {\bf \Psi}^{(3)})= m^2 \Psi^{(3)}\,.
\end{equation}
These lead to (\ref{vd11A}-\ref{vd14}), when $m=0$ provided that the $\Psi^{(3)}$ is chosen as a superposition of a vector (the electric field) and an axial vector (the magnetic field).\footnote{We can continue writing down equations for higher spins in a similar fashion.} When $\chi =0$ we recover the common-used Maxwell equations.

Otherwise, we can start with ($c=\hbar=1$)\footnote{The question of both explicite and implicite dependences of the fields on the time
(and, hence, the "whole-partial derivative" )has been studied in~\cite{vd1Brownstein,vd1Dvoeglazov5}.}
\begin{equation}
\frac{\partial {\bf E}}{\partial t} = curl {\bf B}\,,\quad \frac{\partial {\bf B}}{\partial t} = -curl {\bf E}\,.
\end{equation}
Then,
\begin{eqnarray}
\frac{\partial ({\bf E}+i{\bf B})}{\partial t} - curl ({\bf B}-i{\bf E})&=&0\,,\\
\frac{\partial ( {\bf E} -i{\bf B})}{\partial t} -curl ({\bf B} + i{\bf E}) &=&0\,.
\end{eqnarray}
In the component form:
\begin{eqnarray}
\frac{\partial ({\bf E}+i{\bf B})^i}{\partial t} +i\epsilon^{ijk} \partial_j ({\bf E}+i{\bf B})^k &=&0\,,\\
\frac{\partial ( {\bf E} -i{\bf B})^i}{\partial t} - i\epsilon^{ijk}\partial_j ({\bf E} - i{\bf B})^k &=&0\,.
\end{eqnarray}
Since the spin-1 matrices can be presented in the form: $({\bf S}^i)^{jk}= -i \epsilon^{ijk}$,
we  have 
\begin{eqnarray}
\frac{\partial ({\bf E}+i{\bf B})^i}{\partial t} + ({\bf S}\cdot \nabla )^{ik} ({\bf E}+i{\bf B})^k &=&0\,,\\
\frac{\partial ( {\bf E} -i{\bf B})^i}{\partial t} -  ({\bf S}\cdot \nabla )^{ik} ({\bf E} - i{\bf B})^k &=&0\,.
\end{eqnarray}
Finally, on using that $\hat {\bf p} = -i\hbar \nabla$ we have
\begin{equation}
i\frac{\partial \phi}{\partial t} = ({\bf S}\cdot \hat {\bf p}) \phi\,,\quad 
i\frac{\partial \xi}{\partial t} = - ({\bf S}\cdot \hat {\bf p}) \xi\,.
\end{equation} 
In the following we show that these equations can also be considered as the massless limit of the Weinberg $S=1$ 
quantum-field equation.

Meanwhile, we can calculate the determinants  of the above equations, $Det [E \mp ({\bf S}\cdot {\bf p})] =0$, and we can find
that we have both the causal $E=\pm \vert {\bf p}\vert $ and acausal $E=0$ solutions.\footnote{The possible interpretation of the $E=0$ solutions are the stationary fields.} These results  will be useful
in analyzing the spin-1  quantum-field theory below.

\subsection{The Weinberg $2(2S+1)$ Theory for Spin-1}

It is based on the following
postulates~[Wigner,Weinberg]:

\begin{itemize}

\item
The fields transform according to the formula:
\begin{equation}
U [\Lambda, a] \Psi_n (x) U^{-1} [\Lambda, a] = \sum_m D_{nm}
[\Lambda^{-1}] \Psi_m (\Lambda x +a)\,,\label{vd11} \end{equation}
where $D_{nm} [\Lambda]$ is some representation of $\Lambda$; $x^\mu \rightarrow
\Lambda^\mu_{\quad\nu} \,\,x^\nu +a^\mu$, and $U [\Lambda, a]$ is a
unitary operator.

\item
For $(x-y)$ spacelike one has
\begin{equation}
[\Psi_n (x), \Psi_m (y) ]_\pm =0\,\label{vd12}
\end{equation}
for fermion and boson fields, respectively.

\item
The interaction Hamiltonian density is said by S. Weinberg to be a scalar,
and it is constructed out of the creation and annihilation operators for
the free particles described by the free Hamiltonian $H_0$.

\item
The $S$-matrix is constructed as an integral of the $T$-ordering
product of the interaction Hamiltonians by the Dyson's formula.

\end{itemize}

In this talk we shall be mainly interested in the free-field theory.
Weinberg wrote: ``In order to discuss theories with parity conservation it
is convenient to use $2(2S+1)$-component fields, like the Dirac field.
These do obey field equations, which can be derived as\ldots consequences
of (\ref{vd11},\ref{vd12})."\,\footnote{In the $(2S+1)$ formalism fields
obey only the Klein-Gordon equation, according to the Weinberg wisdom.} 
In such a way he proceeds to form the
$2(2S+1)$-component object
$$\Psi =\begin{pmatrix}\Phi_\sigma\\ \Xi_\sigma\end{pmatrix}$$
transforming according to the Wigner rules. They are the following ones
(see also above, Eqs.~\ref{vd1boost0a},\ref{vd1boost0}):
\begin{eqnarray}
\Phi_\sigma ({\bf p}) &=& \exp (+\Theta \,\hat {\bf p} \cdot {\bf S})
\Phi_\sigma ({\bf 0}) \,,\label{vd1wr1}\\
\Xi_\sigma ({\bf p}) &=& \exp (-\Theta \,\hat {\bf p} \cdot {\bf S})
\Xi_\sigma ({\bf 0}) \,\label{vd1wr2}
\end{eqnarray} 
from the zero-momentum frame. $\Theta$ is the boost parameter,
$\tanh \,\Theta =\vert {\bf p} \vert/ E$, \,$\hat {\bf p} =
{\bf p}/ \vert {\bf p} \vert$, ${\bf p}$ is the 3-momentum of the particle,
${\bf S}$ is the angular momentum operator.
For a given representation the matrices ${\bf S}$ can be constructed. In
the Dirac case (the $(1/2,0)\oplus (0,1/2)$ representation) ${\bf S} =
{\bf \sigma}/2$; in the $S=1$ case (the $(1,0)\oplus (0,1)$
representation) we can choose $(S_i)_{jk} = -i\epsilon_{ijk}$, etc. Hence,
we can explicitly calculate (\ref{vd1wr1},\ref{vd1wr2}).

The task is now to obtain relativistic equations for higher spins.
Weinberg uses the following procedure.
Firstly, he defined the scalar matrix
\begin{equation}
\Pi_{\sigma^\prime \sigma}^{(s)} (q) = (-)^{2s} t_{\sigma^\prime
\sigma}^{\quad \mu_1 \mu_2 \ldots \mu_{2s}} q_{\mu_1} q_{\mu_2}\ldots
q_{\mu_{2s}}
\end{equation}
for the $(S,0)$ representation of the Lorentz group ($q_\mu q_\mu =
-m^2$), with the tensor $t$ being defined by [Weinberg,Eqs.(A4-A5)].
Hence,
\begin{equation}
D^{(s)} [\Lambda] \Pi^{(s)} (q) D^{(s)\,\dagger} [\Lambda] = \Pi^{(s)}
(\Lambda q)\label{vd1wein}
\end{equation}
Since at rest we have $[{\bf S}^{(s)}, \Pi^{(s)} (m)] =0$, then according
to the Schur's lemma $\Pi_{\sigma\sigma^\prime}^{\quad (s)} (m) = m^{2s}
\delta_{\sigma \sigma^\prime}$. After the substitution of $D^{(s)}
[\Lambda]$ in Eq.  (\ref{vd1wein}) one has
\begin{equation}
\Pi^{(s)} (q) = m^{2s} \exp (2\Theta \,\hat {\bf q}
\cdot {\bf S}^{(s)})\,.  \end{equation}
One can construct the analogous
matrix for the $(0,S)$ representation by the same procedure:
\begin{equation} \overline{\Pi}^{(s)} (q) = m^{2s} \exp (- 2\Theta
\hat{\bf q}\cdot {\bf S}^{(s)}) \,.  \end{equation}
Finally, by the direct
verification one has in the coordinate representation
\begin{eqnarray} \overline{\Pi}_{\sigma\sigma^\prime} (-i\partial)
\Phi_{\sigma^\prime} =m^{2s} \Xi_\sigma\,,\\ \Pi_{\sigma\sigma^\prime}
(-i\partial) \Xi_{\sigma^\prime} =m^{2s} \Phi_\sigma\,, \end{eqnarray}
provided that $\Phi_\sigma ({\bf 0})$ and $\Xi_\sigma ({\bf 0})$ are
indistinguishable.\footnote{Later, this fact has been incorporated in the
Ryder book~\cite{vd1Ryder}. Truely speaking, this is an additional postulate.
It is possible that the zero-momentum-frame $2(2S+1)$-component objects
(the 4-spinor in the $(1/2,0)\oplus (0,1/2)$ representation, the bivector
in the $(1,0)\oplus (0,1)$ representation, etc.) are connected by an
arbitrary phase factor~\cite{vd1Dv-ff}.}

As a result one has
\begin{equation}
[ \gamma^{\mu_1 \mu_2 \ldots \mu_{2s}} \partial_{\mu_1} \partial_{\mu_2}
\ldots \partial_{\mu_{2s}} +m^{2s} ] \Psi (x) = 0\,,
\end{equation}
with the Barut-Muzinich-Williams covariantly-defined
matrices 
\cite{vd1Bar-Muz,vd1Sankar}. For the spin-1 they are:
\begin{eqnarray}
&&\gamma_{44} =\begin{pmatrix}0&1\\ 1&0\end{pmatrix}\,,\quad
\gamma_{i4}=\gamma_{4i} = \begin{pmatrix}0&iS_i\\
-iS_i & 0\end{pmatrix}\,,\\
&&\gamma^{ij} = \begin{pmatrix}0&\delta_{ij} -S_i S_j - S_j S_i\\
\delta_{ij} -S_i S_j - S_j S_i & 0\end{pmatrix}\, .
\end{eqnarray}
Later Sankaranarayanan and Good considered another version of
this 
theory~\cite{vd1Sankar} (see also~\cite{vd1Ahluwalia2}). For the $S=1$ case they introduced the
Weaver-Hammer-Good sign operator, ref.~\cite{vd1Weaver}, $m^{2} \rightarrow
m^{2}\, (i\partial/\partial t)/E$, which led to the different parity
properties of an antiparticle with respect to a {\it boson} particle.
Next,  Tucker and Hammer {\it et al}~\cite{vd1TuckerHammer} introduced another higher-spin
equations. In the spin-1 case it is:
\begin{equation} 
[\gamma_{\mu\nu}
\partial_\mu \partial_\nu + \partial_\mu \partial_\mu -2m^2 ] \Psi^{(s=1)}
= 0\,  
\end{equation}
(Euclidean metric is now used). In fact, they added the Klein-Gordon
equation to the Weinberg equation. One can add the
Klein-Gordon equation with arbitrary multiple factor to the Weinberg
equation. So, we can study the generalized Weinberg-Tucker-Hammer equation
($S=1)$, which is written ($p_\mu = -i\partial/\partial x^\mu$):
\begin{equation}
[\gamma_{\alpha\beta}p_\alpha p_\beta +A p_\alpha p_\alpha +Bm^2 ]
\Psi =0\,.
\end{equation}
It has solutions with relativistic dispersion relations $E^2 -{\bf
p}^2 = m^2$, ($c=\hbar=1$) provided that
\begin{equation}
{B\over A+1} = 1\,, \qquad \mbox{or} \qquad{B\over A-1} =1\,.\label{vd1se}
\end{equation}
This can be proven by considering the algebraic equation 
$Det [\gamma_{\alpha\beta} p_\alpha p_\beta +A p_\alpha p_\alpha +Bm^2 ]
=0$. It is  of the 12th order in $p_\mu$. Solving it with respect to
energy one obtains the conditions (\ref{vd1se}). {\bf Unlike the Maxwell equations there are
NO any $E=0$ solutions.}

The solutions in the momentum representation have been explicitly presented by~\cite{vd1Ahluwalia2}:
\begin{eqnarray}
u_{+1} ({\bf p})&=&
\begin{pmatrix}m+\left [ (2p_z^2+p_{+} p_{-}) / 2(E+m)\right ]\\
                      {p_z p_{+}/{\sqrt 2}(E+m)}\\
              { p_{+}^2/ 2(E+m) }\\
               p_z\\
                   {p_{+}/{\sqrt 2}}\\
                   0\end{pmatrix}\,,\\
u_{0}({\bf p}) &=& \begin{pmatrix}{p_z p_{-}/{\sqrt 2}(E+m)}\\
                      m+\left [ {p_{+} p_{-}/(E+m) }\right ]\\
                       -{p_z p_{+}/{\sqrt 2}(E+m)}\\
                       {p_{-}/{\sqrt 2}}\\
                          0\\
                        {p_{+}/{\sqrt 2}}\end{pmatrix}\,,\\
u_{-1}({\bf p})&=&\begin{pmatrix} { p_{-}^2/ 2(E+m) }\\
                             -{p_z p_{-}/{\sqrt 2}(E+m)}\\
                   m+\left [ {(2p_z^2+p_{+} p_{-})/ 2(E+m)}\right]\\
                      0\\
                      {p_{-}/{\sqrt 2}}\\
                   -p_z\end{pmatrix}\,,\label{vd1uv}
\end{eqnarray}
and
\begin{equation}
v_\sigma ({\bf p}) =\gamma_5 u_\sigma ({\bf p}) =\begin{pmatrix}0&1\\
1&0\end{pmatrix} U_\sigma ({\bf p})
\end{equation}
in the standard representation of $\gamma_{\mu\nu}$ matrices.
If the 6-component $v ({\bf p})$ are defined in such way, we inevitably would get 
the additional energy-sign operator~\cite{vd1Weaver,vd1Sankar} $\epsilon = i\partial_t /E =\pm 1$ 
in the dynamical equation,
and the different parities of the corresponding boson and antiboson, $\hat P u_\sigma ({\bf p})
= + u_\sigma ({\bf p})$ and $\hat P v_\sigma ({\bf p}) = - v_\sigma ({\bf p})$.

\section{The Construction of Field Operators}

The method for constructions of field operators has been given in~\cite{vd1Bogoliubov}:\footnote{In this book a bit different  notation for positive- (negative-) energy solutions has been used
comparing with the general accepted one.}
\begin{equation}
\phi (x) = \frac{1}{(2\pi)^{3/2}}\int dk  e^{ikx}\tilde \phi (k)\,.
\end{equation}
From the Klein-Gordon equation we know:
\begin{equation}
(k^2 -m^2) \tilde \phi (k) =0\,.\label{vd1KG1}
\end{equation}
Thus,
\begin{equation}
\tilde \phi (k) = \delta (k^2 - m^2) \phi (k)\,.
\end{equation}
Next,
\begin{eqnarray}
&&\phi (x) = \frac{1}{(2\pi)^{3/2}} \int d k \, e^{ikx}\delta (k^2 -m^2) (\theta (k_0)+ 
\theta (-k_0)) 
\phi (k) =\nonumber\\
&=& \frac{1}{(2\pi)^{3/2}} \int d k \left [ e^{ikx} \delta (k^2 -m^2) \phi^+ (k)
+ e^{-ikx} \delta (k^2 -m^2) \phi^- (k)\right ],\nonumber\\
\end{eqnarray}
where
\begin{equation}
\phi^+ (k) =\theta (k_0) \phi (k)\,,\mbox{and}\,\, \phi^- (k)=\theta (k_0) \phi (-k)\,.
\end{equation}
\begin{eqnarray}
\phi^+ (x) &=& \frac{1}{(2\pi)^{3/2}}\int \frac{d^3 {\bf k}}{2E_k}  e^{+ikx} \phi^+ (k)\,,\\
\phi^- (x) &=& \frac{1}{(2\pi)^{3/2}}\int \frac{d^3 {\bf k}}{2E_k}  e^{-ikx} \phi^- (k)\,.
\end{eqnarray}

In the spinor case (the $(1/2,0)\oplus (0,1/2)$ representation space) we have more components.
Instead of the equation (\ref{vd1KG1}) we have 
\begin{equation}
(\hat k + m) \psi (k) \vert_{k^2 =m^2} =0\,.
\end{equation}
However, again
\begin{equation}
\psi (x) = \frac{1}{(2\pi)^{3/2}} \int d k \, e^{ikx}\delta (k^2 -m^2) (\theta (k_0)+ 
\theta (-k_0)) 
\psi (k)\,,
\end{equation}
and
\begin{equation}
\psi (x)={1\over (2\pi)^3} \int \frac{d^3 {\bf k}}{2E_k} \left [ e^{ikx}\theta(k_0)  
\psi (k)   + e^{-ikx} \theta (k_0) \psi (-k) \right ]\,,
\end{equation}
where $k_0 =E =\sqrt{{\bf k}^2 +m^2}$ is positive in this case. Hence:
\begin{equation}
(\hat k +m)\psi^+ ({\bf k}) =0\,,\quad
(-\hat k + m)\psi^- ({\bf k}) =0\,.
\end{equation}

{\bf Everything is OK? However, please note that the momentum-space Dirac equations
$(\hat k - m) u =0$, $(\hat k +m) v=0$
have solutions $k_0=\pm \sqrt{{\bf k}^2 +m^2}$, both for $u-$ and $v-$ spinors.
This can be checked by calculating the determinants. Usually, one chooses 
$k_0=E=\sqrt{{\bf k}^2 +m^2}$ in the $u-$ and in the $v-$. This is because 
on the classical level (better to say, on the first quantization level) 
the negative-energy $u-$ can be transformed in the positive-energy $v-$, 
and vice versa. This is not precisely so, if we go to the secondary quantization level.
The introduction of creation/annihilation noncommutating operators  gives us more possibilities
in constructing generalized theory even on the basis of the Dirac equation.}

Various-type field operators are possible in the $(1/2,1/2)$ representation. 
During the calculations below we have to present $1=\theta (k_0) +\theta (-k_0)$
( as previously ) in order to get positive- and negative-frequency parts. 
\begin{eqnarray}
&&A_\mu (x) = {1\over (2\pi)^3} \int d^4 k \,\delta (k^2 -m^2) e^{+ik\cdot x}
A_\mu (k) =\nonumber\\
&=& {1\over (2\pi)^3} \sum_{\lambda}^{}\int d^4 k \delta (k_0^2 -E_k^2) e^{+ik\cdot x}
\epsilon_\mu (k,\lambda) a_\lambda (k) =\nonumber\\
&=&{1\over (2\pi)^3} \int {d^4 k \over 2E} [\delta (k_0 -E_k) +\delta (k_0 +E_k) ] 
[\theta (k_0) +\theta (-k_0) ]\nonumber\\
&&e^{+ik\cdot x}
A_\mu (k) = {1\over (2\pi)^3} \int {d^4 k \over 2E} [\delta (k_0 -E_k) +\delta (k_0 +E_k) ] \nonumber\\
&&\left
[\theta (k_0) A_\mu (k) e^{+ik\cdot x}  + 
\theta (k_0) A_\mu (-k) e^{-ik\cdot x} \right ]  =\\
&=&{1\over (2\pi)^3} \int {d^3 {\bf k} \over 2E_k} \theta(k_0)  
[A_\mu (k) e^{+ik\cdot x}  + A_\mu (-k) e^{-ik\cdot x} ]
=\nonumber\\
&=&{1\over (2\pi)^3} \sum_{\lambda}^{}\int {d^3 {\bf k} \over 2E_k}   
[\epsilon_\mu (k,\lambda) a_\lambda (k) e^{+ik\cdot x}  + \epsilon_\mu (-k,\lambda) 
a_\lambda (-k) e^{-ik\cdot x} ].\nonumber
\end{eqnarray}

{\bf In general, due to theorems for integrals and for distributions the presentation $1=\theta (k_0)
+ \theta (-k_0)$ is possible because we use this in the integrand. However, remember, that
we have the $k_0=E=0$ solution of the Maxwell equations.\footnote{Of course, the same procedure can be applied in the construction of the quantum field operator for $F_{\mu\nu}$.} Moreover, it has the experimental confirmation (for instance, the stationary electromagnetic field $curl {\bf B}=0$). Meanwhile the $\theta$ function is NOT defined in $k_0=0$. Do we not loose this solution in the above construction
of the quantum field operator? Mathematicians did not answer me in a straightforward way.}

Moreover, we should transform the second part to $\epsilon_\mu^\ast (k,\lambda) b_\lambda^\dagger (k)$ as usual. In such a way we obtain the charge-conjugate states.\footnote{In the cirtain basis 
it is considered that the charge conjugation operator is just the complex conjugation operator for 4-vectors $A_\mu$.} Of course, one can try to get $P$-conjugates or $CP$-conjugate states too. 

In the Dirac case we should assume the following relation in the field operator:
\begin{equation}
\sum_{\lambda}^{} v_\lambda (k) b_\lambda^\dagger (k) = \sum_{\lambda}^{} u_\lambda (-k) a_\lambda (-k)\,.\label{vd1dcop}
\end{equation}
We know that~\cite{vd1Ryder,vd1Itzykson}
\begin{eqnarray}
\bar u_\mu (k) u_\lambda (k) &=& +m \delta_{\mu\lambda}\,,\\
\bar u_\mu (k) u_\lambda (-k) &=& 0\,,\\
\bar v_\mu (k) v_\lambda (k) &=& -m \delta_{\mu\lambda}\,,\\
\bar v_\mu (k) u_\lambda (k) &=& 0\,,
\end{eqnarray}
but we need $\Lambda_{\mu\lambda} (k) = \bar v_\mu (k) u_\lambda (-k)$.
By direct calculations,  we find
\begin{equation}
-mb_\mu^\dagger (k) = \sum_{\nu}^{} \Lambda_{\mu\lambda} (k) a_\lambda (-k)\,.
\end{equation}
Hence, $\Lambda_{\mu\lambda} = -im ({\bf \sigma}\cdot {\bf n})_{\mu\lambda}$
and 
\begin{equation}
b_\mu^\dagger (k) = i({\bf\sigma}\cdot {\bf n})_{\mu\lambda} a_\lambda (-k)\,.
\end{equation}
Multiplying (\ref{vd1dcop}) by $\bar u_\mu (-k)$ we obtain
\begin{equation}
a_\mu (-k) = -i ({\bf \sigma} \cdot {\bf n})_{\mu\lambda} b_\lambda^\dagger (k)\,.
\end{equation}
Thus, the  above equations  are self-consistent.

In the $(1,0)\oplus (0,1)$ representation we have somewhat different situation. Namely,
\begin{equation}
a_\mu (k) = [1-2({\bf S}\cdot {\bf n})^2]_{\mu\lambda} a_\lambda (-k)\,. 
\end{equation}
This signifies that in order to construct the Sankaranarayanan-Good field operator (which was used by Ahluwalia, Johnson and Goldman~\cite{vd1Ahluwalia2}, it satisfies 
$[\gamma_{\mu\nu} \partial_\mu \partial_\nu - {(i\partial/\partial t)\over E} 
m^2 ] \Psi =0$, we need additional postulates.

We can set for the 4-vector field operator:
\begin{equation}
\sum_{\lambda}^{} \epsilon_\mu (-k,\lambda) a_\lambda (-k) = 
\sum_{\lambda}^{} \epsilon_\mu^\ast (k,\lambda) b_\lambda^\dagger (k)\,,
\label{vd1expan}
\end{equation}
multiply both parts by $\epsilon_\nu [\gamma_{44}]_{\nu\mu}$, and use the normalization conditions for polarization vectors.

However, in the $({1\over 2}, {1\over 2})$ representation we can also expand
(apart the equation (\ref{vd1expan})) in the different way:
\begin{equation}
\sum_{\lambda}^{} \epsilon_\mu (-k, \lambda) a_\lambda (-k) =
\sum_{\lambda}^{} \epsilon_\mu (k, \lambda) a_\lambda (k)\,.
\end{equation}
From the first definition we obtain (the signs $\mp$
depends on the value of $\sigma$):
\begin{equation}
b_\sigma^\dagger (k) = \mp \sum_{\mu\nu\lambda}^{} \epsilon_\nu (k,\sigma) 
[\gamma_{44}]_{\nu\mu} \epsilon_\mu (-k,\lambda) a_\lambda (-k)\,,
\end{equation}
or
\begin{eqnarray}
&&b_\sigma^\dagger (k) = \\
&&{E_k^2 \over m^2} \begin{pmatrix}1+{{\bf k}^2\over E_k^2}&\sqrt{2}
{k_r \over E_k}&-\sqrt{2} {k_l \over E_k}& -{2k_3 \over E_k}\\
-\sqrt{2} {k_r \over E_k}&-{k_r^2 \over {\bf k}^2}& -{m^2k_3^2\over E_k^2 {\bf k}^2}
+{k_r k_l \over E_k^2} & {\sqrt{2} k_3 k_r \over {\bf k}^2}\\
\sqrt{2} {k_l \over E_k}&-{m^2 k_3^2 \over E_k^2 {\bf k}^2} + {k_r k_l \over E_k^2}& -{k_l^2\over {\bf k}^2} & -{\sqrt{2} k_3 k_l \over {\bf k}^2}\\
{2k_3 \over E_k}&{\sqrt{2}k_3 k_r \over {\bf k}^2}& -{\sqrt{2} k_3 k_l\over {\bf k}^2} & {m^2 \over E_k^2} -{2 k_3 \over {\bf k}^2}\end{pmatrix}
\begin{pmatrix}a_{00} (-k)\\ a_{11} (-k)\\
a_{1-1} (-k)\\ a_{10} (-k)\end{pmatrix}\,.\nonumber
\end{eqnarray}

From the second definition $\Lambda^2_{\sigma\lambda} = \mp \sum_{\nu\mu}^{} \epsilon^{\ast}_\nu (k, \sigma) [\gamma_{44}]_{\nu\mu}
\epsilon_\mu (-k, \lambda)$ we have:
\begin{eqnarray}
a_\sigma (k) =  \begin{pmatrix}-1&0&0&0\\
0&{k_3^2 \over {\bf k}^2}& {k_l^2\over {\bf k}^2} & {\sqrt{2} k_3 k_l \over {\bf k}^2}\\
0&{k_r^2 \over {\bf k}^2}& {k_3^2\over {\bf k}^2} & -{\sqrt{2} k_3 k_r \over {\bf k}^2}\\
0&{\sqrt{2}k_3 k_r \over {\bf k}^2}& -{\sqrt{2} k_3 k_l\over {\bf
    k}^2} & 1-{2 k_3^2 \over {\bf k}^2}\end{pmatrix}\begin{pmatrix}a_{00} (-k)\\ a_{11} (-k)\\
a_{1-1} (-k)\\ a_{10} (-k)\end{pmatrix}.
\end{eqnarray}
It is the strange case: the field operator will only destroy particles (like in the $(1,0)\oplus (0,1)$ case). Possibly, we should think about modifications of the Fock space in this case, or introduce several field operators for the $({1\over 2}, {1\over 2})$ representation.

{\bf However,  other way is possible: to construct the left- and right- parts of the $(1,0)\oplus (0,1)$ field operator 
separately each other. In this case the commutation relations may be more complicated.}

Finally, going back to the rest $(S,0)\oplus (0,S)$ objects. {\bf Bogoliubov
constructs them introducing the products with $delta$ functions like $\delta (k_0 -m)$. 
Then, he makes the boost of the "spinors" only, and changes by hand the $\delta$ to
$\delta (k^2 -m^2)$ (where we already have $k_0 =E=\sqrt{{\bf k}^2 +m^2}$). Mathematicians did not answer me, how can it be possible to make the boost
of the $\delta$ functions consistently in such a way.}


The conclusion is: we still have few questions unsolved in the bases of the quantum field theory, which open a room for generalized theories.

\section*{Acknowledgements}
I am grateful to Prof. Z. Oziewicz (organizer) and all participants of the VIII International Workshop
"Graph-Operads-Logics-Category Theory". I would like to mention
Profs. G. Quznetsov and R. Santilli for useful information.

\title{The Bargmann-Wigner Formalism for Spin 2 Fields\thanks{Accepted for the presentation at 
the MG12, Paris, July 2009.}}
\author{V.V. Dvoeglazov}
\institute{%
Universidad de Zacatecas, A. P. 636, Suc. 3 Cruces\\
Zacatecas, Zac. 98062, M\'exico\\
E-mail: valeri@fisica.uaz.edu.mx\\
http://fisica.uaz.edu.mx/\~{}valeri}

\titlerunning{The Bargmann-Wigner Formalism for Spin 2 Fields}
\authorrunning{V.V. Dvoeglazov}
\maketitle

\begin{abstract}
We proceed to derive equations for the symmetric tensor of the second rank on the basis of the Bargmann-Wigner formalism in a straightforward way. The symmetric multispinor of the fourth rank is used. It is constructed out of the Dirac 4-spinors. Due to serious problems with the interpretation of the results obtained on using the standard procedure we generalize it and obtain the spin-2 relativistic equations, which are consistent with those given before. The importance of the 4-vector field (and its gauge part) is pointed out. 
\end{abstract}


The spin-2 case can be of some
interest because it is generally believed that the essential features of
gravitons are  obtained from transverse components of the
$(2,0)\oplus (0,2)$  representation of the Lorentz group. Nevertheless,
the question of the redandant components of the higher-spin relativistic
equations has not yet been understood in detail.
We use the procedure
for the derivation  of higher-spin equations:~\cite{vd2bw-hs,vd2Lurie,vd2dvo-wig}
\begin{eqnarray}
&&\left [ i\gamma^\mu \partial_\mu - m \right ]_{\alpha\alpha^\prime}
\Psi_{\alpha^\prime \beta\gamma\delta} = 0\, ,
\left [ i\gamma^\mu \partial_\mu - m \right ]_{\beta\beta^\prime}
\Psi_{\alpha\beta^\prime \gamma\delta} = 0\, ,\\
&&\left [ i\gamma^\mu \partial_\mu - m \right ]_{\gamma\gamma^\prime}
\Psi_{\alpha\beta\gamma^\prime \delta} = 0\, ,
\left [ i\gamma^\mu \partial_\mu - m \right ]_{\delta\delta^\prime}
\Psi_{\alpha\beta\gamma\delta^\prime} = 0\, .
\end{eqnarray} 
The massless limit (if one needs) should be taken in the end of all
calculations.
We proceed expanding the field function in the set of symmetric matrices
(as in the spin-1 case). In the beginning let us use the
first two indices:
$\Psi_{\{\alpha\beta\}\gamma\delta} =
(\gamma_\mu R)_{\alpha\beta} \Psi^\mu_{\gamma\delta}
+(\sigma_{\mu\nu} R)_{\alpha\beta} \Psi^{\mu\nu}_{\gamma\delta}$.
We would like to write
the corresponding equations for functions $\Psi^\mu_{\gamma\delta}$
and $\Psi^{\mu\nu}_{\gamma\delta}$ in the form:
\begin{equation}
{2\over m} \partial_\mu \Psi^{\mu\nu}_{\gamma\delta} = -
\Psi^\nu_{\gamma\delta}\, , 
\Psi^{\mu\nu}_{\gamma\delta} = {1\over 2m}
\left [ \partial^\mu \Psi^\nu_{\gamma\delta} - \partial^\nu
\Psi^\mu_{\gamma\delta} \right ]\, \label{vd2p2}.
\end{equation} 
The constraints are $(1/m) \partial_\mu \Psi^\mu_{\gamma\delta} =0$
and $(1/m) \epsilon^{\mu\nu}_{\quad\alpha\beta}\, \partial_\mu
\Psi^{\alpha\beta}_{\gamma\delta} = 0$.
Next, we present the vector-spinor and tensor-spinor functions as
\begin{equation}
\Psi^\mu_{\{\gamma\delta\}} = (\gamma^\kappa R)_{\gamma\delta}
G_{\kappa}^{\quad \mu} +(\sigma^{\kappa\tau} R )_{\gamma\delta}
F_{\kappa\tau}^{\quad \mu} \, ,
\Psi^{\mu\nu}_{\{\gamma\delta\}} = (\gamma^\kappa R)_{\gamma\delta}
T_{\kappa}^{\quad \mu\nu} +(\sigma^{\kappa\tau} R )_{\gamma\delta}
R_{\kappa\tau}^{\quad \mu\nu},
\end{equation}
i.~e.,  using the symmetric matrix coefficients in indices $\gamma$ and
$\delta$. Hence, the total function is
\begin{eqnarray}
\Psi_{\{\alpha\beta\}\{\gamma\delta\}}
&=& (\gamma_\mu R)_{\alpha\beta} (\gamma^\kappa R)_{\gamma\delta}
G_\kappa^{\quad \mu} + (\gamma_\mu R)_{\alpha\beta} (\sigma^{\kappa\tau}
R)_{\gamma\delta} F_{\kappa\tau}^{\quad \mu} \nonumber\\ 
 & & {} + (\sigma_{\mu\nu} R)_{\alpha\beta} (\gamma^\kappa R)_{\gamma\delta}
T_\kappa^{\quad \mu\nu} + (\sigma_{\mu\nu} R)_{\alpha\beta}
(\sigma^{\kappa\tau} R)_{\gamma\delta} R_{\kappa\tau}^{\quad\mu\nu};
\end{eqnarray}
and the resulting tensor equations are:
\begin{eqnarray}
&&{2\over m} \partial_\mu T_\kappa^{\quad \mu\nu} =
-G_{\kappa}^{\quad\nu}\, ,
{2\over m} \partial_\mu R_{\kappa\tau}^{\quad \mu\nu} =
-F_{\kappa\tau}^{\quad\nu}\, ,\\
&& T_{\kappa}^{\quad \mu\nu} = {1\over 2m} \left [
\partial^\mu G_{\kappa}^{\quad\nu}
- \partial^\nu G_{\kappa}^{\quad \mu} \right ] \, ,
R_{\kappa\tau}^{\quad \mu\nu} = {1\over 2m} \left [
\partial^\mu F_{\kappa\tau}^{\quad\nu}
- \partial^\nu F_{\kappa\tau}^{\quad \mu} \right ].
\end{eqnarray}
The constraints are re-written to
\begin{eqnarray}
&&{1\over m} \partial_\mu G_\kappa^{\quad\mu} = 0\, ,\quad
{1\over m} \partial_\mu F_{\kappa\tau}^{\quad\mu} =0\, ,\\
&& {1\over m} \epsilon_{\alpha\beta\nu\mu} \partial^\alpha
T_\kappa^{\quad\beta\nu} = 0\, ,\quad
{1\over m} \epsilon_{\alpha\beta\nu\mu} \partial^\alpha
R_{\kappa\tau}^{\quad\beta\nu} = 0\, .
\end{eqnarray}
However, we need to make symmetrization over these two sets
of indices $\{ \alpha\beta \}$ and $\{\gamma\delta \}$. The total
symmetry can be ensured if one contracts the function $\Psi_{\{\alpha\beta
\} \{\gamma \delta \}}$ with {\it antisymmetric} matrices
$R^{-1}_{\beta\gamma}$, $(R^{-1} \gamma^5 )_{\beta\gamma}$ and
$(R^{-1} \gamma^5 \gamma^\lambda )_{\beta\gamma}$ and equate
all these contractions to zero (similar to the $j=3/2$ case
considered in Ref.~\cite[p. 44]{vd2Lurie}). We obtain
additional constraints on the tensor field functions:
\begin{eqnarray}
&& G_\mu^{\quad\mu}=0\, , \quad G_{[\kappa \, \mu ]}  = 0\, , \quad
G^{\kappa\mu} = {1\over 2} g^{\kappa\mu} G_\nu^{\quad\nu}\, ,
\label{vd2b1}\\
&&F_{\kappa\mu}^{\quad\mu} = F_{\mu\kappa}^{\quad\mu} = 0\, , \quad
\epsilon^{\kappa\tau\mu\nu} F_{\kappa\tau,\mu} = 0\, ,\\
&& T^{\mu}_{\quad\mu\kappa} =
T^{\mu}_{\quad\kappa\mu} = 0\, ,\quad
\epsilon^{\kappa\tau\mu\nu} T_{\kappa,\tau\mu} = 0\, ,\\
&& F^{\kappa\tau,\mu} = T^{\mu,\kappa\tau}\, ,\quad
\epsilon^{\kappa\tau\mu\lambda} (F_{\kappa\tau,\mu} +
T_{\kappa,\tau\mu})=0\, ,\\
&& R_{\kappa\nu}^{\quad \mu\nu}
= R_{\nu\kappa}^{\quad  \mu\nu} = R_{\kappa\nu}^{\quad\nu\mu}
= R_{\nu\kappa}^{\quad\nu\mu}
= R_{\mu\nu}^{\quad  \mu\nu} = 0\, , \\
&& \epsilon^{\mu\nu\alpha\beta} (g_{\beta\kappa} R_{\mu\tau,
\nu\alpha} - g_{\beta\tau} R_{\nu\alpha,\mu\kappa} ) = 0\, \quad
\epsilon^{\kappa\tau\mu\nu} R_{\kappa\tau,\mu\nu} = 0\, .\label{vd2f1}
\end{eqnarray} 
Thus, we  encountered with
the known difficulty of the theory for spin-2 particles in
the Minkowski space.
We explicitly showed that all field functions become to be equal to zero.
Such a situation cannot be considered as a satisfactory one (because it
does not give us any physical information) and can be corrected in several
ways.

We have to modify the formalism~\cite{vd2dv-ps}. The field function is now presented as
\begin{equation}
\Psi_{\{\alpha\beta\}\gamma\delta} =
\alpha_1 (\gamma_\mu R)_{\alpha\beta} \Psi^\mu_{\gamma\delta} +
\alpha_2 (\sigma_{\mu\nu} R)_{\alpha\beta} \Psi^{\mu\nu}_{\gamma\delta}
+\alpha_3 (\gamma^5 \sigma_{\mu\nu} R)_{\alpha\beta}
\widetilde \Psi^{\mu\nu}_{\gamma\delta}\, ,
\end{equation}
with
\begin{eqnarray}
\Psi^\mu_{\{\gamma\delta\}} &=& \beta_1 (\gamma^\kappa R)_{\gamma\delta}
G_\kappa^{\quad\mu} + \beta_2 (\sigma^{\kappa\tau} R)_{\gamma\delta}
F_{\kappa\tau}^{\quad\mu} +\beta_3 (\gamma^5 \sigma^{\kappa\tau}
R)_{\gamma\delta} \widetilde F_{\kappa\tau}^{\quad\mu} \, ,\\
\Psi^{\mu\nu}_{\{\gamma\delta\}} &=&\beta_4 (\gamma^\kappa
R)_{\gamma\delta} T_\kappa^{\quad\mu\nu} + \beta_5 (\sigma^{\kappa\tau}
R)_{\gamma\delta} R_{\kappa\tau}^{\quad\mu\nu} +\beta_6 (\gamma^5
\sigma^{\kappa\tau} R)_{\gamma\delta}
\widetilde R_{\kappa\tau}^{\quad\mu\nu}\\
\widetilde \Psi^{\mu\nu}_{\{\gamma\delta\}} &=&\beta_7 (\gamma^\kappa
R)_{\gamma\delta} \widetilde T_\kappa^{\quad\mu\nu} + \beta_8
(\sigma^{\kappa\tau} R)_{\gamma\delta}
\widetilde D_{\kappa\tau}^{\quad\mu\nu}
+\beta_9 (\gamma^5 \sigma^{\kappa\tau} R)_{\gamma\delta}
D_{\kappa\tau}^{\quad\mu\nu}
\end{eqnarray}
Hence, the function $\Psi_{\{\alpha\beta\}\{\gamma\delta\}}$
can be expressed as a sum of nine terms:
\begin{eqnarray}
&& \Psi_{\{\alpha\beta\}\{\gamma\delta\}} =  \nonumber\\
&& \alpha_1 \beta_1 (\gamma_\mu R)_{\alpha\beta} (\gamma^\kappa
R)_{\gamma\delta} G_\kappa^{\quad\mu} +\alpha_1 \beta_2
(\gamma_\mu R)_{\alpha\beta} (\sigma^{\kappa\tau} R)_{\gamma\delta}
F_{\kappa\tau}^{\quad\mu} + \nonumber\\
&+&\alpha_1 \beta_3 (\gamma_\mu R)_{\alpha\beta}
(\gamma^5 \sigma^{\kappa\tau} R)_{\gamma\delta} \widetilde
F_{\kappa\tau}^{\quad\mu} 
+ \alpha_2 \beta_4 (\sigma_{\mu\nu}
R)_{\alpha\beta} (\gamma^\kappa R)_{\gamma\delta} T_\kappa^{\quad\mu\nu}
+\nonumber\\
&+&\alpha_2 \beta_5 (\sigma_{\mu\nu} R)_{\alpha\beta} (\sigma^{\kappa\tau}
R)_{\gamma\delta} R_{\kappa\tau}^{\quad \mu\nu}
+ \alpha_2
\beta_6 (\sigma_{\mu\nu} R)_{\alpha\beta} (\gamma^5 \sigma^{\kappa\tau}
R)_{\gamma\delta} \widetilde R_{\kappa\tau}^{\quad\mu\nu} +\nonumber\\
&+&\alpha_3 \beta_7 (\gamma^5 \sigma_{\mu\nu} R)_{\alpha\beta}
(\gamma^\kappa R)_{\gamma\delta} \widetilde
T_\kappa^{\quad\mu\nu}+
\alpha_3 \beta_8 (\gamma^5
\sigma_{\mu\nu} R)_{\alpha\beta} (\sigma^{\kappa\tau} R)_{\gamma\delta}
\widetilde D_{\kappa\tau}^{\quad\mu\nu} +\nonumber\\
&+&\alpha_3 \beta_9
(\gamma^5 \sigma_{\mu\nu} R)_{\alpha\beta} (\gamma^5 \sigma^{\kappa\tau}
R)_{\gamma\delta} D_{\kappa\tau}^{\quad \mu\nu}\, .
\label{vd2ffn1}
\end{eqnarray}
The corresponding dynamical
equations are given by the set
\begin{eqnarray}
&& {2\alpha_2
\beta_4 \over m} \partial_\nu T_\kappa^{\quad\mu\nu} +{i\alpha_3
\beta_7 \over m} \epsilon^{\mu\nu\alpha\beta} \partial_\nu
\widetilde T_{\kappa,\alpha\beta} = \alpha_1 \beta_1
G_\kappa^{\quad\mu}\,; \label{vd2b}\\
&& \  \nonumber\\
&&{2\alpha_2 \beta_5 \over m} \partial_\nu
R_{\kappa\tau}^{\quad\mu\nu} +{i\alpha_2 \beta_6 \over m}
\epsilon_{\alpha\beta\kappa\tau} \partial_\nu \widetilde R^{\alpha\beta,
\mu\nu} +{i\alpha_3 \beta_8 \over m}
\epsilon^{\mu\nu\alpha\beta}\partial_\nu \widetilde
D_{\kappa\tau,\alpha\beta} - \nonumber\\
&-&{\alpha_3 \beta_9 \over 2}
\epsilon^{\mu\nu\alpha\beta} \epsilon_{\lambda\delta\kappa\tau}
D^{\lambda\delta}_{\quad \alpha\beta} = \alpha_1 \beta_2
F_{\kappa\tau}^{\quad\mu} + {i\alpha_1 \beta_3 \over 2}
\epsilon_{\alpha\beta\kappa\tau} \widetilde F^{\alpha\beta,\mu}\,; \\
&& \  \nonumber\\
&& 2\alpha_2 \beta_4 T_\kappa^{\quad\mu\nu} +i\alpha_3 \beta_7
\epsilon^{\alpha\beta\mu\nu} \widetilde T_{\kappa,\alpha\beta}
=  {\alpha_1 \beta_1 \over m} (\partial^\mu G_\kappa^{\quad \nu}
- \partial^\nu G_\kappa^{\quad\mu})\,; \\
&& \  \nonumber\\
&& 2\alpha_2 \beta_5 R_{\kappa\tau}^{\quad\mu\nu} +i\alpha_3 \beta_8
\epsilon^{\alpha\beta\mu\nu} \widetilde D_{\kappa\tau,\alpha\beta}
+i\alpha_2 \beta_6 \epsilon_{\alpha\beta\kappa\tau} \widetilde
R^{\alpha\beta,\mu\nu}-\nonumber\\
&&- {\alpha_3 \beta_9\over 2} \epsilon^{\alpha\beta\mu\nu}
\epsilon_{\lambda\delta\kappa\tau} D^{\lambda\delta}_{\quad \alpha\beta}
= {\alpha_1 \beta_2 \over m} (\partial^\mu F_{\kappa\tau}^{\quad \nu}
-\partial^\nu F_{\kappa\tau}^{\quad\mu} ) + \nonumber\\
&&+{i\alpha_1 \beta_3 \over 2m}
\epsilon_{\alpha\beta\kappa\tau} (\partial^\mu \widetilde
F^{\alpha\beta,\nu} - \partial^\nu \widetilde F^{\alpha\beta,\mu} )\, .
\label{vd2f}
\end{eqnarray}
The essential constraints have been given in Ref.~\cite{vd2Dvoeglazov-aaca}.
They are  the results of contractions of the field function (\ref{vd2ffn1})
with three antisymmetric matrices, as above. Furthermore,
one should recover the relations (\ref{vd2b1}-\ref{vd2f1}) in the particular
case when $\alpha_3 = \beta_3 =\beta_6 = \beta_9 = 0$ and
$\alpha_1 = \alpha_2 = \beta_1 =\beta_2 =\beta_4
=\beta_5 = \beta_7 =\beta_8 =1$.
As a discussion we note that in such a framework we already have physical
content because only certain combinations of field functions
would be equal to zero. In general, the fields
$F_{\kappa\tau}^{\quad\mu}$, $\widetilde F_{\kappa\tau}^{\quad\mu}$,
$T_{\kappa}^{\quad\mu\nu}$, $\widetilde T_{\kappa}^{\quad\mu\nu}$, and
$R_{\kappa\tau}^{\quad\mu\nu}$,  $\widetilde
R_{\kappa\tau}^{\quad\mu\nu}$, $D_{\kappa\tau}^{\quad\mu\nu}$, $\widetilde
D_{\kappa\tau}^{\quad\mu\nu}$ can  correspond to different physical states
and the equations above describe oscillations one state to another.
Furthermore, from the set of equations (\ref{vd2b}-\ref{vd2f}) one
obtains the {\it second}-order equation for symmetric traceless tensor of
the second rank ($\alpha_1 \neq 0$, $\beta_1 \neq 0$):
${1\over m^2} \left [\partial_\nu
\partial^\mu G_\kappa^{\quad \nu} - \partial_\nu \partial^\nu
G_\kappa^{\quad\mu} \right ] =  G_\kappa^{\quad \mu}$ .
After the contraction in indices $\kappa$ and $\mu$ this equation is
reduced to the set
$\partial_\mu G_{\quad\kappa}^{\mu} = F_\kappa,
{1\over m^2} \partial_\kappa F^\kappa = 0$ ,
i.~e.,  to the equations connecting the analogue of the energy-momentum
tensor and the analogue of the 4-vector potential. 
Further investigations may provide additional foundations to
``surprising" similarities of gravitational and electromagnetic
equations in the low-velocity limit.

\title{New Light on Dark Matter from the LHC}
\author{J. Ellis}
\institute{%
Theory Division, CERN, CH--1211 Gen\`eve 23, Switzerland; \\ 
Theoretical Physics and Cosmology Group, Department of Physics, King's College London, London WC2R 2LS, UK\\
~~~~\\
{\tt CERN-PH-TH/2010-258~~~~~~~~~~~~~~~~~~~~KCL-PH-TH/2010-31}\\
\resizebox{1.5cm}{!}{\includegraphics{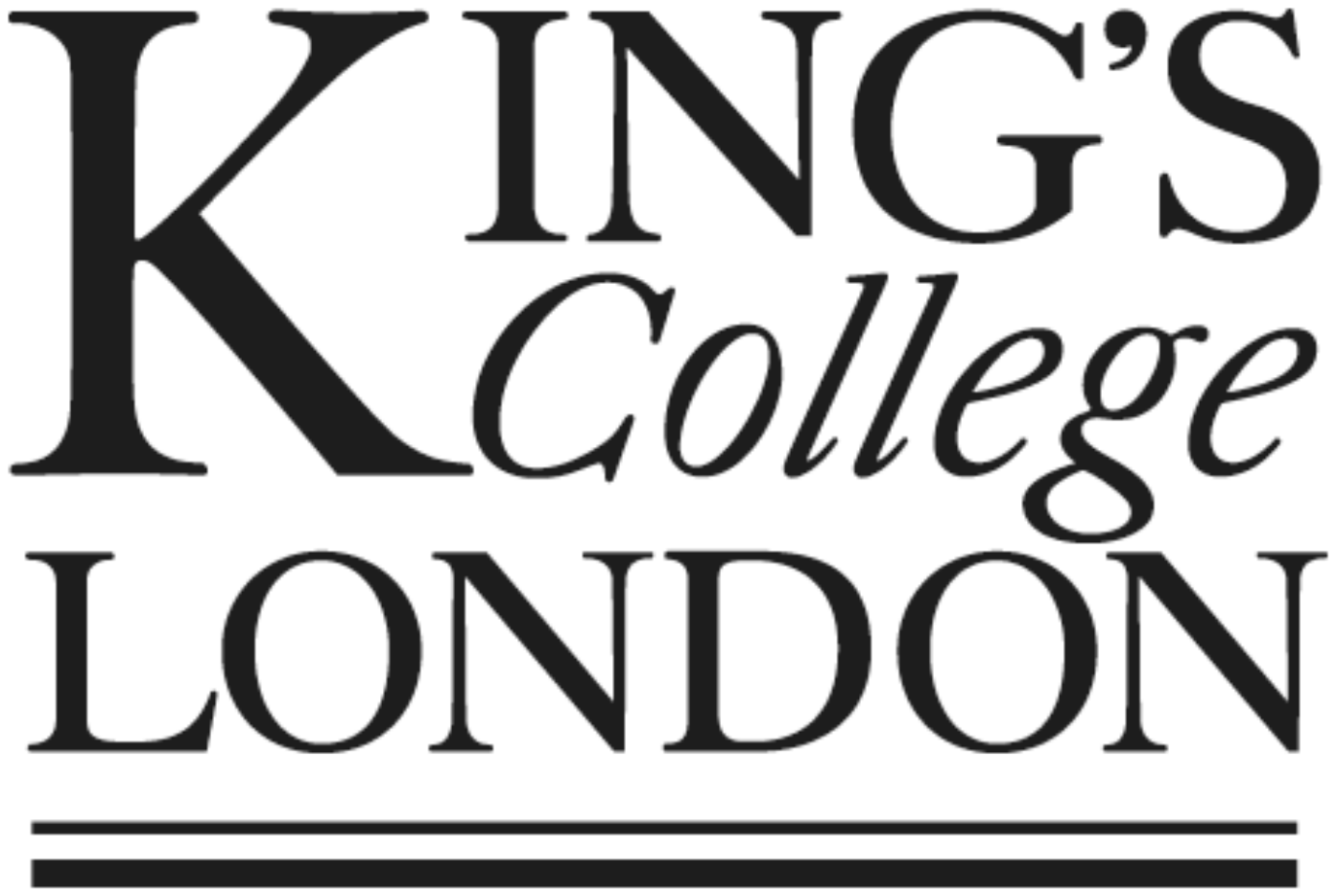}}~~~~~
}

\authorrunning{J. Ellis}
\titlerunning{New Light on Dark Matter from the LHC}
\maketitle

\begin{abstract}
The prospects for detecting a candidate
supersymmetric dark matter particle at the LHC are reviewed, and compared
with the prospects for direct and indirect searches for astrophysical dark matter,
on the basis of a frequentist analysis of the preferred regions of the Minimal
supersymmetric extension of the Standard Model with universal soft
supersymmetry breaking (the CMSSM) and a model with equal but non-universal
supersymmetry-breaking contributions to the Higgs masses (the NUHM1).
LHC searches may have good chances to observe supersymmetry in the near future
- and so may direct searches for astrophysical dark matter particles.
\end{abstract}

\section{Introduction}\label{JEs:intro}

There is a standard list of open questions beyond the Standard Model
of particle physics~\cite{JECOlex}, which includes the following.
(1) What is the origin of particle masses and, in particular, are they due to a Higgs boson?
(2) Why are there so many different types of standard matter particles, notably three neutrino species?
(3) What is the dark matter in the Universe?
(4) How can we unify the fundamental forces?
(5) Last but certainly not least, how may we construct a quantum theory of gravity?
Each of these questions will be addressed, in some way, by experiments at the LHC,
though answers to all of them are not guaranteed!

The central topic of this talk is, of course, question (3) concerning dark matter.
Certainly there are many candidate particles, ranging in mass from axions to
Wimpzillas. However, many candidates fall within the general category of WIMPs 
(weakly-interacting massive particles) weighing between $\sim 100$ and
$\sim 1000$~GeV and hence possibly accessible to the LHC. These include
the lightest Kaluza-Klein particle (LKP)
in some scenarios with extra dimensions~\cite{JELKP}, the lightest T-odd particle (LTP)
in some little Higgs scenarios~\cite{JELTP}, and the lightest supersymmetric particle (LSP)
in supersymmetric models in which R-parity is conserved~\cite{JELSP}.

Historically, the LSP was the first of these WIMP candidates, and personally
I still find the LSP the best motivated,
since there are so many reasons to favour supersymmetry at the TeV scale~\cite{JECOlex}.
It would help the Higgs boson do its job [(1) above], by cancelling the quadratically-divergent
contributions to its mass, and thereby stabilizing the electroweak mass scale~\cite{JEhierarchy}.
Further, supersymmetry predicts the appearance of a Higgs boson at a mass
$\sim 130$~GeV or below, as indicated by the precision electroweak data~\cite{JElightH}.
Supersymmetry at the TeV scale would also aid in the grand unification of the
strong, weak and electromagnetic interactions~\cite{JEGUTs} by enabling their strengths to
evolve to a common value at some high-energy GUT scale [(4) above]. Moreover, supersymmetry is
apparently essential in stringy attempts to construct a quantum theory of
gravity  [(5) above]. However, as Feynman surely would have said, you would not give
five arguments for supersymmetry if you had one good argument, so let us focus on that: the
LSP is an excellent candidate for dark matter  [(3) above]~\cite{JELSP}, as we now discuss.

\section{Supersymmetric Models}\label{JEs:models}

We work within the framework of the minimal supersymetric extension
of the Standard Model (MSSM), in which the known particles are
accompanied by simple supersymmetric partners and there are two
Higgs doublets, with a superpotential coupling denoted by $\mu$ and
a ratio of Higgs v.e.v.s denoted by $\tan \beta$~\cite{JEMSSM}. The bugbear of the MSSM
is supersymmetry breaking, which occurs generically through scalar
masses $m_0$, gaugino fermion masses $m_{1/2}$, trilinear soft
scalar couplings $A_0$ and bilinear soft scalar couplings $B_0$.
In our ignorance about them, the total number of parameters in the MSSM 
exceeds 100! For simplicity,
it is often assumed that these parameters are universal at the scale of
grand unification, so that there are single soft supersymmetry-breaking
parameters $m_0, m_{1/2}, A_0$, a scenario called the constrained
MSSM (CMSSM)~\footnote{I emphasize that the CMSSM is
not to be confused with minimal supergravity (mSUGRA), which
imposes a specific relationship between the trilinear and bilinear
couplings: $B_0 = A_0 - m_0$ as well as a relationship between
the scalar and gravitino masses: $m_0 = m_{3/2}$. These apparently
innocuous extra assumptions affect drastically the nature of the LSP,
and the allowed regions of parameter space~\cite{JEVCMSSM}.}. 
However, this assumption is not strongly motivated
by either fundamental theory or phenomenology. Moreover, as discussed below,
even if $m_0, m_{1/2}, A_0$ are universal, this may be true at some
scale different from the GUT scale~\cite{JEEOS2,JEEMO}.

What happens if the soft supersymmetry-breaking parameters are
not universal? Upper limits on flavour-changing neutral interactions
disfavour models in which different sfermions with the same internal
quantum numbers, e.g., the $ {\tilde d}, {\tilde s}$ squarks have different masses~\cite{JEEN}.
But what about squarks with different internal quantum numbers, or
squarks and sleptons? Various GUT models impose some relations between them,
e.g., the ${\tilde d_R}$ and ${\tilde e_L}$ scalar masses are universal in SU(5) GUTs, 
as are the ${\tilde d_L}, {\tilde u_L}, {\tilde u_R}$ and ${\tilde e_R}$ scalar masses,
and all are equal in SO(10) GUTs. However, none of these arguments rules out
non-universal supersymmetry-breaking scalar masses for the Higgs multiplets,
so one may also consider such non-universal Higgs models (NUHM) with either
one or two additional parameters (NUHM1, NUHM2). Who knows where
string models may finish up among or beyond these possibilities?

The LSP is stable in many supersymmetric models because of a conserved
quantity known as $R$ parity, which may be expressed in terms of baryon
number $B$, lepton number $L$ and spin $S$ as $R \equiv (-1)^{2S - L + 3B}$. 
It is easy to check that all Standard Model particles have $R = +1$ and their
supersymmetric partners have $R = - 1$. The multiplicative conservation of $R$
implies that sparticles must be produced in pairs that heavier sparticles must
decay into lighter sparticles, and that the LSP is stable, because it has no
legal decay mode. It should lie around in the Universe today, as a
supersymmetric relic from the Big Bang~\cite{JELSP}.

In such a scenario, the LSP could have no strong or electromagnetic interactions~\cite{JELSP},
since otherwise it would bind to ordinary matter and be detectable in 
anomalous heavy nuclei, which have been looked for, but not seen.
Possible weakly-interacting scandidates include {\it a priori} the sneutrinos -
which have been excluded by LEP and by direct astrophysical searches for
dark matter, the lightest neutralino $\chi$ - a mixture of the spartners of the
$Z, \gamma$ and neutral Higgs boson, and the gravitino - the supersymmetric
partner of the graviton, which would be a nightmare for astrophysical detection,
but a potential bonanza for collider experiments. Here we concentrate on the neutralino
option, whose classical signature is en event with missing transverse momentum
carried away by invisible dark matter particles. This signature is shared by other
WIMP candidates for dark matter, such as the LKP~\cite{JELKP} and LTP~\cite{JELTP}, though the nature
and kinematics of the visible stuff accompanying the dark matter particles is
model-dependent.

\section{Constraining Supersymmetry}\label{JEs:constraints}

There are significant lower limits on the possible masses of supersymmetric
particles from LEP, which requires any charged sparticle to weigh more than
about 100~GeV~\cite{JELEPsusy}, and the Tevatron collider, which has not found any squarks
or gluinos lighter than about 400~GeV~\cite{JETevatron}. There are also important indirect
constraints implied by the LEP lower limit on the Higgs mass of 114.4~GeV~\cite{JELEPH}, and the
agreement of the Standard Model prediction for $b \to s \gamma$ decay
with experimental measurements. The only possible experimental discrepancy
with a Standard Model prediction is for $g_\mu - 2$~\cite{JEE821}, though the significance
of this discrepancy is still uncertain, as discussed in the following paragraph. 
However, there is one clear discrepancy with the Standard Model of particles, namely the density
of dark matter, which cannot be explained without physics beyond the Standard
Model, such as supersymmetry. The fact that the dark matter density is
constrained to within a range of a few percent~\cite{JEWMAP}:
\begin{equation}
\Omega_{DM} \; = \; 0.111 \pm 0.006
\label{JEWMAP}
\end{equation}
constrains some combination of the parameters of any dark matter model also to within a few percent,
as we shall see shortly in the case of supersymmetry, but the same would be
true in other models.

The calculation of the Standard Model prediction for $g_\mu - 2$ requires an estimate of
the contribution from hadronic vacuum polarization diagrams, that may be obtained either
from $e^+ e^- \to$~hadrons data, or from $\tau \to \nu +$~hadrons decays. Historically,
there has been poor consistency between the $e^+ e^-$ and $\tau$ estimates (though
both differ substantially from the experimental measurement), and
the consistency between different $e^+ e^-$ experiments has not always been
excellent. Since the $\tau$ estimate requires an isospin correction, the $e^+ e^-$ 
estimate is more direct and generally preferred. Accordingly, in the following results are
shown assuming a discrepancy~\cite{JEoldDavier}
\begin{equation}
\Delta ( g_\mu - 2 ) \; = \; (30.2 \pm 8.8) \times 10^{-10}
\label{JEDeltag-2}
\end{equation}
calculated from $e^+ e^-$ data
to be explained by physics beyond the Standard Model, such as supersymmetry.
Very recently, re-evaluations of the $e^+ e^-$ and $\tau$ data
have yielded $\Delta ( g_\mu - 2 ) = (28.7 \pm 8.0) \times 10^{-10}$ 
and $(19.5 \pm 8.3) \times 10^{-10}$~\cite{JEnewDavier},
corresponding to discrepancies of 3.6 and 2.4 $\sigma$, respectively. The
results shown below would differ very little if the newer $e^+ e^-$ estimate were used.
For comparison, some results from dropping the $g_\mu - 2$ constraint altogether
are also shown, and using the $\tau$ decay estimate would give intermediate
results closer to the $e^+ e^-$ estimate.

\begin{figure}
  \includegraphics[width=.45\textwidth]{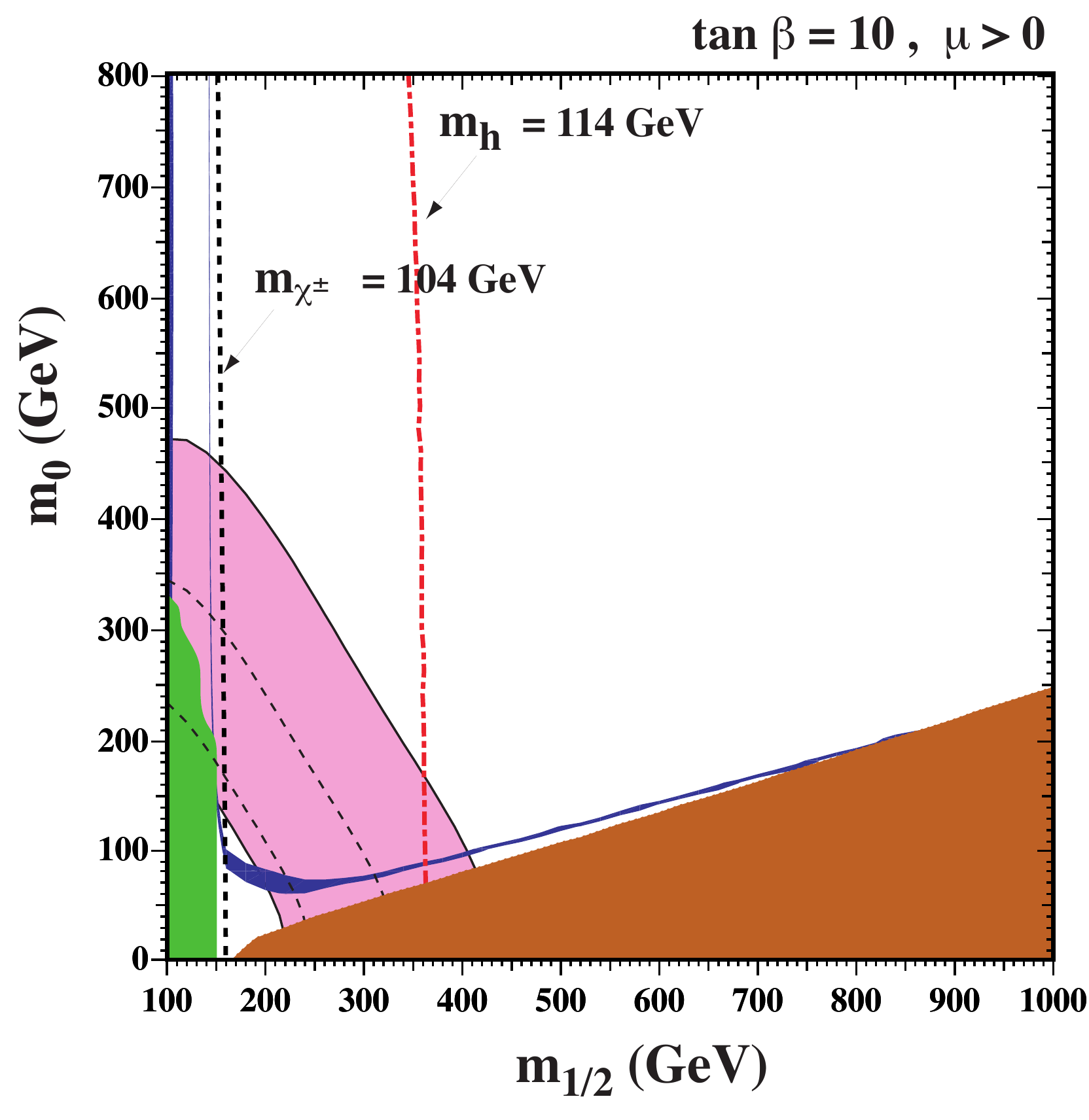}
    \includegraphics[width=.45\textwidth]{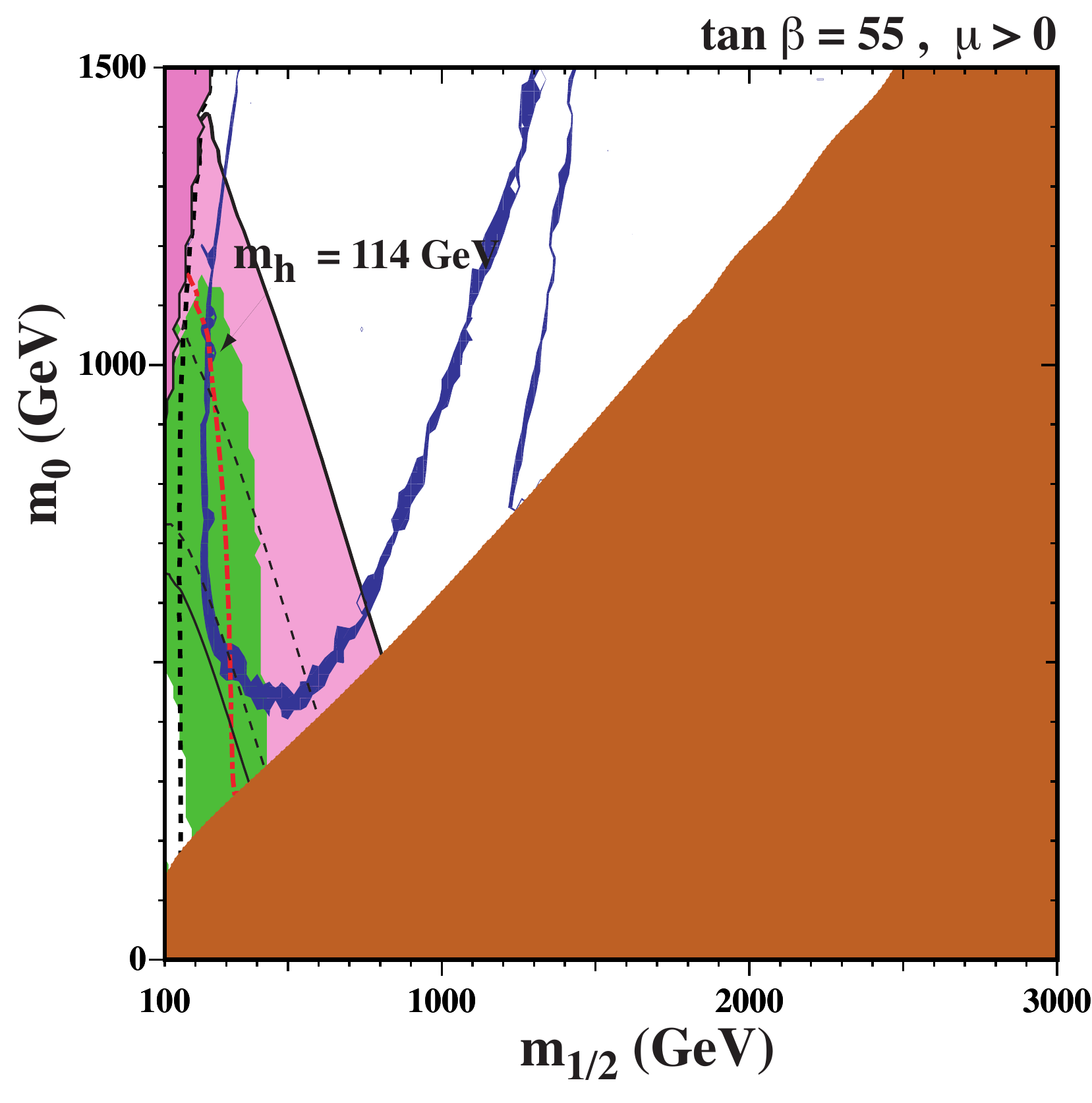}\\
      \includegraphics[width=.45\textwidth]{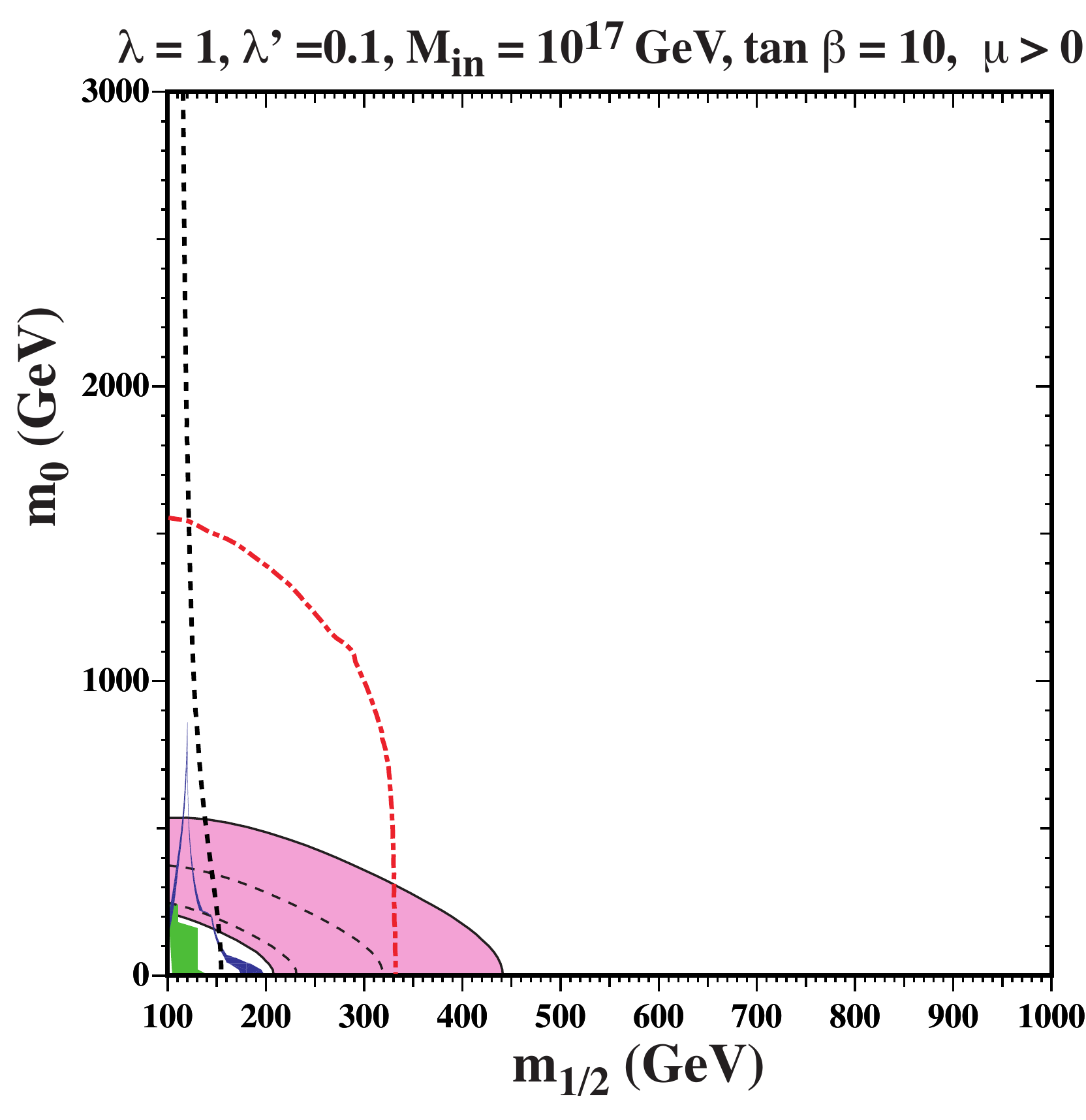}
            \includegraphics[width=.45\textwidth]{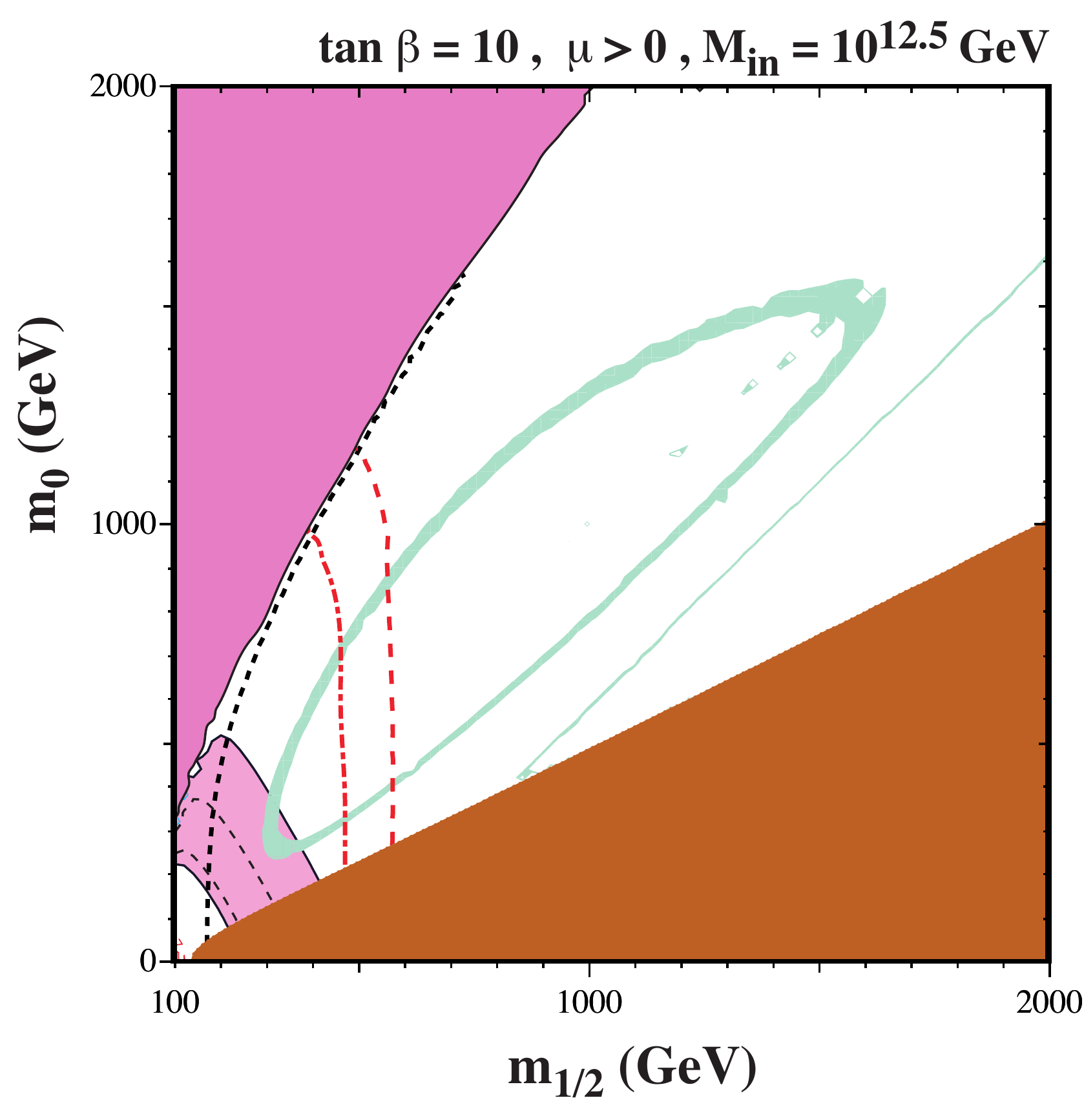}
  \caption{\it The $(m_{1/2}, m_0)$ planes for  (upper left) the CMSSM with $\tan \beta = 10$ and (upper right)
  $\tan \beta = 55$~\protect\cite{JEEOSS}, (lower left) assuming SU(5) universality at 
  $10^{17}$~GeV with representative
  choices of the quartic GUT Higgs couplings~\protect\cite{JEEMO}, 
  and (lower right) assuming scalar mass
  universality at $10^{12.5}$~GeV~\protect\cite{JEEOS2}, 
  all assuming  $\mu > 0, A_0 = 0, m_t = 173.1$~GeV and
$m_b(m_b)^{\overline {MS}}_{SM} = 4.25$~GeV. The near-vertical (red)
dot-dashed lines are the contours $m_h = 114$~GeV~\protect\cite{JELEPH}, 
and the near-vertical (black) dashed
line is the contour $m_{\chi^\pm} = 104$~GeV~\protect\cite{JELEPsusy}.  The medium (dark
green) shaded region is excluded by $b \to s
\gamma$, and the dark (blue) shaded area is the cosmologically
preferred region~\protect\cite{JEWMAP}. In the dark
(brick red) shaded region, the LSP is the charged lighter stau slepton. The
region allowed by the E821 measurement of $g_\mu -2$ at the 2-$\sigma$
level, is shaded (pink) and bounded by solid black lines, with dashed
lines indicating the 1-$\sigma$ ranges.}
  \label{JEfig:CMSSM}
  
\end{figure}

Fig.~\ref{JEfig:CMSSM} demonstrates the impacts of the various theoretical,
phenomenological, experimental and cosmological constraints in $(m_{1/2}, m_0)$
planes under different scenarios with $\mu >0$, assuming that the LSP is the
lightest neutralino, $\chi$. The top panels are for the CMSSM 
with $A_0 = 0$ and (left) $\tan \beta = 10$, (right) $\tan \beta = 55$, two values
that bracket the plausible range~\cite{JEEOSS}. In both cases, we see narrow WMAP-compliant
strips clinging near the boundaries of the (brown) charged LSP region at low $m_0$, 
where LSP-slepton coannihilation is important, and the (pink) region at high $m_{1/2}$
where electroweak symmetry is not broken consistently, called the focus-point strip.
When $\tan \beta = 55$, we also see a diagonal funnel at large $m_{1/2}$ and $m_0$
due to rapid annihilation through direct-channel heavy Higgs poles. In the lower
left panel, also for $\tan \beta = 10$, it is assumed that the scalar masses $m_0$
and the gaugino masses $m_{1/2}$ are universal at the scale $10^{17}$~GeV~\cite{JEEMO},
instead of the GUT scale as in the CMSSM. We see that the coannihilation strip
has shrunk into the region forbidden by the LEP Higgs limit, and the fixed-point
strip has disappeared to larger $m_0$. On the other hand, if $m_0$ universality
is assumed instead to hold at $10^{12.5}$~GeV, as in the bottom right panel, the
coannihilation, fixed-point and funnel regions merge to form an atoll away from
the boundaries of parameter space~\cite{JEEOS2}. In what follows, the standard CMSSM and
the NUHM1 model will be studied, but these panels emphasize that this involves
a dicey assumption.

\section{Global Supersymmetric Fits}

Within the general CMSSM and NUHM frameworks, is it possible to find
a preferred region of supersymmetric masses? To answer this question,
we adopted a frequentist approach and constructed a global likelihood function
using precision electroweak data, the LEP Higgs mass limit (allowing for 
theoretical uncertainties), the cold dark matter density, $b \to s \gamma$
and $B_s \to \mu^+ \mu^-$ constraints and (optionally) $g_\mu - 2$~\cite{JEMC1,JEMC2,JEMC3}.

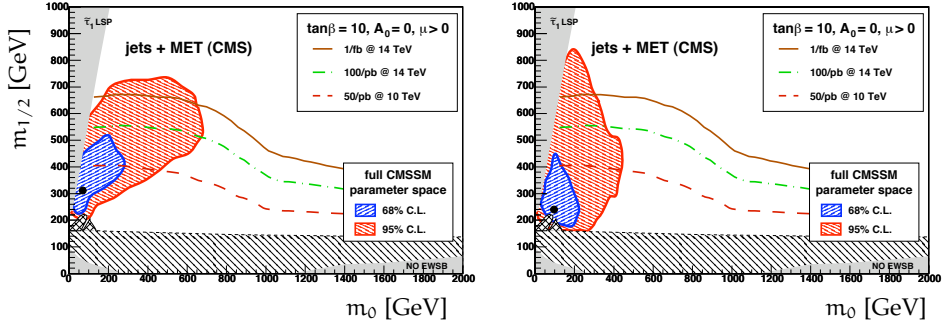
\begin{figure*}[htb!]
\begin{center}
\vspace{-10cm}
\begin{picture}(300,400)
  \put(  -15,   0){ \resizebox{5.75cm}{!}{\includegraphics{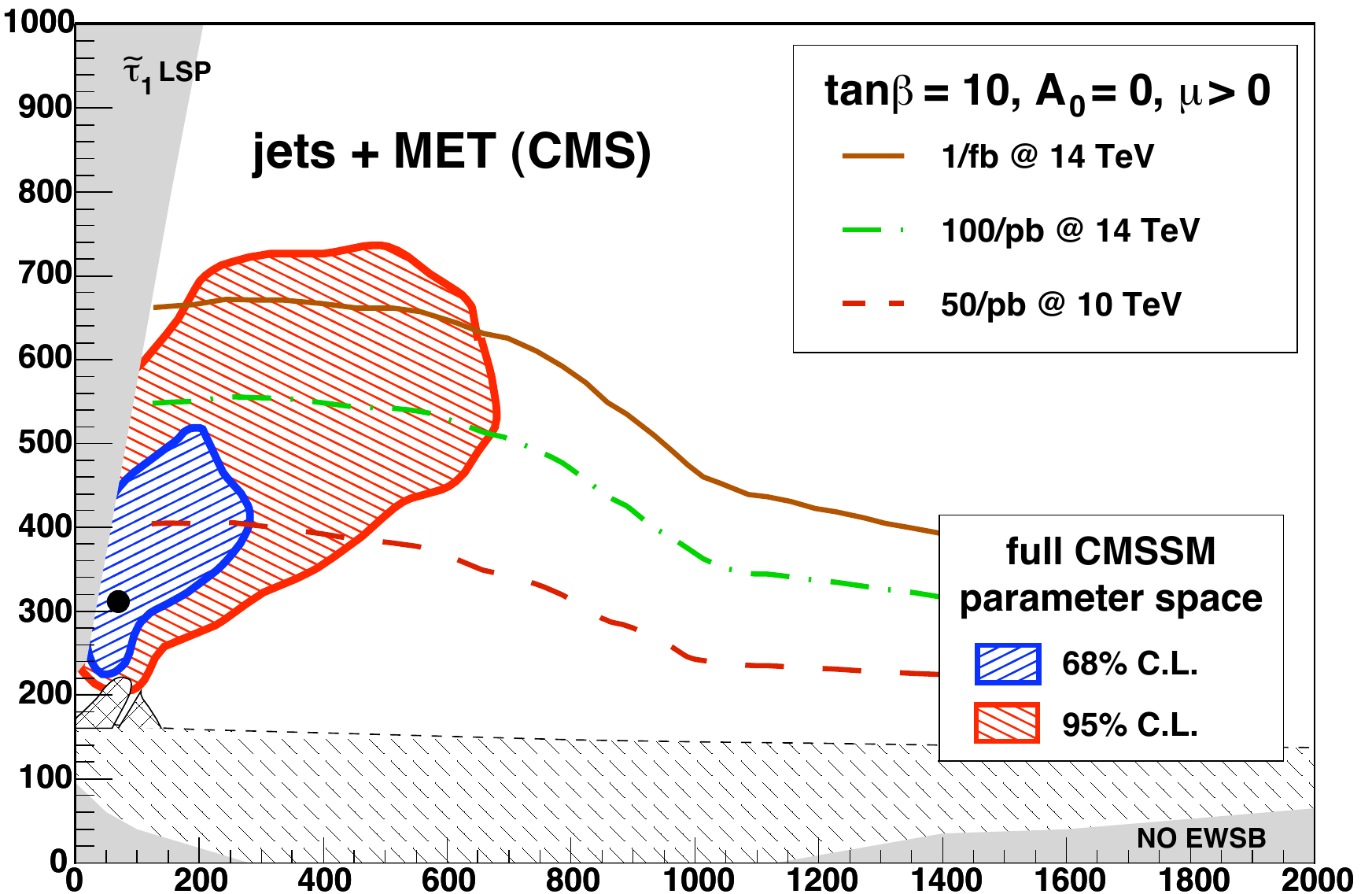}}}
  \put(100,   -12){$m_0$ [GeV]}
  \put( -20,   50){\begin{rotate}{90}$m_{1/2}$ [GeV]\end{rotate}}
  \put(  160,     0){ \resizebox{5.75cm}{!}{\includegraphics{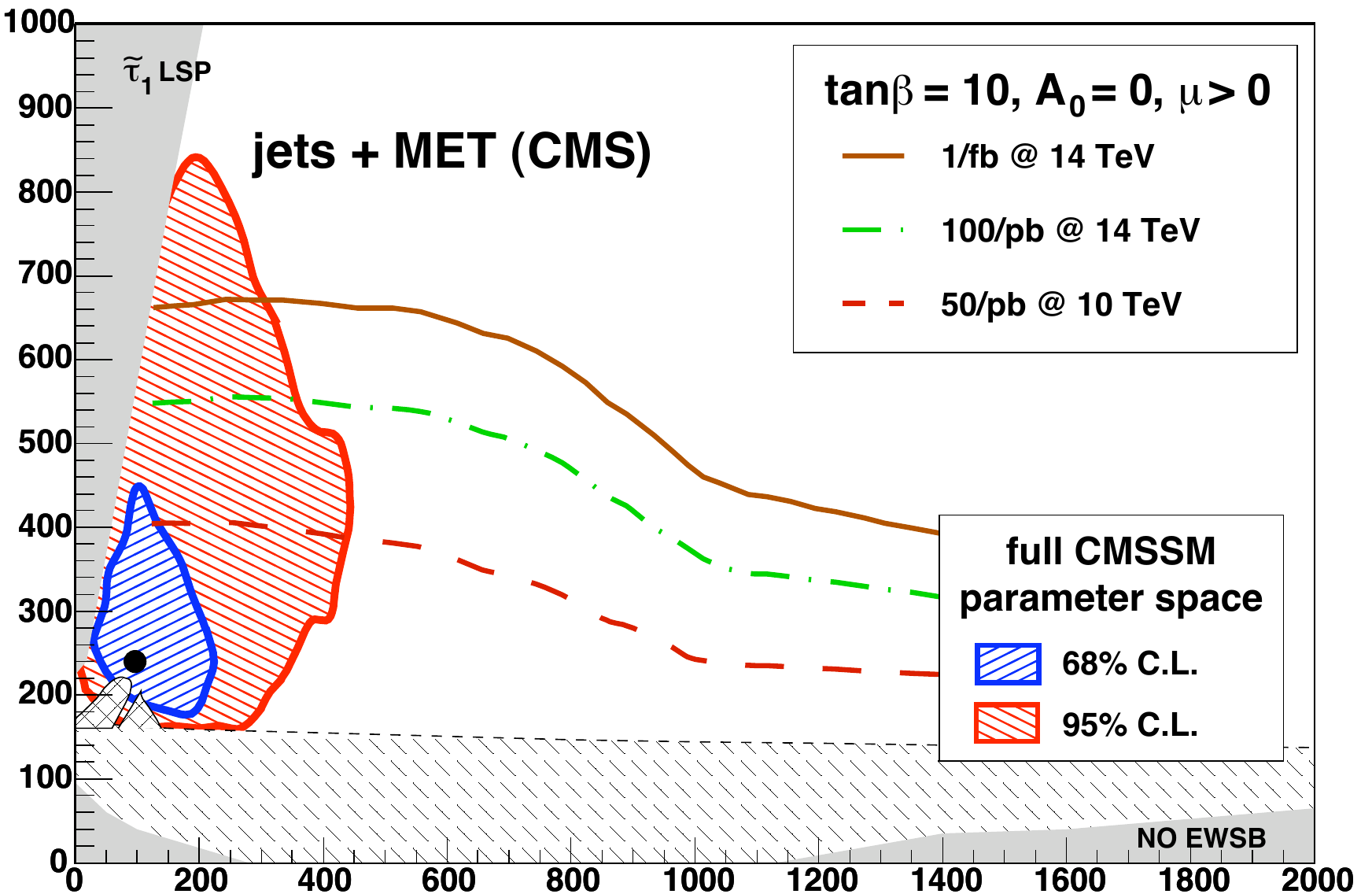}}  }
  \put(280,   -12){$m_0$ [GeV]}
\end{picture}
\end{center}
\caption {\it The $(m_0, m_{1/2})$ planes for (left) the CMSSM and (right) the NUHM1. 
The dark shaded area at low $m_0$ and high $m_{1/2}$ is
  excluded due 
  to a scalar tau LSP, and the light shaded areas at low $m_{1/2}$ do not
  exhibit electroweak symmetry breaking. The nearly horizontal line at
  $m_{1/2} \approx 160$~GeV in the lower panel 
  has $m_{\tilde \chi_1^\pm} = 103$~GeV, and the area
  below is excluded by LEP searches. Just above this contour at low $m_0$
  in the lower panel is the region that is
  excluded by trilepton searches at the Tevatron.
  Shown in each plot is the best-fit point, indicated by a filled
  circle, and the 
  68 (95)\%~C.L.\ contours from our fit as dark grey/blue (light
  grey/red) overlays~\protect\cite{JEMC2}.
  Also shown are 5-$\sigma$ discovery contours at the LHC with
  the indicated luminosities and centre-of-mass energies.
} 
\label{JEfig:planes}
\end{figure*}

\begin{figure*}[htb!]
\resizebox{6.5cm}{!}{\includegraphics{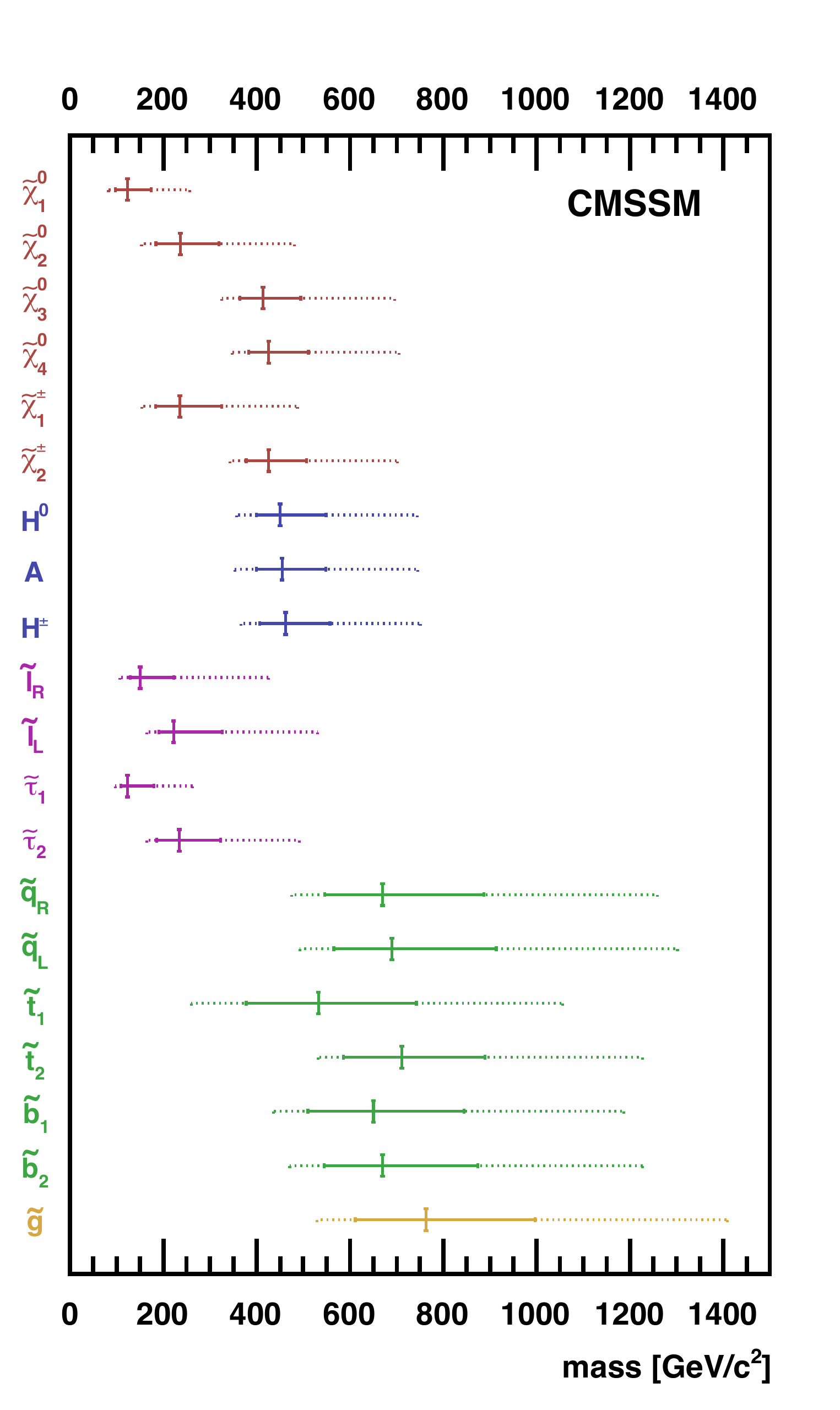}}
\resizebox{6.5cm}{!}{\includegraphics{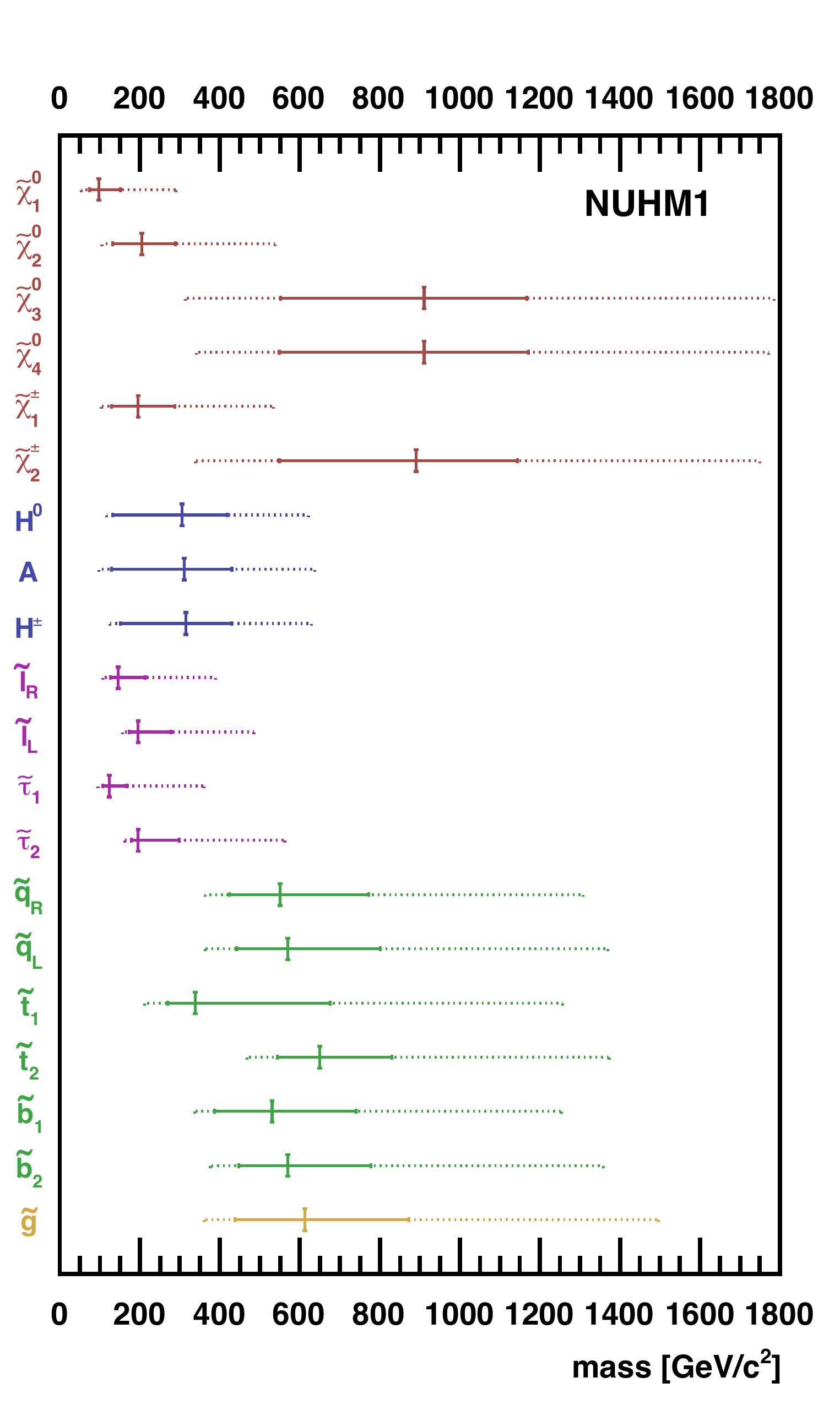}}
\caption{\it Spectra in the CMSSM (left) and the NUHM1 (right). The vertical
solid lines indicate the best-fit values, the horizontal solid lines
are the 68\% C.L.\
ranges, and the horizontal dashed lines are the 95\% C.L.\ ranges for the
indicated mass parameters~\protect\cite{JEMC3}. 
}
\label{JEfig:spectra}
\end{figure*}

\begin{figure*}[htb!]
{\resizebox{6.5cm}{!}{\includegraphics{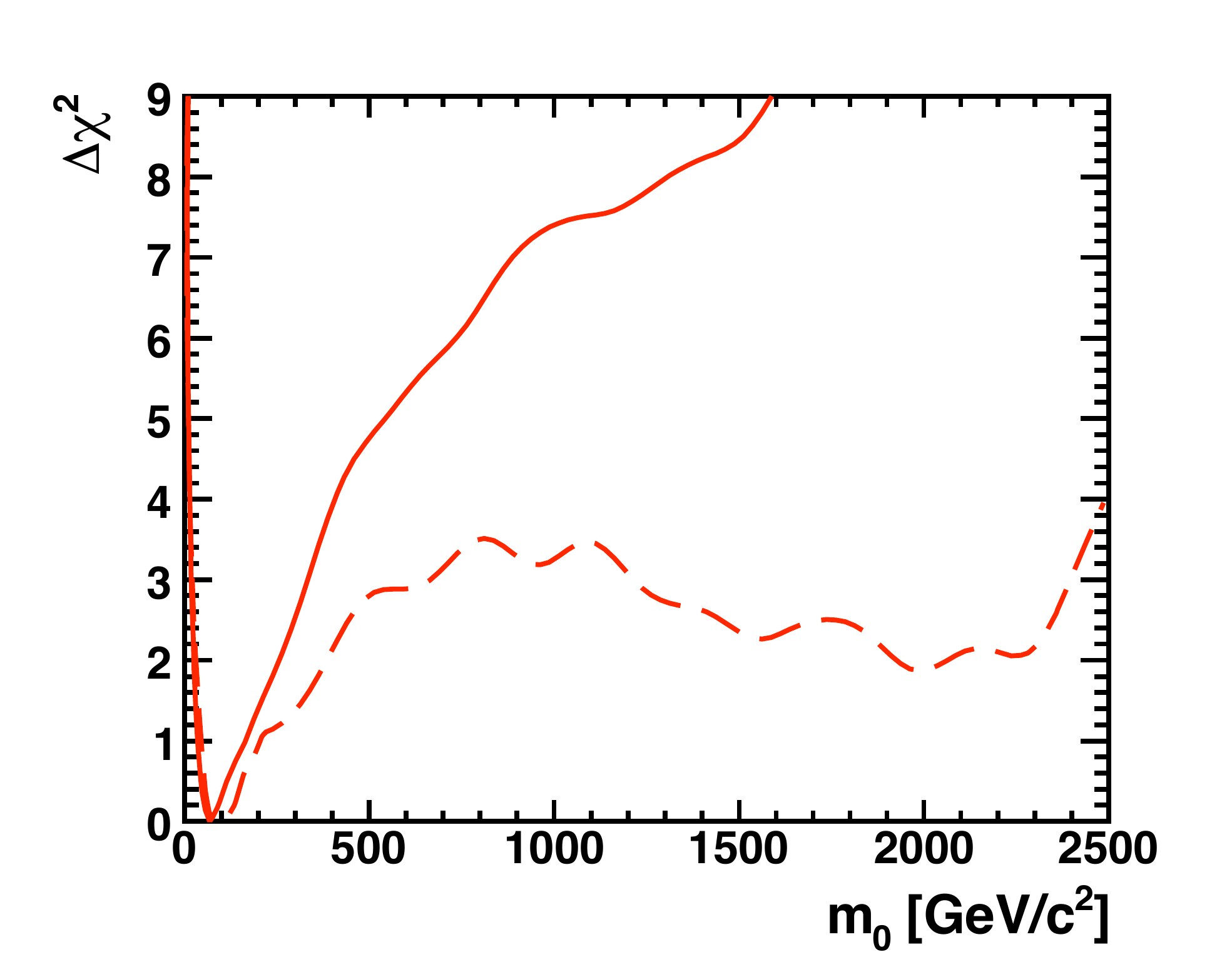}}}  
{\resizebox{6.5cm}{!}{\includegraphics{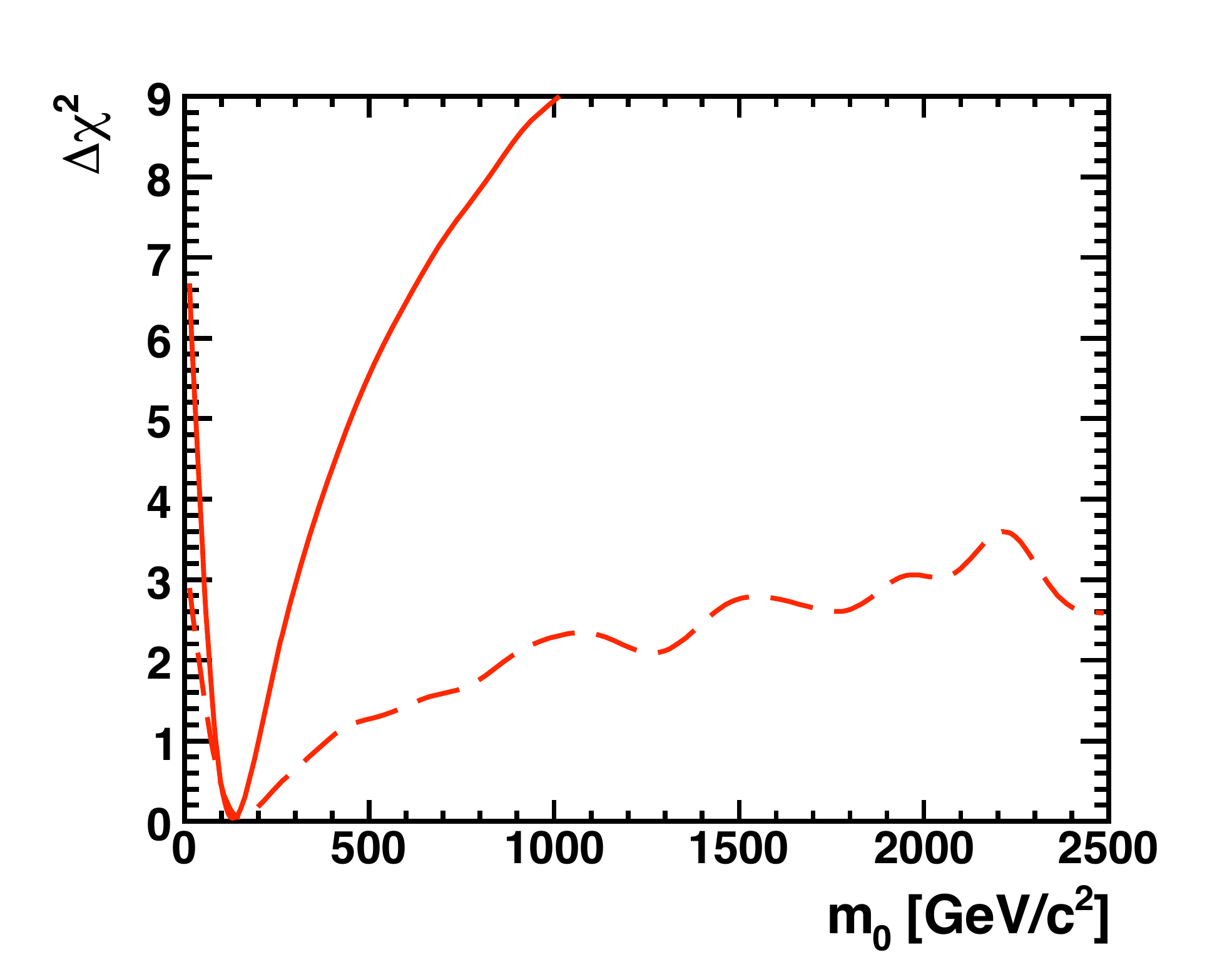}}}
\caption{\it The likelihood functions for $m_0$ in the CMSSM (left plot) and
  in the NUHM1 (right plot). The $\chi^2$ values are shown including
  (excluding) the $g_\mu - 2$ constraint as the solid (dashed) curves~\protect\cite{JEMC3}.
}
\label{JEfig:drop}
\end{figure*}

In both the CMSSM and the NUHM1 we found that small $m_{1/2}$ and $m_0$
in the coannihilation region are preferred, with the focus-point region
disfavoured. The best-fit points, 68\% and 95\% CL regions in the $(m_0, m_{1/2})$
planes of the CMSSM and NUHM1 are shown in Fig.~\ref{JEfig:planes}~\cite{JEMC2}, and
the corresponding spectra are shown in Fig.~\ref{JEfig:spectra}~\cite{JEMC3}. The favoured
areas of the planes shown in Fig.~\ref{JEfig:planes} are quite sensitive to the
treatments of the constraints, particularly $g_\mu - 2$ and $b \to s \gamma$~\cite{JEMC2}.
In the extreme case when the $g_\mu - 2$ constraint is dropped entirely, as
in Fig.~\ref{JEfig:drop}, large values of $m_0$ are no longer strongly
disfavoured, although the other constraints still show some slight preference
for small $m_0$~\cite{JEMC3}.

Fig.~\ref{JEfig:planes} also shows the expected sensitivity of the LHC for a
discovery of supersymmetry with 5-$\sigma$ significance for varying LHC
energies and luminosities. We see that there may be a fair chance to discover 
supersymmetry even in early LHC data. However, at the 95\% CL,
supersymmetry might still lie beyond the reach of the LHC with 1/fb of
data at 14~TeV, as could also be inferred from the 95\% CL ranges
in Fig.~\ref{JEfig:spectra}.

\section{Detecting Supersymmetric Dark Matter}

Several strategies for the detection of WIMP dark matter particles
such as the LSP have been proposed, including the direct search
for scattering on nuclei in the laboratory~\cite{JEGW}, the search for energetic
neutrinos produced by WIMP annihilations in the core of the Sun or Earth~\cite{JEnus},
the search for energetic photons produced by WIMP annihilations
in the galactic centre or elsewhere in the Universe~\cite{JEgammas}, and the searches
for positrons, antiprotons, etc., produced by WIMP annihilations in the
galactic halo~\cite{JEpbars}.

\begin{figure*}[htb!]
\resizebox{6.5cm}{!}{\includegraphics{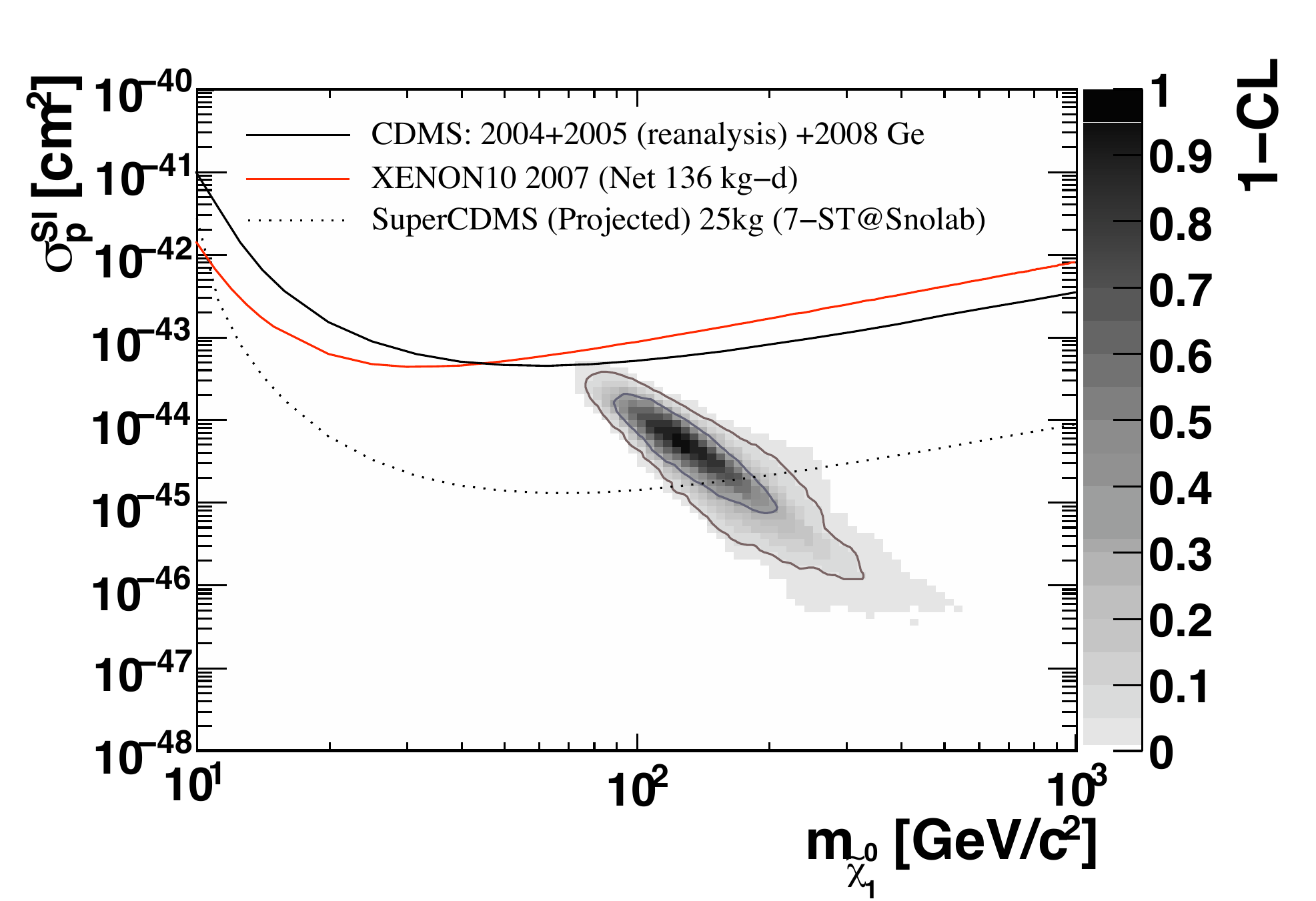}}
\resizebox{6.5cm}{!}{\includegraphics{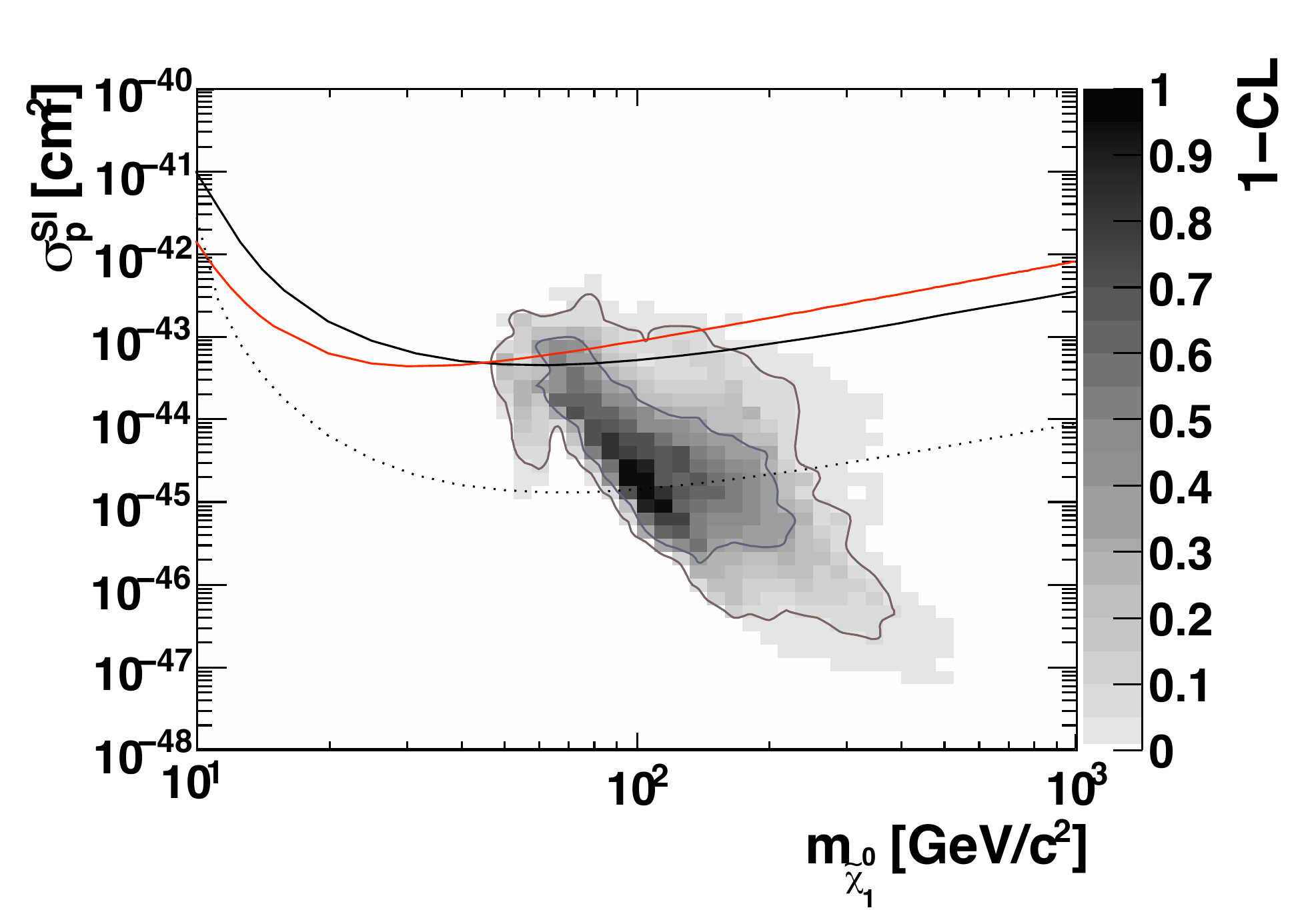}}
\caption{\it The correlation between the spin-independent dark matter
scattering cross section and $m_\chi$
in the CMSSM (left panel) and in the NUHM1 (right panel). The solid lines~\protect\cite{JEDMtool}
are the experimental upper limits from CDMS~\protect\cite{JECDMS} and Xenon10\protect\cite{JEXe10}, 
The dashed line~\protect\cite{JEDMtool} indicates the projected sensitivity of the SuperCDMS 
experiment~\protect\cite{JEsuperCDMS}: that of Xenon100 may be similar.
}
\label{JEfig:scattering}
\end{figure*}

As seen in Fig.~\ref{JEfig:scattering}, within the global fits to
supersymmetric model parameters discussed earlier, our predictions for the direct
nuclear scattering rates in the CMSSM and NUHM1 lie somewhat
below the sensitivities of the CDMS and Xenon10 experiments, 
but within reach of planned upgrades of these experiments~\cite{JEMC3}. 
Subsequently, the CDMS~II~\cite{JECDMSII} and
Xenon100~\cite{JEXe100} experiments have announced results with somewhat
improved sensitivity. In particular, the CDMS II experiment reported
two events with relatively low recoil energies (corresponding possibly to
the scattering of a WIMP weighing $< 30$~GeV) where less than one event
was expected~\cite{JECDMSII}, but this hint was not confirmed by the Xenon100 experiment
in its initial 11-day test run~\cite{JEXe100}. (Nor have possible signals in the DAMA/LIBRA~\cite{JEDAMA} and 
CoGeNT experiments~\cite{JECoGeNT} been confirmed by either CDMS or Xenon100.) It is
expected that updated Xenon100 results with much greater sensitivity will be
announced soon, reaching significantly into the scattering rates expected
within our global fits. (Though it
should be noted that these predictions assume one particular value
for the spin-independent scattering matrix element, which is a significant
source of uncertainty in the predictions~\cite{JEEOSavage1}.)

\begin{figure*}
\begin{center}
\resizebox{11cm}{!}{\includegraphics{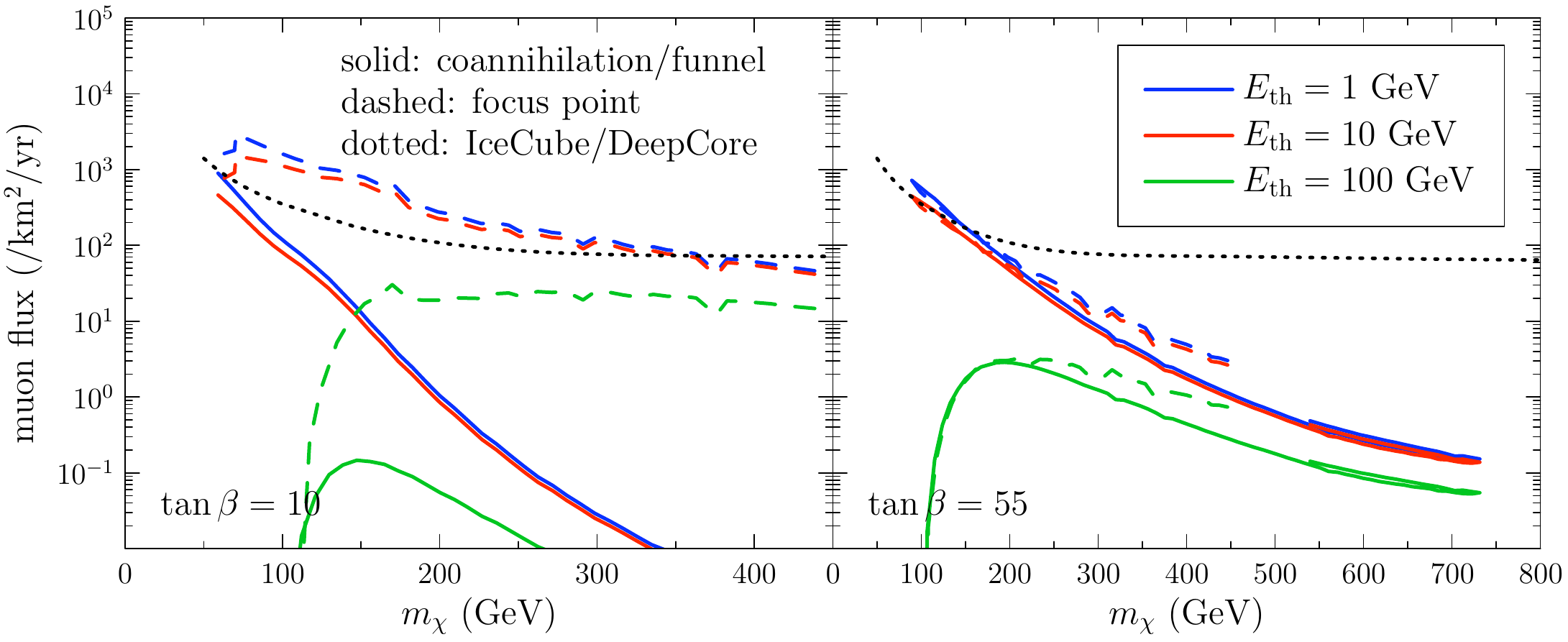}} 
\end{center}
 \caption{\it   
    The CMSSM muon fluxes though a detector calculated for $A_0 = 0$
    and (left) $\tan \beta = 10$, (right) $\tan \beta = 55$,
    along the WMAP strips in the coannihilation/funnel regions (solid)
    and the focus-point region (dashed)~\protect\cite{JEEOSavage}.
    Fluxes are shown for muon energy thresholds of (top to bottom)
    1~GeV, 10~GeV, and 100~GeV.
     Also shown is a conservative estimate of the sensitivity of the
    IceCube/DeepCore detector (dotted)~\protect\cite{JEICDC}, normalized to a muon threshold
    of 1~GeV, for a particular hard annihilation spectrum that is a rough
    approximation to that expected in CMSSM models.
    }
  \label{JEfig:nuflux}
\end{figure*}

The next most promising strategy for indirect detection of dark matter
may be the search for energetic neutrinos emitted by WIMP annihilations
in the core of the Sun~\cite{JEnus}. It is often assumed that the annihilation rate is
in equilibrium with the WIMP capture rate, but this is not true in general
in the CMSSM~\cite{JEEOSavage}. Nor is spin-dependent scattering the dominant
mechanism for LSP capture by the Sun, as is often assumed:
spin-independent scattering on heavier elements inside the Sun may
also be important, even dominant~\cite{JEEOSavage}. As seen in Fig.~\ref{JEfig:nuflux},
in a general survey of the CMSSM
parameter space~\cite{JEEOSavage}, we find significant portions of the focus-point strips,
and some parts of the coannihilation strips, where the flux of energetic 
neutrinos due to LSP annihilations may be detectable in the IceCube/DeepCore
experiment~\cite{JEICDC}.

\section{The Start-up of the LHC}

The LHC made its first collisions on November 29th, 2009, and its first 7-TeV
collisions on March 30th, 2010. Much jubilation, but where are the Higgs boson
and supersymmetry, you may ask. It should be recalled that the total
proton-proton cross section for producing a new particle weighing $\sim 1$~TeV
is $\sim 1$/TeV$^2$, possibly suppressed even further by small couplings
$\sim \alpha^2$, whereas the total cross section $\sim 1/m_\pi^2$, so that
the `interesting' new physics signal is likely to occur in $ \sim 10^{12}$ of
the collisions. This is like looking for a needle in $\sim 100,000$ haystacks!

So far the LHC experiments have seen only a few $\times 10^{12}$
collisions. The missing $E_T$ distribution agrees perfectly with simulations over 
more than 6 orders of magnitude~\cite{JEMET}, and there is no sign yet of an excess of
events that might be due to the production and escape of dark matter
particles, whether they be LSPs, LKPs, LTPs, or whatever. Moreover,
the kinematics of the events with missing $E_T$ is exactly what one
would expect from mismeasured QCD events and other Standard Model
backgrounds: no signs yet of new physics beyond the Standard Model.

\begin{figure*}[htb!]
\begin{center}
\resizebox{8cm}{!}{\includegraphics{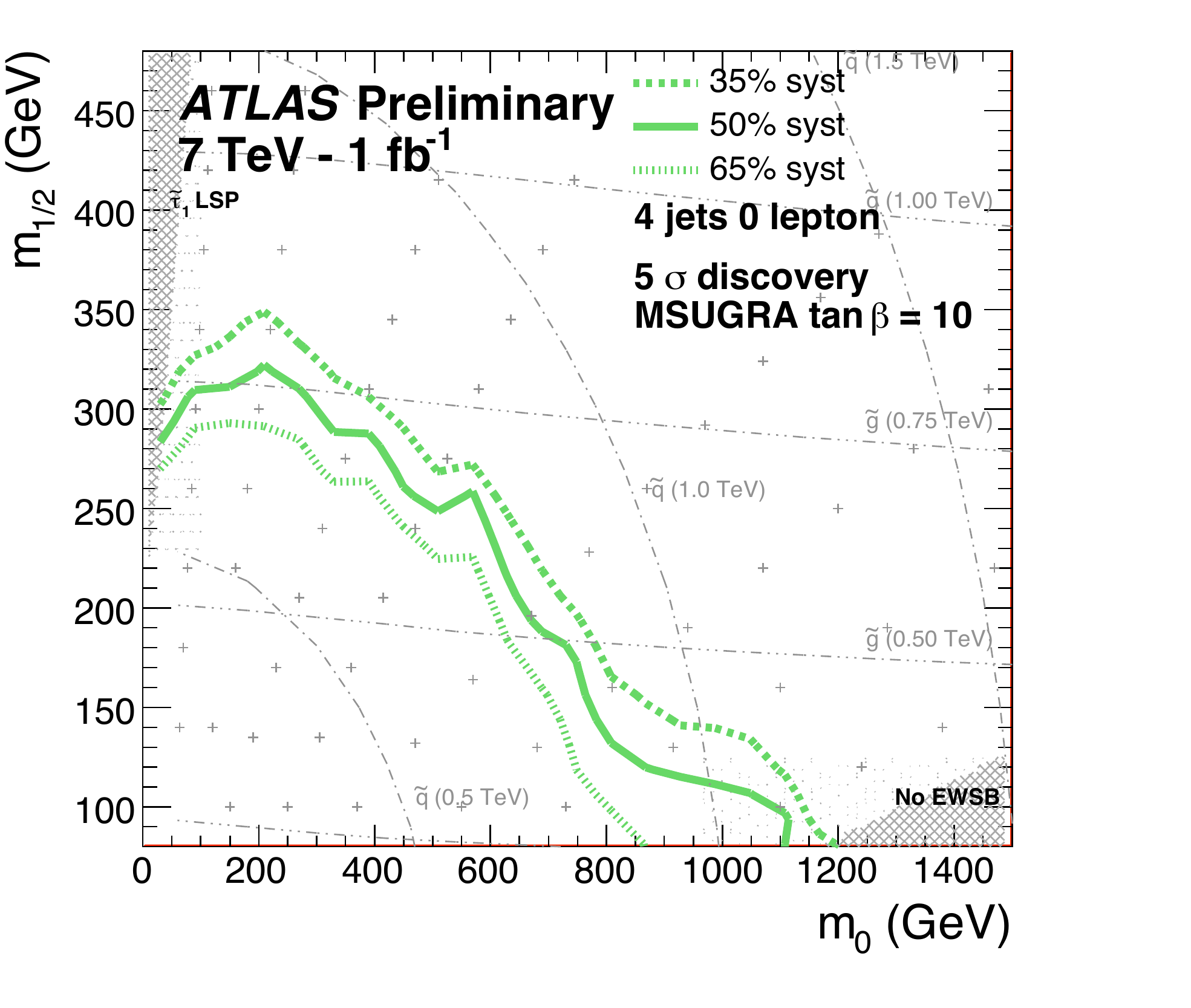}}
\end{center}
\caption{\it The sensitivity in the $(m_0, m_{1/2})$ plane
of the ATLAS experiment for a 5-$\sigma$ discovery of a
supersymmetric signal with 1/fb at 7 TeV in the centre of mass.
}
\label{JEfig:7TeV}
\end{figure*}

The results of our frequentist likelihood analysis were compared in
Fig.~\ref{JEfig:planes} with the estimated sensitivity of the LHC at or
close to its design energy. In 2010 it has been operating at $\sim 7$
TeV and accumulating $\sim 50$/pb of integrated luminosity, which is
sufficient to extend the reach for supersymmetry beyond the Tevatron.
The centre-of-mass energy may be increased in 2011 to 8 or 9 TeV,
and the objective is to accumulate $\sim 1$/fb of integrated
luminosity. Fig.~\ref{JEfig:7TeV} shows the estimated sensitivity of 
supersymmetry searches with the ATLAS experiment~\cite{JEATLAS7} using 1/fb of data at 7 TeV.
Comparing with Fig.~\ref{JEfig:planes}, we see that the best-fit points in the
CMSSM and NUHM1 should lie within reach. There are significant prospects
for soon getting some interesting news about supersymmetry, one way or
the other.

\section{A Conversation with Mrs. Thatcher}

In 1982, Mrs. Thatcher, the British Prime Minister at the time, visited CERN,
and I was introduced to her as a theoretical physicist. ``What exactly do you do?",
she asked in her inimitably intimidating manner. ``I think of things for experimentalists to look
for, and then I hope they find something different", I responded. Somewhat
predictably, Mrs. Thatcher asked ``Wouldn't it be better if they found what
you predicted?" My response was that "If they found exactly what the theorists
predicted, we would not be learning so much". In much the same spirit, I hope
(and indeed expect) that the LHC will become most famous for discovering
something that I did NOT discuss in this talk - as long as it casts light on dark matter!

\title{Extra Dimensional Metric Reversal Symmetry and its Prospect for 
Cosmological Constant and Zero-point Energy Problems, Automatic 
Pauli-Villars-like Regularization, and an Interesting Kaluza-Klein 
Spectrum }
\author{R. Erdem\thanks{E-mail:recaierdem@iyte.edu.tr}}
\institute{%
Department of Physics, \\
{\.{I}}zmir Institute of Technology \\ 
G{\"{u}}lbah{\c{c}}e K{\"{o}}y{\"{u}}, Urla, {\.{I}}zmir 35430, 
Turkey}

\titlerunning{Extra Dimensional Metric Reversal Symmetry\ldots}
\authorrunning{R. Erdem}
\maketitle

\begin{abstract}
The 
metric reversal symmetry was introduced in the context of cosmological 
constant problem. Besides proposing a solution to the cosmological 
constant problem the metric reversal symmetry has also provided a 
framework for solution of the zero-point energy problem, an automatic 
Pauli-Villars-like regularization, and an interesting Kaluza-Klein 
spectrum with interesting phenomenological implications. In this talk I 
give a brief overall summary and discussion of these topics with their 
potential implications.  
\end{abstract}


\section{Introduction: Metric reversal symmetry and the cosmological 
constant problem}

In this talk I will consider a symmetry that may be called metric 
reversal symmetry, in particular, the extra dimensional representations of 
this symmetry. •This symmetry, first, was introduce in
\cite{re1Erdem1}
as a possible solution to 
cosmological constant (CC) problem \cite{re1Weinberg} in a classical setting 
in extra dimensions. Below I define the metric reversal symmetry
and mention its use for CC problem. In the following sections I review 
my recent studies on 
the use of this symmetry at quantum level, namely, zero-point energy 
problem, an 
interesting Kaluza-Klein spectrum, an automatic Pauli-Villars-like 
regularization, and some of their possible implications.

Metric reversal is defined by
\begin{equation}
ds^2\,=\,g_{AB}dx^Adx^B\,\rightarrow\,-ds^2 \label{re1a1}
\end{equation}
This transformation has two realizations: The first 
\cite{re1Erdem1,re1Nobbenhuis,re1tHooft} is
\begin{equation}
x^A\,\rightarrow\,i\,x^A~,~~g_{AB}\,\rightarrow\,g_{AB} \label{re1a2}
\end{equation}
The second \cite{re1Bonelli,re1Erdem2,re1DuffKalk} is
\begin{equation}
x^A\,\rightarrow\,x^A~,~~g_{AB}\,\rightarrow\,-\,g_{AB}\label{re1a3}
\end{equation}
Metric reversal symmetry (MRS) may be imposed at the level of the equations 
of motion (EM) (e.g. Einstein equations) by requiring the equations be 
covariant under MRS or at the level of the action by requiring the action be 
invariant under MRS. MRS forbids a CC in any dimension if it is imposed at 
the level of EM \cite{re1Nobbenhuis,re1tHooft}.
•MRS can be only classically viable if it is introduced at the level of EM.
On the other hand MRS can be extended to the quantum domain if it 
is imposed to the action.
So I prefer to impose MRS at the level of action functional.
The gravitational action                                  
\begin{equation}
S_R = \frac{1}{16\pi\,G}\int \sqrt{(-1)^S g} \,R \,d^Dx \label{re1a4}
\end{equation}
is invariant under either of (\ref{re1a2}) or (\ref{re1a3}) only in                
\begin{equation}                                                     
D=2(2n+1)~~~,~~~~n=0,1,2,3,....~~~.\label{re1a5grav}
\end{equation}                                                       
while the CC action                 
\begin{equation}                                                 
S_C = \frac{1}{8\pi\,G}\int \sqrt{g} \,\Lambda \,d^Dx \label{re1a7act}.
\end{equation}	                                                
is forbidden in $2(2n+1)$ dimensions.                           

So if our space is taken to be 2(2n+1) dimensional (or if the gravitation 
and the CC reside on a 2(2n+1) dimensional subspace of a larger space) 
then the cosmological constant (CC) is forbidden. In this framework the 
accelerated expansion of the universe may be attributed to a small 
breaking of MRS or to an alternative mechanism (such as quintessence, 
modified gravity etc.) if the symmetry is taken to be exact. Another point 
to to be mentioned is that two realizations of MRS are not equivalent in 
matter sector while they are wholly equivalent in the gravitational 
sector. For example $F_{AB}F^{AB}$ is odd under (\ref{re1a2}) while it is 
even under (\ref{re1a3}). So two realizations of MRS may be considered to be 
two different symmetries after the introduction of matter. As we shall see 
in the next section this point may be used to construct a 
model that solves zero-point energy problem as well. The details of these 
points and some other less major points may be found in 
\cite{re1Erdem1,re1Erdem2,re1Erdem3}.

\section{Metric Reversal symmetry and zero-point energy problem}
Quantization results in by-product energies that survive even in the absence 
of any particle. These energies (i.e. zero-point energies (ZPE)) are some 
kind of vacuum energy. They emerge as zero modes of harmonic oscillators 
or fields 
in quantum theory. The total ZPE associated with a particle is constant, 
and is found as the sum over the contributions due to different momenta, 
and is naively infinite. However ZPE is eliminated by subtracting ZPE 
from total energy. In the quantum field theory (QFT) in flat 
space this elimination (normal ordering) has no physical effect because 
changing the energy by a constant does not change the physical results. 
However in QFT in curved space this naive elimination of ZPE is not 
well-defined because gravity couples to all energies. So subtraction process 
affects the physical out-come. Moreover after normal ordering a non-zero 
vacuum energy remains and it is proportional to particle masses. So even the 
renormalized zero-point energy of electron gives a vacuum energy density that 
is $10^{36}$ times the observed energy density of the universe. This may be 
called ZPE problem. Renormalized ZPE may be identified by CC
or as a different kind of vacuum energy depending on how
the infinities are regulzarized in renormalization procedure. Moreover CC 
may get classical field theoretic contributions such as vacuum expectation 
of scalar feilds. Therefore it is better to consider ZPE problem 
separately. 

In an attempt for a solution for ZPE one must take the following points 
into account: CC problem as well should be addressed. Therefore the 
dimension of the subspace we live in should be D = 2(2n+1). 
Another point is that It is easier to impose the symmetry so that ZPE 
vanishes instead of trying to make it small. This requires the symmetry  be 
exact while the scale factor, $a(t)$ in Robertson-Walker metric breaks the 
symmetry generated by (\ref{re1a2}).
Therefore both realizations of MRS 
should be used so that the realization of MRS generated by (\ref{re1a3}) is kept 
intact to impose ZPE vanish while the realization of MRS generated by 
(\ref{re1a2})is broken. The following is the summary of a model that 
satisfies these criteria \cite{re1Erdem4}.

Consider a space consisting of the sum of 2(2n+1) and 2(2m+1) (e.g 6 and
2) dimensional subspaces with the metric
\begin{eqnarray}
ds^2&=&
g_{AB}dx^A\,dx^B
\,+\,g_{A^\prime B^\prime}dx^{A^\prime}\,dx^{B^\prime} \nonumber \\
&=&
\Omega_z(z)[
g_{\mu\nu}(x)
\,dx^{\mu}dx^\nu\,+\,
\tilde{g}_{ab}(y)\,dy^{a}dy^b]
\,+\,
\Omega_y(y)\tilde{g}_{A^\prime B^\prime}(z)\,dz^{A^\prime}dz^{B^\prime}\nonumber\\
\label{re1b1} \\
&&\Omega_y(y)\,=\,\cos{k|y|})
~~,~~~\Omega_z(z)\,=\,\cos{k^\prime|z|} \label{re1b1a} \\
A,B&=&0,1,2,3,5,....N~~,~~~N=2(2n+1) \nonumber \\
A^\prime,B^\prime&=&1^\prime,2^\prime,,....N^\prime~~,~~~N^\prime=2(2m+1)
\nonumber \\
&&~~~~~~~~~~~\mu\nu=0,1,2,3~,~~~a,b=1,2,...,N-4~~,~~~n,m=0,1,2,3......~~.
\nonumber
\end{eqnarray}
The usual four dimensional space is embedded in the first space
$g_{AB}dx^A\,dx^B$. 

I assume that the gravitational sector is invariant under both realizations 
of MRS, that is, under
\begin{eqnarray}
&&ds^2
\,\rightarrow
\,-\,ds^2~~~\mbox{as}~~~~
x^A\,\rightarrow\,i\,x^A\,,~~
x^{A^\prime}\,\rightarrow\,i\,x^{A^\prime}\nonumber \\
&&g_{AB}\,\rightarrow\,g_{AB}\,,~~
g_{A^\prime B^\prime}\,\rightarrow\,g_{A^\prime B^\prime}
\label{re1b2} \\
&&\Rightarrow~~\Omega_z\,\rightarrow\,\Omega_z
\,,~~\Omega_y\,\rightarrow\,\Omega_y \nonumber \\
&&g_{\mu\nu}\,\rightarrow\,g_{\mu\nu}
\,,~~\tilde{g}_{ab}\,\rightarrow\,\tilde{g}_{ab}
\,,~~\tilde{g}_{A^\prime B^\prime}\,\rightarrow\,\tilde{g}_{A^\prime
B^\prime} \label{re1b3}
\end{eqnarray}
and
\begin{eqnarray}
&&ds^2
\,\rightarrow
\,-\,ds^2~~~\mbox{as}~~~~
ky\,\rightarrow\,\pi\,-\,ky~,~~
k^\prime z\,\rightarrow\,\pi\,-\,k^\prime z \nonumber \\
&&x^A\,\rightarrow\,x^A\,,~~
x^{A^\prime}\,\rightarrow\,x^{A^\prime}
\label{re1b4} \\
&&\Rightarrow~~\Omega_z\,\rightarrow\,-\Omega_z
\,,~~\Omega_y\,\rightarrow\,-\Omega_y \nonumber \\
&&g_{\mu\nu}\,\rightarrow\,g_{\mu\nu}
\,,~~\tilde{g}_{ab}\,\rightarrow\,\tilde{g}_{ab}
\,,~~\tilde{g}_{A^\prime B^\prime}\,\rightarrow\,\tilde{g}_{A^\prime
B^\prime} ~~.\label{re1b5}
\end{eqnarray}
Note that the requirements of the homogeneity and isotropy of the 
4-dimensional space together with the equations (\ref{re1b2}-\ref{re1b5}) 
set $g_{\mu\nu}$ to the Minkowski metric $\eta_{\mu\nu}=diag(1,-1,-1,-1)$. 
Later (\ref{re1b3}) will be broken by a small amount in the 
matter sector to accommmodate the cosmic expansion.

The gravitational action is taken to be 
\begin{eqnarray}
S_R &=& \frac{1}{16\pi\,\tilde{G}}\int 
\,dV\,\tilde{R}^2 
\label{re1ca1} \\
dV&=&dV_1\,dV_2~,~~dV_1\,=\,\sqrt{g(-1)^S} \,d^Nx 
~,~~dV_2\,=\,
\sqrt{g^\prime
(-1)^{S^\prime}} 
\,d^{N^\prime}x^\prime \nonumber \\
\label{re1ca2} \\
\tilde{R}&=&R(x,x^\prime)+R^\prime(x,x^\prime) \label{re1ca3}
\end{eqnarray}
where the meaning of the primed and 
the unprimed quantities is evident from (\ref{re1b2}).
After integration over extra dimensions $S_R$ becomes
\begin{eqnarray} S_R &=& \frac{M^{N+N^\prime-4}}{16\pi\,\tilde{G}}\int 
\sqrt{(-1)^S g} 
\sqrt{(-1)^{S^\prime} g^\prime} 
\,2\,R(x)\,R^\prime(x^\prime) 
\,d^Nx
\,d^{N^\prime}x^\prime \nonumber \\
&=& 
 \frac{1}{16\pi\,G}
\int 
\sqrt{(-1)^S g} 
\,R(x)\, \,d^Nx
\label{re1ca10} 
\end{eqnarray}
where
\begin{equation}
 \frac{1}{16\pi\,G}\,=\, 
 M_{pl}^2(\frac{M}{M_{pl}})^2M^{N+N^\prime-6}
\frac{1}{16\pi\,\tilde{G}}\int 
\sqrt{(-1)^{S^\prime} g^\prime} 
\,2\,R^\prime(x^\prime) 
\,d^Dx^\prime \label{re1ca11}
\end{equation}
which is the usual Einstein-Hilbert action. The cosmological constant term is 
still forbidden by either realization of MRS.

Now we consider the subject that is the heart of this section, namely, 
the zero-point energies induced by the matter sector. Here we consider 
only the kinetic term of a scalar field here since this is enough to give the 
essential points of the formulation. The other details and consideration of 
the other fields can be found in \cite{re1Erdem4}. Consider the kinetic part of 
the Lagrangian, ${\cal L}_{Mk}$ for a 
scalar field (in 
the space given in (\ref{re1b2})
\begin{eqnarray}
&&{\cal L}_{\phi\,k} \,=\,
{\cal L}_{\phi\,k1}\,+\, 
{\cal L}_{\phi\,k2} 
\label{re1cba1} \\
&&{\cal L}_{\phi\,k1}\,=\, 
\frac{1}{2}g^{AB}\partial_A\phi\partial_B\phi~~,~~~
{\cal L}_{\phi\,k2} 
\,=\,\frac{1}{2}g^{A^\prime B^\prime}
\partial_{A^\prime}\phi\partial_{B^\prime}\phi
\label{re1cba2}
\end{eqnarray}
For simplicity I take $g_{\mu\nu}=\eta_{\mu\nu}$. Then the corresponding 
action is
\begin{eqnarray}
S_{Mk} &=& 
\int 
\,dV\,{\cal L}_{Mk} 
\nonumber \\
&=&
\frac{1}{2}\int 
\,\sqrt{(-1)^S g}\sqrt{(-1)^{S^\prime} g^\prime} \,d^Dx\,d^Dx^\prime
[\frac{1}{2}g^{AB}\partial_A\phi\partial_B\phi\,+\,
\frac{1}{2}g^{A^\prime B^\prime}
\partial_{A^\prime}\phi\partial_{B^\prime}\phi] \nonumber \\
&=&
\frac{1}{2}\int\,d^4x\,dy_1dy_2dz_1dz_2\,\Omega_z^3\Omega_y\,
\{\Omega_z^{-1}[\eta^{\mu\nu}\partial_\mu\phi\partial_\nu\phi\, 
-\,(\frac{\partial\phi}{\partial y_1})^2
\,-\,(\frac{\partial\phi}{\partial y_2})^2] \nonumber \\
&&
\,-\,\Omega_y[
(\frac{\partial\phi}{\partial z_1})^2
\,+\,
(\frac{\partial\phi}{\partial z_2})^2]\,
\} \nonumber \\
&=&
\frac{1}{2}LL^\prime\int\,d^4x\,\int_0^L\int_0^{L^\prime}dydz\,
\cos^3
{k^\prime z}
\cos{k y}
\{\cos^{-1}{k^\prime z}
[\eta^{\mu\nu}\partial_\mu\phi\partial_\nu\phi\, 
\,-\,(\frac{\partial\phi}{\partial y})^2] \nonumber \\
&&\,-\,\cos^{-1}{k y}
(\frac{\partial\phi}{\partial z})^2
\} 
\label{re1cba3}
\end{eqnarray}
where $y=y_2$, $z=z_2$ is adopted.

$\phi$ may be Fourier decomposed as
\begin{eqnarray}
&&\phi\,=\,
\phi_{AA}(x,y,z)\,+\,
\phi_{AS}(x,y,z)
\,+\,\phi_{SA}(x,y,z)
\,+\,\phi_{SS}(x,y,z)
\nonumber \\
\phi_{AA}(x,y,z)
&=&\sum_{n,m} \,\phi^{AA}_{n,m}(x)\,
\sin{(n\,ky)}
\,\sin{(m\,k^\prime z)}
\label{re1cb6} \\
\phi_{AS}(x,y,z)&=&\sum_{n,m} \,\phi^{AS}_{n,m}(x)\,
\sin{(n\,ky)}
\,\cos{(m\,k^\prime z)}
\label{re1cb7} \\
\phi_{SA}(x,y,z)&=&\sum_{n,m} \,\phi^{SA}_{n,m}(x)\,
\cos{(n\,ky)}
\,\sin{(m\,k^\prime z)}
\label{re1cb8} \\
\phi_{SS}(x,y,z)&=&\sum_{n,m} \,\phi^{SS}_{n,m}(x)\,
\cos{(n\,ky)}
\,\cos{(m\,k^\prime z)}
\label{re1cb9} \\
&&
k=\frac{\pi}{L}~,~
k^\prime=\frac{\pi}{L^\prime}~,~~
0\leq\,y\,\leq\,L 
~,~~0\leq\,z\,\leq\,L^\prime~~,~~~n,m=0,1,2,..... \nonumber
\end{eqnarray}
where we have used 
$k=\frac{\pi}{L}$,
$k^\prime=\frac{\pi}{L^\prime}$ since
$0\leq\,y\,\leq\,L$,
$0\leq\,z\,\leq\,L^\prime$. After replacing this expansion in (\ref{re1cba3}) 
and requiring the action be invariant under extra dimensional parity one 
obtains
\begin{eqnarray}
S_{Mk} &=& 
\frac{1}{2}LL^\prime\int\,d^4x\,\{
\eta^{\mu\nu}\sum_{n,m,r,s}\partial_\mu( 
\,\phi_{n,m}(x)\,)\,\partial_\nu(\,\phi_{r,s}(x)\,) \nonumber \\
&&\qquad\qquad\times\,\int_0^L\,dy\,
\cos{k y}\, 
\sin{(n\,k|y|)}
\sin{(r\,k|y|)}\nonumber\\
&&\qquad\qquad\qquad
\int_0^{L^\prime}\,dz\,
\cos^2{k^\prime z}
\sin{(m\,k^\prime|z|)})
\,\sin{(s\,k^\prime|z|)}) \nonumber \\
&&\,-k^2\,\,
\sum_{n,m,r,s}\,nr
\,\phi_{n,m}(x)\,\phi_{r,s}(x) 
\,\int_0^L\,dy\,
\cos{k y}\, 
\,\cos{(n\,k|y|)}
\cos{(r\,k|y|)} \nonumber \\
&&\qquad\qquad\times\,\int_0^{L^\prime}\,dz\,
\cos^2{k^\prime z}
\sin{(m\,k^\prime|z|)})
\,\sin{(s\,k^\prime|z|)}) \} \nonumber \\
&&\,-k^{\prime 2}\,\,
\sum_{n,m,r,s}\,ms
\,\phi_{n,m}(x)\,\phi_{r,s}(x) 
\,\int_0^L\,dy\,
\sin{(n\,k|y|)}
\sin{(r\,k|y|)} \nonumber \\
&&\qquad\qquad\times\,\int_0^{L^\prime}\,dz\,
\cos^3{k^\prime z}
\cos{(m\,k^\prime|z|)})
\,\cos{(s\,k^\prime|z|)})= \nonumber \\
\end{eqnarray}
\begin{eqnarray}
&=&
\frac{1}{32}(LL^\prime)^2\int\,d^4x\,\{
\eta^{\mu\nu}\sum_{r,s}
\nonumber \\
&&
\partial_\mu 
\bigl[\,\phi_{r-1,s-2}(x)\,+\,\phi_{r-1,s+2}(x)
\,-\,\phi_{r-1,-s-2}(x)\,-\,\phi_{r-1,2-s}(x)\nonumber\\
&& \,+\,2\phi_{r-1,s}(x)
-\,2\,\phi_{r-1,-s}(x) \,+\,\phi_{r+1,s-2}(x)\,\nonumber\\
&&+\,(\,\phi_{r+1,s+2}(x)
\,-\,\phi_{r+1,-s-2}(x)
 \,-\,\phi_{r+1,2-s}(x) \,\nonumber\\
&&+\,2\phi_{r+1,s}(x) 
-\,2\phi_{r+1,-s}(x)\,-\,\phi_{-r-1,s-2}(x)\,)\nonumber \\
&&\,-\,\phi_{-r-1,s+2}(x)\,+\,
\phi_{-r-1,-s-2}(x)\,+\,\phi_{-r-1,2-s}(x)
-\,2\phi_{-r-1,s}(x)\,\nonumber\\
&& +\,2\phi_{-r-1,-s}(x)
\,-\,\phi_{1-r,s-2}(x)\,-\,\phi_{1-r,s+2}(x)
\,+\,\phi_{1-r,-s-2}(x) \nonumber \\
&&+\,\phi_{1-r,2-s}(x)\,-\,2\phi_{1-r,s}(x)\,+\,2\phi_{1-r,-s}(x)
\,\bigr] \partial_\nu(\,\phi_{r,s}(x)\,) \nonumber \\
&&\,-k^2\,\,
\sum_{r,s}\,r
\bigl[\,(r-1)(\,\phi_{r-1,s-2}(x)\,-\,
\phi_{1-r,s-2}(x)\,)\,\nonumber\\
 &&+\,(r-1)(\phi_{r-1,s+2}(x)
\,-\,\phi_{1-r,s+2}(x)\,)\nonumber\\
&&-\,(r-1)(\,\phi_{r-1,-s-2}(x)\,-\,\phi_{1-r,-s-2}(x)\,)
\,\nonumber\\
&&-\,(r-1)(\,\phi_{r-1,2-s}(x)\,-\,\phi_{1-r,2-s}(x)\,) 
+\,2(r-1)(\,\phi_{r-1,s}(x)\,-\,\phi_{1-r,s}(x)\,)
\,\nonumber\\
 &&-\,2(r-1)(\,\phi_{r-1,-s}(x)\,-
\,\phi_{1-r,-s}(x)\,) \nonumber\\
&& +\,(r+1)(\,\phi_{r+1,s-2}(x)\,-\,\phi_{-r-1,s-2}(x)\,)\nonumber\\
&&\,+\,(r+1)(\,\phi_{r+1,s+2}(x)
\,-\,\phi_{-r-1,s+2}(x)\,)\nonumber\\
&&-\,(r+1)(\,\phi_{r+1,-s-2}(x)\,-\,\phi_{-r-1,-s-2}(x)\,)\nonumber\\
&&\,-\,(r+1)(\,\phi_{r+1,2-s}(x)\,-\,\phi_{-r-1,2-s}(x)\,)
+\,2(r+1)(\,\phi_{r+1,s}(x)\,-\,\phi_{-r-1,s}(x)\,)\nonumber\\
&&\,-\,2(r+1)(\,\phi_{r+1,-s}(x)\,-\,\phi_{-r-1,-s}(x)\,)\,\bigr]
\phi_{r,s}(x) \nonumber \\
&&\,-\frac{1}{2}k^{\prime 2}\,\,\sum_{r,s}\,s
\,\bigl[\,(s-3)(\,\phi_{r,s-3}(x)\,-\,\phi_{r,3-s}(x)\,)\nonumber\\
&&\,+\,(s+3)(\,\phi_{r,s+3}(x)\,-\,\phi_{r,-s-3}(x)\,) 
+\,3(s-1)(\,\phi_{r,s-1}(x)\,-\,\phi_{r,1-s}(x)\,)\nonumber\\
&&\,+\,3(s+1)(\,\phi_{r,s+1}(x)\,-\,\phi_{r,-s-1}(x)\,)
+\,(3-s)(\,\phi_{-r,s-3}(x)\,-\,\phi_{-r,3-s}(x)\,)\nonumber\\
&&\,+\,(s+3)(\,\phi_{-r,-s-3}(x)\,-\,\phi_{-r,s+3}(x)\,)\nonumber\\
&&+\,3(1-s)(\,\phi_{-r,s-1}(x)\,-\,\phi_{-r,1-s}(x)\,)\nonumber\\
&&\,-\,3(s+1)(\,\phi_{-r,s+1}(x)\,-\,\phi_{-r,-s-}(x)\,)\,\bigr]
\,\phi_{r,s}(x) \} . 
\label{re1cba5} 
\end{eqnarray}

We notice that the odd modes are coupled to even modes and vica versa. 
In fact this is a result of the invariance under 
$ky^a\rightarrow\,\pi-ky^a$, $k^\prime\,z^a\rightarrow\,\pi-k^\prime\,z^a$ 
that enforces the coupling of even and odd modes to compensate the minus 
coming from the volume element in the action. 

This makes the 
energy-momentum tensor $T_{\mu\nu}$ be of the same form in the extra 
dimensional Fourier modes. The replacement of the expansion of the modes in 
terms of creation and annihilation operators
\begin{equation}
\phi_{n,m}(x)\,=\,\sum_{\vec{k}}\,[\,
a_{n,m}(\vec{k})\,\zeta(t)
e^{i\vec{k}.\vec{x}}
\,+\,a_{n,m}^\dagger(\vec{k})\,\zeta^*(t)
e^{-i\vec{k}.\vec{x}}\,]
\label{re1d3}
\end{equation}
in the energy momentum tensor 
\begin{equation}
T_\mu^\nu=\sum_{m,n,r,s}
\partial_\mu\phi_{n,m}(x)\,
\partial^\nu\phi_{r,s}(x)\, -\,g_\mu^\nu{\cal L}
\end{equation}
results in the terms of the form
\begin{eqnarray}
<0|T^\nu_\mu|0>&\propto&
<0|\,a_{n,m}a_{r,s}^\dagger|0>\,=\,0~,~~
<0|\,a_{r,s}^\dagger 
a_{r,s}|0>\,=\,0 
\label{re1d2} \\
&&~~~~n\neq r~~~~\mbox{and/or}
~~~~m\neq s \nonumber
\end{eqnarray}
(because $a_{r,s}|0>\,=\,0$, and $[\,a_{n,m}, a_{r,s}^\dagger\,]\,=\,0$ 
for $n\neq r$ and/or $m\neq s$) .
In other words there is no contribution to vacuum energy density due to 
zero-point energies in this scheme. This solves the zero point-energy problem.

\section{Metric reversal symmetry and an interesting Kaluza-Klein spectrum}

In this section we shall consider a model where all except a finite number 
of 
Kaluza-Klein modes (i.e. the extra dimenional Fourier modes) are screened by 
the conformal factor in the metric. Note that both the form of 
the conformal factor and the form of the 
mixing of the Kaluza-Klein modes are determined by MRS. 
The details of the analysis given here may be found in \cite{re1Erdem5}.
In this scheme
it is not enough to produce a mode to in order to detect it 
have high enough energies to produce the mode but it is also necessary to 
have them high enough momenta relative to the detector (to expose to the 
sizes smaller than the extra dimension(s)). 
Therefore it has interesting phenomenological implications.

Adopt the following 5-dimensional space
\begin{eqnarray}
ds^2\;=\;\cos{k\,z}\,[\,\eta_{\mu\nu}(x)\,dx^\mu dx^\nu
-\,dz^2\,]
~~~~~~~~\mu,\nu\,=\,0,1,2,3
\label{re1c1}
\end{eqnarray}
where the extra dimension is taken to be compact and have the size $L$,
and $k\,=\,\frac{2\pi}{L}$. Consider fermions with the action
\begin{eqnarray}
S_f &=& \int (\cos{kz})^{\frac{5}{2}}\,{\cal L}_f\,d^4x\,dz
\nonumber \\
&=&
\int (\cos{kz})^2\,
\,i\bar{\chi}\gamma^{a}(\,\partial_a
\,+\,\frac{1}{16}\tan{kz}\,[\,\gamma_a\;,\;\gamma_{5}]\,)\chi
\;d^4x\,dz\;+\,H.C.
\label{re1c2} \\
&&\{\gamma^a,
\gamma^b\}\,=\,2\eta^{ab}~~,~~(\eta^{ab})=diag(1,-1,-1,-1,-1)
\nonumber
\end{eqnarray}
where $H.C.$ stands for Hermitian conjugate, and the term with the 
coefficient $\frac{1}{16}$ 
is the spin connection term.

We impose the following symmetries on the action
\begin{eqnarray}
k\,z\,&\rightarrow&\,\pi\,+\,k\,z . \label{re1c4}\\
x^a\,&\rightarrow&\,-x^a~~~a=0,1,2,3,4 \label{re1c6a} \\
\chi_n(x)
~&\rightarrow&~\epsilon^{\lambda_n}\,{\cal CPT}\,
\chi_n(-x)~~~~~,~~~~~\lambda_n=\frac{i}{2}(-1)^{\frac{n}{2}}
\label{re1c6b}
\end{eqnarray}
where the superscript $a$ refers to the tangent space, $\epsilon$ is some 
constant, and ${\cal CPT}$
denotes the usual 4-dimensional CPT operator (acting on the spinor part
of the field). We also impose anti-periodic boundary conditions in the 
extra dimension i.e. $\chi(x,z)=-\chi(x,z+2\pi\,L)$.

The extra dimensional Fourier expansion of 
$\chi$ in the light of invariance under 
\begin{eqnarray}
\chi&=&\chi_{\cal A}\,+\,\chi_{\cal S} \label{re1a4aa} \\
\chi_{\cal A}\left(x,z\right)
&=&
\sum_{|n|=1}^{\infty}
 \,\tilde{\chi}^{\cal A}_{|n|}\left(x\right)\,
\,\sin{\left(\frac{1}{2}|n|\,k z\right)} 
\label{re1a4a1} \\
\chi_{\cal S}\left(x,z\right)
&=&
\chi_0\left(x\right)\,+\,\sum_{|n|=1}^{\infty}\,
\tilde{\chi}^{\cal S}_{|n|}(x)\,\,\cos{\left(\frac{1}{2}|n|\,k  z\right)} 
\label{re1a4b1} \\
\tilde{\chi}^{{\cal A}({\cal S})}_{|n|}\left(x\right)
&=&\,\chi^{{\cal A}({\cal S})}_n\left(x\right)\,-(+)\,\chi^{{\cal A}({\cal 
S})}_{-n}(x) 
\nonumber
\end{eqnarray}
where $n$ are odd integers (due to anti-boundary conditions in the 
$z$-direction), the absolute value signs in $|n|$ is used to emphasize the 
positiveness of $n$ (due to the symmetry $x^a\rightarrow\,-x^a$). 

After replacing (\ref{re1a4aa}) in (\ref{re1c2}) we obtain
\begin{eqnarray}
&&\sum_{r,s=0}^\infty \int \,d^4x
\,i\bar{\chi}_{\left(2|r|+1\right)}
\gamma^{\bar{\mu}}\partial_{\bar{\mu}}\chi_{\left(2|s|+1\right)}
\times\,2\, \int\,dz\,\left(\cos{kz}\right)^2\,\nonumber\\
&&\qquad\left[\cos{\frac{2|r|+1}{2}kz}\,
\cos{\frac{2|s|+1}{2}kz}
\,-\,\sin{\frac{2|r|+1}{2}kz}\,\sin{\frac{2|s|+1}{2}kz}\,
\right]\,+\,H.C.\nonumber \\
&=&\sum_{r,s=0}^\infty \int \,d^4x
\;i\bar{\chi}_{\left(2|r|+1\right)}\gamma^{\bar{\mu}}\partial_{\bar{\mu}}
\chi_{\left(2|s|+1\right)}\,\nonumber\\
 &&\qquad\qquad\int_0^{L}\,dz\,
\left(\cos{2kz}+1\right)\,
\cos{\left(|r|+|s|+1\right)kz}\,\,+\,H.C. \nonumber \\
&=&\frac{1}{2}\sum_{r,s=0}^\infty \,\int \,d^4x
\,i\bar{\chi}_{\left(2|r|+1\right)}\gamma^{\bar{\mu}}\partial_{\bar{\mu}}
\chi_{\left(2|s|+1\right)} 
\,\int_0^{L}\,dz\,\left[\,\cos{\left(|r|+|s|-1\right)kz}\,\right]
\,+\,H.C.\nonumber\\
\label{re1a8}
\end{eqnarray}
where $2r+1=4l+1$, $2s+1=4p+3$ (l,p=0,1,2,....) or vica versa.
The non-zero contribution to (\ref{re1a8}) are due to the terms where the 
arguments of the cosine functions are zero (or multiples of $2\pi$) i.e. 
the modes that satisfy
\begin{eqnarray}
|r|+|s|-1=0~~~\Rightarrow~~~~~r=0~~,~~s=1
~~~~\mbox{or}~~~~
s=1~~,~~r=0 \label{re1a10}
\end{eqnarray}
Therefore the result of the integration in (\ref{re1a8}) is
\begin{eqnarray}
&&\frac{L}{2}\int \,d^4x
\left[\,i\bar{\chi}_{1}\gamma^{\bar{\mu}}\partial_{\bar{\mu}}\chi_{3}
\,+\,i\bar{\chi}_{3}\gamma^{\bar{\mu}}\partial_{\bar{\mu}}\chi_{1}
\,\right]\,+\,H.C.
\label{re1a11}\\
&&=\frac{1}{2}L
\int \,d^4x
\left[\,i\bar{\psi}\gamma^{\bar{\mu}}\partial_{\bar{\mu}}\psi
\,-\,i\bar{\tilde{\psi}}\gamma^{\bar{\mu}}\partial_{\bar{\mu}}\tilde{\psi} 
\,\right]\,+\,H.C.
\label{re1a12} \\
&&\psi\,=\,
\frac{1}{\sqrt{2}}\left(\,\chi_1\,+\,\chi_3\,\right)
~~,~~\tilde{\psi}\,=\,
\frac{1}{\sqrt{2}}\left(\,\chi_1\,-\,\chi_3\,\right) \label{re1a12a}
\end{eqnarray}
This means that at scales larger than the size of the extra dimension 
(which effectively corresponds to integration over the extra dimension) only 
one fermion and one ghost fermion is observed. The other modes are only 
observed at smaller scales while they are screened at large extra dimensional 
length scales due to the screening because of the (the cosine) form of 
the conformal factor. These modes can be observed only in 
interactions with energies higher than the inverse size of the extra 
dimension. 
At distances greater than the size 
of the extra dimension(s) even when they are already excited they seem 
hidden (unless they have high relative momenta when they interact in the 
detector). 
In other words these modes 
behave like a strange form of dark matter. The experimental predictions of 
this model are quite different from the usual Kaluza-Klein prescription 
and need further study. In high energy colliders the signature of these modes 
would be a sudden increase in the strength of the interactions and a high 
correlation between the interacting particles. In my opinion the 
phenomenological implications of this scheme deserves a separate and
detailed study by its own.  

\section{Metric reversal symmetry and an automatic Pauli-Villars-like 
regularization}

In the usual Kaluza-Klein scheme, Kaluza-Klein tower is an additional source 
of infinites that should be regulated. This property of compact extra 
dimensions is one of the major problems of quantum field theory in extra 
dimensions. On the other we will see
that in the spaces 
with metric reversal symmetry (MRS) there is the possibility 
of an automatic, Pauli-Villars-like on contary to the generic extra 
dimensional spaces. Below I summarize a model of this type. The details of 
this model may be found in \cite{re1Erdem6}.

Consider the following 7-dimensional space ($\mu,\nu\,=\,0,1,2,3$)
\begin{equation}
ds^2\;=\;g_{\mu\nu}(x)\,dx^\mu 
dx^\nu\,-\,
\cos^2{k_2y_2}\,[\,
\,dy_1^2\,+\,\cos^2{k_3y_3}
dy_2^2\,+\,dy_3^2\,]
\label{re1a1dup} 
\end{equation}
where the extra dimensions are compact and have the 
sizes $L_1$, $L_2$, $L_3$, and $k_1\,=\,\frac{2\pi}{L_1}$, 
$k_2\,=\,\frac{2\pi}{L_2}$, $k_3\,=\,\frac{2\pi}{L_3}$. 
Assume the symmetry 
\begin{eqnarray}
&&x^a\;\rightarrow\;-\,x^a~~,~~~~
a=0,1,2,3,5 \label{re1a3a} \\
&&x^b\;\rightarrow\;-\,x^b
~~,~~~~
b=0,1,2,3,6 
\label{re1a3b}
\end{eqnarray}
where $x^5=y_1$, $x^6=y_2$, $x^7=y_3$; and anti-periodic boundary conditions 
are adopted  for the 5th and 6th directions while  
periodic boundary condition is adopted for the 7th direction for the field 
$\chi$ i.e. $\chi(x,z)=-\chi(x,z+L)$ for $z=y_1$, $L=L_1$ or $z=y_2$, 
$L=L_2$ while $\chi(x,y_3)=\chi(x,y_3+L_3)$. Then the Fourier expansion of 
a field $\chi$ is 
\begin{eqnarray}
\chi(x,z)&=&\sum_{n=1}^{\infty}
\{\,f_{|n|}[
\cos{(\frac{|n|kz}{2})}+\sin{(\frac{|n|kz)}{2})}] \nonumber\\
&&+g_{|n|}[
\cos{(\frac{|n|kz)}{2})}-\sin{(\frac{|n|kz)}{2})}]\}
\chi_{|n|}(x) 
\label{re1a5} 
\end{eqnarray}
where 
$z=y_1,y_2$, $k=k_1,k_2$,  
$a_{|n|}$, $b_{|n|}$, $f_{|n|}$, $g_{|n|}$ are some constants. 
Even and odd $n$ correspond to periodic and anti-periodic 
boundary conditions \cite{re1conformal}, respectively.
The modes $\chi_n$ are taken to transform
under (\ref{re1a3a}) and (\ref{re1a3b}) as 
\begin{eqnarray}
&&\varphi_{n,m,r}(x)\,\rightarrow\,\xi^{\lambda_n}{\cal 
CPT}\,\varphi_{n,m,r}(-x)~~~~
\mbox{as}~~~~x^a\,\rightarrow\,-x^a \label{re1aa6}
\\
&&\varphi_{n,m,r}(x)\,\rightarrow\,\xi^{\lambda_m}{\cal 
CPT}\,\varphi_{n,m,r}(-x)~~~~\mbox{as}~~~~x^b\,\rightarrow\,-x^b 
\label{re1ab6} \\
&&\varphi_{n,m,r}(x)\,\rightarrow\,\xi^{\lambda_n+\lambda_m}{\cal 
CPT}\,\varphi_{n,m,r}(-x) \nonumber \\
&&~\mbox{as}~~x^a\,\rightarrow\,-x^a 
~,~x^b\,\rightarrow\,-x^b 
\label{re1a6} \\
&&\lambda_n=\frac{i}{2}(-1)^{\frac{n}{2}}
~~~\lambda_m=\frac{i}{2}(-1)^{\frac{m}{2}}~~~~~a=0,1,2,3,5~~;~~b=0,1,2,3,6 
\nonumber 
\end{eqnarray}
( where $n$, $m$, $r$ are the modes corresponding to $y_1$, $y_2$, 
$y_3$ directions, respectively; $\xi$ is some constant {\it other than 1 
or -1}, and ${\cal CPT}$ denotes the part of (4-dimensional) CPT 
transformation acting on the spinor part of the field.
I also impose the symmetry
\begin{eqnarray}
&&k_{1}y_{1}\;\rightarrow\;
k_{1}y_{1}\,+\,\pi \label{re1a4a} \\
&&k_{2}y_{2}\;\rightarrow\;
k_{2}y_{2}\,+\,\pi
\label{re1a4b}
\end{eqnarray}

In the light of the above observations I consider the following action \begin{eqnarray}
S_{fk1} 
&=& \int \,
\;d^4x\,\,d^3y\,
\cos^3{k_2y_2}\,\cos{k_3y_3}\frac{1}{2}
[{\cal L}_{fk11}\,+\,
{\cal L}_{fk12}]
\;+\,H.C. \label{re1a7}\\
{\cal L}_{fk11}\,&=&\,
\frac{i}{4}[(\bar{\chi}_{(1)}\gamma^{\mu}\,\partial_\mu\chi_{(3)} 
+\bar{\chi_{(1)}}^P\gamma^{\mu}\,\partial_\mu\chi_{(3)}^P) 
\,+\,y_1\rightarrow-y_1] \label{re1a7a} \\
{\cal L}_{fk12}\,&=&\,
\frac{i}{4}[(\bar{\chi}\gamma^{\mu}\,\partial_\mu\chi^P
-\bar{\chi}^P\gamma^{\mu}\,\partial_\mu\chi)+(y_1\rightarrow-y_1)] .
\label{re1a7b}
\end{eqnarray}

After inserting the explicit form of $\chi$ and imposing the symmetries  
(\ref{re1a3a}), (\ref{re1a3b}), (\ref{re1a6}), (\ref{re1a4b}) one finds 
\begin{eqnarray}
{\cal L}_{fk1}&=&
\frac{1}{2}[{\cal L}_{fk11}\,+\,
{\cal L}_{fk12}]\nonumber \\
&=&
\sum_{n_1,m_1=1}^\infty 
\,A_{n1,m1}^{(1,3)}
\,i\bar{\chi}_{n_1}(x,y)\gamma^\mu\partial_\mu\chi_{m_1}(x,y)
\,\cos{\frac{n_1+m_1}{2}k_1y_1}\,+\,H.C.\nonumber \\
&& \label{re1a7cc} 
\end{eqnarray}
where $y=y_2,y_3$.
The spectrum at the scales larger than the size of the extra dimensions 
may be found by integration of $[{\cal L}_{fk11}\,+\,
{\cal L}_{fk12}]$
over the extra dimensions.
\begin{eqnarray}
S_{fk1} 
&=& \int \,\;d^4x\,\,d^2y\,\cos^3{k_2y_2}\,\cos{k_3y_3}
\sum_{n_1,m_1=1}^\infty \,A_{n1,m1}^{(1,3)}
\,i\bar{\chi}_{n_1}(x,y)\gamma^\mu\partial_\mu\chi_{m_1}(x,y) \nonumber \\
&&\times\,\int\,dy_1\,\cos{\frac{n_1+m_1}{2}k_1y_1}\,+\,H.C.\;=\,0 
\label{re1a7c} \\
&&A_{n1,m1}^{(1,3)}\,=\,
(f_{n1}^*g_{m1}+g_{n1}^*f_{m1}+f_{n1}^*f_{m1}-g_{n1}^*g_{m1}) \nonumber 
\end{eqnarray}
The upper index $*$ denotes complex conjugate, $H.C.$ stands for 
Hermitian conjugate, and $f_n$, $g_n$'s 
are those given in (\ref{re1a5}). The subscripts $(1)$, $(3)$, and the 
superscripts $(1,3)$ above refer to the modes with $n=4p+1$ and $n=4p+3$, 
respectively, where $p=0,1,2,.....$. The $y_1\rightarrow-y_1$ terms in 
the above equations stands for the term obtained from the 
previous one by replacing $y_1$'s in that term by $-y_1$ and insures the 
invariance of the Lagrangian ${\cal L}_{fk1}$ under (\ref{re1a3a}). 
The values of $n_1$, $m_1$ in (\ref{re1a7b},\ref{re1a7c}) are fixed by the 
requirement of invariance under 
(\ref{re1a4a}), (\ref{re1aa6}), and are given by
\begin{equation}  
n_1\,=\,4l_1 \,+\,1~,~~ m_1\,=\,4p_1\,+\,3~~~~\mbox{or vica versa}~~~~~~~~
l_1,p_1=0,1,2,....... 
\label{re1a7d}
\end{equation}

It is evident that (\ref{re1a7c}) gives zero 
because $\int_0^{L_1} \,\cos{\frac{n_1+m_1}{2}k_1y_1}\,dy_1=0$
since $n_1+m_1\neq\,0$. Hence there are no observable fermions at scales 
larger than the sizes of the extra dimensions. Therefore an additional action 
must be introduced to accaount for the usual fermions while $S_{fk1}$ may be 
used for a Pauli-Villars-like regularization as we shall see. Assume that on 
the hyper-surface, $y_3=y_1$ the symmetry (\ref{re1a3a}) 
(and (\ref{re1aa6}) ) is broken by a small amount while there is an unbroken 
symmetry under the separate (and simultaneous) applications of 
(\ref{re1a4a1}), (\ref{re1a4b1}), and under the simultaneous application of 
(\ref{re1a3a}) and (\ref{re1a3b}) (and (\ref{re1aa6}) and (\ref{re1ab6} ). Consider the 
following action that obeys these conditions
\begin{eqnarray}
&&S_{fk2}\,=\epsilon\,\int\,\delta(k_3y_3-k_1y_1)
\cos^3{k_2y_2}\,\cos{k_3y_3} 
\frac{1}{2}[{\cal L}_{fk21}\,+\,
{\cal L}_{fk22}]
\;+\,H.C. \label{re1a9}\\
&&
{\cal L}_{fk21}\,=\,
\frac{i}{8}
[(\bar{\chi}_{(1,3)}\gamma^\mu\,\partial_\mu\chi_{(1,3)}+ 
\bar{\chi}_{(1,3)}^{P1,P2}\gamma^\mu\,\partial_\mu\chi_{(1,3)}^{P1,P2}-
\bar{\chi}_{(1,3)}^{P1}\gamma^\mu\,\partial_\mu\chi_{(1,3)}^{P1}-\nonumber\\
&&\bar{\chi}_{(1,3)}^{P2}\gamma^\mu\,\partial_\mu\chi_{(1,3)}^{P2})
+\,(y_{1,2}\rightarrow-y_{1,2})] 
\label{re1a9a} \\
&&{\cal L}_{fk22}\,=\,\nonumber\\
&&\frac{i}{8}[(\bar{\chi}_{(1,3)}\gamma^{\mu}\,\partial_\mu\chi_{(1,3)}^{P1}
+\bar{\chi}_{(1,3)}^{P1}\gamma^{\mu}\,\partial_\mu\chi_{(1,3)}
-\bar{\chi}_{(1,3)}^{P2}\gamma^{\mu}\,\partial_\mu\chi_{(1,3)}^{P1,P2}
-\bar{\chi}_{(1,3)}^{P1,P2}\gamma^{\mu}\,\partial_\mu\chi_{(1,3)}^{P2}
\nonumber \\
&&+\bar{\chi}_{(1,3)}\gamma^{\mu}\,\partial_\mu\chi_{(1,3)}^{P2} 
+\bar{\chi}_{(1,3)}^{P2}\gamma^{\mu}\,\partial_\mu\chi_{(1,3)}
-\bar{\chi}_{(1,3)}^{P1}\gamma^{\mu}\,\partial_\mu\chi_{(1,3)}^{P1,P2}
-\bar{\chi}_{(1,3)}^{P1,P2}\gamma^{\mu}\,\partial_\mu\chi_{(1,3)}^{P1}
\nonumber \\
&&+\bar{\chi}_{(1,3)}^{P1}\gamma^{\mu}\,\partial_\mu\chi_{(1,3)}^{P2}
+\bar{\chi}_{(1,3)}^{P2}\gamma^{\mu}\,\partial_\mu\chi_{(1,3)}^{P1} 
+\bar{\chi}_{(1,3)}\gamma^{\mu}\,\partial_\mu\chi_{(1,3)}^{P1,P2}
+\bar{\chi}_{(1,3)}^{P1,P2}\gamma^{\mu}\,\partial_\mu\chi_{(1,3)}) 
\nonumber \\
&&+(y_1\rightarrow-y_1)] \label{re1a9b}
\end{eqnarray}
where $\epsilon\,<<\,1$ is some constant that accounts for the breaking of 
the symmetry (\ref{re1a3a}) by a small amount. The superscripts $P1$, $P2$ 
refer to the $\chi$'s transformed under (\ref{re1a4a1}), (\ref{re1a4b1}), 
respectively. After replacing the Fourier expansion of $\chi$ and integrating over extra dimensions one obtains
\begin{eqnarray}
S_{fk2}\,&=&\,
\frac{\epsilon L_1L_2L_3}{4\pi}
(f_{1}^*g_{1}+g_{1}^*f_{1}+f_{1}^*f_{1}-g_{1}^*g_{1})\nonumber\\
&&(f_{3}^*g_{3}+g_{3}^*f_{3}+f_{3}^*f_{3}-g_{3}^*g_{3})
\int \,d^4x
\,i\bar{\chi}_{13}\gamma^\mu\partial_\mu\chi_{13}
\label{re1a9four} 
\end{eqnarray}
In other words only the mode $\chi_{13}$ is observed at large scales. If we take $n_3=0$ to be the lowest lying mode in $y_3$ direction then the usual fermions (i.e. the zero mode) are identified by $\chi_{130}$.

Although only $S_{fk2}$ is relavant on the brane $y_1=y_3$ and at large scales both of $S_{fk1}$ and $S_{fk2}$ are relavant on the brane. One must consider small patches in extra dimensional space to regulariza the affect of the delta function. Therefore we integrate ${\cal L}_{fk1}$ and $[{\cal L}_{fk21}+{\cal L}_{fk22}]$ on the patch 
\begin{equation} 
-\Delta\leq\,u\,\leq\Delta~,~~~ v\leq\,v^\prime\,\leq\,v+\Delta^\prime~~~,
~~~~u=k_1y_1-k_3y_3~,~~v=k_1y_1+k_3y_3 \label{re1b1dup}
\end{equation} 
The result of the integration may be expressed as
\begin{equation}
\int\,d^4x\,
{\cal L}_{eff}
\end{equation}
where
\begin{eqnarray}
{\cal L}^{eff}_{fk2} &=&
\frac{i}{2}\lim_{x^\prime\rightarrow x}
\partial_\mu\;\left(
\,\bar{\chi}_{130}(x^\prime),
\bar{\chi}_{310}(x^\prime)
\bar{\chi}_{330}(x^\prime)
\,\right)\,{\bf \tilde{M}}\,
\gamma^{\mu}
\left( 
\begin{array}{c}
\chi_{130}(x)\\
\chi_{310}(x) \\
\chi_{330}(x)
\end{array}
\right)
\label{re1ap3} 
\end{eqnarray}
here
\begin{eqnarray}
{\bf \tilde{M}}&=&\left(
\begin{array}{ccc}
\tilde{{\cal A}}&\tilde{{\cal B}}&\tilde{{\cal C}} \\
\tilde{{\cal B}}&
\tilde{{\cal D}}&0 \\
\tilde{{\cal C}}&0&0
\end{array}
\right) \label{re1ap4} 
\end{eqnarray}
where
\begin{eqnarray}
\tilde{{\cal A}}&\simeq&
\,\epsilon\cos^3{k_2y_2}
\sum_{p_1,s_1=0}^\infty
\tilde{A}_{p_1s_1}^{(1,1)}\,\tilde{T}_{p_1,s_1}^{(1,3)}(y_1) 
\sum_{p_2,s_2=0}^\infty
\tilde{A}_{p_2s_2}^{(3,3)}
\,\cos{[2(p_2+s_2)+1]k_2y_2} \nonumber \\
&&\label{re1ap4a} \\
\tilde{{\cal B}}&\simeq&
\cos^3{k_2y_2}
\sum_{p,s=0}^\infty A_{ps}^{(1)}(y_2)
\,T_{p,s}(y_1) 
\label{re1ap4b} \\
\tilde{{\cal C}}&\simeq&
\cos^3{k_2y_2}
\sum_{p,s=0}^\infty A_{ps}^{(3)}(y_2)
\,T_{p,s}(y_1) 
\label{re1a4c} \\
\tilde{{\cal D}}&\simeq&
\,\epsilon\cos^3{k_2y_2}
\sum_{p_1,s_1=0}^\infty
\tilde{A}_{p_1s_1}^{(3,3)}\,
\tilde{T}_{p_1,s_1}^{(3,1)}(y_1) 
\sum_{p_2,s_2=0}^\infty
\tilde{A}_{p_2s_2}^{(1,1)}
\,\cos{[2(p_2+s_2)+3]k_2y_2} \nonumber \\
\label{re1ap4d} 
\end{eqnarray}
here 
\begin{eqnarray}
&&
\tilde{T}_{p_1,s_1}^{(1,3)}(y_1)\,=\,
\frac{\Delta^\prime}{2}\{
\frac{\cos{(p_1+s_1+1)(k_1y_1+k_3y_3)}}{p_1+s_1+1}\nonumber\\
&&\qquad\qquad+\frac{\cos{(p_1+s_1)(k_1y_1+k_3y_3)}}{p_1+s_1}\} \label{re1ap4e1} \\
&&\tilde{T}_{p_1,s_1}^{(3,1)}(y_1)\,=\,
\frac{\Delta^\prime}{2}\{\frac{\cos{(p_1+s_1+2)(k_1y_1+k_3y_3)}}{p_1+s_1+2}\nonumber\\
&&\qquad\qquad+\frac{\cos{(p_1+s_1+1)(k_1y_1+k_3y_3)}}{p_1+s_1+1}\} \label{re1ap4e2} \\
&& T_{p,s}(y_1)\,=\,
\frac{\Delta^\prime}{(p+s)(p+s+1)}[\,
\sin{(p+s)\Delta}\,\cos{(p+s+1)(k_1y_1+k_3y_3)} \nonumber \\
&&\qquad\qquad+\sin{(p+s+1)\Delta}\,\cos{(p+s)(k_1y_1+k_3y_3)}
\label{re1ap4e3}
\end{eqnarray}
Note that $\Delta^\prime<<2\pi$ is employed in (\ref{re1ap4d}) and 
(\ref{re1ap4e1}-\ref{re1ap4e3}) since  
$\Delta$ and $\Delta^\prime$ should be taken as small as possible 
because my aim is to study point-wise as much as possible (while without 
causing any ambiguity due to the delta function on the brane).
Therefore  provided that $\epsilon\ll 1$
\begin{eqnarray}
\tilde{{\bf M}}\simeq
\left(
\begin{array}{ccc}
0&\tilde{{\cal B}}&\tilde{{\cal C}} \\
\tilde{{\cal B}}&\tilde{{\cal D}}&0 \\
\tilde{{\cal C}}&0&0
\end{array}
\right) \label{re1ap44} 
\end{eqnarray}
Hence the conclusions about the spectrum of the fields 
at the points $k_1y_1\,\neq\,k_3y_3$ essentialy remain the same at the 
points $k_1y_1\,=\,k_3y_3$ (or at the points $k_1y_1\,\simeq\,k_3y_3$). 
The diagonalization of ${\bf M}$ in (\ref{re1ap44}) results in 
\begin{eqnarray}
{\cal L}_{eff} \,\simeq\,i
B(y)[\bar{\psi}_1(x)\gamma^\mu\,\partial_\mu\psi_1(x)
\,-\,\bar{\psi}_2(x)\gamma^\mu\,\partial_\mu\psi_2(x)\,] \label{re1a20} 
\end{eqnarray}
\begin{eqnarray}
\psi_1&=&\frac{1}{2\sqrt{2}}
[\chi_{130}+
(\cos{\theta}\chi_{310}-\sin{\theta}\chi_{330})] \label{re1a20a} \\
\psi_2&=&\frac{1}{2\sqrt{2}}
[\chi_{130}-
(\cos{\theta}\chi_{310}-\sin{\theta}\chi_{330})]
\label{re1a20adup} \\
&&\tan{\theta}=\frac{{\cal B}}{{\cal C}}~~~,~~~
B(y)=\sqrt{({\cal B}^2+{\cal C}^2)}~~,~~~y=y_1,y_2 \label{re1a20b}
\end{eqnarray}
Hence the spectrum at scales smaller than the size of the extra 
dimension has a fermion and a ghost fermion coupled to each standard model 
fermion that appears at scales greater than the size of the extra 
dimensions.
There is another state $\psi_3\,=\,
\sin{\theta}\chi_{310}+\cos{\theta}\chi_{330}$ but this does not 
contribute to (\ref{re1a20}). So it is an auxiliary field. Although  
sign of the kinetic term of $\psi_2$ in (\ref{re1a20}) is opposite of a usual 
fermion (and so it is a ghost-like field) it does not suffer from the 
problems of the usual ghosts. $\psi_1$ or $\psi_2$ in (\ref{re1a20}) can not 
be introduced or removed from (\ref{re1a20}) because (\ref{re1a20}) follows from 
the couplings of $\chi_{130}$, $\chi_{310}$, $\chi_{330}$. So $\psi_1$, 
$\psi_2$ form a single system. For example in this case $\psi_1$, $\psi_2$ 
may be considered as the components of a single field with a 8-component 
spinor and the gamma matrices given by 
$\gamma^\mu\odot\tau_3$ where $\odot$ 
denotes tensor product and $\tau_3$ is the third Pauli matrix. This solves 
the problem of negative norm for $\psi_2$ because there is single norm 
i.e. that of the system composed of $\psi_1$, $\psi_2$. Moreover 
since $\psi_1$ and $\psi_2$ have the same internal space properties and 
they form a single system they may be assigned the same 4-momentum with 
positive energy, and this solves the negative energy problem of 
$\psi_2$. However the extension of this argument to the fields other than 
the fermions is not straightforward and requires additional study.

Eq.(\ref{re1a20}) implies an automatic regularization. The fermion ghost 
fermion pair at smaller scales naturally introduces a cut-off for the loop calculations. This may be seen better as follows: At scales larger than the size of the extra dimensions the relevant field is $\chi_{130}(x)$ and its propagator is
\begin{equation}
D(p)\,=\,\frac{i}{\not{p}+m} \label{re1a22a}
\end{equation}
(where $m$ is the mass of the field at scales larger than the sizes of 
extra dimensions (e.g. induced by Higgs mechanism)) while at smaller scales 
the relevant fields are $\psi_1$ and $\psi_2$ with the effective propagator
\begin{eqnarray}
D_{eff}(p)&=&D_1(p)\,+\,D_2(p)\sim
\frac{i}{B^\prime(\not{p}+m_1)}
-\frac{i}{B^\prime(\not{p}+m_2)} \nonumber \\
\,&=&\,i\frac{m_2-m_1}
{B^\prime(\not{p}+m_1)
(\not{p}+m_2)} \label{re1a22b} \\
B^\prime\,&=&\,N\,B(y)\cos^3{k_2y_2}\cos{k_3y_3} \nonumber
\end{eqnarray}
where $m_1$, $m_2$, in general, may depend on $y_1$, $y_2$, and I have 
assumed for sake of generality that $\psi_1$, $\psi_2$ may 
have two different effective masses at scales smaller than the size of 
extra dimensions that 
may be induced by spin connection terms, Higgs mechanism, or some other 
mechanism. For $m_1= m_2$ this equivalent to finite renormalization while 
for $m_1\neq\,m_2$ it is equivalent to Pauli-Villars regularization 
\cite{re1Pauli-Villars} at propagator level.

\section{Conclusion}
We have seen that metric reversal symmetry gives interesting results for a 
wide range of issues and problems in physics, namely, cosmological 
constant problem, zero-point energy problem, regularization of extra 
dimensional quantum field models, and Pauli-Villars regularization. We 
have also found a an interesting Kaluza- Klein spectrum that may give 
interesting signatures in accelerator experiments and quite 
different, non-conventional dark matter-like spectrum. Therefore the next 
step in this direction may be more realistic models of this form and 
a detailed study of their 
phenomenological implications.




\newcommand{\ahggh}{\mbox{$SU(3)$}}
\title{Masses and Mixing Matrices of Families of Quarks and
Leptons within a $SU(3)$ Flavor Symmetry Model with Radiative
Corrections and Dirac See-saw Mechanisms}
\author{A. Hern\'andez-Galeana}
\institute{%
Departamento de F\'{\i}sica,   Escuela Superior de
F\'{\i}sica y Matem\'aticas, I.P.N., \\
U. P. "Adolfo L\'opez Mateos". C. P. 07738, M\'exico, D.F.,
M\'exico.}

\titlerunning{Masses and Mixing Matrices of Families of Quarks\ldots}
\authorrunning{A. Hern\'andez-Galeana}
\maketitle

\begin{abstract}
I report new results on the study of fermion masses and quark
mixing within a $SU(3)$ flavor symmetry model, where ordinary
heavy fermions, top and bottom quarks and tau lepton become
massive at tree level from {\bf Dirac See-saw} mechanisms
implemented by a new heavy family of $SU(2)_L$ weak singlet vector-like
fermions, while light fermions get masses from one loop
radiative corrections mediated by the massive $SU(3)$ gauge
bosons. A recent quantitative analysis shows the existence of a
low energy space parameter which is able to accommodate the quark
and charged lepton masses as well as the quark mixing angle
$(V_{CKM})_{12}=0.2253$ with the gauge boson masses in the TeV
scale and a vector-like D quark of the order of $900\:GeV $. These
predictions may be tested at the LHC. Furthermore, the above
scenario enable us to suppress simultaneously the tree level
$\Delta F=2$ processes for $K^o-\bar{K^o}$ and $D^o-\bar{D^o}$
meson mixing mediated by these extra horizontal gauge bosons
within current experimental bounds.
\end{abstract}

\section{ Introduction }

The known hierarchical spectrum of quark masses and mixing as well
as the charged lepton masses have suggested to many model building
theorists that light fermion masses could be generated from
radiative corrections\cite{ahg2earlyradm}, while those of the top and
bottom quarks as well as that of the tau lepton are generated at
tree level. This may be understood as a consequence of the
breaking of a symmetry among families ( a horizontal symmetry ).
This symmetry may be discrete \cite{ahg2modeldiscrete}, or continuous,
\cite{ahg2modelcontinuous}. The radiative generation of the light
fermions may be mediated by scalar particles as it is proposed,
for instance, in references \cite{ahg2modelrad,ahg2medscalars} and the
author in \cite{ahg2prd2007}, or also through vectorial bosons as it
happens for instance in "Dynamical Symmetry Breaking" (DSB) and
theories like " Extended Technicolor " \cite{ahg2DSB}.

In this article we deal with the problem of fermion masses and
quark mixing within an extension of the SM introduced by the
author\cite{ahg2albinosu32004} which includes a $SU(3)$ gauged flavor
symmetry commuting with the SM group. In a previous
report\cite{ahg2albinosu32009} we showed that this model has the
ingredients to accommodate a realistic spectrum of charged fermion
masses and quark mixing. We introduce a hierarchical mass
generation mechanism in which the light fermions obtain masses
through one loop radiative corrections, mediated by the massive
bosons associated to the $\ahggh$ family symmetry that is
spontaneously broken, while the masses for the top and bottom
quarks as well as for the tau lepton, are generated at tree level
by the implementation of "Dirac See-saw" mechanisms implemented by
the introduction of a new generation of $SU(2)_L$ weak singlets
vector-like fermions. Recently, some authors have pointed out
interesting features regarding the possibility of the existence of
a sequential fourth generation, see for instance
\cite{ahg2fourthge}. Theories and models with extra matter may also provide
interesting scenarios for present cosmological problems, such as candidates
for the nature of the Dark Matter (\cite{ahg2normaapproach},\cite{ahg2khlopov}).
This is the case of an extra
generation of vector-like matter, both from theoretical and
current experiments. Due to the fact that the vector-like quarks
do not couple to the $W$ boson, the mixing of one $U$ and $D$
vector-like quarks with the SM quarks yield an extended $4\times
4$ non-unitary CKM quark mixing matrix. It has pointed out for
some authors \cite{ahg2vector-like-SU(2)-weak-singlets} that these
type of vector-like fermions are weakly constrained from
Electroweak Precison Data (EWPD) because they do not break
directly the custodial symmetry, then current experimental
constraints on vector-like matter come from the direct production
bounds and their implications on flavor physics. See ref.
\cite{ahg2vector-like-SU(2)-weak-singlets} for further details on
constraints for $SU(2)_L$ singlet vector-like fermions.

\vspace{3mm}

We report in this article a TeV energy space parameter
solution which accounts for the known quark and charged lepton
masses, for the $(V_{CKM})_{12}=0.2253$ Cabibbo mixing angle,
which predict a vector-like $D$ quark of several hundred GeV's and
simultaneously suppress within current bounds the $\Delta F=2$
processes for $K^o-\bar{K^o}$ and $D^o-\bar{D^o}$ meson mixing
mediated by the horizontal gauge bosons.

\section{Model with $SU(3)$ flavor symmetry}

\subsection{Fermion content}

We define the gauge group symmetry $G\equiv SU(3) \otimes G_{SM}$
, where $SU(3)$ is a flavor symmetry among families and
$G_{SM}\equiv SU(3)_C \otimes SU(2)_L \otimes U(1)_Y$ is the
"Standard Model" gauge group of elementary particles. The content
of fermions assumes the ordinary quarks and leptons assigned under
G as: $\Psi_q^o = ( 3 , 3 , 2 , \frac{1}{3} )_L  \;,\; \Psi_l^o =
( 3 , 1 , 2 , -1 )_L \;,\;\Psi_u^o = ( 3 , 3, 1 , \frac{4}{3} )_R
\;,\; \Psi_d^o = (3, 3 , 1 , -\frac{2}{3} )_R \;,\; \Psi_e^o = (3
, 1 , 1,-2)_R $, where the last entry corresponds to the
hypercharge $Y$, and the electric charge is defined by $Q = T_{3L}
+ \frac{1}{2} Y$. The model also includes two types of extra
fermions: Right handed neutrinos $\Psi_\nu^o = ( 3 , 1 , 1 , 0
)_R$, and the $SU(2)_L$ singlet vector-like fermions

 \begin{eqnarray}
U_{L,R}^o= ( 1 , 3 , 1 , \frac{4}{3} )  \qquad , \qquad D_{L,R}^o
= ( 1 , 3 , 1 ,- \frac{2}{3} )  \label{ahg2eq1} \\ N_{L,R}^o= ( 1 , 1 , 1 , 0 )
\qquad , \qquad E_{L,R}^o= ( 1 , 1 , 1 , -2 ) \label{ahg2eq2}\end{eqnarray}

The above fermion content and its assignment under the group G
make the model anomaly free. After the definition of the gauge
symmetry group and the assignment of the ordinary fermions in the
canonical form under the standard model group and in the
fundamental $3$-representation under the $SU(3)$ family symmetry,
the introduction of the right-handed neutrinos is required to
cancel anomalies\cite{ahg2T.Yanagida1979}. The $SU(2)_L$ weak singlets
vector-like fermions have been introduced to give masses at tree
level only to the heaviest family of known fermions through Dirac
See-saw mechanisms. These vector like fermions play a crucial role
to implement a hierarchical spectrum for quarks and charged lepton
masses together with the radiative corrections.

\section{Spontaneous Symmetry breaking}

 The "Spontaneous Symmetry Breaking" (SSB) is proposed to be achieved in the
form:

\begin{equation} G \stackrel{\Lambda_1}{\longrightarrow} SU(2)\otimes G_{SM}
\stackrel{\Lambda_2}{\longrightarrow} G_{SM}
\stackrel{\Lambda_3}{\longrightarrow} SU(3)_C \otimes U(1)_Q  \label{ahg2eq3}\end{equation}

\vspace{1mm} \noindent In order for the model to have the
possibility to be consistent with the known low energy physics,
here $\Lambda_1$, $\Lambda_2$ and $\Lambda_3$ are the scales of
SSB.

\subsection{Electroweak symmetry breaking}

To achieve the spontaneous breaking of the electroweak symmetry to
$U(1)_Q$,  we introduce the scalars: $\Phi = ( 3 , 1 , 2 , -1 )$
and $\Phi^{\prime} = ( 3 , 1 , 2 , +1 )$, with the VEV´s: $\langle
\Phi \rangle^T = ( \langle \Phi_1 \rangle , \langle \Phi_2 \rangle
, \langle \Phi_3 \rangle )$ , $\langle \Phi^{\prime} \rangle^T = (
\langle \Phi^{\prime}_1 \rangle , \langle \Phi^{\prime}_2 \rangle
, \langle \Phi^{\prime}_3 \rangle )$, where $T$ means transpose,
and

\begin{equation} \qquad \langle \Phi_i \rangle = \frac{1}{\sqrt[]{2}} \left(
\begin{array}{c} v_i
\\ 0  \end{array} \right) \qquad , \qquad
\langle \Phi^{\prime}_i \rangle = \frac{1}{\sqrt[]{2}} \left(
\begin{array}{c} 0
\\ V_i  \end{array} \right) \:.\end{equation}

\noindent Assuming $(v_1, v_2, v_3) \neq (V_1, V_2, V_3)$ with
$v_1^2+v_2^2+v_3^2=V_1^2+V_2^2+V_3^2 $, the contributions from
$\langle \Phi \rangle$ and $\langle \Phi^{\prime} \rangle$ yield
the $W$ gauge boson mass $\frac{1}{2} g^2 (v_1^2+v_2^2+v_3^2)
W^{+} W^{-} $. Hence, if we define as usual $M_W=\frac{1}{2} g v$,
we may write $ v=\sqrt{2} \sqrt{v_1^2+v_2^2+v_3^2} \thickapprox
246$ GeV.

\subsection{$SU(3)$ flavor symmetry breaking}

To implement a hierarchical spectrum for charged fermion masses,
and simultaneously to achieve the SSB of $SU(3)$, we introduce the
scalar fields: $\eta_i,\;i=1,2,3$, transforming under the gauge
group as $(3 , 1 , 1 , 0)$ and taking the "Vacuum Expectation
Values" (VEV's): \begin{equation} \langle \eta_3 \rangle^T = ( 0 , 0,
{\cal{V}}_3) \quad , \quad \langle \eta_2 \rangle^T = ( 0 ,
{\cal{V}}_2,0) \quad , \quad \langle \eta_1 \rangle^T = (
{\cal{V}}_1,0,0) \:. \end{equation}

\noindent The above scalar fields and VEV's break completely the
$SU(3)$ flavor symmetry. The corresponding $SU(3)$ gauge bosons
are defined in Eq.(\ref{ahg2eq12}) through their couplings to
fermions. To simplify computations, we impose a $SU(2)$ global
symmetry in the gauge boson masses. Actually this $SU(2)$
global symmetry plays the important role of a $SU(2)$ custodial
symmetry to suppress tree level FCNC in $\Delta F=2$ processes.
So, we assume ${\cal{V}}_1={\cal{V}}_2 \equiv
{\cal{V}}$ in order to cancel mixing between $Z_1$ and $Z_2$
horizontal gauge bosons. Thus, a natural hierarchy among the VEV´s
consistent with the proposed sequence of SSB in Eq.(\ref{ahg2eq3}) is
$ {\cal{V}}_3\:>>\:{\cal{V}} \; \gg
\;\sqrt{v_1^2+v_2^2+v_3^2}=\frac{v}{\sqrt{2}}\simeq
\frac{246\:\text{GeV}}{\sqrt{2}} \backsimeq 173.9 \:\text{GeV}
\approx \;m_t $. Hence, neglecting tiny contributions from
electroweak symmetry breaking, we obtain the gauge bosons
masses

\begin{multline} g_H^2 \left\{ \frac{1}{2} ({\cal{V}})^2
[\:Z_1^2+(Y_1^1)^2+(Y_1^2)^2\:] + \frac{1}{6}\:[\:2
({\cal{V}}_3)^2+({\cal{V}})^2\:]
 \:Z_2^2          \right. \\ \left. + \frac{1}{4}
(\:({\cal{V}}_3)^2+({\cal{V}})^2\:) [\:(Y_2^1)^2+(Y_2^2)^2
+(Y_3^1)^2+(Y_3^2)^2\:] \right\}
\end{multline}

\noindent Thus, we may define the horizontal boson masses
\begin{equation}
\begin{array}{rcl}(M_{Z_1})^2=(M_{Y_1^1})^2=(M_{Y_1^2})^2 & = & M_1^2
\equiv g_H^2 {{\cal{V}}}^2  \:,\\
(M_{Y_2^1})^2=(M_{Y_2^2})^2=(M_{Y_3^1})^2=(M_{Y_3^2})^2 & = & M_2^2
\equiv \frac{g_H^2}{2} ({{\cal{V}}_3}^2+{{\cal{V}}}^2 )   \\
(M_{Z_2})^2 & = & 4/3 M_2^2 - 1/3 M_1^2 \end{array} \:, \label{ahg2eq7}\end{equation}

\noindent with the hierarchy $ M_{Z_2} \gtrsim M_2 > M_1 \gg M_W$. Note that the
$SU(2)$ global symmetry in the gauge boson masses together with the
hierarchy of scales in the SSB yield a spectrum of $SU(3)$
gauge boson masses without mixing in quite good approximation.

\section{ Fermion masses}

\subsection{Dirac See-saw mechanisms}

Now we describe briefly the procedure to get the masses for
fermions. The analysis is presented explicitly for the charged
lepton sector, with a completely analogous procedure for the $u$
and $d$ quark sectors. With the fields of particles introduced in
the model, we may write the gauge invariant Yukawa couplings: \begin{equation}
h\bar{\Psi}_l^o \Phi^\prime E_R^o \;\;+\;\; h_3 \bar{\Psi}_e^o
\eta_3 E_L^o \;\;+\;\; h_2 \bar{\Psi}_e^o \eta_2 E_L^o \;\;+\;\;
h_1 \bar{\Psi}_e^o \eta_1 E_L^o \;\;+\;\ M \bar{E}_L^o E_R^o \;\;+
h.c \label{ahg2eq8}\end{equation}

\noindent where $M$ is a free mass parameter, because its mass
term is gauge invariant, and $h$, $h_1$, $h_2$ and $h_3$ are
Yukawa coupling constants. When the involved scalar fields acquire
VEV's we get, in the gauge basis ${\Psi^{o}_{L,R}}^T = ( e^{o} ,
\mu^{o} , \tau^{o}, E^o )_{L,R}$, the mass terms $\bar{\Psi}^{o}_L
{\cal{M}}^o \Psi^{o}_R + h.c $, where

\begin{equation} {\cal M}^o = \begin{pmatrix} 0 & 0 & 0 & h \:v_1\\ 0 & 0 & 0 & h \:v_2\\
0 & 0 & 0 & h \:v_3\\ - h_1 {\cal{V}} & - h_2 {\cal{V}} & h_3
{\cal{V}}_3 & M \end{pmatrix} \equiv \begin{pmatrix} 0 & 0 & 0 & a_1\\ 0 & 0 & 0 & a_2\\
0 & 0 & 0 & a_3\\ - b_1 & - b_2 & b_3 & c
\end{pmatrix} \;. \end{equation}

\noindent Notice that ${\cal{M}}^o$ has the same structure of a
See-saw mass matrix, but in this case for Dirac fermion masses.
So, we call ${\cal{M}}^o$ a {\bf "Dirac See-saw"} mass matrix.
${\cal{M}}^o$ is diagonalized by applying a biunitary
transformation $\Psi^{o}_{L,R} = V^{o}_{L,R} \;\chi_{L,R}^o$. The
orthogonal matrices $V^{o}_L$ and $V^{o}_R$ are obtained
explicitly in the Appendix A. From $V_L^o$ and $V_R^o$, and using
the relationships defined in this Appendix, one computes \begin{eqnarray}
{V^{o}_L}^T {\cal{M}}^{o} \;V^{o}_R =Diag(0,0,-
\sqrt{\lambda_-},\sqrt{\lambda_+})   \label{ahg2eq10}\\
                                  \nonumber   \\
{V^{o}_L}^T {\cal{M}}^{o} {{\cal{M}}^{o}}^T \;V^{o}_L = {V^{o}_R}^T
{{\cal{M}}^{o}}^T {\cal{M}}^{o} \;V^{o}_R =
Diag(0,0,\lambda_-,\lambda_+)  \:.\label{ahg2eq11}\end{eqnarray}

\noindent where $\lambda_-$ and $\lambda_+$ are the nonzero
eigenvalues defined in Eqs.(\ref{ahg2eq45}-\ref{ahg2eq46}). We see from
Eqs.(\ref{ahg2eq10},\ref{ahg2eq11}) that at tree level the See-saw
mechanism yields two massless eigenvalues associated to the light
fermions: $\sqrt{\lambda_+}$ for the fourth very heavy fermion,
and $\sqrt{\lambda_-}$ of the order of the heaviest ordinary
fermion top, bottom and tau  mass.

\subsection{One loop contribution to fermion masses}

\begin{figure}[htp]
\begin{center}
\includegraphics[width=12cm]{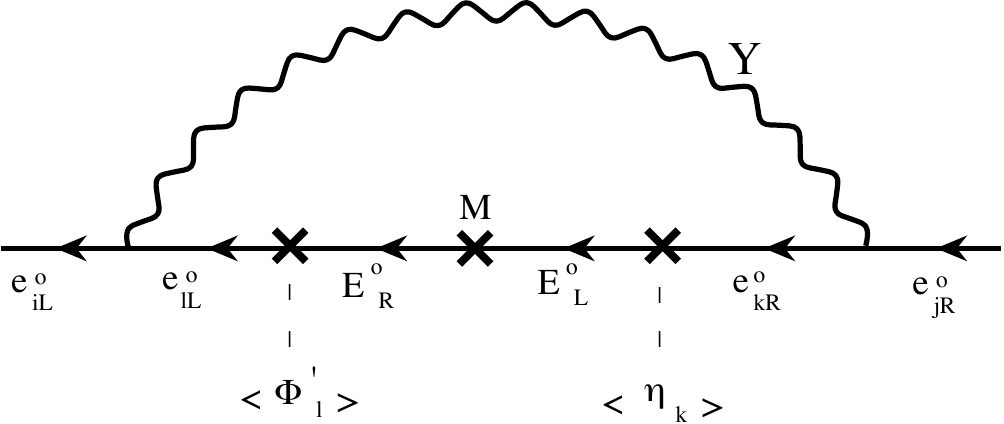}
\end{center}
\caption{\label{ahg2Fig1}Generic one loop diagram contribution to the mass term
$m_{ij} \:{\bar{e}}_{iL}^o e_{jR}^o$.}
\end{figure}

Subsequently, the masses for the light fermions arise through one
loop radiative corrections. After the breakdown of the electroweak
symmetry we can construct the generic one loop mass diagram of
Fig.~\ref{ahg2Fig1}. The vertices in this diagram come from the $SU(3)$ flavor
symmetry interaction Lagrangian
\begin{multline} i {\cal{L}}_{int} = \frac{g_{H}}{2} \left\{ (\bar{e^{o}}
\gamma_{\mu} e^{o}- \bar{\mu^{o}} \gamma_{\mu} \mu^{o})Z_1^\mu +
\frac{1}{\sqrt{3}}(\bar{e^{o}} \gamma_{\mu} e^{o}+ \bar{\mu^{o}}
\gamma_{\mu} \mu^{o} - 2 \bar{\tau^{o}}
\gamma_{\mu} \tau^{o})Z_2^\mu    \right.              \\
+ (\bar{e^{o}} \gamma_{\mu} \mu^{o}+ \bar{\mu^{o}} \gamma_{\mu}
e^{o})Y_1^{1 \mu}+(-i\bar{e^{o}} \gamma_{\mu}
\mu^{o}+ i\bar{\mu^{o}} \gamma_{\mu} e^{o})Y_1^{2 \mu}   \\
+(\bar{e^{o}} \gamma_{\mu} \tau^{o}+ \bar{\tau^{o}} \gamma_{\mu}
e^{o})Y_2^{1 \mu}+(-i\bar{e^{o}} \gamma_{\mu}
\tau^{o}+ i\bar{\tau^{o}} \gamma_{\mu} e^{o})Y_2^{2 \mu}  \\
+ \left. (\bar{\mu^{o}} \gamma_{\mu} \tau^{o}+ \bar{\tau^{o}}
\gamma_{\mu} \mu^{o})Y_3^{1 \mu}+(-i\bar{\mu^{o}} \gamma_{\mu}
\tau^{o}+ i\bar{\tau^{o}} \gamma_{\mu} \mu^{o})Y_3^{2 \mu}
\right\} \:,\label{ahg2eq12}\end{multline}

\noindent where $g_H$ is the $\ahggh$ coupling constant, $Z_1$, $Z_2$
and $Y_i^j\;,i=1,2,3\;,j=1,2$ are the eight gauge bosons. The
crosses in the internal fermion line mean tree level mixing, and
the mass $M$ generated by the Yukawa couplings in Eq.(\ref{ahg2eq8})
after the scalar fields get VEV's. The one loop diagram of Fig.~\ref{ahg2Fig1}
gives the generic contribution to the mass term $m_{ij}
\:{\bar{e}}_{iL}^o e_{jR}^o$

\begin{equation} c_Y \frac{\alpha_H}{\pi} \sum_{k=3,4} m_k^o
\:(V_L^o)_{ik}(V_R^o)_{jk} f(M_Y, m_k^o) \qquad , \qquad \alpha_H
\equiv \frac{g_H^2}{4 \pi} \end{equation}

\noindent  where $M_Y$ is the gauge boson mass, $c_Y$ is a factor
coupling constant, Eq.(12), $m_3^o=-\sqrt{\lambda_-}$ and
$m_4^o=\sqrt{\lambda_+}$ are the See-saw mass eigenvalues,
Eq.(\ref{ahg2eq10}), and $f(a,b)=\frac{a^2}{a^2-b^2}
\ln{\frac{a^2}{b^2}}$. Using again the results of Appendix A, we
compute 

\begin{equation} \sum_{k=3,4} m_k^o \:(V_L^o)_{ik}(V_R^o)_{jk} f(M_Y,
m_k^o)= \frac{a_i \:\beta_j \:M}{\lambda_+ - \lambda_-}\:
F(M_Y,\sqrt{\lambda_-},\sqrt{\lambda_+}) \:,\end{equation}

\noindent with $F(M_Y,\sqrt{\lambda_-},\sqrt{\lambda_+})\equiv
\frac{M_Y^2}{M_Y^2 - \lambda_+} \ln{\frac{M_Y^2}{\lambda_+}} -
\frac{M_Y^2}{M_Y^2 - \lambda_-} \ln{\frac{M_Y^2}{\lambda_-}}\:$,
$\beta_1= -b_1$, $\beta_2= -b_2$ and $\beta_3= b_3$. Adding up all
the one loop $\ahggh$ gauge boson contributions, we get in the gauge
basis the mass terms $\bar{\Psi^{o}_L} {\cal{M}}_1^o  \:\Psi^{o}_R
+ h.c.$,

\vspace{1mm}

\begin{equation} {\cal{M}}_1^o = \left( \begin{array}{ccrc} R_{11} & R_{12} & R_{13}  & 0\\
R_{21} & R_{22} & R_{23} & 0\\ R_{31} & R_{32} & R_{33} & 0\\
0 & 0 & 0 & 0
\end{array} \right) \:\frac{\alpha_H}{\pi}\;,
\end{equation}

\vspace{1mm}

\begin{equation} R_{11}=-\frac{1}{4} F_1 (m_{11} + 2 m_{22}) - \frac{1}{12}
F_{Z_2} m_{11} + \frac{1}{2} F_2 m_{33}  \:,\nonumber\end{equation}
\vspace{1mm}
\begin{equation} R_{22}=-\frac{1}{4} F_1 (2 m_{11} + m_{22}) -
\frac{1}{12} F_{Z_2} m_{22} + \frac{1}{2} F_2 m_{33} \:,\nonumber\end{equation}
\vspace{1mm}
\begin{equation} R_{12}=(\frac{1}{4} F_1 - \frac{1}{12} F_{Z_2}
)m_{12}\quad ,\quad R_{21}=(\frac{1}{4} F_1 - \frac{1}{12} F_{Z_2}
)m_{21} \;,\end{equation} \vspace{1mm}
\begin{equation} R_{33}=\frac{1}{3} F_{Z_2} m_{33} -
\frac{1}{2} F_2 (m_{11} + m_{22}) \quad ,\quad  R_{13}= -
\frac{1}{6} F_{Z_2} m_{13} \:,\nonumber            \end{equation} \vspace{1mm}
\begin{equation} R_{31}= \frac{1}{6} F_{Z_2} m_{31} \quad ,\quad R_{23}= -
\frac{1}{6} F_{Z_2} m_{23}  \quad ,\quad R_{32}= \frac{1}{6} F_{Z_2}
m_{32} \;, \nonumber \end{equation}

\vspace{1mm} \noindent Here, $F_{1,2} \equiv
F(M_{1,2},\sqrt{\lambda_-},\sqrt{\lambda_+})$ and $F_{Z2} \equiv
F(M_{Z_2},\sqrt{\lambda_-},\sqrt{\lambda_+})$, with $M_1 \:,\:M_2$
and $M_{Z_2}$ being the horizontal boson masses defined in
Eq.(\ref{ahg2eq7}),

\begin{equation} m_{ij}=\frac{a_i \:b_j \:M}{\lambda_+ - \lambda_-} = \frac{a_i
\:b_j}{a \:b} \:\sqrt{\lambda_-}\:c_{\alpha} \:c_{\beta} \:,\end{equation}

\noindent and $\cos\alpha \equiv c_{\alpha}\:,\;\cos\beta \equiv
c_{\beta}\:,\;\sin\alpha \equiv s_{\alpha}\:,\;\sin\beta \equiv
s_{\beta}$, as defined in the Appendix. Therefore, up to one loop
corrections we obtain the fermion masses

\begin{equation} \bar{\Psi}^{o}_L {\cal{M}}^{o} \:\Psi^{o}_R + \bar{\Psi^{o}_L}
{\cal{M}}_1^o \:\Psi^{o}_R = \bar{\chi_L^o} \:{\cal{M}}_1
\:\chi_R^o \:,\end{equation}

\vspace{1mm} \noindent with ${\cal{M}}_1\equiv  \left[ Diag(0,0,-
\sqrt{\lambda_-},\sqrt{\lambda_+})+ {V_L^o}^T {\cal{M}}_1^o
\:V_R^o \right]$; explicitly:

\begin{equation} {\cal{M}}_1= \left( \begin{array}{rrrr} q_{11}&q_{12}&c_\beta \:q_{13}&s_\beta \:q_{13} \\
q_{21}& q_{22} & c_\beta \:q_{23} & s_\beta \:q_{23}\\
c_\alpha \:q_{31}& c_\alpha \:q_{32} & -\sqrt{\lambda_-}+c_\alpha
c_\beta
\:q_{33} & c_\alpha s_\beta \:q_{33} \\
s_\alpha \:q_{31}& s_\alpha \:q_{32} & s_\alpha c_\beta \:q_{33} &
\sqrt{\lambda_+}+s_\alpha s_\beta \:q_{33}
\end{array} \right) \;,\label{ahg2eq19}\end{equation}

\noindent where the mass entries $q_{ij}\: ;i,j=1,2,3$ are written
as:

\begin{equation} q_{11}=c_2 \left[ \frac{H}{q} - u q (\frac{\Delta}{2}+J)
\right] \qquad  , \qquad q_{12}=- \frac{a_3}{a} c_1 \epsilon
\frac{H}{q}  \end{equation}

\begin{equation} q_{21}=\frac{b_3}{b} c_1 \epsilon \frac{H}{q} \qquad  , \qquad
q_{22}=c_1 \frac{H}{q} \qquad  , \qquad q_{23}=-
\frac{b^\prime}{b} c_1 \epsilon \frac{H}{q} \end{equation}

\begin{equation} q_{31}=c_2 \left[ \frac{a^\prime}{a_3} \frac{1}{q} \left( H -
\frac{1}{2} u q^2 \Delta \right) + \frac{b^\prime}{b_3} J \right]
\qquad  , \qquad q_{32}=- \frac{a^\prime}{a} c_1 \epsilon
\frac{H}{q}  \end{equation}

\vspace{3mm}

\begin{equation} q_{13}= - c_2 \left[ \frac{b^\prime}{b_3} \frac{1}{q} \left( H
- \frac{1}{2} u q^2 \Delta \right) + \frac{a^\prime}{a_3} J
\right] \end{equation}

\vspace{3mm}

\begin{equation} q_{33}=c_2 \left\{ -u H+J+\frac{1}{6} u^2 q^2 \Delta
-\frac{1}{3} \left[ u^2 q^2 F_1 + \left(
1+\frac{{a^\prime}^2}{a_3^2}+\frac{{b^\prime}^2}{b_3^2} \right)
\right] F_{Z_2} \right\} \end{equation}

\vspace{3mm}

\begin{equation} c_1=\frac{1}{2} c_\alpha c_\beta \frac{a_3 \:b_3}{a\:b}
\frac{\alpha_H}{\pi} \quad , \quad c_2= \frac{a_3 \:b_3}{a\:b}
\:c_1 \quad , \quad u=\frac{\eta_+}{a_3\:b_3} \quad , \quad
\epsilon=\frac{\eta_-}{\eta_+}        \nonumber \end{equation}

\begin{equation} \eta_-=a_1 \:b_2 - a_2 \:b_1 \; , \; \eta_+=a_1 \:b_1 + a_2
\:b_2 \; , \; \frac{a^\prime \:b^\prime}{a_3 \:b_3}=u\:q \:, \end{equation}

\begin{equation} q=\sqrt{1+\epsilon^2} \quad , \quad H=F_2 - u \:F_1 \quad ,
\quad J=F_{Z_2} - u \:F_2 \quad , \quad \Delta=F_{Z_2} - F_1\:.
\nonumber \end{equation}

\vspace{4mm}

\vspace{2mm} \noindent The diagonalization of ${\cal{M}}_1$,
Eq.(19), gives the physical masses for fermions in each sector u,
d and e, using a new biunitary transformation
$\chi_{L,R}^o=V_{L,R}^{(1)} \;\psi_{L,R}$; \;$\bar{\chi_L^o}
\;{\cal{M}}_1\;\chi_R^o= \bar{\psi_L} \:{V_L^{(1)}}^T
{\cal{M}}_1\: V_R^{(1)} \:\psi_R $, with ${\Psi_{L,R}}^T = ( f_1 ,
f_2 , f_3 , F )_{L,R}$ being the mass eigenfields, that is \begin{equation}
{V^{(1)}_L}^T {\cal{M}}_1 \:{\cal M}_1^T \;V^{(1)}_L =
{V^{(1)}_R}^T {\cal M}_1^T \:{\cal{M}}_1 \;V^{(1)}_R =
Diag(m_1^2,m_2^2,m_3^2,M_F^2) \:,\end{equation}

\noindent $m_1^2=m_e^2$, $m_2^2=m_\mu^2$, $m_3^2=m_\tau^2$ and
$M_F^2=M_E^2$ for charged leptons. Therefore, the final
transformation from massless to mass fermions eigenfields in this
scenario reads

\begin{equation} \Psi_L^o = V_L^{o} \:V^{(1)}_L \:\Psi_L \qquad \mbox{and}
\qquad \Psi_R^o = V_R^{o} \:V^{(1)}_R \:\Psi_R \end{equation}

\subsection{Quark Mixing and non-unitary $( V_{CKM} )_{4\times 4}$ }

The interaction of quarks ${f_{uL}^o}^T=(u^o,c^o,t^o)_L$ and
${f_{dL}^o}^T=(d^o,s^o,b^o)_L$ to the $W$ charged gauge boson
is\footnote{Recall that vector like quarks, Eqs.(\ref{ahg2eq1},\ref{ahg2eq2}), are $SU(2)_L$
weak singlets, and so, they do not couple to $W$ boson in the
interaction basis.} \begin{equation} \bar{f^o}_{u L} \gamma_\mu f_{d L}^o
{W^+}^\mu = \bar{\psi}_{u L}\;{V_{u L}^{(1)}}^T\;[(V_{u
L}^o)_{3\times 4}]^T \;(V_{d L}^o)_{3\times 4} \;V_{d
L}^{(1)}\;\gamma_\mu \psi_{d L} \;{W^+}^\mu \:,\end{equation}

\noindent hence, the non-unitary $V_{CKM}$ of dimension $4\times
4$ is identified as

\begin{equation} (V_{CKM})_{4\times 4}\equiv {V_{u L}^{(1)}}^T\;[(V_{u
L}^o)_{3\times 4}]^T \;(V_{d L}^o)_{3\times 4} \;V_{d L}^{(1)}
\:.\end{equation}

\begin{equation} V^o \equiv [(V_{u L}^o)_{3\times 4}]^T \;(V_{d L}^o)_{3\times
4} = \begin{pmatrix} \Omega_{11} & - \frac{s_o}{\sqrt{1+r_u^2}}
& c_\alpha^d \: \Omega_{13} & s_\alpha^d \: \Omega_{13} \\
\frac{s_o}{\sqrt{1+r_d^2}}   & c_o &  \frac{s_o r_d
c_\alpha^d}{\sqrt{1+r_d^2}} &
     \frac{s_o r_d  s_\alpha^d}{\sqrt{1+r_d^2}} \\
c_\alpha^u \: \Omega_{31}  & -  \frac{s_o r_u
c_\alpha^u}{\sqrt{1+r_u^2}} & c_\alpha^u \: c_\alpha^d \: \Omega_{33}
& c_\alpha^u \: s_\alpha^d \: \Omega_{33}  \\
s_\alpha^u \: \Omega_{31}  & -  \frac{s_o r_u
s_\alpha^u}{\sqrt{1+r_u^2}} & s_\alpha^u \: c_\alpha^d \:
\Omega_{33}   & s_\alpha^u \: s_\alpha^d \: \Omega_{33}
\end{pmatrix}  \:,\end{equation}

\vspace{3mm}

\begin{equation} \Omega_{11}=\frac{r_u r_d + c_o}{\sqrt{(1+r_u^2)(1+r_d^2)}}
\quad , \quad \Omega_{13}=\frac{r_d c_o -
r_u}{\sqrt{(1+r_u^2)(1+r_d^2)}} \end{equation}

\begin{equation} \Omega_{31}=\frac{r_u c_o - r_d}{\sqrt{(1+r_u^2)(1+r_d^2)}}
\quad , \quad  \Omega_{33}=\frac{r_u r_d c_o +
1}{\sqrt{(1+r_u^2)(1+r_d^2)}} \end{equation}

\begin{equation} s_o=\frac{v_1}{v^\prime}\:\frac{V_2}{V^\prime} -
\frac{v_2}{v^\prime}\:\frac{V_1}{V^\prime}  \quad , \quad
c_o=\frac{v_1}{v^\prime}\:\frac{V_1}{V^\prime} +
\frac{v_2}{v^\prime}\:\frac{V_2}{V^\prime} \end{equation}

\begin{equation} c_o^2+s_o^2=1 \quad , \quad r_u=(\frac{a^\prime}{a_3})_u \quad
, \quad r_d=(\frac{a^\prime}{a_3})_d        \end{equation}

\vspace{2mm} \noindent $V_i , \;v_i\;,i=1,2$ are related to (e,d)
and (u,$\nu$) fermion sectors respectively.

\section{Preliminary Numerical results}

Using the strong hierarchy for quarks and charged leptons masses
and the results in\cite{ahg2prd2007}, we report here the magnitudes of
quark masses and mixing coming from the analysis of a small region
of the parameter space in this model. For this numerical analysis
we used the input global parameters $\frac{\alpha_H}{\pi}=0.2$,
$M_1=1$ TeV and $M_2=400$ TeV.

\vspace{1mm}

\subsection{Sector d:}

Parameter space: $(\sqrt{\lambda_-})_d= 5.9884$ GeV,
$(\sqrt{\lambda_+})_d =890.886$ GeV, $r_d=0.015$, $u_d=0.2896$,
$\epsilon_d=1.5$, $s_\alpha^d=0.02$, and $s_\beta^d=0.3185$, lead to
the down quark masses: $m_d=5.5$ MeV, $m_s=120$ MeV, $m_b=4.2$
GeV, {\bf $M_D=890.899$ GeV}, and the mixing matrix

\small
\begin{equation} V_{d L}^{(1)}= \left(
\begin{array}{rrrr}
0.5547  &- 0.8318  &-0.0141  &2.24 \times 10^{-5} \\
0.8315  &0.5549  &-0.0225 &  3.56 \times 10^{-5} \\
0.0265&0.0007 &0.9996 & 6.75 \times 10^{-4}\\
-6.0 \times 10^{-5}& - 1.60 \times 10^{-6}& -6.74 \times 10^{-4} &
1
\end{array} \right) \:.\label{ahg2eq35}\end{equation}
\normalsize

\subsection{Sector u:}

Parameter space: $(\sqrt{\lambda_-})_u= 294.377$ GeV,
$(\sqrt{\lambda_+})_u =208.776$ TeV, $r_u=.004$, $u_u=0.9746$,
$\epsilon_u=0$, $s_\alpha^u=.01$ and $s_\beta^u=0.1396$ yield the
up quark masses $m_u=2.5$ MeV, $m_c=1.2$ GeV, $m_t=172$ GeV, $M_U=208.776$ TeV, and
the mixing

\begin{equation} V_{u L}^{(1)}=
\begin{pmatrix}
0.9999&0  &- 0.0035 &  4.10 \times 10^{-7} \\
0 & 1 &0 & 0\\
0.0035 & 0 &0.9999 & 8.26 \times 10^{-5} \\
-7.03 \times 10^{-7} & 0& -8.26 \times 10^{-5} & 1
\end{pmatrix}  \:.\end{equation}

\subsection{$(V_{CKM})_{4\times 4}$}

The See-saw $V^o$ contribution using the parameters;
$s_o=-0.6853$, $s_\alpha^d=0.02$, $s_\beta^d=0.3185$,
$s_\alpha^u=0.01$, $s_\beta^u=0.1396$  reads

\begin{equation} V^o=
\begin{pmatrix}
0.7281  &0.6853  &0.0069  &1.38 \times 10^{-4} \\
-0.6853  &0.7281  &- 0.0102 &  -2.05 \times 10^{-4} \\
- 0.0120&0.0027 &0.9996 & 0.0199\\
- 1.20 \times 10^{-4}& 2.74 \times 10^{-5}& 9.99 \times 10^{-3} &
1.99 \times 10^{-4}
\end{pmatrix} \end{equation}

The above up and down quark mixing matrices $V_{u L}^{(1)}$,
$V_{d L}^{(1)}$ and $V^o$, defined by the
See-saw mixing angles $s_\alpha^d$, $s_\beta^d$, $s_\alpha^u$,
$s_\beta^u$, and the values of parameters $r_u$ and $r_d$, yield
the non-unitary quark mixing

\vspace{4mm}

\begin{equation} (V_{CKM})_{4 \times 4} =
\begin{pmatrix}
.9741  &- 0.2253  &-0.0152  & 2.57 \times 10^{-4} \\
0.2250  &0.9742  &-0.0169 &  - 2.01 \times 10^{-4} \\
0.0186&0.0130 &0.9994 & 0.0206\\
2.23 \times 10^{-4}& 1.23 \times 10^{-4}& 0.0100 &
2.08 \times 10^{-4}
\end{pmatrix} \label{ahg2eq38}\end{equation}

\vspace{4mm}

\noindent Notice that the $(V_{CKM})_{3 \times 3}$ sub-matrix is nearly a
unitary mixing matrix, which is consistent with the measured values for quark mixing.
The entries $(V_{CKM})_{13}$, $(V_{CKM})_{23}$, $(V_{CKM})_{31}$ and $(V_{CKM})_{32}$ lie within the known
 orders of magnitude. However, a more detailed numerical analysis
 is needed to fit them within the allowed ranges reported in the PDG \cite{ahg2PDG2010}.

\subsection{Charged Leptons:}

For this sector, the parameter space: $(\sqrt{\lambda_-})_e=
6.7322$ GeV, $(\sqrt{\lambda_+})_e =2000$ TeV, $r_e=r_d=0.015$,
$u_e=1.0672$, $\epsilon_e=0$, $s_\alpha^e=9.8 \times  10^{-5}$ and
$s_\beta^e=0.0343$, reproduce the known charged lepton masses:
$m_e=0.511$ MeV , $m_\mu=105.658$ MeV and $m_\tau=1776.82$ MeV and $M_E \thickapprox 2000$ TeV

\subsection{FCNC's in $K^o-\bar{K^o}$ meson mixing}

\begin{figure}
\begin{center}
\includegraphics[width=0.8\columnwidth]{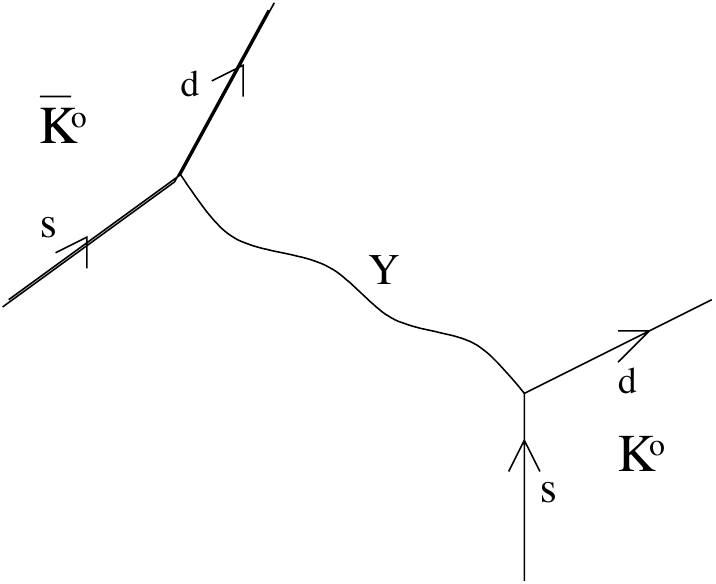}
\caption{\label{ahg2Fig2}Tree level contribution to  $K^o-\bar{K^o}$ from the light $SU(2)$ horizontal gauge bosons.}
\end{center}
\end{figure}

The $SU(3)$ horizontal gauge bosons contribute to new FCNC's, in
particular they mediate $\Delta F=2$ processes at tree level. Here
we compute their leading contribution to $K^o-\bar{K^o}$ meson
mixing. In the previous scenario the up quark sector does not
contribute to $(V_{CKM})_{12}$, and hence the effective
hamiltonian from the tree level diagrams, Fig.~\ref{ahg2Fig2}, mediated by the
$SU(2)$ horizontal gauge bosons of mass $M_1$ to the ${\cal
O}_{LL}(\Delta S=2)=(\bar{d}_L \gamma_\mu s_L)(\bar{d}_L
\gamma^\mu s_L)$ operator, is given by

\begin{equation} {\cal H }_{eff} = C_{\bar{d} s}\:{\cal O}_{LL} \quad , \quad
C_{\bar{d} s} \approx \frac{g_H^2}{4}  \frac{1}{M_1^2}
\frac{r_d^4}{(1+r_d^2)^2} (s_{12}^d)^2 \:,\end{equation}

\noindent and then contribute to the $K^o-\bar{K^o}$ mass
difference as

\begin{equation} \Delta m_K \approx \frac{2 \pi^2}{3}   \frac{\alpha_H}{\pi}
\frac{r_d^4}{(1+r_d^2)^2} (s_{12}^d)^2 \frac{F_K^2}{M_1^2}
B_K(\mu) M_K \:.\end{equation}

\noindent Using the input values: $r_d=0.015$, $\frac{\alpha_H}{\pi} =0.2 $ ,
$s_{12}^d=0.8318$, $F_K=160$ MeV, $M_K=497.614$ MeV and $B_K=0.8$, one gets

\begin{equation} \Delta m_K \thickapprox 4 \times {10}^{-13}\:\text{MeV} \; ,
\end{equation}

\noindent which is an order of magnitude lower than
the current experimental
bound\cite{ahg2PDG2010}, $(\Delta m_K)_{Exp} = M_{K_L} - M_{K_S}
\thickapprox 3.48 \times {10}^{-12}\:\text{MeV}$. The quark
mixing alignment in Eqs.(\ref{ahg2eq35} - \ref{ahg2eq38}) avoids tree
level contributions to $D^0-\bar{D^o}$ mixing mediated by the
$SU(2)$ horizontal gauge bosons.

\section{Conclusions}

We have reported a low energy parameter space within the $SU(3)$
flavor symmetry model extension, which combines tree level "Dirac
See-saw" mechanisms and radiative corrections to implement a
successful hierarchical spectrum for charged fermion masses and quark mixing. We
write explicitly the right predicted values for quark and charged
lepton masses and the quark mixing $(V_{CKM})_{12}=0.2253$, see Section
5, as a result of a particular space parameter with a lower scale
for the horizontal gauge bosons of $1$ TeV and a D vector-like
quark mass prediction of the order of $900$ GeV. So, these extra
particles introduced in the model are within the LHC expectations,
while simultaneously being consistent with current bounds on FCNC
in $K^o-\bar{K^o}$ and $D^o-\bar{D^o}$ meson mixing.

\section*{Appendix A: Diagonalization of the generic Dirac See-saw mass matrix}

\begin{equation} {\cal M}^o=
\begin{pmatrix} 0 & 0 & 0 & a_1\\ 0 & 0 & 0 & a_2\\ 0 & 0 & 0 &
a_3\\ - b_1 & - b_2 & b_3 & c \end{pmatrix} \end{equation}

\vspace{1mm} \noindent Using a biunitary transformation
$\Psi^{o}_L = V^{o}_L \;\chi_L^o$ and  $\Psi^{o}_R = V^{o}_R
\;\chi_R^o $ to diagonalize ${\cal{M}}^o$, the orthogonal matrices
$V^{o}_L$ and $V^{o}_R$ may be written explicitly as

\vspace{4mm}

\begin{equation} V^{o}_L = \left( \begin{array}{ccrr} \frac{v_1 v_3}{v^\prime
v}  & - \frac{v_2}{v^\prime}    & \frac{v_1}{v} \cos\alpha &
\frac{v_1}{v} \sin\alpha\\
\frac{v_2 v_3}{v^\prime v}  & \frac{v_1}{v^\prime}    &
\frac{v_2}{v} \cos\alpha &
\frac{v_2}{v} \sin\alpha\\
- \frac{v^\prime}{v} &   0  & \frac{v_3}{v} \cos{\alpha}
& \frac{v_3}{v} \sin{\alpha}\\
0 & 0 & -\sin{\alpha} & \cos{\alpha}
\end{array} \right) \end{equation}

\vspace{2mm}

\begin{equation} V^{o}_R = \left( \begin{array}{ccrr} \frac{b_1 b_3}{b^\prime
b} & - \frac{b_2}{b^\prime} & - \frac{b_1}{b} \cos{\beta} & -
\frac{b_1}{b} \sin{\beta}\\
\frac{b_2 b_3}{b^\prime b} & \frac{b_1}{b^\prime} & -
\frac{b_2}{b} \cos{\beta} & -
\frac{b_2}{b} \sin{\beta}\\
 \frac{b^\prime}{b}& 0 & \frac{b_3}{b} \cos{\beta} &
\frac{b_3}{b} \sin{\beta}\\ 0 & 0 & -\sin{\beta} & \cos{\beta}
\end{array} \right) \end{equation}

\vspace{1mm} \noindent where $a^\prime=\sqrt{a_1^2+a_2^2}\;, \;
b^\prime=\sqrt{b_1^2+b_2^2} \;, \;a=\sqrt{{a^\prime}^2+a_3^2} \; ,
\; b=\sqrt{{b^\prime}^2+b_3^2} \;,$

\vspace{1mm}

\begin{equation} \lambda_{\pm } = \frac{1}{2} \left( B \pm \sqrt{B^2 -4D} \right) \label{ahg2eq45}
\end{equation}

\vspace{1mm} \noindent are the nonzero eigenvalues of
${\cal{M}}^{o} {{\cal{M}}^{o}}^T$ (${{\cal{M}}^{o}}^T
{\cal{M}}^{o}$), with

\vspace{1mm} \begin{eqnarray} B = a^2 + b^2 + c^2 =
\lambda_{-}+\lambda_{+}\quad &, \quad D= a^2
b^2=\lambda_{-}\lambda_{+} \;,\label{ahg2eq46}\end{eqnarray}

\vspace{1mm}

 \begin{eqnarray} \cos{\alpha} =\sqrt{\frac{\lambda_+ -
a^2}{\lambda_+ - \lambda_-}} \qquad , \qquad \sin{\alpha} =
\sqrt{\frac{a^2
- \lambda_-}{\lambda_+ - \lambda_-}} \:,\nonumber \\
                                     \\
\cos{\beta} =\sqrt{\frac{\lambda_+ - b^2}{\lambda_+ - \lambda_-}}
\qquad , \qquad \sin{\beta} = \sqrt{\frac{b^2 -
\lambda_-}{\lambda_+ - \lambda_-}} \:.\nonumber \end{eqnarray}

\vspace{1mm}

\begin{eqnarray} \cos{\alpha}\: \cos{\beta}=
\frac{c\:\sqrt{\lambda_+}}{\lambda_+ - \lambda_-} \quad , \quad
\cos{\alpha} \:\sin{\beta}=
\frac{b\:c^2\:\sqrt{\lambda_+}}{(\lambda_+ - b^2)(\lambda_+ -
\lambda_-)}        \nonumber \\
                             \\
\sin{\alpha} \:\sin{\beta}= \frac{c\:\sqrt{\lambda_-}}{\lambda_+ -
\lambda_-} \quad , \quad \sin{\alpha} \:\cos{\beta}=
\frac{a\:c^2\:\sqrt{\lambda_+}}{(\lambda_+ - a^2)(\lambda_+ -
\lambda_-)} \nonumber \end{eqnarray}

\vspace{3mm} \noindent Note that in the space parameter $ a^2 \ll
c^2 \:,\:b^2 \; , \; \frac{\lambda_-}{\lambda_+} \ll 1$, so that we
may approach the eigenvalues as

\begin{equation} \lambda_- \approx \frac{D}{B} \approx \frac{a^2\:b^2}{c^2+b^2}
\qquad , \qquad \lambda_+ \approx c^2+b^2+a^2 -
\frac{a^2\:b^2}{c^2+b^2} \end{equation}

\section*{Acknowledgments}

I wish to thank all the organizers and colleagues, in particular
to Maxim Y. Khlopov, N.S. Mankoc-Borstnik, H.B. Nielsen and G.
Moultaka for very useful comments, discussions, and for this
stimulating Workshop at Bled, Slovenia. The author is thankful for
partial support from the "Instituto Polit\'ecnico Nacional",
(Grants from EDI and COFAA) and "Sistema Nacional de
Investigadores" (SNI) in Mexico.

\def\Journal#1#2#3#4{{#1} {\bf #2}, #3 (#4)}
\def\NCA{ Nuovo Cimento}
\def\RNC{ Rivista Nuovo Cimento}
\def\NIM{ Nucl. Instrum. Methods}
\def\NIMA{{ Nucl. Instrum. Methods} A}
\def\NPB{{ Nucl. Phys.} B}
\def\PLB{{ Phys. Lett.}  B}
\def\PRL{ Phys. Rev. Lett.}
\def\PRD{{ Phys. Rev.} D}
\def\ZPC{{ Z. Phys.} C}
\def\GaC{ Gravitation and Cosmology}
\def\GaCS{{ Gravitation and Cosmology} Supplement}
\def\JETP{ JETP}
\def\JETPL{ JETP Lett.}
\def\PAN{ Phys.Atom.Nucl.}
\def\CQG{ Class. Quantum Grav.}
\def\APJ{ Astrophys. J.}
\def\SCI{ Science}
\def\MPLA{{ Mod. Phys. Lett.}  A}
\def\IJTP{ Int. J. Theor. Phys.}
\def\NJP{ New J. of Phys.}
\def\JHEP{ JHEP}
\def\EPHJ{ Eur.Phys.J}
\def\BWP{ Bled Workshops in Physics}
\def\s{{\,\rm s}}
\def\g{{\,\rm g}}
\def\eV{\,{\rm eV}}
\def\keV{\,{\rm keV}}
\def\MeV{\,{\rm MeV}}
\def\GeV{\,{\rm GeV}}
\def\TeV{\,{\rm TeV}}
\def\sv{\left<\sigma v\right>}
\def\cm{{\,\rm cm}}
\def\kpc{{\,\rm kpc}}
\title{Dark Atoms of the Universe: Towards OHe Nuclear Physics}
\author{M.Yu. Khlopov${}^{1,2,3}$, A.G. Mayorov ${}^{1}$ and E.Yu.  Soldatov${}^{1}$}
\institute{%
${}^{1}$National Research Nuclear University "Moscow Engineering Physics Institute", 115409 Moscow, Russia \\
${}^{2}$ Centre for Cosmoparticle Physics "Cosmion" 115409 Moscow, Russia \\
${}^{3}$ APC laboratory 10, rue Alice Domon et L\'eonie Duquet \\75205
Paris Cedex 13, France}

\titlerunning{Dark Atoms of the Universe: Towards OHe Nuclear Physics}
\authorrunning{M.Yu. Khlopov, A.G. Mayorov and E.Yu.  Soldatov}
\maketitle

\begin{abstract}

The nonbaryonic dark matter of the Universe is assumed to consist of
new stable particles.
 A specific case is possible, when new stable particles bear ordinary
 electric charge and bind in heavy "atoms" by ordinary Coulomb
 interaction. Such possibility is severely restricted by
 the constraints on anomalous isotopes of light elements that
 form positively charged heavy species with ordinary electrons.
 The trouble is avoided, if stable particles $X^{--}$ with charge
 -2 are in excess over their antiparticles (with charge +2) and there are no stable
 particles  with charges +1 and -1. Then primordial helium, formed in Big Bang
 Nucleosynthesis, captures all $X^{--}$ in
 neutral "atoms" of O-helium (OHe).  Schrodinger equation for system of nucleus
 and OHe is considered and reduced to an equation of relative motion in a spherically symmetrical potential,
 formed by the Yukawa tail of nuclear scalar isoscalar attraction potential, acting on He beyond the nucleus,
 and dipole Coulomb repulsion between the nucleus and OHe at small distances between nuclear
 surfaces of He and nucleus. The values of coupling strength and mass of $\sigma$-meson,
 mediating scalar isoscalar nuclear potential, are rather uncertain. Within these
 uncertainties and in the approximation of rectangular potential wells and wall
 we find a range of these
 parameters, at which  the sodium nuclei have a few keV binding energy with OHe.
 The result also strongly depend on the precise value of parameter
 $d_o$ that determines the size of nuclei.
 At nuclear parameters, reproducing DAMA results, OHe-nucleus bound states can exist only for
 intermediate nuclei, thus excluding direct comparison with these results in detectors, containing
 very light (e.g. $^3He$) and heavy nuclei (like Xe).

\end{abstract}
\section{Introduction}
\label{contribution:mk1DarkAtoms}
Ordinary matter around us consists of neutral atoms, in which
electrically charged nuclei are bound with electrons. Ordinary
matter is luminous because of electron transitions in atoms. It is
stable owing to stability of its constituents. Electron is the
lightest charged particle. It is stable due to conservation of
electromagnetic charge that reflects local gauge U(1) invariance.
Electromagnetic charge is the source of the corresponding U(1) gauge
field, electromagnetic field. Nuclei are stable because of stability
of nucleons. The lightest nucleon - proton - is the lightest baryon
and stable due to conservation of baryon charge. There is no gauge
field related with baryon charge. Therefore there are two examples
of stable charged particles of the ordinary matter: protected by
gauge symmetry and protected by conserved charge. This excursus in
known physics can give us some idea on possible constituents of dark
atoms, maintaining the dark matter of the Universe.

According to the modern cosmology, the dark matter, corresponding to
$25\%$ of the total cosmological density, is nonbaryonic and
consists of new stable particles. Such particles (see e.g.
\cite{mk1rbook,mk1rCosmoarcheology,mk1rBled07} for review and reference) should
be stable, saturate the measured dark matter density and decouple
from plasma and radiation at least before the beginning of matter
dominated stage. The easiest way to satisfy these conditions is to
involve neutral elementary weakly interacting particles. However it
is not the only particle physics solution for the dark matter
problem and more evolved models of self-interacting dark matter are
possible. In particular, new stable particles may possess new U(1)
gauge charges and bind by Coulomb-like forces in composite dark
matter species. Such dark atoms would look nonluminous, since they
radiate invisible light of U(1) photons. Historically mirror matter
(see \cite{mk1rbook,mk1rOkun} for review and references) seems to be the
first example of such a nonluminous atomic dark matter. In the
studies of new particles Primordial Black holes can play the role of
important theoretical tool (see \cite{mk1rpbh} for review and
references), which in particular can provide constraints on
particles with hidden gauge charges \cite{mk1rDai:2009hx}.

Glashow's tera-helium \cite{mk1rGlashow} has offered a new solution for
dark atoms of dark matter. Tera-U-quarks with electric charge +2/3
formed stable (UUU) +2 charged "clusters" that formed with two -1
charged tera-electrons E neutral [(UUU)EE] tera-helium "atoms" that
behaved like Weakly Interacting Massive Particles (WIMPs). The main
problem for this solution was to suppress the abundance of
positively charged species bound with ordinary electrons, which
behave as anomalous isotopes of hydrogen or helium. This problem
turned to be unresolvable \cite{mk1rFargion:2005xz}, since the model
\cite{mk1rGlashow} predicted stable tera-electrons $E^-$ with charge -1.
As soon as primordial helium is formed in the Standard Big Bang
Nucleosynthesis (SBBN) it captures all the free $E^-$ in positively
charged $(He E)^+$ ion, preventing any further suppression of
positively charged species. Therefore, in order to avoid anomalous
isotopes overproduction, stable particles with charge -1 (and
corresponding antiparticles) should be absent, so that stable
negatively charged particles should have charge -2 only.

Elementary particle frames for heavy stable -2 charged species are
provided by: (a) stable "antibaryons" $\bar U \bar U \bar U$ formed
by anti-$U$ quark of fourth generation \cite{mk1rQ,mk1rI,mk1rlom,mk1rKhlopov:2006dk}
(b) AC-leptons \cite{mk1rKhlopov:2006dk,mk1r5,mk1rFKS}, predicted in the
extension \cite{mk1r5} of standard model, based on the approach of
almost-commutative geometry \cite{mk1rbookAC}.  (c) Technileptons and
anti-technibaryons \cite{mk1rKK} in the framework of walking technicolor
models (WTC) \cite{mk1rSannino:2004qp}. (d) Finally, stable charged
clusters $\bar u_5 \bar u_5 \bar u_5$ of (anti)quarks $\bar u_5$ of
5th family can follow from the approach, unifying spins and charges
\cite{mk1rNorma}. Since all these models also predict corresponding +2
charge antiparticles, cosmological scenario should provide mechanism
of their suppression, what can naturally take place in the
asymmetric case, corresponding to excess of -2 charge species,
$X^{--}$. Then their positively charged antiparticles can
effectively annihilate in the early Universe.

If new stable species belong to non-trivial representations of
electroweak SU(2) group, sphaleron transitions at high temperatures
can provide the relationship between baryon asymmetry and excess of
-2 charge stable species, as it was demonstrated in the case of WTC
\cite{mk1rKK,mk1rKK2,mk1runesco,mk1riwara}.

 After it is formed
in the Standard Big Bang Nucleosynthesis (SBBN), $^4He$ screens the
$X^{--}$ charged particles in composite $(^4He^{++}X^{--})$ {\it
O-helium} ``atoms''
 \cite{mk1rI}.
 For different models of $X^{--}$ these "atoms" are also
called ANO-helium \cite{mk1rlom,mk1rKhlopov:2006dk}, Ole-helium
\cite{mk1rKhlopov:2006dk,mk1rFKS} or techni-O-helium \cite{mk1rKK}. We'll call
them all O-helium ($OHe$) in our further discussion of their
cosmological effects, following the guidelines of \cite{mk1rI2}.

In all these forms of O-helium, $X^{--}$ behaves either as lepton or
as specific "heavy quark cluster" with strongly suppressed hadronic
interaction. Therefore O-helium interaction with matter is
determined by nuclear interaction of $He$. These neutral primordial
nuclear interacting objects contribute to the modern dark matter
density and play the role of a nontrivial form of strongly
interacting dark matter \cite{mk1rStarkman,mk1rMcGuire:2001qj}.

Here after a brief review of the qualitative picture of OHe
cosmological evolution \cite{mk1rI,mk1rFKS,mk1rKK,mk1runesco,mk1rKhlopov:2008rp} we
concentrate on some open questions in the properties of these dark
atoms and their interaction with matter. This analysis is used in
our second contribution to explain the puzzles of dark matter
searches \cite{mk1rDMDA}.

\section{Some features of O-helium Universe}

Following \cite{mk1rI,mk1rlom,mk1rKhlopov:2006dk,mk1rKK,mk1runesco,mk1riwara,mk1rI2} consider
charge asymmetric case, when excess of $X^{--}$ provides effective
suppression of positively charged species.

In the period $100{\,\rm s} \le t \le 300{\,\rm s}$  at $100 \keV\ge T \ge T_o=
I_{o}/27 \approx 60 \keV$, $^4He$ has already been formed in the
SBBN and virtually all free $X^{--}$ are trapped by $^4He$ in
O-helium ``atoms" $(^4He^{++} X^{--})$. Here the O-helium ionization
potential is\footnote{The account for charge distribution in $He$
nucleus leads to smaller value $I_o \approx 1.3 \MeV$
\cite{mk1rPospelov}.} \begin{equation} I_{o} = Z_{x}^2 Z_{He}^2 \alpha^2 m_{He}/2
\approx 1.6 \MeV,\label{mk1rIO}\end{equation} where $\alpha$ is the fine structure
constant,$Z_{He}= 2$ and $Z_{x}= 2$ stands for the absolute value of
electric charge of $X^{--}$.  The size of these ``atoms" is
\cite{mk1rI,mk1rFKS} \begin{equation} R_{o} \sim 1/(Z_{x} Z_{He}\alpha m_{He}) \approx 2
\cdot 10^{-13} \cm \label{mk1rREHe} \end{equation} Here and further, if not
specified otherwise, we use the system of units $\hbar=c=k=1$.

Due to nuclear interactions of its helium constituent with nuclei in
the cosmic plasma, the O-helium gas is in thermal equilibrium with
plasma and radiation on the Radiation Dominance (RD) stage, while
the energy and momentum transfer from plasma is effective. The
radiation pressure acting on the plasma is then transferred to
density fluctuations of the O-helium gas and transforms them in
acoustic waves at scales up to the size of the horizon.

At temperature $T < T_{od} \approx 200 S^{2/3}_3\eV$ the energy and
momentum transfer from baryons to O-helium is not effective
\cite{mk1rI,mk1rKK} because $$n_B \sv (m_p/m_o) t < 1,$$ where $m_o$ is the
mass of the $OHe$ atom and $S_3= m_o/(1 \TeV)$. Here \begin{equation} \sigma
\approx \sigma_{o} \sim \pi R_{o}^2 \approx
10^{-25}\cm^2\label{mk1rsigOHe}, \end{equation} and $v = \sqrt{2T/m_p}$ is the
baryon thermal velocity. Then O-helium gas decouples from plasma. It
starts to dominate in the Universe after $t \sim 10^{12}{\,\rm s}$  at $T
\le T_{RM} \approx 1 \eV$ and O-helium ``atoms" play the main
dynamical role in the development of gravitational instability,
triggering the large scale structure formation. The composite nature
of O-helium determines the specifics of the corresponding dark
matter scenario.

At $T > T_{RM}$ the total mass of the $OHe$ gas with density $\rho_d
= (T_{RM}/T) \rho_{tot} $ is equal to
$$M=\frac{4 \pi}{3} \rho_d t^3 = \frac{4 \pi}{3} \frac{T_{RM}}{T} m_{Pl}
(\frac{m_{Pl}}{T})^2$$ within the cosmological horizon $l_h=t$. In
the period of decoupling $T = T_{od}$, this mass  depends strongly
on the O-helium mass $S_3$ and is given by \cite{mk1rKK}\begin{equation} M_{od} =
\frac{T_{RM}}{T_{od}} m_{Pl} (\frac{m_{Pl}}{T_{od}})^2 \approx 2
\cdot 10^{44} S^{-2}_3 \g = 10^{11} S^{-2}_3 M_{\odot}, \label{mk1rMEPm}
\end{equation} where $M_{\odot}$ is the solar mass. O-helium is formed only at
$T_{o}$ and its total mass within the cosmological horizon in the
period of its creation is $M_{o}=M_{od}(T_{od}/T_{o})^3 = 10^{37}
\g$.

On the RD stage before decoupling, the Jeans length $\lambda_J$ of
the $OHe$ gas was restricted from below by the propagation of sound
waves in plasma with a relativistic equation of state
$p=\epsilon/3$, being of the order of the cosmological horizon and
equal to $\lambda_J = l_h/\sqrt{3} = t/\sqrt{3}.$ After decoupling
at $T = T_{od}$, it falls down to $\lambda_J \sim v_o t,$ where $v_o
= \sqrt{2T_{od}/m_o}.$ Though after decoupling the Jeans mass in the
$OHe$ gas correspondingly falls down
$$M_J \sim v_o^3 M_{od}\sim 3 \cdot 10^{-14}M_{od},$$ one should
expect a strong suppression of fluctuations on scales $M<M_o$, as
well as adiabatic damping of sound waves in the RD plasma for scales
$M_o<M<M_{od}$. It can provide some suppression of small scale
structure in the considered model for all reasonable masses of
O-helium. The significance of this suppression and its effect on the
structure formation needs a special study in detailed numerical
simulations. In any case, it can not be as strong as the free
streaming suppression in ordinary Warm Dark Matter (WDM) scenarios,
but one can expect that qualitatively we deal with Warmer Than Cold
Dark Matter model.

Being decoupled from baryonic matter, the $OHe$ gas does not follow
the formation of baryonic astrophysical objects (stars, planets,
molecular clouds...) and forms dark matter halos of galaxies. It can
be easily seen that O-helium gas is collisionless for its number
density, saturating galactic dark matter. Taking the average density
of baryonic matter one can also find that the Galaxy as a whole is
transparent for O-helium in spite of its nuclear interaction. Only
individual baryonic objects like stars and planets are opaque for
it.

\section{Signatures of O-helium dark matter in the Galaxy}
The composite nature of O-helium dark matter results in a number of
observable effects, which we briefly discuss following
\cite{mk1runesco}.
\subsection{Anomalous component of cosmic rays}
O-helium atoms can be destroyed in astrophysical processes, giving
rise to acceleration of free $X^{--}$ in the Galaxy.

O-helium can be ionized due to nuclear interaction with cosmic rays
\cite{mk1rI,mk1rI2}. Estimations \cite{mk1rI,mk1rMayorov} show that for the number
density of cosmic rays $ n_{CR}=10^{-9}\cm^{-3}$ during the age of
Galaxy a fraction of about $10^{-6}$ of total amount of OHe is
disrupted irreversibly, since the inverse effect of recombination of
free $X^{--}$ is negligible. Near the Solar system it leads to
concentration of free $X^{--}$ $ n_{X}= 3 \cdot 10^{-10}S_3^{-1}
\cm^{-3}.$ After OHe destruction free $X^{--}$ have momentum of
order $p_{X} \cong \sqrt{2 \cdot M_{X} \cdot I_{o}} \cong 2 \GeV
S_3^{1/2}$ and velocity $v/c \cong 2 \cdot 10^{-3} S_3^{-1/2}$ and
due to effect of Solar modulation these particles initially can
hardly reach Earth \cite{mk1rKK2,mk1rMayorov}. Their acceleration by Fermi
mechanism or by the collective acceleration forms power spectrum of
$X^{--}$ component at the level of $X/p \sim n_{X}/n_g = 3 \cdot
10^{-10}S_3^{-1},$ where $n_g \sim 1 \cm^{-3}$ is the density of
baryonic matter gas.

At the stage of red supergiant stars have the size $\sim 10^{15}
\cm$ and during the period of this stage$\sim 3 \cdot 10^{15} \s$,
up to $\sim 10^{-9}S_3^{-1}$ of O-helium atoms per nucleon can be
captured \cite{mk1rKK2,mk1rMayorov}. In the Supernova explosion these OHe
atoms are disrupted in collisions with particles in the front of
shock wave and acceleration of free $X^{--}$ by regular mechanism
gives the corresponding fraction in cosmic rays. However, this
picture needs detailed analysis, based on the development of OHe
nuclear physics and numerical studies of OHe evolution in the
stellar matter.

If these mechanisms of $X^{--}$ acceleration are effective, the
anomalous low $Z/A$ component of $-2$ charged $X^{--}$ can be
present in cosmic rays at the level $X/p \sim n_{X}/n_g \sim
10^{-9}S_3^{-1},$ and be within the reach for PAMELA and AMS02
cosmic ray experiments.

In the framework of Walking Technicolor model the excess of both
stable $X^{--}$ and $Y^{++}$ is possible \cite{mk1rKK2}, the latter
being two-three orders of magnitude smaller, than the former. It
leads to the two-component composite dark matter scenario with the
dominant OHe accompanied by a subdominant WIMP-like component of
$(X^{--}Y^{++})$ bound systems. Technibaryons and technileptons can
be metastable and decays of $X^{--}$ and $Y^{++}$ can provide
explanation for anomalies, observed in high energy cosmic positron
spectrum by PAMELA and in high energy electron spectrum by FERMI and
ATIC.

\subsection{Positron annihilation and gamma lines in galactic
bulge} Inelastic interaction of O-helium with the matter in the
interstellar space and its de-excitation can give rise to radiation
in the range from few keV to few  MeV. In the galactic bulge with
radius $r_b \sim 1 \kpc$ the number density of O-helium can reach
the value $n_o\approx 3 \cdot 10^{-3}/S_3 \cm^{-3}$ and the
collision rate of O-helium in this central region was estimated in
\cite{mk1rI2}: $dN/dt=n_o^2 \sigma v_h 4 \pi r_b^3 /3 \approx 3 \cdot
10^{42}S_3^{-2} \s^{-1}$. At the velocity of $v_h \sim 3 \cdot 10^7
\cm/\s$ energy transfer in such collisions is $\Delta E \sim 1 \MeV
S_3$. These collisions can lead to excitation of O-helium. If 2S
level is excited, pair production dominates over two-photon channel
in the de-excitation by $E0$ transition and positron production with
the rate $3 \cdot 10^{42}S_3^{-2} \s^{-1}$ is not accompanied by
strong gamma signal. According to \cite{mk1rFinkbeiner:2007kk} this rate
of positron production for $S_3 \sim 1$ is sufficient to explain the
excess in positron annihilation line from bulge, measured by
INTEGRAL (see \cite{mk1rintegral} for review and references). If $OHe$
levels with nonzero orbital momentum are excited, gamma lines should
be observed from transitions ($ n>m$) $E_{nm}= 1.598 \MeV (1/m^2
-1/n^2)$ (or from the similar transitions corresponding to the case
$I_o = 1.287 \MeV $) at the level $3 \cdot 10^{-4}S_3^{-2}(\cm^2 \s
\MeV ster)^{-1}$.

It should be noted that the nuclear cross section of the O-helium
interaction with matter escapes the severe constraints
\cite{mk1rMcGuire:2001qj} on strongly interacting dark matter particles
(SIMPs) \cite{mk1rStarkman,mk1rMcGuire:2001qj} imposed by the XQC experiment
\cite{mk1rXQC}. Therefore, a special strategy of direct O-helium  search
is needed, as it was proposed in \cite{mk1rBelotsky:2006fa}.

\section{O-helium interaction with nuclei}
The evident consequence of the O-helium dark matter is its
inevitable presence in the terrestrial matter, which appears opaque
to O-helium and stores all its in-falling flux. After they fall down
terrestrial surface, the in-falling $OHe$ particles are effectively
slowed down due to elastic collisions with matter.In underground
detectors, $OHe$ ``atoms'' are slowed down to thermal energies and
give rise to energy transfer $\sim 2.5 \cdot 10^{-4} \eV A/S_3$, far
below the threshold for direct dark matter detection. It makes this
form of dark matter insensitive to the severe CDMS constraints
\cite{mk1rAkerib:2005kh}. However, $OHe$ induced processes in the matter
of underground detectors can result in observable effects. These
effects, considered in a separate contribution \cite{mk1rDMDA}, strongly
depend on the details of the OHe interaction with nuclei, which we
consider here.

\subsection{Structure of $X^{--}$ atoms with nuclei}
The properties of OHe interaction with matter are determined first
of all by the structure of OHe atom that follows from the general
analysis of the bound states of $X^{--}$ with nuclei.

Consider a simple model \cite{mk1rPospelov}, in which the nucleus is
regarded as a sphere with uniform charge density and in which the
mass of the $X^{--}$ is assumed to be much larger than that of the
nucleus. Spin dependence is also not taken into account so that both
the particle and nucleus are considered as scalars. Then the
Hamiltonian is given by
\begin{equation}
    H=\frac{p^2}{2 A m_p} - \frac{Z Z_x \alpha}{2 R} + \frac{Z Z_x \alpha}{2 R} \cdot (\frac{r}{R})^2,
\end{equation}
for short distances $r<R$ and
\begin{equation}
    H=\frac{p^2}{2 A m_p} - \frac{Z Z_x \alpha}{R},
\end{equation}
for long distances $r>R$, where $\alpha$ is the fine structure
constant, $R = d_o A^{1/3} \sim 1.2 A^{1/3} /(200 MeV)$ is the
nuclear radius, $Z$ is the electric charge of nucleus and $Z_x=2$ is
the electric charge of negatively charged particle $X^{--}$. Since
$A m_p \ll M_X$ the reduced mass is $1/m= 1/(A m_p) + 1/M_X \approx
1/(A m_p)$.

For small nuclei the Coulomb binding energy is like in hydrogen atom
and is given by
\begin{equation}
    E_b=\frac{1}{2} Z^2 Z_x^2 \alpha^2 A m_p.
\end{equation}

For large nuclei $X^{--}$ is inside nuclear radius and the harmonic
oscillator approximation is valid for the estimation of the binding
energy
\begin{equation}
    E_b=\frac{3}{2}(\frac{Z Z_x \alpha}{R}-\frac{1}{R}(\frac{Z Z_x \alpha}{A m_p R})^{1/2}).
\label{mk1rpotosc}
\end{equation}

For the intermediate regions between these two cases with the use of
trial function of the form $\psi \sim e^{- \gamma r/R}$ variational
treatment of the problem \cite{mk1rPospelov} gives
\begin{equation}
    E_b=\frac{1}{A m_p R^2} F(Z Z_x \alpha A m_p R ),
\end{equation}
where the function $F(a)$ has limits
\begin{equation}
    F(a \rightarrow 0) \rightarrow \frac{1}{2}a^2  - \frac{2}{5} a^4
\end{equation}
and
\begin{equation}
    F(a \rightarrow \infty) \rightarrow \frac{3}{2}a  - (3a)^{1/2},
\end{equation}
where $a = Z Z_x \alpha A m_p R$. For $0 < a < 1$ the Coulomb model
gives a good approximation, while at $2 < a < \infty$ the harmonic
oscillator approximation is appropriate.

In the case of OHe $a = Z Z_x \alpha A m_p R \le 1$, what proves its
Bohr-atom-like structure, assumed in our earlier papers
\cite{mk1rI,mk1rlom,mk1rKhlopov:2006dk,mk1rKK,mk1runesco,mk1riwara,mk1rI2}. However, the size of
He, rotating around $X^{--}$ in this Bohr atom, turns out to be of
the order and even a bit larger than the radius $r_o$ of its Bohr
orbit, and the corresponding correction to the binding energy due to
non-point-like charge distribution in He is significant.

Bohr atom like structure of OHe seems to provide a possibility to
use the results of atomic physics for description of OHe interaction
with matter. However, the situation is much more complicated. OHe
atom is similar to the hydrogen, in which electron is hundreds times
heavier, than proton, so that it is proton shell that surrounds
"electron nucleus". Nuclei that interact with such "hydrogen" would
interact first with strongly interacting "protonic" shell and such
interaction can hardly be treated in the framework of perturbation
theory. Moreover in the description of OHe interaction the account
for the finite size of He, which is even larger than the radius of
Bohr orbit, is important. One should consider, therefore, the
analysis, presented below, as only a first step approaching true
nuclear physics of OHe.

\subsection{Potential of O-helium interaction with nuclei}

Our explanation \cite{mk1runesco,mk1riwara,mk1rBled09} of the results of
DAMA/NaI \cite{mk1rBernabei:2003za} and DAMA/LIBRA
\cite{mk1rBernabei:2008yi} experiments is based on the idea that OHe,
slowed down in the matter of detector, can form a few keV bound
state with nucleus, in which OHe is situated \textbf{beyond} the
nucleus. Therefore the positive result of these experiments is
explained by reaction
\begin{equation}
A+(^4He^{++}X^{--}) \rightarrow [A(^4He^{++}X^{--})]+\gamma
\label{mk1rHeEAZ}
\end{equation}
with nuclei in DAMA detector.

In our earlier studies \cite{mk1runesco,mk1riwara,mk1rBled09} the conditions
were found, under which both sodium and iodine nuclei have a few keV
bound states with OHe, explaining the results of DAMA experiments by
OHe radiative capture to these levels. Here we extend the set of our
solutions by the case, when the results of DAMA experiment can be
explained by radiative OHe capture by sodium only and there are no
such bound states with iodine and Tl.

Schrodinger equation for OHe-nucleus system is reduced (taking apart
the equation for the center of mass) to the equation of relative
motion for the reduced mass
\begin{equation}
            m=\frac{Am_p m_o}{Am_p+m_o},
            \label{mk1rm}
 \end{equation}
where $m_p$ is the mass of proton and $m_o\approx M_X+4m_p$ is the
mass of OHe. Since $m_o \approx M_X \gg A m_p$, center of mass of
Ohe-nucleus system approximately coincides with the position of
$X^{--}$.

In the case of orbital momentum \emph{l}=0 the wave functions depend
only on \emph{r}.

The approach of \cite{mk1runesco,mk1riwara,mk1rBled09} assumes the following
picture: at the distances larger, than its size, OHe is neutral,
being only the source of a Coulomb field of $X^{--}$ screened by
$He$ shell
\begin{equation}
U_c= \frac{Z_{X} Z \alpha  \cdot F_X(r)}{r}, \label{mk1repotem}
\end{equation}
where $Z_{X}=-2$ is the charge of $X^{--}$, $Z$ is charge of
nucleus, $F_X(r)=(1+r/r_o) exp(-2r/r_o)$ is the screening factor of
Coulomb potential (see e.g.\cite{mk1rLL3}) of $X^{--}$ and $r_o$ is the
size of OHe. Owing to the negative sign of $Z_{X}=-2$, this
potential provides attraction of nucleus to OHe.

Then helium shell of OHe starts to feel Yukawa exponential tail of
attraction of nucleus to $He$ due to scalar-isoscalar nuclear
potential. It should be noted that scalar-isoscalar nature of He
nucleus excludes its nuclear interaction due to $\pi$ or $\rho$
meson exchange, so that the main role in its nuclear interaction
outside the nucleus plays $\sigma$ meson exchange, on which nuclear
physics data are not very definite. The nuclear potential depends on
the relative distance between He and nucleus and we take it in the
form
\begin{equation}
U_n=-\frac{A_{He} A g^2 exp{(-\mu
|\vec{r}-\vec{\rho}|)}}{|\vec{r}-\vec{\rho}|}. \label{mk1repotnuc}
\end{equation}
Here $\vec{r}$ is radius vector to nucleus, $\vec{\rho}$ is the
radius vector to He in OHe, $A_{He}=4$ is atomic weight of helium,
$A$ is atomic weight of nucleus, $\mu$ and $g^2$ are the mass and
coupling of $\sigma$ meson - mediator of nuclear attraction.

Strictly speaking, starting from this point we should deal with a
three-body problem for the system of He, nucleus and $X^{--}$ and
the correct quantum mechanical description should be based on the
cylindrical and not spherical symmetry. In the present work we use
the approximation of spherical symmetry and take into account
nuclear attraction beyond the nucleus in a two different ways: 1)
nuclear attraction doesn't influence the structure of OHe, so that
the Yukawa potential (\ref{mk1repotnuc}) is averaged over $|\vec{\rho}|$
for spherically symmetric wave function of He shell in OHe; 2)
nuclear attraction changes the structure of OHe so that He takes the
position $|\vec{\rho}|=r_o$, which is most close to the nucleus. Due
to strong attraction of He by the nucleus the second case (which is
labeled "b" in successive numerical calculations) seems more
probable. In the lack of the exact solution of the problem we
present both the results, corresponding to the first case (which are
labeled "m" in successive numerical calculations), and to the second
case (which is labeled "b") in order to demonstrate high sensitivity
of the numerical results to choice of parameters.

In the both cases nuclear attraction results in the polarization of
OHe and the mutual attraction of nucleus and OHe is changed by
Coulomb repulsion of $He$ shell. Taking into account Coulomb
attraction of nucleus by $X^{--}$ one obtains dipole Coulomb barrier
of the form
\begin{equation}
U_d=\frac{Z_{He} Z \alpha r_o}{r^2}. \label{mk1repotdip}
\end{equation}

When helium is completely merged with the nucleus the interaction is
reduced to the oscillatory potential (\ref{mk1rpotosc}) of $X^{--}$ with
homogeneously charged merged nucleus with the charge $Z+2$, given by
\begin{equation}
    E_m=\frac{3}{2}(\frac{(Z+2) Z_x \alpha}{R}-\frac{1}{R}(\frac{(Z+2) Z_x \alpha}{(A+4) m_p R})^{1/2}).
\label{mk1rpotoscEq}
\end{equation}

\begin{figure}
    \begin{center}
        \includegraphics[width=12cm]{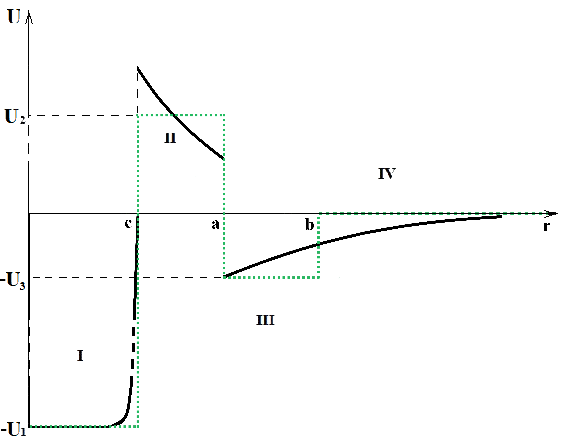}
        \caption{The approximation of rectangular well for potential of OHe-nucleus system.}\label{mk1rpic1}
    \end{center}
\end{figure}

To simplify the solution of Schrodinger equation we approximate the
potentials (\ref{mk1repotem})-(\ref{mk1rpotoscEq}) by a rectangular potential
that consists of a potential well with the depth $U_1$ at $r<c=R$,
where $R$ is the radius of nucleus, of a rectangular dipole Coulomb
potential barrier $U_2$ at $R \le r<a=R+r_o+r_{he}$, where $r_{he}$
is radius of helium nucleus, and of the outer potential well $U_3$,
formed by the Yukawa nuclear interaction (\ref{mk1repotnuc}) and
residual Coulomb interaction (\ref{mk1repotem}). The values of $U_1$ and
$U_2$ were obtained by the averaging of the (\ref{mk1rpotosc}) and
(\ref{mk1repotdip}) in the corresponding regions, while $U_3$ was equal
to the value of the nuclear potential (\ref{mk1repotnuc}) at $r=a$ and
the width of this outer rectangular well (position of the point b)
was obtained by the integral of the sum of potentials
(\ref{mk1repotnuc}) and (\ref{mk1repotem}) from $a$ to $\infty$.
 It leads to the approximate potential, presented on Fig. \ref{mk1rpic1}.

Solutions of Schrodinger equation for each of the four regions,
indicated on Fig. \ref{mk1rpic1}, are given in textbooks (see
e.g.\cite{mk1rLL3}) and their sewing determines the condition, under
which a low-energy  OHe-nucleus bound state appears in the region
III.
\subsection{Low energy bound state of O-helium with nuclei}
\begin{figure}
    \begin{center}
        \includegraphics[width=12cm]{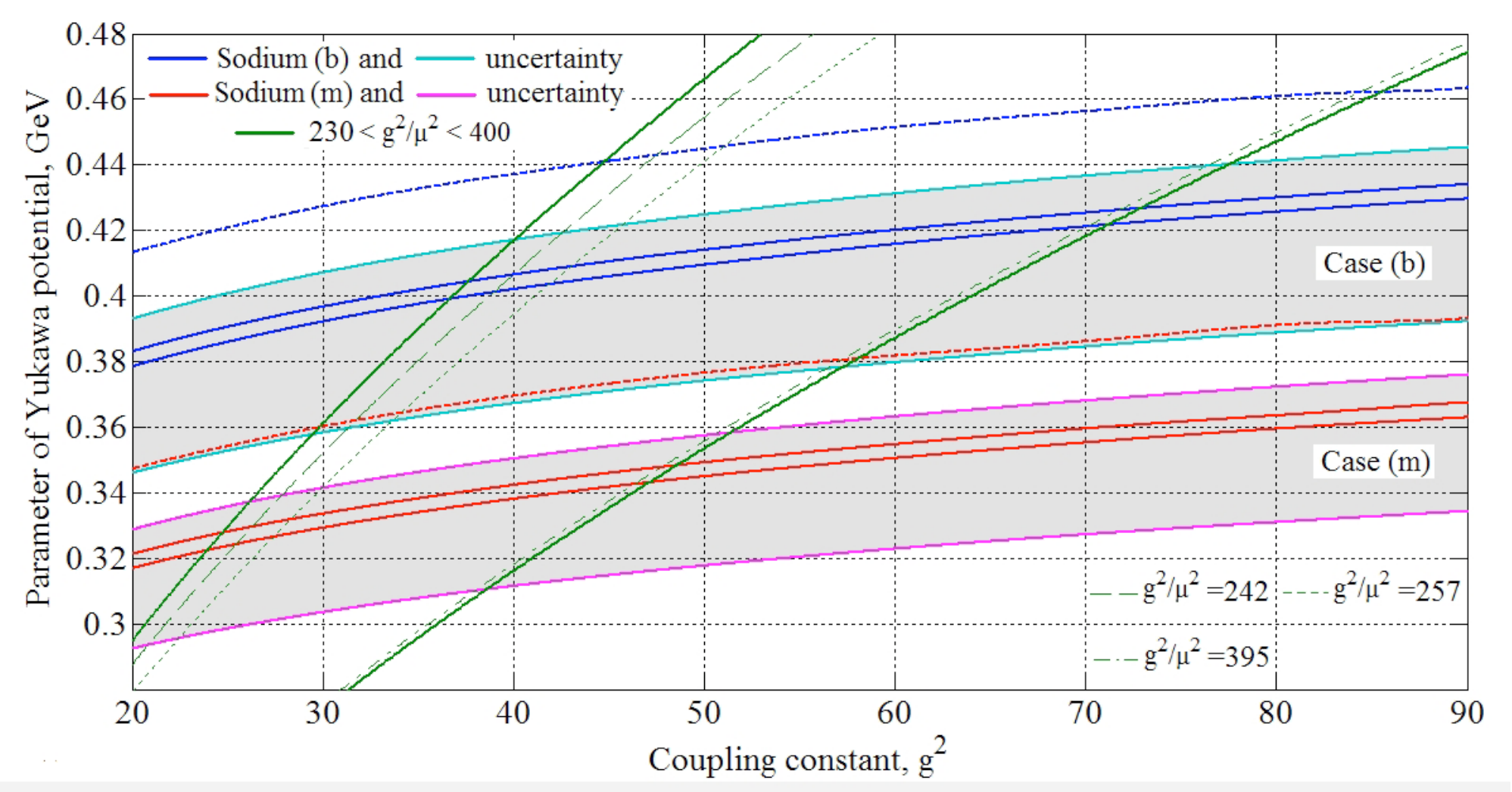}
        \caption{
        The region of parameters $\mu$ and $g^2$, for which Na-OHe system has a level in the interval 4 keV.
        Two lines determine at $d_o=1.2/(200 \MeV)$ the region of parameters, at which the bound
        system of this element with OHe has a 4 keV level.
        In the region between the two strips the energy of level is below 4 keV.
        There are also indicated the range of $g^2/\mu^2$ (dashed lines) as well as their preferred values
        (thin lines) determined in \cite{mk1rnuclear} from parametrization of the relativistic ($\sigma-\omega$) model for nuclear matter.
        The uncertainty in the determination of parameter $1.15/(200 \MeV)<d_o <1.3/(200 \MeV)$ results in the
        uncertainty of $\mu$ and $g^2$ shown by the shaded regions surrounding the lines.
        The case of nuclear Yukawa potential $U_{3m}$, averaged over the orbit of He in OHe, corresponds to the
        lower lines and shaded region, while the upper lines and shaded region around them illustrate
        the case of nuclear Yukawa potential $U_{3b}$ with the position
        of He most close to the nucleus at $\rho=r_o$.}\label{mk1rNa}
    \end{center}
\end{figure}

The energy of this bound state and its existence strongly depend on
the parameters $\mu$ and $g^2$ of nuclear potential (\ref{mk1repotnuc}).
On the Fig. \ref{mk1rNa} the regions of these parameters, giving 4 keV
energy level in OHe bound state with sodium are presented. Radiative
capture to this level can explain results of DAMA/NaI and DAMA/LIBRA
experiments with the account for their energy resolution
\cite{mk1rDAMAlibra}. The lower shaded region on Fig. \ref{mk1rNa}
corresponds to the case of nuclear Yukawa potential $U_{3m}$,
averaged over the orbit of He in OHe, while the upper region
corresponds to the case of nuclear Yukawa potential $U_{3b}$ with
the position of He most close to the nucleus at $\rho=r_o$.The
result is also sensitive to the precise value of $d_o$, which
determines the size of nuclei $R=d_o A^{1/3}$. The two narrow strips
in each region correspond to the experimentally most probable value
$d_o=1.2/(200 \MeV)$. In these calculations the mass of OHe was
taken equal to $m_o=1 TeV$, however the results weakly depend on the
value of $m_o>1 TeV$.

It is interesting that the values of $\mu$ on Fig. \ref{mk1rNa} are compatible with the results of recent experimental measurements of mass of sigma meson \cite{mk1rsigma}.

\subsection{Energy levels in other nuclei}
\begin{figure}
\begin{center}
        \includegraphics[width=12cm]{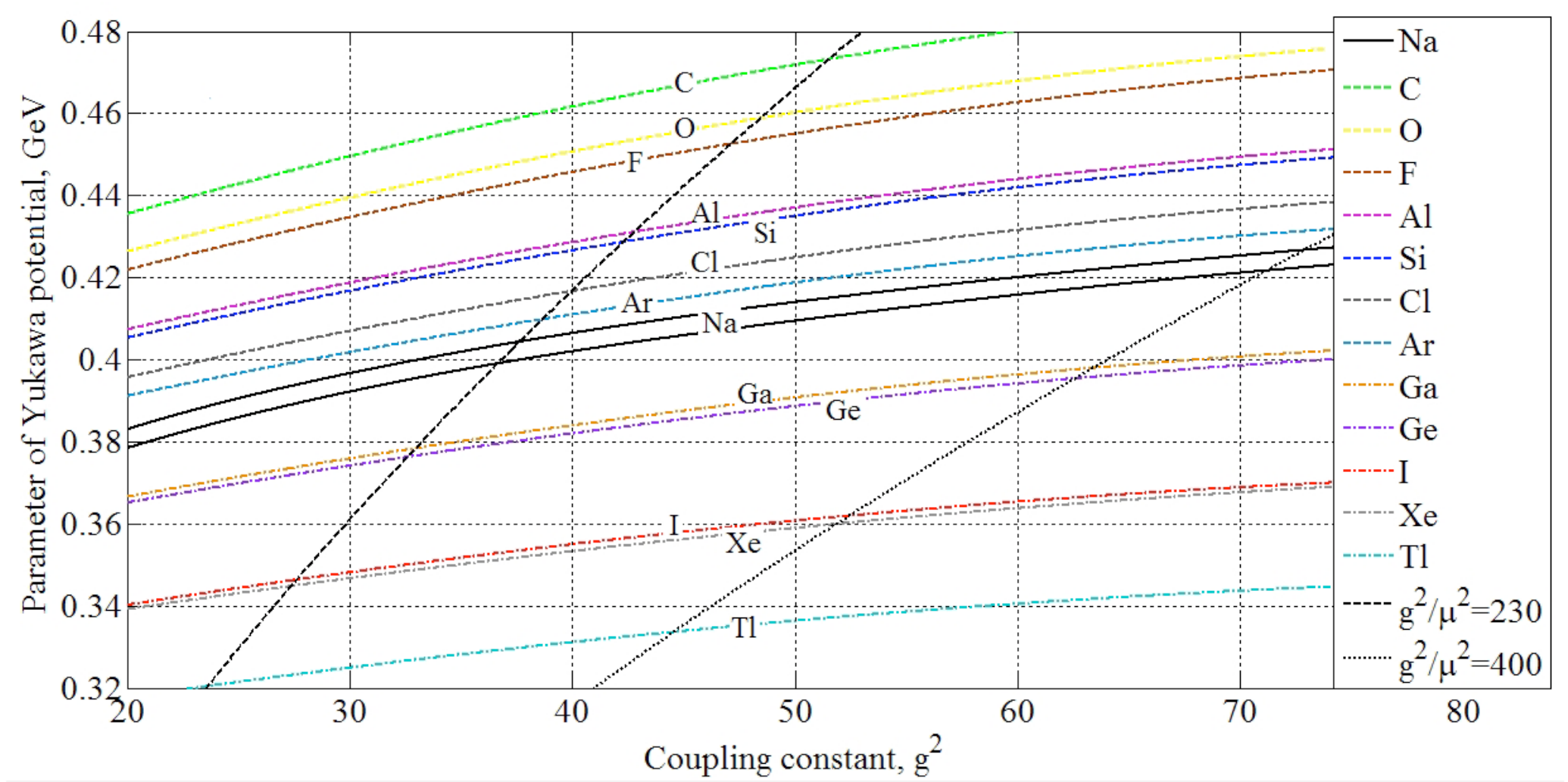}
        \caption{Existence of low energy bound states in OHe-nucleus system
        in the case b for nuclear Yukawa potential $U_{3b}$ with the position of
        He most close to the nucleus at $\rho=r_o$.
        The lines, corresponding to different nuclei, show the upper limit
        for nuclear physics parameters $\mu$ and $g^2$, at which these bound states
        are possible. The choice of parameters corresponding to 4 keV OHe-Na
        bound state, excludes region below Na line.}\label{mk1rExb}
    \end{center}
\end{figure}

\begin{figure}
\begin{center}
        \includegraphics[width=12cm]{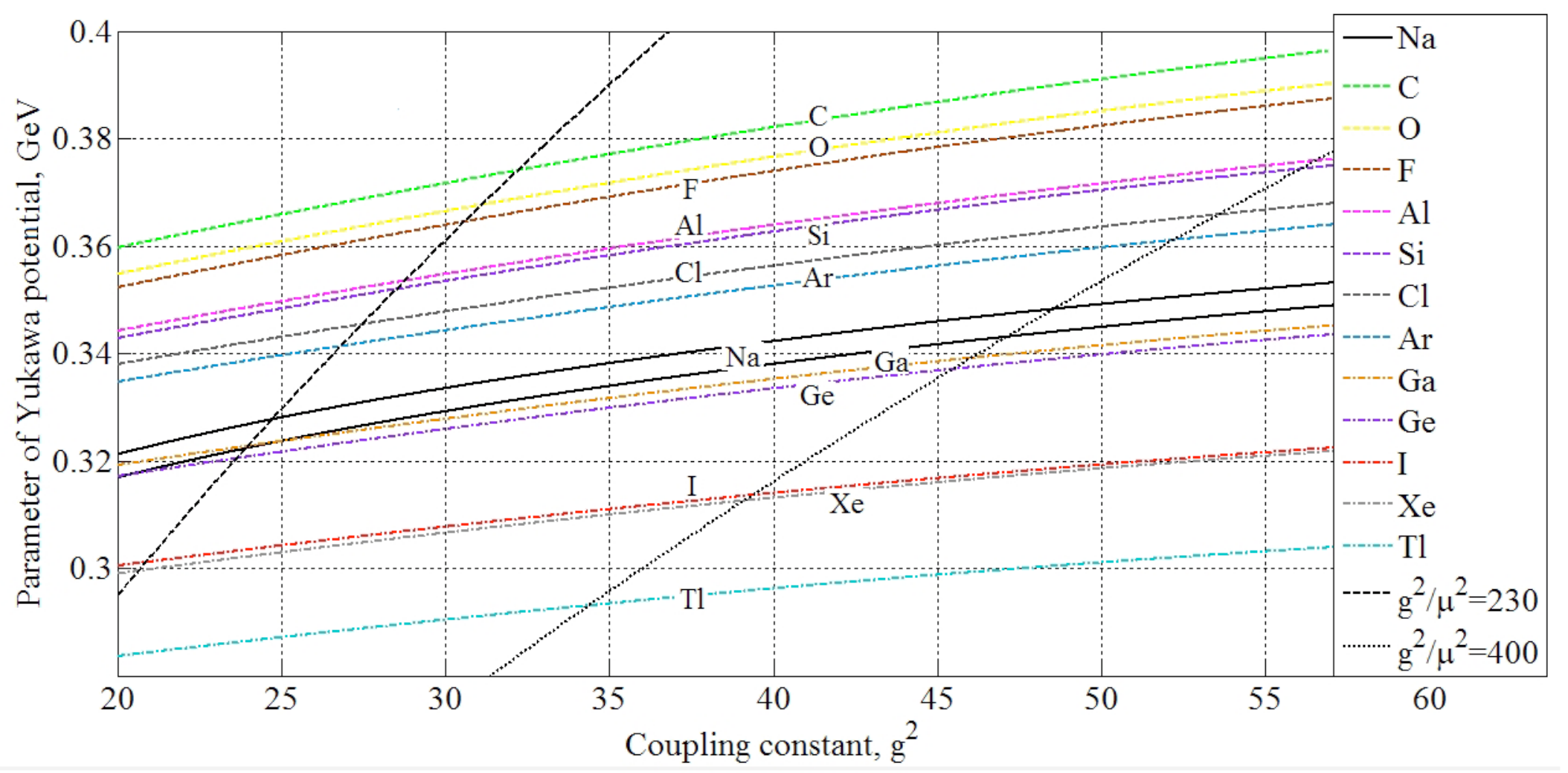}
        \caption{Existence of low energy bound states in OHe-nucleus system
        in the case m for nuclear Yukawa potential $U_{3m}$, averaged over the orbit of He in OHe.
        The lines, corresponding to different nuclei, show the upper limit
        for nuclear physics parameters $\mu$ and $g^2$, at which these bound states
        are possible. The choice of parameters corresponding to 4 keV OHe-Na
        bound state, excludes region below Na line.}\label{mk1rExm}
    \end{center}
\end{figure}

The important qualitative feature of the presented solution is the
restricted range of intermediate nuclei, in which the OHe-nucleus
state beyond nuclei is possible. For the chosen range of nuclear
parameters, reproducing the results of DAMA/NaI and DAMA/LIBRA, we
can calculate the binding energy of OHe-nucleus states in nuclei,
corresponding to chemical composition of set-ups in other
experiments. It turns out that  there are no such states for light
and heavy nuclei. In the case of nuclear Yukawa potential $U_{3b}$,
corresponding to the position of He most close to the nucleus at
$\rho=r_o$, the range of nuclei with bound states with OHe
corresponds to the part of periodic table between B and Ti. This
result is stable independent on the used scheme of numerical
calculations. The upper limits on the nuclear parameters $\mu$ and
$g^2$, at which there exists OHe-nucleus bound state are presented
for this case on Fig.\ref{mk1rExb}. In the case of potential $U_{3m}$,
averaged over the orbit of He in OHe, there are no OHe bound states
with nuclei, lighter than Be and heavier than Ti. However, the
results are very sensitive to the numerical factors of calculations
and the existence of OHe-Ge and OHe-Ga bound states at a narrow
window of parameters $\mu$ and $g^2$ turns to be strongly dependent
on these factors so that change in numbers smaller than 1\% can give
qualitatively  different result for Ge and Ga. The results for the
case (m) are shown on Fig.\ref{mk1rExm}.
Both for the cases (b) and (m) there is a stable conclusion that
there are no OHe-nucleus bound states with Xe, I and Tl.

\begin{figure}
    \begin{center}
        \includegraphics[width=12cm]{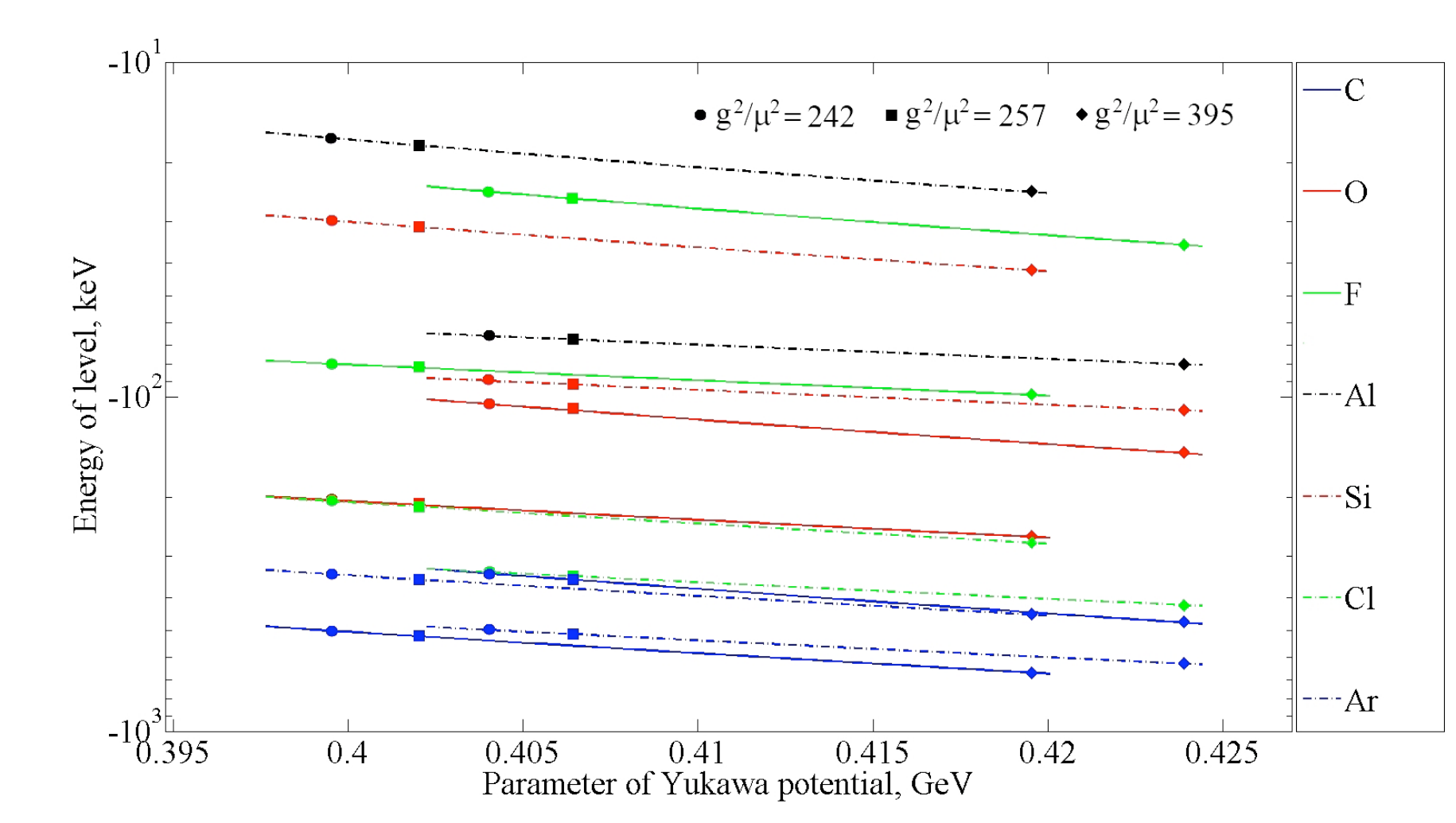}
        \caption{Energy levels in OHe bound system with carbon,
        oxygen, fluorine, argon, silicon, aluminium and chlorine for the case of the nuclear
Yukawa potential $U_{3b}$. The predictions are given for the range
of $g^2/\mu^2$ determined in \cite{mk1rnuclear} from parametrization of
the relativistic ($\sigma-\omega$) model for nuclear matter. The
preferred values of $g^2/\mu^2$ are indicated by the corresponding
marks (squares or circles).}\label{mk1relementsb}
    \end{center}
\end{figure}

\begin{figure}
\begin{center}
        \includegraphics[width=12cm]{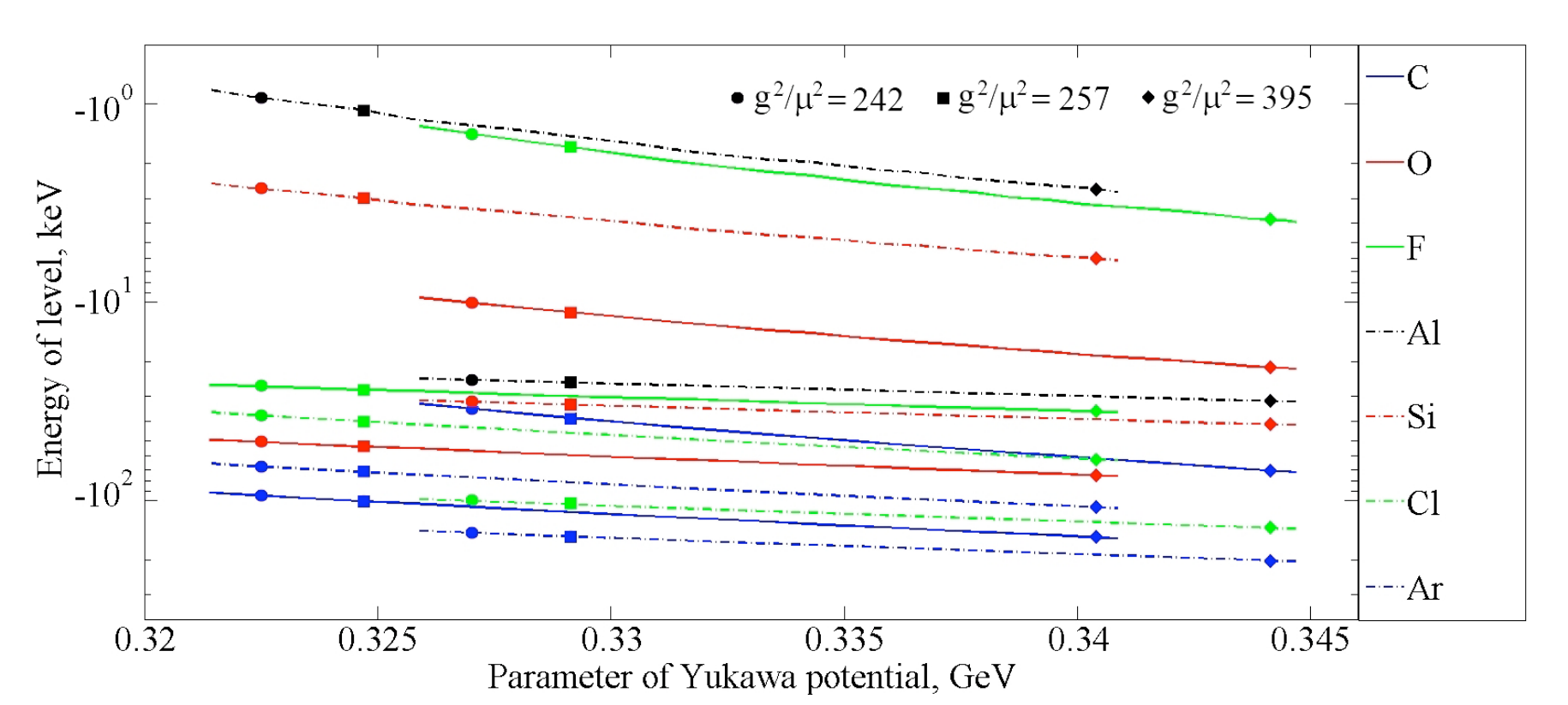}
        \caption{Energy levels in OHe bound system with carbon,
        oxygen, fluorine, argon, silicon, aluminium and chlorine for the case of the nuclear
Yukawa potential $U_{3m}$. The predictions are given for the range
of $g^2/\mu^2$ determined in \cite{mk1rnuclear} from parametrization of
the relativistic ($\sigma-\omega$) model for nuclear matter. The
preferred values of $g^2/\mu^2$ are indicated by the corresponding
marks (squares or circles).}\label{mk1relementsm}
    \end{center}
\end{figure}

For the experimentally preferred value $d_o=1.2/(200 \MeV)$ the
results of calculation of the binding energy of OHe-nucleus systems
for carbon, oxygen, fluorine, argon, silicon, aluminium and chlorine
are presented on Fig. \ref{mk1relementsb} for the case of the nuclear
Yukawa potential $U_{3b}$ and on Fig. \ref{mk1relementsm} for the case
of the potential $U_{3m}$. The difference in these results
demonstrates their high sensitivity to the choice of parameters.

\section{Conclusions}


To conclude, the results of dark matter search in experiments
DAMA/NaI and DAMA/LIBRA can be explained in the framework of
composite dark matter scenario without contradiction with negative
results of other groups. This scenario can be realized in different
frameworks, in particular in Minimal Walking Technicolor model or in
the approach unifying spin and charges and contains distinct
features, by which the present explanation can be distinguished from
other recent approaches to this problem \cite{mk1rEdward} (see also
review and more references in \cite{mk1rGelmini}).

Our explanation is based on the mechanism of low energy binding of
OHe with nuclei. We have found that within the uncertainty of
nuclear physics parameters there exists their range at which OHe
binding energy with sodium is equal to 4 keV and there is no such
binding with iodine and thallium.


With the account for high sensitivity of our results to the values
of uncertain nuclear parameters and for the approximations, made in
our calculations, the presented results can be considered only as an
illustration of the possibility to explain effects in underground
detectors by OHe binding with intermediate nuclei. However, even at
the present level of our studies we can make a conclusion that
effects of such binding should strongly differ in detectors with the
content, different from NaI, and can be absent in detectors with
very light (e.g. $^3He$) and heavy nuclei (like xenon and probably
germanium). Therefore test of results of DAMA/NaI and DAMA/LIBRA
experiments by other experimental groups can become a very
nontrivial task.



\section*{Acknowledgments}


We would like to thank Norma Manko\v c Bor\v stnik and all the
participants of Bled Workshop for stimulating discussions.





\title{Can  the Matter-Antimatter Asymmetry be Easier to Understand
  Within the "Spin-charge-family-theory",
Predicting Twice Four Families and Two Times $SU(2)$ Vector Gauge and Scalar Fields?}
\author{N.S. Manko\v c Bor\v stnik}
\institute{%
Department of Physics, FMF, University of Ljubljana,\\
 Jadranska 19, SI-1000 Ljubljana, Slovenia}

\titlerunning{Can  the Matter-Antimatter Asymmetry\ldots}
\authorrunning{N.S. Manko\v c Bor\v stnik}
\maketitle
 
\begin{abstract}
This contribution is an attempt to try to understand the matter-antimatter asymmetry in the universe
within the  {\it spin-charge-family-theory}~\cite{snmb2bnorma,snmb2bpikanorma}  if assuming that  
transitions in non equilibrium processes among instanton vacua and complex phases in  
mixing matrices are the sources of the matter-antimatter asymmetry, as 
studied in the literature~\cite{snmb2bgross,snmb2brubakovshaposhnikov,snmb2bdinekusenko,snmb2btapeiling} 
for several proposed theories. 
The {\it spin-charge-family-theory} is, namely, very   
promising in showing the right way beyond  the {\it standard model}.  
It predicts  families and their mass matrices, explaining  the origin of the charges
and of the gauge fields.  It predicts that there are, after the  universe passes through two 
$SU(2)\times U(1)$ phase transitions,
in which the symmetry  breaks from $SO(1,3) \times SU(2) \times SU(2) \times U(1) \times SU(3)$ first 
to $SO(1,3) \times SU(2)  \times U(1) \times SU(3)$ and then to 
   $SO(1,3) \times U(1) \times SU(3)$,  twice decoupled four families. The upper four families gain masses 
   in the first phase transition, while the second four families gain masses at the electroweak break. 
To these two breaks of symmetries the scalar non Abelian fields, the (superposition of the) 
gauge fields of the operators generating  families, contribute.
The lightest of the upper four families 
is stable (in comparison with the life of the universe) and is therefore a candidate for 
constituting the dark matter. The heaviest of the lower four families should be seen at the LHC or 
at somewhat higher energies.  
\end{abstract}

\section{Introduction}
\label{snmb2bintroduction}
\label{contribution:snmb2baryon}
The {\it theory unifying spin and charges and predicting families} ({\it spin-charge-family-theory}) 
assumes that spinors carry in $d \ge 4$ ($d= 1 + 13$ is studied) only two kinds of the spin. 
The Dirac kind  $\gamma^{a}$  manifests after  several appropriate breaks of the 
starting symmetry as the spin and all the charges. 
The second kind called  $\{ \gamma^a$ ($\{ \gamma^a, \tilde{\gamma}^b\}_{+}=0$) generates families. 
 Accordingly there 
 are in $d \ge 4$, besides the vielbeins, also the two kinds of the spin connection fields,  which 
 are the gauge fields of the corresponding operators  $S^{ab}$ and $\tilde{S}^{ab}$.  Those 
 connected with $S^{ab}$ manifest in $d=(1+3)$ as the vector gauge fields, while those 
 connected with $\tilde{S}^{ab}$ manifest as the scalar fields and determine on the tree level the 
 mass matrices.

Let me make a short review  of the so far made predictions of the 
{\it spin-charge-family-theory}:

\begin{itemize}

\item   The {\it spin-charge-family-theory} has the explanation for 
        the appearance of the internal degrees of freedom -- the  spin   
        and the charges while unifying them under the assumption that the  
        universe went through several phase transitions which cause the appropriate breaks 
        of the starting symmetry. Then the fact that the right handed (with respect to SO(1,3)) 
        fermions are weak chargeless, while the left handed ones carry the weak charge emerges, 
        as well as that there exist leptons (singlets with respect to the 
        colour charge) and quarks (triplets with respect to the colour charge)~\cite{snmb2bnorma,snmb2bpikanorma}. 
        
\item   The theory  explains the appearance of massless families at 
        the low energy regime under the 
        assumption that there are breaks which leave the massless fermions of only one 
        handedness~\cite{snmb2bhnd}. Assuming that breaks of symmetries affect 
        the whole internal space --- the space defined by both kinds of the Clifford algebra objects ---         
        it predicts in the energy regime close below $10^{16}$ GeV 
        eight massless families. The manifested symmetry is (assumed to be) at this stage 
        $SO(1,3) \times SO(4) \times U(1) \times SU(3)$. 
        The next  break of  the symmetry of the universe to $SO(1,3) \times SU(2) \times 
        U(1) \times SU(3)$ leaves four families massless~\cite{snmb2bpikanorma}, while  the vacuum 
        expectation values of superposition of the  starting fields which manifest in $(1+3)$ 
        as  scalar fields, 
        make the upper four families  and the corresponding gauge fields massive. 
        After the electroweak break also the lower four 
        families become massive due to the vacuum expectation values of superposition of 
        the starting fields, together with the weak bosons. %
       
\item   The theory predicts the fourth family, which will be observed 
        at the LHC or at somewhat higher energies~\cite{snmb2bgmdn}, 
        and the fifth stable family (with no mixing matrix elements couplings 
        to the lower four families in comparison with the age of the universe), the baryons and 
        neutrinos of which are the candidates to form the dark matter.

\item   The masses of this fifth family members are according to the so far made rough 
        estimations~\cite{snmb2bpikanorma,snmb2bgmdn} larger than a few TeV and smaller than $10^{10}$  TeV. 
        The members of the family have approximately the same mass, 
        at least on the tree level~\cite{snmb2bnormaproc2010talk}. 
        
\item   The studies~\cite{snmb2bgn}  of the history of the stable fifth family members in the evolution 
	of the universe and  of their interactions with the ordinary matter in the DAMA's and the 
	CDMS's experiments done so far lead to the prediction that the masses of the fifth family 
	members, if they constitute the dark matter, 
        are  a few hundred TeV, independent of the fifth family fermion-antifermion asymmetry. 
        The Xe experiment looks like to be in disagreement, but careful analyses show that one 
        should wait for further data~\cite{snmb2bdiscussGN} to make the final conclusion. 
        
        The lightest fifth family baryon is, in the case that all the quarks have approximately 
        (within a hundred GeV) the same mass~\cite{snmb2bgn},  the fifth 
        family neutron, due to the attractive electromagnetic interaction. The difference in the 
        weak interaction can be  for large enough masses neglected.
         
\item	The fermion asymmetry  in the {\it approach} has not yet really been studied.

\item   The studies~\cite{snmb2bgn}  of the evolution of this stable fifth family members rely  
        on my rough estimations~\cite{snmb2bgn} of the behaviour of the  coloured  fifth family 
        objects (single quarks and antiquarks  or coloured pairs of quarks or of antiquarks)  
        during the colour phase transition. 
        These estimations namely suggest that the coloured objects either annihilate with the 
        anti-objects or they form  colourless neutrons and antineutrons and correspondingly decouple 
        from the plasma soon after the colour phase transition starts, due to the very strong 
        binding energy of the fifth family baryons (with respect to the first family baryons) 
        long enough before the first family quarks start to form the baryons. 
        These estimations should be followed by more accurate studies. %

\item   The so far done studies 
        suggest strongly that the number density of the fifth 
        family neutrinos (of approximately the same mass as the fifth family quarks and leptons), 
        which also contribute to the dark matter, is pretty much reduced due to the 
        neutrino-antineutrino annihilation closed below the electroweak break.  
        The weak annihilation cross section is expected to play much stronger role 
        for neutrinos than for strongly bound fifth family quarks in the fifth family neutron 
        (due to the huge binding energy of the fifth family quarks), what also remains to be proved. 

\item   The estimations~\cite{snmb2bgmdn} of the properties of the lower four families on the tree level 
        call for the calculations beyond the tree level, which should hopefully demonstrate, that 
        the loop corrections (in all orders) bring the main differences in the properties of 
        the family members. These calculations are in progress~\cite{snmb2bAN}.

        \end{itemize}
 
 Although we can say that the  {\it spin-charge-family-theory} looks  very promising as 
 the right way to  explain where do the assumptions of the {\it standard model} originate, there are 
 obviously many not yet studied, or at least far from being
 carefully enough studied open problems. 
 Many a problem is common to 
 all the theories, like the first family baryon asymmetry, which I am going to discuss within the 
 {\it spin-charge-family-theory}  in this contribution.
  Some of the problems are common  to all the theories  assuming more than so far observed $(1+3)$ 
  dimensions, in particular the {\it spin-charge-family-theory} shares some problems with 
  all the Kaluza-Klein-like theories. We are trying  to solve them first on toy models~\cite{snmb2bhnd}.  
 
 The main new step in the {\it spin-charge-family-theory} --- the explanation of the appearance of 
 families by assuming that both existing kinds of the Clifford algebra objects should be used
 to treat correctly 
 the fermion degrees of freedom --- limits very much the 
 properties of families and their members. The simple starting action in $d= (1+13)$, which in $d=(1+3)$ 
 demonstrates the mass matrices, namely  fixes to high extent the fermion properties after 
 the breaks of symmetries. Therefore this proposal might soon be studied accurately enough to show whether 
 it is the right theory or not.

 This contribution is an attempt to try to understand  what can the {\it spin-charge-family-theory}  
 say about the  fermion-antifermion asymmetry when taking into account the 
 proposals of the references~\cite{snmb2brubakovshaposhnikov,snmb2bdinekusenko,snmb2btapeiling} (and 
 of the works cited therein). These works study the soliton solutions of non Abelian 
 gauge fields  with many different vacua 
  and evaluate fermion number nonconservation due to possible transitions among different  
  vacua in non equilibrium processes 
 during the phase  transitions through which the universe passed. In such processes  
 fermion (and also antifermion) currents are  not conserved since $CP$ is not nonconserved. 
 To the $CP$ nonconservation  also the   complex matrix elements determining the transitions 
 among families contribute and  consequently influence the first family fermion-antifermion   
  asymmetry. 
 
 Since the {\it spin-charge-family-theory} predicts below the unification scale of 
 all the charges two kinds of  phase transitions 
 (first from $SO(1,3) \times SU(2) \times SU(2) \times U(1) \times SU(3)$ to 
 $SO(1,3) \times SU(2) \times U(1) \times SU(3)$, in which the upper four families gain masses and so do 
 the corresponding vector gauge fields, and then from 
 $SO(1,3) \times SU(2) \times U(1) \times SU(3)$  to 
 $SO(1,3) \times U(1) \times SU(3)$, in which the lower four families and the 
 corresponding gauge fields gain masses), in which besides the vector gauge fields
 also  the scalar gauge fields (the gauge fields of $\tilde{S}^{ab}$ and also of $S^{ab}$ 
 with the scalar index with respect to $(1+3)$) contribute,  
 the fermion-antifermion asymmetry might very probably have for the stable fifth family  an opposite sign  
 than for the first family.

 It might therefore  be that the existence of 
  two kinds of four families, together with two kinds of the vector gauge fields and two kinds of the 
  scalar fields help to 
  easier understanding the first family fermion-antifermion asymmetry.

 Although I am studying the fermion asymmetry, together with the discrete symmetries, in   
 the {\it spin-charge-family-theory} for quite some time (not really intensively), this contribution 
 is  stimulated by the question of M.Y. Khlopov~\cite{snmb2bMYN}, since he is 
 proposing the scenario, in which my stable fifth family members should  manifest an 
 opposite fermion asymmetry than the 
 first family members, that is antifermion-fermion asymmetry. While in the case that the fifth family 
 members have masses around 100 TeV or higher and the neutron is the lightest baryon and 
 neutrino the lightest  lepton~\cite{snmb2bgn}  the fifth family baryon asymmetry plays no role  
 (since in this case the fifth family neutrons and neutrinos as well as their antiparticles 
 interact weakly enough  among  themselves and with the ordinary matter that the 
 assumption that they constitute the dark matter is in agreement  with the observations). Maxim~\cite{snmb2bm} 
 claims that the fifth family members with the quark masses not higher than  10 TeV are also the candidates 
 for the dark matter, provided that $\bar{u}_5 \bar{u}_5 \bar{u}_5$ is the lightest antibaryon and   that 
 there is an excess of antibaryons over the baryons in the fifth family case.

 \section{A short overview of the theory unifying spin and charges and explaining families}
 \label{snmb2bapproach}

  In this section I briefly repeat the main ideas of the {\it spin-charge-family-theory}. 
   I kindly ask the reader to learn more about this theory in the references~\cite{snmb2bnorma,snmb2bpikanorma}  
   as well as in my talk presented in this proceedings and in the references therein. 
 
I am proposing a simple action in $d=(1+13)$-dimensional space. Spinors carry two kinds of the spin 
(no charges).\\ 
i. The Dirac spin, described by $\gamma^a$'s, defines the spinor representation in $d=(1+ 13)$.  
After the break of the starting symmetry  $SO(1,13)$ (through $SO(1,7) \times 
SO(6)$) to the symmetry of the {\it standard model}  in $d=(1+3)$ ($SO(1,3)\times U(1)\times SU(2)\times SU(3)$)  
it defines  the hyper charge ($U(1)$),  the weak charge ($SU(2)$, with the 
left handed representation of $SO(1,3)$ manifesting naturally the weak charge and the right 
handed ones appearing as the weak singlets) and  the colour charge ($SU(3)$). \\
ii. The second kind of the spin~\cite{snmb2bnorma},  
described by $\tilde{\gamma}^a$'s ($\{\tilde{\gamma}^a, \tilde{\gamma}^b\}_{+}= 2 \, \eta^{ab}$) and  
anticommuting with the Dirac $\gamma^a$ ($\{\gamma^a, \tilde{\gamma}^b\}_{+}=0$),  
defines the families of spinors.\\ 
Accordingly spinors interact with the two kinds of the spin connection fields and the vielbeins. 

We have
\begin{eqnarray}
&& \{ \gamma^a, \gamma^b\}_{+} = 2\eta^{ab} =  
\{ \tilde{\gamma}^a, \tilde{\gamma}^b\}_{+},\quad
\{ \gamma^a, \tilde{\gamma}^b\}_{+} = 0,\nonumber\\
&&S^{ab}: = (i/4) (\gamma^a \gamma^b - \gamma^b \gamma^a), \quad
\tilde{S}^{ab}: = (i/4) (\tilde{\gamma}^a \tilde{\gamma}^b 
- \tilde{\gamma}^b \tilde{\gamma}^a),\quad  \{S^{ab}, \tilde{S}^{cd}\}_{-}=0.\nonumber\\
\label{snmb2bsnmb:tildegclifford}
\end{eqnarray}
The action
\begin{eqnarray}
S            \,  &=& \int \; d^dx \; E\;{\mathcal L}_{f} +  
\nonumber\\  
               & & \int \; d^dx \; E\; (\alpha \,R + \tilde{\alpha} \, \tilde{R})\,,
               \end{eqnarray}
\begin{eqnarray}
{\mathcal L}_f &=& \frac{1}{2}\, (E\bar{\psi} \, \gamma^a p_{0a} \psi) + h.c.\,, 
\nonumber\\
p_{0a }        &=& f^{\alpha}{}_a p_{0\alpha} + \frac{1}{2E}\, \{ p_{\alpha}, E f^{\alpha}{}_a\}_-, 
\nonumber\\  
   p_{0\alpha} &=&  p_{\alpha}  - 
                    \frac{1}{2}  S^{ab} \omega_{ab \alpha} - 
                    \frac{1}{2}  \tilde{S}^{ab}   \tilde{\omega}_{ab \alpha}\,,                   
\nonumber\\ 
R              &=&  \frac{1}{2} \, \{ f^{\alpha [ a} f^{\beta b ]} \;(\omega_{a b \alpha, \beta} 
- \omega_{c a \alpha}\,\omega^{c}{}_{b \beta}) \} + h.c. \;, 
\nonumber\\
\tilde{R}      &=& \frac{1}{2}\,   f^{\alpha [ a} f^{\beta b ]} \;(\tilde{\omega}_{a b \alpha,\beta} - 
\tilde{\omega}_{c a \alpha} \tilde{\omega}^{c}{}_{b \beta}) + h.c.\;, 
\label{snmb2bwholeaction}
\end{eqnarray}
manifests ($f^{\alpha [a} f^{\beta b]}= f^{\alpha a} f^{\beta b} - f^{\alpha b} f^{\beta a}$) 
after the break of symmetries all the known 
gauge fields and the scalar fields, and the mass matrices. 
To see the manifestation of the covariant momentum and the mass matrices we rewrite 
formally the action for a Weyl spinor in $d=(1+13)$  as follows  
\begin{eqnarray}
{\mathcal L}_f &=&  \bar{\psi}\gamma^{m} (p_{m}- \sum_{A,i}\; g^{A}\tau^{Ai} A^{Ai}_{m}) \psi 
+ \nonumber\\
               & &  \{ \sum_{s=7,8}\;  \bar{\psi} \gamma^{s} p_{0s} \; \psi \}  + \nonumber\\
               & & {\rm the \;rest}, 
\label{snmb2bfaction}
\end{eqnarray}
where $m=0,1,2,3$ with
\begin{eqnarray}
\tau^{Ai} = \sum_{a,b} \;c^{Ai}{ }_{ab} \; S^{ab},
\nonumber\\ 
\{\tau^{Ai}, \tau^{Bj}\}_- = i \delta^{AB} f^{Aijk} \tau^{Ak}.
\label{snmb2btau}
\end{eqnarray}
All the charges and the spin of one family are determined by $S^{ab}$, 
with $S^{ab}$ as the only internal degree of freedom of one family  
(besides the family quantum number, determined by $\tilde{S}^{ab}$),  
manifesting after the breaks at the low energy regime as the generators of the observed 
groups~(Eq.~(\ref{snmb2btau})) $U(1), SU(2)$ and $SU(3)$, for $A=1,2,3$, respectively.

The breaks of the starting symmetry from $SO(1,13)$ to the symmetry $SO(1,7) \times SU(3) \times U(1)$ 
and further to $SO(1,3) \times SU(2) \times SU(2) \times U(1) \times SU(3) $ 
are assumed to leave all the low lying families of spinors massless. 
There are eight such massless families ($2^{8/2-1}$) 
before further breaks. 

Accordingly the first row of the action in Eq.~(\ref{snmb2bfaction}) manifests the effective {\it standard model} 
fermions part of the action before the weak break, while the second part manifests, 
after the appropriate breaks of 
symmetries (when $\omega_{ab \sigma}$  and $\tilde{\omega}_{ab \sigma}$, $\sigma \in (5,6,7,8),$ 
fields gain the nonzero vacuum expectation values 
on the tree level)  the mass matrices.

 The generators $\tilde{S}^{ab}$ take care of the families, transforming each member of one family 
 into the corresponding member of another family, due to the fact that 
 $\{S^{ab}, \tilde{S}^{cd}\}_{-}=0$ (Eq.(\ref{snmb2bsnmb:tildegclifford})). 
 
 Using the technique~\cite{snmb2bsnmb:hn02hn03} and analysing the vectors as the eigenvectors of the 
 {\it standard model} groups we present vectors in the space of charges and spins in terms of 
 projectors and nilpotents as can be learned in Appendix, in the references~\cite{snmb2bnorma,snmb2bpikanorma} and 
 also in my talk in the Proceedings of Bled workshop 2010. 
 
 I present  in Table~\ref{snmb2bTable I.} the eightplet (the representation of $SO(1,7)$ of quarks of a 
 particular colour charge ($\tau^{33}=1/2$, 
 $\tau^{38}=1/(2\sqrt{3})$), and $U(1)$ charge ($\tau^{4}=1/6$) and on Table~\ref{snmb2bTable Il.} the 
 eightplet of the corresponding (colour chargeless) leptons. 
 %
 %
 \begin{table}
 \begin{center}
 \begin{tabular}{|r|c||c||c|c||c|c|c||r|r|}
 \hline
 i&$$&$|^a\psi_i>$&$\Gamma^{(1,3)}$&$ S^{12}$&$\Gamma^{(4)}$&
 $\tau^{13}$&$\tau^{23}$&$Y$&$Q$\\
 \hline\hline
 && ${\rm Octet},\;\Gamma^{(1,7)} =1,\;\Gamma^{(6)} = -1,$&&&&&&& \\
 && ${\rm of \; quarks}$&&&&&&&\\
 \hline\hline
 1&$ u_{R}^{c1}$&$ \stackrel{03}{(+i)}\,\stackrel{12}{(+)}|
 \stackrel{56}{(+)}\,\stackrel{78}{(+)}
 ||\stackrel{9 \;10}{(+)}\;\;\stackrel{11\;12}{[-]}\;\;\stackrel{13\;14}{[-]} $
 &1&$\frac{1}{2}$&1&0&$\frac{1}{2}$&$\frac{2}{3}$&$\frac{2}{3}$\\
 \hline 
 2&$u_{R}^{c1}$&$\stackrel{03}{[-i]}\,\stackrel{12}{[-]}|\stackrel{56}{(+)}\,\stackrel{78}{(+)}
 ||\stackrel{9 \;10}{(+)}\;\;\stackrel{11\;12}{[-]}\;\;\stackrel{13\;14}{[-]}$
 &1&$-\frac{1}{2}$&1&0&$\frac{1}{2}$&$\frac{2}{3}$&$\frac{2}{3}$\\
 \hline
 3&$d_{R}^{c1}$&$\stackrel{03}{(+i)}\,\stackrel{12}{(+)}|\stackrel{56}{[-]}\,\stackrel{78}{[-]}
 ||\stackrel{9 \;10}{(+)}\;\;\stackrel{11\;12}{[-]}\;\;\stackrel{13\;14}{[-]}$
 &1&$\frac{1}{2}$&1&0&$-\frac{1}{2}$&$-\frac{1}{3}$&$-\frac{1}{3}$\\
 \hline 
 4&$ d_{R}^{c1} $&$\stackrel{03}{[-i]}\,\stackrel{12}{[-]}|
 \stackrel{56}{[-]}\,\stackrel{78}{[-]}
 ||\stackrel{9 \;10}{(+)}\;\;\stackrel{11\;12}{[-]}\;\;\stackrel{13\;14}{[-]} $
 &1&$-\frac{1}{2}$&1&0&$-\frac{1}{2}$&$-\frac{1}{3}$&$-\frac{1}{3}$\\
 \hline
 5&$d_{L}^{c1}$&$\stackrel{03}{[-i]}\,\stackrel{12}{(+)}|\stackrel{56}{[-]}\,\stackrel{78}{(+)}
 ||\stackrel{9 \;10}{(+)}\;\;\stackrel{11\;12}{[-]}\;\;\stackrel{13\;14}{[-]}$
 &-1&$\frac{1}{2}$&-1&$-\frac{1}{2}$&0&$\frac{1}{6}$&$-\frac{1}{3}$\\
 \hline
 6&$d_{L}^{c1} $&$\stackrel{03}{(+i)}\,\stackrel{12}{[-]}|
 \stackrel{56}{[-]}\,\stackrel{78}{(+)}
 ||\stackrel{9 \;10}{(+)}\;\;\stackrel{11\;12}{[-]}\;\;\stackrel{13\;14}{[-]} $
 &-1&$-\frac{1}{2}$&-1&$-\frac{1}{2}$&0&$\frac{1}{6}$&$-\frac{1}{3}$\\
 \hline
 7&$ u_{L}^{c1}$&$\stackrel{03}{[-i]}\,\stackrel{12}{(+)}|
 \stackrel{56}{(+)}\,\stackrel{78}{[-]}
 ||\stackrel{9 \;10}{(+)}\;\;\stackrel{11\;12}{[-]}\;\;\stackrel{13\;14}{[-]}$
 &-1&$\frac{1}{2}$&-1&$\frac{1}{2}$&0&$\frac{1}{6}$&$\frac{2}{3}$\\
 \hline
 8&$u_{L}^{c1}$&$\stackrel{03}{(+i)}\,\stackrel{12}{[-]}|\stackrel{56}{(+)}\,\stackrel{78}{[-]}
 ||\stackrel{9 \;10}{(+)}\;\;\stackrel{11\;12}{[-]}\;\;\stackrel{13\;14}{[-]}$
 &-1&$-\frac{1}{2}$&-1&$\frac{1}{2}$&0&$\frac{1}{6}$&$\frac{2}{3}$\\
 \hline\hline
 \end{tabular}
 \end{center}
 \caption{\label{snmb2bTable I.} The 8-plet of quarks - the members of $SO(1,7)$ subgroup of the 
 group $SO(1,13)$, 
 belonging to one Weyl left 
 handed ($\Gamma^{(1,13)} = -1 = \Gamma^{(1,7)} \times \Gamma^{(6)}$) spinor representation of 
 $SO(1,13)$. 
 It contains the left handed weak charged quarks and the right handed weak chargeless quarks 
 of a particular 
 colour $(1/2,1/(2\sqrt{3}))$. Here  $\Gamma^{(1,3)}$ defines the handedness in $(1+3)$ space, 
 $ S^{12}$ defines the ordinary spin (which can also be read directly from the basic vector, both
 vectors  with both spins, $\pm \frac{1}{2}$, are presented), 
 $\tau^{13}$ defines the third component of the weak charge, $\tau^{23}$ the third component 
 of the $SU(2)_{II}$ charge, 
 $\tau^{4}$ (the $U(1)$ charge) defines together with the
 $\tau^{23}$  the hyper charge ($Y= \tau^4 + \tau^{23}$), $Q= Y + \tau^{13}$ is the 
 electromagnetic charge. 
 The reader can find the whole Weyl representation in the ref.~\cite{snmb2bPortoroz03}.}
 \end{table}
 %
 %
 
 %
 %
 \begin{table}
 \begin{center}
 \begin{tabular}{|r|c||c||c|c||c|c|c||r|r|}
 \hline
 i&$$&$|^a\psi_i>$&$\Gamma^{(1,3)}$&$ S^{12}$&$\Gamma^{(4)}$&
 $\tau^{13}$&$\tau^{23}$&$Y$&$Q$\\
 \hline\hline
 && ${\rm Octet},\;\Gamma^{(1,7)} =1,\;\Gamma^{(6)} = -1,$&&&&&&& \\
 && ${\rm of \; quarks}$&&&&&&&\\
 \hline\hline
 1&$ \nu_{R}$&$ \stackrel{03}{(+i)}\,\stackrel{12}{(+)}|
 \stackrel{56}{(+)}\,\stackrel{78}{(+)}
 ||\stackrel{9 \;10}{(+)}\;\;\stackrel{11\;12}{(+)}\;\;\stackrel{13\;14}{(+)} $
 &1&$\frac{1}{2}$&1&0&$\frac{1}{2}$&$0$&$0$\\
 \hline 
 2&$\nu_{R}$&$\stackrel{03}{[-i]}\,\stackrel{12}{[-]}|\stackrel{56}{(+)}\,\stackrel{78}{(+)}
 ||\stackrel{9 \;10}{(+)}\;\;\stackrel{11\;12}{[-]}\;\;\stackrel{13\;14}{[-]}$
 &1&$-\frac{1}{2}$&1&0&$\frac{1}{2}$&$0$&$0$\\
 \hline
 3&$e_{R}$&$\stackrel{03}{(+i)}\,\stackrel{12}{(+)}|\stackrel{56}{[-]}\,\stackrel{78}{[-]}
 ||\stackrel{9 \;10}{(+)}\;\;\stackrel{11\;12}{[-]}\;\;\stackrel{13\;14}{[-]}$
 &1&$\frac{1}{2}$&1&0&$-\frac{1}{2}$&$-1$&$-1$\\
 \hline 
 4&$ e_{R} $&$\stackrel{03}{[-i]}\,\stackrel{12}{[-]}|
 \stackrel{56}{[-]}\,\stackrel{78}{[-]}
 ||\stackrel{9 \;10}{(+)}\;\;\stackrel{11\;12}{[-]}\;\;\stackrel{13\;14}{[-]} $
 &1&$-\frac{1}{2}$&1&0&$-\frac{1}{2}$&$-1$&$-1$\\
 \hline
 5&$e_{L}$&$\stackrel{03}{[-i]}\,\stackrel{12}{(+)}|\stackrel{56}{[-]}\,\stackrel{78}{(+)}
 ||\stackrel{9 \;10}{(+)}\;\;\stackrel{11\;12}{[-]}\;\;\stackrel{13\;14}{[-]}$
 &-1&$\frac{1}{2}$&-1&$-\frac{1}{2}$&0&$-\frac{1}{2}$&$-1$\\
 \hline
 6&$e_{L} $&$\stackrel{03}{(+i)}\,\stackrel{12}{[-]}|
 \stackrel{56}{[-]}\,\stackrel{78}{(+)}
 ||\stackrel{9 \;10}{(+)}\;\;\stackrel{11\;12}{[-]}\;\;\stackrel{13\;14}{[-]} $
 &-1&$-\frac{1}{2}$&-1&$-\frac{1}{2}$&0&$-\frac{1}{2}$&$-1$\\
 \hline
 7&$ \nu_{L}$&$ \stackrel{03}{[-i]}\,\stackrel{12}{(+)}|
 \stackrel{56}{(+)}\,\stackrel{78}{[-]}
 ||\stackrel{9 \;10}{(+)}\;\;\stackrel{11\;12}{[-]}\;\;\stackrel{13\;14}{[-]}$
 &-1&$\frac{1}{2}$&-1&$\frac{1}{2}$&0&$-\frac{1}{2}$&$0$\\
 \hline
 8&$\nu_{L}$&$\stackrel{03}{(+i)}\,\stackrel{12}{[-]}|\stackrel{56}{(+)}\,\stackrel{78}{[-]}
 ||\stackrel{9 \;10}{(+)}\;\;\stackrel{11\;12}{[-]}\;\;\stackrel{13\;14}{[-]}$
 &-1&$-\frac{1}{2}$&-1&$\frac{1}{2}$&0&$-\frac{1}{2}$&$0$\\
 \hline\hline
 \end{tabular}
 \end{center}
 \caption{\label{snmb2bTable Il.} The 8-plet of leptons - the members of $SO(1,7)$ subgroup of the 
 group $SO(1,13)$, 
 belonging to one Weyl left 
 handed ($\Gamma^{(1,13)} = -1 = \Gamma^{(1,7)} \times \Gamma^{(6)}$) spinor representation of 
 $SO(1,13)$. 
 It contains the colour chargeless left handed weak charged leptons and the right handed weak 
 chargeless leptons. The rest of notation is the same as in Table~\ref{snmb2bTable Il.}.  
 }
 \end{table}

 In both tables the vectors are chosen to be the eigenvectors of the operators of 
 handedness $\Gamma^{(n)}$,  
 the generators $\tau^{13}, \, \tau^{23}, \,\tau^{33}$  $ \tau^{38}$,  $Y= \tau^{4} + \tau^{23}$ and
 $Q= Y + \tau^{13}$. They are also 
 eigenvectors of the corresponding $\tilde{S}^{ab}$, $\tilde{\tau}^{Ai}, A=1,2,3$ and $\tilde{Y}, \tilde{Q}$. 
One easily sees that the right handed vectors (with respect to $SO(1,3)$ )  are weak ($SU(2)_{I}$) 
chargeless and are doublets with respect to the second $SU(2)_{II}$, while the left handed are  weak 
charged and singlets with respect to $SU(2)_{II}$. 

The generators $\tilde{S}^{ab}$ transform each member of a family into the same member of other 
$2^{\frac{8}{2}-1}$ families. 
 The eight families of the first 
 member of the eightplet of quarks from Table~\ref{snmb2bTable I.}, for example, that is of the right 
 handed $u$-quark 
 of the spin $\frac{1}{2}$,  are presented in the left column of Table~\ref{snmb2bTable II.}. 
 The corresponding right handed neutrinos, belonging to eight different families, are presented 
 in the right column of the same table. The $u$-quark member of the eight families and the $\nu$ 
 members of the same eight families
 are generated by $\tilde{S}^{cd}$, $c,d \in \{0,1,2,3,5,6,7,8\}$ from any starting family.
 \begin{table}
 \begin{center}
 \begin{tabular}{|r||c||c||c||c||}
 \hline
 $I_R$ & $u_{R}^{c1}$&
 $ \stackrel{03}{[+i]}\,\stackrel{12}{(+)}|\stackrel{56}{(+)}\,\stackrel{78}{[+]}||
 \stackrel{9 \;10}{(+)}\:\; \stackrel{11\;12}{[-]}\;\;\stackrel{13\;14}{[-]}$ & 
 $\nu_{R}$&
 $ \stackrel{03}{[+i]}\,\stackrel{12}{(+)}|\stackrel{56}{(+)}\,\stackrel{78}{[+]}||
 \stackrel{9 \;10}{(+)}\;\;\stackrel{11\;12}{(+)}\;\;\stackrel{13\;14}{(+)}$ 
 \\
 \hline
  $II_R$ & $u_{R}^{c1}$&
  $ \stackrel{03}{[+i]}\,\stackrel{12}{(+)}|\stackrel{56}{[+]}\,\stackrel{78}{(+)}||
  \stackrel{9 \;10}{(+)}\;\;\stackrel{11\;12}{[-]}\;\;\stackrel{13\;14}{[-]}$ & 
  $\nu_{R}$&
  $ \stackrel{03}{(+i)}\,\stackrel{12}{[+]}|\stackrel{56}{(+)}\,\stackrel{78}{[+]}||
  \stackrel{9 \;10}{(+)}\;\;\stackrel{11\;12}{(+)}\;\;\stackrel{13\;14}{(+)}$ 
 \\
 \hline
 $III_R$ & $u_{R}^{c1}$&
 $ \stackrel{03}{(+i)}\,\stackrel{12}{[+]}|\stackrel{56}{(+)}\,\stackrel{78}{[+]}||
 \stackrel{9 \;10}{(+)}\;\;\stackrel{11\;12}{[-]}\;\;\stackrel{13\;14}{[-]}$ & 
 $\nu_{R}$&
 $ \stackrel{03}{(+i)}\,\stackrel{12}{[+]}|\stackrel{56}{[+]}\,\stackrel{78}{(+)}||
 \stackrel{9 \;10}{(+)}\;\;\stackrel{11\;12}{(+)}\;\;\stackrel{13\;14}{(+)}$ 
 \\
 \hline
 $IV_R$ & $u_{R}^{c1}$&
 $ \stackrel{03}{(+i)}\,\stackrel{12}{[+]}|\stackrel{56}{[+]}\,\stackrel{78}{(+)}||
 \stackrel{9 \;10}{(+)}\;\;\stackrel{11\;12}{[-]}\;\;\stackrel{13\;14}{[-]}$ & 
 $\nu_{R}$&
 $ \stackrel{03}{[+i]}\,\stackrel{12}{(+)}|\stackrel{56}{[+]}\,\stackrel{78}{(+)}|| 
 \stackrel{9 \;10}{(+)}\;\;\stackrel{11\;12}{(+)}\;\;\stackrel{13\;14}{(+)}$ 
 \\
 \hline\hline\hline
 $V_R$ & $u_{R}^{c1}$&
 $ \stackrel{03}{(+i)}\,\stackrel{12}{(+)}|\stackrel{56}{(+)}\,\stackrel{78}{(+)} ||
 \stackrel{9 \;10}{(+)}\;\;\stackrel{11\;12}{[-]}\;\;\stackrel{13\;14}{[-]}$ & 
 $\nu_{R}$&
 $ \stackrel{03}{(+i)}\,\stackrel{12}{(+)}|\stackrel{56}{(+)}\,\stackrel{78}{(+)} ||
 \stackrel{9 \;10}{(+)}\;\;\stackrel{11\;12}{(+)}\;\;\stackrel{13\;14}{(+)}$ 
 \\
 \hline
 $VI_R$ & $u_{R}^{c1}$&
 $ \stackrel{03}{(+i)}\,\stackrel{12}{(+)}|\stackrel{56}{[+]}\,\stackrel{78}{[+]}|| 
 \stackrel{9 \;10}{(+)}\;\;\stackrel{11\;12}{[-]}\;\;\stackrel{13\;14}{[-]}$ & 
 $\nu_{R}$&
 $ \stackrel{03}{(+i)}\,\stackrel{12}{(+)}|\stackrel{56}{[+]}\,\stackrel{78}{[+]}||
 \stackrel{9 \;10}{(+)}\;\;\stackrel{11\;12}{(+)}\;\;\stackrel{13\;14}{(+)}$ 
 \\
 \hline
 $VII_R$ & $u_{R}^{c1}$&
 $\stackrel{03}{[+i]}\,\stackrel{12}{[+]}|\stackrel{56}{(+)}\,\stackrel{78}{(+)}|| 
 \stackrel{9 \;10}{(+)}\;\;\stackrel{11\;12}{[-]}\;\;\stackrel{13\;14}{[-]}$ & 
 $\nu_{R}$&
 $\stackrel{03}{[+i]}\,\stackrel{12}{[+]}|\stackrel{56}{(+)}\,\stackrel{78}{(+)}||
 \stackrel{9 \;10}{(+)}\;\;\stackrel{11\;12}{(+)}\;\;\stackrel{13\;14}{(+)}$ 
 \\
 \hline
 $VIII_R$ & $u_{R}^{c1}$&
 $ \stackrel{03}{[+i]}\,\stackrel{12}{[+]}|\stackrel{56}{[+]}\,\stackrel{78}{[+]}|| 
 \stackrel{9 \;10}{(+)}\;\;\stackrel{11\;12}{[-]}\;\;\stackrel{13\;14}{[-]}$ & 
 $\nu_{R}$&
 $ \stackrel{03}{[+i]}\,\stackrel{12}{[+]}|\stackrel{56}{[+]}\,\stackrel{78}{[+]}|| 
 \stackrel{9 \;10}{(+)}\;\;\stackrel{11\;12}{(+)}\;\;\stackrel{13\;14}{(+)}$ 
 \\
 \hline 
 \end{tabular}
 \end{center}
 \caption{\label{snmb2bTable II.} Eight families of the right handed $u_R$ quark with the spin $\frac{1}{2}$, 
  the colour charge $\tau^{33}=1/2$, $\tau^{38}=1/(2\sqrt{3})$ and of the colourless right handed  
  neutrino $\nu_R$ of the spin $\frac{1}{2}$ are presented in the left and in the right column, 
  respectively.
  $S^{ab}, a,b \in \{0,1,2,3,5,6,7,8\}$ transform $u_{R}^{c1}$ of the spin $\frac{1}{2}$ and the 
  chosen colour $c1$ to all the members of the same colour: to the right handed $u_{R}^{c1}$ 
  of the spin $-\frac{1}{2}$, 
  to the left $u_{L}^{c1}$ of both spins ($\pm \frac{1}{2}$), to the right handed $d_{R}^{c1}$ of both spins 
  ($\pm \frac{1}{2}$) and to the left handed $ d_{L}^{c1}$ of both spins ($\pm \frac{1}{2}$). They transform 
  equivalently the right handed   neutrino $\nu_R$ of the spin $\frac{1}{2}$ to the right handed 
  $\nu_R$ of the spin ($-\frac{1}{2}$), to  $\nu_L$ of both spins, to $e_R$ of both spins and to 
  $e_L$ of both spins. $\tilde{S}^{ab}, a,b \in \{0,1,2,3,5,6,7,8\}$ transform a chosen member of one family 
  into the same member of all the eight families.}
 \end{table}

Let us present also the quantum numbers of  the families from Table~\ref{snmb2bTable II.}. 
In Table~\ref{snmb2bTable IV.} 
the handedness of the families
$\tilde{\Gamma}^{(1+3)}(= -4i \tilde{S}^{03} \tilde{S}^{12})$, 
$\tilde{S}^{03}_{L}, \tilde{S}^{12}_L$, $\tilde{S}^{03}_{R}, \tilde{S}^{12}_R$ (the diagonal matrices of 
$SO(1,3)$ ), $\tilde{\tau}^{13}$ 
(of one of the two $SU(2)_{I}$), $\tilde{\tau}^{23}$ (of the second 
$SU(2)_{II}$) are presented. 

%
 \begin{table}
 \begin{center}
 \begin{tabular}{|r||r|r|r|r|r|r|r|r|r|r|r||}
 \hline
$i$ & $\tilde{\Gamma}^{(1+3)}$& $\tilde{S}^{03}_{L}$&$ \tilde{S}^{12}_L$& $\tilde{S}^{03}_{R}$& 
$\tilde{S}^{12}_R$& $\tilde{\tau}^{13}$ & $\tilde{\tau}^{23}$&$\tilde{\tau}^{4}$ & 
$\tilde{Y}'$&$\tilde{Y}$&$\tilde{Q}$ \\
\hline
\hline
$1$& $-1$ &$ - \frac{i}{2}$ &$ \frac{1}{2}$& $0$& $0$& $  \frac{1}{2}$& $0$&$-\frac{1}{2}$&$0$&$-\frac{1}{2}$&$0$\\ 
\hline
$2$& $-1$ &$ -\frac{i}{2}$ &$  \frac{1}{2}$& $0$& $0$& $-\frac{1}{2}$& $0$&$-\frac{1}{2}$&$0$&$-\frac{1}{2}$&$-1$\\
\hline
$3$& $-1$ &$ \frac{i}{2}$ &$ - \frac{1}{2}$& $0$& $0$& $ \frac{1}{2}$& $0$&$-\frac{1}{2}$&$0$&$-\frac{1}{2}$&$0$\\
\hline
$4$& $-1$ &$ \frac{i}{2}$ &$ - \frac{1}{2}$& $0$& $0$& $-\frac{1}{2}$& $0$&$-\frac{1}{2}$&$0$&$-\frac{1}{2}$&$-1$\\
\hline\hline
$5$& $1 $ & $0$ & $0$& $\frac{i}{2}$ &$ \frac{1}{2}$& $0 $& $ \frac{1}{2}$&$-\frac{1}{2}$&$\frac{1}{2}$&$0$&$0$\\ 
\hline
$6$& $1 $ & $0$ & $0$& $\frac{i}{2}$ &$  \frac{1}{2}$& $0 $& $-\frac{1}{2}$&$-\frac{1}{2}$&$-\frac{1}{2}$&$-1$&$-1$\\
\hline
$7$& $1 $ & $0$ & $0$& $-\frac{i}{2}$ &$- \frac{1}{2}$& $0 $& $ \frac{1}{2}$&$-\frac{1}{2}$&$\frac{1}{2}$&$0$&$0$\\
\hline
$8$& $1 $ &  $0$& $0$& $- \frac{i}{2}$ &$-\frac{1}{2}$& $0 $& $-\frac{1}{2}$&$-\frac{1}{2}$&$-\frac{1}{2}
$&$-1$&$-1$\\
\hline
\hline
 \end{tabular}
 \end{center}
 \caption{\label{snmb2bTable IV.}  The quantum numbers of each member of the eight families presented in 
 Table~\ref{snmb2bTable II.} are presented: The handedness of the families 
 $\tilde{\Gamma}^{(1+3)}= -4i \tilde{S}^{03} \tilde{S}^{12}$, the left and right handed $SO(1,3)$ 
 quantum numbers 
$\tilde{S}^{03}_{L}, \tilde{S}^{12}_L$, $\tilde{S}^{03}_{R}, \tilde{S}^{12}_R$ (of $SO(1,3)$ group in the 
$\tilde{S}^{mn}$ sector), $\tilde{\tau}^{13}$ 
 of  $SU(2)_{I}$, $\tilde{\tau}^{23}$ of the second 
$SU(2)_{II}$, $\tilde{\tau}^4$, 
$\tilde{Y}'= \tilde{\tau}^{23} -  
\tilde{\tau}^4 \, \tan\tilde{\theta}_2$, taking $\tilde{\theta}^2=0$, $\tilde{Y}
=\tilde{\tau}^{4} + \tilde{\tau}^{23}$, 
$\tilde{Q}= \tilde{\tau}^{4} + \tilde{S}^{56}$.  See also the ref.~\cite{snmb2bnormaproc2010talk}.
}
\end{table}

We see in Table~\ref{snmb2bTable IV.} that four of the eight families are singlets with respect to 
one of the two $SU(2)$ ($SU(2)_{I}$) groups determined by $\tilde{S}^{ab}$ and doublets with respect to the 
second $SU(2)$ ($SU(2)_{II}$), while the remaining four families are doublets with respect to the first 
$SU(2)_{I}$ and singlets with respect to the second $SU(2)_{II}$. When the first break  
appears, to which besides the vielbeins also the spin connections  contribute, we expect that 
if only one of the two $SU(2)$ subgroups of $SO(1,7) \times U(1)$ breaking into
$SO(1,3) \times SU(2)\times U(1)$ contributes in the break~\cite{snmb2bnormaproc2010talk}, namely that of the charges 
$\tilde{\tau}^{2 i}$, together with $\tilde{N}^{i}_{-} $, there will be four families massless and mass 
protected after this break, namely those, 
which are singlets with respect to $\vec{\tilde{\tau}}^{2}$ and with respect to $\tilde{N}^{i}_{-} $
(Table~\ref{snmb2bTable IV.}), while for the other four families  the vacuum expectation values of the 
scalars (particular combinations of vielbeins $f^{\sigma}{}_{s}$, and spin connections 
$\tilde{\omega}_{abs}, s \in \{5,8\}$) will take care of the mass matrices on the tree level and beyond  
the tree level. 

 \subsection{Discrete symmetries of the theory unifying spin and charges and explaining families}
\label{snmb2bcpt} 

Let us define the discrete operators of the parity ($ P$) and of the charge conjugation 
($C$).
\begin{eqnarray}
\label{snmb2bcp}
P &=& \gamma^0 \, \gamma^8 \, I_{x}\, , \nonumber\\
C &=& \Pi_{Im\, \gamma^a}\, \gamma^a \, K\,.
\end{eqnarray}
$K$ means complex conjugation, while in our choice of matrix representation of the $\gamma^a$ 
matrices $\Pi_{Im\, \gamma^a}\, \gamma^a \,= \gamma^2 \gamma^5 \gamma^7  \gamma^9 \gamma^{11} \gamma^{13}$.

One can easily check that $P$ transforms  the $ u_{R}^{c1}$ from the first row in Table~\ref{snmb2bTable I.} 
into the $ u_{L}^{c1}$ of the seventh row in the same table. The $CP$ transforms the fermion 
states of table~\ref{snmb2bTable I.} into the corresponding states of antifermions: $ u_{R}^{c1}$ 
from the first row in table~\ref{snmb2bTable I.} with the spin $\frac{1}{2}$, weak chargeless
and of the colour charge
($(\frac{1}{2}, \frac{1}{2 \sqrt{3}})$) into a right handed antiquark $ \bar{u}_{R}^{\bar{c1}}$, 
weak charged and of the colour charge ($(-\frac{1}{2}, -\frac{1}{2 \sqrt{3}})$) as presented 
in table~\ref{snmb2bTable anti}.
 \begin{table}
 \begin{center}
 \begin{tabular}{|r|c||c||c|c||c|c|c||r|r|}
 \hline
 i&$$&$|^a\psi_i>$&$\Gamma^{(1,3)}$&$ S^{12}$&$\Gamma^{(4)}$&
 $\tau^{13}$&$\tau^{23}$&$Y$&$Q$\\
 \hline\hline
 && ${\rm Octet},\;\Gamma^{(1,7)} =-1,\;\Gamma^{(6)} = 1,$&&&&&&& \\
 && ${\rm of \; antiquarks}$&&&&&&&\\
 \hline\hline
 1&$ \bar{u}_{R}^{\bar{c1}}$&$ \stackrel{03}{[-i]}\,\stackrel{12}{[-]}|
 \stackrel{56}{[-]}\,\stackrel{78}{(+)}
 ||\stackrel{9 \;10}{[-]}\;\;\stackrel{11\;12}{(+)}\;\;\stackrel{13\;14}{(+)} $
 &1&$-\frac{1}{2}$&-1&$-\frac{1}{2}$&$0$&$-\frac{1}{6}$&$-\frac{2}{3}$\\
 \hline 
 2&$\bar{u}_{R}^{\bar{c1}}$&$\stackrel{03}{(+i)}\,\stackrel{12}{(+)}|
 \stackrel{56}{[-]}\,\stackrel{78}{(+)}
 ||\stackrel{9 \;10}{[-]}\;\;\stackrel{11\;12}{(+)}\;\;\stackrel{13\;14}{(+)}$
 &1&$ \frac{1}{2}$&-1&$-\frac{1}{2}$&$0$&$-\frac{1}{6}$&$-\frac{2}{3}$\\
 \hline
 3&$\bar{d}_{R}^{\bar{c1}}$&$\stackrel{03}{[-i]}\,\stackrel{12}{[-]}|
 \stackrel{56}{(+)}\,\stackrel{78}{[-]}
 ||\stackrel{9 \;10}{[-]}\;\;\stackrel{11\;12}{(+)}\;\;\stackrel{13\;14}{(+)}$
 &1&$-\frac{1}{2}$&-1&$ \frac{1}{2}$&$0$&$-\frac{1}{6}$&$\frac{1}{3}$\\
 \hline 
 4&$ \bar{d}_{R}^{\bar{c1}} $&$\stackrel{03}{(+i)}\,\stackrel{12}{(+)}|
 \stackrel{56}{(+)}\,\stackrel{78}{[-]}
 ||\stackrel{9 \;10}{[-]}\;\;\stackrel{11\;12}{(+)}\;\;\stackrel{13\;14}{(+)} $
 &1&$\frac{1}{2}$&-1&$\frac{1}{2}$&$0$&$-\frac{1}{6}$&$\frac{1}{3}$\\
 \hline
 5&$\bar{d}_{L}^{\bar{c1}}$&$\stackrel{03}{(+i)}\,\stackrel{12}{[-]}|
 \stackrel{56}{(+)}\,\stackrel{78}{(+)}
 ||\stackrel{9 \;10}{[-]}\;\;\stackrel{11\;12}{(+)}\;\;\stackrel{13\;14}{(+)}$
 &-1&$-\frac{1}{2}$&1&$0$&$\frac{1}{2}$&$\frac{1}{3}$&$\frac{1}{3}$\\
 \hline
 6&$\bar{d}_{L}^{\bar{c1}} $&$\stackrel{03}{[-i])}\,\stackrel{12}{(+)}|
 \stackrel{56}{(+)}\,\stackrel{78}{(+)}
 ||\stackrel{9 \;10}{[-]}\;\;\stackrel{11\;12}{(+)}\;\;\stackrel{13\;14}{(+)} $
 &-1&$\frac{1}{2}$&1&$0$&$\frac{1}{2}$&$\frac{1}{3}$&$\frac{1}{3}$\\
 \hline
 7&$ \bar{u}_{L}^{\bar{c1}}$&$\stackrel{03}{(+i)}\,\stackrel{12}{[-]}|
 \stackrel{56}{[-]}\,\stackrel{78}{[-]}
 ||\stackrel{9 \;10}{[-]}\;\;\stackrel{11\;12}{(+)}\;\;\stackrel{13\;14}{(+)}$
 &-1&$-\frac{1}{2}$&1&$0$&$-\frac{1}{2}$&$-\frac{2}{3}$&$-\frac{2}{3}$\\
 \hline
 8&$\bar{u}_{L}^{\bar{c1}}$&$\stackrel{03}{[-i]}\,\stackrel{12}{(+)}|
 \stackrel{56}{[-]}\,\stackrel{78}{[-]}
 ||\stackrel{9 \;10}{[-]}\;\;\stackrel{11\;12}{(+)}\;\;\stackrel{13\;14}{(+)}$
 &-1&$ \frac{1}{2}$&1&$0$&$-\frac{1}{2}$&$-\frac{2}{3}$&$-\frac{2}{3}$\\
 \hline\hline
 \end{tabular}
 \end{center}
 \caption{\label{snmb2bTable anti} The 8-plet of antiquarks  to the quarks 
 obtained from Table~\ref{snmb2bTable I.} by the $CP$ ($= \gamma^2 \gamma^5 \gamma^7  
 \gamma^9 \gamma^{11} \gamma^{13}K \gamma^0 \gamma^8 \, I_{x}$) conjugation.} 
 \end{table}
 \section{The fermion-antifermion asymmetry within the theory unifying spin and charges 
 and explaining families}
 \label{snmb2bmatterasymmetry}

 As said in the abstract, I shall here  follow the ideas from
 the references~\cite{snmb2bgross,snmb2brubakovshaposhnikov,snmb2bdinekusenko,snmb2btapeiling}. The difference from the 
 studies there in here is, as explained, in the number of  families (there are two decoupled 
 groups of four families and consequently two stable families), in the number of 
 gauge fields contributing to the phase transitions and in the types of the gauge 
 fields contributing to phase transitions.
 
 Let us assume that the fermion-antifermion asymmetry is zero, when the expanding 
 universe cools down to the 
 temperature below the unification scale of the spin and the charges that is to the 
 temperature below, let say, $10^{16}$ TeV, when there are eight massless families, 
 manifesting  the symmetry $SO(1,3) \times SU(2) \times SU(2) \times U(1) \times SU(3)$, and 
 distinguishing among themselves in the quantum numbers defined by $\tilde{S}^{ab}$.

 Then we must investigate, how much do the following processes contribute to  the fermion-antifermion 
 asymmetry  in non equilibrium thermal processes in the expanding universe:   
 
 \begin{itemize}
  
 \item The nonconservation of currents on the quantum level
 due to the triangle anomalies~\cite{snmb2btapeiling,snmb2brubakovshaposhnikov,snmb2bdinekusenko}, 
 which are responsible for 
 $P$ and $CP$ nonconservation  
 \begin{eqnarray}
 \label{snmb2bnonconsrvedcurrents}
 \partial^{m} \, j^{A i \,\alpha  (i)}_{ m} &=& \frac{(g^{A})^2}{8 \pi^2}  \; \frac{1}{2} \,
 \varepsilon_{mnpr}\, F^{Ai\, mn} F^{Ai \, pr}.
 \end{eqnarray}
 Here $j^{A i \,\alpha }_{ m} $ stays  for the currents of fermions (and antifermions), which carry 
 a particular charge denoted by a charge group $A$, in our case $A=4$ means the $U(1)$ 
 charge originating in $SO(6)$,   $A=3$ means the $SU(3)$ (colour) 
 charge, 
 $A=2_I$ means the weak $SU(2)_I$ charge of the left handed doublets, while $A=2_{II}$  stays 
  for the $SU(2)_{II}$
 charge of the right handed singlets before the $SU(2)_{II}$ break, 
 $A=1$ stays for the actual $U(1)$ charge (the {\it standard model} like hyper charge 
 after the $SU(2)_{II}$ break and the electromagnetic one after the weak break).

 In my case also the fields, which look like scalar fields in $d=(1+3)$, $\tilde{A}^{\tilde{A}i}_{s}$, 
 $s,t \in{5,6,\cdots}$, and to which the fermions are 
 coupled, contribute. 
 
 All the fermions and antifermions, which are coupled to a particular gauge 
 field $A^{Ai}_{m}$ and 
 in my case  also $\tilde{A}^{\tilde{A}i}_{s}$ contribute to the current 
 \begin{eqnarray}
 \label{snmb2bcurrent}
 j^{A i \,\alpha (i)}_m = \psi^{Ai\, \alpha (i)\dagger}\, \gamma^0 \gamma^m \,\psi^{Ai \, \alpha (i)}. 
 \end{eqnarray}
 $(i) \in  \{1,8\}$ enumerates families, in my case twice four families which are distinguishable 
 by the quantum numbers originating in $\tilde{S}^{ab}$, namely, after the break of $SU(2)_I$  
 the lower four families, which are 
 doublets with respect to $\tilde{N}^{i}_{+}$ and $\tilde{\tau}^{I\,i}$ and singlets with respect to 
 $\tilde{N}^{i}_{-}$ and $\tilde{\tau}^{II\,i}$,  stay massless, 
 while the upper four families are doublets 
 with respect to  $\tilde{N}^{i}_{-}$ and $\tilde{\tau}^{IIi}$ and singlets with respect to 
 $\tilde{N}^{i}_{+}$ and $\tilde{\tau}^{Ii}$. After the electroweak break all the eight families 
 become massive, but the upper four families have no mixing matrix elements since the way of breaking 
 leaves all the 
 $\omega_{msa}$ and $\tilde{\omega}_{msa}$, with $m=0,1,2,3; s=5,6,\cdots$, equal to zero. 
  $\alpha$  distinguishes the multiplets in each family, in my case of the two $SU(2)$ 
  gauge groups $\alpha$ distinguishes the $SU(2)_I$ doublets, that is one colour singlet and 
  one colour triplet, 
  and the $SU(2)_{II}$ doublets, again one colour singlet and one colour triplet.
  $A^{Ai}_{m}$ are the corresponding gauge fields, with tensors  
 $F^{A}_{mn} =  \tau^{A i} \,F^{A i}_{mn}$ and $F^{A i}_{mn}= A^{A i}_{n,m} - A^{A i}_{m,n} +
 g^A \, f^{Ai j k} \, A^{A j}_{m}\, A^{A k}_{n} $. (The scalar fields 
 $\tilde{A}^{\tilde{A}i}_{s}$ define tensors  
  $\tilde{F}^{\tilde{A}i\, st}= \tilde{A}^{\tilde{A}i}_{t,s} -
  \tilde{A}^{\tilde{A}i}_{s,t} + g^{\tilde{}A} \, f^{A\, ijk} \tilde{A}^{\tilde{A}j}_{s}\,
 \tilde{A}^{\tilde{A}k}_{t}$.)   
 
 The nonconserved currents affect the fermions and antifermions. (In the later case the 
 $\tau^{A i}$ are replaced by $\bar{\tau}^{A i}$, both fulfilling the same 
 commutation relations $\{\tau^{A i}, \tau^{B j} \}_{-} = 
 i \delta^{AB} \, f^{A \,ijk} \, \tau^{A k} $, $\{\bar{\tau}^{A i}, \bar{\tau}^{B j} \}_{-} = 
 i \delta^{AB} \, f^{A \,ijk} \, \bar{\tau}^{A k}$). One obtains $\bar{\tau}^{A i}$ from 
$\tau^{A i}$ by the 
 $CP$ transformation $P= \gamma^0 \gamma^8 \,I_{x}$, while $C= \prod_{ Im \gamma^a}\,\gamma^a \, K$
 (See~\ref{snmb2bcpt}).

 \item The nonconservation of the fermion numbers originating in the 
 complex phases of  the mixing matrices of the two times $4 \times 4$ mass matrices for 
 each member of a family, after the two successive breaks causes two phase transitions 
 when the symmetry $SU(2) \times SU(2) \times U(1)$ breaks first to $SU(2) \times U(1)$ 
 and finally to  $U(1)$ and the two types of gauge 
 fields manifest their masses while the two groups of four with the 
 mixing matrices decoupled families gain nonzero mass matrices in the first break the 
 upper four families and in the second break the lower four families.  
 
 \end{itemize}

I am following here the references~\cite{snmb2bgross,snmb2brubakovshaposhnikov,snmb2bdinekusenko,snmb2btapeiling}.
The nonconservation of currents may be expected whenever the  non-Abelian  gauge fields 
 manifest a non trivial structure of  vacua, originating 
in  the {\it instanton} solutions of the Euclidean non-Abelian gauge theories in $(1+3)-$dimensional 
space, that is in $A^{A}_{m}$, which  fulfil the boundary condition 
$\lim_{r \to \infty} \;\tau^{Ai}\,A^{Ai}_{m} = U^{-1} \partial_m U$, summed over $i$ for a particular 
gauge group $A$ (and 
similarly might be that the fields 
$ \lim_{\rho \to \infty} \;\tilde{\tau}^{\tilde{A}i}\,\tilde{A}^{\tilde{A}i}_{s}= 
U^{-1} \partial_s U$, with $r=\sqrt{(x^0)^2 + \vec{x}^2}$ and $\rho = \sqrt{\sum_s}\, (x^s)^2$, 
for a particular $\tilde{A}$ , contribute as well, where the effect of the triangle anomalies 
in the case of scalar gauge 
fields depending on $x^{\sigma}, \sigma = 5,6,7,8$ and the corresponding meaning of the winding numbers  
distinguishing among the different vacua in this case might be negligible and should be studied). 
The vacua  distinguish among themselves in the topological quantum numbers $n_{A}$ ($ n_{\tilde{A}}$), 
determined by  a particular choice of $U$
\begin{eqnarray}
\label{snmb2bna}
n_{A} = \frac{(g^{A})^2 }{16 \pi^2} \, \int d^4 x \, \varepsilon_{mnpr}\, Tr (F^{Amn} F^{Apr})= 
\frac{(g^{A})^2}{32 \pi^2} \, \int d^{4} x \partial_{m} K^{A\, m}, 
\end{eqnarray}
where $K^{A}_m= \sum_{i}4 \varepsilon_{mnpr} \,( A^{Ai}_{n} \partial_{p} A^{A}_{r}+ 
\frac{2}{3}\, g^A f^{Aijk} A^{Ai}_{n} A^{Aj}_{p}A^{Ak}_{r}).$ 
(Similarly also the topological quantum number $n_{\tilde{A}}$ might be non negligible.)

Instanton solutions fulfilling 
the boundary condition $\lim_{r \to \infty}\; \tau^{Ai}\, A^{Ai}_{m}  = U^{-1} \partial_m U$ 
for a particular gauge group $A$ 
(or $\lim_{\rho \to \infty}\; \tilde{\tau}^{\tilde{A}i}\,\tilde{A}^{\tilde{A}i}_{s} = 
U^{-1} \partial_m U$ for a particular $\tilde{A}$), each with its own $U$ for a particular 
$A$ (or $\tilde{A}$), connect vacua  
$|n_{A}>$ with different winding numbers~\footnote{In the ref.~\cite{snmb2btapeiling,snmb2bgross} the 
{\it instanton} field 
$\tau^{Ai}\, A^{Ai}_{m}= \frac{r^2}{r^2 + \lambda^2} \,U^{-1} \partial_m U$, with $U = 
\frac{x^0 + i\vec{\sigma}^{A} \cdot \vec{x}}{r}$, $r^2 = (x^0)^2 + \vec{x}^2,$ is presented. 
Operators $\vec{\sigma}^{A}$ 
are the three Pauli matrices,  used  to denote the $SU(2)$ gauge group in this case: $\vec{\tau}^{A}=
\frac{\vec{\sigma}^{A}}{2}$, $\{\tau^{Ai}, \tau^{Aj}\}_{-} = i \varepsilon^{ijk} \tau^{Ak}$. 
The corresponding 
action  $ \int d^{4}x\, \frac{1}{2} \varepsilon_{mnpr} \, F^{Ai\, mn} F^{Ai pr}= \frac{8 \pi^2}{g^2}$, 
while  $U = 
\frac{x^0 + i\vec{\sigma}^{A} \cdot \vec{x}}{r}$, defines the $n=1$ vacuum state. 
}
$n_{A}$ (and correspondingly for $n_{\tilde{A}}$).
The true vacuum $|\theta^{A}>$ is for each $A$ (let it count also $\tilde{A}$) in a stationary 
situation a superposition of the vacua, 
determined by the time 
independent gauge transformation~\cite{snmb2bgross} ${\cal T}$, 
${\cal T} |\theta^{A}> = e^{i\theta^{A}} |\theta^{A}>$, 
where $\theta^{A}$ is a parameter, which weights  
 the contribution of a vacuum to the effective Lagrange density 
${\cal L}_{eff}= {\cal L} + \sum_{A} \, \frac{\theta^{A}}{16 \pi^2} \, 
F^{Ai \,mn} \frac{1}{2} \varepsilon_{mnpr}
F^{Ai\, pr}$, for a particular gauge field. ${\cal T}$ acts as the raising operator for the handedness 
(chirality).  
The second term of the effective Lagrange density ${\cal L}_{eff}$  violates parity $P$ and  then also $CP$.  
The vacuum state with the definite handedness has also a definite topological quantum number. 
In the presence of the massless fermions all the vacua  $|\theta^{A}>$, for each $A$,  are equivalent.

The fermion currents~(Eq.(\ref{snmb2bcurrent})) are not conserved in  processes, for which 
 the gauge fields are such  that the corresponding winding number $n_A$ of Eq.~(\ref{snmb2bna}) is nonzero.
 Correspondingly also the fermion (and antifermion numbers), carrying the corresponding charge, are 
 not conserved 
\begin{eqnarray}
\label{snmb2bnafermion}
\Delta n_{A i \,\alpha  (i)}= n_{A}. 
\end{eqnarray}
The fermion number of all the fermions interacting with the same  non-Abelian gauge field with nonzero 
winding number, either of a vector or of a scalar type (whose contribution should be studied and 
hopefully understood),  changes 
in such processes for the {\it same amount}: Any member of a family, interacting with the particular 
field and therefore also the corresponding members of each family, either a quark or  a lepton member 
of doublets, change for the same amount, before the  breaks or after the breaks 
(in my case first from $SO(1,3) \times SU(2) \times SU(2) \times U(1) \times SU(3)$ to
$SO(1,3) \times SU(2)  \times U(1) \times SU(3)$ and finally to $SO(1,3)  \times U(1) \times SU(3)$) 
of the symmetries.

For a baryon three quarks are needed. It is the conservation of the colour charge which 
requires  that  the lepton number and the baryon number ought to be conserved 
separately as long as the charge group is a global symmetry. The transformations, which allow 
rotations of a lepton to a quark or opposite, conserve the fermion number, but not the lepton and not the 
baryon number.

  Instanton solutions of the non-Abelian gauge fields, which connect different vacua (see the 
refs.~\cite{snmb2btapeiling}, page 481, and~\cite{snmb2brubakovshaposhnikov}, page 6), are characterized by the 
 highest value of the {\it instanton} field between the two vacua, that is by the {\it sphaleron} energy. 

The question arises, can the {\it instanton} solutions be responsible for the baryon asymmetry of the universe?
The authors of the papers~\cite{snmb2brubakovshaposhnikov,snmb2bdinekusenko} discuss and evaluate the 
probability for tunnelling from one vacuum to the other at low energy regime and also at the energies of 
{\it sphalerons}. 
When once the  system of gauge fields is in one vacuum the probability for the 
transition to another vacuum depends not only on the {\it sphalerons} height (energy)  
but also on the temperature. If the temperature is low, then the 
transition is negligible. At the temperature above the phase transition 
(the authors~\cite{snmb2brubakovshaposhnikov}
discuss the electroweak phase transition starting at around $100$ GeV, while in my case there is also 
the $SU(2)_{II}$ phase transition at around $10^{16}$ GeV or slightly below) when the fermions 
are massless and the
expansion rate of the universe is much slower that the rate of nonconservation of the fermion number,
and in the case of non equilibrium processes in phase transitions, the fermion number nonconservation 
can be large. 
The authors conclude that more precise  evaluations (treating several models) of the probability 
that in a non thermal equilibrium   phase  transition and below it the fermion  
number would not be conserved   
due to  transitions to vacua with different winding numbers 
in the amount as observed  for the (first family) baryon number excess in the universe are needed.

What can be concluded about the fermion number asymmetry, caused by the transitions of gauge fields 
to different vacua, in my case, where at energies above the 
$ SU{2}_{II}$ phase transition there are eight families of massless fermions, with the charges manifesting 
the symmetries first of $SU(2)_I \times SU(2)_{II} \times U(1)$ and correspondingly with the two 
kinds of the vector gauge 
$SU(2)$ fields which both might demonstrate  the vacua with different winding numbers? In addition 
also the scalar gauge fields might contribute with their even more rich vacua (if they do that at all). 
The phase transitions caused first by the break of the symmetry $SU(2)_I \times SU(2)_{II} \times U(1)$ to 
$SU(2)_{II} \times U(1)$, when the upper four families gain masses (and the  corresponding gauge 
vector fields become massive)
and then by the final break to $U(1)$, with the $\tilde{S}^{ab}$ sector causing the masses 
in both transitions and may be 
also taking care of the richness of vacua with different winding numbers, 
might show up after a careful study as a mechanism 
for generating the fermion-antifermion (or the antifermion-fermion) asymmetry. Although I do not 
yet see, 
how do the non equilibrium processes in the first order phase transitions decide about the 
excess of either fermions or  of antifermions.

So, is it in my case possible that the two successive non equilibrium phase transitions 
leave the excess of antifermions in the case of the upper four families and the excess  
of fermions in the lower four families? Or there is a negligible excess of either fermions or 
antifermions in the upper four families? We saw in the ref.~\cite{snmb2bgn} that an excess of either 
fermions or antifermions is not important for massive enough (few $100$ TeV)  stable 
fifth family members. The excess of fermions over antifermions is certainly what universe made 
a choice of for the lower four families, whatever the reason for this fact is. Can this be easier 
understood  within the {\it spin-charge-family-theory}?
All these need a careful study.

The fermion number nonconservation  originates also in the complex phases of the mass mixing matrices 
of  each of the two groups of four family members. It  might be that the vacua, triggered by  
{\it instanton} solutions of the gauge vector and scalar fields, and the mass matrices, determined 
on the tree level by the 
vacuum expectation values of the scalar gauge fields in the $\tilde{S}^{ab}$ sector, are connected  
(since in the {\it instanton} case also the 
scalar fields, the gauge fields of charges originating in $\tilde{S}^{ab}$ might exhibit the 
{\it instanton} solutions).

\section{Conclusion}
\label{snmb2bconclusion}

In this contribution I pay attention to the origin of baryon asymmetry of our universe within the 
{\it spin-charge-family-theory} under the assumption that  the asymmetry is caused {\bf i.} by the 
{\it instanton} solutions of the non-Abelian gauge fields which determine vacua with different 
winding numbers and {\bf ii.} by the complex matrix elements of the mixing matrices. 

The {\it spin-charge-family-theory} namely assumes besides the Dirac Clifford algebra objects also the 
second ones $\tilde{\gamma^a}$ as a necessary mechanism (or better a mathematical tool) 
which should be used in order that we consistently describe both: spin and charges, as well as families.
The second kind is namely responsible for generating families, defining the equivalent 
representations with respect to the Dirac one. Correspondingly there are 
besides  the two kinds of the vector gauge fields, the  $SU(2)_{I}$ and $SU(2)_{II}$, also the scalar 
gauge fields, the two $SU(2)$ from $SO(4)$ and the two $SU(2)$ from $SO(1,3)$, the superposition of the 
gauge fields of 
$\tilde{S}^{ab} (= \frac{i}{4} (\tilde{\gamma^a} \tilde{\gamma^b} - \tilde{\gamma^b}\tilde{\gamma^a}))$,
 which might contribute to vacua  with different winding numbers (what has to be studied). The scalar fields,  
originating in the $\tilde{S}^{ab}$ charges, are responsible  with their  vacuum expectation values 
(and in loop corrections)
for the mass matrices of fermions after the breaks of symmetries.

The theory predicts twice four families (which 
differ in the family quantum numbers in the way that the upper four families are doublets 
with respect to $\tilde{\tau}^{II \, i}$ and $\tilde{N}^{i}_{-}$, while the lower four families are 
doublets with respect to $\tilde{\tau}^{I \, i},$ and $\tilde{N}^{i}_{+}$) which all  are 
massless above the last two phase transitions. 

What should be clarified in the {\it spin-charge-family-theory} is whether the predicted twice four  
families (rather than once three families of the {\it standard model}) and the fact that there are gauge 
fields belonging to two kinds of generator ($S^{ab}$ and $\tilde{S}^{ab}$) make the baryon number 
asymmetry easier to be understood within these two phenomena --- the {\it instanton} responsibility for the
fermion number nonconservation and the complex matrix elements of the mixing matrices responsibility 
for the fermion number nonconservation.

The manifestation of the {\it instanton} gauge vector and scalar fields in the determination of the 
properties of the vacuum might be correlated with 
the vacuum expectation values of the scalar fields defining the mass matrices of  twice the four families.
 Both manifestations appear in possibly non equilibrium phase transitions of the expanding universe, 
which cause breaking of particular symmetries and also the 
fermion number nonconservation. In this contribution I just follow the way 
suggested by the ref.~\cite{snmb2brubakovshaposhnikov} 
and by the authors cited in this reference, while taking into account the requirement of the 
{\it spin-charge-family-theory}. The fermion number nonconservation obviously ended in the   
excess of (what we call) fermions for the lower four families, while for the upper four 
families we have to see whether there is the  
excess of either the stable fifth family  fermions or antifermions.  
To answer these questions a careful study is needed. It even might be that there was at the non equilibrium 
phase transitions the same  excess of antifermions for the upper four families as it is of fermions 
for the  lower four families, while later the complex matrix elements in the mixing matrices change 
this equality drastically. But yet it must be understood the origin of both sources of the 
fermion number nonconservation.

\section*{Appendix: Some useful relations}
\label{snmb2bsabprop}

The following Cartan subalgebra set of the algebra $S^{ab}$ (for both sectors) is chosen:
\begin{eqnarray}
S^{03}, S^{12}, S^{56}, S^{78}, S^{9 \;10}, S^{11\;12}, S^{13\; 14}\nonumber\\
\tilde{S}^{03}, \tilde{S}^{12}, \tilde{S}^{56}, \tilde{S}^{78}, \tilde{S}^{9 \;10}, 
\tilde{S}^{11\;12}, \tilde{S}^{13\; 14}.
\label{snmb2bcartan}
\end{eqnarray}
%
A left handed ($\Gamma^{(1,13)} =-1$) eigen state of all the members of the 
Cartan  subalgebra 
\begin{eqnarray}
&& \stackrel{03}{(+i)}\stackrel{12}{(+)}|\stackrel{56}{(+)}\stackrel{78}{(+)}
||\stackrel{9 \;10}{(+)}\stackrel{11\;12}{(-)}\stackrel{13\;14}{(-)} |\psi \rangle = \nonumber\\
&&\frac{1}{2^7} 
(\gamma^0 -\gamma^3)(\gamma^1 +i \gamma^2)| (\gamma^5 + i\gamma^6)(\gamma^7 +i \gamma^8)||
\nonumber\\
&& (\gamma^9 +i\gamma^{10})(\gamma^{11} -i \gamma^{12})(\gamma^{13}-i\gamma^{14})
|\psi \rangle .
\label{snmb2bstart}
\end{eqnarray}
represent the $u_R$-quark with spin up and of one colour.

$ \tilde{S}^{ab} $  generate families from the starting $u_R$ quark
In particular $\tilde{S}^{03}(= \frac{i}{2}
[\stackrel{03}{\tilde{(+i)}} \stackrel{12}{\tilde{(+)}} +
\stackrel{03}{\tilde{(-i)}} \stackrel{12}{\tilde{(+)}} +
\stackrel{03}{\tilde{(+i)}} \stackrel{12}{\tilde{(-)}}+
\stackrel{03}{\tilde{(-i)}} \stackrel{12}{\tilde{(-)}}])$  applied on 
a right handed $u_R$-quark with spin up and a particular colour generate a state which is again 
 a right handed $u$-quark of the same colour.
\begin{eqnarray}
\stackrel{03}{\tilde{(-i)}}\stackrel{12}{\tilde{(-)}} &&
\stackrel{03}{(+i)}\stackrel{12}{(+)}| \stackrel{56}{(+)} \stackrel{78}{(+)}||
\stackrel{9 10}{(+)} \stackrel{11 12}{(-)} \stackrel{13 14}{(-)}=\nonumber\\
&&\stackrel{03}{[\,+i]} \stackrel{12}{[\,+\,]}| \stackrel{56}{(+)} \stackrel{78}{(+)}||
\stackrel{9 10}{(+)} \stackrel{11 12}{(-)} \stackrel{13 14}{(-)},
\end{eqnarray}
where 
\begin{eqnarray}
\stackrel{ab}{(\pm i)}         &=& 
\frac{1}{2}\, ( \gamma^a \mp  \gamma^b), 
\stackrel{ab}{(\pm 1)} =          \frac{1}{2} \,( \gamma^a \pm i\gamma^b),\nonumber\\
\stackrel{ab}{[\pm i]}& =& \frac{1}{2} (1 \pm   \gamma^a \gamma^b), \quad
\stackrel{ab}{[\pm 1]} = \frac{1}{2} (1 \pm i \gamma^a \gamma^b), \nonumber\\
\stackrel{ab}{\tilde{(\pm i)}} &=& 
\frac{1}{2}  (\tilde{\gamma}^a \mp  \tilde{\gamma}^b), \quad
\stackrel{ab}{\tilde{(\pm 1)}} = 
\frac{1}{2}  (\tilde{\gamma}^a \pm i\tilde{\gamma}^b), \nonumber\\ 
\stackrel{ab}{\tilde{[\pm i]}} &=& \frac{1}{2} (1 \pm \tilde{\gamma}^a \tilde{\gamma}^b), \quad
\stackrel{ab}{\tilde{[\pm 1]}} = \frac{1}{2} (1 \pm i \tilde{\gamma}^a \tilde{\gamma}^b). 
\label{snmb2bdeftildefun}
\end{eqnarray}

We present below some useful relations which are easy to derive~\cite{snmb2bpikanorma}. 

\begin{eqnarray}
\label{snmb2brelations}
\stackrel{ab}{(k)} \stackrel{ab}{(k)}& =& 0, \;   \stackrel{ab}{(k)} \stackrel{ab}{(-k)}
= \eta^{aa}  \stackrel{ab}{[k]}, \; 
\stackrel{ab}{[k]} \stackrel{ab}{[ k]} =   \stackrel{ab}{[k]}, \nonumber\\
\stackrel{ab}{[k]} \stackrel{ab}{[-k]} &=& 0, 
\stackrel{ab}{(k)} \stackrel{ab}{[ k]} = 0,\quad \; 
\stackrel{ab}{[k]} \stackrel{ab}{( k)}  = \stackrel{ab}{(k)}, \nonumber\\ 
\stackrel{ab}{(k)} \stackrel{ab}{[-k]} &=& \stackrel{ab}{(k)}\, ,
\quad \; \stackrel{ab}{[k]} \stackrel{ab}{(-k)} =0.   
\end{eqnarray}
\begin{eqnarray}
\stackrel{ab}{\tilde{(k)} }  \stackrel{ab}{(k)}& =& 0, 
\quad \;
 \stackrel{ab}{\tilde{(-k)}} \stackrel{ab}{(k)}= 
-i \eta^{aa}                 \stackrel{ab}{[k]},\nonumber\\ 
 \stackrel{ab}{\tilde{( k)}} \stackrel{ab}{[k]}&=& 
i  \stackrel{ab}{(k)},\quad
 \stackrel{ab}{\tilde{( k)}} \stackrel{ab}{[-k]} = 0.
\label{snmb2bgraphbinomsfamilies}
\end{eqnarray}
\begin{eqnarray}
N^{\pm}_{+}         &=& N^{1}_{+} \pm i \,N^{2}_{+} = 
 - \stackrel{03}{(\mp i)} \stackrel{12}{(\pm )}\,, \quad N^{\pm}_{-}= N^{1}_{-} \pm i\,N^{2}_{-} = 
  \stackrel{03}{(\pm i)} \stackrel{12}{(\pm )}\,,\nonumber\\
\tilde{N}^{\pm}_{+} &=& - \stackrel{03}{\tilde{(\mp i)}} \stackrel{12}{\tilde{(\pm )}}\,, \quad 
\tilde{N}^{\pm}_{-}= 
  \stackrel{03} {\tilde{(\pm i)}} \stackrel{12} {\tilde{(\pm )}}\,,\nonumber\\ 
\tau^{1\pm}         &=& (\mp)\, \stackrel{56}{(\pm )} \stackrel{78}{(\mp )} \,, \quad   
\tau^{2\pm}=            (\mp)\, \stackrel{56}{(\mp )} \stackrel{78}{(\mp )} \,,\nonumber\\ 
\tilde{\tau}^{1\pm} &=& (\mp)\, \stackrel{56}{\tilde{(\pm )}} \stackrel{78}{\tilde{(\mp )}}\,,\quad   
\tilde{\tau}^{2\pm}= (\mp)\, \stackrel{56}{\tilde{(\mp )}} \stackrel{78}{\tilde{(\mp )}}\,.
\end{eqnarray}
 %

%
 
 %
\title{Mass Matrices of Twice Four Families of Quarks\\ and Leptons,
 Scalars and Gauge Fields as Predicted by\\
 the {\it Spin-charge-family-theory}
\thanks{On line lecture  of N.~S.~M.~B. available at \\%
http://viavca.in2p3.fr/what\_comes\_beyond\_the\_standard\_models\_xiii.html} }
\author{N.S. Manko\v c Bor\v stnik}
\institute{%
Department of Physics, FMF, University of Ljubljana,\\
Jadranska 19, SI-1000 Ljubljana, Slovenia}

\titlerunning{Mass Matrices of Twice Four Families\ldots}
\authorrunning{N.S. Manko\v c Bor\v stnik}
\maketitle

\begin{abstract}
The theory unifying spin and charges and predicting families,  proposed~\cite{snmb3tgnorma,snmb3tgpikanorma} as  
a new way to explain
the assumptions of the {\it standard model of the electroweak and colour interactions}, 
predicts at the low energy regime two (by the mixing matrices  decoupled) groups of four families. 
All of them are massless before the two final breaks and identical with respect to the charges and 
spin. They differ only in the family quantum number. 
The fourth family of the lower group of four families (three of them known) is predicted to be  possibly
observed at the LHC or at somewhat higher energies and the stable of the higher four families -- the 
fifth family -- is the candidate to constitute the dark matter. 
In this paper the properties of the fields -- spinors, gauge fields and scalar fields -- before 
and after each of the two last  successive breaks, leading to the (so far) observed quarks, 
leptons and gauge fields, are analysed as they 
follow from the {\it spin-charge-family-theory}. 
\end{abstract}

\section{Introduction } 
\label{contribution:snmb1twogroups}
The {\it theory unifying spin and charges and predicting 
families}~\cite{snmb3tgnorma,snmb3tgpikanorma,snmb3tgnproc2007,snmb3tgdark2010,snmb3tggmdn,snmb3tggn} (to be called the {\it spin-charge-family-theory})
seems  promising in 
explaining the assumptions of the {\it standard model of the electroweak and colour 
interactions}, in particular since it is proposing  the mechanism for generating families.  It assumes in 
$d= (1+ (d-1))$, $d=14$,  a  simple starting action for spinors and the gauge fields: The spinors 
carry only two kinds of the spin (no charges), namely the one postulated by Dirac 80 years ago and the second 
kind proposed by the author of the paper (there is no  third kind of the spin),   and 
interact correspondingly with only the vielbeins and the two kinds of the spin connection fields.  
The action for the corresponding gauge fields contains the curvature in the first power for both kinds of the 
spin connection fields.

After several breaks 
of the starting symmetry the starting action~\ref{snmb3tgwholeaction}  manifests at the low energy 
regime before the final 
(the electroweak) break the effective action (in which only the left handed quarks and leptons curry the 
weak charge while the right handed ones are weak chargeless) with four (rather than three already observed) 
massless families and with four additional, from the lower four families  
decoupled, massive  families (differing from the lower four families in the quantum numbers defined by the 
second kind of spin). The  stable  family of the massive four families -- the fifth family -- is 
accordingly predicted to be the candidate for the dark matter constituent. 
The effective action manifests before the electroweak break (the break of $SU(2)_{I} \times U(1)_{I}$ 
into $U(1)$) 
the massless gauge fields of the 
{\it standard model} and the gauge fields which obtain masses at the higher scale due to   
vielbeins and the two kinds of the spin connection fields which manifest as the scalar  (with respect to 
$SO(1,3)$) fields 
when $SU(2)_{II} \times SU(2)_{I} \times U(1)_{II} \times SU(3)$ breaks  into the standard model group 
$SU(2)_{I} \times U(1)_{I} \times SU(3)$ before the electroweak break.
This break manifests besides in the massive $SU(2)$ and massless $U(1)$ gauge fields also in 
the mass matrices 
of the upper four families, caused by the vacuum expectation values of some superposition of 
the scalar fields originating in all the gauge fields of the action. The upper four families 
are doublets with respect to 
$SU(2)_{II}$  (whose generators of the infinitesimal transformations are $\tilde{\tau}^{2i}$, expressible 
with $\tilde{S}^{st}$, $s,t=\in (5,6,7,8)$) as well as with respect to $\tilde{N}^{i}_{-}$ of $SO(1,3)$  
(expressible with $\tilde{S}^{m,n}$, $m,n \in (0,1,2,3)$) and singlets with respect to the group $SU(2)_I$  
(with $\tilde{\tau}^{1i}$ as generators) and with respect to $\tilde{N}^{i}_{+}$ of $SO(1,3)$. 
To the final -- the electroweak -- break the vielbeins and the two kinds of the spin connection fields,  
with the quantum numbers different from the quantum numbers of the "upper scalar field", contribute, 
 manifesting the properties 
of the  {\it standard model} Higgs. The gauge weak fields obtain masses and so do the lower four families, 
which are singlets with respect to $\tilde{\tau}^{2i}$  and $\tilde{N}^{i}_{-}$, while 
they are doublets with respect to $\tilde{\tau}^{1i}$  and $\tilde{N}^{i}_{+}$.

The fourth family  might be seen (as the so far made rough estimations show~\cite{snmb3tgpikanorma,snmb3tggmdn}) 
at the LHC or at somewhat higher energies.  
The stable fifth family members forming neutral (with respect to the colour and electromagnetic 
charge) baryons and the fifth family neutrinos are predicted to explain the origin of the 
dark matter~\cite{snmb3tggn}. 

Since due to the {\it spin-charge-family-theory} the break of the symmetry,  caused by the nonzero 
expectation values of those vielbeins 
and  both kinds of the spin connection fields which are connected with the break,   
and manifesting in the masses of the corresponding gauge fields  and  
in the masses of the higher four families, occurs 
below $10^{13}$ GeV and pretty much above the 
masses of the fourth family and accordingly pretty much above the scale of the electroweak break,  
and since in this theory  the unification scale of the electroweak and colour 
interactions seems to be below the scale of the break of the supersymmetry, the 
{\it spin-charge-family-theory} 
does not speak for  observing
the {\it supersymmetric particles} at the LHC.

Although the estimations of the properties of families done so far are very approximate~\cite{snmb3tggmdn,snmb3tggn},
yet the predictions  give a hope that the starting assumptions  of the {\it spin-charge-family-theory}
 are the right ones: i. Both existing  Clifford algebra operators determine 
properties of fermions. The Dirac $\gamma^a$'s  manifest in the low energy regime  
the spin and all the charges of fermions (like in the Kaluza-Klein[like] 
theories~\footnote{The Kaluza-Klein[like] theories have difficulties with 
(almost) masslessness of the spinor fields at the low energy regime. In the ref.~\cite{snmb3tgdhn} we 
are proposing possible solutions to these kind of difficulties.}), giving to left handed spinors the 
weak charge and leaving the right handed spinors weak chargeless. The second kind, 
forming the equivalent representations with 
respect to the Dirac one, is responsible for the appearance of families. 
ii. Fermions carrying only 
the corresponding two kinds of the spin (no charges) interact with the corresponding gravitational 
fields -- the vielbeins and (the two kinds of) the spin connections. The spin connections  connected with the 
Dirac's gammas manifest at the low energy regime all the known gauge fields, those connected with the second 
kind of gammas are correspondingly responsible, together with the vielbeins, for the masses of 
gauge fields and of fermions. iii. The assumed starting 
action for spinors and  gauge fields in $d$-dimensional space is simple.

The project to come from the starting action through spontaneous (nonperturbative) breaks of 
symmetries to the effective action at low (measurable) energy regime is very demanding. Although 
one easily sees that a part of the starting action manifests, after the breaks of symmetries, at the tree 
level the mass matrices of the families with the observed properties of the three lowest families, 
and that a part of the vielbeins 
together with the two kinds of the spin connection fields manifests the scalar fields, taking care 
of the masses of the gauge bosons, yet  
several proofs are still needed besides those done so far~\cite{snmb3tgsnmb:hn02hn03}
to guarantee that the {\it spin-charge-family-theory} does lead to the measured 
effective action of the {\it standard model} as well as very demanding calculations  in addition 
to the rough estimations~\cite{snmb3tgpikanorma,snmb3tggmdn,snmb3tggn} done so far are needed to show that predictions 
agree also with the measured values of the masses and mixing matrices of the so  far observed fermions.

The {\it spin-charge-family-theory} predicts two kinds of the contributions to the mass matrices. 
One kind distinguishes on the 
tree level only among the members of one family (among the u-quark, d-quark, neutrino and 
electron, the left and right handed), while the other kind distinguishes only among the families. 
Beyond the tree level both kinds  start to contribute coherently and 
a detailed study should manifest the drastic differences in properties of quarks and leptons: 
in their masses and mixing matrices~\cite{snmb3tgnproc2007,snmb3tgdark2010}.

In this work the mass matrices,  the two groups of the scalar fields giving masses to gauge 
fields to which they interact, and the gauge fields  are studied. We shall follow  all these  fields 
through the final two successive breaks, keeping in mind   
the {\it standard model} assumptions. 
The properties of the fields and the consequences on the so far observed phenomena are discussed,  
paying attention to the contributions of the scalar and gauge 
fields when going beyond the tree level. 

The concrete evaluations and discussions about the properties of the mass matrices beyond the 
tree level are done in the common contribution with A. Hern\'andez-Galeana~\cite{snmb3tgalbinonorma2010}.

\section{Action of the   theory unifying  spin and charges and predicting families}

The  {\it spin-charge-family-theory} assumes that the spinor carries 
in $d(=(1 + 13))$-dimensional 
space  two kinds of the spin, no charges~\cite{snmb3tgnorma,snmb3tgpikanorma,snmb3tgnproc2007}:\\ 
i. The Dirac spin, described by $\gamma^a$'s, defines the  spinor representations in $d=(1+ 13)$. 
After the break of the starting symmetry 
$SO(1,13)$ (through $SO(1,7) \times 
SO(6)$) to the symmetry of the {\it standard model}  in 
$d=(1+3)$ ($SO(1,3)\times U(1)\times SU(2)\times SU(3)$)  
it defines  the hyper charge ($U(1)$),  the weak charge ($SU(2)$, with the 
left handed representation of $SO(1,3)$ manifesting naturally the weak charge and the right 
handed ones appearing as the weak singlets) and  the colour charge ($SU(3)$). \\
ii. The second kind of the spin~\cite{snmb3tgsnmb:hn02hn03},  
described by $\tilde{\gamma}^a$'s ($\{\tilde{\gamma}^a, \tilde{\gamma}^b\}_{+}= 2 \, \eta^{ab}$) and  
anticommuting with the Dirac $\gamma^a$ ($\{\gamma^a, \tilde{\gamma}^b\}_{+}=0$),  
defines the families of spinors. 

There is no  third kind of the Clifford algebra objects. The appearance of the two kinds of the 
Clifford algebra objects can be understood as follows:
If the Dirac one corresponds to the multiplication of any spinor object $B$ (any product of the Dirac 
$\gamma^a$'s, which represents a spinor state when being applied on a spinor vacuum state $|\psi_0>$) 
from the left hand side, the second kind of the Clifford objects  can be  understood 
(up to a factor, determining the Clifford evenness ($n_B=2k$) or oddness ($n_B=2k+1$) of the object  
$B$ as the multiplication of the object from the right hand side
\begin{eqnarray}
 \tilde{\gamma}^a B \, |\psi_0>: = i(-)^{n_B} B \gamma^a\, |\psi_0>, 
 \label{snmb3tgBt}
 \end{eqnarray}
 with $|\psi_0>$ determining the 
spinor vacuum state. 
Accordingly we have
\begin{eqnarray}
&& \{ \gamma^a, \gamma^b\}_{+} = 2\eta^{ab} =  
\{ \tilde{\gamma}^a, \tilde{\gamma}^b\}_{+},\quad
\{ \gamma^a, \tilde{\gamma}^b\}_{+} = 0,\nonumber\\
&&S^{ab}: = (i/4) (\gamma^a \gamma^b - \gamma^b \gamma^a), \quad
\tilde{S}^{ab}: = (i/4) (\tilde{\gamma}^a \tilde{\gamma}^b 
- \tilde{\gamma}^b \tilde{\gamma}^a),\quad  \{S^{ab}, \tilde{S}^{cd}\}_{-}=0.\nonumber\\
\label{snmb3tgsnmb:tildegclifford}
\end{eqnarray}
The  {\it spin-charge-family-theory} proposes a simple action for a Weyl 
spinor which carries in $d=(1+13)$  two kinds of the  
spin (no charges) and for  the corresponding gauge fields  
\begin{eqnarray}
S            \,  &=& \int \; d^dx \; E\;{\mathcal L}_{f} +  
\nonumber\\  
               & & \int \; d^dx \; E\; (\alpha \,R + \tilde{\alpha} \, \tilde{R}),
               \end{eqnarray}
\begin{eqnarray}
{\mathcal L}_f &=& \frac{1}{2}\, (E\bar{\psi} \, \gamma^a p_{0a} \psi) + h.c., 
\nonumber\\
p_{0a }        &=& f^{\alpha}{}_a p_{0\alpha} + \frac{1}{2E}\, \{ p_{\alpha}, E f^{\alpha}{}_a\}_-, 
\nonumber\\  
   p_{0\alpha} &=&  p_{\alpha}  - 
                    \frac{1}{2}  S^{ab} \omega_{ab \alpha} - 
                    \frac{1}{2}  \tilde{S}^{ab}   \tilde{\omega}_{ab \alpha},                   
\nonumber\\ 
R              &=&  \frac{1}{2} \, \{ f^{\alpha [ a} f^{\beta b ]} \;(\omega_{a b \alpha, \beta} 
- \omega_{c a \alpha}\,\omega^{c}{}_{b \beta}) \} + h.c. \;, 
\nonumber\\
\tilde{R}      &=& \frac{1}{2}\,   f^{\alpha [ a} f^{\beta b ]} \;(\tilde{\omega}_{a b \alpha,\beta} - 
\tilde{\omega}_{c a \alpha} \tilde{\omega}^{c}{}_{b \beta}) + h.c.\;. 
\label{snmb3tgwholeaction}
\end{eqnarray}
Here~\footnote{$f^{\alpha}{}_{a}$ are inverted 
vielbeins to 
$e^{a}{}_{\alpha}$ with the properties $e^a{}_{\alpha} f^{\alpha}{\!}_b = \delta^a{\!}_b,\; 
e^a{\!}_{\alpha} f^{\beta}{\!}_a = \delta^{\beta}_{\alpha} $. 
Latin indices  
$a,b,..,m,n,..,s,t,..$ denote a tangent space (a flat index),
while Greek indices $\alpha, \beta,..,\mu, \nu,.. \sigma,\tau ..$ denote an Einstein 
index (a curved index). Letters  from the beginning of both the alphabets
indicate a general index ($a,b,c,..$   and $\alpha, \beta, \gamma,.. $ ), 
from the middle of both the alphabets   
the observed dimensions $0,1,2,3$ ($m,n,..$ and $\mu,\nu,..$), indices from 
the bottom of the alphabets
indicate the compactified dimensions ($s,t,..$ and $\sigma,\tau,..$). 
We assume the signature $\eta^{ab} =
diag\{1,-1,-1,\cdots,-1\}$.} 
$f^{\alpha [a} f^{\beta b]}= f^{\alpha a} f^{\beta b} - f^{\alpha b} f^{\beta a}$. 
The action~(Eq.(\ref{snmb3tgwholeaction})) manifests after the break of symmetries all the known 
gauge fields and the scalar fields, and the mass matrices. 
To see the manifestation of the covariant momentum and the mass matrices we rewrite 
formally the action for a Weyl spinor in $d=(1+13)$  as follows  
\begin{eqnarray}
{\mathcal L}_f &=&  \bar{\psi}\gamma^{m} (p_{m}- \sum_{A,i}\; g^{A}\tau^{Ai} A^{Ai}_{m}) \psi 
+ \nonumber\\
               & &  \{ \sum_{s=7,8}\;  \bar{\psi} \gamma^{s} p_{0s} \; \psi \}  + \nonumber\\
               & & {\rm the \;rest}, 
\label{snmb3tgfaction}
\end{eqnarray}
where $m=0,1,2,3$ with
\begin{eqnarray}
\tau^{Ai} = \sum_{a,b} \;c^{Ai}{ }_{ab} \; S^{ab},
\nonumber\\ 
\{\tau^{Ai}, \tau^{Bj}\}_- = i \delta^{AB} f^{Aijk} \tau^{Ak}.
\label{snmb3tgtau}
\end{eqnarray}
All the charges and the spin of one family are determined by $S^{ab}$, 
with $S^{ab}$ as the only internal degree of freedom of one family  
(besides the family quantum number, determined by $\tilde{S}^{ab}$). 
Index $A$ enumerates all possible vector fields corresponding to the spinor 
charges  defined by the infinitesimal generators $\tau^{Ai}$ of the corresponding Lie groups, 
expressed as the superposition of  $S^{ab}$ (Eq.~(\ref{snmb3tgtau})). 
At the low energy regime we expect that $A$ will stay for the groups $U(1), SU(2)$ and $SU(3)$. 
The breaks of the starting symmetry from $SO(1,13)$ to the symmetry $SO(1,7) \times SU(3) \times U(1)$ 
and further to $SO(1,3) \times SU(2) \times SU(2) \times U(1) \times SU(3) $ 
are assumed to leave all the low lying families of spinors massless~\footnote{We proved that it is possible to 
have massless fermions after a (particular) break if we start with massless fermions and 
assume particular boundary conditions after the break or  the "effective 
two dimensionality" cases~\cite{snmb3tgsnmb:hn02hn03}}. After the break of $SO(1,13)$ to 
$SO(1,7) \times SU(3) \times U(1)$ there are eight such families ($2^{8/2-1}$), all left 
handed with respect to $SO(1,13)$. 

Accordingly the first row of the action in Eq.~(\ref{snmb3tgfaction}) manifests the effective {\it standard model} 
fermions part of the action, while the second part manifests, after the appropriate breaks of 
symmetries when $\omega_{ab \sigma}$  and $\tilde{\omega}_{ab \sigma}$, $\sigma \in (5,6,7,8),$ 
fields gain the nonzero values 
on the tree level -- the vacuum expectation values -- the mass matrices. We shall 
comment this part in sections~\ref{snmb3tgyukawandhiggs},~\ref{snmb3tgyukawatreefbelow}. The third part in the third 
row stays for all the rest, which at low energies is expected to be negligible, or might 
 slightly influence the mass matrices beyond the tree level.

 The generators $\tilde{S}^{ab}$ take care of the families, transforming each member of one family 
 into the corresponding member of another family, due to the fact that 
 $\{S^{ab}, \tilde{S}^{cd}\}_{-}=0$ (Eq.(\ref{snmb3tgsnmb:tildegclifford})). 
 
 Correspondingly also the action for vielbeins and spin connections of the two kinds, with 
 the Lagrange densities $\alpha\,E\, R$ and $\tilde{\alpha}\,E \, \tilde{R}$, manifests at 
 the low energy regime, after the 
 breaks of the starting symmetry, as the known vector fields -- the gauge fields of $U(1), \,SU(2),\, SU(3)$ and 
 the ordinary gravity -- playing the role of also the scalar (Higgs) fields, which bring masses to the 
 gauge field, 
 as we shall see in  subsection~\ref{snmb3tggaugefieldaction}.

\section{Spinor action after  breaks}
\label{snmb3tgspinoraction}

We shall follow the spontaneous breaks (expected to be highly nonperturbative) of the starting symmetry,  
from  $SO(1,13)$ to  $SO(1,7) \times U(1)\times SU(3)$  and further to $SO(1,3) \times 
SO(4) \times U(1) \times SU(3)$ with eight families of  massless spinors ($2^{\frac{8}{2}-1}$), 
 when massless 
spinors  manifest the $U(1), \,SU(3)$ and the two $SU(2)$ charges, one $SU(2)$ left handed 
with respect to  $SO(1,3)$ 
(we shall call it $SU(2)_{I}$) and  the other $SU(2)$
right handed with respect to  $SO(1,3)$ (we shall call it $SU(2)_{II}$).  All these properties 
concern $S^{ab}$.  The corresponding gauge fields of $U(1), \,SU(3)$ and the two $SU(2)$ are massless. 
Then we follow all these fields (the spinor, gauge vector and scalar fields) 
through the break of one of the two $SU(2)$ 
($SU(2)_{II} \times U(1)_{II} $ to $U(1)_{I}$) when four of the families obtain the nonzero mass matrices 
due to the nonzero vacuum expectation values of those scalar fields - 
vielbeins and 
the second kind  of the 
spin connection fields, the gauge fields of 
$\tilde{S}^{ab}$ (indeed of $\tilde{\tau}^{2i}$ so that the scalar fields are $ \tilde{A}^{2i}_{\sigma}$), 
and of $\tilde{N}^{i}_{-}$ (which are the generators of the infinitesimal transformations of 
one of the $SU(2)$ in the $SO(1,3)$ group), 
with the scalar fields 
$ \tilde{A}^{\tilde{N}_{-}i}_{\sigma}$ - contributing to the break of 
$SU(2)_{II}$. The nonzero vacuum expectation values of the same scalar fields  
appearing at this break  
take care of masses of those gauge fields $A^{2i}_{a}$
which carry the  $SU(2)_{II} $ charge (namely $\tau^{2i}$).

The masslessness of the remaining four families are guaranteed by the fact 
that they are singlets with respect to 
$SU(2)_{II}$  in the $\tilde{S}^{ab}$ sector, 
that is with respect to $\tilde{\tau}^{2i}$, 
(they are  singlets also with respect to $\tilde{N}^{i}_{-}$), while they are doublets 
with respect to $\tilde{\tau}^{1i}$, that is with respect to the second $SU(2)_{I}$  
(they are  doublets also with respect to $\tilde{N}^{i}_{+}$). The gauge fields of $SU(2)_{I}$, 
which are singlets with respect to $SU(2)_{II}$, stay at this break massless. 
The weak charged left handed 
and the weak chargeless right handed (with respect to $SO(1,3)$) spinors are correspondingly still 
mass protected unless finally the electroweak symmetry breaks ($SU(2)_{I} \times U(1)_{I}$) to  the symmetry of 
$SO(1,3) \times U(1) \times SU(3)$. 
To this break 
again the vielbeins and now both kinds of the spin connection 
fields (the gauge fields of $S^{ab}$ and of $\tilde{S}^{ab}$), those which manifest the nonzero 
expectation values, contribute.


We shall study  the properties of the gauge and scalar fields and of the spinors
mass matrices  
for both groups of four families, taking into account the contribution of the dynamical fields   
beyond the tree level.

The technique~\cite{snmb3tgsnmb:hn02hn03} offers an easy way to keep track of the symmetry 
properties of spinors after breaks.  

Following the refs.~\cite{snmb3tgsnmb:hn02hn03} we define the vectors, that is the nilpotents 
($\,\stackrel{ab}{(k)}{}^2= 0$) and projectors ($\,\stackrel{ab}{[k]}{}^2 = \stackrel{ab}{[k]}$),  
\begin{eqnarray}
\stackrel{ab}{(\pm i)}: &=& \frac{1}{2}(\gamma^a \mp  \gamma^b),  \; 
\stackrel{ab}{[\pm i]}: = \frac{1}{2}(1 \pm \gamma^a \gamma^b), \quad
{\rm for} \,\; \eta^{aa} \eta^{bb} = -1, \nonumber\\
\stackrel{ab}{(\pm )}: &= &\frac{1}{2}(\gamma^a \pm i \gamma^b),  \; 
\stackrel{ab}{[\pm ]}: = \frac{1}{2}(1 \pm i\gamma^a \gamma^b), \quad
{\rm for} \,\; \eta^{aa} \eta^{bb} =1,
\label{snmb3tgsnmb:eigensab}
\end{eqnarray} 
which are the eigen vectors  of $S^{ab}$ as well as of $\tilde{S}^{ab}$ as follows
\begin{eqnarray}
S^{ab} \stackrel{ab}{(k)} =  \frac{k}{2} \stackrel{ab}{(k)}, \quad 
S^{ab} \stackrel{ab}{[k]} =  \frac{k}{2} \stackrel{ab}{[k]}, \quad
\tilde{S}^{ab} \stackrel{ab}{(k)}  = \frac{k}{2} \stackrel{ab}{(k)},  \quad 
\tilde{S}^{ab} \stackrel{ab}{[k]}  =   - \frac{k}{2} \stackrel{ab}{[k]}.\;\;
\label{snmb3tgsnmb:eigensabev}
\end{eqnarray}
One should recognize that $\gamma^a$ transform   
$\stackrel{ab}{(k)}$ into  $\stackrel{ab}{[-k]}$, while 
$\tilde{\gamma}^a$ transform  $\stackrel{ab}{(k)}$ 
into $\stackrel{ab}{[k]}$ 
\begin{eqnarray}
\gamma^a \stackrel{ab}{(k)}= \eta^{aa}\stackrel{ab}{[-k]},\; 
\gamma^b \stackrel{ab}{(k)}= -ik \stackrel{ab}{[-k]}, \; 
\gamma^a \stackrel{ab}{[k]}= \stackrel{ab}{(-k)},\; 
\gamma^b \stackrel{ab}{[k]}= -ik \eta^{aa} \stackrel{ab}{(-k)},\;\;\;
\label{snmb3tgsnmb:graphgammaaction}
\end{eqnarray}
\begin{eqnarray}
\tilde{\gamma^a} \stackrel{ab}{(k)} = - i\eta^{aa}\stackrel{ab}{[k]},\;
\tilde{\gamma^b} \stackrel{ab}{(k)} =  - k \stackrel{ab}{[k]}, \;
\tilde{\gamma^a} \stackrel{ab}{[k]} =  \;\;i\stackrel{ab}{(k)},\; 
\tilde{\gamma^b} \stackrel{ab}{[k]} =  -k \eta^{aa} \stackrel{ab}{(k)}. \; 
\label{snmb3tgsnmb:gammatilde}
\end{eqnarray}
Correspondingly it is easy to show how do the equivalent representations (with respect to $S^{ab}$), 
formed by $\tilde{S}^{ab}$,  manifest. Defining  the basis vectors in the internal space of 
spin degrees of freedom 
in $d=(1+13)$  as products of the projectors and nilpotents from Eq.~(\ref{snmb3tgsnmb:eigensab}) 
on the spinor vacuum state, 
the representation of one Weyl spinor with respect to $S^{ab}$  manifests, after the 
above discussed breaks, the spin and all the charges of one family, while $\tilde{S}^{ab}$ 
determines families, in agreement with the {\it standard model} assumption that before 
the electroweak break of symmetry the three known families (in my case there are four rather than 
three families) are identical up to the family 
quantum number. 

While the {\it standard model} assumes the existence and the number of 
families (the so far measured number of families), the simple starting action 
(Eq.~(\ref{snmb3tgwholeaction})) of the  {\it spin-charge-family-theory} -- by assuming that both existing 
Clifford algebra objects  $\gamma^a$'s and $\tilde{\gamma}^a$'s  
determine properties of spinors 
-- offers 
the explanation for the origin of families and consequently predicts the number of families  
which might be observed at the low energy regime. 
Since there exist two Clifford algebra objects, the assumption to use both of them seems needed, 
otherwise we would have to explain why one kind of gammas manifest in the theoretical description of spinors 
and the other does not.

We arrange the products of nilpotents and projectors to be the eigenvectors of 
the Cartan subalgebra $S^{03}, S^{12}, S^{56}, S^{78}, S^{9\,10}, S^{11\,12}, S^{13\,14}$ and, 
at the same time,  they are also the
eigenvectors of the corresponding  $\tilde{S}^{ab}$.  
We analyse these states in terms of the subgroups 
$SO(1,3)$, $U(1)$, $SU(2)$ and $SU(3)$ to manifest that one representation of the group $SO(1,13)$ 
manifests with respect to $S^{ab}$, after the break of $SO(1,13)$ to $SO(1,7) \times SU(3) \times U(1)$ 
and further to $SO(1,3) \times SU(2) 
\times U(1) \times SU(3)  $,  all the members of one family, 
namely the left handed weak charged quarks and leptons and the right handed weak chargeless  
quarks and leptons. The generators $\tilde{S}^{ab}$ transform each member of one octet of 
$SO(1,7)$, that is  $ 2^{\frac{8}{2}-1}$ vectors, into the same member of one of  the $2^{\frac{8}{2}-1}$ 
eight families of quarks and leptons.

We analyse one Weyl spinor of $SO(1,13)$ in terms of the subgroup 
$SO(1,3)$ ($  SU(2) \times SU(2) $)%
\begin{eqnarray}
\label{snmb3tgso13}
\vec{N}_{\pm}=\frac{1}{2} (S^{23}\pm i S^{01},S^{31}\pm i S^{02}, S^{12}\pm i S^{03} ),
\end{eqnarray}
the subgroup  
 $SU(2)\times$ $SU(2)$ of $SO(4)$
 \begin{eqnarray}
 \label{snmb3tgso4}
 \vec{\tau}^{1}=\frac{1}{2} (S^{58}-  S^{67}, \,S^{57} + S^{68}, \,S^{56}-  S^{78} )\nonumber\\
 \vec{\tau}^{2}=\frac{1}{2} (S^{58}+  S^{67}, \,S^{57} - S^{68}, \,S^{56}+  S^{78} ),
 \end{eqnarray}
 and the subgroup 
 $SU(3) \times U(1)$,  originating in $SO(6)$,    
  \begin{eqnarray}
 \label{snmb3tgso6}
 \vec{\tau}^{3}: = &&\frac{1}{2} \,\{  S^{9\;12} - S^{10\;11} ,\,
  S^{9\;11} + S^{10\;12} ,\, S^{9\;10} - S^{11\;12} ,\nonumber\\
 && S^{9\;14} -  S^{10\;13} ,\,  S^{9\;13} + S^{10\;14} ,\,
  S^{11\;14} -  S^{12\;13},\nonumber\\
 && S^{11\;13} +  S^{12\;14} ,\, 
 \frac{1}{\sqrt{3}} ( S^{9\;10} + S^{11\;12} - 
 2 S^{13\;14})\},\nonumber\\
 \tau^{4}: = &&-\frac{1}{3}(S^{9\;10} + S^{11\;12} + S^{13\;14}).
 \end{eqnarray}
 %
 
 One finds that $N^{\pm}_{+}= N^{1}_{+} \pm i \,N^{2}_{+} = 
 - \stackrel{03}{(\mp i)} \stackrel{12}{(\pm )}$, $N^{\pm}_{-}= N^{1}_{-} \pm i\,N^{2}_{-} = 
  \stackrel{03}{(\pm i)} \stackrel{12}{(\pm )}$. 
 We shall rename $\vec{N}_{\pm}$ into $\vec{N}_{L}$ for $(+)$ and $\vec{N}_{R}$ for $(-)$. 
 All the operators have their equivalent ones in the $\tilde{S}^{mn}$ sector, which follow when  
  replacing $S^{mn}$ with $\tilde{S}^{mn}$: $\vec{\tilde{N}}_{\pm}=\frac{1}{2} (\tilde{S}^{23}\pm i 
 \tilde{S}^{01},  \tilde{S}^{31}\pm i \tilde{S}^{02}, \tilde{S}^{12}\pm i \tilde{S}^{03} )$, while 
 $\tilde{N}^{\pm}_{+}= - \stackrel{03}{\tilde{(\mp i)}} \stackrel{12}{\tilde{(\pm )}}$,
 $\tilde{N}^{\pm}_{-}= 
  \stackrel{03} {\tilde{(\pm i)}} \stackrel{12} {\tilde{(\pm )}}$. We rename  $\vec{\tilde{N}}_{\pm} $ into 
$\vec{\tilde{N}}_{L}$ for $(+)$ and into $\vec{\tilde{N}}_{R}$ for $(-)$.

 We also find that  $\tau^{1\pm}= (\mp)\, \stackrel{56}{(\pm )} \stackrel{78}{(\mp )} $ and  
 $\tau^{2\pm}= (\mp)\, \stackrel{56}{(\mp )} \stackrel{78}{(\mp )} $. 
 In the  $\tilde{S}^{st}$ sector we find equivalently $\vec{\tilde{\tau}}^{(1)}=
  \frac{1}{2} (\tilde{S}^{58} -  \tilde{S}^{67}, \,\tilde{S}^{57} + \tilde{S}^{68}, \,
 \tilde{S}^{56} -  \tilde{S}^{78} )$ and  $\vec{\tilde{\tau}}^{(2)}=
  \frac{1}{2} (\tilde{S}^{58} -  \tilde{S}^{67}, \,\tilde{S}^{57} + \tilde{S}^{68}, \,
 \tilde{S}^{56} -  \tilde{S}^{78} )$, 
 while $\tilde{\tau}^{1\pm}= (\mp)\, \stackrel{56}{\tilde{(\pm )}} \stackrel{78}{\tilde{(\mp )}} $ and  
 $\tilde{\tau}^{2\pm}= (\mp)\, \stackrel{56}{\tilde{(\mp )}} \stackrel{78}{\tilde{(\mp )}} $.

 In the $\tilde{S}^{st}; \, s,t \in  \{9,10,11,12,13,14\}$ sector we have equivalent 
 operators, if exchanging 
 $S^{ab}$ with $\tilde{S}^{ab}$, $\vec{\tau}^{3}$ with $\vec{\tilde{\tau}}^{3}$ and 
 $\tau^{4}$ with $\tilde{\tau}^{4}$, following the above prescriptions for the group $SO(1,3)$ and $SO(4)$. 
 We present some useful relations in Appendix.
 
 After the break of one of  $SU(2)$ ($SU(2)_{II} \times U(1)_{II} $, we shall call $SU(2)_{I}=SU(2)_{W}$,
 into $SU(2)_{W} \times U(1)$ we determine the 
 properties of spinors in terms of 
  $\tau^{1i}$ and the hypercharge $Y=
  \frac{1}{2}\,(S^{56}+  S^{78})+ \tau^{4}$ and equivalently in the tilde sector in terms of
  $\tilde{\tau}^{1i}$ and 
 $\tilde{Y}$. 
 
 The eightplet follows for the $u$ and $d$ quarks (each of them in three colour charges) 
 as well as for colourless leptons. Each eightplet includes the left handed weak charged 
 and the right  handed weak chargeless members. 
%

I present  in Table~\ref{snmb3tgTable I.} the eightplet (the representation of $SO(1,7)$ of quarks of a 
particular colour charge ($\tau^{33}=1/2$, 
$\tau^{38}=1/(2\sqrt{3})$), and $U(1)$ charge ($\tau^{4}=1/6$) and on Table~\ref{snmb3tgTable Il.} the 
eightplet of the corresponding (colour chargeless) leptons. 
%
%
\begin{table}
\begin{center}
\begin{tabular}{|r|c||c||c|c||c|c|c||r|r|}
\hline
i&$$&$|^a\psi_i>$&$\Gamma^{(1,3)}$&$ S^{12}$&$\Gamma^{(4)}$&
$\tau^{13}$&$\tau^{23}$&$Y$&$Q$\\
\hline\hline
&& ${\rm Octet},\;\Gamma^{(1,7)} =1,\;\Gamma^{(6)} = -1,$&&&&&&& \\
&& ${\rm of \; quarks}$&&&&&&&\\
\hline\hline
1&$ u_{R}^{c1}$&$ \stackrel{03}{(+i)}\,\stackrel{12}{(+)}|
\stackrel{56}{(+)}\,\stackrel{78}{(+)}
||\stackrel{9 \;10}{(+)}\;\;\stackrel{11\;12}{[-]}\;\;\stackrel{13\;14}{[-]} $
&1&$\frac{1}{2}$&1&0&$\frac{1}{2}$&$\frac{2}{3}$&$\frac{2}{3}$\\
\hline 
2&$u_{R}^{c1}$&$\stackrel{03}{[-i]}\,\stackrel{12}{[-]}|\stackrel{56}{(+)}\,\stackrel{78}{(+)}
||\stackrel{9 \;10}{(+)}\;\;\stackrel{11\;12}{[-]}\;\;\stackrel{13\;14}{[-]}$
&1&$-\frac{1}{2}$&1&0&$\frac{1}{2}$&$\frac{2}{3}$&$\frac{2}{3}$\\
\hline
3&$d_{R}^{c1}$&$\stackrel{03}{(+i)}\,\stackrel{12}{(+)}|\stackrel{56}{[-]}\,\stackrel{78}{[-]}
||\stackrel{9 \;10}{(+)}\;\;\stackrel{11\;12}{[-]}\;\;\stackrel{13\;14}{[-]}$
&1&$\frac{1}{2}$&1&0&$-\frac{1}{2}$&$-\frac{1}{3}$&$-\frac{1}{3}$\\
\hline 
4&$ d_{R}^{c1} $&$\stackrel{03}{[-i]}\,\stackrel{12}{[-]}|
\stackrel{56}{[-]}\,\stackrel{78}{[-]}
||\stackrel{9 \;10}{(+)}\;\;\stackrel{11\;12}{[-]}\;\;\stackrel{13\;14}{[-]} $
&1&$-\frac{1}{2}$&1&0&$-\frac{1}{2}$&$-\frac{1}{3}$&$-\frac{1}{3}$\\
\hline
5&$d_{L}^{c1}$&$\stackrel{03}{[-i]}\,\stackrel{12}{(+)}|\stackrel{56}{[-]}\,\stackrel{78}{(+)}
||\stackrel{9 \;10}{(+)}\;\;\stackrel{11\;12}{[-]}\;\;\stackrel{13\;14}{[-]}$
&-1&$\frac{1}{2}$&-1&$-\frac{1}{2}$&0&$\frac{1}{6}$&$-\frac{1}{3}$\\
\hline
6&$d_{L}^{c1} $&$\stackrel{03}{(+i)}\,\stackrel{12}{[-]}|
\stackrel{56}{[-]}\,\stackrel{78}{(+)}
||\stackrel{9 \;10}{(+)}\;\;\stackrel{11\;12}{[-]}\;\;\stackrel{13\;14}{[-]} $
&-1&$-\frac{1}{2}$&-1&$-\frac{1}{2}$&0&$\frac{1}{6}$&$-\frac{1}{3}$\\
\hline
7&$ u_{L}^{c1}$&$\stackrel{03}{[-i]}\,\stackrel{12}{(+)}|
\stackrel{56}{(+)}\,\stackrel{78}{[-]}
||\stackrel{9 \;10}{(+)}\;\;\stackrel{11\;12}{[-]}\;\;\stackrel{13\;14}{[-]}$
&-1&$\frac{1}{2}$&-1&$\frac{1}{2}$&0&$\frac{1}{6}$&$\frac{2}{3}$\\
\hline
8&$u_{L}^{c1}$&$\stackrel{03}{(+i)}\,\stackrel{12}{[-]}|\stackrel{56}{(+)}\,\stackrel{78}{[-]}
||\stackrel{9 \;10}{(+)}\;\;\stackrel{11\;12}{[-]}\;\;\stackrel{13\;14}{[-]}$
&-1&$-\frac{1}{2}$&-1&$\frac{1}{2}$&0&$\frac{1}{6}$&$\frac{2}{3}$\\
\hline\hline
\end{tabular}
\end{center}
\caption{\label{snmb3tgTable I.} The 8-plet of quarks - the members of $SO(1,7)$ subgroup of the 
group $SO(1,13)$, 
belonging to one Weyl left 
handed ($\Gamma^{(1,13)} = -1 = \Gamma^{(1,7)} \times \Gamma^{(6)}$) spinor representation of 
$SO(1,13)$. 
It contains the left handed weak charged quarks and the right handed weak chargeless quarks 
of a particular 
colour $(1/2,1/(2\sqrt{3}))$. Here  $\Gamma^{(1,3)}$ defines the handedness in $(1+3)$ space, 
$ S^{12}$ defines the ordinary spin (which can also be read directly from the basic vector, both
vectors  with both spins, $\pm \frac{1}{2}$, are presented), 
$\tau^{13}$ defines the third component of the weak charge, $\tau^{23}$ the third component 
of the $SU(2)_{II}$ charge, 
$\tau^{4}$ (the $U(1)$ charge) defines together with the
$\tau^{23}$  the hyper charge ($Y= \tau^4 + \tau^{23}$), $Q= Y + \tau^{13}$ is the 
electromagnetic charge. 
The reader can find the whole Weyl representation in the ref.~\cite{snmb3tgPortoroz03}.}
\end{table}
%
%

%
%
\begin{table}
\begin{center}
\begin{tabular}{|r|c||c||c|c||c|c|c||r|r|}
\hline
i&$$&$|^a\psi_i>$&$\Gamma^{(1,3)}$&$ S^{12}$&$\Gamma^{(4)}$&
$\tau^{13}$&$\tau^{23}$&$Y$&$Q$\\
\hline\hline
&& ${\rm Octet},\;\Gamma^{(1,7)} =1,\;\Gamma^{(6)} = -1,$&&&&&&& \\
&& ${\rm of \; quarks}$&&&&&&&\\
\hline\hline
1&$ \nu_{R}$&$ \stackrel{03}{(+i)}\,\stackrel{12}{(+)}|
\stackrel{56}{(+)}\,\stackrel{78}{(+)}
||\stackrel{9 \;10}{(+)}\;\;\stackrel{11\;12}{(+)}\;\;\stackrel{13\;14}{(+)} $
&1&$\frac{1}{2}$&1&0&$\frac{1}{2}$&$0$&$0$\\
\hline 
2&$\nu_{R}$&$\stackrel{03}{[-i]}\,\stackrel{12}{[-]}|\stackrel{56}{(+)}\,\stackrel{78}{(+)}
||\stackrel{9 \;10}{(+)}\;\;\stackrel{11\;12}{[-]}\;\;\stackrel{13\;14}{[-]}$
&1&$-\frac{1}{2}$&1&0&$\frac{1}{2}$&$0$&$0$\\
\hline
3&$e_{R}$&$\stackrel{03}{(+i)}\,\stackrel{12}{(+)}|\stackrel{56}{[-]}\,\stackrel{78}{[-]}
||\stackrel{9 \;10}{(+)}\;\;\stackrel{11\;12}{[-]}\;\;\stackrel{13\;14}{[-]}$
&1&$\frac{1}{2}$&1&0&$-\frac{1}{2}$&$-1$&$-1$\\
\hline 
4&$ e_{R} $&$\stackrel{03}{[-i]}\,\stackrel{12}{[-]}|
\stackrel{56}{[-]}\,\stackrel{78}{[-]}
||\stackrel{9 \;10}{(+)}\;\;\stackrel{11\;12}{[-]}\;\;\stackrel{13\;14}{[-]} $
&1&$-\frac{1}{2}$&1&0&$-\frac{1}{2}$&$-1$&$-1$\\
\hline
5&$e_{L}$&$\stackrel{03}{[-i]}\,\stackrel{12}{(+)}|\stackrel{56}{[-]}\,\stackrel{78}{(+)}
||\stackrel{9 \;10}{(+)}\;\;\stackrel{11\;12}{[-]}\;\;\stackrel{13\;14}{[-]}$
&-1&$\frac{1}{2}$&-1&$-\frac{1}{2}$&0&$-\frac{1}{2}$&$-1$\\
\hline
6&$e_{L} $&$\stackrel{03}{(+i)}\,\stackrel{12}{[-]}|
\stackrel{56}{[-]}\,\stackrel{78}{(+)}
||\stackrel{9 \;10}{(+)}\;\;\stackrel{11\;12}{[-]}\;\;\stackrel{13\;14}{[-]} $
&-1&$-\frac{1}{2}$&-1&$-\frac{1}{2}$&0&$-\frac{1}{2}$&$-1$\\
\hline
7&$ \nu_{L}$&$ \stackrel{03}{[-i]}\,\stackrel{12}{(+)}|
\stackrel{56}{(+)}\,\stackrel{78}{[-]}
||\stackrel{9 \;10}{(+)}\;\;\stackrel{11\;12}{[-]}\;\;\stackrel{13\;14}{[-]}$
&-1&$\frac{1}{2}$&-1&$\frac{1}{2}$&0&$-\frac{1}{2}$&$0$\\
\hline
8&$\nu_{L}$&$\stackrel{03}{(+i)}\,\stackrel{12}{[-]}|\stackrel{56}{(+)}\,\stackrel{78}{[-]}
||\stackrel{9 \;10}{(+)}\;\;\stackrel{11\;12}{[-]}\;\;\stackrel{13\;14}{[-]}$
&-1&$-\frac{1}{2}$&-1&$\frac{1}{2}$&0&$-\frac{1}{2}$&$0$\\
\hline\hline
\end{tabular}
\end{center}
\caption{\label{snmb3tgTable Il.} The 8-plet of leptons - the members of $SO(1,7)$ subgroup of the 
group $SO(1,13)$, 
belonging to one Weyl left 
handed ($\Gamma^{(1,13)} = -1 = \Gamma^{(1,7)} \times \Gamma^{(6)}$) spinor representation of 
$SO(1,13)$. 
It contains the colour chargeless left handed weak charged leptons and the right handed weak 
chargeless leptons. The rest of notation is the same as in Table~\ref{snmb3tgTable Il.}.  
}
\end{table}

In both tables the vectors are chosen to be the eigenvectors of the operators of 
handedness $\Gamma^{(n)}$,  
the generators $\tau^{13}, \, \tau^{23}, \,\tau^{33}$  $ \tau^{38}$,  $Y= \tau^{4} + \tau^{23}$ and
$Q= Y + \tau^{13}$. They are also 
eigenvectors of the corresponding $\tilde{S}^{ab}$, $\tilde{\tau}^{Ai}, A=1,2,3$ and $\tilde{Y}, \tilde{Q}$. 

The  tables  for the two additional choices of the colour charge of quarks 
 follow from the Table~\ref{snmb3tgTable I.} by changing the colour part of 
 the states~\cite{snmb3tgPortoroz03}, by applying $\tau^{3i}$ on the states of Table~\ref{snmb3tgTable Il.}.  .

 The generators $\tilde{S}^{ab}$ take care of the families, transforming each member of one family 
 into the same member of another family, due to the fact that 
 $\{S^{ab}, \tilde{S}^{cd}\}_{-}=0$ (Eq.(\ref{snmb3tgsnmb:tildegclifford})). The eight families of the first 
 member of the eightplet of quarks from Table~\ref{snmb3tgTable I.}, for example, that is of the right 
 handed $u$-quark 
 of the spin $\frac{1}{2}$,  are presented in the left column of Table~\ref{snmb3tgTable II.}. 
 The corresponding right handed neutrinos, belonging to eight different families, are presented 
 in the right column of the same table. The $u$-quark member of the eight families and the $\nu$ 
 members of the same eight families
 are generated by $\tilde{S}^{cd}$, $c,d \in \{0,1,2,3,5,6,7,8\}$ from any starting family.
 \begin{table}
 \begin{center}
 \begin{tabular}{|r||c||c||c||c||}
 \hline
 $I_R$ & $u_{R}^{c1}$&
 $ \stackrel{03}{[+i]}\,\stackrel{12}{(+)}|\stackrel{56}{(+)}\,\stackrel{78}{[+]}||
 \stackrel{9 \;10}{(+)}\:\; \stackrel{11\;12}{[-]}\;\;\stackrel{13\;14}{[-]}$ & 
 $\nu_{R}$&
 $ \stackrel{03}{[+i]}\,\stackrel{12}{(+)}|\stackrel{56}{(+)}\,\stackrel{78}{[+]}||
 \stackrel{9 \;10}{(+)}\;\;\stackrel{11\;12}{(+)}\;\;\stackrel{13\;14}{(+)}$ 
 \\
 \hline
  $II_R$ & $u_{R}^{c1}$&
  $ \stackrel{03}{[+i]}\,\stackrel{12}{(+)}|\stackrel{56}{[+]}\,\stackrel{78}{(+)}||
  \stackrel{9 \;10}{(+)}\;\;\stackrel{11\;12}{[-]}\;\;\stackrel{13\;14}{[-]}$ & 
  $\nu_{R}$&
  $ \stackrel{03}{(+i)}\,\stackrel{12}{[+]}|\stackrel{56}{(+)}\,\stackrel{78}{[+]}||
  \stackrel{9 \;10}{(+)}\;\;\stackrel{11\;12}{(+)}\;\;\stackrel{13\;14}{(+)}$ 
 \\
 \hline
 $III_R$ & $u_{R}^{c1}$&
 $ \stackrel{03}{(+i)}\,\stackrel{12}{[+]}|\stackrel{56}{(+)}\,\stackrel{78}{[+]}||
 \stackrel{9 \;10}{(+)}\;\;\stackrel{11\;12}{[-]}\;\;\stackrel{13\;14}{[-]}$ & 
 $\nu_{R}$&
 $ \stackrel{03}{(+i)}\,\stackrel{12}{[+]}|\stackrel{56}{[+]}\,\stackrel{78}{(+)}||
 \stackrel{9 \;10}{(+)}\;\;\stackrel{11\;12}{(+)}\;\;\stackrel{13\;14}{(+)}$ 
 \\
 \hline
 $IV_R$ & $u_{R}^{c1}$&
 $ \stackrel{03}{(+i)}\,\stackrel{12}{[+]}|\stackrel{56}{[+]}\,\stackrel{78}{(+)}||
 \stackrel{9 \;10}{(+)}\;\;\stackrel{11\;12}{[-]}\;\;\stackrel{13\;14}{[-]}$ & 
 $\nu_{R}$&
 $ \stackrel{03}{[+i]}\,\stackrel{12}{(+)}|\stackrel{56}{[+]}\,\stackrel{78}{(+)}|| 
 \stackrel{9 \;10}{(+)}\;\;\stackrel{11\;12}{(+)}\;\;\stackrel{13\;14}{(+)}$ 
 \\
 \hline\hline\hline
 $V_R$ & $u_{R}^{c1}$&
 $ \stackrel{03}{(+i)}\,\stackrel{12}{(+)}|\stackrel{56}{(+)}\,\stackrel{78}{(+)} ||
 \stackrel{9 \;10}{(+)}\;\;\stackrel{11\;12}{[-]}\;\;\stackrel{13\;14}{[-]}$ & 
 $\nu_{R}$&
 $ \stackrel{03}{(+i)}\,\stackrel{12}{(+)}|\stackrel{56}{(+)}\,\stackrel{78}{(+)} ||
 \stackrel{9 \;10}{(+)}\;\;\stackrel{11\;12}{(+)}\;\;\stackrel{13\;14}{(+)}$ 
 \\
 \hline
 $VI_R$ & $u_{R}^{c1}$&
 $ \stackrel{03}{(+i)}\,\stackrel{12}{(+)}|\stackrel{56}{[+]}\,\stackrel{78}{[+]}|| 
 \stackrel{9 \;10}{(+)}\;\;\stackrel{11\;12}{[-]}\;\;\stackrel{13\;14}{[-]}$ & 
 $\nu_{R}$&
 $ \stackrel{03}{(+i)}\,\stackrel{12}{(+)}|\stackrel{56}{[+]}\,\stackrel{78}{[+]}||
 \stackrel{9 \;10}{(+)}\;\;\stackrel{11\;12}{(+)}\;\;\stackrel{13\;14}{(+)}$ 
 \\
 \hline
 $VII_R$ & $u_{R}^{c1}$&
 $\stackrel{03}{[+i]}\,\stackrel{12}{[+]}|\stackrel{56}{(+)}\,\stackrel{78}{(+)}|| 
 \stackrel{9 \;10}{(+)}\;\;\stackrel{11\;12}{[-]}\;\;\stackrel{13\;14}{[-]}$ & 
 $\nu_{R}$&
 $\stackrel{03}{[+i]}\,\stackrel{12}{[+]}|\stackrel{56}{(+)}\,\stackrel{78}{(+)}||
 \stackrel{9 \;10}{(+)}\;\;\stackrel{11\;12}{(+)}\;\;\stackrel{13\;14}{(+)}$ 
 \\
 \hline
 $VIII_R$ & $u_{R}^{c1}$&
 $ \stackrel{03}{[+i]}\,\stackrel{12}{[+]}|\stackrel{56}{[+]}\,\stackrel{78}{[+]}|| 
 \stackrel{9 \;10}{(+)}\;\;\stackrel{11\;12}{[-]}\;\;\stackrel{13\;14}{[-]}$ & 
 $\nu_{R}$&
 $ \stackrel{03}{[+i]}\,\stackrel{12}{[+]}|\stackrel{56}{[+]}\,\stackrel{78}{[+]}|| 
 \stackrel{9 \;10}{(+)}\;\;\stackrel{11\;12}{(+)}\;\;\stackrel{13\;14}{(+)}$ 
 \\
 \hline 
 \end{tabular}
 \end{center}
 \caption{\label{snmb3tgTable II.} Eight families of the right handed $u_R$ quark with the spin $\frac{1}{2}$, 
  the colour charge $\tau^{33}=1/2$, $\tau^{38}=1/(2\sqrt{3})$ and of the colourless right handed  
  neutrino $\nu_R$ of the spin $\frac{1}{2}$ are presented in the left and in the right column, 
  respectively.
  $S^{ab}, a,b \in \{0,1,2,3,5,6,7,8\}$ transform $u_{R}^{c1}$ of the spin $\frac{1}{2}$ and the 
  chosen colour $c1$ to all the members of the same colour: to the right handed $u_{R}^{c1}$ 
  of the spin $-\frac{1}{2}$, 
  to the left $u_{L}^{c1}$ of both spins ($\pm \frac{1}{2}$), to the right handed $d_{R}^{c1}$ of both spins 
  ($\pm \frac{1}{2}$) and to the left handed $ d_{L}^{c1}$ of both spins ($\pm \frac{1}{2}$). They transform 
  equivalently the right handed   neutrino $\nu_R$ of the spin $\frac{1}{2}$ to the right handed 
  $\nu_R$ of the spin ($-\frac{1}{2}$), to  $\nu_L$ of both spins, to $e_R$ of both spins and to 
  $e_L$ of both spins. $\tilde{S}^{ab}, a,b \in \{0,1,2,3,5,6,7,8\}$ transform a chosen member of one family 
  into the same member of all the eight families.}
 \end{table}

To have an overview over the properties of the members of  one (any one of the eight) family let us 
present in Table~\ref{snmb3tgTable III.} 
the quantum numbers of particular members  of any of the eight families: The handedness 
$\Gamma^{(1+3)}(= -4i S^{03} S^{12}$), 
$\,S^{03}_{L}, \,S^{12}_L$, $\,S^{03}_{R}, \,S^{12}_R$, $\tau^{13}$ (of the weak $SU(2)_{W}$), $\tau^{23}$ 
(of $SU(2)_{II}$), the hyper charge $Y=\tau^{4} + \tau^{23}$, the electromagnetic charge 
$Q$ and the $SU(3)$ status, 
that is, whether the member is a member of the triplet  (the quark with the one of the charges 
$\{(\frac{1}{2}, \frac{1}{2 \sqrt{3}}),
(-\frac{1}{2},  \frac{1}{2 \sqrt{3}} ),(0, - \frac{1}{ \sqrt{3}})\}$) or the colourless lepton. 
%
%
 \begin{table}
 \begin{center}
 \begin{tabular}{|c||c|c|c|c|c|c|c|c|c|c|c||}
 \hline
& $\Gamma^{(1+3)}$& $S^{03}_{L}$&$ S^{12}_L$& $S^{03}_{R}$& $S^{12}_R$& 
$\tau^{13}$ & $\tau^{23}$ & $Y$ & $Q$ & $SU(3)$&$Y'$\\
\hline
\hline
$u_{Li}$& $-1$ &$ \mp \frac{i}{2}$ &$ \pm \frac{1}{2}$& $0$& $0$& $\frac{1}{2}$& $0$&$\frac{1}{6}$& 
$\frac{2}{3}$&{\rm triplet}& $-\frac{1}{6}\, \tan^{2}\theta_{2}$\\ 
\hline
$d_{Li}$& $-1$ &$ \mp \frac{i}{2}$ &$ \pm \frac{1}{2}$& $0$& $0$& $-\frac{1}{2}$& $0$&$\frac{1}{6}$& 
$-\frac{1}{3}$&{\rm triplet}& $-\frac{1}{6}\, \tan^{2}\theta_{2}$\\
\hline
$\nu_{Li}$& $-1$ &$ \mp \frac{i}{2}$ &$ \pm \frac{1}{2}$& $0$& $0$& $\frac{1}{2}$& $0$&$-\frac{1}{2}$& 
$0$&{\rm singlet} & $\frac{1}{2}\, \tan^{2}\theta_{2}$\\
\hline
$e_{Li}$& $-1$ &$ \mp \frac{i}{2}$ &$ \pm \frac{1}{2}$& $0$& $0$& $-\frac{1}{2}$& $0$&$-\frac{1}{2}$& 
$-1$&{\rm singlet}& $\frac{1}{2}\, \tan^{2}\theta_{2}$\\
\hline\hline
$u_{Ri}$& $1$ &$0$& $0$&$\pm \frac{i}{2}$ &$ \pm \frac{1}{2}$& $0 $& $ \frac{1}{2}$&$\frac{2}{3}$& 
$\frac{2}{3}$&{\rm triplet}& $\frac{1}{2}\,(1- \frac{1}{3} \tan^{2}\theta_{2})$\\ 
\hline
$d_{R
i}$& $1$ &$0$& $0$&$\pm \frac{i}{2}$ &$ \pm \frac{1}{2}$& $0 $& $-\frac{1}{2}$&$-\frac{1}{3}$& 
$-\frac{1}{3}$&{\rm triplet} & $-\frac{1}{2}\,(1+ \frac{1}{3} \tan^{2}\theta_{2})$\\
\hline
$\nu_{Ri}$&$1$&$0$& $0$&$\pm \frac{i}{2}$ &$ \pm \frac{1}{2}$& $0 $& $ \frac{1}{2}$& $0$& 
$0$&{\rm singlet} & $\frac{1}{2}\,(1+  \tan^{2}\theta_{2})$\\
\hline
$e_{Ri}$& $1$ &$0$& $0$&$\pm \frac{i}{2}$ &$ \pm \frac{1}{2}$& $0 $& $-\frac{1}{2}$& $-1$& 
$-1$&{\rm singlet} & $-\frac{1}{2}\,(1-  \tan^{2}\theta_{2})$\\
\hline
\hline
 \end{tabular}
 \end{center}
 \caption{\label{snmb3tgTable III.}  The quantum numbers of the members -- quarks and leptons, left and right handed -- 
 of any of the eight families ($i \in\{1,\cdots,8\}$) are presented: The handedness 
 $\Gamma^{(1+3)}= -4i S^{03} S^{12}$, 
$S^{03}_{L}, S^{12}_L$, $S^{03}_{R}, S^{12}_R$, $\tau^{13}$ of the weak $SU(2)_{I}$, $\tau^{23}$ of the second 
$SU(2)_{II}$, the hyper charge $Y=\tau^{4} + \tau^{23}$, the electromagnetic charge $Q$,  the $SU(3)$ status, 
that is, whether the member is a triplet -- the quark with the one of the charges determined by 
$\tau^{33}$ and $\tau^{38}$, that is one of $\{(\frac{1}{2}, \frac{1}{2 \sqrt{3}}),
(-\frac{1}{2},  \frac{1}{2 \sqrt{3}} ),(0, - \frac{1}{ \sqrt{3}})\}$ -- or a singlet, and the charge 
$Y'= {\tau^{23}- \tau^4 \,\tan^{2}\theta_{2}}$.  }
 \end{table}

Let us present also the quantum numbers of  the families from Table~\ref{snmb3tgTable II.}. 
In Table~\ref{snmb3tgTable IV.} 
the handedness of the families
$\tilde{\Gamma}^{(1+3)}(= -4i \tilde{S}^{03} \tilde{S}^{12})$, 
$\tilde{S}^{03}_{L}, \tilde{S}^{12}_L$, $\tilde{S}^{03}_{R}, \tilde{S}^{12}_R$ (the diagonal matrices of 
$SO(1,3)$ ), $\tilde{\tau}^{13}$ 
(of one of the two $SU(2)_{I}$), $\tilde{\tau}^{23}$ (of the second 
$SU(2)_{II}$) are presented. 

%
 \begin{table}
 \begin{center}
 \begin{tabular}{|r||r|r|r|r|r|r|r|r|r|r|r||}
 \hline
$i$ & $\tilde{\Gamma}^{(1+3)}$& $\tilde{S}^{03}_{L}$&$ \tilde{S}^{12}_L$& $\tilde{S}^{03}_{R}$& 
$\tilde{S}^{12}_R$& $\tilde{\tau}^{13}$ & $\tilde{\tau}^{23}$&$\tilde{\tau}^{4}$ & 
$\tilde{Y}'$&$\tilde{Y}$&$\tilde{Q}$ \\
\hline
\hline
$1$& $-1$ &$ - \frac{i}{2}$ &$ \frac{1}{2}$& $0$& $0$& $  \frac{1}{2}$& $0$&$-\frac{1}{2}$&$0$&$-\frac{1}{2}$&$0$\\ 
\hline
$2$& $-1$ &$ -\frac{i}{2}$ &$  \frac{1}{2}$& $0$& $0$& $-\frac{1}{2}$& $0$&$-\frac{1}{2}$&$0$&$-\frac{1}{2}$&$-1$\\
\hline
$3$& $-1$ &$ \frac{i}{2}$ &$ - \frac{1}{2}$& $0$& $0$& $ \frac{1}{2}$& $0$&$-\frac{1}{2}$&$0$&$-\frac{1}{2}$&$0$\\
\hline
$4$& $-1$ &$ \frac{i}{2}$ &$ - \frac{1}{2}$& $0$& $0$& $-\frac{1}{2}$& $0$&$-\frac{1}{2}$&$0$&$-\frac{1}{2}$&$-1$\\
\hline\hline
$5$& $1 $ & $0$ & $0$& $\frac{i}{2}$ &$ \frac{1}{2}$& $0 $& $ \frac{1}{2}$&$-\frac{1}{2}$&$\frac{1}{2}$&$0$&$0$\\ 
\hline
$6$& $1 $ & $0$ & $0$& $\frac{i}{2}$ &$  \frac{1}{2}$& $0 $& $-\frac{1}{2}$&$-\frac{1}{2}$&$-\frac{1}{2}$&$-1$&$-1$\\
\hline
$7$& $1 $ & $0$ & $0$& $-\frac{i}{2}$ &$- \frac{1}{2}$& $0 $& $ \frac{1}{2}$&$-\frac{1}{2}$&$\frac{1}{2}$&$0$&$0$\\
\hline
$8$& $1 $ &  $0$& $0$& $- \frac{i}{2}$ &$-\frac{1}{2}$& $0 $& $-\frac{1}{2}$&$-\frac{1}{2}$&$-\frac{1}{2}
$&$-1$&$-1$\\
\hline
\hline
 \end{tabular}
 \end{center}
 \caption{\label{snmb3tgTable IV.}  The quantum numbers of each member of the eight families presented in 
 Table~\ref{snmb3tgTable II.} are presented: The handedness of the families 
 $\tilde{\Gamma}^{(1+3)}= -4i \tilde{S}^{03} \tilde{S}^{12}$, the left and right handed $SO(1,3)$ 
 quantum numbers (Eq.~(\ref{snmb3tgso13})
$\tilde{S}^{03}_{L}, \tilde{S}^{12}_L$, $\tilde{S}^{03}_{R}, \tilde{S}^{12}_R$ (of $SO(1,3)$ group in the 
$\tilde{S}^{mn}$ sector), $\tilde{\tau}^{13}$ 
 of  $SU(2)_{I}$ , $\tilde{\tau}^{23}$ of the second 
$SU(2)_{II}$, $\tilde{\tau}^4$ (Eq.~(\ref{snmb3tgso6})), $\tilde{Y}'= \tilde{\tau}^{23} -  
\tilde{\tau}^4 \, \tan\tilde{\theta}_2$, taking $\tilde{\theta}^2=0$, $\tilde{Y}
=\tilde{\tau}^{4} + \tilde{\tau}^{23}$, 
$\tilde{Q}= \tilde{\tau}^{4} + \tilde{S}^{56}$. 
}
\end{table}

We see in Table~\ref{snmb3tgTable IV.} that four of the eight families are singlets with respect to 
one of the two $SU(2)$ ($SU(2)_{I}$) groups determined by $\tilde{S}^{ab}$ and doublets with respect to the 
second $SU(2)$ ($SU(2)_{II}$), while the rest four families are doublets with respect to the first 
$SU(2)_{I}$ and singlets with respect to the second $SU(2)_{II}$. When the first break 
spontaneously (and nonperturbatively) 
appears, to which besides the vielbeins also the spin connections  contribute, we expect that 
if only one of the two $SU(2)$ subgroups of $SO(1,7) \times U(1)$ breaking into
$SO(1,3) \times SU(2)\times U(1)$ contributes in the break~\cite{snmb3tgnproc2007}, namely that of charges 
$\vec{\tilde{\tau}}^{2}$, together with $\tilde{N}^{i}_{-} $, there will be four families massless and mass 
protected after this break, namely those, 
which are singlets with respect to $\vec{\tilde{\tau}}^{2}$ and with respect to $\tilde{N}^{i}_{-} $
(Table~\ref{snmb3tgTable IV.}), while for the other four families  the vacuum expectation values of the 
scalars (particular combinations of vielbeins $f^{\sigma}{}_{s}$, and spin connections 
$\tilde{\omega}_{abs}, s \in \{5,8\}$) will take care of the mass matrices on the tree level and beyond  
the tree level.

Making a choice that vacuum expectation value of $\tilde{A}^{4}_{\pm}=0$ and $\tilde{\theta}^2=0$ 
at the first break, it follows 
that the lower four families do not obtain any contribution from the  $\tilde{\omega}_{abs}$ 
to the mass matrices, and stay accordingly massless, until the second (weak) break 
occurs.

Since $\vec{\tau}^{1}|\psi_{R}>=0$ (while the mass term connects $\psi_{L}$ with 
$\psi_{R}$: $\psi^{\dagger}_{L} \,\gamma^0 \gamma^s \,p_{0s} \, \psi_{R}$) and since we expect 
after the first break that the lower four families stay massless (like it is the case 
in the {\it standard model} where the three families are massless before the electroweak break), it must be 
that all the Yukawa couplings of the type $\vec{A}^{2}_{s}=0, s=7,8,$ and $A^{4}_{s}=0$ 
must be zero on the tree level. Since $\psi^{\dagger}_{L} \,\gamma^0 \gamma^s \,p_{0s}\, \psi_{R}, 
s=5,6$ would mix states with different electromagnetic charges, these terms are forbidden. (It 
remains to show, how does this happen.)

The spinor action of Eq.(\ref{snmb3tgfaction}) will in this case manifest the masslessness of the lowest four families, 
while the higher four families will obtain the masses, to which the term 
$ \sum_{s=7,8}\;  \bar{\psi} \gamma^{s} p_{0s} \; \psi$ will contribute on the tree level when 
manifesting particular properties, as it will be discussed in section~\ref{snmb3tgyukawatreefbelow}. 
The tree level contributions would in this case offer the same mass for all the members of one family.
The contributions beyond the tree level distinguish between the $u_i$ and the $d_i, \, i=V,VI,
VII,VIII,$  quarks and between and  $\nu_i$ and $e_i$.

\section{Scalar and gauge fields in $d=(1+3)$ after  breaks}
\label{snmb3tgyukawandhiggs}

What manifests in the  {\it spin-charge-family-theory} as scalar fields in $d=(1+3)$-dimensional space  
after a particular break of a symmetry (which occurs nonadiabatically and spontaneously in a 
kind of a phase transition) are the vielbeins  $e^s{}_{\sigma} $  
\begin{displaymath}
\label{snmb3tge}
 e^a{}_{\alpha} = 
\left( \begin{array}{c c} 
\delta^{m}{}_{\mu}  & 0 \\
 
 0 &  e^{s}{}_{\sigma} \\ 
\end{array}\right)
\end{displaymath}
in a strong correlation with the spin connection fields of both kinds, $\tilde{\omega}_{st \sigma}$  and  
 with $\omega_{ab \sigma} $, with  
indices $s,t, \sigma \in \{5,6,7,8\}$. 

The gauge fields then correspondingly appear as
\begin{displaymath}
\label{snmb3tgeg}
 e^a{}_{\alpha} = 
\left( \begin{array}{c c} 
\delta^{m}{}_{\mu}  & 0 \\
 e^{s}{}_{\mu}= e^{s}{}_{\sigma} E^{\sigma}{}_{Ai} A^{Ai}_{\mu}  &  \;\; e^{s}{}_{\sigma} \\ 
\end{array}\right), 
\end{displaymath}
with 
$ E^{\sigma}{}_{Ai} =  \tau_{Ai} \, x^{\sigma},$ 
where $A^{Ai}_{\mu} $ are the gauge fields, corresponding to (all possible) Kaluza-Klein 
charges $\tau^{Ai}$, manifesting in $d=(1+3)$.
Since the space symmetries  include only $S^{ab}$  ($M^{ab}= L^{ab} + S^{ab}$)  and not
$\tilde{S}^{ab}$, 
there are no vector gauge fields of the type 
$e^{s}{}_{\sigma} \tilde{E}^{\sigma}{}_{Ai} \tilde{A}^{Ai}_{\mu}$, 
with $\tilde{E}^{\sigma}{}_{Ai} =  \tilde{\tau}_{Ai} \, x^{\sigma}$.  
The gauge fields of 
$\tilde{S}_{ab}$ manifest in $d=(1+3)$ only as scalar fields.

The vielbeins and spin connection fields from Eq.~(\ref{snmb3tgwholeaction}) 
($\int  d^dx \, E\; (\alpha \,R + \tilde{\alpha} \, \tilde{R})$) are  
manifesting in $d=(1+3)$ in the following effective action, if no gravity is assumed in $d=(1+3)$  
($e^m{}_{\mu}= \delta^{m}{}_{\mu} $)  
\begin{eqnarray}
\label{snmb3tgvbscstc2}
 S_{b} &=& \int \; d^{(1+3)}x \;\{-\frac{\varepsilon^{A}}{4}\, F^{Ai mn}\, F^{Ai}{}_{mn}  + 
 \frac{1}{2}\, (m^{Ai})^2\,\, A^{Ai}_m A^{Ai\, m} + {\rm scalar \; terms}\, \},\nonumber\\
\end{eqnarray}
where  $m^{Ai}$ of those gauge fields of the charges $\tau^{Ai}$,  which 
symmetries are unbroken,  are zero. Nonzero masses  correspond 
to the broken symmetries, to which  $\tilde{\omega}_{ab\sigma}$, $e^{s}{}_{\sigma}$ and 
$\omega_{ab \sigma}$ contribute. 

In the breaking procedures, when 
$SO(1,7) \times U(1) \times SU(3)$ 
and further to $SO(1,3)  \times SU(2) \times SU(2) \times SU(3) \times U(1)$, all the members 
of the eight families 
of quarks and leptons stay massless (as discussed above) and so do the corresponding gauge 
fields~\cite{snmb3tgdhn}.

Detailed studies of the appearance of breaks of symmetries as follow from 
the starting action, the corresponding manifestation of masses of the 
gauge fields involved in these breaks, as well as the appearance of the nonzero 
vacuum expectation values of the fields which manifest in the mass matrices of the families 
involved in particular breaks is under 
considerations and will be presented elsewhere, when it will be finished. 
In this paper I  just assume that the final two breaks manifest in mass matrices on the tree level, on 
massive gauge fields and in the two types of 
scalar fields, with the symmetries dictated by the breaks and discuss the properties 
of the gauge fields, of the mass matrices beyond the tree level and of the scalar fields. 


%
\subsection{Properties of the scalar fields contributing to the breaks of symmetries}
\label{snmb3tgscalarsub}

I have not yet derived how do the two breaks occur nonperturbatively, and how in details do 
the gauge scalar fields of the two kinds, $\tilde{\omega}_{abs}$ and $\omega_{abs}$, together 
with the vielbeins $f^{\sigma}{}_{s}$ cause the break of symmetries and the corresponding 
phase transitions. Although the symmetries of the gauge scalar fields and the symmetries of 
their vacuum expectation values are known as discussed  above, yet their values (numbers) 
are not known. Also the way how the scalar fields contribute to the masses of the gauge fields 
waits  to be studied.

Let me therefore, only to clarify what the scalar fields are doing, assume in this section that in 
the two successive breaks the two kinds of the scalar fields manifest as the Higgs field does 
in the {\it standard model}. These two scalar fields then obviously differ in all the quantum numbers, 
each of these two fields being coupled to one of the two gauge vector fields.

Let in the break from $SO(1,3)  \times SU(2) \times SU(2) \times U(1) \times SU(3) $ to 
$SO(1,3)  \times SU(2) \times U(1) \times SU(3)$ the scalar field named $\Phi_{II}$ manifest, while in the 
electroweak break (that is in the break from $SO(1,3)  \times SU(2) \times U(1) \times SU(3)$ to 
$SO(1,3)  \times U(1) \times SU(3)$) the scalar field named $\Phi_{I}$ manifests. 
Both scalar fields are assumed 
to be in the fundamental representations with respect to the two $SU(2)$ in both internal spaces, $S^{ab}$ 
and $\tilde{S}^{ab}$. These here assumed scalar fields are indeed manifestation of particular scalar 
forms of vielbeins and spin connections 
with the Einstein index $\sigma \in \{5,6,7,8\}$, which, as already said, 
need to be studied.

Then when the break from $SO(1,3)  \times SU(2) \times SU(2) \times SU(3) 
\times U(1)$ to 
$SO(1,3)  \times SU(2) \times U(1) \times SU(3)$  occurs, I assume that it is $\Phi_{II}$ from 
Table~\ref{snmb3tgTable V.}, which 
manifests the nonzero vacuum expectation value. The scalar field $\Phi_{II}$ is in Table~\ref{snmb3tgTable V.}
expressed in terms of nilpotents and projectors, and manifests as a four vector,  
 a two vector with respect to $S^{ab}$ and a two vector 
with respect to $\tilde{S}^{ab}$.   

%
 \begin{table}
 \begin{center}
 \begin{tabular}{|r||c|r|r|r|r|r||r|r|r|r||}
 \hline
&$\Phi_i$ & $\tau^{13}$& $\tau^{23}$& $\tau^{4}$&$ Y$ & $Q$&$\tilde{\tau}^{13}$& $\tilde{\tau}^{23}$& 
  $\tilde{Y}$&$\tilde{Q}$  \\
\hline
\hline
$ \Phi_{II\,1}$& $\stackrel{56}{(+)}\,\stackrel{78}{(+)}
$&$0$ &$ \frac{1}{2}$ &$ \frac{1}{2}$&$1$& $1$&$0$&$ \frac{1}{2}$&$\frac{1}{2}$&$\frac{1}{2}$\\ 
\hline\hline\hline
$ {\bf \Phi_{II\,2}}$& ${\bf \stackrel{56}{[-]}\,\stackrel{78}{[-]}}  
$&$0$&$-\frac{1}{2}$ &$ \frac{1}{2}$&$0$& $0$&$0$&$ \frac{1}{2}$&$\frac{1}{2}$&$\frac{1}{2}$\\
\hline\hline\hline
$ \Phi_{II\,3}$&$\stackrel{56}{[+]}\,\stackrel{78}{[+]}
$& $0$ &$ \frac{1}{2}$ &$ \frac{1}{2}$&$1$& $1$&$0$&$-\frac{1}{2}$&$-\frac{1}{2}$&$-\frac{1}{2}$\\
\hline
$\Phi_{II \,4}$&$\stackrel{56}{(-)}\,\stackrel{78}{(-)}
$& $0$ &$-\frac{1}{2} $ &$ \frac{1}{2}$&$0$& $0$&$0$&$-\frac{1}{2}$&$-\frac{1}{2}$&$-\frac{1}{2}$\\
\hline\hline
 \end{tabular}
 \end{center}
 \caption{\label{snmb3tgTable V.}  One possible choice for the $SU(2)\times SU(2)$ part of the 
 scalar four vector $\Phi_{II}$,  which is assumed to manifest the (not yet carefully 
 enough studied) contribution of all the scalar gauge fields to the properties of the gauge fields at 
 the first break (from $SO(1,3)\times SU(2)\times SU(2)\times U(1)\times SU(3)$ to 
 $SO(1,3)\times SU(2)\times U(1)\times SU(3)$), is presented together with the corresponding 
 quantum numbers. $\Phi_{II}$ is  in the fundamental 
 representations with respect to  these two groups in the $S^{st}$ and $\tilde{S}^{st}$ sector, 
 $s,t \in \{5,6,7,8\}$. Here $Y= \tau^4 + \tau^{23}$, $\tilde{\tau}^4 =0$, 
 $\tilde{Y}= \tilde{\tau}^4 + \tilde{\tau}^{23}=  \tilde{\tau}^{23}=\tilde{Q}$ and 
 $Q= Y + \tau^{13}$. 
 $\vec{\tau}^{1}$ defines the infinitesimal generators of the weak $SU(2)_{I}$ group and 
 $\vec{\tau}^{2}$ of the second $SU(2)_{II}$ group. It is $\Phi_{II \,2}$ which manifests nonzero 
 vacuum expectation value. 
 }
\end{table}

I assume, as can be read from table~\ref{snmb3tgTable V.},  therefore $\Phi_{II}$ is a weak  
($SU(2)_{I}$) singlet and $SU(2)_{II}$ doublet 
(that means that it is 
in the fundamental 
representations with respect to $SU(2)_{I}$ and $SU(2)_{II}$). It is also a doublet with respect to 
the corresponding $SU(2)_{I}$ and $SU(2)_{II}$ in the internal space of $\tilde{S}^{ab}$. 
I assume that $\Phi_{II \,2}$ gains a nonzero vacuum expectation value 
$<0|\Phi_{II\,2}|0>=
\frac{v_{II}}{\sqrt{2}}$.

Since the break concerns the $SU(2)_{II}$ symmetry, the corresponding gauge fields 
manifest as the massive fields.
It then follows from Table~\ref{snmb3tgTable V.} that $\vec{\tau}^{1}\, \Phi_{II\,2}= 0$, 
$\vec{\tau}^{2}\, \Phi_{II\,2}\ne 0$, $\tau^{4} \,\Phi_{II\,2}\, \ne 0$, 
$Y \, \Phi_{II\,2}=0$, $Q \, \Phi_{II\,2}=0$,  $\vec{\tilde{\tau}}^{1}\, \Phi_{II\,2}= 0$,  
$\vec{\tilde{\tau}}^{2}\, \Phi_{II\,2}\ne 0$, 
$\tilde{Y} \Phi_{II\,2} \ne 0$. 
The gauge fields 
\begin{eqnarray}
\label{snmb3tgmphi2}
A^{2\pm}_m, \;\;A^{Y'}_m,\;\;
\end{eqnarray}
manifest as  massive fields, 
while $A^{Y}_m$ 
stays massless. Here $Y'= \tau^{23}- \tau^{4} \tan^2\theta_{2}$
and $\tilde{Y}'= \tilde{\tau}^{23}- \tilde{\tau}^{4}\tan^2\tilde{\theta}_{2}$.

In section~\ref{snmb3tgyukawatreefbelow} the appearance of the mass matrices for the upper four families after the break 
of $SO(1,3)  \times SU(2) \times SU(2) \times U(1) \times SU(3)$ to 
$SO(1,3)  \times SU(2) \times U(1) \times SU(3)$ is discussed, while the lower four families stay 
massless.
The gauge fields $A^{2\pm}_m, \;\;A^{Y'}_m$ contribute beyond the tree level to the mass terms and so do 
in higher loop corrections also the scalar fields.

The second break from 
$SO(1,3)  \times SU(2) \times U(1) \times SU(3)$ to $SO(1,3)  \times U(1) \times SU(3)$, which 
occurs at the weak scale, is caused by a nonzero vacuum 
expectation value of 
scalar fields, in which vielbeins and spin connections of both kinds contribute. 
Let us, as at the case of the first  ($SU(2)_{II}$)
break, suggest the properties of the supplement of all the scalar (as before with respect to $SO(1,3)$) 
gauge fields which influence the properties of the corresponding  gauge  bosons 
of the $SU(2)_{I}$  and $U(1)$ groups. Above I named this supplement $\Phi_{I}$. It  
manifests  the  {\it standard model} Higgs field, with the nonzero vacuum expectation value 
$<0|\Phi_{I\,2}|>$.
One possible choice for the quantum numbers of the scalar field $\Phi_{I}$ is presented in 
Table~\ref{snmb3tgTable VI.}.  
 \begin{table}
 \begin{center}
 \begin{tabular}{|r||c|r|r|r|r|r||r|r|r|r||}
 \hline
&$\Phi_i$ & $\tau^{13}$& $\tau^{23}$& $\tau^{4}$&$ Y$ & $Q$&$\tilde{\tau}^{13}$& $\tilde{\tau}^{23}$
&$\tilde{Y}$&  $\tilde{Q}$  \\
\hline
$\Phi_{I\,1}$& $\stackrel{56}{[+]}\,\stackrel{78}{[-]}
$& $  \frac{1}{2}$&$0$ &$ \frac{1}{2}$&$\frac{1}{2}$&$1$&$ -\frac{1}{2}$&$0$& $0$&$-\frac{1}{2}$\\ 
\hline\hline\hline
${\bf \Phi_{I\,2}}$& ${\bf \stackrel{56}{(-)}\,\stackrel{78}{(+)}}
$& $- \frac{1}{2}$&$0$ &$ \frac{1}{2}$&$\frac{1}{2}$&$0$&$- \frac{1}{2}$&$0$& $0$&$-\frac{1}{2}$\\ 
\hline\hline\hline 
$\Phi_{I\,3}$& $\stackrel{56}{(+)}\,\stackrel{78}{(-)}
$& $  \frac{1}{2}$&$0$ &$ \frac{1}{2}$&$\frac{1}{2}$&$1$&$  \frac{1}{2}$&$0$& $0$&$\frac{1}{2}$\\ 
\hline
$\Phi_{I\,4}$& $\stackrel{56}{[-]}\,\stackrel{78}{[+]}
$& $ -\frac{1}{2}$&$0$ &$ \frac{1}{2}$&$\frac{1}{2}$&$0$&$  \frac{1}{2}$&$0$& $0$&$\frac{1}{2}$\\ 
\hline\hline 
 \end{tabular}
 \end{center}
 \caption{\label{snmb3tgTable VI.}  One possible choice for the $SU(2)\times SU(2)$ part of the 
 scalar four vector $\Phi_{I}$, which is assumed to manifest the (not yet carefully 
 enough studied) contribution of all the scalar gauge fields to the properties  of 
 the gauge $SU(2)_{I}$ and 
 $U(1)$ fields  at the second  $SU(2)_{I}$, that is at the weak, break 
 (from $SO(1,3)\times SU(2)\times U(1)\times SU(3)$ to $SO(1,3)\times U(1)\times SU(3)$). 
 The scalar field $\Phi_{I}$ is in the fundamental 
 representations with respect to  these two $SU(2)$ groups in the $S^{st}$ and $\tilde{S}^{st}$ sector, 
 $s,t \in \{5,6,7,8\}$.
 }
\end{table}

 $\Phi_{I}$ is  in the fundamental representations with respect to  the two groups 
$SU(2)\times SU(2)$  in the $S^{st}$ and $\tilde{S}^{st}$ sector, 
 $s,t \in \{5,6,7,8\}$. In table~\ref{snmb3tgTable VI.} 
 $Y= \tau^4 + \tau^{23}$, $\tilde{\tau}^4 =0$, $\tilde{Y}= \tilde{\tau}^4 + \tilde{\tau}^{23}= 
 \tilde{\tau}^{23}=\tilde{Q}$ and 
 $Q= Y + \tau^{13}$.  
 $\vec{\tau}^{1}$ defines the infinitesimal generators of the weak $SU(2)_{I}$ group and 
 $\vec{\tau}^{2}$ of the second $SU(2)_{II}$ group. It is $\Phi_{I \,2}$ which manifests a nonzero 
 vacuum expectation value. 
$\Phi_{I}$ is assumed to be a weak  ($SU(2)_{I}$) doublet and $SU(2)_{II}$ singlet 
(that means that it is 
in the fundamental 
representations with respect to $SU(2)_{I}$ and $SU(2)_{II}$). It is also a doublet with respect to 
the corresponding $SU(2)_{I}$ and $SU(2)_{II}$ in the internal space of $\tilde{S}^{ab}$. 
Let  a nonzero vacuum expectation value be $<0|\Phi_{I\,2}|0>=
\frac{v_{I}}{\sqrt{2}}$.

Since the break concerns the $SU(2)_{I}$($ \equiv SU(2)_{W} $) symmetry, the corresponding gauge fields 
manifest as the massive fields.
It  follows from Table~\ref{snmb3tgTable VI.} that $\vec{\tau}^{1}\, \Phi_{I\,2} \ne 0  $, 
$\vec{\tau}^{2}\, \Phi_{I\,2}= 0$, $\tau^{4} \,\Phi_{I\,2}\, \ne 0$, 
$Y \, \Phi_{I\,2} \ne 0$, $Q \, \Phi_{I\,2}=0$,  $\vec{\tilde{\tau}}^{1}\, \Phi_{I\,2}\ne 0$,  
$\vec{\tilde{\tau}}^{2}\, \Phi_{I\,2}= 0$, 
$\tilde{Y} \Phi_{I\,2} = 0$. 
The weak gauge fields 
\begin{eqnarray}
\label{snmb3tgmphi22}
W^{2\pm}_m, \;\;Z^{Q'}_m \equiv Z_m,\;\;
\end{eqnarray}
manifest as  massive fields, as it should to be in  agreement with the experimental data 
while the electromagnetic gauge field $A^{Q}_m \equiv A_m$ 
stays massless. Here $Q'= \tau^{13}- Y\, \tan^2\theta_{1}$
and $\tilde{Y}'= \tilde{\tau}^{13}- \tilde{Y} \,\tan^2\tilde{\theta}_{1}$.

In section~\ref{snmb3tgyukawatreefbelow} the appearance of the mass matrices for the lower four families 
after the break of $SO(1,3)  \times SU(2)  \times U(1) \times SU(3)$ to 
$SO(1,3)   \times U(1) \times SU(3)$ is discussed.  
After this last break the lower four families 
become massive. 
 The vector and scalar gauge fields 
 contribute beyond the tree level 
 to the mass matrices of the lower and the upper four families, what will be discussed in 
 section~\ref{snmb3tgyukawatreefbelow}.

Let me point out once more that the mass term 
in Eq.~(\ref{snmb3tgfaction}) does not need  a scalar field which would 
"dress" the weak chargeless spinors into the weak charged one as it is assumed in the {\it standard model}, 
since the operator 
$ \sum_{s=7,8}\;   \gamma^{s} p_{0s}$ from Eq.~(\ref{snmb3tgfaction}),   when it applies   on a right 
handed weak chargeless family 
member, transforms the right handed weak chargeless member to the corresponding left handed weak 
charged one  of the same family, as 
it can be seen also in Table~\ref{snmb3tgTable I.}. 

The proposed scalar fields are here only to demonstrate, as we pointed out above, what is the role
  of the scalar fields originating in vielbeins and the two kinds of the spin connection 
fields ($f^{s}{}_{\sigma}, \omega_{st\sigma}, \tilde{\omega}_{ab\sigma}$) following from the starting 
action after the breaks and what needs to be studied. Not 
necessarily will the result confirm the above assumptions. The choices I made are namely 
not the only possible. But it also may happen that the idea of the scalar field as 
a (collective in the  {\it spin-charge-family-theory}) degree of freedom, 
suggested by  the {\it standard model}, 
is not what is happening and that when the masses of the gauge fields and 
the mass matrices of the families  will be derived from the  {\it spin-charge-family-theory}, there will 
be no degrees of freedom, which would behave  
as scalar fields.  I leave these open problems for further studies.

\subsection{The gauge fields after the breaks}
\label{snmb3tggaugefieldaction}

After the break of $SO(1,3)\times SU(2)_{I} \times SU(2)_{II} \times U(1) 
\times SU(3)$ into $SO(1,3)\times SU(2)_{I} \times U(1) \times SU(3)$, the gauge fields~\cite{snmb3tgnproc2007} 
$A^{2\pm}_{m}$ 
as well as one superposition 
of $A^{23}_{m}$ and $A^{4}_{m}$ ($A^{Y'}_m$) 
become massive, while  another superposition ($A^{Y}_m$) and the gauge fields $\vec{A}^{1}_{m}$ 
stay massless, due to the 
charges (and consequently  the interaction properties) of  the nonzero expectation value of the 
scalar field $\Phi_{II 2}$ as assumed in Table~\ref{snmb3tgTable V.} and explained in the previous 
subsection~\ref{snmb3tgscalarsub}.

The fields $A^{Y'}_{m}$  and $A^{2 \pm}_{m}$, manifesting 
as massive fields, and $A^{Y}_{m}$ which stay massless,  are defined as the superposition of the 
old ones as follows
%
%
\begin{eqnarray}
\label{snmb3tgnewfieldssab} 
A^{23}_{m}  &=& A^{Y}_{m} \sin \theta_2 + A^{Y'}_{m} \cos \theta_2, \nonumber\\
A^{4}_{m} &=& A^{Y}_{m} \cos \theta_2 - A^{Y'}_{m} \sin \theta_2, \nonumber\\
A^{2\pm}_m &=& \frac{1}{\sqrt{2}}(A^{21}_m \mp  i A^{22}_m),
\end{eqnarray}
for $m=0,1,2,3$ and a particular value of $\theta_2$. 
The massive  scalar fields $A^{Y'}_{s}$, $A^{2\pm}_{s}$, $A^{Y'}_{s}$  contribute through the 
terms $\tau^{2-}\tau^{2 +} u_{R}= u_{R} $   to the fermion 
mass matrices below the tree level.

The corresponding operators for the charges of these new gauge fields are then 
\begin{eqnarray}
\label{snmb3tgnewoperatorssab}
Y&=& \tau^{4}+ \tau^{23}, \quad Y'= \tau^{23} - \tau^{4} \tan^{2} \theta_{2}, 
\quad \tau^{2\pm} = \tau^{21}\pm i \tau^{22}. 
\end{eqnarray}
The new coupling constants become  $g^{Y}= g^{4} \cos \theta_{2}$, $g^{Y'}= g^{2} \cos \theta_{2}$,  
while $A^{2\pm}_m $ have a coupling constant $\frac{g^2}{\sqrt{2}}$. (In nonperturbative phase transitions, 
as the two $SU(2)$ phase transitions where the vielbeins and the two kinds of the spin connection fields 
contribute are,  one must be careful when making a choice of the 
coupling constants $g$.)


We made the assumption that the symmetries in the $\tilde{S}^{ab}$ and $S^{ab}$ sector are broken 
simultaneously. 
The nonzero expectation value of $\Phi_{II\, 2}$ means indeed   
nonzero expectation values of the scalar fields $\tilde{A}^{2i}_{s} $ contributing  to the masses  
of the gauge  bosons and causing that appropriate superposition of the 
gauge fields manifest. 
We have for the scalar fields correspondingly
\begin{eqnarray}
\label{snmb3tgnewfieldstsab} 
\tilde{A}^{23}_{s}  &=& \tilde{A}^{\tilde{Y}}_{s} \sin \tilde{\theta}_2 + \tilde{A}^{\tilde{Y}'}_{s} 
\cos \tilde{\theta}_2, \nonumber\\
\tilde{A}^{4}_{s} &=& \tilde{A}^{\tilde{Y}}_{s} \cos \tilde{\theta}_2 - \tilde{A}^{\tilde{Y}'}_{s} 
\sin \tilde{\theta}_2, \nonumber\\
\tilde{A}^{2\pm}_s &=& \frac{1}{\sqrt{2}}(\tilde{A}^{21}_s \mp  i \tilde{A}^{22}_s),
\end{eqnarray}
for $s= 7,8$ and a particular value of  $\tilde{\theta}_2$. These scalar fields, having a nonzero 
vacuum expectation values, manifest mass matrices of the upper four families on the 
tree level.

The corresponding new  operators are then 
\begin{eqnarray}
\label{snmb3tgnewoperatorstsab}
\tilde{Y}&=& \tilde{\tau}^{4}+ \tilde{\tau}^{23}, \quad \tilde{Y}'= \tilde{\tau}^{23} - 
\tilde{\tau}^{4} \tan^{2} \tilde{\theta}_{2}, 
\quad \tilde{\tau}^{2\pm} = \tilde{\tau}^{21}\pm i \tilde{\tau}^{22}. 
\end{eqnarray}
The new coupling constants are correspondingly $\tilde{g}^{\tilde{Y}}= \tilde{g}^{4} \cos \tilde{\theta}_{2}$,
$\tilde{g}^{\tilde{Y}'}= \tilde{g}^{2} \cos \tilde{\theta}_{2}$,  
while $\tilde{A}^{2\pm}_a $ have a coupling constant $\frac{\tilde{g}^2}{\sqrt{2}}$.

At the weak scale break, when $SO(1,3)\times SU(2) \times U(1) \times SU(3)$ breaks into 
$SO(1,3) \times U(1) \times SU(3)$,  
the vacuum expectation value of  the scalar field manifests 
as $\Phi_{I 2}$ and are again (assumed to be) triggered by the (scalar, with respect to $d=1+3$ ) 
vielbeins $e^{s}{}_{\sigma}$ and the two kinds of the spin connection fields, 
$\tilde{\omega}_{ab\sigma}$ and $\omega_{st\sigma}$, this time the  
 break concerns $SU(2)_{I}(\equiv SU(2)_{W}$.
The assumed properties of the scalar field as presented in Table~\ref{snmb3tgTable VI.} manifest  
 (in the {\it standard model}-like way) the new massive gauge fields  
 $Z_{m}, W^{1\pm}_m$  
and the massless 
$A_{m}$, 
 \begin{eqnarray}
 \label{snmb3tgnewfieldsweaksab}
 A^{13}_{m} &=& A_{m} \sin \theta_1 + Z_{m} \cos \theta_1,\nonumber\\ 
 A^{Y}_{a}  &=& A_{m} \cos \theta_1 -  Z_{m} \sin \theta_1,\nonumber\\ 
 W^{\pm}_m &=& \frac{1}{\sqrt{2}}(A^{11}_m \mp i  A^{12}_m).
 \end{eqnarray}
  The corresponding operators for charges are  
 \begin{eqnarray}
 \label{snmb3tgnewoperatorsweaksab}
 Q  &=&  \tau^{13}+ Y = S^{56} +  \tau^{4},\nonumber\\
 Q' &=& -Y \tan^2 \theta_1 + \tau^{13}, \nonumber\\
 \tau^{1\pm}&=& \tau^{11} \pm i\tau^{12},
 \end{eqnarray}
 and the new coupling constants are $e = 
 g^{Y} \cos \theta_1$, $g' = g^{1}\cos \theta_1$  and $\tan \theta_1 = 
 \frac{g^{Y}}{g^1} $, everything in agreement with the {\it standard model}. 
 
 Similarly as in the break $SU(2)_{II}$ new scalar fields manifest after the break also in 
 the $\tilde{S}^{ab}$ sector 
 \begin{eqnarray}
 \label{snmb3tgnewfieldsweaktildesab}
 \tilde{A}^{13}_{s} &=& \tilde{A}_{s} \sin \tilde{\theta}_1 + 
 \tilde{Z}_{s} \cos \tilde{\theta}_1,\nonumber\\ 
 \tilde{A}^{\tilde{Y}}_{s} &=& \tilde{A}_{s} \cos \tilde{\theta}_1 -  
 \tilde{Z}_{s} \sin \tilde{\theta}_1, \nonumber\\
 \tilde{W}^{\pm}_{s} &=& \frac{1}{\sqrt{2}}(\tilde{A}^{11}_{s} \mp i  \tilde{A}^{12}_{s}).
 \end{eqnarray}
 The corresponding new operators follow  
 \begin{eqnarray}
 \label{snmb3tgnewoperatorsweaktildesab}
 \tilde{Q}  &=&  \tilde{\tau}^{13}+ \tilde{Y} = \tilde{S}^{56} +  \tilde{\tau}^{4},\nonumber\\
 \tilde{Q'} &=& -\tilde{Y} \tan^2 \tilde{\theta}_1 + \tilde{\tau}^{13},\nonumber\\
 \tilde{\tau}^{1\pm}&=& \tilde{\tau}^{11} \pm i\tilde{\tau}^{12}, 
 \end{eqnarray}
 with the new coupling constants $\tilde{e} = 
 \tilde{g}^{Y}\cos \tilde{\theta}_1$, $\tilde{g'} = 
 \tilde{g}^{1}\cos \tilde{\theta}_1$  and $\tan \tilde{\theta}_1 = 
\frac{\tilde{g}^{Y}}{\tilde{g}^1} $.

The Lagrange density for the gauge and the assumed scalar fields (simulating the scalar fields
following from the starting action after the appropriate breaks of symmetries and the 
two final breaks) with the nonzero vacuum 
expectation values of $\Phi_{II 2}$ and $\Phi_{I 2}$ would be 
\begin{eqnarray}
\label{snmb3tgvbscstc}
 S_{b} &=& \int \; d^{(1+3)}x \;\{-\frac{\varepsilon^{A}}{4}\, F^{Ai mn}\,F^{Ai}{}_{mn}  
 \nonumber\\ 
&& \;\;+ \sum_{\pi=I,II}\,[(p_{0m} \Phi_{\pi})^{\dagger}(p_{0}{}^{m} \Phi_{\pi}) - 
V (\Phi^{\dagger}_{\pi} \Phi_{\pi})\},\nonumber\\
&& \;\;  p_{0m} = p_m - \tau^{Ai} \,A^{Ai}{}_{m}. 
\end{eqnarray}
\section{Mass matrices on the tree and beyond the tree  level}
\label{snmb3tgyukawatreefbelow}

In the  two subsections~(\ref{snmb3tgscalarsub},~\ref{snmb3tggaugefieldaction}) of the section~\ref{snmb3tgyukawandhiggs} the 
properties of the scalar and gauge fields after the two successive breaks as it might appear due 
to the  {\it spin-charge-family-theory} were discussed. It remains to study 
the properties of the fermion fields after each of these two breaks, first on the tree level 
and then beyond the tree level.

It is the second term of the fermion action, presented in Eq.~(\ref{snmb3tgfaction}),  
which manifests after the breaks of  symmetries resulting in the  nonzero vacuum expectation values of 
the fields contributing in $p_{0s}$ as  mass matrices. 
The operator $\gamma^s p_{0s}$
can be rewritten in terms of the nilpotents as follows 
\begin{eqnarray}
\label{snmb3tgYaction1}
{\mathcal L}_Y &=& 
\psi^{\dagger} \,\gamma^0\,(\stackrel{78}{(+)} p_{0+} +  \stackrel{78}{(-)} p_{0-} \, \psi,\nonumber\\
p_{0 \pm} &=& p_{07}\mp i p_{08},\quad p_{0s}= p_{s} - \frac{1}{2}\, \tilde{S}^{ab}\,\tilde{\omega}_{ab s} 
- \frac{1}{2}\, S^{ab}\,\omega_{ab s}, 
\end{eqnarray}
with $s=7,8$. It will be assumed that the main part to the mass matrices the vacuum expectation values 
of the $\frac{1}{2}\, \tilde{S}^{ab}\,\tilde{\omega}_{ab s}$  contribute and that  accordingly 
the dynamical term $p_s$ is negligible, while only those scalar components of 
$\frac{1}{2}\, S^{ab}\,\omega_{ab s}$ are nonzero, which do conserve the electromagnetic charges in the 
mass terms. 

The nonzero expectation values of the scalar fields  (the superposition of 
$f^{\sigma}_{s} \tilde{\omega}_{ab \sigma} $)
$\vec{\tilde{A}}^{2}_{s}$  cause the nonzero  mass matrices on the tree level 
and although the left handed 
members of the upper four families do carry the weak charge and the right handed do not, 
the mass protection is lost: The mass term of Eq.~(\ref{snmb3tgfaction}) is after the break 
$SU(2)_{II} \times U(1)_{II}$, due to nonzero vacuum expectation values of 
$\vec{\tilde{A}}^{2}_{s}$, different from 
zero and it is expected to be of the order of the scale of the break of  $SU(2)_{II} \times U(1)_{II}$.

Since the lower four families are singlets with respect to $\vec{\tilde{\tau}}^{2}$ 
and with respect to $\vec{\tilde{N}}_R$, as can be seen in Table~\ref{snmb3tgTable IV.}, 
all the Yukawa couplings originating in $\vec{\tilde{\tau}}^{2} \, \vec{\tilde{A}}^{2}_{s}$ and 
$\vec{\tilde{N}}_{R}\, \tilde{A}^{\tilde{N}_R}_{s}$, $s=7,8$, are equal to zero for 
the lower four families. 

Since $\gamma^0 \gamma^s\, \tau^{1i} $ on any right handed member of any 
family is equal to zero, and since all the mass matrix elements of  the lower four 
families are zero after the break of 
$SO(1,3)\times SU(2)_{I} \times SU(2)_{II} \times U(1)_{II} \times SU(3)$ to $SO(1,3)
\times SU(2)_{I} \times U(1)_{I} \times SU(3)$, which manifests in nonzero vacuum expectation values    
of $\vec{\tilde{A}}^{2}_{s}$ and  $\tilde{A}^{\tilde{N}_R}_{s}$ (but not in nonzero 
vacuum expectation values of $\vec{\tilde{A}}^{1}_{s}$ and  $\tilde{A}^{\tilde{N}_
L}_{s}$ ), and consequently in non zero mass matrices of the upper four families,  
the lower four families stay massless (as long as the vacuum expectation values of 
$\tilde{A}^{1i}_{s}$  and $\tilde{A}^{4}_{s}$, as well as $A^{13}_s$ and $A^{4}_s$ stay equal 
to zero)~\footnote{
The nonzero vacuum expectation values of either $A^{2i}_s$ or $A^{1i}_s$ would cause nonconservation 
of the electromagnetic charge, since $\tau^{2 +}$ transforms  
 $u_R$ into $d_R$, and $\tau^{2 -}$ $d_R$ into $u_R$, while $\tau^{1-}$ transforms  
 $u_L$ into $d_L$, and $\tau^{1 +}$ $d_L$ into $u_L$ and correspondingly 
 is true for leptons. Therefore only nonzero 
 values of $A^{23}_s$ or $A^{13}_s$ make sense. Similarly would the scalar fields, which 
 correspond to the operators $N^{\pm}_{R}, N^{\pm}_{L}$, cause that the mass terms would not 
 conserve the spin.}. 
We shall require for simplicity that scalars  $\omega_{abs}, s =7,8,$ stay after the break of 
$SU(2)_{II} \times U(1)_{I}$  equal to zero.
After this assumption  it follows that the mass matrices are on the tree level the same 
for  $u$-quarks and $\nu$ and 
$d$-quarks and $e$, since $\tilde{S}^{ab}$ do not distinguish among the members of one 
family (they only distinguish among different families).  This is, however, not true, when 
going beyond the tree level, since the massive gauge fields, like $ A^{Y'}_{m}, \, A^{2\pm}_{m}$, 
interacting with family members beyond the tree level together with the scalar 
dynamical fields, make the mass matrices different for a different  family member.

Table~\ref{snmb3tgTable VII.} represents the mass matrix elements on the tree level for the upper 
four families after the first break. 
The notation $\tilde{a}^{\tilde{A}i}_{\pm}=$ $-\tilde{g}^{\tilde{A}i}\, \tilde{A}^{\tilde{A}i}_{\pm
}$ is used. The sign $(\pm)$ distinguishes  between the values of the $u$-quarks and $d$-quarks and 
between the values of $\nu$ and $e$.
 \begin{table}
 \begin{center}
\begin{tabular}{|r||c|c|c|c|c|c|c|c||}
\hline
 &$ I $&$ II $&$ III $&$ IV $&$ V $&$ VI $
 &$ VII $&$ VIII$\\
\hline\hline
$I \;\;\; $ & $ \quad 0 \quad $ & $ \quad 0 \quad $ & $ \quad 0 \quad $ & $ \quad 0 \quad$
          & $ 0 $ & $ 0 $ & $ 0 $ & $ 0 $\\
\hline
$II\;\; $ & $ 0 $ & $ 0 $ & $ 0 $ & $ 0 $
          & $ 0 $ & $ 0 $ & $ 0 $ & $ 0 $\\
\hline
$III\;$ & $ 0 $ & $ 0 $ & $ 0 $ & $ 0 $
          & $ 0 $ & $ 0 $ & $ 0 $ & $ 0 $\\ 
\hline
$IV\;\, $ & $ 0 $ & $ 0 $ & $ 0 $ & $ 0 $
          & $ 0 $ & $ 0 $ & $ 0 $ & $ 0 $\\
\hline\hline
$ V \;\; $ & $ 0 $ & $ 0 $ & $ 0 $ & $ 0 $ & 
$  \frac{1}{2}\, (\tilde{a}^{23}_{\pm} + \tilde{a}^{\tilde{N}^{3}_{R}}_{\pm})$ & 
$  - \tilde{a}^{2-}_{\pm}$ & $ -\tilde{a}^{\tilde{N}_{R}^{+}}_{\pm} $ & $ 0 $\\
\hline
$ VI \;\,$ & $ 0 $ & $ 0 $ & $ 0 $ & $ 0 $ &
$ 
-\tilde{a}^{2+}_{\pm}$ & 
$ \frac{1}{2}(-\tilde{a}^{23}_{\pm } + \tilde{a}^{\tilde{N}^{3}_{R}}_{\pm}) $ & 
$ 0 $ & $- \tilde{a}^{\tilde{N}_{R}^{+}}_{\pm}   $\\ 
\hline
$VII \;$ & $ 0 $ & $ 0 $ & $ 0 $ & $ 0 $ & $   - \tilde{a}^{\tilde{N}_{R}^{-}}_{\pm} $ & $ 0 $ &
$  \frac{1}{2}\,( \tilde{a}^{23}_{\pm}  - \tilde{a}^{\tilde{N}^{3}_{R}}_{\pm}) $ & 
$  -\tilde{a}^{2-}_{\pm}$ \\
\hline
$VIII$ & $ 0 $ & $ 0 $ & $ 0 $ & $ 0 $ & $ 0 $ & $ - \tilde{a}^{\tilde{N}_{R}^{-}}_{\pm}  $ &  
$ - \tilde{a}^{2+}_{\pm}$ & 
$ - \frac{1}{2}\,( \tilde{a}^{23}_{\pm} + \tilde{a}^{\tilde{N}^{3}_{R}}_{\pm})$ \\
\hline\hline
\end{tabular}
 \end{center}
 \caption{\label{snmb3tgTable VII.}  The mass matrix for the eight families of quarks and 
 leptons after the break of $SO(1,3)\times SU(2)_{I}  \times SU(2)_{II} \times U(1)_{II} \times SU(3)$ 
 to $SO(1,3) \times  SU(2)_{I} \times U(1)_{I} \times SU(3)$. The contribution comes 
 from the term $\tilde{S}^{ab} \,\tilde{\omega}_{abs}$ in $p_{0s}$ in Eq.(\ref{snmb3tgYaction1}). 
 The notation $\tilde{a}^{\tilde{A}i}_{\pm}$ stays for 
 $-\tilde{g}^{\tilde{A}i}\, \tilde{A}^{\tilde{A}i}_{\pm}$,  $(\mp)$ distinguishes $u_{i}$ from
 $d_{i}$ and $\nu_{i}$ from $e_{i}$.
 }
\end{table} 

All the matrix elements can be expressed in terms of $\tilde{\omega}_{abs}$.
Since after the break $SU(2)_{II} \times U(1)_{II}$ 
$\vec{\tilde{A}}^{\tilde{N}_{L}}_{s}=0$ and  
$\vec{\tilde{A}}^{1}_{s}= 0$, then $\vec{\tilde{A}}^{\tilde{N}_{R}}_{s}= 2(\tilde{\omega}_{23s}, 
\tilde{\omega}_{13s},
\tilde{\omega}_{12s})$ and $\vec{\tilde{A}}^{2}_{s}= 2(\tilde{\omega}_{58s}, 
\tilde{\omega}_{57s}, \tilde{\omega}_{56s})$,
$s=7,8$. 

The masses of the lowest of the higher four family were evaluated in ref.~\cite{snmb3tggn} from the cosmological 
and direct measurements, when assuming that baryons of this stable family (with no mixing matrix  
to the lower four families) 
constitute the dark matter. 


The lower four families, which stay massless after the break of the one of the two 
$SU(2)$ subgroups of $SO(4)$ ($SU(2)_{II}\times U{1}_{II}$), obtain masses when the second 
$SU(2)_{I} \times U(1)_{I}$  
break occurs at the electroweak scale, manifesting in nonzero vacuum expectation values 
of the scalar fields $\tilde{A}^{1i}_{s}$  and $\tilde{A}^{4}_{s}$, as well as $A^{13}_s$ and $A^{4}_s$, 
and also in the masses of the gauge fields  $W^{\pm}_{m}$ and $Z_m$. We  simulated the appearance 
of the masses of the vector gauge fields, in order to illustrate assumptions  of the 
{\it standard model}, by the scalar $\Phi_{I }$ and its nonzero vacuum expectation value.

Like in the case of the upper four families, also here is the mass matrix contribution from 
the nonzero vacuum expectation values of $f^{\sigma}{}_{s}\, \tilde{\omega}_{ab \sigma}$ 
on the tree level   
the same for the quarks and  the leptons ($(\pm)$ distinguish between the values of the $u$-quarks 
and $d$-quarks and 
between the values of $\nu$ and $e$), while the contribution from   
$f^{\sigma}{}_{s}\, \omega_{ab \sigma}$, which do not contradict the observed properties (do conserve the 
electromagnetic charge), contribute on the tree level only the constant times the identity matrix. 
 This diagonal contribution, the same for all the four lower families, 
comes from terms like  $Q' A^{Q'}_s, s=7,8$.  

Beyond the tree level, however, the difference among the members of one family 
in all the mass matrix elements emerges, through the contribution of the gauge and the scalar fields.

Table~\ref{snmb3tgTable VIII.} represents the contribution of $\tilde{g}^{\tilde{A}i}\tilde{\tau}^{\tilde{A}i}\,
\tilde{A}^{\tilde{A}i}_{\pm}$ to the mass matrix 
elements on the tree level for the lower  
four families after the weak break, the contribution from the  terms like 
$Q' \,A^{Q'}_{s}$,  which are diagonal and 
equal for all the families, but distinguish among the members of one family are not present.  
The notation $\tilde{a}^{\tilde{A}i}_{\pm}=$ $-\tilde{g}^{\tilde{A}i}\, \tilde{A}^{\tilde{A}i}_{\pm}$ is used. 
 \begin{table}
 \begin{center}
\begin{tabular}{|r||c|c|c|c||}
\hline
 &$ I $&$ II $&$ III $&$ IV $\\
\hline\hline
$I \;\;$& $  \frac{1}{2}\, (\tilde{a}^{13}_{\pm} + \tilde{a}^{\tilde{N}^{3}_{L}}_{\pm})$ & 
$   \tilde{a}^{1+}_{\pm}$ & $ \tilde{a}^{\tilde{N}_{L}^{+}}_{\pm} $& $0$ \\
\hline
$II\;$ &  $ \tilde{a}^{1-}_{\pm} $ & 
$ \frac{1}{2}( -\tilde{a}^{13}_{\pm } + \tilde{a}^{\tilde{N}^{3}_{L}}_{\pm}) $&$0$& 
$ \tilde{a}^{\tilde{N}_{L}^{+}}_{\pm}   $ \\ 
\hline
$III$ & $ \tilde{a}^{\tilde{N}_{L}^{-}}_{\pm} $ & $0$&  
$  \frac{1}{2}\,( \tilde{a}^{13}_{\pm}  - \tilde{a}^{\tilde{N}^{3}_{L}}_{\pm}) $ & 
 $\tilde{a}^{1+}_{\pm}$ \\
\hline
$IV$ & $0$&$  \tilde{a}^{\tilde{N}_{L}^{-}}_{\pm}  $ &  
$  \tilde{a}^{1-}_{\pm}$ &
$ - \frac{1}{2}\,( \tilde{a}^{13}_{\pm} + \tilde{a}^{\tilde{N}^{3}_{L}}_{\pm})$ \\
\hline\hline
\end{tabular}
 \end{center}
 \caption{\label{snmb3tgTable VIII.}  The mass matrix for the lower four  families of quarks and 
 leptons after the electroweak break. Only the contributions coming  from the terms 
 $\tilde{S}^{ab} \,\tilde{\omega}_{abs}$ in $p_{0s}$ in Eq.(\ref{snmb3tgYaction1}) are presented. 
 The notation $\tilde{a}^{\tilde{A}i}_{\pm}$ stays for $-\tilde{g}\, \tilde{A}^{\tilde{A}i}_{\pm}$, 
 where $(\mp)$ stays to distinguish  between the values of the $u$-quarks and $d$-quarks and 
between the values of $\nu$ and $e$. 
 The terms coming from $S^{ss'}\, \omega_{ss'\,t}$ are not presented here. They are the same for 
 all the families,  but distinguish among the family members.  
 }
\end{table} 

After the weak break again all the matrix elements are expressible  
in terms of $\tilde{\omega}_{abs}$ since $\vec{\tilde{A}}^{1i}_s = (\tilde{\omega}_{58 s}- 
\tilde{\omega}_{67 s}, \tilde{\omega}_{57 s}+ \tilde{\omega}_{68 s}, \tilde{\omega}_{56 s}- 
\tilde{\omega}_{78 s})$, while $|\frac{\tilde{A}^{1i}_s}{\tilde{A}^{2i}_s}|\ll1$.


The mass matrices of the lower four families were studied and evaluated in the ref.~\cite{snmb3tggmdn} 
under the assumption 
that if going beyond the tree level the differences in the mass matrices of different 
family members start to manifest. The symmetry properties of the mass matrices from 
Table~\ref{snmb3tgTable VIII.}  were assumed while fitting the matrix elements to the experimental data 
for the three observed families within the accuracy of the experimental data.

\subsection{Mass matrices  beyond the tree level}
\label{snmb3tgbreakbelowtree}

The contribution of $\tilde{S}^{ab} \, \omega_{abs}$ to the mass matrix elements in Eq.~(\ref{snmb3tgYaction1}) 
is for  each family member the same, while terms like $Q' A^{Q'}_s$ and $Q A^{Q}_s$, if nonzero, 
distinguish among family members, contributing only to the diagonal terms the same value for all  
the families. When going beyond the tree level the massive gauge  and scalar fields start 
to contribute to each of mass matrix elements differently for different family members. 
It is my prediction (and hope) that these contributions (in all the orders below the tree level) 
are generating the 
mass matrix elements which then manifest  the masses and the mixing matrices for the so far 
observed three families and predict properties for the fourth family members. 
In the ref.~\cite{snmb3tgalbinonorma2010} we are studying the contributions from the matrix elements 
below the tree level in one and two loop corrections to the tree level. 

After the break of $SU(2)_{II} \times U(1)_{II}$ all the massive gauge fields ($A^{Y'}_m, A^{2  \pm}_m$), 
as well as the massive scalar dynamical fields ($\tilde{A}^{2i}_s, \tilde{A}^{\tilde{N}_R\,i}_{s} $),
contribute  in the loop corrections to the mass matrices of  the upper four families. 
 Equivalently also  the corresponding massive gauge fields and the massive scalar dynamical fields 
 contribute to the mass matrices 
  of the lower four families below the tree level.  Some of these loop corrections 
  also influence (slightly) the mass matrices of the upper four families.

Let me suggest that, for example, that after the weak break (for which $Q' \nu_R =0$, 
$Q' e_R = \tan^2 \theta^1$, 
 $Q' u_R =-\frac{2}{3}\, \tan^2 \theta^1 $ and
$Q' d_R =\frac{1}{3}\, \tan^2 \theta^1 $) the  mass matrices  -- after taking into account 
the loop corrections (in which the dynamical scalar fields 
and massive gauge fields  contribute) -- are expressed  as a sum of matrices, each 
one "belonging" to the operator ${Q'}$ on the power $k$ or ${Y'}$  on the power $k$ or  
${Q'^k}$  ${Y'^{k'}}$
\begin{eqnarray}
\label{snmb3tgqprime}
M=\sum_{k=0}^{\infty}\, {Q'}^k \, M_{ Q'\,k} +  \sum_{k=1}^{\infty}\, {Y'}^k \, M_{ Y'\,k} 
+ \sum_{k=1, k'=1}^{\infty}\, {Q'}^k \,{Y'}^{k'} \, M_{Q' Y'\,k k'}! 
\end{eqnarray}
Then the mass matrices would manifest explicitly
the differences  among  family members.  ${Q'}^k $ and  ${Y'}^k $ stay for the operators on the power $k$ and 
$M_{ QO\,k}$  for the matrices. 

Since there are no terms below the tree level, which would contribute to the 
matrix elements which are zero on the tree level, the mass matrices keep the 
zeros also when going below the tree level. 

More about the mass matrices below the tree level can be found in 
the ref.~\cite{snmb3tgalbinonorma2010}.

\section{Conclusions}
\label{snmb3tgconclusions}

The  {\it spin-charge-family-theory}~\cite{snmb3tgnorma,snmb3tgpikanorma,snmb3tgnproc2007,snmb3tgdark2010,snmb3tggmdn,snmb3tggn} is,  
by proposing the mechanism for generating families of quarks and leptons and consequently  
predicting  the existence of families, the number of families at (sooner or later) observable 
energies, the mass matrices for each of the family member
(and correspondingly the masses and the mixing matrices of families of quarks 
and leptons), 
offering the way beyond the {\it standard model of 
the electroweak and colour interactions}. It predicts the fourth family to 
be possibly measured at the LHC or at some higher energies and the fifth family which is, 
since it is decoupled in the mixing matrices
from the lower four families and it is correspondingly stable, the 
candidate to form the dark matter~\cite{snmb3tggn}.

In the ref.~\cite{snmb3tgpikanorma,snmb3tgnproc2007,snmb3tggmdn} we made a rough estimation of  the properties 
of the gauge fields and 
quarks and leptons for the lower four families on the tree level as predicted by the 
 {\it spin-charge-family-theory}.   
We took into account that going beyond the tree level the mass matrices for different family members 
differ in all the matrix elements  keeping the symmetry  evaluated on 
the tree level. We fitted the matrix elements to the existing experimental data 
for the observed three families within the experimental accuracy and for a chosen mass 
of each of the fourth family member. We predict the  elements of the mixing matrices for 
the fourth family 
members as well as the weakly measured matrix elements of the three observed families. 
In the ref.~\cite{snmb3tggn} we evaluated the masses of the stable fifth family  (belonging  
to  the upper four families)  
under the assumption that  neutrons and neutrinos of this stable fifth family form the dark matter. 
We study the properties of the fifth family  neutrons, their freezing out of  the cosmic plasma 
during the evolution of the universe, as well as  their interaction among themselves and 
with the ordinary matter in the direct experiments.

In this paper we study properties of the gauge fields, scalar fields and the eight 
families of quarks and leptons as they  follow from the  {\it spin-charge-family-theory}
on the tree level and 
also below the tree level  after the last two successive breaks, trying to understand better 
what could happen during these two breaks and after it.  We made several assumptions,
since we are not (yet) able to  evaluate correctly how do these two final breaks occur and 
what does trigger them. 

In the first break several scalar (with respect to $SO(1,3)$) fields 
$f^{\sigma}{}_{s}\, \tilde{\omega}_{abs}$, breaking 
one of the two $SU(2)$ symmetries ($SU(2)_{I} \times SU(2)_{II}\times U(1)_{II}$ into 
$SU(2)_{I}\times U(1)_{I}$), gain 
nonzero vacuum expectation values, causing the nonzero mass matrices of the upper four families on the 
tree level and bringing masses to some of the gauge fields. These scalar fields manifest, from the 
point of view of the vector gauge fields carrying the quantum numbers of $\vec{\tau}^{2}$, 
something  like a scalar field in the {\it standard model} sense  for the 
upper four families, while they leave the lower four families massless, since the second $SU(2)_{I}$ 
symmetry stays unbroken. (I presented a possible manifestation of the influence of the scalar fields 
contributing to this break and influencing properties of the gauge vector fields as doublet field 
with respect to the two $SU(2)_{II}$, one in the $S^{ab}$ and the other in the $\tilde{S}^{ab}$).

In the second break, that is in the electroweak break, several other combinations of 
$f^{\sigma}{}_{s}\, \tilde{\omega}_{ab\sigma}$  together with some of the 
scalars  $f^{\sigma}{}_{s}\, \omega_{s't\sigma}$ contribute 
with their nonzero 
vacuum expectation values to the 
mass matrices of the lower four families and  to the masses of the gauge fields $W^{\pm}_m$ 
and $Z_m$. I simulated the influence of the scalar fields on the 
properties of the gauge fields with the scalar (Higgs) field to understand better 
 the {\it standard model} assumptions.  This scalar field 
 is assumed to be a doublet with respect to the two $SU(2)_{I}$, one in the $S^{ab}$ and the other in 
 the $\tilde{S}^{ab}$).

The properties of the scalar fields, of the (twice four) family mass matrices and of the gauge 
fields (of the two $SU(2)$  and the corresponding $U(1)$ symmetries) should be 
 calculated from the simple starting action of the  {\it spin-charge-family-theory}~\ref{snmb3tgwholeaction}. 
 In this paper we were  not yet able to  do these very demanding calculations, which should answer 
 why and how the successive breaks occur, first from $SO(1,13)$ to $SO(1,7) \times U(1) \times SU(3)$, 
 leaving eight  massless families  and further to $SO(1,3) \times SU(2) \times U(1) \times SU(3)$, 
 with four massless and four massive families and the corresponding gauge fields, and finally to 
 massive four lower families, decoupled in the mixing matrices from the higher four families, the 
 three lowest of them with the properties of the observed ones and  to the massive weak bosons and massless 
 photon. 
 
Let me tell all the assumptions I made in this paper and in the references cited in this paper.\\

i. The properties of all the fields, fermionic and bosonic, in $d=(1+3)$ follow from the 
simple starting actions for massless spinors which interact with the vielbeins and the two kinds 
of the spin connection fields, the gauge fields of $S^{ab}$  and $\tilde{S}^{ab}$ and for the corresponding 
vielbeins in spin connection fields. The operators $S^{ab}$ are expressible with the Dirac's $\gamma^a$'s operator, 
and obviously manifest in $d=(1+3)$ the spin and all the charges of fermions, while  $\tilde{S}^{ab}$ are determined 
 with the second kind of the Clifford algebra objects, $\tilde{\gamma}^{a}$'s, 
which I am proposing for the description of families of fermions. The corresponding 
Lagrange densities for the gauge fields are assumed to be  linear in the curvature. 
(The breaks in ordinary space and the corresponding conserved symmetries require the appearance 
of the gauge vector fields in $d=(1+3)$, while there is no  gauge vector fields in $d=(1+3)$ 
originating in  $\tilde{S}^{ab}$.)\\

ii. Breaks from $SO(1,13)$, with only one massless spinor in $d=(1+13)$, to $SO(1,7) \times U(1) 
\times SU(3) $ and further to $SO(1,3) \times SU(2) \times SU(2) \times U(1) \times SU(3) $ leave 
eight massless families and also the corresponding massless gauge fields.\\

iii. Each break manifests in the vacuum expectation values of particular superposition of 
$f^{\sigma}{}_{s} \tilde{\omega}_{ab \sigma} $ and $f^{\sigma}{}_{s} \omega_{ab \sigma} $, such that 
are in agreement with the observations. They are scalar fields with respect to $d=(1+3)$. \\

iv. The properties of the gauge vector fields in $d=(1+3)$ should follow directly from the starting 
Lagrange density due to interaction with those scalar fields which gain nonzero vacuum expectation 
values after each of the two breaks. I simulated in this paper these interactions with the appearance of 
the two scalar fields of the kind as 
assumed in the {\it standard model}, which are doublets with respect to  one of the two $SU(2)$ symmetries 
in the $S^{ab}$ and $\tilde{S}^{ab}$ sector -- that one which breaks. \\

v. The breaks occur simultaneously in both sectors $S^{ab}$ and $\tilde{S}^{ab}$. The  
responsibility  for a particular break have the scalar fields, that is the gauge fields of 
$S^{ab}$, $\tilde{S}^{ab}$ and the spin connections. \\

vi. Only that part of the simple starting action manifests in $d=(1+3)$ as the mass term, which is
in agreement with the observations. \\

The evaluations of the mass matrix elements  
below the tree level is presented in the contribution~\cite{snmb3tgalbinonorma2010} in this 
proceedings and also in the paper which is in preparation. 
The evaluations will hopefully  help to understand within the 
 {\it spin-charge-family-theory} why there are 
so great differences in the masses and mixing matrices within the observed members of 
the families. It should also make more accurate predictions for the properties of the fourth family, as well 
as for all the upper four families.

\section*{Appendix: Some useful relations}
\label{snmb3tgsabprop}

The following Cartan subalgebra set of the algebra $S^{ab}$ (for both sectors) is chosen:
\begin{eqnarray}
S^{03}, S^{12}, S^{56}, S^{78}, S^{9 \;10}, S^{11\;12}, S^{13\; 14}\nonumber\\
\tilde{S}^{03}, \tilde{S}^{12}, \tilde{S}^{56}, \tilde{S}^{78}, \tilde{S}^{9 \;10}, 
\tilde{S}^{11\;12}, \tilde{S}^{13\; 14}.
\label{snmb3tgcartan}
\end{eqnarray}
%
A left handed ($\Gamma^{(1,13)} =-1$) eigen state of all the members of the 
Cartan  subalgebra 
\begin{eqnarray}
&& \stackrel{03}{(+i)}\stackrel{12}{(+)}|\stackrel{56}{(+)}\stackrel{78}{(+)}
||\stackrel{9 \;10}{(+)}\stackrel{11\;12}{(-)}\stackrel{13\;14}{(-)} |\psi \rangle = \nonumber\\
&&\frac{1}{2^7} 
(\gamma^0 -\gamma^3)(\gamma^1 +i \gamma^2)| (\gamma^5 + i\gamma^6)(\gamma^7 +i \gamma^8)||
\nonumber\\
&& (\gamma^9 +i\gamma^{10})(\gamma^{11} -i \gamma^{12})(\gamma^{13}-i\gamma^{14})
|\psi \rangle .
\label{snmb3tgstart}
\end{eqnarray}
represent the $u_R$-quark with spin up and of one colour.

$ \tilde{S}^{ab} $  generate families from the starting $u_R$ quark
In particular $\tilde{S}^{03}(= \frac{i}{2}
[\stackrel{03}{\tilde{(+i)}} \stackrel{12}{\tilde{(+)}} +
\stackrel{03}{\tilde{(-i)}} \stackrel{12}{\tilde{(+)}} +
\stackrel{03}{\tilde{(+i)}} \stackrel{12}{\tilde{(-)}}+
\stackrel{03}{\tilde{(-i)}} \stackrel{12}{\tilde{(-)}}])$  applied on 
a right handed $u_R$-quark with spin up and a particular colour generate a state which is again 
 a right handed $u$-quark of the same colour.
\begin{eqnarray}
\stackrel{03}{\tilde{(-i)}}\stackrel{12}{\tilde{(-)}} &&
\stackrel{03}{(+i)}\stackrel{12}{(+)}| \stackrel{56}{(+)} \stackrel{78}{(+)}||
\stackrel{9 10}{(+)} \stackrel{11 12}{(-)} \stackrel{13 14}{(-)}=\nonumber\\
&&\stackrel{03}{[\,+i]} \stackrel{12}{[\,+\,]}| \stackrel{56}{(+)} \stackrel{78}{(+)}||
\stackrel{9 10}{(+)} \stackrel{11 12}{(-)} \stackrel{13 14}{(-)},
\end{eqnarray}
where 
\begin{eqnarray}
\stackrel{ab}{(\pm i)}         &=& 
\frac{1}{2}\, ( \gamma^a \mp  \gamma^b), 
\stackrel{ab}{(\pm 1)} =          \frac{1}{2} \,( \gamma^a \pm i\gamma^b),\nonumber\\
\stackrel{ab}{[\pm i]}& =& \frac{1}{2} (1 \pm   \gamma^a \gamma^b), \quad
\stackrel{ab}{[\pm 1]} = \frac{1}{2} (1 \pm i \gamma^a \gamma^b), \nonumber\\
\stackrel{ab}{\tilde{(\pm i)}} &=& 
\frac{1}{2}  (\tilde{\gamma}^a \mp  \tilde{\gamma}^b), \quad
\stackrel{ab}{\tilde{(\pm 1)}} = 
\frac{1}{2}  (\tilde{\gamma}^a \pm i\tilde{\gamma}^b), \nonumber\\ 
\stackrel{ab}{\tilde{[\pm i]}} &=& \frac{1}{2} (1 \pm \tilde{\gamma}^a \tilde{\gamma}^b), \quad
\stackrel{ab}{\tilde{[\pm 1]}} = \frac{1}{2} (1 \pm i \tilde{\gamma}^a \tilde{\gamma}^b). 
\label{snmb3tgdeftildefun}
\end{eqnarray}

We present below some useful relations which are easy to derive~\cite{snmb3tgpikanorma}. 

\begin{eqnarray}
\label{snmb3tgrelations}
\stackrel{ab}{(k)} \stackrel{ab}{(k)}& =& 0, \;   \stackrel{ab}{(k)} \stackrel{ab}{(-k)}
= \eta^{aa}  \stackrel{ab}{[k]}, \; 
\stackrel{ab}{[k]} \stackrel{ab}{[ k]} =   \stackrel{ab}{[k]}, \nonumber\\
\stackrel{ab}{[k]} \stackrel{ab}{[-k]} &=& 0, 
\stackrel{ab}{(k)} \stackrel{ab}{[ k]} = 0,\quad \; 
\stackrel{ab}{[k]} \stackrel{ab}{( k)}  = \stackrel{ab}{(k)}, \nonumber\\ 
\stackrel{ab}{(k)} \stackrel{ab}{[-k]} &=& \stackrel{ab}{(k)} ,
\quad \; \stackrel{ab}{[k]} \stackrel{ab}{(-k)} =0.   
\end{eqnarray}
\begin{eqnarray}
\stackrel{ab}{\tilde{(k)} }  \stackrel{ab}{(k)}& =& 0, 
\quad \;
 \stackrel{ab}{\tilde{(-k)}} \stackrel{ab}{(k)}= 
-i \eta^{aa}                 \stackrel{ab}{[k]},\nonumber\\ 
 \stackrel{ab}{\tilde{( k)}} \stackrel{ab}{[k]}&=& 
i  \stackrel{ab}{(k)},\quad
 \stackrel{ab}{\tilde{( k)}} \stackrel{ab}{[-k]} = 0.
\label{snmb3tggraphbinomsfamilies}
\end{eqnarray}
The relations
\begin{eqnarray}
N^{\pm}_{+}         &=& N^{1}_{+} \pm i \,N^{2}_{+} = 
 - \stackrel{03}{(\mp i)} \stackrel{12}{(\pm )}\,, \quad N^{\pm}_{-}= N^{1}_{-} \pm i\,N^{2}_{-} = 
  \stackrel{03}{(\pm i)} \stackrel{12}{(\pm )}\,,\nonumber\\
\tilde{N}^{\pm}_{+} &=& - \stackrel{03}{\tilde{(\mp i)}} \stackrel{12}{\tilde{(\pm )}}\,, \quad 
\tilde{N}^{\pm}_{-}= 
  \stackrel{03} {\tilde{(\pm i)}} \stackrel{12} {\tilde{(\pm )}}\,,\nonumber\\ 
\tau^{1\pm}         &=& (\mp)\, \stackrel{56}{(\pm )} \stackrel{78}{(\mp )} \,, \quad   
\tau^{2\pm}=            (\mp)\, \stackrel{56}{(\mp )} \stackrel{78}{(\mp )} \,,\nonumber\\ 
\tilde{\tau}^{1\pm} &=& (\mp)\, \stackrel{56}{\tilde{(\pm )}} \stackrel{78}{\tilde{(\mp )}}\,,\quad   
\tilde{\tau}^{2\pm}= (\mp)\, \stackrel{56}{\tilde{(\mp )}} \stackrel{78}{\tilde{(\mp )}}\,,
\end{eqnarray}
 are already derived in the text.

\title{Bohmian Quantum Mechanics \\ or \\ 
What Comes {\sl Before} the Standard Model}
\author{G. Moultaka}
\institute{%
Laboratoire de Physique Th\'eorique \& Astroparticules \\
Universit\'e Montpellier 2, CNRS/IN2P3, Montpellier, France \\
gilbert.moultaka@univ-montp2.fr}

\titlerunning{Bohmian Quantum Mechanics\ldots}
\authorrunning{G. Moultaka}
\maketitle

\begin{abstract}
We give a short account of an old and somewhat unfashionable approach to quantum mechanics, arguing though
for its potential to provide a new kind of Higgsless physics beyond the Standard Model.
\end{abstract}

\section{Introduction}
\label{contribution:moultaka}
\begin{small}
\begin{quotation} { {\it [...But why then had Born not told me of this 'pilot wave'? If only to point out what was wrong with it?
Why did von Neumann not consider it? More extraordinarily, why did people go on producing 'impossibility' proofs,
after 1952, and as recently as 1978? When even Pauli, Rosenfeld, and Heisenberg, could produce no more 
devastating criticism of Bohm's version than to brand it as 'metaphysical' and 'ideological'? Why is the pilot wave
picture ignored in text books? Should it not be taught, not as the only way, but as an antidote to the prevailing
complacency? To show that vagueness, subjectivity, and indeterminism, are not forced on us by experimental facts,
but by deliberate theoretical choice? ...]}} \end{quotation} \end{small}

\noindent
When reading those lines written by John S. Bell twenty eight years ago  \cite{gm1bell1}, one feels, if one knows about
the Bohmian quantum mechanics, somewhat bewildered by their topicality... and why is it that, even today, an 
overwhelming majority of physicists is still being raised on the belief that quantum indeterminism 
has been thrust on us, a long time ago, by the objective observation of the natural phenomena, the 
'experimental facts', to the extent that the only choice left is to incorporate it once and for all in our natural way 
of thinking, and to carve it in the abstract manipulation of our formalism?  Of course, some still worry about
conceptual issues in quantum mechanics, and there seems to be an increasing awareness nowadays that the
celebrated 'Copenhagen interpretation' is only one among various other approaches which are equally sound
for the description of the quantum phenomena \cite{gm1laloe}. Nevertheless, a majority sees the whole issue as merely 
'philosophical', and from there very unlikely to play any significant role in the modern quest for 
a unified description of the fundamental laws of Nature. There is a simple reason for that, for why
bother about as many different approaches, including the Bohmian quantum mechanics (BQM),
 if at the end of the day they all lead to {\sl exactly} the same predictions for the experimentally observable 
phenomena? Then, if not for epistemological reasons, only a few would see 
an interest in learning and using BQM even though, as we will see, it has by far less conceptual
problems (at least in the non-relativistic domain) than the orthodox interpretation of quantum mechanics.
In fact, it would be a progress if this were the only reason to ignore BQM, rather than the erroneous\footnote{as shown in 
\cite{gm1bell2} p. 1-13, p. 14-21 and in \cite{gm1bell1}} and somewhat
widespread belief that BQM has been proved theoretically wrong by von Neumann's theorem \cite{gm1vonneumann}, 
and experimentally wrong through the tested violation of Bell's inequalities \cite{gm1aspectetal}!\\

The present talk aims at a concise introduction to the essential features of BQM. However, the emphasis will be put on 
the fact that BQM {\sl does} have the seeds for being not just {\sl `yet another interpretation predicting merely the 
same physical phenomena as the orthodox interpretation'}. It is this latter aspect  which could in our opinion make the 
BQM a particularly interesting starting point for speculations about new physics, whether beyond the standard model (BSM)
of particle physics, beyond the standard cosmology, or even at the most fundamental level questioning the structure
of space-time \cite{gm1bohmhiley}.

\section{A historical burden}

\noindent
It would still be a long way to go from the above claim to a well defined formulation such as for instance a 
BQM-based BSM physics, that would hoist it to a level comparable to the fashionable BSM speculations such as Grand 
Unification, Supersymmetry, Extra-dimensions, etc. or even the less standard ones some of which have been presented in 
this conference.\footnote{The word 'speculation' should not have here any belittling connotation; it only serves to 
stress that up to now none of the BSM scenarios have met a firm experimental support, except perhaps
for the neutrino sector. However, strictly speaking, even the latter does not need any BSM physics
for its description, at least not more than for the other massive fermions of the model, notwithstanding
the issue of Dirac or Majorana nature of the massive neutrinos.}  Indeed, as we will recall very briefly, 
historical as well as sociological factors discouraged a sustained development of physical theories in the BQM 
language that would have put the latter at a level comparable to that for the orthodox quantum mechanics, 
and in particular as concerns the Relativistic Quantum Field Theories (QFT).

It is at the celebrated 1927 Solvay congress that the main ideas were presented for the first time by Louis de Broglie,
of what  David Bohm will take up several years  later and develop into a self-consistent theory \cite{gm1bohm12,gm1bohm3}, 
BQM (known also as the de Broglie-Bohm pilot-wave quantum mechanics). Initially interested in developing his principle 
of 'double solution for the Schr\"odinger equation', but running short of time for the Solvay congress,
 de Broglie had to content himself with a less sophisticated approach postulating separately a wave and a particle.
But as it goes, his approach was severely criticized by Pauli which made him rally Bohr and others in what will become
the orthodox quantum mechanics.\footnote{A very interesting account of the history of de Broglie's pilot-wave can be found in ref. \cite{gm1cushing}.}

\section{BQM in a nutshell}

\noindent
Let us illustrate the idea in the simplest system, a point particle of mass $m$ feeling a potential energy $V(\vec{x})$.
We assume here for simplicity zero spin and no non-abelian internal degrees of freedom.  
In the quantum realm we associate
with this particle a scalar complex-valued wave function

\begin{equation}
\psi(\vec{x},t) \equiv R(\vec{x},t) e^{i S(\vec{x},t)/\hbar} \label{gm1eq:wavef}
\end{equation}

\noindent
governed by the Schr\"odinger equation

\begin{equation}
i \hbar \frac{\partial}{\partial t} \psi(\vec{x},t) = \hat{H} \psi(\vec{x},t) = (-\frac{\hbar^2}{2 m}
\vec{\nabla}_{\vec{x}}^2 + V(\vec{x})) \psi(\vec{x},t)  \label{gm1eq:schroed}
\end{equation}

\noindent
In eq.(\ref{gm1eq:wavef}), $R$ and $S$ are real-valued functions which are uniquely defined for a given $\psi$,
up to an additive $2 \pi n \hbar$ constant to $S$ where $n$ is an arbitrary relative integer.\footnote{Without loss of 
generality we take $\psi$ to be properly normalized so that $\int |\psi|^2 d^3\vec{x} = 1$
when integrating over the whole space volume.}   
As in the orthodox interpretation, we will assume the {\sl probability postulate}, namely that 
$ R(\vec{x}, t)^2 (=|\psi|^2)$ gives the probability  density of {\sl finding} the particle at a given instant $t$ 
in an  infinitesimal volume $d^3 \vec{x}$ around a given space 
point $\vec{x}$. The analogy, however, stops here. None of the other postulates of the orthodox 
interpretation (the measurement process, the wave function collapse,...)  will be assumed hereafter. Instead, one 
simply adds the {\sl postulate} that well-defined trajectories in space and time are still assigned to the point particle,
 governed by  the following dynamical equation:   
  
\begin{equation}
\vec{p}\equiv m \frac{d}{dt}\vec{x}(t)=\vec{\nabla}S(\vec{x}(t),t) \label{gm1eq:particle}
\end{equation}

\noindent
where now $\vec{x}(t)$ is the actual trajectory of the particle and $\vec{p}$ its momentum. The above equation is thus
essentially deterministic since it gives a physical meaning to a simultaneous knowledge of the position and 
velocity of the particle at any instant of time $t$. Furthermore, the occurrence of $S$ in Eq.(\ref{gm1eq:particle})
implies that the wave function $\psi$ plays a dynamical role since it affects the equation of motion of the
particle, {\sl i.e.}  $\psi$ is {\sl guiding} the particle, whence the name 'pilot-wave'. One could then worry
about the coherence of this approach altogether, where $\psi$ encodes in the same time a probabilistic 
as well as a deterministic information! A first tentative answer to this worry lies in that the above-mentioned 
features are encoded in two independent parts of $\psi$, namely the modulus $R$ and the phase
$S$. However there is much more to it since  $R$ and $S$ are dynamically 
correlated through Eq.(\ref{gm1eq:schroed}) as we will shortly discuss, so that one still has to understand
how can the two apparently exclusive features live together. To see what is happening we first note that
plugging Eq.(\ref{gm1eq:wavef}) in Eq.(\ref{gm1eq:schroed}) and identifying
the real and imaginary parts one finds,      
   
\begin{eqnarray}
&&
\displaystyle
{\frac{\partial S}{\partial t} + \frac{ (\vec{\nabla} S)^2}{2 m} + V +\;\;
 {\fbox{
 {$\displaystyle -\frac{\hbar^2}{2 m} \frac{\vec{\nabla}^2 R}{R}$ }}}= 0} \label{gm1eq:schroed1}\\
\nonumber \\
&& \displaystyle {\frac{\partial R^2}{\partial t} + \vec{\nabla} \cdot (R^2 \frac{\vec{\nabla}
S}{m}) = 0 \label{gm1eq:schroed2}}
\end{eqnarray}

\noindent
It is noteworthy that in the limit $\hbar \to 0$ one retrieves from Eq.(\ref{gm1eq:schroed1}) the classical
Hamilton-Jacobi equation, where $S$ is now identified as Hamilton's principle function. This 
clarifies the origin of the postulated equation (\ref{gm1eq:particle}) as being nothing else but the classical
relation giving the momentum in terms of the derivative of a generating function of canonical
transformations. The crucial point appears then as the fact of having retained relation (\ref{gm1eq:particle}) even when
$\hbar \ne 0$, thus pushing further the well-known close connection between the Hamilton-Jacobi theory and the dual
particle and wave mechanics.\footnote{see for instance, Goldstein's Classical Mechanics, {\sl Addison-Wesley},
chap. 9.} Furthermore, from equations (\ref{gm1eq:schroed1}) and (\ref{gm1eq:particle} ) one finds readily a generalized
Newton's law, 
$m \frac{d^2 \vec{x}}{dt^2}  =  - \vec{\nabla} ( V(\vec{x}) + U(\vec{x}))$, where 
$U= - \frac{\hbar^2}{2 m} \frac{\vec{\nabla}^2 R}{R}$. One thus sees that there is effectively, 
on top of $ - \vec{\nabla} V$, 
an extra force $ - \vec{\nabla} U$ given by a potential energy of {\sl quantum mechanical} nature acting on the 
particle. The interpretation of Eq.(\ref{gm1eq:schroed1}) as being a quantum generalization of the Hamilton-Jacobi 
equation becomes now evident with $V + U$ being the total potential energy,  $\frac{ (\vec{\nabla} S)^2}{2 m}$
the kinetic energy {\sl c.f.} Eq.(\ref{gm1eq:particle}), and $-\frac{\partial S}{\partial t}$ the total energy of the 
particle${}^5$. It follows that the real function $R(\vec{x}, t)$ determines, not only the probability of finding
the particle (the probability postulate stated above), but also the quantum force guiding the particle. 
So the question asked previously of how can the probabilistic and deterministic features coexist in the same
theory, becomes even sharper as they both emerge from the same function. This strongly suggests giving up either on 
the deterministic feature or on the probabilistic feature as being fundamental. In the standard quantum mechanics one 
gives up on the former; here we will rather give up on the latter and assume that the probabilistic feature is emerging
from classical statistical considerations: in realistic settings the initial conditions, here the initial position 
of the point particle, are always endowed with uncertainties so that a probability density distribution $P(\vec{x}, t)$
has to be provided. Thus the real content of the {\sl probability postulate} in BQM is to assume that the classical
probability distribution is given by 

\begin{equation} 
P(\vec{x}, t) = |\psi|^2. \label{gm1eq:probapost} 
\end{equation} 

\noindent
But then its content is no more equivalent to that
of the orthodox quantum mechanics. Indeed, in the latter there would be no meaning asking {\sl why} do we adopt such
a postulate, for it is simply a founding postulate, while in BQM it is legitimate to wonder why a classical 
uncertainty which is in principle reducible at will by improving the precision on the initial conditions for the
point particle, should be constrained by the wave function? We will see that there is an elegant and self-consistent
 answer to this question which is provided by the theory itself. Before discussing this point let us first summarize
the main assumptions of BQM and their consequences:
 
\begin{itemize}
\item[-a-] the wave function $\psi$ satisfies the Schr\"odinger equation
\item[-b-] the particle is guided by the wave through 
$\frac{d \vec{x}}{dt}  = \frac{1}{m} \vec{\nabla} S_{|\vec{x}= \vec{x}(t)}$
\item[-c-] in practice, the initial position of
the particle is never known with absolute precision $\rightarrow$ assume a statistical ensemble with probability density
$P(\vec{x}, t)= |\psi(\vec{x}, t)|^2$
\end{itemize}

\noindent
The principle result \cite{gm1bohm12} is that assumptions -a- \& -b- \& -c- lead exactly
to the same predictions as the orthodox quantum mechanics, including the uncertainty principle, and answer by 
themselves all the criticism de Broglie had to face during the 1927 Solvay congress.\footnote{Actually Bohm sent his papers to Pauli in 1952 but never
got any answer back.} In particular the crucial {\sl reduction of the wave packet following measurement} postulate,
made by von Neumann, can be in fact derived from these assumptions and thus becomes a dynamical consequence of the 
measurement process. Moreover, essentially all the conceptual issues which arise within the orthodox interpretation
(double-slit experiment, delayed-choice, EPR,...) are completely resolved in the BQM approach through the effect
of the quantum potential $- \frac{\hbar^2}{2 m} \frac{\vec{\nabla}^2 R}{R}$, \cite{gm1bell2}. We will not dwell further here on these 
features of BQM (see e.g. \cite{gm1holland} for an extensive account), but only stress that as a matter of fact
 there is still even today no experiment that can distinguish between BQM and the orthodox quantum mechanics,
and thus no objective reason to reject the former in favour of the latter; experiment does {\sl not} tell us by itself 
that indeterminism is indeed a fundamental feature of the natural  phenomena rather than a phenomenological by-product 
of an underlying causal behaviour, albeit classically statistical or chaotic,  and Occam's razor is almost helpless 
when having to choose between a conceptually fuzzy but technically simple 
and familiar explanation, and a conceptually clear but  mathematically somewhat involved one. \\

What we find more interesting to explore is the possibility of weakening the postulates of BQM in such a way 
that the known quantum phenomena, and the effective equivalence between BQM and orthodox quantum mechanics are
retrieved in a given limit, but where experimental deviations leading to new physical phenomena can occur
away from this limit. [Such an approach is of course a common place in the realm of Beyond the Standard Model physics, 
except for the fact of being perhaps much more upstream in the present case.] Let us thus come back to the question
we have asked previously concerning the status of the $P(\vec{x}, t)= |\psi(\vec{x}, t)|^2$ relation in postulate -c-.
We should first note that requiring this relation to hold for any  $t$ conceals an important consistency
between -a- and -b-. Indeed, since the point particle follows a deterministic trajectory,  the time evolution of a 
statistical ensemble of such particles with a given initial density distribution $P(\vec{x}, t_0)$ at some time $t=t_0$
will give uniquely the value of $P$ at any other time $t > t_0$, determined by the actual trajectories of all the 
particles of the ensemble. So -c- will be consistent only if the postulated relation between $P$ and whatever other
function is a fixed point of the dynamical evolution. This is indeed what happens in BQM thanks to equations 
(\ref{gm1eq:schroed2}) and (\ref{gm1eq:particle}); one can prove \cite{gm1bohm3} that if $P(\vec{x}, t_0)= |\psi(\vec{x}, t_0)|^2$ 
is satisfied at one given time $t_0$ then $P(\vec{x}, t)= |\psi(\vec{x}, t)|^2$ becomes valid for all $t > t_0$
due to the dynamical trajectories of the particle, thus showing some robustness in assumption -c-.\footnote{
To reach this result one uses the similarity between eq.(\ref{gm1eq:schroed2}) and the general continuity equation
for probability densities, $\frac{\partial P}{\partial t} + \vec{\nabla} \cdot (P \frac{d}{dt}\vec{x}(t)) = 0$,
together with eq.(\ref{gm1eq:particle}), 
to show that the ratio $P/|\psi|^2 \equiv f(\vec{x}, t)$ is a constant of motion when $\vec{x}$ is replaced by the
actual particle trajectory, i.e. $\frac{d}{dt} f(\vec{x}(t), t)= \frac{\partial f}{\partial t} +  
\frac{d}{dt}\vec{x}(t)\cdot \vec{\nabla} f = 0$.} Nonetheless, one still feels somewhat uncomfortable about the
peculiarity of the initial condition, for why should one require $P(\vec{x}, t_0) = |\psi(\vec{x}, t_0)|^2$
to hold in the first place?

\section{BSM physics from BQM?} 

\noindent
What happens if the initial conditions are such that  $P(\vec{x}, t_0) \neq |\psi(\vec{x}, t_0)|^2$?
As initially anticipated by Bohm \cite{gm1bohm3} and established on rather more firm grounds later on
(see for e.g. \cite{gm1bohmhiley1}, \cite{gm1valentini}), a statistical ensemble governed by the dynamical assumptions
-a- and -b- at the {\sl microscopic} level and evolving from an initial time $t_0$ where
$P(\vec{x}, t_0) \neq |\psi(\vec{x}, t_0)|^2$, will reach {\sl statistical  equilibrium} at a later time $t$
only when the equality $P(\vec{x}, t) \neq |\psi(\vec{x}, t)|^2$ is reached. This is a very profound result
in the realm of BQM as it allows to reinterpret the quantum phenomena as we know them as corresponding 
to the physical properties of some system in statistical equilibrium, and thus to predict deviations from 
quantum mechanics  when the system is out-of-equilibrium! These features can be elegantly described by resorting
to an adapted definition of the celebrated Boltzmann $H$ function and proving a corresponding $H$-theorem.  
More specifically, following \cite{gm1valentini}, one considers a system of $N$ interacting particles, with space coordinates 
$(X_1,X_2, ..., X_N) \equiv \vec{X}$, where the interactions among those particles are given by a classical potential 
$V(\vec{X})$ as well as the dynamical assumptions -a- and -b-, and defining $f(\vec{X}, t)$ by  
$P(\vec{X}, t)= f(\vec{X}, t) \times |\psi(\vec{X}, t)|^2 $, one introduces a notion of entropy $S=-H$ with 
$H = \int d\Sigma |\psi|^2 f \; \log f$ 
where $d\Sigma$ is a configuration space (not a phase space) volume element. An $H$-theorem then follows
 for the fine-grained function H, $\frac{{dH}}{dt} = 0$, as well as for its coarse-grained version $\bar{H}$
(when $N$ is large enough), $\frac{d\bar{H}}{dt} \leq 0$. Thus $\bar{H}$ decreases with time to its minimum value 
(maximum of the entropy) where the system reaches statistical equilibrium. As shown in \cite{gm1valentini} this
minimum is attained when $\bar{f}$, the ratio of the coarse-grained $P(\vec{X}, t)$ and  $ |\psi(\vec{x}, t_0)|^2$, 
reaches $1$. It thus becomes clear in what sense assumption -c- in BQM can be viewed not as a postulate but as
an emerging property once one considers the coarse-grained probability distribution  $\overline{P(\vec{X}, t)}$ 
rather than $P(\vec{X}, t)$ and the coarse grained square of the wave-function $\overline{|\psi|^2}$ rather than 
$|\psi|^2$ in a system with a large number of interacting particles such as our Universe: when the elements of such a 
large system interact with each other for a sufficiently long time, 
$\overline{P(\vec{X}, t)}$ and $\overline{|\psi(\vec{X}, t)|^2}$ will evolve tending to equal each other
when $\bar{H}$ reaches its minimum at some time $t=T$, and $\overline{P(\vec{X}, t)}= \overline{|\psi(\vec{X}, t)|^2}$
for any $t > T$ even if they started very different at some initial time $t=t_0$. Moreover, once the statistical 
equilibrium is reached any subsystem (including the subsystems made of one single particle) will also satisfy  
$\overline{P(\vec{X'}, t)}= \overline{|\psi(\vec{X'}, t)|^2}$ where $\vec{X'}$ is the configuration space of the 
subsystem, \cite{gm1valentini}, so that all ordinary quantum phenomena are guaranteed.

This set of results provides a physically coherent understanding of why it is difficult, if not impossible, to 
distinguish in practice between the orthodox and the BQM approaches to quantum phenomena, since in terms of the
BQM approach the Universe as a whole has had enough time to reach today the 'quantum equilibrium' in the sense 
defined above. Nonetheless, this does also open an interesting window for potential new physics: for one thing,
any deviation from $\overline{P} = \overline{|\psi|^2}$ leads to deviations from the Heisenberg uncertainty principle
and concomitantly allows for instantaneous signals (see, \cite{gm1bohmhiley}, \cite{gm1valentini}, \cite{gm1bohmhiley1} ) which
can have far reaching consequences on the interplay between Lorentz invariance and quantum effects,
and violations thereof\footnote{As 
emphasized by Bell, in experiments of the EPR type Lorentz invariance is rescued by the uncertainty principle. It should
thus not come as a surprise that the non-local features related to EPR would lead to violation of Lorentz invariance
once the uncertainty principle is modified.}; for another, one expects from the previous discussion that deviations 
from $\overline{P} = \overline{|\psi|^2}$ can indeed occur in transient regimes towards statistical equilibrium  or in 
local fluctuations around this equilibrium. Effects of the first kind can be relevant to very early universe physics,
before the Universe as a whole has settled in the 'quantum equilibrium', while the second kind can be relevant to 
out-of-equilibrium regimes for {\sl local} subsystems.\\

In the last part of this talk we will speculatively argue that there are physical systems where effects of this second 
type can possibly occur in a {\sl natural} way; namely systems described by massive Yang-Mills theories such as in 
the Standard Model of electroweak interactions. As stated in section 2, relativistic field theories have not been
developed in the BQM approach to a level comparable to that of the familiar relativistic QFT. Without such a 
development, the foregoing discussion might indeed look very speculative. For instance, a version of the 
full-fledged Standard Model in the BQM language is not available. Still, the existing formulations (see for e.g.
\cite{gm1holland} and part II of \cite{gm1bohmhiley2}), in particular the photon field and abelian gauge theories in the 
BQM language, lend themselves to generalization to the non-abelian case, at least far enough to formulate our argument. 
Let us state it here in words: in the standard formulation it is common knowledge that pure massive Yang-Mills theories
suffer from bad high-energy behaviour and from the lack of renormalizability of quantum effects due to the longitudinal 
component of the spin-1 massive field. The physical consequence is that processes involving scattering of this
longitudinal component violate unitarity, thus signaling that the theory is sick as it stands. Technically, this
comes about due to the quantum field theoretic version of the orthodox quantum mechanical postulate 
Eq.(\ref{gm1eq:probapost}),
{\sl relating a scattering probability density ${\cal P}$ to the square of the corresponding Green's function $G$.}
  As we know, the standard remedy is to add new physics in such a way to cancel the high energy behaviour and restore 
unitarity, the most fashionable being to postulate a scalar Higgs particle. But it is crucial that any postulated 
remedy, whether extending the Higgs sector, or assuming some alternative dynamical sector, seeks for the same aim, the 
restoration of unitarity, that is modifying the Green's function so that the total probability does not exceed one. 

This rational of unitarity restoration does not need to hold anymore when the BQM approach is adopted.
Indeed, the relation between ${\cal P}$ and the square of $G$ can now be violated, akin to the
violation of Eq.(\ref{gm1eq:probapost}) if the system is out of statistical 'quantum equilibrium', as discussed previously.
In this case there would be no unitarity violation if ${\cal P}$ and $G$ are sufficiently decorrelated from one another.
 Hence, while in the orthodox approach the scattering of longitudinal $W$ or $Z$ bosons requires 
the onset of new particles or interactions at the TeV scale, in the BQM approach Nature has the choice between
the latter solution keeping the system in 'quantum equilibrium', or instead putting the system locally out of 
'quantum equilibrium' which we would lead as new physics to typical Lorentz invariance and uncertainty principle 
violations.

%

\section*{Acknowledgements}
I would like to thank the organizers as well as all the participants at this meeting for the very friendly 
as well as scientifically critical state of mind. 


\newcommand\hbnmnlnf{ln \, f}
\author{H.B. Nielsen${}^1$ and M. Ninomiya${}^{2,3}$}
\title{Backward Causation in Complex Action Model\,---
  Superdeterminism and Transactional Interpretations\thanks{OIQP-10-04 \\
RIKEN-TH-189}}
\institute{%
${}^1$The Niels Bohr Institute,
University of Copenhagen,\\
Copenhagen $\phi$, DK2100, Denmark\\
${}^2$Okayama Institute for Quantum Physics,\\  
Kyoyama 1, Okayama 700-0015, Japan \\
${}^3$Theoretical Physics Laboratory,\\ 
The Institute of Physics and Chemical Research (RIKEN),\\ 
Wako, Saitama 351-0198, Japan}

\authorrunning{H.B. Nielsen and M. Ninomiya}
\titlerunning{Backward Causation in Complex Action Model}
\maketitle
\begin{abstract}
 It is shown that the transactional interpretation of quantum mechanics being referred 
back to Feynman-Wheeler time reversal symmetric radiation theory is 
reminiscent of our complex action model. In our model 
the initial conditions are, in principle, even calculable. Thus the model 
philosophically points towards superdeterminism, but really the problem associated with Bell's theorem is solved in our complex action model by removing 
the necessity for signals to travel faster than the speed of light for consistency.

As previously published our model predicts that the LHC (Large Hadron Collider) should have a 
failure before producing as many Higgs-particles as 
would have been produced by the SSC (Superconducting Super Collider) accelerator. 
In the present article, we point out that a card game to decide 
whether to restrict the operation of the LHC, which we proposed as means of testing our model will be a success under all circumstances. 
 \end{abstract}


\section{Introduction}\label{hbn1mn:intro}
\label{contribution:mnhbn}
In our previous articles \cite{hbn1mnsearch} we proposed that one should use 
the LHC machine to look for 
backward causation effects. 
Indeed, we proposed a model \cite{hbn1mnsearch,hbn1mn3,hbn1mn4,hbn1mn5} 
in which the realized history of the 
universe was selected so as to minimize a certain functional 
of the history, 
the functional being the imaginary part of the action $S_I$\,[history], 
which only exists in our model.
In general, it is assumed in science that there is no prearrangement \cite{hbn1mnr1} 
of initial conditions determining the occurrence or nonoccurrence of several events.
However, in a BBC radio broadcast J.~Bell proposed a solution to the problem 
of Einstein-Podlosky-Rosen ``super-determinism'' \cite{hbn1mn6}. 
Furthermore, Cramer has been developing a transactional interpretation of 
quantum mechanics \cite{hbn1mnCramer}, 
which involves Feynman-Wheeler's radiation theory\cite{hbn1mnWheeler_Feynman_interaction} 
that has backward causation 
in its formalism. 

Also, one of the present authors (H.B.N.) and his group 
previously proposed models that are nonlocal in time (and space) \cite{hbn1mn7,hbn1mn8,hbn1mn9}.
Similar backward causation effects have also been proposed 
in connection with the hypothesis that, for example, humanity would cause 
a new vacuum to appear, (a ``vacuum bomb,'') devised 
by one of the present authors (H.B.N.) and collaborators \cite{hbn1mn10}. 

Our proposal is to test whether there are such 
prearrangements in nature, that is, prearrangements that prevent 
producing Higgs particles, such as LHC and SSC, from being functional. 
Our model, based on the imaginary part of the action \cite{hbn1mnsearch,hbn1mn11}, 
begins with a series of not completely convincing, 
but still suggestive, assumptions that lead to the 
prediction that large machines producing Higgs particles should turn out not 
to work in the history of the universe actually 
being realized.

The two main points of the present article are the following:
\begin{itemize}
\item[A)] To argue that our model with imaginary action [1-4] is a very 
natural type of model if one decides to go for Bell's proposal of 
superdeterminism [6] or the transactional interpretation.
\item[B)] To argue that by performing an experiment to test our 
model involving a card game to determine the placing of restrictions on the running of the LHC 
can only be successful. 
\end{itemize}

These points will be explained below: \\
A)~~~~The point of superdeterminism is to overcome the problem of quantum mechanics associated with Bell's theorem 
and the very general assumptions of locality. 
This is done by discarding the assumption that experimentalists in each of the 
entanglement connected objects have ``free will'' when choosing quantity 
to measure. Rather, the idea is that they could NOT have chosen 
to measure anything different from what they indeed decided to measure. 
In other words it is as if the decision of these experimentalists was fixed; 
thus it may even be something that is, in principle, calculable.
At least, it should not be argued that other possibilities 
could have been chosen for their measurements other than those they actually chose. 
In this case, of course, the Einstein-Podolsky-Rosen theorem or 
equivalently Bell theorem has no validity. If you only measure whatever you 
measure, it is only the element of reality that corresponds to that which is 
relevant, and no paradoxes occur.

Since, in the classical approximation, our model leads to a model 
from the initial condition {\em everything} can be determined 
in principle --- but not in practice --- purely by calculation only using the 
coupling constants including their imaginary parts as the input, 
this is an clear 
example of superdeterminism (at least classically). We argue in 
section~\ref{hbn1mn:sec2} that something like our model is strongly called for
to overcome the problems of measurement theory, as was also repeated 
in the Proceedings at the Vaxjo conference by one of us \cite{hbn1mn14}. 
\\
B)~~~~The plan behind the practical experiment that we proposed 
was to produce some random 
numbers --- partly by drawing cards and partly by quantum random number 
generation --- and then to translate these random numbers in accordance with 
the rules of the game, into some restrictions on the luminosity or 
energy or both of the LHC. Thus, the LHC might, for instance, only be 
allowed to operate up to a certain beam energy.
I.~Stewart \cite{hbn1mn11} proposed that pauses in operation should be determined by random numbers.

The idea is merely to require any restriction on the LHC to have a 
probability $p$ that is deemed, through the rules of the game, 
to be very small. 
The probability $p$ for a ``close the LHC'' card should be \,
$p\sim 10^{-6}$ \cite{hbn1mnsearch}. 

It is clear that even a small probability of a restriction 
being enforced on the LHC, 
its luminosity or beam energy 
results in an artificially imposed risk for the LHC project.

It is, 
however, 
the main focus of the present paper to point out 
(as was briefly stated in a previous article \cite{hbn1mnsearch}) 
that 
even though our proposed project of restricting the LHC 
based on random numbers seems to give rise to a loss, 
in fact, whatever happens seems 
--- initially at least --- to be a gain i.e., a success!

In the present article it is demonstrated that a success in this sense is guaranteed to be the result, 
seemingly with almost 100\% certainty (but in reality not quite 100\%).

The main point is that the occurrence of a restriction with a probability of $10^{-6}$ resulting from such a card game is 
already evidence for our model, which would thus be discovered by such 
a card draw.

Of course, the whole exercise of making the proposed card game experiment 
for search for some backward causation of the type we propose would be futile 
if such an effect was already excluded by earlier experiments. 
If indeed we look for completely disastrous bad luck in any attempt 
to produce even a single Higgs particle, presumably the Tevatron of 
FNAL in Chicago would present a counterexample. 
Although not a single Higgs particle has been convincingly recognized to have 
been produced at the Tevatron, one expects that according to theoretical 
expectations based on, for example Standard Model, several thousand Higgs particles would already have been produced, 
although even that is in sufficient number for a discovery, 
since only exclusions of mass regions have so far been found.
However the LHC accelerator as well as the SSC would, if working, produce many 
more Higgs particles in the long run than the Tevatron. 
Thus, it is certainly a possibility that the effect caused backward particles 
to be achieved in the LHC and the SSC canceled in 1993, 
while it was insignificant in case of the Tevatron.

It has also been proposed that the mere observation of cosmic rays with 
sufficient energy for the Higgs particles to be produced even on fixed target 
should represent an argument against the possibility of the effect 
we propose to investigate. 
However, if such an effect existed, 
we predict that the amount of cosmic rays with such 
energies would be reduced by the backward causation effect, one might imagine 
that sources of cosmic rays might be directed so that their radiation is sent 
in the direction of regions with a low density of stars and planets so as to 
avoid the production of Higgs particle on Earth. But we do not have sufficient 
data to measure whether there should --- say 300,000 Higgs particles --- 
be any statistically effect of this type, 
because our understanding of the amount 
of cosmic rays without there being such an effect is insufficient.
\section{Relation between Superdeterminism and our complex action model}
\label{hbn1mn:sec2}
We have already remarked in the introduction that 
in our complex action model the imaginary part of the action determines 
the initial conditions. 
If we denote the complex action 
\begin{equation}
  S\left[\mathrm{path}\right]=
    S_R \left[\mathrm{path}\right]+iS_I \left[\mathrm{path}\right],
\end{equation}
where $S_R$ and $S_I$ are the real and imaginary parts, respectively, 
approximately speaking, among all the solutions to the classical equations 
of motion obtained by ignoring the imaginary part using the approximation 
\begin{equation}
   \delta S_R =0, 
\end{equation}
the solution with the minimal (i.e., most negative) $S_I[\mathrm{sol}]$ is 
that which we are living through (i.e., our current history). 
That is to say the statement in our complex action model for our history of 
the universe to be selected is 
\begin{equation}
   S_I \, [sol\,]\, \mathrm{will\ be\ minimal}.
\end{equation}
The fact that we have such a formula -- wherein the mathematical expression for 
$S_I$ in terms of the fields (development) is very similar and analogous to 
the usual Standard Model action expression, except that the coefficients 
deviate -- for the realized history $sol$ means that even the initial 
conditions (contained in $sol$) are \emph{calculable} in principle, 
although not in practice.
With such a model as ours, in which one can thus calculate ``everything'' 
in principle, one can especially imagine calculating the choice of 
EPS-type experiment that experimentalists 
would perform on particles that were separated in this experiment. 
Let us remind the reader that in an EPS or Bell theorem experiment 
a pair of quanta(=\,particles) are produced in a correlated (entangled) state, 
and these particles successively separate 
and travel run in different directions. 
Two different experimentalists at a distance then attempt to detect 
and measure the properties of the particles. 
It is important that these experimentalists, if they have ``free will'', can 
decide which properties to measure, for example 
the spin component along a particular direction, or the momentum or position, 
\emph{after} the particles are widely separated. 
Of course, in our complex action model, similar to in usual 
deterministic models, this free will is an illusion, 
since using our model we could, in principle, even have calculated what 
quantities the experimentalists decided to measure.

 The problem with quantum mechanics associated with 
Bell's theorem is that under mild assumptions, mainly that no signal 
can travel arbitrarily fast from one measurement place to another 
so as to communicate the experimentalist's choice, the quantum mechanical 
statistical predictions are not consistent. 
 If one takes the point of view that we only need to consider the truly realized 
situation, i.e. choice of experiments used to measure the particles,  
and that we can totally ignore the illusion possibilities associated with 
apparent free will of the experimentalists, then 
Bell's theorem falls away and quantum mechanics can be seen Bell's theorem 
without any problems. 

 Thus, we may logically state that our model even strengthens the application 
of superdeterminism to avoid the problem of Bell's theorem. 
We claim that our model 
strengthens for superdeterminism because it even makes the initial state, 
and thus the deterministically determined decision of the experimentalists, 
calculable, so that the requirement that we should be allowed to vary 
at least the initial conditions in deriving Bell's theorem would 
no longer hold in our model. 

 However, we shall argue in the next section that in spite of 
our model strengthening the grounds for superdeterminism, 
it is in fact another feature of our model that removes 
the problem associated with Bell's theorem.
 In fact, once we have effective backward causation 
--- so that, for instance, the potential switching on of the SSC to produce 
many Higgs bosons can \textit{backwardly} cause 
the Congress of the United States to stop funding --- 
then the rule that a signal cannot move with arbitrarily high speed is 
no longer required.
 The signal could instead move slowly along in the future and then go 
backward in time using the feature of backward causation in our model. 

\section{Analysis of how our model solves the problem of the EPS-Bell theorem}

As Bell himself was aware\,---\,and why he thus did not like superdeterminism as 
a means of resolving the problem of quantum mechanics\, ---\, 
there can be many small details that are 
very difficult to control, and these details can influence experimentalists' 
decisions. Such details indicate, for example, the reasoning in their brains 
and may well represent their ``free will''. We believe that a 
superdeterminism solution that solves the problem of Bell's theorem 
associated with quantum mechanics 
by postulating that even such ``free will'' 
\,--- \,though simulating details that lead 
forward in time to the experimentalist's decisions\,---\, 
can be somehow integrated 
and calculated by the particle choosing its property when measured 
is a somewhat unreasonable philosophy. 
 How indeed should a particle B at site B where a measurement is 
made ``know'' and ``understand'' in detail the contemplations 
of the experimentalists at the other site A\,?\ 
 It sounds more reasonable to assume that such detailed calculations 
of the experimentalists of team A cannot be calculated 
at observation site B. Upon supplementing quantum mechanics with locality, 
etc. by such a reasonable extra postulate, the loophole in the problem of Bell's theorem 
is applied to quantum mechanics would be closed. However, with such an extra reasonable assumption, 
quantum mechanics would be truly in crisis. 
 Our strengthening of quantum mechanics by making the initial state ``calculable'' in principle will 
a priori not help much against the reasonable assumption suggested. 
 Thus, regarding the true superdeterminism solution to the problem of Bell's theorem 
associated with quantum mechanics, our model is not particularly helpful. 

 There is, however, \textit{another way} in which our model may more realistically help to solve the problem of Bell's theorem associated with quantum mechanics. Since the formula for $S_I \mathrm{[history]_{minimal}}$ in the integral form, 
\begin{eqnarray}
   S_I \left[\mathrm{history}\right]
   =\int_{\mathrm{all~times}}L_I \left(\mathrm{history}(t)\right)dt, 
\end{eqnarray}
has contributions from through time, from the beginning to the end of time, it also 
includes contributions from the future. 
 In order that our model has a chance of being viable, we must of course 
hope or speculate that, for example, because of the special conditions during 
times of inflation of the universe, the contributions to $S_I [\mathrm{history}]$ from 
$L_I \left(\mathrm{history}(t)\right)$ for time $t$ in eras of e.g inflation were 
by far the most important, so that what happened in era of inflation 
dominated the selection of which history was to be realized 
(i.e. the one we are living through now). 
Only with such an assumption of the inflation-era contribution to the integral 
$S_I =\int L_I dt$ \ dominating the selection of the initial 
conditions (the solution) will our normal experience agree with the second law of 
thermodynamics, meaning that only the start of the history was strongly organized 
in the sense of having low entropy and essentially nothing being prearranged 
by the fine tuning of initial conditions destined to make future events happen. 
 However, in our complex action model there should be at least some 
attempts to make such prearrangements, 
meaning that in our model there are in principle, 
events that occur that cannot be predicted statistically using conventional theory, so that one 
might denote them as ``miracles'', or ``antimiracles'' in the case of an undesirable event. 
 Usually, however, we expect that the contributions from an era such as the 
inflation era shortly after the ``Big Bang'' (if there was a Big Bang) 
would dominate. 

 When, however, we consider a quantum experiment involving a measurement, 
the result of which can seemingly, independently of the initial conditions, 
produce different measured values with finite nonzero probabilities, 
it becomes plausible that the future contribution 
$\int_{t_\mathrm{exp}}^{\infty }L_I dt$ 
may become important where $t_\textrm{ext}$ denotes the time a which the experiment is performed. 
In our model, we therefore suppose that the outcome of a quantum experiment 
is not a priori purely fortuitous or accidental, but actually depends on 
the (future) contribution to the imaginary part of the action 
$\int_{t_\mathrm{exp}}^{\infty }L_I dt$. 
 That is to say, we expect that the outcome of the measurement is the result 
that minimizes the contribution of 
$\int_{t_\mathrm{exp}}^{\infty }L_I dt$\, 
to
$S_I$ depending on this outcome. 

 If we have, as we now assume about our model, a theory in which 
the outcome of a quantum measurement is selected by minimizing the integral 
$\int_{t_\mathrm{exp}}^{\infty }L_I dt$  
extending into the distant future of the world, 
then the problem associated with 
signals traveling between the sites A and B in EPS-type 
experiments faster than light or arbitrarily fast loses its interest. 
Namely, to avoid problems with Bell's theorem, 
instead of requiring signals faster than light, signals that reach the future 
of the particles being measured are required, because it is the future of the particle 
(roughly speaking) that determines the result of the quantum measurement. 
 Actually, it is not so much the future of the particle itself as the results 
of the measurement propagated into the future 
that affects the integral 
$\int_{t_\mathrm{exp}}^{\infty }L_I dt$ 
to be minimized, which gives us the result of the measurement.
 This picture of measurement results being determined 
by minimizing the future part of the imaginary part of the action $S_I$, 
$\int_{t_\mathrm{exp}}^{\infty }L_I dt$, might be 
roughly described as information first going forward in time, where it 
contributes to $\int_{t_\mathrm{exp}}^{\infty }L_I dt$, then 
causing what we call backward causation, i.e. an effect backward in time 
that determines the measurement results. 
 
 To clarify how our model treats the problem of Bell's theorem, 
we hypothetically imagine that two measurement sites, A and B, 
are kept isolated throughout the future. Then the measurement results, 
determined in our model (in a complicated way) from the future integral 
contribution $\int_{t_\mathrm{exp}}^{\infty }L_I dt$, 
cannot be correlated.
In other words, the strange correlation 
(or any correlation) 
between the measurement results assigning sites A and B have a common 
future, which contradict Bell's theorem can contribute to $\int_{t_\mathrm{exp}}^{\infty }L_I dt$ \,
and thereby enable the minimization of this integral, providing the correlation. 

 This ``explanation'' of the violation of Bell's theorem in our model 
by the future contribution to $S_I$, i.e., 
$\int_{t_\mathrm{exp}}^{\infty }L_I dt$, 
determines the measurement results is superior to genuine 
superdeterminism, because it does not require the complicated contemplations 
of the experimentalist teams to be ``known'' and ``understood'' by any particles. 

 In fact, the usual Copenhagen interpretation (or Born) rule is approximately 
 reproduced in our model by making the approximation 
\begin{equation}
    \mid \mathrm{B}(t)\left\rangle \right\langle \mathrm{B}(t)\mid \sim 1, 
\end{equation}
where ket $\mid \mathrm{B}(t)\rangle$ and its bra $\langle \mathrm{B}(t)\mid$ is 
given by the functional integral over the exponentiated action from the era after $t_{\mathrm{exp}}$ 
\begin{eqnarray}
   S_{\mathrm{after}\,t_\mathrm{exp}} = \int_{t_\mathrm{exp}}^{\infty} Ldt,
\end{eqnarray}
i.e., we previously defined the following expression in a basis consisting of basis vectors 
$|\vec{q}\rangle$, 
\begin{eqnarray}
  \left\langle \vec{q}\mid \mathrm{B}(t)\right\rangle \hat{=} 
  \int_{\mathrm{with~conditions~path}(t)=\vec{q}} 
  e^{\frac{i}{\hbar}S_{\mathrm{after}(t)}(\mathrm{path})} 
  &&D\mathrm{path}. \\
  &&{\tiny (\mathrm{paths~from~path}~t_{\mathrm{exp}}~\mathrm{to}~\infty )} \nonumber
\end{eqnarray}

 We thus see that although everything, even what actually occurs and is 
measured, is superdetermined in our model in the sense that it is 
in principle calculable, the main way in which our model can be claimed 
to solve the problem associated with Bell's theorem is that, by having dependence 
on the future via the integral 
$\int_{t_\mathrm{exp}}^{\infty }L_I dt$, 
information or a signal sent backward in time is obtained to determine the outcome of the measurement.
If such backward signaling is allowed, the locality principle, 
formally implemented by the (complex) action of the form 
\begin{eqnarray}
   S\left[\mathrm{path}\right]=
   \int L\left(\mathrm{path}(x), \partial \mathrm{path}(x)\right)\sqrt{g}dx,    
\end{eqnarray}
where the Lagrangian density 
$L\left(\mathrm{path}(x), \partial \mathrm{path}(x)\right)$ 
only depends on the field development called the path in the infinitesimal 
neighborhood of the space-time point $x$, i.e., on $\mathrm{path}(x)$ 
and its first derivative $\partial \mathrm{path}(x)$, 
can still be compatible with arbitrarily fast information transfer. 
It is by means of backward causation via the 
$\int_{t_\mathrm{exp}}^{\infty }L_I dt$ dependence that our model 
--- in a reasonably plausible way --- circumvents one of the assumptions 
behind Bell's theorem and thus its problem associated with quantum mechanics. 
\section{The transactional interpretation}

There is another proposal regarding the interpretation of quantum mechanics, 
is even more similar to ours than the previously discussed superdeterminism, 
and that is the transactional interpretation. 
In fact, the transactional interpretation has an interesting 
common feature with our model: formal influence from the future. 

In the transactional interpretation, this formal backward causation or 
influence from the future is clearly alluded to by the fact 
that the transactional interpretation has such effects in as far as 
it is based on the Feynman-Wheeler electrodynamics. 
In the Feynman-Wheeler theory of electrodynamics, the usual boundary 
conditions used to derive electromagnetic radiation in terms of 
retarded waves are replaced by a time-reversal-invariant boundary condition.
This Feynman-Wheeler postulate is that an electrically charged object 
sends out both a retarded contribution \textit{and} an advanced contribution to electromagnetic fields so that the total emission is time-reversal-invariant. 
This means that, formally, fields propagate \textit{both} backward 
and forward in time. Thus, Feynman-Wheeler theory formally has an influence 
from the future built into it. It is nontrivial to argue on the basis of their theory that, in practice, we seemingly obtain only the retarded waves, and the 
argument is invalid in some cosmological scenarios. 
Namely the argument is based on a discussion in which an absorber of the light is 
strongly needed. 

Therefore, when the transactional interpretation is based on the effect of charged matter on the wave 
function for the photon (essentially the electromagnetic field) 
in the way proposed by Feynman and Wheeler 
it appears a priori as if the transactional interpretation 
contains an influence from the future. 
However, it is claimed by proponents of the transactional interpretation 
such as Cramer 
\cite{hbn1mnCramer}
that one can distinguish strong and weak principles 
of causality. The weak principle of causality, which merely claims that a 
cause shall come before its effect in the case of macroscopic observations 
and observer--to-- observer communication. 
However, Cramer claims that at present there is no present experimental 
evidence for a 
causal principle stronger than the weak principle. 
This interpretation even allows the possibility of backward causation on the microscopic level, 
since strong causality does not hold. 
Another point suggesting the validity of our model of complex action is the occurrence 
of \textit{two} wave functions: OW (offer wave) and CW (confirmation wave). 
This occurrence has similarities to the $\left\langle q|A(t) \right\rangle$ and  
$\left\langle B(t)|q \right\rangle$ wave functions in our functional 
integral based on the complex action model, defined by the following ``half time" 
functional integrals: 
\begin{eqnarray}
   \left\langle q|A(t) \right\rangle =
   \int_{\mathrm{{with~boundary \atop conditions} \atop q'=path(t).}} 
   e^{\frac{i}{\hbar} \int_{-\infty \mathrm{(beginning)}}^{t} \mathrm{L(path,\partial path)}dt} \mathrm{Dpath_{(half)}},
\end{eqnarray}
and
\begin{eqnarray}
   \left\langle B(t)|q' \right\rangle =
   \int e^{\frac{i}{\hbar} \int_{t}^{\infty } 
   \mathrm{L(path,\frac{dpath}{dt}}dt} \mathrm{Dpath_{(half)}}.
\end{eqnarray}

\subsection{Further review of the transactional interpretation}
As far as we understand, the point of the transactional model is that 
echoes of advanced waves responding to retarded and advanced waves considered 
in a pedagogical way finally lead to a total field that obeys the following. 
\begin{itemize}
  \item[a)] The usual type of boundary conditions of no waves before emission and no waves after absorption.
  \item[b)] Some quantization conditions, for example, that the energy is supposedly given by a Planck quantization rule.
\end{itemize}

We must consider the fact that the field in the nonzero region becomes concentrated 
along a narrow track in space(time) connecting the emitter to the absorber. 
If this is a correct interpretation of the transactional model, then 
the direction of motion of the emitted photon is, from the start, geared 
to reach the absorber. 
However, this means that it is indeed strongly influenced by the future. 
This is, of course, expected in a model based on backward causality 
containing Feynman-Wheeler theory. This means that, at least in principle, 
the influence from the future has been incorporated into the transactional interpretation scheme. 
However, it may turn out to not exist macroscopically in the scheme. 
\subsection{Is our complex action model equivalent to the transactional interpretation?}
Although both our complex action model and the transactional interpretation 
model are characterized by an influence from the future, 
they are not exactly the same, since we have different details regarding the influence 
from the future. In fact, in principle there exist a series of parameters 
in our model in the form of the imaginary part of the action 
$S_I =\int L_I dt$ to be chosen before we have a definite model, while in 
the transactional interpretation model one uses the Feynman-Wheeler time-reversal-symmetric emission rule ($\sim$ boundary condition) to determine 
\textit{how} the future influences the past.

However, in a general way we may obtain a very close correspondence between the two models. 
Presumably the best way to obtain such a correspondence 
is to apply the second quantized theory in the language of field theory in the 
functional integral taken as a fundamental principle in our model.

That is to say, we take our abstract ``path" to represent possible development 
of all the fields (for simplicity we only consider bosonic fields 
$\psi (X^{\mu })$). 
This means that the phase space --- in our model --- 
is a space of infinitely many dimensions, the coordinates of which are 
partly the fields $\psi_i (\vec{X})$ and partly their conjugate fields. 

A crucial feature in our complex action model and 
the use of an action integral over all time (including all the past and 
all the future) is that the initial conditions, or rather the solution to 
the equations of motion, determine the field (i.e. the field is in principle calculable). 
This classical solution, obtained by minimizing the imaginary action 
$S_I$, describes a path throughout time. 
It thus even makes it possible in principle to calculate the outcome 
of quantum experiments (in our complex action model). 
Considering this model with the fields as variables describing 
the path, we can thus use our model to obtain --- up to a few small splittings of 
the track --- (in principle but not in practice) a classical 
solution to the field development. 
However, such a classical field development is from the viewpoint of quantization 
a specific development of the wave function. 
This wave function can now be considered, as it is in the transactional 
interpretation, as simply an ordinary (i.e., well-defined classical) field! 
In this way, our complex action model, considered as a theory 
of fields, provides exactly the same picture as the transactional interpretation. 

\subsection{How does single quantization occur about?}
When making such a wave function or, more precisely, when the fields become 
classical solutions, one might wonder how to 
quantize the energy of a photon in terms of $\hbar \omega $, where $\omega$ 
is the frequency.

In our model unless we allow there to typically be 
not only selected classical solutions, but also a discrete 
series of solution close to each other. 
In the case of a photon being transmitted from an emitter to an absorber 
over a long (space) distance, these similar but different 
classical solutions. which are relevant and contribute to our functional integral, 
will be solutions within a number of oscillations 
of the field from the emitter to the absorber. However each extra 
oscillation of the field will give an extra phase factor in our ``fundamental" 
functional integral. These different ``neighboring" routes will only add up 
constructively provided the total phase difference between the contributions 
from the different classical (field) solutions happen to be 
(at least approximately) zero. 
Such a condition for constructive interference between contributions in the 
Wentzel-Dirac-Feynman integral from slightly deviating 
classical solutions could lead to a quantization rule in a single quantized 
language.

It appears that in the transactional interpretation the quantization of energy 
and momentum is imposed as an extra condition without any explanation behind it. 

This would correspond to our model if one could carry out quantization without 
having our functional integral at a more fundamental level. Namely, 
the phase factor from the functional integration 
would  have no place in the picture. One would have to include  
as a kind of Bohr quantization condition as an addition. 

\subsection{What can be concluded from the close connection of our model with the transactional interpretation?}
One can consider the close coincidence of our complex action model 
with the transactional interpretation from two opposing viewpoints: 
One can accept transactional interpretation and show that it is a model 
``of our type", thus supporting such models. Alternatively one can 
accept our model and say that can be used to derive not exactly the conventional transactional interpretation model, but a transactional-interpretation-type model. 
The latter does not necessarily have the Feynman-Wheeler specific way of sending 
equal-strength waves backward and forward in time. But in the model 
the important 
``philosophical" aspects would be exactly the same: The wave functions 
(in the single-particle picture) could be considered as ordinary ($\sim $classical) 
fields, and there would be an influence from the future so that a particle 
would be guided in the right direction from the start as in, for example, 
Renninger's negative-result experiment.


\section{What we need to solve the EPR problems}

Even though it is not so much the superdeterminism in our model (in the 
sense of everything being calculable in principle) that makes it 
compatible with Bell's theorem 
and quantum mechanics as its lack of information 
when only going forward in time 
that causes the compatibility of our model with quantum mechanics. 
Locally our model is nevertheless supported by the problems of quantum 
mechanics. 

 We can generally state that any theory with backward causation 
such as our model would clearly be potentially able to circumvent the problems of Bell's theorem 
by forward and backward motion in time, allowing an effect/a signal to effectively travel 
faster than light. 
Such potential theories with backward causation would be able to solve 
the problems of Bell's theorem. This type of ``theory'' can be claimed 
to be supported by the problems of Bell's theorem associated with quantum mechanics. 
 This fact makes it particularly important to look for any backward causation 
effects whenever possible. Since we so far have only 
weak, if any, evidence for even very rare prearranged events, it appears that 
daily life should exhibit extremely little backward causation according to any 
viable physics theory. However, the physical systems with much higher energy per particle than 
in daily life, we may have looked less carefully for prearranged 
events (miracles).
It is therefore suggested that, e.g., to look for a possible way 
of avoiding the problems associated with Bell's theorem, one should investigate  each new accelerator for 
prearranged events. 

 If a prearrangement governing the world (e.g., via the initial conditions) 
was made to arrange or avoid the occurrence of some phenomenon 
due to high-energy accelerators, the easiest 
(least miraculous) way to arrange or avoid such a phenomenon might be 
to favor or disfavor the building of the accelerator itself. 

 As an example of something disfavored by our complex action model, 
it could be that there is a special type of particle which, if produced will 
contribute to, for example, reducing the probability that the accelerator produces. 
If so, then an accelerator producing this type of particle 
\,---\,particularly if in large numbers\,---\, 
should be prearranged so as not to operate for a long time in the mode producing 
many such particles. In our model, we suggest that a type of particle 
that has such an advance effect thus leading to bad luck in the running of 
accelerators is the Higgs boson, because we consider that the term 
$\ldots+m_h^2\mid_I \cdot \mid\phi_{H} \mid^{2}+\ldots $  
in the imaginary part of the Lagrangian density 
\begin{eqnarray}
   L_I (x)=\ldots+m_h^2\mid_I \cdot \mid\phi_{H} \mid^{2}+\ldots 
\end{eqnarray}
is dominant from a dimensional argument. 
 The imaginary part of the Lagrangian density $L_I (x)$ is, of course, 
the space-time density for the imaginary part of the action 
\begin{eqnarray}
   S_I \left[\mathrm{path}\right]=
   \int_{\mathrm{over~all~space~time~incl.} \atop \mathrm{past}~\textit{{and}}~\mathrm{future}}
   L_I \left(x,\mathrm{path}(x)\right)\sqrt{g}\,d^4 x.
\end{eqnarray}

 Our dimensional argument is that if the natural units was the Planck unit, 
the natural value for the quantity \,$m_h^2\mid_I$, having the dimension of mass square, 
would be the square of Planck mass $m_{\mathrm{PL}}\sim 10^{19}GeV^{2}$, i.e., 
$m_h^2 \mid_I \sim \left(10^{19}GeV\right)^2 \sim 10^{38}GeV^{2}=10^{32}TeV^{2}$, 
which is tremendously large from the viewpoint of LHC physics. 

 If an accelerator indeed has the potential of producing many of these 
bad-luck-producing new particles, we might observe it by statistically 
investigating 
whether the accelerators have had good or bad luck technically and politically. 
 Here, the reader should immediately think of the world's largest 
accelerator the SSC which was canceled in 1993 by Congress.

 As we have already suggested in earlier papers, it might be difficult to perform 
a clear statistical investigation of the potential bad luck unless one carries 
out a very clear experiment involving betting in a card game preferably combined with quantum 
random numbers to decide whether a certain accelerator 
which, of course, we propose to be the LHC should operate and at what luminosity 
and energy.   

\section{A card game to determine LHC restrictions can only be a success!}

If such a card game is executed, two possibilities may occur.
\begin{enumerate}
\item[1)] A card combination of the most common type is drawn, leading to no restrictions.
      Then the LHC can operate without any restrictions which is a positive result happy 
      because an argument against our theory has been found at close to zero expense.
      Our theory has been disproved, 
      or at least the probability that it is correct has been drastically reduced. 
      This is a very good scenario!
\item[2)] A card combination corresponding to a restriction is drawn. 
      Although, it is a significant loss that the LHC cannot operate full capacity, 
      our backward causation theory or a similar theory has been proved.
      If backward causation really exist, 
      it would be so interesting that it might be counted as a greater discovery than that of finding supersymmetric partners or the Higgs boson.
      This would be a fantastic discovery made with the LHC! 
      Moreover the restriction drawn did not involve the total closure of LHC, 
	it would be likely that 
      the Higgs boson and perhaps the supersymmetric partners 
      would be found quickly even if 
      the statistics might initially be slightly worse than hoped for. 
\end{enumerate}

 It would be a wonderful achievement for CERN and the LHC to find 
\textit{back\-ward causation} 
by simply obeying probably very mild restriction.
 We should remember that the rules of our card game should 
result in a milder restriction having a much higher chance 
of being drawn than a very strong restriction such as the closure of the LHC.

There is, however, a small chance of 
a true loss, even though it will not be initially noticed. 
It is possible, although not likely, 
that a random number game will lead to a restriction even if our model, 
and any model with backward causation, is incorrect.
In this case, we would have had a bad bargain:
not only would we lose the full applicability of the LHC 
but we would also have obtained, by a statistical fluctuation, the wrong impression 
that a backward-causation-containing model was indeed true without 
this actually being the case. 
We should certainly arrange probability of a restriction $p$ 
to be sufficiently low 
to make this undesirable case extremely unlikely.

One would, according to this argument, initially suspect 
that it would be more profitable not to perform our random number 
LHC-restricting experiment 
because
if our theory was correct the LHC would, in any case, be closed or restricted somehow 
by prearranged bad luck,
which happened to the SSC, 
for which Congress in the U.S.A. terminated economic support.
However, 
we argue that it would be more agreeable to have LHC stopped 
or restricted by a random number game rather than by an unfortunate event 
such as the political withdrawal of support. 
The main reason why artificially caused random number withdrawal is 
preferential is that we would, in this case, obtain more solid support 
for our, or a similar, model being true than for the same restriction 
occurring through another event.

To see that 
the truth of our theory of imaginary action determined 
by history  would be more convincingly shown 
if we had a card or random number closure rather than a ``normal'' failure, 
we could contemplate
how much more convincing our theory would be today 
if the SSC had been closed after a random number experiment 
rather than mainly for economic reasons or perhaps because of 
the collapse of the Soviet Union, 
which made the competition to build large accelerators 
costing 60 billion dollars no longer worthwhile. 

It is sometimes explained that the SSC \cite{hbn1mn12} 
encountered bad luck because of various mistakes or accidents, 
but had its bad luck been due to a card game, such ideas would not matter. 
Everything is an accident, 
but we would know the probabilities of bad luck very reliably. 
Thus, if a card game had been set up so that the probability of closing was 
sufficiently small, we would have been sure that the closing of 
the SSC was due to a (anti)miracle. 

In the following, we shall present a numerical example to 
illustrate formally that a more reliable knowledge of the truth of 
our theory is obtained by a random number experiment. 
This comes under the discussion of point 2 among the reasons for conducting 
our proposed experiment in the following section.

\section{Reasons for conducting our proposed experiment}

The reasons for conducting the card game experiment are as follows, 
\begin{enumerate}
\item[1)] To obtain greater conviction about the truth of our theory, provided it is true of course.
\item[2)] To possibly avoid the advance effects of backward causation. 
\end{enumerate} 

In formula we should estimate averages for measuring these two benefits. 

\subsection{Greater conviction of the truth of our model} 

For reason 1)\,---\,the conviction that our theory 
is indeed correct\,---\,we need some measurements. 
Both the result of the card game and the failure of the LHC 
for other reasons are statistical events. 
However, while we have very reliable ideas about what probability 
$p$ 
to assign to a given class of card combinations, 
our assignment of a reliable value probability of failure 
$f$ 
for other reasons is very difficult and has a huge uncertainty. 
Therefore, if LHC fails for a reason other than a random number game, 
we could still not be truly convinced that our theory is correct 
even though we could comment on the 
remarkable fact that we had written about the failure 
while it appeared that the LHC was still likely to operate. 

\subsection{Miraculousity and estimating the amount of evidence for our model}

Why the difference between obtaining support for 
our model by a ``natural'' failure of the LHC and a by failure 
caused merely by a ``restrict the LHC'' 
card being drawn in a card game, 
gives rise to an important difference in reliability 
in our model? 
 To understand that we shall give a slightly formal illustration using a statistical model 
that is not very precise but is appropriate for illustrating our point. 

 If, in our model, the failure of the LHC 
occurs merely through a series of small coincidental bad-luck events that 
can easily happen, 
then the number and unlikeliness of elements in this series of 
bad luck events must be proportional to 
$  - \hbnmnlnf = | \hbnmnlnf| $
, where
$f$
is the probability of failure. 
We could call this quantity
$ - \hbnmnlnf $ 
the ``miraculousity'' of failure in a seemingly natural way. 
This concept of 
miraculousity
is a measure for how many 
``submiracles''
must occur. 
Examples of submiracles, include  
``the person on watch has drunk a bit too much'', 
``the connection between superconducting cables 
  having too high resistance'', 
``The accident is in a relatively inaccessible part of the tunnel directly 
under one of Jura mountains''.

Now if we set up a card or quantum mechanically based 
random number generator so that 
``restrict LHC''
occurs with probability  $p$ , 
it must generate 
--- by the selection of the realized history in our model ---
a number of adjusted accidents (or submiracles) 
proportional to 
$ -ln\, p =|ln\, p| $ .
Essentially, if our theory is true, whether the failure 
of the LHC will arise via a card or a quantum random number game 
or via a natural reason will depend on which of the two alternative 
miraculousities, 
$ -ln\, p $
or 
$ -ln\, f $, 
is the smallest. 
There will, of course, be a preference for lower miraculousity: 
the less miraculous of the two alternative set of events 
leading to failure will 
be more likely to occur. 
This would require fewer submiracles. 

We can define $f$ so that indeed 
$-ln\, f$ 
gives a measure of the miraculousity, but it is very difficult, 
even for people building the LHC, to convincingly determine 
what to accept or predict about this miraculousity defined by 
$ -ln\, f =|ln\, f| $ .
At best, one can predict it with an appreciable uncertainty. 
That is to say, we obtain, from some simulation 
--- say by a Monte Carlo method or  
simply theoretically --- a probability distribution for 
miraculousity $|ln\, f|$. 
 To illustrate our point of estimating the degree of 
conviction obtained in the case of a 
``natural'' and\,/\,or ``normal'' 
failure, we can assume that the calculation of probability by (computer) 
simulation of the political and technical procedures related to 
CERN and LHC lead to a Gaussian distribution for the miraculousity 
$-ln\, f$.
That is to say, we assume the probability distribution
  \begin{eqnarray}
     && P\left(| ln\, f | \right) d\,|ln\, f|  =  \nonumber \\ 
     && \approx \frac{1}{\sigma \sqrt{2\pi  }}\> \mathrm{exp} 
        \Big( \frac{1}{2\sigma^2}\left(|ln \, f|-\ |ln \, f_0 |\right)^2 
        \Big) \ d\,|ln\, f|.\nonumber \\ 
      \label{hbn1mnGauss}
  \end{eqnarray}
Here, $\sigma$ is the spread of the distribution for the logarithm of 
$f$ , i.e., 
the miraculousity. 

Now let us consider the degree of remarkableness of the failure 
depending on whether it is due to the card 
or quantum random number game 
or to a ``normal'' failure, i.e., other reasons 
such as a meteor or a bad electrical connection between the superconductors. 

In the case of a card or quantum random number game, the number of 
sub-miracles in the card or quantum packing is proportional to 
$-ln\,p$, 
where $p$ is the arranged probability say by the game rules. 

However, if there is instead a ``normal'' failure due to, for example, 
the decision of some 
members of a cabinet, then we will tend 
to believe that the true miraculousity 
$-ln\, f = |ln\,f|$
for that failure is indeed at the low end of the estimated Gaussian 
distribution.
In other words, we would expect that the ``true'' probability 
for failure $f$ is rather high, i.e., 
$f>f_0$, 
or presumably even 
$f \gg f_0 $. 

Let us evaluate the expected probability for a seemingly 
``normal'' failure, i.e., not caused by a card game, etc. 
This expected normal probability of failure is 
\begin{eqnarray}
\left\langle f \right\rangle
  =\int ^{\infty }_{-\infty }\frac{1}{\sigma \sqrt{2\pi }}\cdot f 
   \cdot \mathrm{exp} \left(-\frac{1}{2\sigma ^2}(ln\, f-ln\, f_0)^2 \right)
d\,|ln\, f|
\end{eqnarray}
(We imagine that the miscalculation due to including the region 
$f>1$ is negligible, but one could of course improve accuracy if needed). 

  We write 
$f=e^{-|ln\,f|}$. 
We had hoped to expect ``normal'' failure to occur 
with the probability 
\begin{eqnarray}
\left\langle f \right\rangle 
    &&=\int ^{\infty }_{-\infty }\frac{1}{\sigma \sqrt{2\pi }}
       \cdot \mathrm{exp} \left(-\frac{1}{2\sigma ^2}(|ln\, f|-|ln\, f_0|)^2 
       -|ln\, f| \right)d\,|ln\, f| \nonumber\\
    &&=\int ^{\infty }_{-\infty }\frac{1}{\sigma \sqrt{2\pi }} \cdot \mathrm{exp} 
       \biggl(-\frac{1}{2\sigma ^2}\left[ \left(|ln\, f|-|ln\, f_0|
       +\sigma ^2\right)^2  \right. \nonumber \\
    && \qquad \qquad \qquad \qquad \qquad \qquad \qquad \quad      
         \Bigl. -\sigma^4 +2\sigma^2 |ln\,f_o| \Bigr] \biggr) d\,|ln\, f| \nonumber\\
    &&=\mathrm{exp}\left( \frac{\sigma^2}{2}-|ln\, f_o|\right) =f_o\,e^{\sigma^2 /2}. 
\end{eqnarray}

 Hence, the remarkability or apparent miraculousness of the 
outcome that the LHC should fail seemingly by a ``normal'' accident 
\,---\,such as a political decision\,---\,
is not the ``miraculousity'' corresponding to the most likely value for 
$f$, i.e., $-ln\,f_o =|ln\, f_o |$, but rather to the 
$\mathrm{``remarkableness"}=-ln\left\langle f \right\rangle =|ln\left\langle f \right\rangle\!|=|ln\,f_o|-\frac{\sigma^2}{2}$. 

It is this correction by the term 
$-\frac{\sigma^2}{2}$ 
that results in there being less conviction of our model being true 
if the failure of the LHC appears to be due to a ``normal'' failure 
than if a failure is caused by a card or quantum random number game. 
One should keep in mind that, according to our model, 
which of the two types of reason for failure mainly depends on the relative sizes of 
$-ln\, f$ and $-ln\, p$ . 

 In this way, our theory would be more convincing 
if the failure of the LHC was found to be due to a card game than by a 
``normal'' failure.
 It would thus be scientifically profitable if we could provoke a 
card game failure instead of a ``normal'' failure; we would have 
the possibility of determining whether our model was correct. 
 In the case of our model being wrong, the card game 
would only add to the total probability of failure of the LHC, 
making a card game a risk and an unprofitable action.

 Should our theory be right, the failure of the LHC would be guaranteed 
to have $\frac{2}{3}$ probability, and in that case, the probability of total 
failure would not change greatly whether we perform 
a card game or not.
 In that case, we would simply transfer some of the probability of failure 
from ``normal'' failure to failure due to the card game or a similar exercise.

If we place an economic value on the degree of confidence 
we would obtain if our model was indeed true, depending on 
whether one failure or another actually occurs, 
we could express this benefit in the form 
\begin{eqnarray}
  b_{1)}&&=c\cdot \mathrm{``remarkableness''} \nonumber \\
        &&=c\cdot
        \left\{ \begin{array}{l}
           |ln\,p| \ \mathrm{if\ game\ failure} \\
           |ln\left\langle f \right\rangle\!|=|ln\,f_o|-\frac{\sigma^2}{2}
           \ \mathrm{if\ ``normal''\ failure}. 
        \end{array} \right.  
\end{eqnarray}

In the case of our theory being correct, which occurs with probability 
$r$, we estimated that the LHC would be stopped with $\frac{2}{3}$ 
probability \cite{hbn1mnsearch}; Hence this benefit can be calculated as an average, 
\begin{eqnarray}
  \left\langle b_{1)}\right\rangle 
     && =c\cdot \mathrm{``remarkableness''} \nonumber \\
     && =c \left\langle \Biggl( \frac{p}{f+p} |ln\,p|+\frac{f}{f+p}
         \left(|ln\, f_o |- \frac{\sigma^2}{2} \right) \Biggr) r\frac{2}{3}
         \right\rangle_{\mathrm{Gauss}},
\end{eqnarray}
where the average $\left\langle \cdots \right\rangle_{\mathrm{Gauss}}$ 
is simply the average over distribution~\ref{hbn1mnGauss} .

 For instance, in the limit of a very small probability $p$ assigned to 
a random number restricting the LHC, we would obtain 
\begin{eqnarray}
\left\langle b_{1)}\right\rangle \approx 
    c\left(|ln\, f_o |- \frac{\sigma^2}{2} \right) r\cdot\frac{2}{3}
    +cp\left\langle \frac{1}{f} \right\rangle \left(|ln\,p|-|ln\,f_o | +
    \frac{\sigma^2}{2}\right) r\> \frac{2}{3}+ \ldots . \label{hbn1mnlowp}
\end{eqnarray}
If, on the other hand, we set $p\gg \left\langle f \right\rangle$, 
we would obtain 
\begin{eqnarray}
\left\langle b_{1)}\right\rangle \approx  \Biggl(|ln\,p|+
\frac{\left\langle f \right\rangle}{p} \left(|ln\,f_o |- \frac{\sigma^2}{2}
-|ln\,p|\right) \Biggr) r\cdot \frac{2}{3}. 
\end{eqnarray}

It is important to notice that, as the previous discussion suggested, 
the correction term in \ref{hbn1mnlowp}
will, for sufficiently small $p$, give increasing $p$ benefit with increasing; 
$p$ 
so it would be beneficial in terms of this benefit 
$b_{1)}$ to attain an increase in the confidence of our knowledge 
that $p$ is not zero in our model.

\section{Avoiding undesirable backwardly caused events}

In our earlier paper, 
in our estimates of whether it would be profitable to perform our card game 
or random number game experiment, we included the consideration 
that if we indeed have backward causation resulting in the LHC becoming inoperable, 
then these prearrangements may have adverse side effects 
and, a priori, perhaps also possible. 
 The side effects of backward causation might end up being huge in much the same way 
as the famous forward causation effect of the butterfly in the 
``butterfly effect''. However, in the same way that it is difficult to predict 
whether the effects are good or bad when the butterfly beats its wings 
in a particular way, 
it is difficult to know if the pre-arrangements set up to prevent the LHC  
from working are good or bad.
 If we consider possibilities such as the closure of CERN or an earthquake 
in Geneva, 
we may judge the effect to be bad, but if we consider even earlier or more distant 
prearrangements, it becomes increasingly difficult to estimate 
whether the effect is good or bad.
For instance, it is possible that a major factor behind the SSC
being terminated by Congress was 
the collapse of the Soviet Union \cite{hbn1mn13}. 
 This was a huge backward causation effect but it is difficult to evaluate 
it being good or bad. 
 Thus, it would have been difficult to evaluate in advance whether 
our card game would have been profitable had our theory been known 
at the time of the collapse of the Soviet Union.

In the previous articles \cite{hbn1mnsearch}, 
we defined $d$ as the cost of the excess damage arising 
when a ``normal'' failure of the LHC occurs. 

We imagine that huge backward causation effects occurring 
very remotely  
from the LHC are probably averaged out to zero, 
similar to the effects of the butterfly wing in the distant future. 
 Hence, the important contributions to the cost of damage 
 $d$ are close in time (and space) to the LHC itself. In our previous study
 we very roughly estimated that 
$d \approx 10 \cdot $ ``cost of LHC'' $\simeq 10 \cdot 3.3 \cdot 10^9$ CHF
$=3.3 \cdot 10^{10}$ CHF. 

 In the case of failure due to a card game, there may also be huge effects, 
but the evaluation of whether the overall effect is good or bad would be 
completely opaque.
 Only the performance of the actual experiment may have a predictable 
average effect. Therefore, in the case of such an artificial failure, 
the effect is equally likely to be good or bad and the obvious loss 
because of the  restriction imposed by the  
$d_\mathrm{\,rest.loss}$ card that was drawn. 

We should ensure that the latter damage is almost certainly minimal 
by assigning mild restrictions to be much more likely outcomes than 
heavy restrictions. 

The damage done or, by switching the sign, the (negative) benefit, is 
\begin{eqnarray}
-b_{2)}=d\cdot\frac{2}{3}r\cdot \frac{f}{f+p}+d_\mathrm{\,rest.loss} \cdot
\Biggl( p\left(1-\frac{2}{3}r\right)+\frac{2}{3}r\cdot \frac{p}{f+p}\Biggr) , 
\end{eqnarray}
where we used the notation $d_\mathrm{\,rest.loss}$ for the cost of the restrictions. 

\section{Conclusion and outlook}

In this article we have discussed two major topics in connection with
our previously proposed model in which the action is assumed to be complex.

The first of these topics could have been considered starting from the 
problems of EPS problem as on page~\pageref{hbn1mn:intro} 
the Bell's theorem, which states that quantum mechanics makes predictions
in the case of entangled particles being measured that are in 
disagreement with seemingly very reasonable assumptions.  There is, however, 
as noticed by Bell himself, a resolution if one
makes use of the fact that, for given initial conditions, the measurements 
performed by 
experimentalists at two significantly separated positions A 
and B in the Bell or EPS experiment are already in principle 
determined by the determinism of at least classical approximation.
This deterministically determined choice of the experiment being performed
makes the need to discuss several simultaneously possible 
choices ( by ``free will'') irrelevant. Regarding our complex action 
model, this point should be emphasized since the initial conditions are 
even, in principle, calculable.

However, we believe that it is NOT this true superdeterminism that 
makes our model with the complex action more able to cope with the problem of 
Bell's theorem, but rather the fact that our model predicts that
\emph{the measurement results depend on events in the future}!
It is this backward causation property of our model that makes
the assumption of no signal traveling faster than the speed of light
a prerequisite for Bell's theorem, unreliable in our model.
The point is that if the future can influence the past by adjusting 
the initial conditions, or in this case, by having a relevant influence
on the outcome of a measurement, then a genuine signal traveling at a velocity 
less than speed of light from A to B is not needed. Instead we can 
have an effect from the future that is influenced by a signal from
A. However, if one can wait for a signal to arrive from the future, 
there is no need for the signal to travel faster than the speed of light, 
because the signal has sufficient time to reach the future.
 Actually we found that our model essentially reproduces a second quantized 
version of, in principle calculable, classical fields that can be identified 
with wavefunctions including echoes from the future in the transactional 
interpretation.
In this respect, our model is essential by identical to the transactional 
interpretation model, although we do not have the exact Feynman-Wheeler 
time-reflection-invariant emission. Rather, the influence from the future 
in our model is determined by parameters in the imaginary part of the action.

Thus, we claimed that the problem with Bell's theorem requires an influence
from a future effect, and thus one should attempt to look for such
backward causation whenever some new field of physics is being 
explored. Using our special model of complex action, the obvious place to 
look for such effects at the present time is in the highest-energy 
accelerators. Thus we should look for such effects 
in the LHC.

We have argued that it would be profitable to perform 
our previously proposed experiment involving the generation of some random numbers 
\,---\,by a drawing card or by a quantum random number generator, 
or even both ways,\,---\,
and letting them determine whether restrictions should be applied to the beam energy, 
the luminosity and/or other parameters. 

The main point was that our theory, referred to as 
``the complex action model'', 
is indeed shown to be correct if the LHC is 
stopped by our proposed game rather than if it failed for some technical 
or political reason. 
The reason for the suggesting that our model is correct if the LHC was 
stopped by decision based on a random number or card game other than by a 
``normal" technical or political failure is that it is very difficult to estimate 
in advance how likely it is for a ``normal'' failure of the LHC to occur. 

The greatest encouragement for performing the experiment in the near future 
is that  
whatever happens as a result of our proposed experiment, the experiment should 
either be a success or cause no harm. 
The point is that in the case of restriction being imposed 
by random numbers, we have, because of the very fact that 
these random numbers were generated, obtained the shocking and 
monumental discovery that ``backward causation'' exists. 
Such a discovery of the future influencing the present and past would 
be monumental. 
Consequently, it would be a fantastic success 
for the LHC to have caused such a discovery!
\section*{Acknowledgements}
The authors wish to thank the Niels Bohr Institute and 
Okayama Institute for Quantum Physics for the hospitality 
extended to them. This work was supported by Grants -- 
in-Aids for Scientific Research, No. 19540324 
and No. 21540290 from the 
Ministry of Education, Culture, Sports, Science and Technology, Japan. 



\cleardoublepage
\thispagestyle{empty}
\vspace*{5cm}
{\bfseries \Large Discussion Section I}\\[1cm]
\addcontentsline{toc}{chapter}{Discussion Section I}
All discussion contributions are arranged alphabetically with respect to the first author's name.

\newpage

\cleardoublepage
\title{Is the Prediction of the "Spin-charge-family-theory" 
that the Fifth Family Neutrons Constitute the Dark Matter in Disagreement with the 
XENON100 Experiment?}
\author{G. Bregar${}^1$, R.F. Lang${}^2$ and N S. Manko\v c Bor\v stnik${}^1$}
\institute{%
${}^1$Department of Physics, FMF, University of
Ljubljana, Jadranska 19, SI-1000 Ljubljana\\
${}^2$Physics Department, Columbia University, New York, New York 10027, USA}

\titlerunning{Is the Prediction of the "Spin-charge-family-theory"\ldots}
\authorrunning{G. Bregar, R.F. Lang and N.S. Manko\v c Bor\v stnik}
\maketitle
 
\begin{abstract}
This discussion is  to clarify what can be concluded from the so far analysed experimental data obtained 
in the XENON100 experiment~\cite{blmbDAprile:2010um} about the prediction of the ref.~\cite{blmbDgn}, which states:
If the DAMA/LIBRA~\cite{blmbDDAMA} experiment measures the fifth family neutrons predicted by the 
"spin-charge-family-theory" (of the author N.S. Manko\v c Bor\v stnik~\cite{blmbDnorma,blmbDpikanorma,blmbDNF}) 
then new  direct 
experiments will in a few years measure the fifth family neutrons as well.
\end{abstract}

\section{Introduction}
\label{blmbDintroduction}
\label{contribution:gbrlsnmb}
The theory {\it unifying spin and charge and predicting families} 
(the {\it spin-charge-family-theory} proposed by N.S. Manko\v c 
Bor\v stnik~\cite{blmbDnorma,blmbDpikanorma,blmbDNF}) predicts the stable fifth family with masses 
of the family members of around a few hundred TeV/$c^2$. The authors G. Bregar and 
N.S. Manko\v c Bor\v stnik analysed in
~\cite{blmbDgn} 
the properties of the members of this family 
in the evolution of Universe up to today, evaluating also their scattering on the 
first family nuclei in Earth and in particular in direct measurements. They conclude: 
The fifth family quarks i. with the masses at around a few hundred TeV/$c^2$ and ii. with
the mass difference between  the $u_5$ and $d_5$ quark not larger then a few hundred GeV,
would decouple from the primordial plasma, forming the fifth family neutrons and antineutrons. Today this   
would constitute  the main part of the dark matter.  If  the DAMA/LIBRA experiment~\cite{blmbDDAMA} 
is measuring the fifth family neutrons, then other direct experiments will in a few years 
confirm the fifth family neutrons as the dark matter constituents.

The CDMS~\cite{blmbDCDMS} experiment has not observed so far any dark matter signal. Also  the XENON100 
experiment~\cite{blmbDAprile:2010um} has not observed any signal which it could be interpreted as 
the dark matter signal. 
In the ref.~\cite{blmbDgn}, the comparison between the DAMA/LIBRA and CDMS experiments 
are done and the conclusion made, that these two experiments are not (at least yet) in disagreement. 

During the Bled Workshop "What comes beyond the standard models" the discussions with 
one of the authors of the XENON100 experiment took place after the talk of N.S. Manko\v c Bor\v stnik, 
in which she reviewed the {\it spin-charge-family-theory} and the  predictions of this theory, 
which follow from the studies done so far. This contribution is the review of the discussions after 
her talk on the subject of the direct measurement experiments and the prediction of 
the {\it spin-charge-family-theory}.

Following the analyses of the fifth family neutron properties from the ref.~\cite{blmbDgn} we estimate 
in this contribution how many events should be observed in the XENON100 experiment in comparison with the 
annual modulated events of the DAMA experiment, if both measure the fifth family neutrons as the 
dark matter constituents.

\section{New results from the Xenon100 experiment}
\label{blmbDxe}

\begin{figure}[!htb]
\begin{center}
\includegraphics[width=0.8\columnwidth,clip,trim=0 0 20 400]{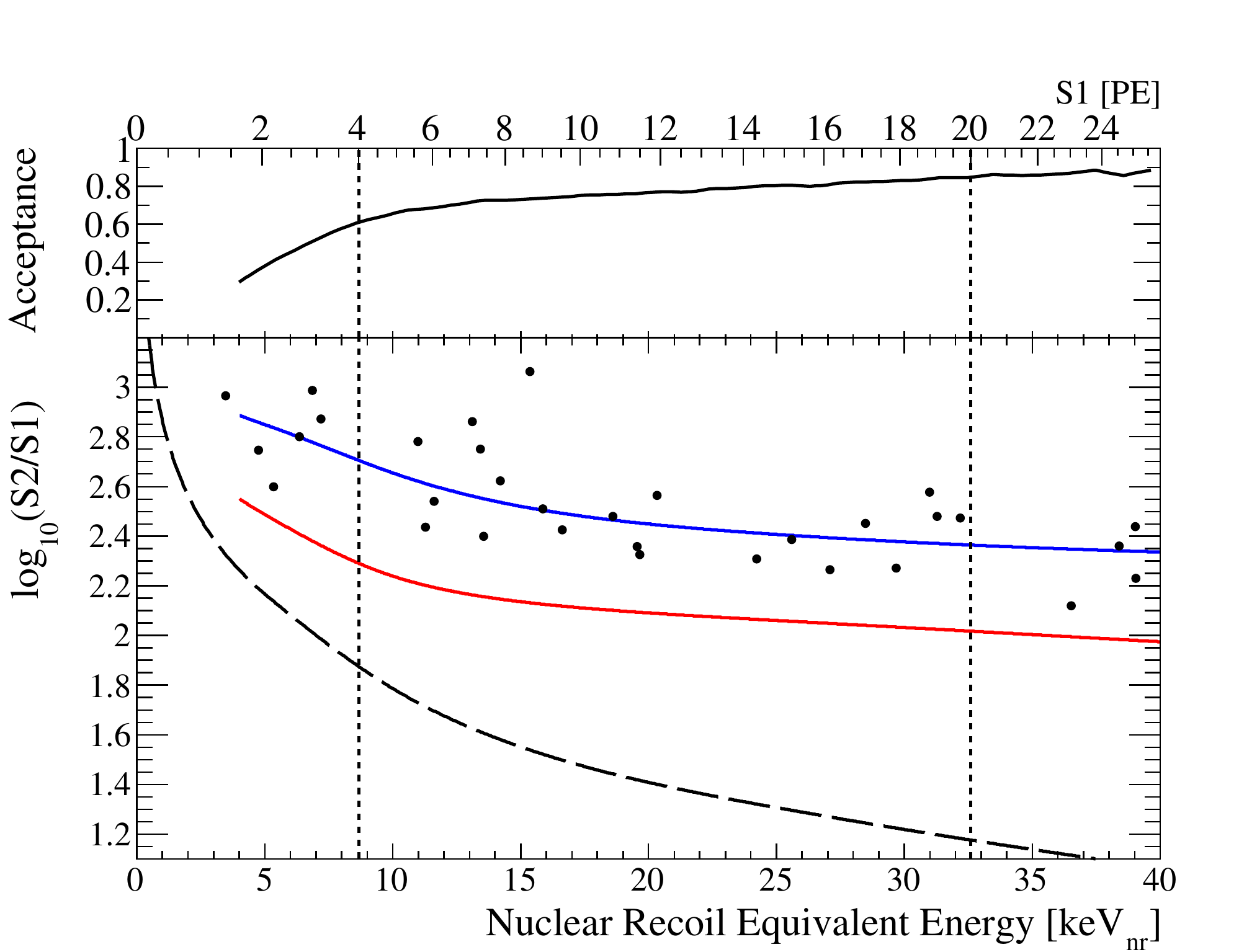}
\caption{Cut acceptance (top, not including 50\% acceptance 
from $\mathrm{S2/S1}$ discrimination) and discrimination 
parameter $\log_{10}(\mathrm{S2/S1})$ (bottom) as functions 
of nuclear recoil energy for events observed in the 40~kg fiducial volume 
during 11.17~live days of the first XENON100 data release~\cite{blmbDAprile:2010um}. 
Coloured lines correspond to the median $\log_{10}(\mathrm{S2/S1})$ 
values of the electronic (blue) and nuclear (red) recoil 
bands. The dark matter search energy window from 
8.7-32.6~keV$_{\mathrm{nr}}$ (vertical, dashed) and the 
S2 software threshold of 300~PE (long dashed) are also shown. 
No nuclear recoil events are observed. Figure taken from ~\cite{blmbDAprile:2010um}.}
\label{blmbDfig:discrimination}
\end{center}
\end{figure}

As the DAMA/LIBRA experiment also  XENON100 experiment is located at the Laboratori Nazionali del 
Gran Sasso, Italy. 
It is a two-phase liquid/gaseous xenon 
time-projection-chamber (TPC) made for the direct detection of the dark matter. 
A first search for the dark matter particles scattering elastically and spin independently  
on 62 kg of liquid xenon detector 
were analysed in the first published results~\cite{blmbDAprile:2010um}. The exposure was 
approximately 11 days and the analysis taking  into account only 
the inner 40~kg target of xenon to reduce the contribution from background radioactivity. In
the energy range of interest for the 
search for the dark matter as indicated on  figure~\ref{blmbDfig:discrimination} 
no events  which could be considered as the result of the dark matter 
constituent scattering on Xe nuclei were found. 
%
%
\section{Short review of the basic idea of the DAMA experiment~\cite{blmbDDAMA}}
\label{blmbDidea}

The basic idea of the DAMA experiment~\cite{blmbDDAMA} is to 
measure the annual modulation of the rate of events in a large NaI/Tl
scintillator. It is expected that the rate should modulate due to the motion 
of  Earth through the cloud of the dark matter particles.  Sun itself 
moves through this cloud as it travels around the center of our galaxy. 
When  Earth and Sun move in the same direction relative to the center 
of our galaxy, the flux of  incoming dark matter particles is increased 
and therefore also the rate of events measured in the DAMA experiment should 
correspondingly increase. The expression for the rate is 
\begin{eqnarray}
\label{blmbDra}
R_A = \, N_A \,  \frac{\rho_{0}}{m_{c_5}} \;
\sigma(A) \, v_S \, \varepsilon_{v_{dmS}}\, \varepsilon_{\rho} \, 
\left( 1 + 
\frac{\varepsilon_{v_{dmES}}}{\varepsilon_{v_{dmS}}} \, \frac{v_{ES}}{v_S}\, \cos \theta
\, \sin \omega t \right)\, , 
\end{eqnarray}
where $\theta=60^o$ is the angle between the plane of rotation of  Earth around 
 Sun and the plane of rotation of  Sun around the center of our galaxy. 
$R_A$ is the number of events per unit time and unit 
active mass of the detector, $\rho_0$ is the local mass density
of the dark matter which is unknown  ( say within the factor of $10$) with 
respect to the average density (see also the ref.~\cite{blmbDgn}) and which we take into account 
with the factor $\varepsilon_{\rho}$, measuring this indeterminacy of the local density of
dark matter.
In the XENON100 experiment the detector  is pure Xe, in the  
DAMA experiment it is NaI(Tl), in the CDMS experiment it is Ge. 
In all  experiments  
cuts are applied on  data to exclude fake events.
The term effective exposure stands for the exposure that takes into 
account the subsequently reduced acceptance of the experiment with respect to a signal.
The probability for a detector to register an event as a dark 
matter collision is influenced also by the energy threshold 
of the detector since the detector is blind below the energy threshold. 
In this analysis  we take into account  this effect by using a factor 
$\varepsilon_{cut}$. 
So if the rate of events taking place in the detector material is $R_A$ and 
the effective exposure is $ex_{{\rm eff}}$,  %
then the average value of events the detector should recognize as the dark matter events
is equal to $R_A \,ex_{{\rm eff}} \varepsilon_{cut}$.
$N_A$ is the number of scatterers per unit mass that have atomic mass $A$.
In the DAMA experiment it is  I  ($A_I=127$), which for heavy enough dark matter clusters
(in the case of the fifth family members of the mass of a few hundred TeV this is the case)
contribute the most ($A_{Na}=23$ is in this case negligible).  
 In this case $N_I=4.0\times 10^{24}\; \textrm{kg}^{-1}$.
In the XENON100 experiment with xenon ($A_{Xe}=131$)
$N_{Xe}=4.6\times 10^{24}\; \textrm{kg}^{-1}$. 
$\sigma(A)$ is the low energy cross section for  the collision  
of the dark matter constituent  and the nuclei of the detector. 
We must point out that this cross
section should scale as $\sigma(A)\propto A^4$ since for the case when dark matter 
constituents are for orders of magnitude more massive than the nuclei the scattering is
coherent which brings a factor $A^2$ while another factor 
$A^2$ comes from the phase space.
$v_S$ is the velocity of the Sun around the center of our galaxy, which is  in the 
region  $100 \; {\rm km/s} < v_S < 270 \; {\rm km/s}$.
$\varepsilon_{v_{dmS}}$ is an indeterminacy factor that tells how much 
the flux of dark matter particles onto a detector differs from the flux
in the case in which the Sun is moving through the cloud of dark matter particles
that are at rest relative to the center of galaxy. In such  (rather artificial)
case the intrinsic motion of dark matter particles does not increase the
flux. This factor depends only on the motion of the dark matter constituents 
and Sun.
$m_{c_5}$ is the mass of a dark matter particle.
 The ratio $\varepsilon_{v_{dmES}}/\varepsilon_{v_{dmS}}$ describes
the indeterminacy of the amplitude of the annual modulation of the rate. It
depends only on the motion of  Sun, Earth and the dark matter. 
$v_{ES}$ is the velocity of  Earth relative to  Sun.
$\omega$ fixes the period to one year. 
A detailed description of the analysis is presented in the already 
mentioned reference~\cite{blmbDgn}.
The amplitude of the annual modulation and the average rate are
connected in the following way
\begin{equation}
\frac{\Delta R_A}{\bar{R}_A}=
\frac{\varepsilon_{v_{dmES}}}{\varepsilon_{v_{dmS}}} \, \frac{v_{ES}}{v_S}\, \cos \theta.
\end{equation}

%
\section{What should the XENON100 experiment observe if  DAMA and XENON100  measure the 
fifth family neutrons with the mass of several hundred TeV/$c^2\,$?}
\label{blmbDcomparison}

In the following we calculate the number of events the XENON100 experiment
should have measured in the first data taking if we assume the events that 
the DAMA experiment measures  in the observed annual modulation are the fifth family 
neutrons predicted by the {\it spin-charge-family-theory}. 
Similar to the CDMS experiment, the XENON100 experiment has so far measured no events which these 
two experiments would recognize as dark matter events. 

From Eq.~(\ref{blmbDra}) and
the properties of $\sigma(A)$ we see that $R_A\propto N_A A^4$.
The number of events  CDMS should have measured is given 
according to the following formula, where $e_{{\rm eff} \, Xe}=172 \; \textrm{kg}\cdot  
\textrm{day}$
is the effective exposure %
\begin{eqnarray}
N&=&e_{{\rm eff} \, Xe} \; \bar{R}_{Xe} \; \varepsilon_{cut\; Xe} \nonumber\\
&=&e_{{\rm eff} \, Xe} \; \varepsilon_{cut\; DAMA} \; \Delta R_{I}
\frac{\varepsilon_{v_{dmS}}}{\varepsilon_{v_{dmES}}}
\frac{v_S}{v_{ES}\cos\theta}
\frac{N_{Xe}}{N_{I}}
\left(\frac{A_{Xe}}{A_{I}} \right)^4
\frac{\varepsilon_{cut\; Xe}}{\varepsilon_{cut\; DAMA}}.
\end{eqnarray}
The experimental value of the measured amplitude of the annual 
modulation of the DAMA experiment~\cite{blmbDDAMA} is
$\Delta R_I \; \varepsilon_{cut\; DAMA}=0.052\; \textrm{event}\; \textrm{day}^{-1}
\textrm{kg}^{-1}$. For the value $v_S=220 \; \textrm{km/s}$ 
we obtain 
\begin{eqnarray}
N &=&
172 \cdot 0.052
\frac{\varepsilon_{v_{dmS}}}{\varepsilon_{v_{dmES}}}
\; 14.7
\; \frac{4.6}{4.0}
\; \left(\frac{131}{127} \right)^4
\;\frac{\varepsilon_{cut\; Xe}}{\varepsilon_{cut\; DAMA}}\nonumber\\
&=&
170
\; \frac{\varepsilon_{v_{dmS}}}{\varepsilon_{v_{dmES}}}
\; \frac{\varepsilon_{cut\; Xe}}{\varepsilon_{cut\; DAMA}}\,.
\end{eqnarray}
Let us assume  the velocity range
 of Sun  within the interval $100 {\rm km}/s < v_S < 270 {\rm} km/s$. 
Then for the Sun's velocities $v_S = (100,\; 170,\; 220,\; 270){\rm km/s}$
we correspondingly obtain the number of events that the XENON100 experiment should measure
\begin{eqnarray}
v_S &=& (100,\; 170,\; 220,\; 270){\rm km/s}, \nonumber\\
   N&=& (77,\;\; 132,\; 170,\; 209)  \frac{\varepsilon_{v_{dmS}}}{\varepsilon_{v_{dmES}}}
\; \frac{\varepsilon_{cut\; Xe}}{\varepsilon_{cut\; DAMA}}\,.
\end{eqnarray}

Comparison among different experiments weakens the dependence on most of the 
uncertainties. What stays is the ratio of the uncertainties about the cuts and the 
uncertainties in the relative velocities  which concern the dark matter properties.
The ratio on the uncertainties about the cuts is determined  mainly on the 
threshold of the detectors and it is not expected to  be 
an order of magnitude away from 1. From extreme case models about  possible  ways of motion 
of the dark matter constituents  in our galaxy, which we performed,  we expect  
the ratio $\frac{\varepsilon_{v_{dmS}}}{\varepsilon_{v_{dmES}}}$, determined  
by the dynamics of the dark matter constituents in the galaxy, to be from 1/3 to 3.
We note that this XENON100 data was taken in November, when 
signals with strong annual modulation are expected to be rather 
weak.

Assuming   simple Poisson statistics one can quote the probabilities  that the XENON100 experiment 
 observes  zero events  
when the average expected  number of events
is $(3,5,10,100)$
 \begin{eqnarray}
 \label{blmbDprobability}
&& (3,\quad \; 5, \quad  \quad 10, \quad \quad \,100) \;\; \quad {\rm expected \,\,\,events}\nonumber\\
&&(5,\quad 0.7,\quad 4 \times10^{-3}, 10^{-42})\; \% \;{\rm probability \; to \; observe\; zero \,\,\,events}. 
\end{eqnarray}

\section{Conclusion}

A new data release from XENON100 is expected very soon, 
with an exposure more than an order of magnitude above 
the published data. This will help to constrain the nature of 
dark matter also in the scenario of the {\it spin-charge-family-theory} 
that is considered here.

\title{Masses and Mixing Matrices of Families of Quarks and
Leptons Within the Approach of\\ N.S. Manko\v c Bor\v stnik {\it
Unifying Spins and Charges and predicting families}}
\author{A. Hern\'andez-Galeana${}^1$ and N.S. Manko\v c Bor\v stnik${}^2$}
\institute{%
${}^1$Departamento de F\'{\i}sica,   Escuela Superior de
F\'{\i}sica y Matem\'aticas, I.P.N., \\
U. P. "Adolfo L\'opez Mateos". C. P. 07738, M\'exico, D.F.,
M\'exico.\\
${}^2$Department of Physics, FMF, University of Ljubljana,
Jadranska 19, SI-1000 Ljubljana, Slovenia}

\titlerunning{Masses and Mixing Matrices of Families\ldots}
\authorrunning{A. Hern\'andez-Galeana and N.S. Manko\v c Bor\v stnik}
\maketitle

\begin{abstract}
The {\it approach unifying spin and charges and predicting families}, proposed by N.S.M.B., 
predicts at the low energy regime two groups of four families, decoupled
in the mixing matrix elements. To the mass matrices there are two kinds of  contributions.
One kind distinguishes on the tree level only among the members of one family, that is among the
$u$-quark, $d$-quark, neutrino and electron, the left and right handed, while
the other kind distinguishes only among the families. Beyond the tree level
both kinds start to contribute coherently and it is expected that a detailed study of the
properties of mass matrices beyond the tree level will explain
a drastic difference in masses and mixing matrices between quarks and leptons.
We report in this contribution on  the analysis of one
loop corrections to the tree level fermion masses and mixing matrices. The loop diagrams are
mediated by gauge bosons and scalar fields.
\end{abstract}

\section{Introduction}

\label{contribution:AN}
The {\it approach unifying spin and charges and predicting families} (hereafter named the
{\it spin-charge-family-theory}~\cite{ansnmbnorma,ansnmbpikanorma,ansnmbNF}), proposed by
N.S.  Manko\v c Bor\v stnik, seems promising to show the
right way beyond the {\it standard model of fermions and bosons}.
The reader is kindly asked to learn more about
this theory in the ref.~\cite{ansnmbNF} in this proceedings (and in the references therein).
Following analyses of the ref.~\cite{ansnmbNF}, we shall here repeat  only those parts,
which are necessary  for the explanation to what conclusions  one 
loop corrections beyond the tree level might lead.

The {\it spin-charge-family-theory} predicts eight massless families of quarks and leptons
before the two successive breaks --
first from $SU(2)_{I} \times SU(2)_{II} \times U(1)_{II} \times  SU(3)$   to
$SU(2)_{I} \times U(1)_{I} \times SU(3)$ and then from $SU(2)_{I} \times U(1)_{I} \times SU(3)$
to $ U(1) \times SU(3)$. Mass matrices originate in a simple starting action: They are determined by
the nonzero vacuum expectation values of the scalar (with respect to $SO(1,3)$) fields,  to which vielbeins
and the two kinds of  spin connection fields contribute. Each of the two breaks is triggered
by different (orthogonal) superposition of  scalar fields.
The mass matrices for eight families appear to be four times four by diagonal matrices, with 
no mixing matrix elements among the upper four and the lower four families. There are, correspondingly,
two (with respect to the life of our Universe) stable families: the fifth and the observed first family.
The fifth family members are candidates to form the dark matter, the fourth family waits to be observed.

After the first break (from $SU(2)_{II} \times SU(2)_{I} \times U(1)_{II} \times  SU(3)$ to
$SU(2)_{I} \times U(1)_{I} \times SU(3)$),  which occurs,
below  $\approx 10^{13}$ GeV, the upper four families become massive. In the second break, which is the
{\it standard model}-like electroweak break, also the lower four families became massive.

Rough estimations made so far~\cite{ansnmbpikanorma,ansnmbgmdn,ansnmbgn} on the tree level, which took into account besides the
elementary particle data also the cosmological  data, show that the stable of the upper four families
might have masses~\cite{ansnmbgn} of the order of $100$ TeV/$c^2$. The contribution~\cite{ansnmbMN} 
discusses also the possibility that the masses are much smaller, of around a few TeV/$c^2$.
For the lower four families~\cite{ansnmbpikanorma,ansnmbgmdn} we were not really able to predict
the masses of the fourth family members,
we only estimated for chosen masses of the fourth family members their mixing matrices.

In this contribution we are studying, following suggestions from the ref.~\cite{ansnmbNF},
properties of the mass matrices of  twice four families,
taking into account the one loop corrections to the tree level
estimations. We namely hope to see already within the one and may be two loops corrections the explanation
for the differences in masses and mixing matrices between quarks and leptons, as well as within quarks and
within leptons.

To the loop corrections the gauge boson fields and the scalar field contribute, as explained in
the ref.~\cite{ansnmbNF}.

Let us write the effective action (it is  from the ref.~\cite{ansnmbNF} as Eq.~(5))
for eight  families of quarks and leptons ($\psi $),
left and right handed. The action is  formally rewritten in a way   to manifest the {\it standard model}
properties
\begin{eqnarray}
{\mathcal L}_f &=&  \bar{\psi}\gamma^{m} (p_{m}- \sum_{A,i}\; g^{A}\tau^{Ai} A^{Ai}_{m}) \psi
+ \nonumber\\
               & &  \{ \sum_{s=7,8}\;  \bar{\psi} \gamma^{s} p_{0s} \; \psi \}  + \nonumber\\
               & & {\rm the \;rest},
\label{ansnmbfaction}
\end{eqnarray}
where $m=0,1,2,3$ with
\begin{eqnarray}
\tau^{Ai} = \sum_{a,b} \;c^{Ai}{ }_{ab} \; S^{ab},
\nonumber\\
\{\tau^{Ai}, \tau^{Bj}\}_- = i \delta^{AB} f^{Aijk} \tau^{Ak}.
\label{ansnmbtau}
\end{eqnarray}
The second row of Eq.~(\ref{ansnmbfaction}) defines the mass matrices. We shall call it ${\mathcal L}_Y$.
We namely have (Eq.~(26)in the ref.~\cite{ansnmbNF})
\begin{eqnarray}
\label{ansnmbYaction1}
{\mathcal L}_Y &=&
\psi^{\dagger} \,\gamma^0\,(\stackrel{78}{(+)} p_{0+} +  \stackrel{78}{(-)} p_{0-} \, \psi,\nonumber\\
p_{0 \pm} &=& p_{07}\mp i p_{08},\quad p_{0s}= p_{s} - \frac{1}{2}\, \tilde{S}^{ab}\,\tilde{\omega}_{ab s}
- \frac{1}{2}\, S^{ab}\,\omega_{ab s},
\end{eqnarray}
with $s=7,8,$  and $\stackrel{78}{(\pm)}= \frac{1}{2}\, (\gamma^7 \pm \gamma^8)$.
$\stackrel{78}{(-)}$ transforms, for example, a righ handed $u_{R}$-quark of a particular colour and spin
to the corresponding left handed $u_{L}$-quark of the same spin and colour.

We see that to the mass terms both, $\tilde{S}^{ab}\,\tilde{\omega}_{ab s}$  (transforming one family
into another) as well as $S^{ab}\,\omega_{ab s}$ might contribute (a superposition
which "sees" the electromagnetic charge  $Q$ and the quantum number $Q'$) whenever the spin connection
scalar fields $\omega_{ab s}$ and $\tilde{\omega}_{ab s}$, each in a  superposition
determined by a particular break, gain nonzero vacuum expectation values.

To the first break (when one of the two $SU(2)$ together with $ U(1)$ breaks,
namely   $ SU(2)_{II} \times U(1)_{II}$
breaks into $U(1)_{I}$) the scalar fields (with respect to $SO(1,3)$) originating in vielbeins and
particular superposition of spin connection fields of $\tilde{S}^{ab}$ contribute.

From the ref.~\cite{ansnmbNF} we read in Table~\ref{snmb3tgTable VIII.} a general shape of mass matrices after the first break,
when  the upper four families gain masses, while the lower four families are still massless.
We present this   table, Table~\ref{ansnmbTable VII.}, also here. To learn the meaning of
the notation $\tilde{a}^{\tilde{A}i}_{\pm}$ 
the reader should look at  Section V. of the ref.~\cite{ansnmbNF}. For  discussions here only the
symmetry of mass matrices is important.
$(\pm)$ distinguishes  between the values of the $u$-quarks and $d$-quarks and
between the values of $\nu$ and $e$.
 \begin{table}
 \begin{center}
\begin{tabular}{|r||c|c|c|c|c|c|c|c||}
\hline
 &$ I $&$ II $&$ III $&$ IV $&$ V $&$ VI $
 &$ VII $&$ VIII$\\
\hline\hline
$I \;\;\; $ & $  \;0 \; $ & $ \;0 \; $ & $ \; 0 \;$ & $ \; 0 \;$
          & $ 0 $ & $ 0 $ & $ 0 $ & $ 0 $\\
\hline
$II\;\; $ & $ 0 $ & $ 0 $ & $ 0 $ & $ 0 $
          & $ 0 $ & $ 0 $ & $ 0 $ & $ 0 $\\
\hline
$III\;$ & $ 0 $ & $ 0 $ & $ 0 $ & $ 0 $
          & $ 0 $ & $ 0 $ & $ 0 $ & $ 0 $\\
\hline
$IV\;\, $ & $ 0 $ & $ 0 $ & $ 0 $ & $ 0 $
          & $ 0 $ & $ 0 $ & $ 0 $ & $ 0 $\\
\hline\hline
$ V \;\; $ & $ 0 $ & $ 0 $ & $ 0 $ & $ 0 $ &
$  \frac{1}{2}\, (\tilde{a}^{23}_{\pm} + \tilde{a}^{\tilde{N}^{3}_{R}}_{\pm})$ &
$  - \tilde{a}^{2-}_{\pm}$ & $ -\tilde{a}^{\tilde{N}_{R}^{+}}_{\pm} $ & $ 0 $\\
\hline
$ VI \;\,$ & $ 0 $ & $ 0 $ & $ 0 $ & $ 0 $ &
$
-\tilde{a}^{2+}_{\pm}$ &
$ \frac{1}{2}(-\tilde{a}^{23}_{\pm } + \tilde{a}^{\tilde{N}^{3}_{R}}_{\pm}) $ &
$ 0 $ & $- \tilde{a}^{\tilde{N}_{R}^{+}}_{\pm}   $\\
\hline
$VII \;$ & $ 0 $ & $ 0 $ & $ 0 $ & $ 0 $ & $   - \tilde{a}^{\tilde{N}_{R}^{-}}_{\pm} $ & $ 0 $ &
$  \frac{1}{2}\,( \tilde{a}^{23}_{\pm}  - \tilde{a}^{\tilde{N}^{3}_{R}}_{\pm}) $ &
$  -\tilde{a}^{2-}_{\pm}$ \\
\hline
$VIII$ & $ 0 $ & $ 0 $ & $ 0 $ & $ 0 $ & $ 0 $ & $ - \tilde{a}^{\tilde{N}_{R}^{-}}_{\pm}  $ &
$ - \tilde{a}^{2+}_{\pm}$ &
$ - \frac{1}{2}\,( \tilde{a}^{23}_{\pm} + \tilde{a}^{\tilde{N}^{3}_{R}}_{\pm})$ \\
\hline\hline
\end{tabular}
 \end{center}
 \caption{\label{ansnmbTable VII.}  The mass matrix for the eight families of quarks and
 leptons after the break of $SO(1,3)\times SU(2)_{I}  \times SU(2)_{II} \times U(1)_{II} \times SU(3)$
 to $SO(1,3) \times  SU(2)_{I} \times U(1)_{I} \times SU(3)$. The contribution comes from a particular
 superposition of  spin connection fields, the gauge fields of $\tilde{S}^{ab}$.
 $(\mp)$ distinguishes $u_{i}$ from
 $d_{i}$ and $\nu_{i}$ from $e_{i}$.
 }
\end{table}

To the second break
(when the remaining  $SU(2)$ contributes, namely   $ SU(2)_{I} \times U(1)_{I}$
breaks into $U(1)$)  besides the scalar fields originating in
vielbeins and in orthogonal (to the first) superposition  of spin connection fields of $\tilde{S}^{ab}$,
also the scalar fields originating in spin connections of $S^{ab}$ contribute.
This is explained in the ref.~\cite{ansnmbNF}.

Table~\ref{ansnmbTable VIII.} represents the mass matrices for the lower
four families on the tree level. Only the contribution of the scalar fields which originate
in the gauge fields of $\tilde{S}^{ab}$ are included.
The contribution from the  terms like
$Q' \,A^{Q'}_{s}$,  which are diagonal and
equal for all the families, but distinguish among the members of one family are not present.
The notation $\tilde{a}^{\tilde{A}i}_{\pm}=$ $-\tilde{g}^{\tilde{A}i}\, \tilde{A}^{\tilde{A}i}_{\pm}$ is used.
 \begin{table}
 \begin{center}
\begin{tabular}{|r||c|c|c|c||}
\hline
 &$ I $&$ II $&$ III $&$ IV $\\
\hline\hline
$I \;\;$& $  \frac{1}{2}\, (\tilde{a}^{13}_{\pm} + \tilde{a}^{\tilde{N}^{3}_{L}}_{\pm})$ &
$   \tilde{a}^{1+}_{\pm}$ & $ \tilde{a}^{\tilde{N}_{L}^{+}}_{\pm} $& $0$ \\
\hline
$II\;$ &  $ \tilde{a}^{1-}_{\pm} $ &
$ \frac{1}{2}( -\tilde{a}^{13}_{\pm } + \tilde{a}^{\tilde{N}^{3}_{L}}_{\pm}) $&$0$&
$ \tilde{a}^{\tilde{N}_{L}^{+}}_{\pm}   $ \\
\hline
$III$ & $ \tilde{a}^{\tilde{N}_{L}^{-}}_{\pm} $ & $0$&
$  \frac{1}{2}\,( \tilde{a}^{13}_{\pm}  - \tilde{a}^{\tilde{N}^{3}_{L}}_{\pm}) $ &
 $\tilde{a}^{1+}_{\pm}$ \\
\hline
$IV$ & $0$&$  \tilde{a}^{\tilde{N}_{L}^{-}}_{\pm}  $ &
$  \tilde{a}^{1-}_{\pm}$ &
$ - \frac{1}{2}\,( \tilde{a}^{13}_{\pm} + \tilde{a}^{\tilde{N}^{3}_{L}}_{\pm})$ \\
\hline\hline
\end{tabular}
 \end{center}
 \caption{\label{ansnmbTable VIII.}  The mass matrix for the lower four  families of quarks and
 leptons after the electroweak break. Only the contributions coming  from the spin connection fields,
 originating in  $\tilde{S}^{ab} $
 are presented.
 $(\mp)$ distinguishes  between the values of the $u$-quarks and $d$-quarks and
between the values of $\nu$ and $e$.
 The terms coming from spin connection fields originating in $S^{ss'}\,$
 are not presented here. They are the same for
 all the families,  but distinguish among the family members.
 }
\end{table}

We shall study in this contribution  the properties of  only the lowest
two families of each  of the two groups of families, neglecting the
nonzero mass matrix elements between the lower and the upper two families, which
on the tree level looks in the {\it spin-charge-family-theory} rather small, in
particular for the upper four families.
 The measured values of the mixing matrices for the observed families supports such an
 assumption for quarks, at least for these first studies, but not for leptons.
Of course, going beyond the tree level can drastically change the values of matrix
elements, what obviously must happen at least for  leptons of the lower four families.


%
\section{Mass matrices beyond the tree level}
\label{ansnmbbeyond}

It seems meaningful~\cite{ansnmbNF}, that is in accordance with the experimental data, to assume
that to the upper four families only $\tilde{\omega}_{abs}$ contribute on the tree level,
while for the lower four families also the fields $\omega_{st t'}$ contribute a diagonal
terms, the same for all the families of a particular member, but different for
different family members.

We shall study in this contribution, as we already say, only mass matrices of the
lowest two families,
that is only two times two matrices, for either the upper or the lower four families,
neglecting the coupling  to the rest of the two families. That is we shall put
the matrix elements $\tilde{a}^{\tilde{N}^{3, \pm}_{L,R}}_{\pm}=0$, in both
tables, Table~\ref{ansnmbTable VII.} and Table~\ref{ansnmbTable VIII.}, since the {\it spin-charge-family-theory}
suggets that these are much smaller than the rest of matrix elements.

Regarding the upper group of families
the $ 2 \times 2 $ mass matrices look accordingly like

\vspace{2mm}

\begin{equation} M_{\pm}^{ \scriptstyle{V,VI}}=
\begin{pmatrix} a_1 & b \\ b & a_2
\end{pmatrix}^{\scriptstyle{ V,VI} }_{\pm}\;,
\end{equation}

\vspace{2mm}

\noindent the same for  $u_i$ and $\nu_i$ {$(-)$} and for $d_i$ and $e_i$ {$(+)$}.

For the lowest two families to the  mass matrix $M_{\pm}^{\scriptstyle{I,II}}$
a different diagonal term $a^{\alpha} \, I_{2\times 2}$ should be added for each of the  family member
$\alpha= u,d,\nu,e$, where $I$ is the identity matrix.

\vspace{2mm}

\begin{equation} M^{\alpha,\scriptstyle{I,II}}= a^{\alpha} \,I_{2 \times 2} +
\begin{pmatrix} a_1 & b \\ b & a_2,
\end{pmatrix}^{\scriptstyle{ I,II} }_{\pm}\;.
\end{equation}

Let us, before going to the loop corrections, diagonalize these tree level matrices.
If we write any of the mass matrices (for any member of either the upper two or
the lower two families)

\vspace{2mm}

\begin{equation}  {\cal
M}_o= \begin{pmatrix} a_1 & b \\ b & a_2
\end{pmatrix}
\end{equation}

\vspace{2mm}

\noindent and any of the two correcponding (left and right handed) vectors as

\vspace{2mm}

\begin{equation} \psi_{L,R}^o=
\begin{pmatrix} \psi_1^o \\ \psi_2^o
\end{pmatrix}_{L,R}\;,
\end{equation}

\vspace{2mm}
\noindent
then a generic  mass term from Eq.(\ref{ansnmbYaction1}) and presented 
in tables~\ref{ansnmbTable VII.},\ref{ansnmbTable VIII.}, only $2 \times 2$ matrices,
can be diagonalized with the  orthogonal matrix
\vspace{2mm}

\begin{equation} V = \begin{pmatrix} \cos\theta & \sin\theta \\ - \sin\theta & \cos\theta      \end{pmatrix} \quad , \quad
\cos\theta = \sqrt{\frac{a_2 - m_1}{m_2 - m_1}} \;\; , \;\; \sin\theta
= \sqrt{\frac{a_1 - m_1}{m_2 - m_1}} 
\end{equation}

\vspace{2mm}
\noindent  with $m_{1,2} = \frac{1}{2}\, [ (a_1 + a_2) \mp \, \sqrt{(a_1 + a_2)^2 - 4\, (a_1 a_2- b^2)} ]$. 
We have

\vspace{2mm}

\begin{equation} \psi_{L,R}^o = V\:\psi_{L,R} \quad ; \quad V^T {\cal M}_o \:V
=M_D =Diag(m_1,m_2)   
\end{equation}

\vspace{2mm} \noindent with $m_1$ and $m_2$  the mass
eigenvalues of ${\cal M}_o$, and

\vspace{2mm}
\begin{equation} \bar{\psi_L^o}\:{\cal M}_o\: \psi_R^o = \bar{\psi_L}\:M_D\:
\psi_R\end{equation}\,.

\vspace{2mm}

\subsection{  Gauge bosons $A_m^{Y^\prime}$ contribution to  one loop corrections }
\label{ansnmbgaugebosons}

What follows can be found in Subsection IV.B. of the ref.~\cite{ansnmbNF}.
Let us first treat the upper two (out of four) families. From the expressions which
will be obtained for the upper two (out of four) families it is not difficult to obtain the
corresponding expressions for the lowest two families.

According to the ref.~\cite{ansnmbNF}
$\psi_R^o$ transforms as $(2, \tau^4)$, $\tau^4=\frac{1}{6}$ ($-\frac{1}{2}$) for quarks (leptons),
under $SU(2)_{II} \times
U(1)_{II}$, while $\psi_L^o$ transforms as $(1,\tau^4 )$ under $SU(2)_{II} \times
U(1)_{II}$.  Therefore the  covariant  momenta contributing to the mass matrices are
\begin{eqnarray}
     p_{o m}  \: \psi_R^o = \{ p_m -
&\bigl[&     g_Y \:Y \:A_m^Y + g_2\: \cos\theta_2 \:Y^\prime
\:A_m^{Y^\prime}+\nonumber\\
           &&\frac{1}{\sqrt{2}} \left( \tau^{2+} A_m^{2+} +
\tau^{2-} A_m^{2-}\right)  \bigr] \, \} \:\psi_R^o\, ,
\end{eqnarray}
 and
\begin{eqnarray}
    p_{o m}  \: \psi_L^o &=&  \{ p_m - g_4 \tau^4 \:A_m^{\tau^4}\}\,
    \: \psi_L^o = \nonumber\\
&=&  \{ p_m  - \left(     g_4 \:\tau^4 \cos\theta_2 \:A_m^Y - g_4\:\tau^4
\sin\theta_2 \:A_m^{Y^\prime}  \right)\,\} \:\psi_L^o\, .
\end{eqnarray}

From these expressions for the covariant derivatives the gauge couplings
to $A_m^{Y^\prime}$ follow

\begin{equation}  g_{u_{iL}}^\prime   \bar{u}_{iL} \gamma^m A_m^{Y^\prime}  u_{iL} +
     g_{u_{iR}}^\prime   \bar{u}_{iR} \gamma^m A_m^{Y^\prime}  u_{iR} \end{equation}

\begin{equation} g_{\nu_{iL}}^\prime \bar{\nu}_{iL} \gamma^m A_m^{Y^\prime} \nu_{iL} +
    g_{\nu_{iR}}^\prime \bar{\nu}_{iR} \gamma^m A_m^{Y^\prime} \nu_{iR} \end{equation}\, ,

\noindent with $i=V,VI$, and where
\begin{equation}  g_{u_{iL}}^\prime = - \frac{1}{6} g_4 \sin\theta_2 \quad ; \quad
     g_{u_{iR}}^\prime = g_2 \cos\theta_2 \: \frac{1}{2} (1 - \frac{1}{3}
\tan^2\theta_2) \;, \label{ansnmbeq16}\end{equation} 

\begin{equation} g_{\nu_{iL}}^\prime = \frac{1}{2} g_4 \sin\theta_2 \quad ; \quad
    g_{\nu_{iR}}^\prime = g_2 \cos\theta_2 \: \frac{1}{2} (1 +
\tan^2\theta_2) \;, \label{ansnmbeq17}\end{equation}

\noindent with $\sin\theta_2$ and $\cos\theta_2$ defined by the relationship

\begin{equation} g^Y=\frac{g^2\:g^4}{\sqrt{(g^2)^2+(g^4)^2}}=g^2\:\sin\theta_2=g^4\:\cos\theta_2 \;,
\end{equation}

\noindent $g^2,\:g^4\:g^Y$ being the couplings of $SU(2)_{II}$, $U(1)_{II}$ and $U(1)_{I}$, respectively.

\begin{figure}
\centering\includegraphics{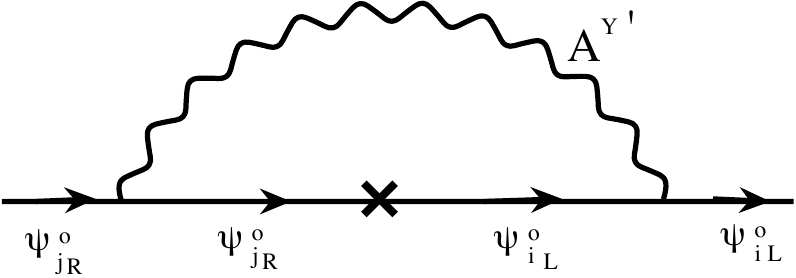}
\caption{\label{ansnmbFig1}One loop contribution from $A_m^{Y^\prime}$
gauge boson.}
\end{figure}

The tree level mass terms and gauge couplings may be
used to draw the one loop diagrams of Fig.~\ref{ansnmbFig1}. We write the
internal fermion lines of these diagrams in terms of the mass
eigenvalues and eigenfields of ${\cal M}_o$. So, the one loop
contribution in the interaction basis reads

\begin{equation} \bar{\psi_{L}^o}\:{\cal M}_1^\prime\: \psi_{R}^o=
\bar{\psi_{L}}\:V^T\:{\cal M}_1^\prime \:V \:\psi_{R} \qquad ;
\qquad {\cal M}_1^\prime=
\begin{pmatrix} w_{11} & w_{12} \\ w_{21} & w_{22}
\end{pmatrix}  \:,\end{equation}

\vspace{2mm}
\noindent where 

\begin{equation} w_{11}= \cos^2\theta\:m_1\:G_1 +  \sin^2\theta\:m_2\:G_2 \;,\end{equation} 

\begin{equation} w_{22}=
\sin^2\theta\:m_1\:G_1 + \cos^2\theta\:m_2\:G_2    \;,\end{equation}
                            
\begin{equation} w_{21}=w_{12}=\cos\theta\:\sin\theta \:(-m_1\:G_1 + m_2\:G_2 ) \; , \end{equation}

\vspace{2mm}

\begin{equation}  G_i \equiv G(M,m_i)=
\frac{g_L^\prime\:g_R^\prime}{16\:\pi^2}\:\frac{M^2}{M^2-m_i^2}\ln{\frac{M^2}{m_i^2}} \quad , \quad 
 M^2=M_{A^{Y^\prime}}^2 \;, \end{equation}

\noindent where the corresponding couplings $(g_L^\prime\:g_R^\prime)_{ui}$ and  $(g_L^\prime\:g_R^\prime)_{\nu i}$
read from from Eqs.(\ref{ansnmbeq16},\ref{ansnmbeq17}).
\vspace{3mm}

Explicit computation yields

\begin{equation} {V}^T\: {\cal M}_1^\prime \:V = \begin{pmatrix} m_1\:G_1 & 0 \\ 0 & m_2\:G_2
\end{pmatrix} \end{equation}

\noindent Hence the one loop contributions from $A^{Y^\prime}$ give corrections to the tree level mass
eigenvalues, but does not correct mixing.


%
\subsection{Scalar fields  contribution to  one loop corrections }
\label{ansnmbscalarbosons}

\begin{figure}
\centering\includegraphics[width=12cm]{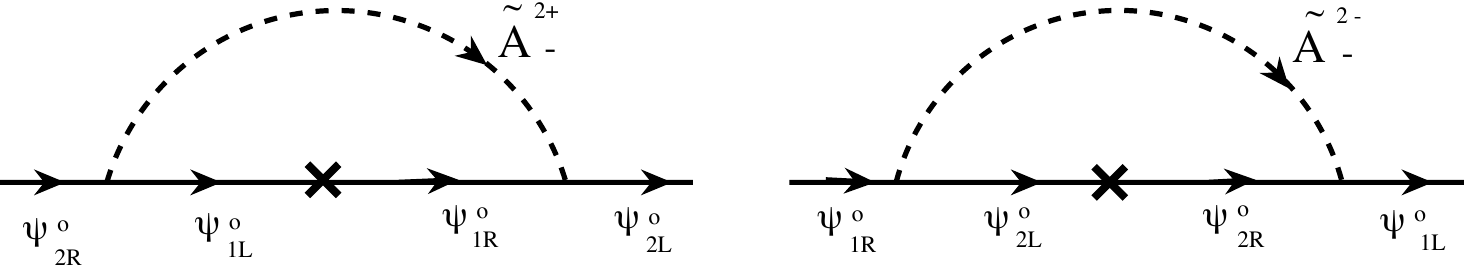} 
\caption{\label{ansnmbFig2}Loop contributions from $\tilde{A}_{-}^{2\pm}$.}
\end{figure}

\begin{figure}
\centering\includegraphics{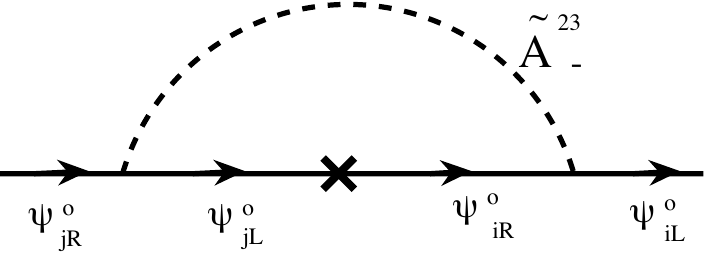} 
\caption{\label{ansnmbFig3}Loop contributions from $\tilde{A}_{-}^{23}$.}
\end{figure}

Following an analogous procedure as in the previous subsection we calculate
in what follows the contribution from the scalar fields, the scalar fields of $\tilde{S}^{ab}$,
to loop corrections.  Couplings of heavy families
$V$ and $VI$ to the scalar tilde fields $\tilde{A}^{2 i}_{\pm}$ are
\begin{multline}  \frac{\tilde{g}_2}{2}\:\left[
\left( \overline{\psi_2^o}_{L,R}\:{\psi_1^o}_{R,L} +
\overline{\psi_1^o}_{L,R}\:{\psi_2^o}_{R,L}
\right)\:\tilde{A}^{21}_{\pm} \right.  \\ + \left( i\:
\overline{\psi_2^o}_{L,R}\:{\psi_1^o}_{R,L} - i\:
\overline{\psi_1^o}_{L,R}\:{\psi_2^o}_{R,L}
\right)\:\tilde{A}^{22}_{\pm}           \\
+ \left. \left( \overline{\psi_1^o}_{L,R}\:{\psi_1^o}_{R,L} -
\overline{\psi_2^o}_{L,R}\:{\psi_2^o}_{R,L}
\right)\:\tilde{A}^{23}_{\pm} \right] 
\end{multline}
Using these scalar couplings and the tree level mass matrices for the  upper two
 family members $\psi_{iR,L}, i=V,VI$, first for $u_{i L,R}$ quarks
 and neutrinos $\nu_{i L,R}$,
 we can draw the one
loop diagrams of Figs.~\ref{ansnmbFig2} and~\ref{ansnmbFig3}.

\noindent It follows then

\begin{equation} \bar{\psi_{L}^o}\: \tilde{\cal M}_1 \: \psi_{R}^o=
\bar{\psi_{L}}\:V^T\:\tilde{\cal M}_1 \:V \:\psi_{R} \qquad ;
\qquad \tilde{\cal M}_1= \begin{pmatrix} \Omega_{11} & \Omega_{12} \\
\Omega_{21} & \Omega_{22}
\end{pmatrix}  \:,\end{equation}

\vspace{2mm} \noindent with

\begin{eqnarray} \Omega_{11}= 2 \left( \sin^2\theta\:\Sigma_1^1 + \cos^2\theta\:\Sigma_2^1
\right) +
\cos^2\theta\:\Sigma_1^3 +  \sin^2\theta\:\Sigma_2^3     \\
                                  \nonumber \\
\Omega_{22}= 2 \left( \cos^2\theta\:\Sigma_1^1 + \sin^2\theta\:\Sigma_2^1 \right)+
\sin^2\theta\:\Sigma_1^3 +  \cos^2\theta\:\Sigma_2^3    \\
                                \nonumber \\
\Omega_{21}=\Omega_{12}=\cos\theta\:\sin\theta \:( \Sigma_1^3 - \Sigma_2^3 ) \end{eqnarray}

\vspace{2mm}

\begin{equation} \Sigma_i^j \equiv m_i\:{\cal G}(M_j,m_i) \; , \; {\cal
G}(M_j,m_i)=
\frac{\tilde{g}_2^2}{4}\:\frac{1}{16\:\pi^2}\:\frac{M_j^2}{M_j^2-m_i^2}\ln{\frac{M_j^2}{m_i^2}}
\end{equation}

\begin{equation} M_1^2= M_{\tilde{A}^{21}}^2=M_{\tilde{A}^{22}}^2 \qquad , \qquad
M_3^2 =M_{\tilde{A}^{23}}^2   \label{ansnmbeq31} \end{equation}

\vspace{2mm}

\noindent Let us remind here that the scalar fields $\tilde{A}^{23}$ and $\tilde{A}^{4}$ 
mix in the breaking $SU(2)_{II}\times U(1)_{II}$ into $U(1)_I$ (Eq.(19) in the ref.\cite{ansnmbNF}),
so that it is consistent to consider $M_1 \neq M_3$ in Eq.(\ref{ansnmbeq31}).

\vspace{2mm}

From explicit computation one gets

\begin{equation}  \Gamma = {V}^T\: \tilde{\cal M}_1    \:V = \begin{pmatrix} \Gamma_{11}  & \Gamma_{12} \\
 \Gamma_{21} & \Gamma_{22} \end{pmatrix} \end{equation}

\vspace{2mm} \noindent where

\begin{eqnarray} \Gamma_{11}= 2 \:\Sigma_2^1 + \Sigma_1^3 - 4 \cos^2\theta\:\sin^2\theta \left(
\Sigma_1^3 - \Sigma_1^1 + \Sigma_2^1 - \Sigma_2^3 \right)    \\
                                       \nonumber \\
\Gamma_{22}= 2 \:\Sigma_1^1 + \Sigma_2^3 + 4 \cos^2\theta\:\sin^2\theta \left(
\Sigma_1^3 - \Sigma_1^1 + \Sigma_2^1 - \Sigma_2^3 \right)  \\
                                       \nonumber   \\
\Gamma_{21}=\Gamma_{12}= 2 \:\cos\theta\:\sin\theta\:(\cos^2\theta - \sin^2\theta)\left( \Sigma_1^3 -
\Sigma_1^1 + \Sigma_2^1 - \Sigma_2^3 \right) \;, \end{eqnarray}

\vspace{2mm} \noindent where we may write

\begin{equation}  \Sigma_1^3 - \Sigma_1^1 + \Sigma_2^1 -
\Sigma_2^3  = \frac{\tilde{g}_2^2}{64\:\pi^2} \:F(M_1,M_3,m_1,m_2) \end{equation}

\vspace{2mm} \noindent with

\begin{multline} F(M_1,M_3,m_1,m_2) = \left\{    m_1 \left[ \frac{M_3^2}{M_3^2-m_1^2}\ln{\frac{M_3^2}{m_1^2}}
 -  \frac{M_1^2}{M_1^2-m_1^2}\ln{\frac{M_1^2}{m_1^2}}    \right] \right. \\
                                                      \\
\left. -   m_2 \left[
\frac{M_3^2}{M_3^2-m_2^2}\ln{\frac{M_3^2}{m_2^2}}
 -  \frac{M_1^2}{M_1^2-m_2^2}\ln{\frac{M_1^2}{m_2^2}}    \right]
 \right\}    \end{multline}

\vspace{2mm} \noindent Finally, taking into account the tree level
contribution, we obtain up to one loop corrections the mass matrix

\begin{equation} \bar{\psi}_L \: \left[ V^T\:{\cal M}_1^\prime \:V + V^T\:
\tilde{\cal M}_1 \:V + Diag(m_1,m_2) \right] \psi_R \equiv
\bar{\psi}_L \: {\cal M}\:\psi_R  \end{equation}

\vspace{2mm}

 \begin{equation} {\cal M}=\begin{pmatrix} M_{11} &  M_{12}
\\  M_{21} &  M_{22} \end{pmatrix}  \:, \label{ansnmboneloopmassterms} \end{equation}

\vspace{2mm} \noindent with the mass matrix elements

\begin{eqnarray} M_{11}= m_1 + m_1\:G_1 +  \Gamma_{11} \quad , \quad M_{22}=
m_2 + m_2\:G_2 +  \Gamma_{22}  \\
                             \nonumber \\
M_{21}= M_{12}=\Gamma_{12} \end{eqnarray}

\vspace{2mm} \noindent Diagonalizing the mass matrix
${\cal M}$, Eq.(\ref{ansnmboneloopmassterms}),  the physical
masses for fermions within one loop correction follow.




\vspace{2mm} \noindent Notice that in the limit $m_1 , m_2 \ll M_1 , M_3 $, which is accomplished for at least the
lower 1 and 2 families, we may approach

\begin{equation} F(M_1,M_3,m_1,m_2) \approx (m_2 -
m_1)\:\ln{\frac{M_1^2}{M_3^2}} \;,\end{equation}

\vspace{2mm} \noindent and

\begin{equation} \Sigma_i^j \approx
\frac{\tilde{g}_2^2}{64\:\pi^2}\:m_i\:\ln{\frac{M_j^2}{m_i^2}} \;,\end{equation}

\vspace{2mm} \noindent So, the mass terms $M_{ij}$ in this limit
may be written explicitly as

\vspace{2mm}

\begin{multline} M_{11} \thickapprox m_1 +
m_1\:\frac{g_L^\prime\:g_R^\prime}{16\:\pi^2}\:\ln{\frac{M^2}{m_1^2}} \\
+ \frac{\tilde{g}_2^2}{64\:\pi^2}\: \left[
2m_2\:\ln{\frac{M_1^2}{m_2^2}} + m_1\:\ln{\frac{M_3^2}{m_1^2}} - 4
\cos^2\theta \sin^2\theta \:(m_2 - m_1)\:\ln{\frac{M_1^2}{M_3^2}} \right] \label{ansnmbM11}
\end{multline}

\vspace{2mm}

\begin{equation} M_{21}=M_{12} \thickapprox \frac{\tilde{g}_2^2}{32\:\pi^2}\:\cos\theta\:
\sin\theta \:(\cos^2\theta - \sin^2\theta)\:(m_2 - m_1)\:\ln{\frac{M_1^2}{M_3^2}} \label{ansnmbM12}
\end{equation}

\vspace{2mm}

\begin{multline} M_{22} \thickapprox m_2 +
m_2\:\frac{g_L^\prime\:g_R^\prime}{16\:\pi^2}\:\ln{\frac{M^2}{m_2^2}} \\
+ \frac{\tilde{g}_2^2}{64\:\pi^2}\: \left[
2m_1\:\ln{\frac{M_1^2}{m_1^2}} + m_2\:\ln{\frac{M_3^2}{m_2^2}} + 4
\cos^2\theta \:\sin^2\theta \:(m_2 - m_1)\:\ln{\frac{M_1^2}{M_3^2}} \right] \label{ansnmbM22}
\end{multline}

\section{Discussion and Conclusions}

The results reported from the analysis of one loop diagrams coming from
$A_m^{Y^\prime}$ gauge boson and the scalar tilde fields
$\tilde{A}^{2 i}$ show that loop corrections contribute to both
fermion masses and mixing. A detailed quantitative analysis is in
progress to find out the spectrum of masses and mixing that we are
able to accommodate taking into account tree level and 
loop corrections. Even without present here some numerical results,
it is important to point out that for the upper two families the main 
differences among family mass matrices, for instance between very heavy U
quarks and neutrinos N, come from loop corrections, while for the lower two 
families the diagonal mass terms, differing for each family member and 
being the same for all families, play an important role together with the loop 
contributions to distinguish for example between up quarks and neutrino mass
matrices within this spin-charge-family theory.

\vspace{2mm} \noindent The expressions for the mass terms $M_{ij}$
as written in Eqs.(\ref{ansnmbM11}-\ref{ansnmbM22}) are useful to obtain the
quantitative corrections to the the tree level masses and mixing
from loop contributions at least for the lower 1 and 2 families.

\title{Can the Stable Fifth Family of the "Spin-charge-family-theory" Proposed by
N.S. Manko\v c Bor\v stnik Fulfil the M.Y. Khlopov Requirements and Form the Fifth Antibaryon
Clusters with Ordinary He Nucleus?}
\author{M. Y. Khlopov${}^{1,2,3}$ and N.S. Manko\v c Bor\v stnik${}^4$}
\institute{%
${}^1$Moscow Engineering Physics Institute (National Nuclear Research University), 115409 Moscow, Russia \\
${}^2$Centre for Cosmoparticle Physics "Cosmion" 115409 Moscow, Russia \\
${}^3$APC laboratory 10, rue Alice Domon et L\'eonie Duquet \\75205 Paris
Cedex 13, France\\
${}^4$Department of Physics, FMF, University of Ljubljana,\\
 Jadranska 19, SI-1000 Ljubljana}

\authorrunning{M. Y. Khlopov and N.S. Manko\v c Bor\v stnik}
\titlerunning{Can the Stable Fifth Family\ldots}
\maketitle

\begin{abstract}
This discussion is to try to clarify whether the dark matter can be
made out of the clusters of the members of the stable fifth family, which
is predicted by the {\it approach unifying spin and charges and
predicting families} proposed by Norma~\cite{mksnmbDnorma,mksnmbDpikanorma} (to be
named as  the {\it spin-charge-family-theory}) in the scenario proposed by Maxim.
Maxim's scenario differs from the  Norma's one published in the ref.~\cite{mksnmbDgn},
in which Gregor and Norma
 evaluated that the fifth family members have the masses of
around a few hundred TeV and that the current quark masses of the
stable fifth family do not differ among themselves more than a few
hundred GeV (see~\cite{mksnmbDgn,mksnmbDMN}). Then independently of the
quark-antiquark fifth family asymmetry the fifth family neutrons and
antineutrons constitute the dark matter, while the contribution of
the fifth family neutrinos is negligible, provided that the fifth
neutrino-antineutrino asymmetry is small enough. Maxim assumes for
his scenario i. that the fifth family quarks are not heavier than a
few TeV, ii. that there is an antiquark-quark asymmetry and iii.
that $\bar{u}_5 \bar{u}_5 \bar{u}_5$ is the lightest baryon. If
these three conditions are fulfilled Maxim~\cite{mksnmbDMaxim} found that
the  fifth family anti$u$-quark  cluster with a charge $-2 $ forming
with  the ordinary He nucleus an electromagnetically neutral object
(Maxim calls it OHe) might be what constitute the dark matter. These
two scenarios are discussed in this contribution.
\end{abstract}

\section{Introduction}
\label{mksnmbDintroduction}

\label{contribution:mksnmbDisc}
The theory unifying spin and charges and predicting families --  the {\it spin-charge-family-theory} --
proposed by N.~S.~Manko\v c Bor\v stnik~\cite{mksnmbDnorma,mksnmbDpikanorma,mksnmbDNF}  is very
promising in showing the right way beyond  the {\it standard model}.

\begin{itemize}
\item{It predicts  families and their mass matrices, explaining  the origin of the charges
and of the gauge fields~\cite{mksnmbDnorma,mksnmbDpikanorma,mksnmbDNF}.}
\item{It predicts that there are, before the  universe passes through the two
$SU(2) \times U(1)$ phase transitions, eight massless families of quarks and leptons~\cite{mksnmbDNF}.}
\item{It predicts  after these two phase transitions,
in which the symmetry  breaks from $SO(1,3) \times SU(2) \times SU(2) \times U(1) \times SU(3)$ first
to $SO(1,3) \times SU(2)  \times U(1) \times SU(3)$ and then to
   $SO(1,3) \times U(1) \times SU(3)$,  twice decoupled four families.
   The upper four families gain masses
   in the first phase transition, while the second four families gain masses at the electroweak break.
To these two breaks of symmetries the scalar non Abelian fields, the (superposition of the)
gauge fields of the operators generating  families, contribute~\cite{mksnmbDNF}.}
\item{The lightest of the upper four families
is stable (in comparison with the life of the universe) and is therefore a candidate for
constituting the dark matter. Their masses are so far only roughly estimated~\cite{mksnmbDgn,mksnmbDMN} to
be around a few hundred TeV/$c^2$.}
\item{The heaviest of the lower four families should be seen at the LHC or
at somewhat higher energies~\cite{mksnmbDpikanorma,mksnmbDgmdn}. The calculations
below the tree level are under considerations~\cite{mksnmbDAN}.}
\end{itemize}

There are still many open questions to be solved in this {\it spin-charge-family-theory}, which are not
directly connected with the question about the dark matter content.  Many a problem is
shared with many other theories.
This discussion  concerns mainly the question which one (if any) of the two scenarios for the
dark matter constituents has a better chance to be the right one.

Each of these two scenarios lies on the assumptions and  rough estimations.
We kindly ask the reader to learn more about the {\it spin-charge-family-theory}, their assumptions
and so far made estimations  in the two contributions of Norma in this proceedings~\cite{mksnmbDNF,mksnmbDNMA} and in the
papers cited therein. Here we will present only those estimations that are relevant for
this discussion. That is that they help to  distinguish between the two proposals for the
constituents of the dark matter, if both originating in the stable fifth family
members of the {\it spin-charge-family-theory}.

The assumptions  (many of them supported by  rough estimations) and
the open questions concerning the Norma's proposal for the dark
matter constituents, which are relevant for this discussion, are
presented below. Many an item concerns also the Maxim's proposal.
When commenting the items, which follow, also comments to some of
the open problems concerning the Maxim's proposal are added. Some
additional discussions and comments about the open questions
concerning  the Maxim's proposal
 are presented in section~\ref{mksnmbDtwoscenarios}.
\begin{enumerate}
\item{The masses of the upper four families lie above the lower four families and below the
unification scale of the three charges~\cite{mksnmbDpikanorma,mksnmbDNF,mksnmbDgn}, say,
below $10^{16}$ GeV/$c^2$.}
\item{The masses of each of the member of the upper four families, and consequently also of the
fifth family members, are approximately the same at least on the tree level~\cite{mksnmbDNF}.}
\item{Close below the weak $SU(2) \times U(1)$ phase transition the number density of the fifth
family neutrinos (with approximately the same masses  as the fifth family electrons and quarks)
is pretty much reduced due to a strong neutrino-antineutrino annihilation~\cite{mksnmbDGN}. The contribution
of the fifth family neutrinos to the dark matter is then smaller (at least not larger) than
the contribution of the
fifth family neutrons, provided that the neutrino-antineutrino asymmetry is small enough. }
\item{In the colour phase transition, which appears at the temperature close below $1$/$k_b$ GeV,
the coloured
fifth family objects either annihilate with the anti-objects or form the colourless neutrons and
anti-neutrons and correspondingly decouple from the plasma at the very beginning of
the colour phase transition due to very strong binding energies of the fifth family
baryons~\cite{mksnmbDgn,mksnmbDGNC}.}
\item{Among the open (not yet studied) questions of the {\it spin-charge-family-theory}  is
what is the origin of the matter-antimatter in the universe and how does the
fermion-antifermion asymmetry manifest in the case of the stable fifth family. }
\item{Some additional open question will be discussed
in section~\ref{mksnmbDtwoscenarios}.}

\end{enumerate}

Comments to the first assumption: This assumption is  mild, provided that the
universe really passed through two separate phase transitions of the kind as follow: First from
$SO(1,3) \times SU(2) \times SU(2) \times U(1) \times SU(3)$
to $SO(1,3) \times SU(2)  \times U(1) \times SU(3)$ and then to
   $SO(1,3) \times U(1) \times SU(3)$.

Comments to the second assumption:  Rough estimations  show~\cite{mksnmbDNF} that at least
on the tree level, but also within one or two loops calculations~\cite{mksnmbDAN}, there is not much possibility for
large mass differences  among the fifth family members. If the difference ($m_{u_{5}}- m_{d_{5}}$)
is not larger than a few hundred GeV/$c^2$ in the case of the few hundred TeV/$c^2$ quark masses,
then the neutron is the stable fifth family
member~\cite{mksnmbDgn,mksnmbDMN} due to repulsive electromagnetic interaction among quarks (and  taking into
account also the weak interaction).
For lighter quark masses   smaller differences among the masses of the members are allowed
if the fifth family neutron should be
the lightest fifth family baryon. For large enough current mass differences, that is for lighter enough
fifth family $u_{5}$-quark with respect to $d_5$-quark, it might happen that
($\bar{u}_5 \bar{u}_5 \bar{u}_5$) be the lightest fifth family antibaryon, although it does not look very
probable.

Comments to the third assumption: Besides the fifth family colourless clusters of quarks also
fifth family neutrinos  contribute to the dark matter. If the masses of these neutrinos are
a few  hundred TeV/$c^2$, approximately the same as the masses of the other fifth family members,
their contribution to the dark matter would not be in agreement with the measured dark matter
density, it would be high to huge. So far done rough estimations (which are still under
consideration) show that the
number density of the fifth family neutrinos  is pretty much reduced due to the
neutrino-antineutrino annihilation (the higher is the mass the stronger is annihilation),
provided that the fifth family neutrino-antineutrino asymmetry is small enough.
The weak annihilation cross section is expected to play much stronger role for neutrinos
than for strongly bound quarks in the neutron (due to the huge binding energy of the fifth
family quarks), which remains to be proved.

A too large fifth family neutrino-antineutrino asymmetry would lead to contradiction
with the experimental data for both proposals for the dark matter constituents, either  the  Norma's
or   the Maxim's ones.

Comments to the fourth assumption: Norma~\cite{mksnmbDgn} has estimated how do in the colour
phase transition behave  heavy  enough (with the masses of a few hundred TeV/$C^2$ or more)
coloured  fifth family objects, that is single quarks or clusters
of quarks. This estimation suggests that the coloured
objects either annihilate with the anti-objects or form the colourless neutrons and
anti-neutrons and correspondingly decouple from the plasma at the very beginning of
the colour phase transition due to  very strong binding energies of the fifth family
baryons (with respect to the first family baryons). This happens long enough before
the first family quarks start to form the baryons, so that there is negligible
amount of colourless clusters with the fifth and first family quarks.

If the masses of the fifth family
quarks would be of the order of hundreds GeV, for example, or lower, the colourless
only fifth family clusters, which would succeed to decouple from plasma, would be very rare.
Mostly they would behave similarly as the first
family coloured objects.
They would either
annihilate during the colour phase transition if there would be no
quark-antiquark asymmetry or form stable colourless objects with the first family
quarks or antiquarks.
This estimation needs to be
followed by more accurate studies to be proved.

Comments to the fifth alinea. In this proceedings there is the contribution of
Norma~\cite{mksnmbDNMA} in which she is making a first step to understand
 the matter-antimatter asymmetry in the universe
within the  {\it spin-charge-family-theory} 
if assuming that
transitions in non equilibrium processes among instanton vacua  as well as complex phases in
mixing matrices are the sources of the fermion-antifermion asymmetry, as
studied in the literature~\cite{mksnmbDgross,mksnmbDrubakovshaposhnikov,mksnmbDdinekusenko,mksnmbDtapeiling}
for several proposed theories. She is pointing out that there are several phenomena,
like: i. There are two kinds of phase transitions of the electroweak type at two
very different temperatures and with two different types of gauge vector fields. ii. There
are two in masses very different stable families. iii. There are  scalar fields, responsible
for generating masses of families at phase transitions, which also might contribute to
the variety of vacua. A deep understanding of  fermion-antifermion asymmetry within the
{\it spin-charge-family-theory} is necessary to see whether the excess of antibaryons
is for the fifth family members acceptable, or even natural, since  the first family
members manifest the excess of baryon number.

It might be, for example, that $\Delta n_{1 \alpha} +\Delta n_{5 \alpha}=0$, for each family
member $\alpha$. That would mean that at some level the total number of fermions of each species
would be conserved. Of course, one should know why does this  happen, if it happens at all.
In addition the same excess of antibaryons of the fifth family and the excess of baryons
of the first family would not be in agreement with the measured excess of the dark matter over
the ordinary matter. But the $CP$  non-conservation due to complex phases in mixing matrices
might during two phase transitions  cure this to the right direction for all the family members.

\section{Which of these two scenarios assuming  that the Norma's fifth family members constitute
the dark matter has a better chance to be the right one?}
\label{mksnmbDtwoscenarios}

Rough estimations of properties of the  fifth family members show that their mas\-ses are on the tree level
very probably a few  $100$ TeV or higher and that all the fifth
family members  have approximately the same mass,
as it is  discussed in section~\ref{mksnmbDintroduction} and in the refs.~\cite{mksnmbDgn,mksnmbDNF,mksnmbDAN,mksnmbDMN}, while
similar estimations  show that the  first family members differ
in their masses already on the tree
level (and the same is true for the remaining three of the lower four family members)~\cite{mksnmbDNF,mksnmbDpikanorma,mksnmbDgmdn}.
Next more sophisticated studies  beyond the tree level will help to better evaluate and consequently
better understand the properties of the lower four families as well as of the fifth family.
It should be found whether the fifth family members have the  masses of the order of a few
hundred TeV, all the quarks the same within the accuracy of  a few hundred GeV.
 If this is not the case and if the masses  might be lower, say
a few TeV as in the scenario of Maxim,
can then be the  $u_5$- quark mass lighter enough  then the mass of the $d_5$ quark in order
that $u_5  u_5 u_5$ would be the lightest baryon.

Studies of the fermion-antifermion asymmetry for  the {\it
spin-charge-family-theo\-ry} as discussed in
section~\ref{mksnmbDintroduction} and in the contribution~\cite{mksnmbDNMA} of
this proceedings should also show, whether there is a possibility
for the antifermion excess for the fifth family members. As we said,
if masses of the fifth family members are of a few $100$ TeV, the
fermion-antifermion asymmetry does not influence much the content of
the fifth family neutrons and antineutrons in the dark
matter~\cite{mksnmbDgn}, provided that additional studies of the behaviour
of  very heavy stable quarks during the colour phase transition of
the plasma (which starts at $Tk_b \approx 1$ GeV) will confirm that
(due to a very large binding energy of heavy quarks in colourless
clusters) (almost) all the coloured fifth family objects either
annihilate or form the colourless clusters long before the light
quarks start to form colourless clusters. It stays to prove that
indeed the amount of clusters made out of the first and the fifth
family members  which might remain after the colour phase transition
are not in contradiction with the experimental data. %
If the masses are much
lower, say below a few TeV, then the fermion-antifermion asymmetry
starts to decide how many of the fifth family baryons  survive the
colour phase transition, similarly as it is the case for the first
family quarks. Besides  that
 many of the  fifth and the first family members (like $u_5u_1d_1$) will constitute common clusters and
 might severe contradict the experimental data.
 It is evaluated in the ref.~\cite{mksnmbDMaxim}
 that the ratio of the number  of mixed  clusters of the kind $\bar{u}_5 u_1$ and
 $\bar{u}_5  \bar{u}_5 \bar{d}_1$
and the ordinary (first family) baryons should  not exceed
  $\sim 10^{-8}$. More detailed studies are  needed to see whether this is indeed happening.

In the ref.~\cite{mksnmbDgn} the authors estimate the scattering amplitudes on the ordinary
 (made mostly of the first family nuclei) matter commenting  the DAMA~\cite{mksnmbDDAMA} and
 CDMS~\cite{mksnmbDCDMS} experiments.
They conclude that, if there are (mostly, since also the fifth family neutrinos contribute as we
have discussed above) fifth family neutrons which constitute the dark matter (the study
of the behaviour of the fifth family members in the expanding universe in the ref.~\cite{mksnmbDgn} speaks for
this possibility) the CDMS  or some other direct experiment will in a few years confirm their estimations.
The new measurements with Xe  looks like to be in disagreement with the DAMA experiment, but the
analyses, presented in this Proceedings~\cite{mksnmbDGLN} suggest that we should wait for the (hopefully
soon coming) analyses of their experiment with higher statistics, which will decide, whether or
not  the DAMA, CDMS and XENON100 experiments measure our fifth family neutrons, whose masses are around
a few hundred TeV.

Once the fifth family neutrons decouple from the primordial plasma, they are not influenced much by the
following phase transitions.


This is not the case for the Maxim's scenario in which the fifth
family members have masses below a few TeV/$c^2$ and the $\bar{u}_5
\bar{u}_5 \bar{u}_5$  is the lightest fully fifth family cluster.
Many of the fifth family quarks which enter into the colour phase
transition will not succeed to form the colourless clusters from
only fifth family quarks or antiquarks before  the lower four
families start to enter in this process of forming colourless
objects. In this case the binding energy of such clusters is  namely
$100$ times smaller than in the case of quarks with a few hundred
TeV/$c^2$ mass. One should study this dynamics very carefully. For
TeV range teraquarks~\cite{mksnmbDFargion} and stable quarks of the 4th
family~\cite{mksnmbDMaxim,mksnmbDQ} the studies were done and might be used also
in the case of {\it spin-charge-family-theory}. Accordingly a
significant fraction of the  fifth family quarks (with masses of a
few TeV/$c^2$)  constitute the colourless clusters with the lower
four family members and because of the nonzero mixing matrix
elements to the lowest first family members, these clusters  would
include at the end the first and  the fifth family members. Further
history of such clusters in the expanding universe after the colour
phase transition was~\cite{mksnmbDMaxim,mksnmbDFargion,mksnmbDQ} and still should be
studied.
These studies take into account the
hadronic recombination processes, like $(\bar{u}_5 u_1) + (u_5 \bar{u}_1 \Rightarrow
 \pi^{0} + (\bar{u}_5 u_5)$,
$(\bar{u}_5 u_1) + (\bar{u}_5 \bar{u}_5 \bar{u}_1) \Rightarrow (\bar{u}_5 \bar{u}_5
\bar{u}_5) + \pi^{0}_{1}$ or
$(\bar{u}_5 u_1) + u_5 u_1 d_{-1} \Rightarrow p+ (\bar{u}_5 u_5)$,  with further decay of $(\bar{u}_5 u_5)$.
These processes reduce the number of the fifth-first family
hadrons, but their amount 
might remain significant even after the nucleosynthesis, in particular
stable $(\bar{u}_5 u_1)$. 

There will be after the colour phase transition also colourless clusters of the fifth family quarks only,
which will be (because of  the assumed antifermion-fermion asymmetry and  the assumption that the
lightest fifth family quark $u_5$ is enough lighter than $d_5$ quark)  the
$\bar{u}_5 \bar{u}_5 \bar{u}_5$.
This object has the electromagnetic charge equal to $-2$ and would probably in the phase transition
when the electromagnetically neutral objects are produced, form the electromagnetically neutral objects with
the He nucleus.
There are two main possibilities:\\
%
i. The heavy colourless object $\bar{u}_5 \bar{u}_5 \bar{u}_5$ will seat in the middle of the
He nucleus. One can easily evaluate the binding energy of this system.
\begin{eqnarray}
\label{mksnmbDu5inthecenter}
E_B= - 2 \frac{Z \alpha_{elm} \hbar c}{ <r_{He}>} \approx  -
\frac{ 4 \cdot 200 {\rm MeV}}{ 137 <r_{He}>/fm }
\approx  – 3 {\rm MeV}.
\end{eqnarray}
ii. The  light He nucleus rotates around the heavy baryon $\bar{u}_5 \bar{u}_5 \bar{u}_5$.
The estimated binding energy of such a configuration is
\begin{eqnarray}
\label{mksnmbDu5outofthecenter}
E_B= - \frac{1}{2n^2} (2Z \alpha_{elm})^2 \, m_{He} c^2
\approx  - \frac{1}{n^2} \,\frac{(2Z)^{2} \, m_{He} c^2}{2 \cdot 137^2}  \approx -2 {\rm MeV} \cdot \frac{1}{n^2}.
\end{eqnarray}
There might be that the superposition of both configurations is the
stable one. Maxim  and his co-authors are commenting both
possibilities in their contribution to this
proceedings~\cite{mksnmbDmaximBled2}. They study the behaviour of these
objects made out of the fifth family quarks, predicted by the {\it
spin-charge-family-theory}, when scattering on the ordinary (first)
family nuclei and if being measured in the direct measurement
experiments of DAMA, CDMS and XENON100.

The properties  of the fifth family members in the Norma's scenario and the behaviour of these
members when decoupling from the cosmic plasma
up to today is studying by Gregor and Norma in the ref.~\cite{mksnmbDgn}. There  also  scattering among themselves
and on the ordinary matter, together with the estimates about all possible measurable
consequences, with the interpretation of the direct measurements experiments on DAMA,CDMS and XENON100
included.

 \section{Conclusions}

 In this contribution we stressed the assumptions on which the Norma's {\it spin-charge-family-theory}
 is built, explaining briefly the properties of  by this theory predicted stable fifth family baryons and
 leptons might have. Although the rough estimations seems to speak for the Norma's prediction
 for the properties of the fifth family baryons and leptons, which have a real chance to explain
 the dark matter origin, with the fifth family members of around a few hundred TeV, it also might
 be that  the fifth family members are much lighter with the masses of around a few  TeV up to
 $10$ TeV. This latter case Maxim assumes and his group made estimation what is happening in the
 evolution after the colour phase transition. Since there are many open problems to be solved,
 none of these to scenarios are excluded. Further theoretical studies, many a study is under
 considerations, will give more clear picture whether  the {\it spin-charge-family-theory}
 is showing the right way beyond the {\it standard model}, and in particular which of the two scenarios,
 if any, is the right one.  Final decision is always on the experiment.

 %
\title{Puzzles of Dark Matter - More Light on Dark Atoms?}
\author{M.Yu. Khlopov${}^{1,2,3}$, A.G. Mayorov${}^{1}$ and E.Yu. Soldatov${}^{1}$}
\institute{%
${}^{1}$National Research Nuclear University "Moscow Engineering Physics Institute", 115409 Moscow, Russia \\
${}^{2}$ Centre for Cosmoparticle Physics "Cosmion" 115409 Moscow, Russia \\
${}^{3}$ APC laboratory 10, rue Alice Domon et L\'eonie Duquet \\75205
Paris Cedex 13, France}

\titlerunning{Puzzles of Dark Matter - More Light on Dark Atoms?}
\authorrunning{M.Yu. Khlopov, A.G. Mayorov and E.Yu. Soldatov}
\maketitle

\begin{abstract}

Positive results of dark matter searches in experiments DAMA/NaI and
DAMA/ LIBRA confronted with results of other groups can imply
nontrivial particle physics solutions for cosmological dark matter.
Stable particles with charge -2, bound with primordial helium in
O-helium "atoms" (OHe), represent a specific nuclear-interacting
form of dark matter. Slowed down in the terrestrial matter, OHe is
elusive for direct methods of underground Dark matter detection
using its nuclear recoil. However, low energy binding of OHe with
sodium nuclei can lead to annual variations of energy release from
OHe radiative capture in the interval of energy 2-4 keV in DAMA/NaI
and DAMA/LIBRA experiments. At nuclear parameters, reproducing DAMA
results, the energy release predicted for detectors with chemical
content other than NaI differ in the most cases from the one in DAMA
detector. Moreover there is no bound systems of OHe with light and
heavy nuclei, so that there is no radiative capture of OHe in
detectors with xenon or helium content.
Due to dipole Coulomb barrier, transitions to more energetic levels
of Na+OHe system with much higher energy release are suppressed in
the correspondence with the results of DAMA experiments. The
proposed explanation inevitably leads to prediction of abundance of
anomalous Na, corresponding to the signal, observed by DAMA.

\end{abstract}
\section{Introduction}
\label{contribution:mk2dm}
In our previous paper \cite{mk2rLevels} we have shown that the set of
conditions for dark matter candidates
\cite{mk2rbook,mk2rCosmoarcheology,mk2rBled07}can be satisfied for new stable
charged particles, if they are hidden in neutral atom-like states.
To avoid anomalous isotopes overproduction, stable particles with
charge -1 (like tera-electrons \cite{mk2rGlashow,mk2rFargion:2005xz}) should
be absent, so that stable negatively charged particles should have
charge -2 only. In the row of possible models, predicting such
particles
\cite{mk2rI,mk2rlom,mk2rKhlopov:2006dk,mk2rQ,mk2rKhlopov:2006dk,mk2r5,mk2rFKS,mk2rKK,mk2rKhlopov:2008rp}
stable charged clusters $\bar u_5 \bar u_5 \bar u_5$ of (anti)quarks
$\bar u_5$ of 5th family from the {\it spin-charge-family-theory}
\cite{mk2rNorma} can also find their place (see \cite{mk2rDiscussion}).

In the asymmetric case, corresponding to excess of -2 charge
species, $X^{--}$, they bind in "dark atoms" with primordial $^4He$
as soon as it is formed in the Standard Big Bang Nucleosynthesis. We
call such dark atoms O-helium ($OHe$) \cite{mk2rI2} and assume that they
are the dominant form of the modern dark matter.

Here we concentrate on effects of O-helium dark matter in
underground detectors. We present qualitative confirmation of the
earlier guess \cite{mk2rI2,mk2rKK2,mk2riwara,mk2runesco} that the positive results
of dark matter searches in DAMA/NaI (see for review
\cite{mk2rBernabei:2003za}) and DAMA/LIBRA \cite{mk2rBernabei:2008yi}
experiments can be explained by O-helium, resolving the controversy
between these results and the results of other experimental groups.

\section{Radiative capture of OHe in the underground detectors}
\subsection{O-helium in the terrestrial matter} The evident
consequence of the O-helium dark matter is its inevitable presence
in the terrestrial matter, which appears opaque to O-helium and
stores all its in-falling flux.

After they fall down terrestrial surface, the in-falling $OHe$
particles are effectively slowed down due to elastic collisions with
matter. Then they drift, sinking down towards the center of the
Earth with velocity \begin{equation} V = \frac{g}{n \sigma v} \approx 80 S_3
A^{1/2} \cm/\s. \label{mk2rdif}\end{equation} Here $A \sim 30$ is the average
atomic weight in terrestrial surface matter, $n=2.4 \cdot 10^{24}/A \cm^{-3}$
is the number density of terrestrial atomic nuclei, $\sigma v$ is the rate
of nuclear collisions, $m_o \approx M_X+4m_p=S_3 \TeV$ is the mass of O-helium,
$M_X$ is the mass of the $X^{--}$ component of O-helium, $m_p$ is the mass of proton and $g=980~ \cm/\s^2$.

Near the Earth's surface, the O-helium abundance is determined by
the equilibrium between the in-falling and down-drifting fluxes.

The in-falling O-helium flux from dark matter halo is
$$
  F=\frac{n_{0}}{8\pi}\cdot |\overline{V_{h}}+\overline{V_{E}}|,
$$
where $V_{h}$-speed of Solar System (220 km/s), $V_{E}$-speed of
Earth (29.5 km/s) and $n_{0}=3 \cdot 10^{-4} S_3^{-1} \cm^{-3}$ is the
local density of O-helium dark matter. For qualitative estimation we
don't take into account here velocity dispersion and distribution of particles
in the incoming flux that can lead to significant effect.

At a depth $L$ below the Earth's surface, the drift timescale is
$t_{dr} \sim L/V$, where $V \sim 400 S_3 \cm/\s$ is given by
Eq.~(\ref{mk2rdif}). It means that the change of the incoming flux,
caused by the motion of the Earth along its orbit, should lead at
the depth $L \sim 10^5 \cm$ to the corresponding change in the
equilibrium underground concentration of $OHe$ on the timescale
$t_{dr} \approx 2.5 \cdot 10^2 S_3^{-1}\s$.

The equilibrium concentration, which is established in the matter of
underground detectors at this timescale, is given by
\begin{equation}
    n_{oE}=\frac{2\pi \cdot F}{V} = n_{0}\frac{n \sigma v}{4g} \cdot
    |\overline{V_{h}}+\overline{V_{E}}|,
\end{equation}
where, with account for $V_{h} > V_{E}$, relative velocity can be
expressed as
$$
    |\overline{V_{o}}|=\sqrt{(\overline{V_{h}}+\overline{V_{E}})^{2}}=\sqrt{V_{h}^2+V_{E}^2+V_{h}V_{E}sin(\theta)} \simeq
$$
$$
\simeq V_{h}\sqrt{1+\frac{V_{E}}{V_{h}}sin(\theta)}\sim
V_{h}(1+\frac{1}{2}\frac{V_{E}}{V_{h}}sin(\theta)).
$$
Here $\theta=\omega (t-t_0)$ with $\omega = 2\pi/T$, $T=1yr$ and
$t_0$ is the phase. Then the concentration takes the form
\begin{equation}
    n_{oE}=n_{oE}^{(1)}+n_{oE}^{(2)}\cdot sin(\omega (t-t_0))
    \label{mk2rnoE}
\end{equation}

So, there are two parts of the signal: constant and annual
modulation, as it is expected in the strategy of dark matter search
in DAMA experiment \cite{mk2rBernabei:2008yi}.

\subsection{Radiative capture of O-helium by sodium}

In the essence, our explanation of the results of experiments
DAMA/NaI and DAMA/LIBRA is based on the idea that OHe, slowed down
in the terrestrial matter and present in the matter of DAMA
detectors, can form a few keV bound state with sodium nuclei, in
which OHe is situated \textbf{beyond} the nucleus. Radiative capture
to this bound state leads to the corresponding energy release and
ionization signal, detected in DAMA experiments.

The rate of radiative capture of OHe by nuclei can be calculated
\cite{mk2riwara,mk2runesco} with the use of the analogy with the radiative
capture of neutron by proton with the account for: i) absence of M1
transition that follows from conservation of orbital momentum and
ii) suppression of E1 transition in the case of OHe. Since OHe is
isoscalar, isovector E1 transition can take place in OHe-nucleus
system only due to effect of isospin nonconservation, which can be
measured by the factor $f = (m_n-m_p)/m_N \approx 1.4 \cdot
10^{-3}$, corresponding to the difference of mass of neutron,$m_n$,
and proton,$m_p$, relative to the mass of nucleon, $m_N$. In the
result the rate of OHe radiative capture by nucleus with atomic
number $A$ and charge $Z$ to the energy level $E$ in the medium with
temperature $T$ is given by
\begin{equation}
    \sigma v=\frac{f \pi \alpha}{m_p^2} \frac{3}{\sqrt{2}} (\frac{Z}{A})^2 \frac{T}{\sqrt{Am_pE}}.
    \label{mk2rradcap}
\end{equation}

Formation of OHe-nucleus bound system leads to energy release of its
binding energy, detected as ionization signal.  In the context of
our approach the existence of annual modulations of this signal in
the range 2-6 keV and absence of such effect at energies above 6 keV
means that binding energy of Na-OHe system in DAMA experiment should
not exceed 6 keV, being in the range 2-4 keV. The amplitude of
annual modulation of ionization signal (measured in counts per day
per kg, cpd/kg) is given by
\begin{equation}
\zeta=\frac{3\pi \alpha \cdot n_o N_A V_E t Q}{640\sqrt{2}
A_{med}^{1/2} (A_I+A_{Na})} \frac{f}{S_3 m_p^2} (\frac{Z_i}{A_i})^2
\frac{T}{\sqrt{A_i m_p E_i}}= a_i\frac{f}{S_3^2} (\frac{Z_i}{A_i})^2
\frac{T}{\sqrt{A_i m_p E_i}}. \label{mk2rcounts}
\end{equation}
Here $N_A$ is Avogadro number, $i$ denotes Na, for which numerical
factor $a_i=4.3\cdot10^{10}$, $Q=10^3$ (corresponding to 1kg of the
matter of detector), $t=86400 \s$, $E_i$ is the binding energy of
Na-OHe system and $n_{0}=3 \cdot 10^{-4} S_3^{-1} \cm^{-3}$ is the
local density of O-helium dark matter near the Earth. The value of
$\zeta$ should be compared with the integrated over energy bins
signals in DAMA/NaI and DAMA/LIBRA experiments and the result of
these experiments can be reproduced for $E_{Na} = 3 \keV$. The
account for energy resolution in DAMA experiments \cite{mk2rDAMAlibra}
can explain the observed energy distribution of the signal from
monochromatic photon (with $E_{Na} = 3 \keV$) emitted in OHe
radiative capture.

At the corresponding values of $\mu$ and $g^2$ there is no binding
of OHe with iodine and thallium \cite{mk2rLevels}.

It should be noted that the results of DAMA experiment exhibit also
absence of annual modulations at the energy of MeV-tens MeV. Energy
release in this range should take place, if OHe-nucleus system comes
to the deep level inside the nucleus. This transition implies
tunneling through dipole Coulomb barrier and is suppressed below the
experimental limits.

\subsection{OHe radiative capture by other nuclei}

For the chosen range of nuclear parameters, reproducing the results
of DAMA/NaI and DAMA/LIBRA, our results  \cite{mk2rLevels} indicate that
there are no levels in the OHe-nucleus systems for heavy nuclei. In
particular, there are no such levels in Xe and most probably in Ge,
what seem to prevent direct comparison with DAMA results in CDMS and
XENON100 experiments. However, even in this case presence of silicon
in the chemical composition of CDMS set-up can provide some
possibility for test of OHe interpretation of these results. The
levels in Si-OHe system were calculated in \cite{mk2rLevels}. The two
sets of solutions were obtained for each of approximation in
description of Yukawa potential:
\begin{itemize}
\item[i] the case (m) for nuclear Yukawa potential $U_{3m}$, averaged over the
orbit of He in OHe,
\item[ii] the case (b) of the nuclear Yukawa potential
$U_{3b}$ with the position of He most close to the nucleus.
\end{itemize}
These two approximations correspond to the larger and smaller
distance effects of nuclear force, respectively, so that the true
picture should be between these two extremes.

For the parameters, reproducing results of DAMA experiment the
predicted energy level of OHe-silicon bound state is generally
beyond the range 2-6 keV, being in the most cases in the range of
30-40 keV or 90-110 keV by absolute value. It makes elusive a
possibility to test DAMA results by search for ionization signal in
the same range 2-6 keV in other set-ups with content that differs
from Na and I. Even in the extreme case (m) of ionization signal in
the range 2-6 keV our approach naturally predicts its suppression in
accordance with the results of CDMS \cite{mk2rKamaev:2009gp}.

It should be noted that strong sensitivity of the existence of the
OHe-Ge bound state to the values of numerical factors \cite{mk2rLevels}
doesn't exclude such state for some window of nuclear physics
parameters. The corresponding binding energy  would be about 450-460
keV, what proves the above statement even in that case.

Since OHe capture rate is proportional to the temperature, it looks
like it is suppressed in cryogenic detectors by a factor of order
$10^{-4}$. However, for the size of cryogenic devices  less, than
few tens meters, OHe gas in them has the thermal velocity of the
surrounding matter and the suppression relative to room temperature
is only $\sim m_A/m_o$. Then the rate of OHe radiative capture in
cryogenic detectors is given by Eq.(\ref{mk2rradcap}), in which room
temperature $T$ is multiplied by factor $m_A/m_o$, and the
ionization signal (measured in counts per day per kg, cpd/kg) is
given by Eq.(\ref{mk2rcounts}) with the same correction for $T$
supplemented by additional factors $2 V_h/V_E$ and
$(A_I+A_{Na})/A_i$, where $i$ denotes Si.
 To illustrate possible effects of OHe in
various cryogenic detectors we give in Tables~\ref{mk2rta1} and
\ref{mk2rta2} energy release, radiative capture rate and counts per day
per kg for the pure silicon for the preferred values of nuclear
parameters.

\begin{table}

\center
\begin{tabular}{|c|c|c|c|c|c|c|}
    \hline
        $g^2/\mu^2, GeV^{-1}$ & 242 & 242 & 257 & 257 & 395 & 395\\
    \hline
        Energy, $keV$ & 2.7 & 31.9 & 3.0 & 33.2 & 6.1 & 41.9\\
    \hline
        $\sigma V \cdot 10^{-33}, cm^3/s$ & 19.3 & 5.6 & 18.3 & 5.5 & 12.8 & 4.9\\
    \hline
        $\xi \cdot 10^{-2}, cpd/kg$ & 10.8 & 3.1 & 10.2 & 3.1 & 7.2 & 2.7\\
    \hline
\end{tabular}
\caption{\label{mk2rta1} Effects of OHe in pure silicon cryogenic detector in the
case m for nuclear Yukawa potential $U_{3m}$, averaged over the
orbit of He in OHe \cite{mk2rLevels}.}

\end{table}

\begin{table}

\center
\begin{tabular}{|c|c|c|c|c|c|c|}
    \hline
        $g^2/\mu^2, GeV^{-1}$ & 242 & 242 & 257 & 257 & 395 & 395\\
    \hline
        Energy, $keV$ & 29.8 & 89.7 & 31.2 & 92.0 & 42.0 & 110.0\\
    \hline
        $\sigma V \cdot 10^{-33}, cm^3/s$ & 5.8 & 3.3 & 5.7 & 3.3 & 4.9 & 3.0\\
    \hline
        $\xi \cdot 10^{-2}, cpd/kg$ & 3.3 & 1.9 & 3.2 & 1.9 & 2.7 & 1.7\\
    \hline
\end{tabular}
\caption{\label{mk2rta2} Effects of OHe in pure silicon cryogenic detector for the
case of the nuclear Yukawa potential $U_{3b}$ with the position of
He most close to the nucleus \cite{mk2rLevels}.}

\end{table}

\section{Conclusions}

The results of dark matter search in experiments DAMA/NaI and
DAMA/LIBRA can be explained in the framework of our scenario without
contradiction with the results of other groups. This scenario can be
realized in different frameworks, in particular, in the extensions
of Standard Model, based on the approach of almost commutative
geometry, in the model of stable quarks of 4th generation that can
be naturally embedded in the heterotic superstring phenomenology, in
the models of stable technileptons and/or techniquarks, following
from Minimal Walking Technicolor model or in the approach unifying
spin and charges. Our approach contains distinct features, by which
the present explanation can be distinguished from other recent
approaches to this problem \cite{mk2rEdward} (see also for review and
more references in \cite{mk2rGelmini}).

The proposed explanation is based on the mechanism of low energy
binding of OHe with nuclei. Within the uncertainty of nuclear
physics parameters there exists a range at which OHe binding energy
with sodium is in the interval 2-4 keV. Radiative capture of OHe to
this bound state leads to the corresponding energy release observed
as an ionization signal in DAMA detector.

OHe concentration in the matter of underground detectors is
determined by the equilibrium between the incoming cosmic flux of
OHe and diffusion towards the center of Earth. It is rapidly
adjusted and follows the change in this flux with the relaxation
time of few minutes. Therefore the rate of radiative capture of OHe
should experience annual modulations reflected in annual modulations
of the ionization signal from these reactions.

An inevitable consequence of the proposed explanation is appearance
in the matter of DAMA/NaI or DAMA/LIBRA detector anomalous
superheavy isotopes of sodium, having the mass roughly by $m_o$
larger, than ordinary isotopes of these elements. If the atoms of
these anomalous isotopes are not completely ionized, their mobility
is determined by atomic cross sections and becomes about 9 orders of
magnitude smaller, than for O-helium. It provides their conservation
in the matter of detector. Therefore mass-spectroscopic analysis of
this matter can provide additional test for the O-helium nature of
DAMA signal. Methods of such analysis should take into account the
fragile nature of OHe-Na bound states, since their binding energy is
only few keV.

With the account for high sensitivity of the numerical results to
the values of nuclear parameters and for the approximations, made in
the calculations, the presented results can be considered only as an
illustration of the possibility to explain puzzles of dark matter
search in the framework of composite dark matter scenario. An
interesting feature of this explanation is a conclusion that the
ionization signal expected in detectors with the content, different
from NaI, should be dominantly in the energy range beyond 2-6 keV.

Moreover, it is shown that in detectors, containing light nuclei
(e.g. helium-3) and heavy nuclei (e.g. xenon) there should be no
bound states with OHe. In the framework of our approach it means
that the physical nature of effects, observed in DAMA/NaI and
DAMA/LIBRA experiments, cannot be probed in XENON10, XENON100
experiments or in the future detectors with He-3 content. Test of
the nature of these results in CDMS experiment should take into
account the difference in energy release and rate of radiative
capture of OHe by silicon as well as in Ge, if OHe-Ge bound state
does exist. The uncertainty in the existence of OHe-Ge bound state
makes problematic direct test of our model in pure germanium
detectors.  It should be noted that the excess of low energy events
reported in the CoGent experiment can be hardly explained by
radiative capture of OHe. Therefore test of results of DAMA/NaI and
DAMA/LIBRA experiments by other experimental groups can become a
very nontrivial task.

It is interesting to note that in the framework of our approach
positive result of experimental search for WIMPs by effect of their
nuclear recoil would be a signature for a multicomponent nature of
dark matter. Such OHe+WIMPs multicomponent dark matter scenarios
that naturally follow from AC model \cite{mk2rFKS} and from models of
Walking technicolor \cite{mk2rKK2} can be also realized as OHe
(dominant)+5th neutrino (sub-dominant) model in the framework of
{\it spin-charge-family-theory} \cite{mk2rNorma} (see
\cite{mk2rDiscussion}).

The presented approach sheds new light on the physical nature of
dark matter. Specific properties of dark atoms and their
constituents are challenging for the experimental search. The
development of quantitative description of OHe interaction with
matter confronted with the experimental data will provide the
complete test of the composite dark matter model.


\title{Families of Spinors in $d=(1+5)$ Compactified on an Infinite Disc with 
the Zweibein Which Makes a Disc Curved on $S^2$ and a Possibility for Masslessness}
\author{D. Lukman${}^1$, N.S. Manko\v c Bor\v stnik${}^1$ and H.B. Nielsen${}^2$} 
\institute{%
${}^1$Department of Physics, FMF, University of Ljubljana,\\
Jadranska 19, SI-1000 Ljubljana, Slovenia\\
${}^2$Department of Physics, Niels Bohr Institute,
Blegdamsvej 17,\\
Copenhagen, DK-2100}

\titlerunning{Families of Spinors in $d=(1+5)$ Compactified\ldots}
\authorrunning{D. Lukman, N.S. Manko\v c Bor\v stnik and H.B. Nielsen}
\maketitle

\begin{abstract} 
In the ref.~\cite{dhnhn} we present the case of a spinor in $d=(1+5)$ compactified on 
an (formally) infinite disc with 
the zweibein which makes a disc curved on $S^2$ and with the spin connection 
field  which allows on such a sphere only one massless spinor state of a particular charge, 
which couples the spinor chirally to the corresponding Kaluza-Klein gauge field. 
In this contribution we include in this toy model also families,
as proposed by the {\it theory unifying spin and charges and 
predicting families}~\cite{dhnnorma92939495,dhnholgernorma20023,dhnNF} proposed by N.S.Manko\v c Bor\v stnik. 
We study possible masslessness 
of spinors and their properties following mostly the assumptions and derivations of the 
ref.~\cite{dhnhn} with the definition of families as 
proposed in the refs.~\cite{dhnnorma92939495,dhnholgernorma20023,dhnNF}.
\end{abstract}

\section{Introduction}
\label{dhnintroduction}

\label{contribution:dhn}
The idea of Kaluza and Klein~\cite{dhnkk} was almost killed by the "no-go theorem" of 
E. Witten~\cite{dhnwitten}  telling that 
these kinds of Kaluza-Klein[like] theories with the gravitational fields only (that is with 
vielbeins and spin connections) have  severe difficulties with obtaining  massless fermions 
chirally coupled to the Kaluza-Klein-type gauge fields in $d=1+3$, as well as with the 
appearance of families,
as required by the {\it standard model}.

In this contribution there are families of spinors in $d=(1+5)$ included in the spinor action 
as proposed by the {\it theory of unifying spin and charges and predicting families} of the author
N.S. Manko\v c Bor\v stnik~\cite{dhnnorma92939495,dhnholgernorma20023,dhnNF} 
(we shall call it the {\it spin-charge-family-theory}). 
The space we use here is the same as in the ref.~\cite{dhnhn}. We  roll up the 
two extra dimensions  by a zweibein to $S^2$ with one point - 
the south pole - excluded. Thus the space is  non compact. The  volume is finite and    
suggests accordingly  that the usually 
expected problem with extra non compact dimensions  
having a continuous spectrum is not present in our model.
Correspondingly  the "no-go theorem" of E. Witten should not be valid (because of a special  
singularity at the south pole)  and we were able in the ref.~\cite{dhnhn}  
to achieve masslessness for an appropriate 
choice of the spin connection  fields, which are the part of the gravitational gauge fields 
as the vielbeins are. We did not need  the presence of the external (additional) gauge fields.

As it is not difficult to recognize, the two dimensional spaces are very special~\cite{dhnmil,dhndeser}.  
Namely, in dimensions higher than two, 
when we have no fermions present and only  the curvature in the first power in the Lagrange 
density, the spin connections are normally determined from the vielbein fields and   
the torsion is zero. In the two dimensional spaces, the vielbeins do not determine 
the spin connection fields.

In the here proposed types of models there is the chance for having chirally mass protected fermions 
in a theory in which the chirally 
protecting effective four dimensional gauge fields are {\it true} Kaluza-Klein[like] fields,  
the degrees of which inherit from the higher dimensional gravitational ones.
We are thus hoping for a revival of true Kaluza-Klein[like] models as candidates for 
phenomenologically viable models!

One of us has been trying for long to develop the  {\it approach unifying spins and charges} 
(N.S.M.B.)~\cite{dhnnorma92939495,dhnholgernorma20023} so that 
spinors which carry in $d\ge 4$ nothing but two kinds of the spin (no charges), would manifest 
in $d=(1+3)$ all the properties assumed by the {\it standard model} and does accordingly 
share with the Kaluza-Klein[like] theories the problem of masslessness of the fermions 
before the final breaks~\cite{dhnNF}. 

In this contribution we take into account that there are two kinds of the Clifford algebra objects. 
Beside the Dirac $\gamma^a$ also $\tilde{\gamma}^a$  of Norma. Correspondingly the 
covariant momentum  
\begin{eqnarray}
p_{0 s} &=& f^{\sigma}{\!}_{s}p_{0 \sigma}, \quad p_{0 \sigma} \psi = 
p_{ \sigma} - \frac{1}{2} S^{t t'} 
\omega_{t t' \sigma} - \frac{1}{2} \tilde{S}^{cd} 
\tilde{\omega}_{cd \sigma} \nonumber\\
a &=& 0,1,2,3,5,6,\quad \alpha=(0),(1),(2),(3),(5),(d),\nonumber\\
 \quad s&=&5,6,\quad \sigma =(5),(6). 
\label{dhncovp}
\end{eqnarray}
While $S^{ab}$ define the spin in and the charge in $(1+3)$, define $\tilde{S}^{ab}$ 
(forming the equivalent representations~\cite{dhnholgernorma20023} with respect to $S^{ab}$, 
$\{\tilde{\gamma}^a, \tilde{\gamma}^b\}_{+}= 2 \, \eta^{ab}$, 
$\{\gamma^a, \tilde{\gamma}^b\}_{+}=0$) the families of spinors.
A spinor carries  in $d\ge 4$  accordingly two kinds of the spin and interacts  with 
two corresponding kinds of the gauge fields, the gauge fields of  $S^{ab} = \frac{i}{4}
\,(\gamma^a \gamma^b - \gamma^b \gamma^a)$ and the gauge fields of 
$\tilde{S}^{ab} = \frac{i}{4} \,(\tilde{\gamma}^a \tilde{\gamma}^b 
- \tilde{\gamma}^b \tilde{\gamma}^a)$.

We make a choice of both kinds of spin connection fields in a similar way 
$f^{\sigma}{}_{s} \tilde{\omega}_{ab\sigma} = f^{\sigma}{}_{s} \omega_{56\sigma} \,
\frac{\tilde{F}_{ab}}{F}$, with $f^{\sigma}{}_{s'}\, \omega_{st \sigma} = i F\, f \, \varepsilon_{st}\; 
  \frac{e_{s' \sigma} x^{\sigma}}{(\rho_0)^2}$, while zweibein is chosen to curve the infinite disc
  on $S^2$, and study whether 
we can make some of families massless, mass protected and chirally coupled through the 
charge defined by $S^{56}$ to the corresponding Kaluza-Klein gauge field.

\section{The action}
\label{dhns:action}

We shall follow to high extend the derivations present in the ref.\cite{dhnhn} pointing out those differences, 
which appear due to the inclusion of families.
Let us  start in the $2(2n+1)$-dimensional space with gravity only,  
described by the action 
  \begin{eqnarray}
         {\cal S} = \alpha \int \; d^d x \, E  (\alpha {\cal\,  R} + \tilde{\alpha} \tilde{{\cal\,  R}}).
  \label{dhnaction}
  \end{eqnarray}
with the two Riemann scalars,  ${\cal R} = {\cal R}_{abcd}\,\eta^{ac}\eta^{bd}$and 
                               $\tilde{{\cal R}} = \tilde{{\cal R}}_{abcd}\,\eta^{ac}\eta^{bd}$,
                               determined by the Riemann tensors
%
$     {\cal R}_{abcd}      = \frac{1}{2}\, f^{\alpha}{}_{[a} f^{\beta}{}_{b]}(\omega_{cd \beta, \alpha} 
- \omega_{ce \alpha} \omega^{e}{}_{d \beta} ) + h.c.\;$, 
$\tilde{{\cal R}}_{abcd}  =  \frac{1}{2}\, f^{\alpha [ a} f^{\beta b ]} \; (\tilde{\omega}_{cd \beta,\alpha} - 
\tilde{\omega}_{c e \alpha} \tilde{\omega}^{e}{}_{d \beta}) + h.c.\;. $
%
with vielbeins $f^{\alpha}{\!}_{a}$~\footnote{$f^{\alpha}{}_{a}$ are inverted 
vielbeins to 
$e^{a}{}_{\alpha}$ with the properties $e^a{}_{\alpha} f^{\alpha}{\!}_b = \delta^a{}_b,\; 
e^a{}_{\alpha} f^{\beta}{}_a = \delta^{\beta}_{\alpha} $. 
Latin indices  
$a,b,..,m,n,..,s,t,..$ denote a tangent space (a flat index),
while Greek indices $\alpha, \beta,..,\mu, \nu,.. \sigma,\tau ..$ denote an Einstein 
index (a curved index). Letters  from the beginning of both the alphabets
indicate a general index ($a,b,c,..$   and $\alpha, \beta, \gamma,.. $ ), 
from the middle of both the alphabets   
the observed dimensions $0,1,2,3$ ($m,n,..$ and $\mu,\nu,..$), indices from 
the bottom of the alphabets
indicate the compactified dimensions ($s,t,..$ and $\sigma,\tau,..$). 
We assume the signature $\eta^{ab} =
diag\{1,-1,-1,\ldots,-1\}$.
}, the gauge fields of the infinitesimal generators of translation, and  spin connections $\omega_{ab\alpha}$
the gauge fields of the $S^{ab}= \frac{i}{4}(\gamma^a \gamma^b - \gamma^b \gamma^a)$ and 
$\tilde{\omega}_{ab\alpha}$
the gauge fields of the $\tilde{S}^{ab}= \frac{i}{4}(\tilde{\gamma}^a \tilde{\gamma}^b - 
\tilde{\gamma}^b \tilde{\gamma}^a)$. 
$[a\,\,b]$ means that the antisymmetrization must be performed over the two indices $a$ and $b$.

We make a choice for $d=2$ of a zweibein, which curves an infinite disc 
(a two dimensional infinite plane with the rotational symmetry around the 
axes perpendicular to the plane) into 
a sphere $S^2$ with the radius $\rho_{0}$ and  with a hole in the southern pole , just as we did 
in the ref.~\cite{dhnhn}
\begin{eqnarray}
e^{s}{}_{\sigma} = f^{-1}
\begin{pmatrix}1  & 0 \\
 0 & 1 
 \end{pmatrix},
f^{\sigma}{}_{s} = f
\begin{pmatrix}1 & 0 \\
0 & 1 \\
\end{pmatrix},
\label{dhnfzwei}
\end{eqnarray}
with 
\begin{eqnarray}
\label{dhnf}
f &=& 1+ (\frac{\rho}{2 \rho_0})^2= \frac{2}{1+\cos \vartheta},\nonumber\\ 
x^{(5)} &=& \rho \,\cos \phi,\quad  x^{(6)} = \rho \,\sin \phi, \quad E= f^{-2}.
\end{eqnarray}
The angle $\vartheta$ is the ordinary azimuthal angle on a sphere. 
The last relation follows  from $ds^2= 
e_{s \sigma}e^{s}{}_{\tau} dx^{\sigma} dx^{\tau}= f^{-2}(d\rho^{2} + \rho^2 d\phi^{2})$.
We use indices $s,t=5,6$ to describe the flat index in the space of an infinite plane, and 
$\sigma, \tau = (5), (6), $ to describe the Einstein index.  
$\phi$ determines the angle of rotations around  the axis through the two poles of a sphere, 
while $\rho = 2 \rho_0 \, \sqrt{\frac{1- \cos \vartheta}{1+ \cos \vartheta}}$. 
The volume of this noncompact sphere is finite, equal to $V= \pi\, (2 \rho_0)^2$.  The symmetry 
of $S^2$ is a symmetry of $U(1)$ group. We  look for  chiral fermions on this sphere, that is 
the fermions of 
only one handedness and accordingly mass protected, without including any extra fundamental gauge 
fields to the action from Eq.(\ref{dhnaction}). 


We make a choice  of the spin connection fields as follows
\begin{eqnarray}
  f^{\sigma}{}_{s'}\, \omega_{st \sigma} &=& i F\, f \, \varepsilon_{st}\; 
  \frac{e_{s' \sigma} x^{\sigma}}{(\rho_0)^2}\, 
  ,\quad s=5,6,\,\,\; \sigma=(5),(6), \nonumber\\
  f^{\sigma}{}_{s'}\, \tilde{\omega}_{st \sigma} &=& i \tilde{F}_{56}\, f \, \varepsilon_{st}\; 
  \frac{e_{s' \sigma} x^{\sigma}}{(\rho_0)^2}\, ,\nonumber\\
  f^{\sigma}{}_{s'} \tilde{A}^{\pm}_\sigma &=& i \tilde{F}^{\pm} \, 
  f \, \frac{e_{s' \sigma} x^{\sigma}}{(\rho_0)^2}\, , 
\label{dhnomegas}
\end{eqnarray}
with 
\begin{eqnarray}
\label{dhnso13}
\vec{\tilde{N}}_{\pm}&=&\frac{1}{2} (\tilde{S}^{23}\pm i \tilde{S}^{01},\tilde{S}^{31}\pm i \tilde{S}^{02}, 
\tilde{S}^{12}\pm i \tilde{S}^{03} )\, \nonumber\\
f^{\sigma}{}_{s}\tilde{A}^{\pm i}_{\sigma} &=& f^{\sigma}{}_{s} \{
(\tilde{\omega}_{23 \sigma} \mp i \tilde{\omega}_{01 \sigma}), 
(\tilde{\omega}_{31 \sigma} \mp i \tilde{\omega}_{02 \sigma}),
(\tilde{\omega}_{12 \sigma} \mp i \tilde{\omega}_{03 \sigma})\}\, .
\end{eqnarray}

The above choices of both kinds of the spin connection fields allow for an interval of values 
of $F$,  $\tilde{F}_{56}$ and 
$\tilde{F}^{\pm}$ 
massless spinors of a particular handedness and charge on such 
$S^2$, as we shall see in sect.~\ref{dhnequations}. 
Accordingly, if we have families of Weyl spinors in $d=(1+5)$-dimensional space, and 
this space breaks into $M^{1+3}$ cross an infinite disc, which by a zweibein is curved on $S^2$,
then at least for a particular choices for the spin connections from Eq.~(\ref{dhnomegas}) we 
know the solutions for the gauge 
fields fulfilling the equations of motion for the action linear in the curvature, 
where the vielbein and  
spin connection guarantee masslessness of spinors in the space $d=(1+3)$. 
This is possible, since for $d=2$ the spin connections and zweibein can in the absence of fermions 
be chosen independently (see the ref.~\cite{dhnhn}).

Let us write down now the Lagrange density for Weyl spinor families 
$${\cal L}_{W} = \frac{1}{2} [(\psi^{\dagger} E \gamma^0 \gamma^a p_{0a} \psi) + 
(\psi^{\dagger} E \gamma^0\gamma^a p_{0 a}
\psi)^{\dagger}],$$ 
leading to  
\begin{eqnarray}
{\cal L}_{W}&=& \psi^{\dagger}\, \gamma^0 \gamma^a E \{f^{\alpha}{}_a p_{\alpha} +
\frac{1}{2E} \{p_{\alpha},f^{\alpha}{}_a E\}_-   -\frac{1}{2} S^{cd}  \omega_{cda}
- \frac{1}{2} \tilde{S}^{cd} 
\tilde{\omega}_{cda}
 \}\psi,\nonumber\\
\label{dhnweylL}
\end{eqnarray}
with $ E = \det(e^a{\!}_{\alpha}) $ and with both kinds of spin connection fields on a disc 
as written in Eq.(~\ref{dhnomegas}).  
%
%
Let us have no gravity in $d=(1+3)$ ($f^{\mu}{}_m = \delta^{\mu}_m$ and  
$\omega_{mn\mu}=0$ for $ m,n=0,1,2,3,\ldots; \;,  \mu =0,1,2,3, \ldots $) and let us make  a choice of  a 
zweibein and spin connection on our disc as written in Eqs.~(\ref{dhnf},\ref{dhnomegas}). 
($S^2$ does not break the rotational symmetry on the disc, it breaks the translational symmetry 
after making a choice of the northern and southern pole.)

Then the  equations of motion for spinors (the Weyl equations) which follow from the Lagrange density 
of Eq.~(\ref{dhnweylL}) are 
\begin{eqnarray}
&&\{E\gamma^0 \gamma^m p_m + E f \gamma^0 \gamma^s \delta^{\sigma}_s  ( p_{0\sigma} 
+  \frac{1}{2 E f}
\{p_{\sigma}, E f\}_- )\}\psi=0,\quad {\rm with} \nonumber\\
&& p_{0\sigma} = p_{\sigma}- \frac{1}{2} S^{st}\omega_{st \sigma} - \frac{1}{2} \tilde{S}^{st} 
\tilde{\omega}_{st \sigma},
\label{dhnweylE}
\end{eqnarray}
with $f$ from  Eq.~(\ref{dhnf})
and with $ \omega_{st \sigma}$ from Eq.~(\ref{dhnomegas}).

\section{Solutions of the equations of motion  for spinors}
\label{dhnequations}

%
%
The  solution  of the equations of motion (\ref{dhnweylE}) for a spinor   
in $(1+5)$-dimensional space, which breaks into  
$M^{(1+3)}$ and a noncompact $S^2$, should be written as a superposition
of all  four ($2^{6/2 -1}=4$) states of a single Weyl representation (one family)  
as well as the superposition of all the families being correlated through the equations of motion.  
(We kindly ask the reader to see the technical details  about how to write 
a Weyl representation in terms of the Clifford algebra objects after making a choice of 
the Cartan subalgebra,  
for which we take: $S^{03}, S^{12}, S^{56}$ in the refs.~\cite{dhnholgernorma20023,dhnhn,dhnNF}.)
There are $2^{6/2 -1}=4$ families. 
In our technique~\cite{dhnholgernorma20023}, where the states are defined as a product of nilpotents 
$\stackrel{ab}{(\pm i)}: = \frac{1}{2}(\gamma^a \mp  \gamma^b)$ and projectors  
$\stackrel{ab}{[\pm i]}: = \frac{1}{2}(1 \pm \gamma^a \gamma^b)$ 
\begin{eqnarray}
\stackrel{ab}{(\pm i)}: &=& \frac{1}{2}(\gamma^a \mp  \gamma^b),  \; 
\stackrel{ab}{[\pm i]}: = \frac{1}{2}(1 \pm \gamma^a \gamma^b), \quad
{\rm for} \,\; \eta^{aa} \eta^{bb} = -1, \nonumber\\
\stackrel{ab}{(\pm )}: &= &\frac{1}{2}(\gamma^a \pm i \gamma^b),  \; 
\stackrel{ab}{[\pm ]}: = \frac{1}{2}(1 \pm i\gamma^a \gamma^b), \quad
{\rm for} \,\; \eta^{aa} \eta^{bb} =1,
\label{dhnsnmb:eigensab}
\end{eqnarray} 
which are the eigen vectors  of $S^{ab}$ as well as of $\tilde{S}^{ab}$ as follows
\begin{eqnarray}
S^{ab} \stackrel{ab}{(k)} =  \frac{k}{2} \stackrel{ab}{(k)}, \quad 
S^{ab} \stackrel{ab}{[k]} =  \frac{k}{2} \stackrel{ab}{[k]}, \quad
\tilde{S}^{ab} \stackrel{ab}{(k)}  = \frac{k}{2} \stackrel{ab}{(k)},  \quad 
\tilde{S}^{ab} \stackrel{ab}{[k]}  =   - \frac{k}{2} \stackrel{ab}{[k]}\;,\nonumber\\
\label{dhnsnmb:eigensabev}
\end{eqnarray}
with the properties that $\gamma^a$ transform   
$\stackrel{ab}{(k)}$ into  $\stackrel{ab}{[-k]}$, while 
$\tilde{\gamma}^a$ transform  $\stackrel{ab}{(k)}$ 
into $\stackrel{ab}{[k]}$ 
\begin{eqnarray}
\gamma^a \stackrel{ab}{(k)}= \eta^{aa}\stackrel{ab}{[-k]},\; 
\gamma^b \stackrel{ab}{(k)}= -ik \stackrel{ab}{[-k]}, \; 
\gamma^a \stackrel{ab}{[k]}= \stackrel{ab}{(-k)},\; 
\gamma^b \stackrel{ab}{[k]}= -ik \eta^{aa} \stackrel{ab}{(-k)},\;\;\;\nonumber\\
\label{dhnsnmb:graphgammaaction}
\end{eqnarray}
\begin{eqnarray}
\tilde{\gamma^a} \stackrel{ab}{(k)} = - i\eta^{aa}\stackrel{ab}{[k]},\;
\tilde{\gamma^b} \stackrel{ab}{(k)} =  - k \stackrel{ab}{[k]}, \;
\tilde{\gamma^a} \stackrel{ab}{[k]} =  \;\;i\stackrel{ab}{(k)},\; 
\tilde{\gamma^b} \stackrel{ab}{[k]} =  -k \eta^{aa} \stackrel{ab}{(k)}. \; \nonumber\\
\label{dhnsnmb:gammatilde}
\end{eqnarray}
the four spinor families, each with four vectors, which are 
 eigen vectors of the chosen Cartan subalgebra with the eigen values $\frac{k}{2}$, 
are  presented with the following four times four products of projections 
$\stackrel{ab}{[k]}$ and nilpotents 
$\stackrel{ab}{(k)}$:

\begin{align}
\label{dhnweylrep}
\varphi^{1 I}_{1} &= \stackrel{56}{(+)} \stackrel{03}{(+i)} \stackrel{12}{(+)}\psi_0,
&\varphi^{1 II}_{1} &= \stackrel{56}{(+)} \stackrel{03}{[+i]} \stackrel{12}{[+]}\psi_0, \nonumber\\
\varphi^{1 I}_{2} &=\stackrel{56}{(+)}  \stackrel{03}{[-i]} \stackrel{12}{[-]}\psi_0,
&\varphi^{1 II}_{2} &= \stackrel{56}{(+)}  \stackrel{03}{(-i)} \stackrel{12}{(-)}\psi_0,\nonumber\\
\varphi^{2 I}_{1} &=\stackrel{56}{[-]}  \stackrel{03}{[-i]} \stackrel{12}{(+)}\psi_0,
&\varphi^{2 II}_{1} &= \stackrel{56}{[-]}  \stackrel{03}{(-i)} \stackrel{12}{[+]}\psi_0,\nonumber\\
\varphi^{2 I}_{2} &=\stackrel{56}{[-]} \stackrel{03}{(+i)} \stackrel{12}{[-]}\psi_0,
&\varphi^{2 II}_{2} &= \stackrel{56}{[-]} \stackrel{03}{[+i]} \stackrel{12}{(-)}\psi_0,\nonumber\\
&&
\nonumber\\
\varphi^{1 III}_{1} &= \stackrel{56}{[+]} \stackrel{03}{[+i]}\stackrel{12}{(+)}\psi_0,
&\varphi^{1 IV}_{1} &= \stackrel{56}{[+]} \stackrel{03}{(+i)} \stackrel{12}{[+]}\psi_0,\nonumber\\
\varphi^{1 III}_{2} &=\stackrel{56}{[+]}  \stackrel{03}{(-i)}\stackrel{12}{[-]}\psi_0,
&\varphi^{1 IV}_{2} &=\stackrel{56}{[+]}  \stackrel{03}{[-i]} \stackrel{12}{(-)}\psi_0,\nonumber\\
\varphi^{2 III}_{1} &=\stackrel{56}{(-)}  \stackrel{03}{(-i)}\stackrel{12}{(+)}\psi_0,
&\varphi^{2 IV}_{1} &=\stackrel{56}{(-)}  \stackrel{03}{[-i]} \stackrel{12}{[+]}\psi_0,\nonumber\\
\varphi^{2 III}_{2} &=\stackrel{56}{(-)} \stackrel{03}{[+i]}\stackrel{12}{[-]}\psi_0,
&\varphi^{2 IV}_{2} &=\stackrel{56}{(-)}  \stackrel{03}{(+i)} \stackrel{12}{(-)}\psi_0,
\end{align}
where  $\psi_0$ is a vacuum  for the spinor state.
One can find from each state of each family all the states of the same family by applying $S^{ab}$. 
Similarly one reaches from each family member
the same  member of all the other families, by the application of $\tilde{S}^{ab}$, or any 
superposition of such $\tilde{S}^{ab}$. 
If we write the operators of handedness in $d=(1+5)$ as $\Gamma^{(1+5)} = \gamma^0 \gamma^1 
\gamma^2 \gamma^3 \gamma^5 \gamma^6$ ($= 2^3 i S^{03} S^{12} S^{56}$), in $d=(1+3)$ 
as $\Gamma^{(1+3)}= -i\gamma^0\gamma^1\gamma^2\gamma^3 $ ($= 2^2 i S^{03} S^{12}$) 
and in the two dimensional space as $\Gamma^{(2)} = i\gamma^5 \gamma^6$ 
($= 2 S^{56}$), we find that all four states are left handed with respect to 
$\Gamma^{(1+5)}$, with the eigen value $-1$, the first two states are right handed and the second two 
 states are left handed with respect to 
$\Gamma^{(2)}$, with  the eigen values $1$ and $-1$, respectively, while the first two are 
left handed 
and the second two right handed with respect to $\Gamma^{(1+3)}$ with the eigen values $-1$ and $1$, 
respectively. 
Taking into account Eq.~(\ref{dhnweylrep}) we may write 
the most general wave function  
$\psi^{(6)}$ obeying Eq.~(\ref{dhnweylE}) in $d=(1+5)$ as
\begin{eqnarray}
\psi^{(6)} = \sum_{i=I,II,III,IV} \, \psi^{(6 i)}= 
\sum_{i=I,II,III,IV}\,( {\cal A}^{i} \,{\stackrel{56}{(+)}}\,\psi^{(4 i)}_{(+)} + 
{\cal B}^{i} \,{\stackrel{56}{[-]}}\, \psi^{(4 i)}_{(-)}), 
\label{dhnpsi6}
\end{eqnarray}
where ${\cal A}^{i}$ and ${\cal B}^{i}$ depend on $x^{\sigma}$, while $\psi^{(4 i)}_{(+)}$ 
and $\psi^{(4 i)}_{(-)}$  determine the spin 
and the coordinate dependent parts of the wave function $\psi^{(6)}$ in $d=(1+3)$ in accordance 
with the definition in Eq.(\ref{dhnweylrep}), for example, 
\begin{eqnarray}
\psi^{(4 I)}_{(+)} &=& \alpha^{I}_+ \; {\stackrel{03}{(+i)}}\, {\stackrel{12}{(+)}} + 
\beta^{I}_+ \; {\stackrel{03}{[-i]}}\, {\stackrel{12}{[-]}}, \nonumber\\ 
\psi^{(4 I)}_{(-)}&=& \alpha^{I}_- \; {\stackrel{03}{[-i]}}\, {\stackrel{12}{(+)}} + 
\beta^{I}_- \; {\stackrel{03}{(+i)}}\, {\stackrel{12}{[-]}}. 
\label{dhnpsi4}
\end{eqnarray}
Using $\psi^{(6)}$ in Eq.~(\ref{dhnweylE}) and separating dynamics in $(1+3)$ and on $S^2$, 
the following relations follow, from which we recognize the mass term $m^{I}$:  
$\frac{\alpha^{i}_{+}}{\alpha^{i}_-} (p^0-p^3) - \frac{\beta^{i}_+}{\alpha^{i}_-} (p^1-ip^2)= m^{i},$ 
$\frac{\beta^{i}_+}{\beta^{i}_-} (p^0+p^3) - \frac{\alpha^{i}_+}{\beta^{i}_-} (p^1+ip^2)= m^{i},$ 
$\frac{\alpha^{i}_-}{\alpha^{i}_+} (p^0+p^3) + \frac{\beta_-}{\alpha_+} (p^1-ip^2)= m^{i},$
$\frac{\beta^{i}_-}{\beta^{i}_+} (p^0-p^3) + \frac{\alpha^{i}_-}{\beta^{i}_+} (p^1-ip^2)= m^{i}.$ 
(One notices that for massless solutions  ($m^{i}=0$)  $\psi^{(4i)}_{(+)}$ 
and $\psi^{(4i)}_{(-)}$, for each $i=  I,II,III,IV,$ 
decouple.) 
Taking the above derivation into account Eq.~(\ref{dhnweylE}) transforms into
\begin{eqnarray}
\label{dhnweylEp}
f \, &&\{(p_{05} + i 2S^{56}\,p_{06}) + \frac{1}{2E}\, \{p_{5} + i 2S^{56}\,p_{6}, Ef\}_{-} \}\, \psi^{(6)}
+ \gamma^0 \gamma^5 \, m^{i} \, \psi^{(6)}=0,\nonumber\\
&&p_{0s}= f^{\sigma}{}_{s} (p_{\sigma} - \frac{1}{2}(S^{ab} \omega_{ab \sigma}
+ \tilde{S}^{ab} \tilde{\omega}_{ab \sigma})).
\end{eqnarray}
Having the rotational symmetry around the axis perpendicular to the plane of the fifth and the sixth 
dimension we require that $\psi^{(6)}$ is the eigen function of the total angular momentum
operator $M^{56}= x^5 p^6-x^6 p^5  + S^{56}= -i \frac{\partial}{\partial \phi} + S^{56}$
\begin{eqnarray}
M^{56}\psi^{(6)}= (n+\frac{1}{2})\,\psi^{(6)}.
\label{dhnmabx}
\end{eqnarray}
Accordingly we write
\begin{eqnarray}
\psi^{(6)}= {\cal N}\, \sum_{i=I,II,III,IV}\, ({\cal A}^{i}_{n}\, \stackrel{56}{(+)}\, \psi^{(4i)}_{(+)}  
+ {\cal B}^{i}_{n+1}\, e^{i \phi}\, \stackrel{56}{[-]}\, \psi^{(4 i)}_{(-)})\, e^{in \phi}.
\label{dhnmabpsi}
\end{eqnarray}

Let us assume first that only $\omega_{56 s}\ne 0$ and $\tilde{\omega_{56 s}}\ne 0$. That means
that  we take in Eq.~(\ref{dhnomegas}) only $F \ne 0 $ and $\tilde{F}_{56} \ne 0$, while we put $\tilde{F}^{\pm} = 0$. 

Then, for $x^{(5)}$ and $x^{(6)}$ from Eq.~(\ref{dhnf})  
and for the zweibein from 
Eqs.(\ref{dhnfzwei},\ref{dhnf}) and the spin connections from Eq.(\ref{dhnomegas}) one obtains 
the two solutions, 
namely $\psi^{6 I}$ and 
$\psi^{6 II}$ both massless and both of the same handedness, 
while $\psi^{6 III}$ and $\psi^{6 IV}$ are massive and have the 
same mass. One can read this from equations
%
\begin{eqnarray}
\label{dhnweylErho}
if \,&& \{ e^{i \phi 2S^{56}}\, [(\frac{\partial}{\partial \rho} + \frac{i\, 2 S^{56}}{\rho} \, 
(\frac{\partial}{\partial \phi}) ) -  \frac{1}{2 \,f} \, \frac{\partial f}{\partial \rho }\, 
(1- 2 F \, 2S^{56} - 2 \tilde{F}_{56}\, 2\tilde{S}^{56} )\,]
\, \}  \psi^{(6)}\nonumber\\
&&+ \gamma^0 \gamma^5 \, m \, \psi^{(6)}=0.
\end{eqnarray}

After taking into account that $S^{56} \stackrel{56}{(+)}= \frac{1}{2} \stackrel{56}{(+)}$, 
$S^{56} \stackrel{56}{[+]}= \frac{1}{2} \stackrel{56}{[+]}$, 
$S^{56} \stackrel{56}{[-]}= -\frac{1}{2} \stackrel{56}{[-]}$,
$S^{56} \stackrel{56}{(-)}= -\frac{1}{2} \stackrel{56}{[-]}$, while 
$\tilde{S}^{56} \stackrel{56}{(+)}= \frac{1}{2} \stackrel{56}{(+)}$, 
$\tilde{S}^{56} \stackrel{56}{[+]}= - \frac{1}{2} \stackrel{56}{[+]}$,
$\tilde{S}^{56} \stackrel{56}{[-]}= \frac{1}{2} \stackrel{56}{[-]}$ and 
$\tilde{S}^{56} \stackrel{56}{(-)}= -\frac{1}{2} \stackrel{56}{[-]}$,
we end up with the equations of motion
 for ${\cal A}^{i}_n$ and ${\cal B}^{i}_{n+1}$ as follows, taking into account that the equations 
 for $i=I$ are the same as for $i=II$ and for $i=III$ are the same as for $i=IV$
\begin{eqnarray}
&&-if \,\{ \,(\frac{\partial}{\partial \rho} + \frac{n+1}{\rho})  -   
  \frac{1}{2\, f} \, \frac{\partial f}{\partial \rho}\, (1+ 2F- 2\tilde{F}_{56})\}  {\cal B}^{I,II}_{n+1} + 
  m^{I,II}\; {\cal A}^{I,II}_n = 0,  
\nonumber\\
&&-if \,\{ \,(\frac{\partial}{\partial \rho} - \quad \frac{n}{\rho}) -   
  \frac{1}{2\, f} \, \frac{\partial f}{\partial \rho}\, (1- 2F - 2\tilde{F}_{56})\}  {\cal A}^{I,II}_{n} + 
  m^{I,II} \; {\cal B}^{I,II}_{n+1} = 0\, ,\nonumber\\
  \\
&&-if \,\{ \,(\frac{\partial}{\partial \rho} + \frac{n+1}{\rho})  -   
  \frac{1}{2\, f} \, \frac{\partial f}{\partial \rho}\, (1+ 2F+ 2\tilde{F}_{56})\}  {\cal B}^{III, IV}_{n+1} + 
  m^{III,IV} \; {\cal A}^{III, IV}_n = 0,  
\nonumber\\
&&-if \,\{ \,(\frac{\partial}{\partial \rho} - \quad \frac{n}{\rho}) -   
  \frac{1}{2\, f} \, \frac{\partial f}{\partial \rho}\, (1- 2F + 2\tilde{F}_{56})\}  {\cal A}^{III, IV}_{n} + 
  m^{III,IV} \; {\cal B}^{III,IV}_{n+1} = 0.\nonumber
\label{dhnequationm56gen1}
\end{eqnarray}
Let us treat first the massless case ($m^{i}=0$). Taking into account that $F\frac{f-1}{f \rho} = 
\frac{\partial}{\partial \rho} \ln f^{\frac{F}{2}}$ and that $E=f^{-2}$, 
%
%
we get correspondingly the solutions
\begin{eqnarray}
{\cal B}^{I,II}_n &=& {\cal B}^{I,II}_0 \,\;\; \rho^{-n} f^{F - \tilde{F}_{56}+1/2}, \nonumber\\
{\cal A}^{I,II}_n &=& {\cal A}^{I,II}_0 \,\;\; \rho^{n} f^{-F - \tilde{F}_{56}+1/2}, \nonumber\\ 
{\cal B}^{III,IV}_n &=& {\cal B}^{III,IV}_0 \, \rho^{-n} f^{ F + \tilde{F}_{56}+1/2}, \nonumber\\
{\cal A}^{III,IV}_n &=& {\cal A}^{III,IV}_0       \,\rho^{n} f^{-F + \tilde{F}_{56}+1/2}.
\label{dhnmasslesseqsol}
\end{eqnarray}
Requiring that only normalizable (square integrable) solutions are acceptable 
\begin{eqnarray}
2\pi \, \int^{\infty}_{0} \,E\, \rho d\rho {\cal A}^{i \star}_{n} {\cal A}^{i}_{n} && < \infty, \nonumber\\
2\pi \, \int^{\infty}_{0} \,E\, \rho d\rho {\cal B}^{i\star}_{n} {\cal B}^{i}_{n} && < \infty, 
\label{dhnmasslesseqsolf}
\end{eqnarray}
$i=I,II,III,IV$,  it follows 
\begin{eqnarray}
&&{\rm for}\; {\cal A}^{I,II}_{n}: -1 < n < 2(F+ \tilde{F}_{56}) , \nonumber\\
&&{\rm for}\; {\cal B}^{I,II}_{n}: 2(F-\tilde{F}_{56}) < n < 1, \nonumber\\
&&{\rm for}\; {\cal A}^{III,IV}_{n}: -1 < n < 2(F - \tilde{F}_{56}) , \nonumber\\
&&{\rm for}\; {\cal B}^{III,IV}_{n}: 2(F+\tilde{F}_{56}) < n < 1, 
\quad n \;\; {\rm is \;\; an \;\;integer}.
\label{dhnmasslesseqsolf1a}
\end{eqnarray}
One immediately sees that for $F=0=\tilde{F}_{56}$ there is no solution for the zweibein from Eq.~(\ref{dhnf}). 

Eq.~(\ref{dhnmasslesseqsolf1a}) tells us that the strengths $F, \tilde{F}_{56}$ of the spin connection fields 
$\omega_{56 \sigma}$ and $\tilde{\omega}_{56 \sigma}$ can make a choice between the 
massless solutions ${\cal A}^{I,II}_n$ and ${\cal B}^{I,II}_n, {\cal A}^{III,IV}_n, {\cal B}^{III,IV}_n$: 
For 
\begin{eqnarray}
0< 2(F+ \tilde{F}_{56}) \le 1,\quad \tilde{F}_{56} \le F
\label{dhnFformassless}
\end{eqnarray}
 the only massless solution are the two left handed spinors with respect to 
$(1+3)$
\begin{eqnarray}
\psi^{(6 \;I,II)m=0}_{\frac{1}{2}} ={\cal N}_0  \; f^{-F- \tilde{F}_{56}+1/2} 
\stackrel{56}{(+)}\psi^{(4 \; I,II)}_{(+)}.
\label{dhnMassless}
\end{eqnarray} 
The solutions~(Eq.\ref{dhnMassless}) are the eigen functions  of $M^{56}$ with the eigen value $1/2$. 
No right handed massless 
solution is allowed. 
For the  particular choice  $2(F + \tilde{F}_{56})=1$ the spin connection fields  
$-S^{56} \omega_{56\sigma} - \tilde{S}^{56} \tilde{\omega}_{56\sigma}$ 
compensates the term $\frac{1}{2Ef} \{p_{\sigma}, Ef \}_- $ and the  left handed spinor
with respect to $d=(1+3)$ becomes a constant with respect to $\rho $ and $\phi$.  

To make one of these two states massive, one must include also the term $\tilde{S}^{mn} \tilde{\omega}_{mns}$.

For $2(F + \tilde{F}_{56})=1$ it is easy to find also the massive solutions  at least for 
$m^{i}, i=I,II$ (see ref.~\cite{dhnhn}) 
of Eq.~(\ref{dhnequationm56gen1}).

If we allow also nonzero values of $\tilde{\omega}_{mn\sigma}$, that is nonzero values 
of $\tilde{F}^+ \ne 0$ and $\tilde{F}^- \ne 0$ for our particular choices of 
$\tilde{\omega}_{mn\sigma}$ (Eq.\ref{dhnomegas}) we might end with only one massless solution.
The system of equations to be solved is then
\begin{eqnarray}
-if \bigl\{ [(\frac{\partial}{\partial\rho} - \frac{n}{\rho}) 
           &-& \frac{1}{2f} \frac{\partial f}{\partial\rho} 
           (1-2F-2\tilde{F}_{56}-2\tilde{F}^{-})]  \mathcal{A}_n^I \nonumber\\
           &-& \frac{1}{2f} \frac{\partial f}{\partial\rho}
           (2\tilde{F}^{-}-2i\tilde{F}^{+}) \mathcal{A}_n^{II}
    \bigr\}  + m^{I} \mathcal{B}_{n+1}^I = 0,
\end{eqnarray}
\begin{eqnarray} 
-if \bigl\{ [(\frac{\partial}{\partial\rho} - \frac{n}{\rho}) 
           &-& \frac{1}{2f} \frac{\partial f}{\partial\rho} 
            (1-2F-2\tilde{F}_{56}+2\tilde{F}^{-})]  \mathcal{A}_n^{II}\nonumber\\
           &-& \frac{1}{2f} \frac{\partial f}{\partial\rho}
           (2\tilde{F}^{-}+2i\tilde{F}^{+}) \mathcal{A}_n^{I}
    \bigr\} + m^{II} \mathcal{B}_{n+1}^{II} = 0,
\end{eqnarray}
\begin{eqnarray}     
-if \bigl\{ [(\frac{\partial}{\partial\rho} + \frac{n+1}{\rho}) 
           &-& \frac{1}{2f} \frac{\partial f}{\partial\rho} 
                (1+2F-2\tilde{F}_{56}-2\tilde{F}^{-})]  \mathcal{B}_{n+1}^{I}\nonumber\\
           &-& \frac{1}{2f} \frac{\partial f}{\partial\rho} 
           (2\tilde{F}^{-}-2i\tilde{F}^{+}) \mathcal{B}_{n+1}^{II}
   \bigr\} + m^{I} \mathcal{A}_{n}^{I} = 0,
\end{eqnarray}
\begin{eqnarray} 
-if \bigl\{ [(\frac{\partial}{\partial\rho} + \frac{n+1}{\rho}) 
           &-& \frac{1}{2f} \frac{\partial f}{\partial\rho} 
              (1+2F-2\tilde{F}_{56}+2\tilde{F}^{-})]  \mathcal{B}_{n+1}^{II}\nonumber\\
           &-& \frac{1}{2f} \frac{\partial f}{\partial\rho} 
              (2\tilde{F}^{-}+2i\tilde{F}^{+}) \mathcal{B}_{n+1}^{I}
    \bigl\} + m^{II} \mathcal{A}_{n}^{II} = 0,
\end{eqnarray}
\begin{eqnarray} 
-if \bigl\{ [(\frac{\partial}{\partial\rho} - \frac{n}{\rho}) 
         &-& \frac{1}{2f} \frac{\partial f}{\partial\rho}
           (1-2F+2\tilde{F}_{56}-2\tilde{F}^{+})]  \mathcal{A}_n^{III}\nonumber\\
         &-& \frac{1}{2f} \frac{\partial f}{\partial\rho}
          (-2\tilde{F}^{+}+2i\tilde{F}^{-}) \mathcal{A}_n^{IV}
    \bigr\} + m^{III} \mathcal{B}_{n+1}^{III} = 0,
\end{eqnarray}
\begin{eqnarray} 
-if \bigl\{ [(\frac{\partial}{\partial\rho} - \frac{n}{\rho}) 
           &-& \frac{1}{2f} \frac{\partial f}{\partial\rho} 
           (1-2F+2\tilde{F}_{56}+2\tilde{F}^{+})]  \mathcal{A}_n^{IV}\nonumber\\
           &-& \frac{1}{2f} \frac{\partial f}{\partial\rho} 
             (-2\tilde{F}^{+}-2i\tilde{F}^{-}) \mathcal{A}_n^{III}
    \bigr\} + m^{IV} \mathcal{B}_{n+1}^{IV} = 0,
\end{eqnarray}
\begin{eqnarray} 
-if \bigl\{ [(\frac{\partial}{\partial\rho} + \frac{n+1}{\rho}) 
          &-& \frac{1}{2f} \frac{\partial f}{\partial\rho} 
               (1+2F+2\tilde{F}_{56}-2\tilde{F}^{+})]  \mathcal{B}_{n+1}^{III}\nonumber\\
           &-& \frac{1}{2f} \frac{\partial f}{\partial\rho} 
             (-2\tilde{F}^{+}+2i\tilde{F}^{-}) \mathcal{B}_{n+1}^{IV}
    \bigl\} + m^{III} \mathcal{A}_{n}^{III} = 0,
\end{eqnarray}
\begin{eqnarray} 
-if \bigl\{ [(\frac{\partial}{\partial\rho} + \frac{n+1}{\rho}) 
           &-& \frac{1}{2f} \frac{\partial f}{\partial\rho} 
                  (1+2F+2\tilde{F}_{56}+2\tilde{F}^{+})]  \mathcal{B}_{n+1}^{IV}\nonumber\\
           &-& \frac{1}{2f} \frac{\partial f}{\partial\rho} 
            (-2\tilde{F}^{+}-2i\tilde{F}^{-}) \mathcal{B}_{n+1}^{III}
    \bigr\} + m^{IV} \mathcal{A}_{n}^{IV} = 0. 
\end{eqnarray}

\section{Conclusions}
\label{dhnconclusion}

We show in this contribution that  one can  escape from the "no-go theorem" of Witten~\cite{dhnwitten}, 
that is one can guarantee the masslessness of spinors and their chiral coupling  to the 
Kaluza-Klein[like] gauge fields when breaking the symmetry from the $d$-dimensional one to 
$M^{(1+3)} \times M^{d-4}$ space, in cases which we call the "effective two dimensionality" 
even without boundaries, as we proposed in the references~\cite{dhnhnkk06} and when there are more then 
one family. 

We assume the Lagrange density with zweibein which curves the infinite disc on $S^2$ and the 
two kinds of spin connection fields, the gauge fields of $S^{ab} = \frac{i}{4}
\,(\gamma^a \gamma^b - \gamma^b \gamma^a)$ and the gauge fields of 
$\tilde{S}^{ab} = \frac{i}{4} \,(\tilde{\gamma}^a \tilde{\gamma}^b 
- \tilde{\gamma}^b \tilde{\gamma}^a)$ as proposed by the {\it spin-charge-family-theory}. 
For a particular choice of the coordinate dependence of 
both spin connection fields and for the intervals of the strengths of these fields we found 
massless spinor solutions of only one handedness and correspondingly mass protected. They 
also couple to the correspondent Kaluza-Klein gauge fields.  

Work is in progress.

\title{Are Superheavy Quark Clusters Viable Candidates for the Dark Matter?
\thanks{Delivered in two talks, by M. Rosina and by N.S. Manko\v{c} Bor\v{s}tnik at the 
Mini workshop Bled 2010, Dressing Hadrons}}
\author{N.S. Manko\v{c} Bor\v{s}tnik${}^{1}$ and M. Rosina${}^{1,2}$}
\institute{%
${}^{1}${Faculty of Mathematics and Physics, University of Ljubljana,
     Jadranska~19, P.O.~Box 2964, 1001 Ljubljana, Slovenia}\\
${}^{2}${J.~Stefan Institute,  1000 Ljubljana, Slovenia}}

\authorrunning{N.S. Manko\v{c} Bor\v{s}tnik and M. Rosina}
\titlerunning{Are Superheavy Quark Clusters Viable Candidates for the Dark Matter?}
\maketitle

\begin{abstract}
The ordinary matter, as we know it, is made mostly of the first family quarks and leptons, 
while the theory together with experiments has proven so far that there are (at least) 
three families. The explanation of the origin of families is one of the most promising 
ways to understand the assumptions of the {\it Standard Model}. 
The {\it Spin-Charge-Family }theory~\cite{mrsnmb1M,mrsnmb1AM} does propose the mechanism 
for the appearance of families which bellow the energy of unification scale of 
the three known charges form two decoupled groups of four families.
The lightest of the upper four families, is predicted~\cite{mrsnmb1BM} to have stable members and
to be the candidate to constitute the dark matter. 
The clustering of quarks from the fifth family into baryons in the evolution of the universe  is discussed.

In this contribution we study how much the electroweak interaction influences the properties of baryons of the fifth family. 
\end{abstract}

\section{Introduction}
\label{mrsnmb1introduction}
\label{contribution:mrsnmb}
The {\it Standard Model} has no explanation for either 
the existence of families and their properties or for the appearance of the scalar field, which 
in the {\it Standard Model} determines the properties of the electroweak gauge fields. A theory 
which would explain the origin of families and the mechanism causing the observed properties 
of the quarks, leptons and gauge fields is needed. The {\it Spin-Charge-Family }theory~\cite{mrsnmb1M,mrsnmb1AM} 
is very promising for this purpose.

The {\it Spin-Charge-Family} theory points out that  there are two kinds of the $\gamma^a$ 
operators, the Dirac ones and the ones observed by one of the authors (SNMB)~\cite{mrsnmb1M,mrsnmb1AM} and called $\tilde{\gamma}^a$,
and it proposes that both should appear in an acceptable theory (or it should be proven that one of these two 
kinds has no application at the observable energy regime).  Since the operators 
$\tilde{\gamma}^a$ and $\gamma^b$ anticommute, 
while the corresponding generators of rotations in $d$-dimensional space commute 
($[ S^{ab},\tilde{S}^{cd}]=0$), both kinds form  equivalent representations with respect to each other. 
If Dirac operators are used to describe  spin and  charges~\cite{mrsnmb1M,mrsnmb1AM}, then
the other kind must be used to describe families, which obviously form an equivalent representations 
with respect to spin and charges. 

The properties of the fifth family quarks and leptons, and corresponding baryons, have been evaluated in ref.~\cite{mrsnmb1BM}, concluding that the fifth family neutron is very probably the most stable nucleon. In this paper, the formation of neutrons and anti-neutrons from the fifth family quarks and anti-quarks in the cooling plasma has been followed in the expanding universe. Their behaviour in the colour phase transition up to the 
present dark matter, as well as the scattering of the fifth family neutrons among themselves and on the ordinary matter has been evaluated.

The purpose of this contribution is to show an example how one can use standard  hadronic 
calculations in order to examine possible higher families and candidates for dark matter. It is also 
a demonstration of how much  the properties of clusters depend on the masses of the objects forming the clusters. 

We shall use the promising unified {\it Spin-Charge-Family} theory~\cite{mrsnmb1M} which has been 
developed by one of the authors (SNMB) in the recent decade. The reader can find details about the 
theory in the references~\cite{mrsnmb1M}, while Sect.~\ref{mrsnmb1normatheory} is a short overview, needed 
for the purpose of this contribution. 

Let us remind the reader about possible prejudices one might have at the first moment against accepting the 
particles which interact with the colour interaction,  as candidates for dark matter. 
We  discuss  these prejudices in order to demonstrate that superheavy quark clusters  are  
legitimate candidates worth exploring, provided they are stable.
\begin{enumerate}
\item
{\em Superheavy quarks are too short-lived.} This is  true for the fourth family predicted by  
the {\it  Spin-Charge-Family} theory, or any other proposal if the mixing matrix elements 
to the lower mass families are not negligible. However, the {\it Spin-Charge-Family} theory~\cite{mrsnmb1M,mrsnmb1BBLM} predicts eight families, with the upper four families (almost) 
decoupled from the lower four families. This makes one of the quarks of the fifth family, 
actually one of possible baryonic clusters,  practically stable.
\item
{\em Either the charged baryon u$_5$u$_5$u$_5$ or the charged baryon d$_5$d$_5$d$_5$ would be the lightest, 
depending on whether u$_5$ or d$_5$ is lighter.} 
Charged clusters cannot, of course, constitute dark matter. Forming the atoms with the first family 
electrons they would have far too large scattering amplitude to be consistent with the  properties of  dark matter. 
However, if one takes into account also the electro-weak interaction between quarks, 
then the neutral baryon n$_5$ = u$_5$d$_5$d$_5$ can very probably be the lightest,  
provided the u-d mass difference is not too large. 
The ref.~\cite{mrsnmb1BM} estimates the allowed differences, here we present the ratio between the 
weak and electromagnetic contributions for different fifth family baryons in more detail  (Sect.~\ref{mrsnmb1calculations}).
\item
{\em Strongly interacting particles have far too large cross section to be ``dark''}. 
The scattering cross section of any neutral cluster due to any interaction depends 
strongly also on the mass of the constituents. The fifth family baryons, interacting with the 
fifth family "nuclear force", have very small cross section if the masses are large enough.
For $m_{5} = 100$ TeV, for example,  the size of the cluster is of the order 10$^{-4}$ fm 
or less and the geometrical cross section as small as 10$^{-10}$ fm$^2$. 
\item 
{\em Did the fifth family quarks and/or their clusters form and survive after the big bang 
and during galaxy formation?} We kindly invite the reader to learn about the history of 
the fifth family clusters in the expanding universe from the paper~\cite{mrsnmb1BM}. In a  
hot plasma, when the temperature $T$ is much higher than the mass of the fifth family 
members, $T>>m_5$, the fifth family members 
behave as massless and are created out of plasma and annihilate back in the thermodynamically equilibrium in the same way as  other fermions and fields, which are massless or have low enough masses. When due to the expansion 
of the universe the temperature lowers bellow the mass of the family members, $T<m_5 $, 
they can be annihilated while the creation starts to be  less and less probable. 
When the temperature  falls bellow the binding energy of the clusters of the fifth family 
quarks they start to form clusters. Once the cluster is formed, it starts to interact with a very small "fifth family nuclear force" 
and survives also the colour phase transition up to now. In \cite{mrsnmb1BM,mrsnmb1Bled} the scattering of the fifth family 
neutrons in the experimental equipment of DAMA~\cite{mrsnmb1DAMA} and CDMS~\cite{mrsnmb1CDMS} is evaluated and 
discussed.
\end{enumerate}

\section{The Spin-Charge-Family theory}
\label{mrsnmb1normatheory} 

In this section  a  short introduction 
to the {\it Spin-Charge-Family} theory~\cite{mrsnmb1M} is presented. Only the  essential things 
are reviewed hoping  to make the reader curious to start thinking  about the differences in the 
hadronic properties of the very heavy fifth family hadrons as compared to the lowest three families.

The {\it Spin-Charge-Family} theory proposes in $d=(1 + (d-1))$ dimensions a very simple starting action   
for spinors  which carry both  kinds of the spin generators ($\gamma^a$  and $\tilde{\gamma}^a$ 
operators) and  for the corresponding gauge fields. Multidimensional spinors unify the spin and electro-weak-colour charge degrees of freedom. A spinor couples in $d=1+13$ 
to vielbeins and (through two kinds of the spin generators) to spin connection fields.  
Appropriate breaking of the starting symmetry leads to the 
left-handed quarks and leptons in $d=(1+3)$ dimensions, which carry the weak charge while the 
right handed ones carry no weak charge. The  {\it Spin-Charge-Family} theory is offering the answers to  
the questions about the origin of families of quarks and leptons, about the explicit 
values of their masses and mixing matrices, predicting the fourth family to be possibly 
seen at the LHC or at somewhat higher energies~\cite{mrsnmb1BBLM},   
as well as about the masses of the scalar  and the weak gauge fields, about the dark 
matter candidates~\cite{mrsnmb1BM}, and about breaking the discrete symmetries.

The simple action in $d=(1+13)$-dimensional space of the {\it Spin-Charge-Family} theory~\cite{mrsnmb1M}
\begin{eqnarray}
S &=& \int \; d^dx \; E\;{\mathcal L}_{f} +  \int \; d^dx \; E\; {\mathcal L}_{g} 
\label{mrsnmb1action}
\end{eqnarray}
contains the Lagrange density for two kinds of gauge fields linear in the curvature 
\begin{eqnarray}
\label{mrsnmb1lgauge}
{\mathcal L}_{g} &= &E  \;(\alpha \, R + \tilde{\alpha}  \tilde{R}), \nonumber\\
R &=& f^{\alpha [a}f^{\beta b]} \;(\omega_{a b \alpha,\beta} - \omega_{c a \alpha}
\omega^{c}{}_{b \beta}),
\tilde{R} = f^{\alpha [a} f^{\beta b]} \;(\tilde{\omega}_{a b \alpha,\beta} - \tilde{\omega}_{c a \alpha}
\tilde{\omega}^{c}{}_{b \beta}),\nonumber\\
\end{eqnarray}
and for a spinor, which carries  in $d=(1+13)$ dimensions two  kinds of the spin represented by the two 
kinds of the Clifford algebra objects~\cite{mrsnmb1M} 
\begin{eqnarray}
\label{mrsnmb1clifford}
S^{ab}&=& \frac{i}{4} (\gamma^a \gamma^b - \gamma^b \gamma^a),\quad   
\tilde{S}^{ab} = \frac{i}{4} (\tilde{\gamma}^a \tilde{\gamma}^b - \tilde{\gamma}^b \tilde{\gamma}^a), \nonumber\\
\{\gamma^a, \gamma^b \}_{+}&=& 2 \eta^{ab}= \{\tilde{\gamma}^a, \tilde{\gamma}^b \}_{+}, \quad 
 \{\gamma^a, \tilde{\gamma}^b \}_{+}=0,\quad \{S^{ab},\tilde{S}^{cd}\}_{-}= 0.
\end{eqnarray}
%
The interaction is only between the vielbeins and the two kinds of spin connection fields 
\begin{eqnarray}
     {\mathcal L}_f &=& \frac{1}{2} (E\bar{\psi} \, \gamma^a p_{0a} \psi) + h.c. \nonumber\\
p_{0a }&=& f^{\alpha}{}_a p_{0\alpha}, \quad  p_{0\alpha} =  p_{\alpha}  - 
                     \frac{1}{2}  S^{ab}   \omega_{ab\alpha} - 
                    \frac{1}{2}   \tilde{S}^{ab}   \tilde{\omega}_{ab\alpha}.
\label{mrsnmb1lspinor}
\end{eqnarray}
This action offers a  real possibility to explain the assumptions of the {\it standard 
model}~\footnote{This is the only theory in the literature to our knowledge, 
which does not explain the appearance of families by just postulating their numbers in one or 
another way, through the choice of a group, for example, 
but by offering the mechanism for generating families.}.

The {\it Spin-Charge-Family }theory  predicts an even number of families, among which is the 
fourth family, which might be seen at 
the LHC~\cite{mrsnmb1M,mrsnmb1BBLM} or at somewhat higher energies and
the fifth family with neutrinos and baryons with masses of several
hundred TeV forming  dark matter~\cite{mrsnmb1BBLM}.  

The action in Eq.~(\ref{mrsnmb1action}) starts with the massless spinor which through two kinds of  spins interacts 
with the two kinds of the spin connection fields.
The Dirac kind of the Clifford 
algebra objects ($\gamma^a$) 
determines, when the group $SO(1,13)$ is analysed with respect to the {\it Standard Model} groups in 
$d=(1+3)$ dimensions, the spin and all charges, manifesting the left handed quarks and leptons  
carrying the weak charge and the right handed weak-neutral quarks and leptons.
Accordingly, the Lagrange density ${\mathcal L}_f$ manifests after the appropriate breaking of symmetries
all the properties of one family of fermions as assumed by the {\it Standard Model}, with 
the three kinds of  charges coupling  fermions to the corresponding three gauge fields
(first term  of Eq.(\ref{mrsnmb1lsmyukawa}).

The second kind ($\tilde{\gamma}^a$) of the 
Clifford algebra objects (defining the equivalent representations with respect to the Dirac one) 
determines families. Accordingly, the spinor Lagrange density, after the spontaneous breaking
of the starting symmetry ($SO(1,13)$ into $SO(1,7)\times U(1)\times SU(3)$ and further    
into $SO(1,3) \times SU(2) \times SU(2) \times U(1) \times SU(3)$) generates
the {\it Standard Model-like} Lagrange density for massless spinors of  (four $+$ four) families 
(defined by $2^{8/2-1}=8$ spinor states for each member of one family). After the first symmetry breaking the upper four families  decouple from the lower four families (in the Yukawa couplings). In the final symmetry breaking 
(leading to $SO(1,3)   \times U(1) \times SU(3)$) the upper four families obtain masses through the mass matrix (the second term  of Eq.(\ref{mrsnmb1lsmyukawa}). The third term $("{\rm the \;rest}")$ is unobservable at low energies 
\begin{eqnarray}
{\mathcal L}_f &=& \bar{\psi}\gamma^{m} (p_{m}- \sum_{A,i}\; g^{A}\tau^{Ai} A^{Ai}_{m}) \psi 
+  \sum_{s=7,8}\;  \bar{\psi} \gamma^{s} p_{0s} \; \psi   +  {\rm the \;rest}. 
\label{mrsnmb1lsmyukawa}
\end{eqnarray}
Here $\tau^{Ai}$ ($= \sum_{a,b} \;c^{Ai}{ }_{ab} \; S^{ab}$) determine the hypercharge ($A=1$), the weak ($A=2$) 
and the colour ($A=3$) charge: $\{\tau^{Ai}, \tau^{Bj}\}_- = i \delta^{AB} f^{Aijk} \tau^{Ak}$, $f^{1ijk}=0$, 
$f^{2ijk}=\varepsilon^{ijk}$, where $f^{3ijk}$ is the $SU(3)$ structure tensor. 

The evaluation of  masses and mixing matrices of the lower four families~\cite{mrsnmb1BBLM} suggests that 
the fifth family masses should be above a few TeV, while evaluations of the breaks of symmetries from the 
starting one (Eq.~\ref{mrsnmb1action}) suggests that these masses should be far bellow $10^{10}$ TeV. 

We have not yet  evaluated a possible fermion number non-conservation in the dynamical history 
of the universe either for the first (the lower four) or for the fifth (the upper four) families. 
However, the evaluation of the history of the fifth family baryons up to today's dark matter does not 
depend much on the matter anti-matter asymmetry, as long as the masses are higher than a few $10$ TeV. 
So our prediction that if DAMA~\cite{mrsnmb1DAMA} really measures the family neutrons, also other direct experiments 
like CDMS~\cite{mrsnmb1CDMS} should in a few years observe the dark matter clusters, does not depend on the 
baryon number non-conservation~\cite{mrsnmb1BM}. 

Following the history of the fifth family members in the expanding universe up to today and 
estimating also the scattering properties of this fifth family on the ordinary matter, the evaluated 
masses of the fifth family quarks,  under the assumption that the lowest mass fifth family baryon 
is the fifth family neutron, are in the interval 
\begin{eqnarray}
\label{mrsnmb1all}
200 \,{\rm TeV} < m_5 <  10^5 \,\mathrm {TeV}.
\end{eqnarray} 
The fifth family neutrino mass $m_{\nu 5}$ is estimated to be in the interval between
a few TeV and  a few hundred TeV.

\section{The superheavy neutron from the fifth family as a candidate for the dark matter}
\label{mrsnmb1calculations}
 
 We want to put limits on u-d quark mass differences so that the neutral baryon $n_5$ appears as the lightest.
 First we calculate the dominant properties of a three-quark cluster~\cite{mrsnmb1BM}, 
 its binding energy and size. For this purpose we assume equal superheavy masses and we 
 realize that in this regime the colour interaction is coulombic (one gluon exchange dominates at these energies ). 
 For three nonrelativistic particles with attractive coulombic interaction we 
 solve the Hamiltonian
 \begin{eqnarray}
 \label{mrsnmb1H}
 H = 3m + \sum_i \frac{\vec{p}_i^2}{2m} - \frac{(\sum_i\vec{p}_i)^2}{6m} 
- \sum_{i<j} \frac{2}{3}\frac{\alpha_{\mathrm{s}}}{r_{ij}}.
\end{eqnarray}

The potential energy of the solution can be conveniently parametrized as
\begin{eqnarray}
 \label{mrsnmb1V}
V_{\mathrm{s}}=-\frac{2}{3} \alpha_{\mathrm{s}} \epsilon, \quad
\epsilon=\langle\sum_{i<j} \frac{1}{r_{ij}}\rangle = 3\eta \alpha_{\mathrm{s}} m_5,
\end{eqnarray}
where $m_5$ is the average mass of quarks in the fifth family. 
The binding energy is then (according to the virial theorem)
\begin{eqnarray}
 \label{mrsnmb1E}
E= \frac{1}{2} \, V_{\mathrm{s}}= -E_{\mathrm{kin}} = - \alpha_{\mathrm{s}}^2 \eta m_5.
\end{eqnarray}
The parameter $\eta$ for a variational solution using Jacobi coordinates and exponential profiles
was calculated in \cite{mrsnmb1BM}: $\eta=0.66$.

The splitting of baryons in the fifth family is caused by the u-d mass difference as well as by 
the potential energy of the electro-weak interaction. In the studied energy range, 
the electro-weak interaction has a coulombic form, determined by the exchange of one photon or one 
massless weak boson, and can be treated as a perturbation. Even if we are far above the electroweak phase transition,
it is convenient to work in the basis using Weinberg mixing of $\gamma$ and Z since this basis is more familiar to 
low energy hadron physicists.

We split the electro-weak interaction in five contributions, electric, Z-exchan\-ge Fermi (=vector), Z-exchange Gamov-Teller (=axial), W-exchange Fermi (=vector), W-exchange Gamov-Teller (=axial)
\begin{eqnarray}
 \label{mrsnmb1M}
M=\sum_i m_i + E + \left(V_{\mathrm{EM}}+ V_{\mathrm{Z}}^{\mathrm{F}}+ V_{\mathrm{Z}}^{\mathrm{GT}}
+ V_{\mathrm{W}}^{\mathrm{F}}+ V_{\mathrm{W}}^{\mathrm{GT}}\right).
\end{eqnarray}
Separate terms are as follows 
\begin{eqnarray}
 \label{mrsnmb1Vterms}
 V_{\mathrm{EM}}&=& \langle\sum_{i<j}Q_iQ_j\rangle\>\alpha_{\mathrm{EM}}\epsilon,\nonumber\\
V_{\mathrm{Z}}^{\mathrm{F}}&=& \langle\sum_{i<j}(\frac{t^0_i}{2}-\sin^2\vartheta_{\mathrm{W}}Q_i)
(\frac{t^0_j}{2}-\sin^2\vartheta_{\mathrm{W}}Q_j)\rangle\> \alpha_{\mathrm{Z}}\epsilon,\;
V_{\mathrm{Z}}^{\mathrm{GT}}= \langle\sum_{i<j}\frac{t^0_i t^0_j}{4}\vec{\sigma}_i\vec{\sigma}_j)\rangle\>
\alpha_{\mathrm{Z}}\epsilon,\nonumber\\
V_{\mathrm{W}}^{\mathrm{F}}&=& \langle\sum_{i<j}\frac{t^-_it^+_j+t^+_it^-_j}{8}\rangle\>\alpha_{\mathrm{W}}\epsilon,\quad V_{\mathrm{W}}^{\mathrm{GT}}= \langle\sum_{i<j}\frac{t^-_it^+_j+t^+_it^-_j}{8}\vec{\sigma}_i\vec{\sigma}_j\rangle\>
\alpha_{\mathrm{W}}\epsilon.
\end{eqnarray}
Here $\vec{t}=\frac{1}{2}\vec{\tau}$ are isospin operators, $t^+=(t_x+t_y)$, and $\vec{\sigma}$ are 
Pauli spin matrices.
Separate terms are evaluated in Table~\ref{mrsnmb1EW}. Note that the vector contributions (also the electromagnetic) are 
the same for N and $\Delta$ baryons while the axial contributions differ dramatically. 
The lowest two lines give the sum of these contributions for the choice of the coupling 
constants given below. The unnecessary decimal places are there if you like to check the 
reproducibility of the results.

In the numerical example we choose the average quark mass $m_5$ = 100 TeV and the corresponding average momentum of each quark $p=\sqrt{2m_5\,E_{\mathrm{kin}}/3} = 5.1$~TeV (see below). At this momentum scale, we read the running coupling constants  from Particle Data Group diagram \cite{mrsnmb1PDG} as 
$\alpha_{\mathrm{s}}=\alpha_3=1/13$, 
$\alpha_{\mathrm{W}}=\alpha_2=1/32$ and $\alpha_1=1/56$. 
The latter gives $\sin^2\vartheta_{\mathrm{W}}= (1+ \frac{5}{3}\frac{\alpha_\mathrm{W}}{\alpha_1})^{-1}= 0.255\approx 1/4 ,\;               \alpha_{\mathrm{EM}}=\alpha_{\mathrm{W}} \sin^2\vartheta_{\mathrm{W}}=1/128 \;$ and
$\alpha_{\mathrm{Z}}=\alpha_{\mathrm{W}}/ \cos^2\vartheta_{\mathrm{W}} = 1/24$.

\begin{table}[hhh]
\centering
\vspace*{4mm}
\begin{tabular}{|l|l|l|l|l|} \hline 
&&&&\\
& uuu & uud & udd & ddd \\ \hline
&&&&\\
$V_{\mathrm{EM}}/\epsilon\alpha_{\mathrm{EM}}$                   & +4/3 & 0     & -1/3 & +1/3 \\ 
&&&&\\
$V_{\mathrm{Z}}^{\mathrm{F}}/\epsilon\alpha_{\mathrm{Z}}$  & +1/48 & -1/48 & 0 & +4/48 \\
&&&&\\
$V_{\mathrm{Z}}^{\mathrm{GT}}(\mathrm{N})/\epsilon\alpha_{\mathrm{Z}}$&&-15/48&-15/48&\\
&&&&\\
$V_{\mathrm{Z}}^{\mathrm{GT}}(\Delta)/\epsilon\alpha_{\mathrm{Z}}$  &-9/48     &+3/48&+3/48&-9/48\\
&&&&\\
$V_{\mathrm{W}}^{\mathrm{F}}/\epsilon\alpha_{\mathrm{W}}$                      &0&+1/4&+1/4&0\\
&&&&\\
$V_{\mathrm{W}}^{\mathrm{GT}}(\mathrm{N})/\epsilon\alpha_{\mathrm{W}}$&&-30/48&-30/48&  \\
&&&&\\
$V_{\mathrm{W}}^{\mathrm{GT}}(\Delta)/\epsilon\alpha_{\mathrm{W}}$         &0&-1/4&-1/4&0\\
\hline
&&&&\\
$V_{\mathrm{EW}}(\mathrm{N})/\epsilon$ && -0.0256 & -0.0273 &\\
&&&&\\
$V_{\mathrm{EW}}(\Delta)/\epsilon$ & +0.0035 & +0.0017 & -0.0000 & -0.0017 \\
\hline
\end{tabular}
\caption{Electro-weak contributions to superheavy baryon masses  \label{mrsnmb1EW} }   
\end{table}

In this example, the binding energy $E=-0.39$ TeV and the average reciprocal distance 
$\langle 1/r_{ij} \rangle = \epsilon/3 = \eta\alpha_{\mathrm{s}}m_5=5.1 \,\mathrm{TeV} 
= 2.6\cdot10^4 \mathrm{fm}^{-1}$.

Finally, we come to our goal to make limits on u-d mass difference such that the neutral baryon remains the lightest.
\begin{enumerate}
\item
$m_{u5}-m_{d5} < (0.0273-0.0017)\epsilon = 0.0256\,\epsilon $ prevents udd$\to$ddd.
\item
$m_{u5}-m_{d5} > (-0.0273+0.0256)\epsilon = -0.0017\,\epsilon $ prevents udd$\to$uud.
\end{enumerate}
For our value of $\epsilon=15.24$ TeV this reads
$$-0.026 \,\mathrm{TeV}\;< m_{u5}-m_{d5} <\; 0.39 \, \mathrm{TeV}.$$
This limits are narrow compared to the mass scale $m_5=100$ TeV, but they are not so narrow if the mass generating mechanism is of order of 100 GeV.

\section{Conclusion}

In this contribution we put light on the hadronic properties of the very heavy stable 
fifth family as predicted by the {\it Spin-Charge-Family} theory, proposed by one of the authors~\cite{mrsnmb1M}.
The evaluations presented in Sect.~\ref{mrsnmb1calculations} were already partially done in~\cite{mrsnmb1BM}.
However, we try to convince the hadron physicists that if  the {\it Spin-Charge-Family} theory is the right way
to explain the assumptions of the {\it Standard Model} 
then the hadron physicists will have a pleasant time to study properties of the clusters forming
dark matter  with their knowledge form the lower three families.

\title{Complex Action Functioning as Cutoff and De Broglie-Bohm Particle}
\author{H.B.~Nielsen${}^1$, N.S. Manko\v c Bor\v stnik${}^2$,
  K. Nagao${}^3$ and G. Moultaka${}^4$}
\institute{%
${}^1$The Niels Bohr Institute, Copenhagen, Denmark\\ hbech@nbi.dk\\%
${}^2$University of Ljubljana, jadranska 19, 1000 Ljubljana, Slovenia \\ norma.mankoc@fmf.uni-lj.si\\%
${}^3$The Niels Bohr Institute, Copenhagen, Denmark\\%
${}^4$University of Montpellier II, 34095 Montpellier, cedex 5, France\\
        gilbert.moultaka@lpta.univ-montp2.fr} 

\titlerunning{Complex Action Functioning as Cutoff and De Broglie-Bohm Particle}
\authorrunning{H.B.~Nielsen, N.S. Manko\v c Bor\v stnik, K. Nagao and G. Moultaka}
\maketitle

\begin{abstract}
The purpose of this discussion contribution is to
 suggest the possibility that the imaginary action model 
could function as a cut off in loop diagrams. We argue also 
that the complex action model of M. Ninomiya and H.B. Nielsen  
has the DeBroglie-Bohm-particle appearing by itself, which is in a way 
already present in the contribution to this conference~\cite{ghnkNNB}. 
 \end{abstract}

\section{Introduction}
In the contribution by M. Ninomiya and H.B. Nielsen to this workshop~\cite{ghnkNNB}
it were suggested that the model of Ninomiya and 
Nielsen~\cite{ghnkown} would lead to improvement in the sense of interpretation 
of quantum mechanics. In the present discussion contribution we shall
estimate how does this model with a complex action lead to that a 
very narrow range of paths come to dominate the Wentzel-Dirac-Feynman-path 
integral, since we use a path integral with 
integration over the whole phase space and not only over the configuration 
space alone as can also be chosen. This may at first looks as to be in contradiction 
with the Heisenberg uncertainty relation, but it should be stressed that the 
metaphysical way in which the  result of Ninomiya and Nielsen 
 about the dominance of some narrow 
classes or some discrete sets of classes of paths is not in contradiction 
with what one can achieve with wave functions or rather cannot achieve.
In fact, Heisenberg's result is the information {\em we can have 
about the quantum system and still be able to use it}, whereas the information
one can obtain out of the model with the extremely narrow region dominating 
the dynamics in the phase space according to 
the pathway dominance range is completely useless to work with. In fact 
this information is what one can claim one has about a particle in the time 
in between preparation and observation, say if one prepares its momentum and 
measures its position. Then one could  metaphysically claim that in the 
period between the preparation and the measurement one has {\em both} the 
prepared momentum and the observed position at the same time. This claim is however 
totally useless and could fundamentally not be tested by further experiments 
because it would, if such an experiment is performed, lead to a 
disturbance of conditions and thus would spoil the correctness of the claim.
In spite of being useless one could however still with good metaphysical 
right uphold that indeed in such a time interval a particle has both momentum 
and position at much more accurate values than allowed by the Heisenberg 
uncertainty principle. If one really takes seriously that in the 
complex action model everything is calculable just from the expression for 
the action, i.e. mainly the coupling constants and the form of the action,
then one  could in principle calculate (but it would be exceedingly hard and in 
practise totally impossible) this narrow range of dominating paths, meaning 
essentially an up to a very little uncertainty classical path. So if we could 
use such unrealistic but possible calculations we would indeed have the 
Heisenberg uncertainty violating prediction! In this sense we must say that 
in principle in our model there is no Heisenberg uncertainty principle at 
the metaphysical level. Such a metaphysical classical state of the system 
is extremely reminiscent of the Bohm-DeBroglie interpretation of quantum 
mechanics about which G. Moultaka has talked at this workshop~\cite{ghnkgilbert}.

Can one find, using the metaphysical way of treating our universe (or any system 
with extremely many degrees of freedom),  with the complex action assuming the phase space 
of coordinates and momenta, the way for cutting away most of the space in a consistent way?  
Does the narrow range of dominating paths (meaning almost a classical path) help to 
make the theories of the Kaluza-Klein-kind renormalizable or at least trustable?

One of the open problems of the Kaluza-Klein[like] theories is, namely, the renormalizability 
of these theories. Even if one studies properties of a system of fermions interacting 
with the gauge gravity fields far bellow the 
quantum gravity regime, yet  is the consistent treatment of the cutoff and correspondingly 
the renormalizability of the approach questionable. We suggest that the complex action 
as proposed by Ninomia and Nielsen~\cite{ghnkown}  play a role of a cut of in loops 
diagrams for the Kaluza-Klein[like] theories. The {\it spin-charges-family-theory}~\cite{ghnknorma}, 
proposed by N.S. Manko\v c Bor\v stnik and presented and discussed in this workshop as a promising 
theory for explaining open questions of the {\it standard model}, is besides proposing the mechanism 
for generating families (and consequently possibly explaining the appearance of the  masses 
and mixing matrices of fermions), unifying the spin and the charges  into only the spin. 
This {\it spin-charges-family-theory} is  namely sharing many a difficulty with the 
Kaluza-Klein[like] theories. One of these difficulties is also the cut off problem. 
We  propose in this contribution that the imaginary 
action might help to make a choice of a trustable cut off.  

What are conditions which the system must fulfil that the complex action model  start to be 
efficient or usable in the sense, that it helps to make a choice of a very narrow part 
of the phase space of momenta and coordinates at least metaphysically? And how could one use it 
when describing systems, like quarks, hadrons, nuclei, atoms, molecules, scattering of 
particles on slits, and so on? How such cases come along with both, complex action model and 
the Bohm-DeBroglie interpretation of quantum 
mechanics.

\section{A typical shape of the phase space distributions corresponding 
to the $|A(t)>$ and $<B(t)|$ states in the complex action model.}

In this section we argue using the Lyapunov-exponent or better 
the Lyapunov-matrix when discussing properties of the universe existing for a very long 
time  that the two states  $|A(t)>$ and $<B(t)|$,  defined in the contribution by H.B. Nielsen 
and M. Ninomiya~\cite{ghnkNNB} or 
in~\cite{ghnkown},  
that the first, $|A(t)>$, may be considered  as a sort of wave functions
describing the state of the universe which is favoured by having low action $S_I$ up to the time
$t$ from the beginning of  the time, and the second, $<B(t)|$, a sort of hidden variable wave function 
 expressing a similar favourite state with respect to  
$S_I$ coming from the time interval between $t$ and the end of the  time.
In fact we define these two wave functions from  the  complex action 
model as a fundamental formulation from the functional path integral 
\begin{equation}
\int exp(\frac{i}{\hbar}*S[path]) \,{\cal D} path, \label{ghnkoriginal}
\end{equation}              
by splitting it up into two factors 
\begin{eqnarray}
<q|A(t)> & = & \int_{with \; \; path(t)=q} exp(\frac{i}{\hbar} 
*S_{-\infty \to t}[path]) \,{\cal D}path, \nonumber \\
<B(t)|q> & = & \int_{with \; \; path(t)=q} exp(\frac{i}{\hbar} 
*S_{t \to \infty}[path]) \,
{\cal D}path.
\end{eqnarray}
Here 
\begin{eqnarray}
S_{-\infty \to t}[path] & = & \int_{-\infty}^{t} L(path(t') )\,dt, \nonumber \\
S_{t \to  \infty}[path] & = & \int_t^{\infty} L(path(t') ) \,dt 
\end{eqnarray}
and the subscript ``$with \;\; path(t) =q$''  means that we only include 
those paths which end at time $t$ with 
representing the configuration  point $q$ in the path way integration.
In the case of $<q|A(t)>$ we only use half paths from the 
beginning of time - symbolized by $-\infty$ to the finite time $t$, while 
in the definition of $<B(t)|q>$ we similarly only use half paths from $t$
to the end of time, symbolized by $\infty$. 
We say that we split up the original functional integral~(\ref{ghnkoriginal}),
because  we immediately see that 
\begin{equation}
<B(t)|A(t)> = \int exp(\frac{i}{\hbar}*S[path]) {\cal D} path,
\end{equation} 
where here the time integration region is from the beginning of 
time to the end of time, although we have delete the index telling this 
so that we have put indeed,
\begin{equation}
S[path] = S_{-\infty \, to \, \infty}[path]. 
\end{equation}

The idea  is to seek to estimate the shape of the distribution in phase 
space describing in the best way the wave packet corresponding to the states
$|A(t)>$ and $<B(t)|$. In a classical approximation one should get the 
state $|A(t)>$ by developing forward to time $t$ a state determined roughly 
in some time prior to $t$ by ``optimising'(minimizing) $S_I$. In thinking of 
such a development during long times we have to have in mind how does the 
development of a series of very close (infinitesimally close) classical 
starting states in phase space develop as time goes on, and this is given 
by a matrix which is a generalization of the Lyapunov exponent. In fact if 
one phase space point $P_2$ deviates from another  infinitesimally close one 
$P_1$ by an infinitesimal vector in phase space ${\vec l}$, this distance 
vector ${\vec l(t)}$ will develop with time exponentially in the sense 
that
\begin{equation}
\vec l(t) = exp({\bf \lambda} * t )\,\vec l(0)
\end{equation}        
where ${\bf \lambda}$ is a matrix with the order being equal to the dimension
of the phase space. If the vector $\vec l(t_{start})$ at the starting time 
$t_{start}$ has components along the subspace of positive eigenvalues for the 
matrix ${\bf \lambda}$, the components in this space will grow up very 
drastically during sufficiently long time, while on the other hand the 
components in the subspace of negative eigenvalues will grow smaller and 
smaller as time passes. If thus at some time the starting state was selected
by the $S_I$ to be in some not especially elongated region and essentially 
just one quantum state (we speculate that this is  the selection 
at some close to Big Bang time), then as time goes on this region will be 
more and more contracted in the  ${\bf \lambda}$ negative eigenvalue 
subspace directions, while it will be expanded in the positive eigenvalue 
directions. After a long time - i.e. when $t$ has become long after the 
era of the strongest influence of  $S_I$ - the region representing the 
most favourite state at time $t$, that is just $|A(t)>$,  becomes very contracted
in the directions corresponding to the negative eigenvalue subspace 
and very elongated in the directions corresponding to the positive eigenvalue subspace.
This means that approximately this region corresponding to the state $|A(t)>$
becomes a surface of dimension as the number of positive eigenvalues of 
${\bf \lambda}$ lying in the phase space, probably not a flat surface but a 
curved smooth one. Similarly - but now we  can say time reversed - we obtain 
that the phase space region corresponding to the state $<B(t)|$ will be a very 
extended surface while strongly   contracted in other directions. For $<B(t)|$ 
we must imagine that the dependence on  the minimization on $S_I$ on what goes on in the 
future - of the time $t$ - determines in some presumably far future which 
state would be most favourable 
and then we must imagine how to develop backwardly (backward to time $t$) this
most favoured state. The development under such 
a backward development is again exponential and given by a metric similar 
to the ${\bf \lambda}$ from before. Now however we develop a negative time
namely from the presumably far future time back to the time $t$. Again some 
eigenvalues shrink under this backward development while others expand 
drastically. We therefore again obtain that the region in the phase space roughly
describing $<B(t)|$ has the shape of a very extended surface. Both surfaces, the surface
for $|A(t)>$ and that for $<B(t)|$, have dimensions presumably about the 
half dimension of the phase space, since they had their dimension given by 
number of respectively negative and positive eigenvalues of matrices of order
of the dimension of phase space. In case they have really this half dimension 
of phase space both, their intersection will be point wise. That is to say 
they will intersect in a discrete set of points typically (if they intersect 
at all).

Such an intersection in one or a few points would mean that our whole model
predicts {\em essentially one or a few classical solutions with very little 
uncertainty to represent the dominant part of the functional integral}!
If we - metaphysically may be - take this dominant region - the overlap
of the $|A(t)>$ region  with the $<B(t)|$ one - in the space of paths
to represent the realized history of the universe, then we have reached 
a picture in which the universe runs through a development in which the 
conjugate variables (i.e. momentum and position) are much more accurately
determined than (formally) the Heisenberg uncertainty principle allows for.
That is to say: The metaphysical picture put forward in our model turns 
out to deliver a classical picture in the sense that there is approximately 
a totally classically development, so that our complex action model makes it
approximately as if there really were a true classical development as one 
would have imagined before quantum mechanics were invented. Well, we have the 
tiny deviation from this picture that there will typically not be only one
such classical development, but rather several although still a discrete set
of them.

\section{A proposal for cutting off by means of the complex action model}

It is the main purpose of the present discussion and contribution to point 
out that we have a hope that the complex action  of H.B. Nielsen and M. Ninomiya 
is offering a "physical" mechanism for a cutoff and correspondingly find 
the "philosophical" support for the higher dimensional Kaluza-Klein-like 
theories which are not renormalizable. One could namely claim: we know a 
mechanism that in principle will cut off the divergences and replace them 
by finite expressions depending on the support of $S_I$ effects of the 
complex action model.


Let us show  how does such a principal cut off mechanism appear in the complex 
action model! 

Let us therefore very shortly remind ourselves how can one get in the complex 
action  the "usual" quantum mechanics. 
The basic approximation  to reproduce the 
(usual) quantum mechanics in the complex action model is that we 
approximate the projection operator on the future-determined state 
$|B(t)>$, the hidden variable state we could call it, by a unit operator
\begin{equation}
|B(t)><B(t)| \approx N * I \label{ghnkBB1}
\end{equation}           
where $N$ is  an unimportant normalization factor and $I$ is the unit 
operator. The argumentation for statistically justifying this 
approximation to be used for making the Born-probability distribution 
so as to obtain the usual expectation value formula from the 
one suggested at first in the complex action model goes with an ergodicity-like 
approximation. The hidden variable state from the future $<B(t)|$ affected by 
$S_I$ is, as we mentioned in the previous section, essentially given by some 
favourable state in a presumably far future extrapolated backward in time 
through a large amount of time. Then if this time is long and the system, the 
universe, is roughly an ergodic system, 
we will argue that all states 
have almost the same practical chance for being the state $<B(t)|$. In this 
way we count that all states in some basis for the Hilbert space are equally 
likely to be in the state $<B(t)|$. 
\begin{eqnarray}
\label{ghnkcutoffO}
<O>_t & = & \frac{\int exp(\frac{i}{\hbar}*S[path])\,O(path(t))\,{\cal D}path }
                 {\int exp(\frac{i}{\hbar}*S[path])\,            {\cal D}path} \nonumber\\
& = & \frac{<B(t)|O|A(t)>}{<B(t)|A(t)>} \nonumber\\
& = & \frac{<A(t)|B(t)><B(t)|O|A(t)>}{<A(t)|B(t)> <B(t)|A(t)>} \nonumber \\
& \approx & \frac{<A(t)|N*1 O|A(t)>}{<A(t)|N*1|A(t)>} \nonumber\\
& = & \frac{<A(t)|O|A(t)>}{<A(t)|A(t)>}. 
\end{eqnarray}
Thus we have justified the approximation~(\ref{ghnkBB1}). 
We hope that since the two states (as a function of a phase space) of a system, one 
describing the developing of the system from the very beginning up to the time $t$ ($A(t)$) 
and the second describing the system from the very end backward up to time $t$ ($B(t)$), define 
as an overlap a  very tiny part of the phase space, the idea is that knowing this phase space, 
that is some almost classical solutions, would help us to make a choice of an appropriate cut off.



\cleardoublepage
\thispagestyle{empty}
\vspace*{5cm}
{\bfseries \Large Discussion Section II}\\[1cm]
\addcontentsline{toc}{chapter}{Discussion Section II}

As one of the Editors, I thank  R. Mirman for his contribution, apologizing him 
that we put his discussions in a special part of the Proceedings. Although I do not share 
his opinion that none of theories originating in higher than $d=(1+3)$-dimensional can have 
any support, or even, are in contradiction with the so far observed phenomena,  I found his 
discussions useful and even very much needed. It would be, namely,  nice if we shall in the near future 
open a special discussion section, in which we would discuss how important are new consistent 
theories with, as Ronald would say, a lot of fantasy, which might help to see the presently 
accepted theories (as quantum mechanics and quantum fields theory and both standard models are) 
which we test in the low energy 
regime, from different point of view, which might bring a lot of new light in understanding  the 
accepted formalism and correspondingly understand better our 
Universe. However, very critical discussions with  arguments against such theories are 
equally useful. In particular, since many research positions are occupied by 
those, working on such theories.  \textit{N.S. Manko\v c Bor\v stnik}

\newpage

\author{R. Mirman}
\title{Where does the Science Go?\thanks{sssbbg@gmail.com}}
\institute{%
14U\\
155 E 34 Street\\
New York, NY  10016
}

\titlerunning{Where does the Science Go?}
\authorrunning{R. Mirman}
\maketitle

\begin{abstract}

Absurdity strengthens belief in nonsense, fantasy, so misleading physicists, officials, appropriators --- 
wasting much time, talent (?) and money. 
This is illustrated for the glaring nonsense of string 
theory and the hugely expensive Higgs fantasy. 

\end{abstract}

\section{Discussion}

Our universe is populated by strange and frightening creatures: Hecatonchires, golems, strings with weird dimensions, little green men, Higgs bosons, and on and on. Shockingly people even believe in them. No matter how strange, implausible, mathematically impossible, there is still intensively held belief, even certanty, in the impossible, the phantasmagoric, the nonexistent. Often their absurdity makes them more attractive, more deeply believed.

Of course we physicists know better than to take such glaring nonsense seriously. And of course so many physicists do, wasting their careers and much public money because the attraction of fantasy is so strong.

Indeed not a few physicists, but thousands, spend their entire lives on fantasy and daydreams. Fantasy, wishful thinking, are so compelling that they become addictive; escape becomes impossible. There are so many examples in current physics that we can only consider two.

The discussions here are not opinion that there can be reasonable disagreements with, but rigorous mathematical results, and properties of reality about which there can be no doubt.  

This is part of the discussion of the reasons for the laws of physics, and the grave misunderstandings of the laws and of physics, Many of these considerations, but far from all, are discussed in greater depth, often with proofs, elsewhere~(\cite{rm1anmb}; \cite{rm1anm2}; \cite{rm1anm3}; \cite{rm1anm4}; \cite{rm1aia}; \cite{rm1agf}; \cite{rm1aml}; \cite{rm1apt}; \cite{rm1aqm}; \cite{rm1acnfr}; \cite{rm1abna}; \cite{rm1abnb}; \cite{rm1abnc}; \cite{rm1aop}; \cite{rm1acg}; \cite{rm1aimp}; \cite{rm1amsi}; \cite{rm1aobn};
\cite{rm1arn}; \cite{rm1acu}). 

The laws have reasons that reason can know, but that unreason tries to hide. 

So on to fantasy. 

There is in "physics" a weird belief that fundamental objects are made of strings. There is not the slightest reason to believe this, it requires space to have some absurd dimension, violently disagreeing with reality, and since it has long been known that a universe is possible only with dimension 3+1 it is mathematically impossible\cite{rm1acu};\cite{rm1abnb};\cite{rm1aimp}. 

String theory has been shown. with mathematical rigor, to be impossible yet physicists, who of course know this, have managed to get billions of dollars of the txpayer's money, to study it and its extensions. What else can this be besides very serious fraud? 

That their theories disagree with Nature never bothers physicists. Since their theories are elected they must be correct, and if Nature disagrees, then clearly Nature must be wrong. A lot of physicists spend their careers fixing Nature so that it agrees with their theories.   

Where did this strange belief in strings come from?

Physicists are very upset about the infinities caused by point particles. Where in the formalism is there even the slightest hint of particles, let alone point particles? Physics is determined by its formalism --- equations --- not by vague pictures physicists pick up in kindergarten (although physicists prefer kindergarten. 

Point particles (even particles) do not exist, there is nothing in the formalism that gives the slightest hint of them. They have the sane status as little green men. Can anyone show the slightest hint, beyond their fantasies, of point particles? What objects are is considered elsewhere\cite{rm1aimp}. 

And what of the infinities? In physicists' favorite approximation scheme (perturbation theory) there are integrals in intermediate terms with lower limits of 0. When the calculation is carried out the result is finite, completely reasonable, and in agreement with reality (perhaps that is why physicists dislike it so strongly) to a huge number of significant figures. Clearly the infinities in the intermediate steps have no meaning, being just a quirk of the way the mathematics is done. 

And if a different approximation scheme were used these questions would never have arisen. However as physicists realize they are so extremely important that their choice of approximation scheme determines the laws of Nature. They really believe as can be seen from the huge number of physicists who spend their entire careers trying to fix Nature so it agrees with their theories. 

There is no problem strings are supposed to solve. 

We then have thousands of physicists wasting their entire careers trying to solve the terrible problems caused by point particles yet not a single one realized there are none (neither particles nor problems). Does anyone really believe that? Or is the fraud deliberate? 

Thus string theory is a completely crackpot theory, with no rationale, violently disagreeing with reality and mathematically impossible.

That is why physicists are so enthusiastic about it, and so enthusiastic about wasting billions of dollars of the taxpayer's money because of it.    

As another example of physicists' belief that if Nature disagrees with their theories then Nature must be wrong --- so spendinng vast amounts of money and time to revise Nature to agree with their theories --- we consider the hugely expensive Higgs fantasy. 

There has been much effort, many wasted careers, and much money, searching for the Higgs boson (providing a rationale for multi-billion dollar accelerators). The range of mass values in which it lies gets smaller and smaller. When it reaches 0 physicists will be left wondering where it is (an important question since they are already arguing who should get the Nobel Prize for its nonexistent discovery). The trivial answer of course is that it exists only in fantasy (including about Nobel Prizes). Since there is a waste of so much taxpayer money, and so many careers that (perhaps) could be put to better use, we must consider why physicists believe so strongly in this mythical object. This tells much about modern ''physics'', and about the sociology of ''science''.

We start with the concept of gauge transformations.

Take an object (preferably not this computer) and drop it. The higher you hold it, the harder it hits (the more damage it does). Thus the potential energy (the energy due to position) becomes kinetic energy (the energy due to motion), which upon striking becomes ...). But the kinetic energy, thus the speed so the danage, is the same whether dropped on the 7th floor or the 290th floor (ignoring the very slight change of gravitational potential). It only depends on the distance between the initial and final points. The potential energy only depends on the change of position; potential energy is independent (here) of the 0 point. It has no meaning, only differences do. It is invariant to change of the 0 point. This transformation between 0 points, is called a gauge transformation, and the independence of the 0 point is gauge invariance.

Sounds dreary, doesn't it?

This considers motion along a line, but space has 3+1 dimensions (3 space and 1 time). Thus gauge transformations refer to the arbitrary addition of a (not completely arbitrary) function to the potential (the physical object, not the field). Electromagnetism (and likewise gravitation) are invariant under gauge transformations. This seems like a boring, highly technical (or  highly boring, technical) point, but it gets physicists very excited. They feel that gauge transformations are so wonderful God must really love them so made all interactions gauge invariant. There is a very minor problem. It isn't true. The interactions between protons and pions are not, nor are any others except for electromagnetism and gravitation (the two, and only two, massless objects).

That their theories disagree with Nature never bothers physicists. They just redesign Nature to fit their theories. That is the reason for the Higgs boson (fantasy). Thus Nature is redesigned to make all massive objects massless, so gauge invariant. And since they elected the fact that all interactions are gauge invariant, thus all objects massless, it must be true. There is not the slightest doubt of that; elected facts are always certainly true. And Nature must agree.

Where does gauge invariance come from, besides its election?

Of course physicists like to generalize from a single case, here masslessness. Electromagnetism and gravitation are massless objects. How can an object be massless? These have 0 rest mass, which means they can never be at rest. It is possible to have a proton, or pion, hanging around, never going anywhere. But not a photon (the ''particle'' of light, properly electromagnetism). That must always be moving. It has energy (thus mass) but not rest mass since it can never be at rest (and likewise gravitation), but certainly not pions or protons).
 
Now why are there gauge transformations? Electrons and photons are spinning like the earth or a top or like dizzy people whose theories may be the result of that. To represent the spin we draw an arrow perpendicular to the plane of the spin, for the earth from the S to the N poles. This gives the plane of the spin (try it for Uranus).

Consider an electron and a photon moving in the same direction (their momentum vectors are parallel) and with their spins along these vectors (thus also parallel). All 4 vectors, both momenta and both spins, are along the same line.

We now perform a rotation leaving the directions of propagation the same, but changing the spin direction of the electron. Unfortunately the spin direction of the photon does not change. The spin of the photon is required (by the mathematics, not God) to lie along its direction of motion. Thus there are transformations acting on the electron, changing its spin direction, that have (seemingly) no effect on the photon. What are these? Of course gauge transformations. They are effectively (necessary) left-over transformations acting on the photon, not able to change its spin direction, as they can for massive objects like electrons or protons (thus being gauge transformations). Thus it is these extra transformations, not God's love as physicists believe, that gives gauge transformations. This has long been known\cite{rm1aimp}, and their properties have long been worked out\cite{rm1aml} Why does this belief that God's love give gauge transformations lead to the mythical Higgs boson (which has been named after Peter Higgs who is by no means mythical)?

This election of all interactions being gauge invariant, thus all objects massless, including us (which seems a little unlikely), requires these objects to ''seem'' to have mass even though we have decided they do not. Thus a new field is introduced which slows particles down, like moving through water or molasses, thus seeming massive. It is much more fun revising Nature than revising our theories. This is the (clearly mythical) Higgs field, and the particle going with it, the Higgs boson.

And these must exist since physicists elected them (and spend huge amounts  of the taxpayers' money looking for them). Isn't spending hundreds of millions (perhaps billions), much more entertaining than thinking a little bit? And certainly physicists, being so extremely important, will not let Nature or mathematics disagree with them. 

All these absurdities are well known, yet physicists waste huge amounts of the taxpayers' money on known nonsense. What else is this besides serious fraud?

\section*{Acknowledgements}

These discussions could not have existed without Norma Manko\v c Bor\v stnik.

\cleardoublepage
\cleardoublepage
\thispagestyle{empty}
\vspace*{5cm}
\begin{center}{\bfseries \Large Virtual Institute of Astroparticle
    Physics\\ Presentation}
\end{center}
\addcontentsline{toc}{chapter}{Virtual Institute of Astroparticle Physics Presentation}
\newpage

\cleardoublepage
\title{VIA Presentation}
\author{M.Yu. Khlopov${}^{1,2,3}$}
\institute{%
${}^{1}$Moscow Engineering Physics Institute (National Nuclear Research University), 115409 Moscow, Russia \\
${}^{2}$Centre for Cosmoparticle Physics "Cosmion" 125047 Moscow, Russia \\
${}^{3}$APC laboratory 10, rue Alice Domon et L\'eonie Duquet \\75205
Paris Cedex 13, France}

\titlerunning{VIA Discussions at XIII Bled Workshop}
\authorrunning{M.Yu. Khlopov}
\maketitle

\begin{abstract}

Virtual Institute of Astroparticle Physics (VIA), integrated in the
structure of Laboratory of Astroparticle physics and Cosmology (APC)
is evolved in a unique multi-functional complex, combining various
forms of collaborative scientific work with programs of education at
distance. The activity of VIA takes place on its website and
includes regular videoconferences with systematic basic courses and
lectures on various issues of astroparticle physics, online
transmission of APC Colloquiums, participation at distance in
various scientific meetings and conferences, library of their
records and presentations, a multilingual forum. VIA virtual rooms
are open for meetings of scientific groups and for individual work
of supervisors with their students. The format of a VIA
videoconferences was effectively used in the program of XIII Bled
Workshop to provide a world-wide participation at distance in
discussion of the open questions of physics beyond the standard
model.

\end{abstract}

\section{Introduction}
Studies in astroparticle physics link astrophysics, cosmology,
particle and nuclear physics and involve hundreds of scientific
groups linked by regional networks (like ASPERA/ApPEC \cite{mkVIAaspera})
and national centers. The exciting progress in these studies will
have impact on the fundamental knowledge on the structure of
microworld and Universe and on the basic, still unknown, physical
laws of Nature (see e.g. \cite{mkVIAbook} for review).

In the proposal \cite{mkVIAKhlopov:2008vd} it was suggested to organize a
Virtual Institute of Astroparticle Physics (VIA), which can play the
role of an unifying and coordinating structure for astroparticle
physics. Starting from the January of 2008 the activity of the
Institute takes place on its website \cite{mkVIAVIA} in a form of regular
weekly videoconferences with VIA lectures, covering all the
theoretical and experimental activities in astroparticle physics and
related topics. The library of records of these lectures, talks and
their presentations is now accomplished by multi-lingual forum. In
2008 VIA complex was effectively used for the first time for
participation at distance in XI Bled Workshop \cite{mkVIAarchiVIA}. Since
then VIA videoconferences became a natural part of Bled Workshops'
programs, opening the room of discussions to the world-wide
audience. Its progress was presented in \cite{mkVIABledVIA}. Here the
current state-of-art of VIA complex, integrated since the end of
2009 in the structure of APC Laboratory, is presented in order to
clarify the way in which VIA discussion of open questions beyond the
standard model took place in the framework of XIII Bled Workshop.
\section{The current structure of VIA complex}
\subsection{The forms of VIA activity}
\begin{figure}
    \begin{center}
        \includegraphics[scale=0.6]{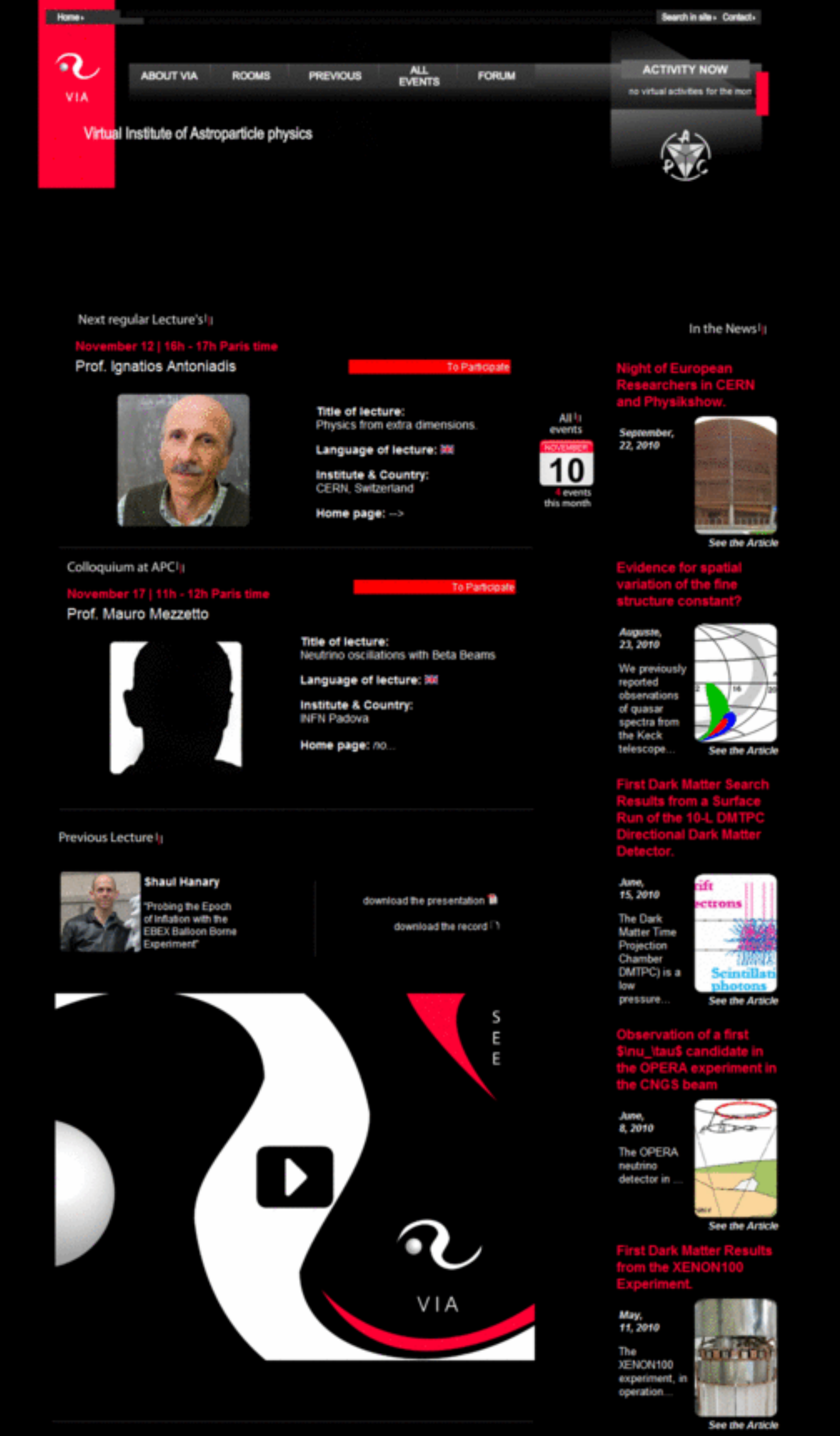}
        \caption{The home page of VIA site}
        \label{mkVIAhomevia}
    \end{center}
\end{figure}

The structure of
VIA complex is illustrated on Fig. \ref{mkVIAhomevia}.
The home page, presented on this figure, contains the information on
VIA activity and menu, linking to directories (along the upper line
from left to right): with general information on VIA (What is VIA),
entrance to VIA virtual lecture hall and meeting rooms (Rooms), the
library of records and presentations of VIA Lectures, records of
online transmissions of Conferences, APC Seminars and Colloquiums
and courses (Previous), Calender of the past and future VIA events
(All events) and Forum. In the end of this line "Activity Now"
provides the direct entrance in the virtual room with current
meeting. In the upper right angle there are links to Google search
engine (Search in site) and to contact information (Contacts). The
announcement of the next VIA lecture and VIA online transmission of
APC Colloquium occupy the main part of the homepage with the record
of the previous VIA lecture below. In the time of the announce event
(VIA lecture or transmitted APC Colloquium) it is sufficient to
click on "to participate" on the announcement and to Enter as Guest
in the corresponding Virtual room. The Calender links to the program
of future VIA lectures and events. The right column on the VIA
homepage lists the announcements of the hot news of Astroparticle
physics.

In 2010 special COSMOVIA tours were undertaken in Switzerland
(Geneve), Belgium (Brussels, Liege) and Italy (Turin, Pisa, Bari,
Lecce) in order to test stability of VIA online transmissions from
different parts of Europe. Positive results of these tests have
proved the stability of VIA system and stimulate this practice to be
continued.

It is assumed that the VIA forum can continue and extend the
discussion of questions that were put in the interactive VIA events.
The Forum is intended to cover the topics: beyond the standard
model, astroparticle physics, cosmology, gravitational wave
experiments, astrophysics, neutrinos. Presently activated in
English, French and Russian with trivial extension to other
languages, the Forum represents a first step on the way to
multi-lingual character of VIA complex and its activity.

One of the interesting forms of Forum activity is the educational
work. Having attended the VIA course of lectures in order to be
admitted to exam students should put on Forum a post with their
small thesis. Professors comments and proposed corrections are put
in a Post reply so that students should continuously present on
Forum improved versions of work until it is accepted as
satisfactory. Then they are admitted to pass their exam. The record
of videoconference with their oral exam is also put in the
corresponding directory of forum. Such procedure provides completely
transparent way of estimation of students' knowledge.

\subsection{VIA lectures and virtual meetings} 
\begin{figure}
    \begin{center}
        \includegraphics[scale=0.55]{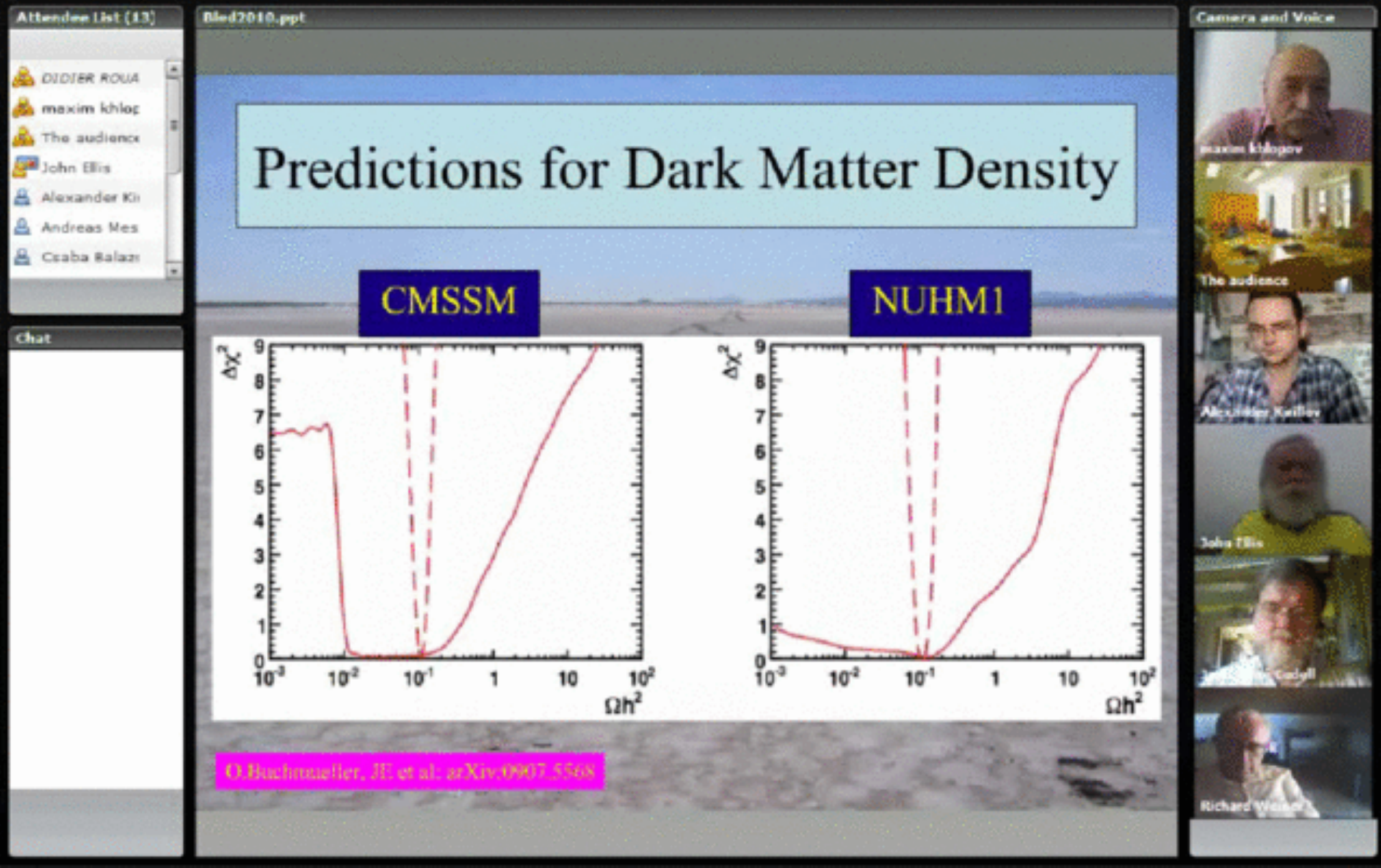}
        \caption{Videoconference Bled-Marburg-Liege-Geneve-Moscow-Australia with lecture by John Ellis,
        which he gave from his office in CERN,
        Switzerland, became a part of the program of
           XIII Bled Workshop.}
        \label{mkVIAellis}
    \end{center}
\end{figure}

First tests of VIA
system, described in \cite{mkVIAKhlopov:2008vd,mkVIAarchiVIA,mkVIABledVIA},
involved various systems of videoconferencing. They included skype,
VRVS, EVO, WEBEX, marratech and adobe Connect. In the result of
these tests the adobe Connect system was chosen and properly
acquired. Its advantages are: relatively easy use for participants,
a possibility to make presentation in a video contact between
presenter and audience, a possibility to make high quality records
and edit them, removing from records occasional and rather rare
disturbances of sound or connection, to use a whiteboard facility
for discussions, the option to open desktop and to work online with
texts in any format. The regular form of VIA meetings assumes that
their time and Virtual room are announced in advance. Since the
access to the Virtual room is strictly controlled by administration,
the invited participants should enter the Room as Guests, typing
their names, and their entrance and successive ability to use video
and audio system is authorized by the Host of the meeting. The
format of VIA lectures and discussions is shown on Fig. \ref{mkVIAellis},
illustrating the talk given by John Ellis from CERN in the framework
of XIII Workshop. The complete record of this talk and other VIA
discussions are available on VIA website \cite{mkVIAVIAbled}.

The ppt or pdf file of presentation is uploaded in the system in
advance and then demonstrated in the central window. Video images of
presenter and participants appear in the right window, while in the
lower left window the list of all the attendees is given. To protect
the quality of sound and record, the participants are required to
switch out their audio system during presentation and to use upper
left Chat window for immediate comments and urgent questions. The
Chat window can be also used by participants, having no microphone,
 for questions and comments during Discussion. In the end of presentation
 the central window can be used for a whiteboard utility
 as well as the whole structure of windows can be changed,
 e.g. by making full screen the window with the images of participants of discussion.

\section{\label{mkVIABled} VIA Sessions at Bled Workshop}
\subsection{The program of discussions}
In the course of Bled Workshop meeting the list of open questions
was stipulated, which was proposed for wide discussion with the use
of VIA facility (see \cite{mkVIAnorma}).

\begin{figure}
    \begin{center}
        \includegraphics[scale=0.55]{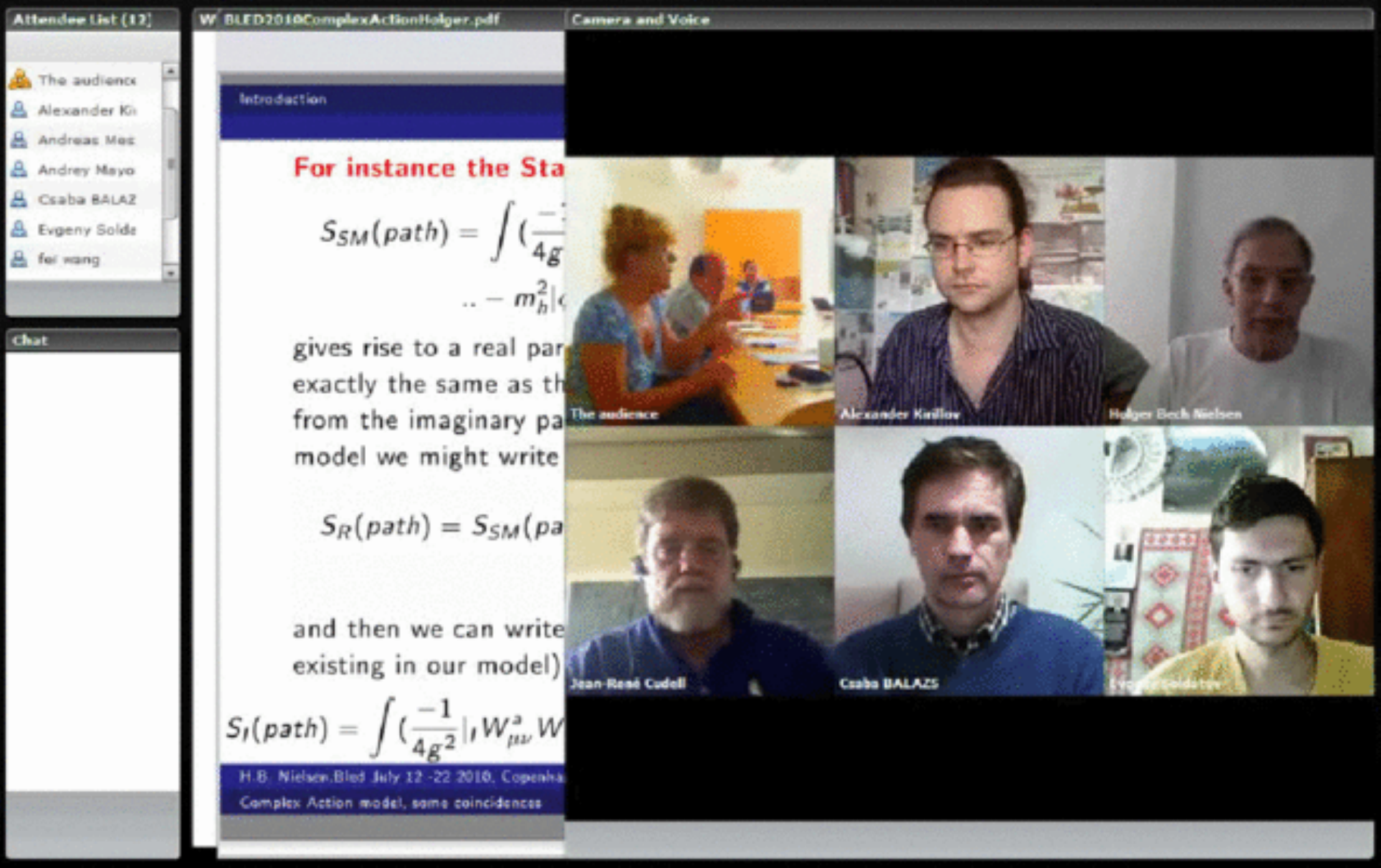}
        \caption{Videoconference with VIA talk by Holger Bech Nielsen}
        \label{mkVIAholger}
    \end{center}
\end{figure}

The list of these questions was put on the VIA site and all the
participants of VIA sessions were invited to address them during VIA
discussions. Some of them were covered in the VIA lectures (see
their records in \cite{mkVIAVIAbled}):
\begin{itemize}
\item[$\bullet$]  "Is the "Approach unifying spins and charges" into only two kinds of the
spin offering the right answers to the open questions of the
standard models?" by Norma Mankoc-Borstnik (in two parts )
\item[$\bullet$] "Arguments for there being Backward Causation, funny
coincidences." by Holger Bech Nielsen (in two parts)
\item[$\bullet$] "Atoms of dark matter from new stable charged particles"
by Maxim Khlopov
\item[$\bullet$] "New LHC Light on Dark Matter" by John Ellis
\end{itemize}
The use of the sensitive audio system KONFTEL 300W \cite{mkVIAkonftel}
supported VIA discussions inside the Bled conference room in the way
most natural for the non-formal atmosphere of Bled Workshops. The
advantage of the interactivity of VIA system has provided distant
participants to share this atmosphere and contribute the discussion.
The important refinement of adobe Connect system is that it is
possible now to use pdf files, what facilitated VIA presentations by
N.S. Manko\v c Bor\v stnik and Holger Bech Nielsen (see and image
form the latter videoconference on Fig. \ref{mkVIAholger}).

\subsection{VIA discussions}
VIA discussion sessions of XIII Bled Workshop have developed from
the first experience at XI Bled Workshop \cite{mkVIABregar:2008zz} and
their more regular practice at XII Bled Workshop \cite{mkVIABledVIA}.
They became a regular part of the Workshop's programme.

\begin{figure}
    \begin{center}
        \includegraphics[scale=0.55]{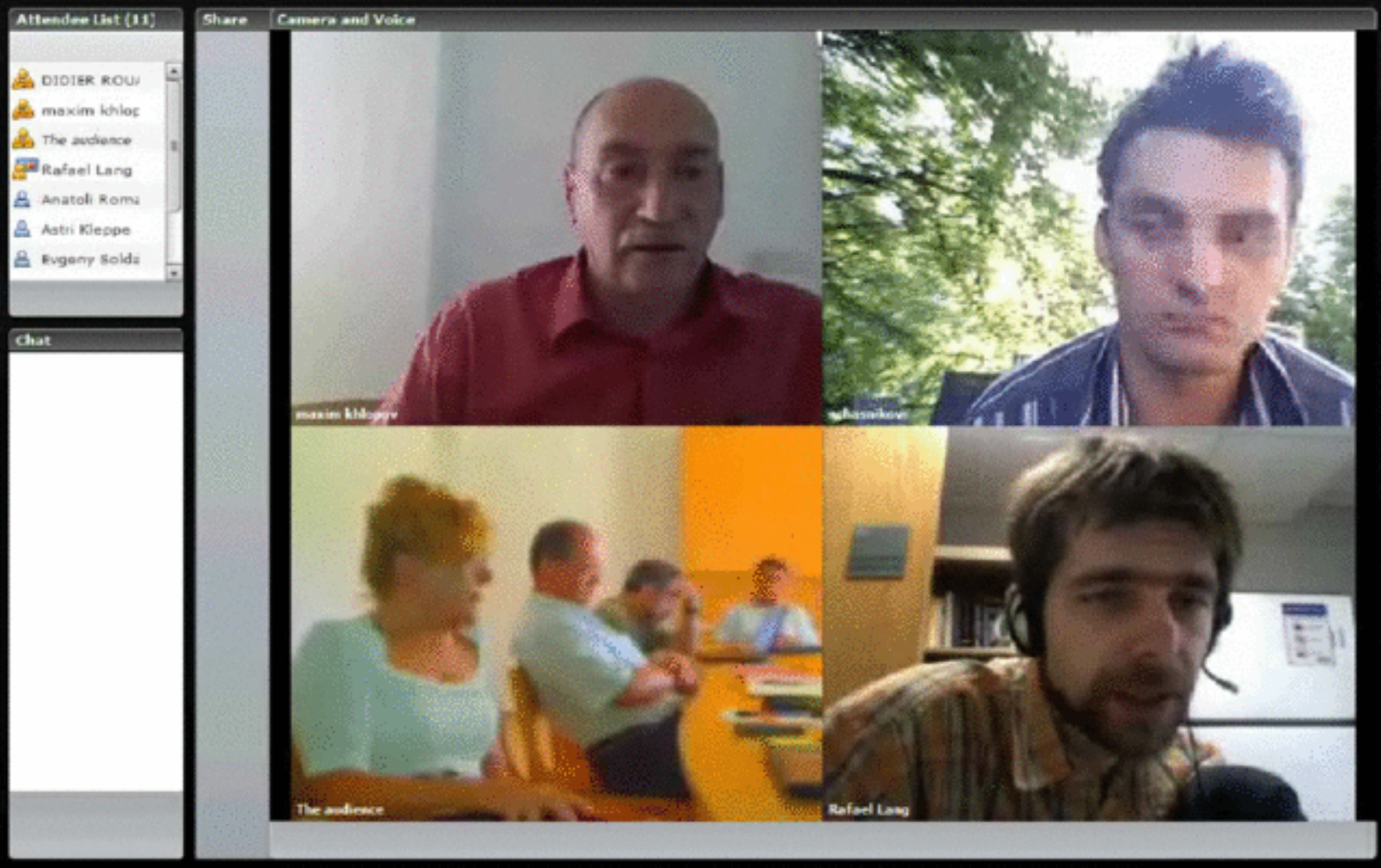}
        \caption{Bled Conference Discussion Bled-Moscow-CERN-Norge-Marburg-Liege-USA}
        \label{mkVIAdm}
    \end{center}
\end{figure}

 In the framework of the program of Bled Workshop John Ellis, staying in his office in CERN,
 gave his talk "New LHC Light on Dark Matter" and took part in the
 successive discussion. VIA sessions were finished by the discussion of puzzles
 of dark matter searches (see \cite{mkVIABled}). N.S. Manko\v c Bor\v stnik presented possible dark matter candidates
 that follow from the approach, unifying spins and charges, and Maxim Khlopov presented composite
 dark matter scenario, mentioning that it can offer the solution for the puzzles of direct
 dark matter searches as well as that it can find physical basis in the above approach.
 The comments by Rafael Lang from his office in USA were very important
 for clarifying the current status of  experimental constraints on the possible properties
 of dark matter candidates (Fig. \ref{mkVIAdm})

VIA sessions provided participation at distance in Bled discussions
for John Ellis (CERN, Switezerland), K.Belotsky, N.Chasnikov,
A.Mayorov and E. Soldatov (MEPhI, Moscow), J.-R. Cudell (Liege,
Belgium), R.Weiner (Marburg, Germany) C. Balasz (Australia) and many
others.

\section{Conclusions}

Current VIA activity is integrated in the structure of APC
laboratory and includes regular weekly videoconferences with VIA
lectures, online transmissions of APC Colloquiums and Seminars, a
solid library of their records and presentations, work of
multi-lingual VIA Internet forum.

The Scientific-Educational complex of Virtual Institute of
Astroparticle physics can provide regular communications between
different groups and scientists, working in different scientific
fields and parts of the world, get the first-hand information on the
newest scientific results, as well as to support various educational
programs on distance. This activity would easily allow finding
mutual interest and organizing task forces for different scientific
topics of astroparticle physics and related topics. It can help in
the elaboration of strategy of experimental particle, nuclear,
astrophysical and cosmological studies as well as in proper analysis
of experimental data. It can provide young talented people from all
over the world to get the highest level education, come in direct
interactive contact with the world known scientists and to find
their place in the fundamental research.

VIA sessions became a natural part of a program of Bled Workshops,
opening the room of discussions of physics beyond the Standard Model
for distant participants from all the world.
\section*{Acknowledgements}
 The initial step of creation of VIA was
 supported by ASPERA. I am grateful to P.Binetruy, J.Ellis and S.Katsanevas for
 permanent stimulating support, to J.C. Hamilton for support in VIA
 integration in the structure of APC laboaratory,
to K.Belotsky, A.Kirillov and K.Shibaev for assistance in
educational VIA program, to A.Mayorov and E.Soldatov for fruitful
collaboration, to M.Pohl, C. Kouvaris, J.-R.Cudell,
 C. Giunti, G. Cella, G. Fogli and F. DePaolis for cooperation in the tests of VIA online
 transmissions in Switzerland, Belgium and Italy and to D.Rouable for help in
 technical realization and support of VIA complex.
 I express my gratitude to N.S. Manko\v c Bor\v stnik, G.Bregar, D. Lukman and all
 Organizers of Bled Workshop for cooperation in the online
 transmission is of VIA Discussion Sessions at XIII Bled Workshop.



\backmatter

\thispagestyle{empty}
\parindent=0pt
\begin{flushleft}
\mbox{}
\vfill
\vrule height 1pt width \textwidth depth 0pt
{\parskip 6pt

{\sc Blejske Delavnice Iz Fizike, \ \ Letnik~11, \v{s}t. 2,} 
\ \ \ \ ISSN 1580-4992

{\sc Bled Workshops in Physics, \ \  Vol.~11, No.~2}

\bigskip

Zbornik 13. delavnice `What Comes Beyond the Standard Models', 
Bled, 12.~-- 22.~julij 2010

Proceedings to the 13th workshop 'What Comes Beyond the Standard Models', 
Bled, July 12.--22.,  2010

\bigskip

Uredili Norma Susana Manko\v c Bor\v stnik, Holger Bech Nielsen in Dragan Lukman 

Publikacijo sofinancira  Javna agencija za knjigo Republike Slovenije 

Brezpla\v cni izvod za udele\v zence 

Tehni\v{c}na urednica Tadeja \v{S}ekoranja

\bigskip

Zalo\v{z}ilo: DMFA -- zalo\v{z}ni\v{s}tvo, Jadranska 19,
1000 Ljubljana, Slovenija

Natisnila NTD v nakladi 150 izvodov

\bigskip

Publikacija DMFA \v{s}tevilka 1799

\vrule height 1pt width \textwidth depth 0pt}
\end{flushleft}


\end{document}